%% file: dissert.tex
\documentclass[twoside,openright,a4paper,11pt,makeidx]{amsbook}
\pdfoutput=1
\input{usepackage.tex}
\begin{document}
\title[Dissertation]{Automated Evaluation of One-Loop Six-Point
Processes\\
for the LHC}
%\author[T.~Reiter]{Thomas~Reiter\\[1cm]Supervisor: Thomas Binoth}
\author[T.~Reiter]{Thomas~Reiter}
\address{The University of Edinburgh\\School of Physics}
\email{Thomas.Reiter@ed.ac.uk}
\date{15\;September 2008}
\frontmatter
\maketitle

\cleardoublepage
\thispagestyle{plain}
\chapter*{Declaration}
\begin{center}
\begin{minipage}{10cm}
I declare that this thesis has been composed by myself
and that the work presented is my own.
This work has not been submitted for any other degree or professional
qualification.

\vspace{2.5cm}
\makeatletter
\hspace{1cm}Edinburgh, \@date
\makeatother
\end{minipage}
\end{center}

\cleardoublepage
% The next couple of lines mime a chapter head
% to give the word "Abstract" a consistent look
% but to avoid that it appears in the contents.
\phantomsection
\addcontentsline{toc}{chapter}{Abstract}
\thispagestyle{plain}
\topskip 7.5pc
\begin{center}
\bfseries\Large Abstract
\end{center}\par
\begin{center}
% Since the abstract is by definition very short
% I reduce the textwidth manually to make it more
% readable. That makes it about 50 characters per
% line. A newspaper column is about 40.
\begin{minipage}{0.7\textwidth}
\input abstract
\end{minipage}
\end{center}
\vfil\null

\cleardoublepage
\pagestyle{plain}
\pdfbookmark[0]{Contents}{the.TOC.label}
\tableofcontents
\pagestyle{fancy}
\mainmatter
\phantomsection
\include{preface}

\include{qcd}

\include{results}
\include{conclusion}
\backmatter
\appendix
\include{appendix-distributions}

\include{appendix-intdetails}

\include{appendix-integrals}

\include{implementation}

% already included in imp-sections.nw: \include{appendix-ffactory}

\let\chaptername\relax
\include{appendix-minor}

% Use the HEP bibliography style
%\bibliographystyle{hep}
\bibliographystyle{hepdoi}
\clearpage
%\addcontentsline{toc}{chapter}{Bibliography}
\phantomsection
\bibliography{dissert}        % bibliography
\clearpage
\include{very-end}
\end{document}

%% file: usepackage.tex
\usepackage[a4paper,pdftex]{geometry}
\usepackage{setspace}
\usepackage{amsmath}
\usepackage{amstext}
\usepackage{amsfonts}
\usepackage{amssymb}
\usepackage{xspace}
\usepackage{perm}
\usepackage{bibunits}

\usepackage[batch]{feynmp}            % feynMF package
\usepackage[pdftex,a4paper]{hyperref} % hyperreferences for PDF output
\usepackage{fancyhdr}                 % defining own headers
\usepackage{acronym}                  % acronyms
\usepackage[pdftex]{graphicx}
\usepackage[usenames]{color}
\usepackage{algorithmic}
\usepackage{algorithm}

\usepackage{fancyvrb}
\usepackage[thinspace,mediumqspace,squaren,textstyle]{SIunits}

\usepackage{subfig}                   % sub-figures
\usepackage[final]{remark}
\makeindex

\input{macros}                        % my own macros
\input{layout}

%% file: macros.tex
\DeclareMathOperator{\Kern}{Ker}
\DeclareMathOperator{\Trace}{tr}
\DeclareMathOperator{\rank}{rk}
\DeclareMathOperator{\diag}{diag}
\DeclareMathOperator{\sgn}{sgn}
\DeclareMathOperator{\shape}{sh}
\newcommand{\Lagrangeian}{{\person{Lagrang}ian\xspace}}
\newcommand{\eps}{\ensuremath{\varepsilon}}
\newcommand{\transposed}{\mathsf{T}\!}
\newcommand{\diff}[1][{}]{{\mathrm{d}}^{#1}\!}
\newcommand{\born}[1]{{#1}^\mathcal{B}}
\newcommand{\virt}[1]{{#1}^\mathcal{V}}
\newcommand{\real}[1]{{#1}^\mathcal{R}}
\newcommand{\phasespace}[1]{\diff{\Phi^{(#1)}}}
\newcommand{\measureF}[1]{F^{(#1)}_\mathrm{J}}
\newcommand{\ME}{\mathcal{M}}
\newcommand{\One}{\mathbb{I}}
\newcommand{\fmslash}[1]{\ensuremath{/\!\!\!{#1}}}
\newcommand{\pslash}[1][{}]{\fmslash{p}_{#1}}
\newcommand{\qslash}[1][{}]{\fmslash{q}_{#1}}
\newcommand{\kslash}[1][{}]{\fmslash{k}_{#1}}
\newcommand{\bra}[1]{\left\langle{#1}\right\vert}
\newcommand{\ket}[1]{\left\vert{#1}\right\rangle}
\newcommand{\braket}[1]{\left\langle{#1}\right\rangle}
\newcommand{\cbraket}[1]{\left[{#1}\right]}

\newcommand{\tr}[2][{}]{\Trace^{#1}\!\left\{{#2}\right\}}
\newcommand{\person}[1]{{\textsc{#1}}}
\newcommand{\Ld}{{\mathcal{L}}}
\newcommand{\Diff}{{\mathcal{D}}}
\newcommand{\Dslash}{\,\fmslash{\!\Diff}}
\newcommand{\Glue}{{\mathcal{A}}}
\newcommand{\Dipole}{{\mathcal{D}}}
\newcommand{\Rset}{{\mathbb{R}}}
\newcommand{\Nset}{{\mathbb{N}}}

\newcommand{\Cset}{{\mathbb{C}}}
\newcommand{\Zset}{{\mathbb{Z}}}

\newcommand{\comma}{\expandafter{,}}
\newcommand{\cd}{\mbox{\textperiodcentered}}
\newcommand{\bul}{\mbox{\bf\textperiodcentered}}
\newcommand{\figref}[1]{Figure~\ref{#1}}

\newcommand{\fortran}{\texttt{Fortran}\xspace}
\newcommand{\fortranXC}{\texttt{Fortran90}\xspace}
\newcommand{\maple}{\texttt{Maple}{\scriptsize\texttrademark}\xspace}
\newcommand{\form}{\texttt{FORM}\xspace}
\newcommand{\qgraf}{\texttt{QGraf}\xspace}

\newcommand{\LieGroup}[2][GL]{%
	\texorpdfstring{\ensuremath{\mathrm{#1}(#2)}}{#1(#2)}}
\newcommand{\gln}[1][N]{\LieGroup[Gl]{#1}}
\newcommand{\sln}[1][N]{\LieGroup[Sl]{#1}}
\newcommand{\son}[1][N]{\LieGroup[SO]{#1}}
\newcommand{\spn}[1][N]{\LieGroup[Sp]{#1}}
\newcommand{\sun}[1][N]{\LieGroup[SU]{#1}}
\newcommand{\Sym}[1][k]{\ensuremath{{\mathcal S}_{#1}}}
\newcommand{\nj}[1]{\ensuremath{{#1}\textendash{}j}}

\newcommand{\cmdvar}[1]{$\langle$\textit{#1}$\rangle$}
\newsavebox{\MYBOX}
\newenvironment{headquote}[1]{%
\begin{flushright}
\begin{minipage}[t]{0.5\textwidth}
\sbox{\MYBOX}{--- \sc #1}\sl}{%
\rm\hfill{\usebox\MYBOX}
\end{minipage}
\end{flushright}
}

\newtheorem{theorem}{Theorem}
\newtheorem{lemma}{Lemma}
\newtheorem{corollary}{Corollary}
\newtheorem{definition}{Definition}

\newcommand{\NWtarget}[2]{\hypertarget{#1}{#2}}
\newcommand{\NWlink}[2]{\hyperlink{#1}{#2}}

\newcommand{\NWtxtMacroRefIn}{Macro referenced in}

\newcommand{\NWtxtDefBy}{Defined by}
\newcommand{\NWtxtRefIn}{Referenced in}

\newcommand{\NWsep}{${\diamond}$}

\definecolor{gray}{rgb}{0.5,0.5,0.5}

%% file: layout.tex
\makeatletter
\def\cleardoublepage{\clearpage\if@twoside \ifodd\c@page\else%
    \hbox{}%
    \thispagestyle{empty}%              % Empty header styles
    \newpage%
    \if@twocolumn\hbox{}\newpage\fi\fi\fi}
\makeatother

\hypersetup{%
	backref=true,
	colorlinks=true,
	linkcolor=black,
	citecolor=black,
   filecolor=black,
	urlcolor=black
}

\makeatletter
\def\maketitle{
   \hypersetup{%
		pdftitle=\@title,%
		pdfauthor={\authors},%
		pdfsubject={PhD Thesis}
   }
   \thispagestyle{empty}
   \vspace{1in}
   \begin{tabular}{rl}
   \parbox[b]{0.8\textwidth}{
   \begin{flushright}
   \Huge\sffamily\slshape School of Physics and Astronomy\\
   \large\rmfamily\itshape{The University of Edinburgh}
   \end{flushright}
   }&\includegraphics[width=0.15\textwidth,bb=0 0 495 498]{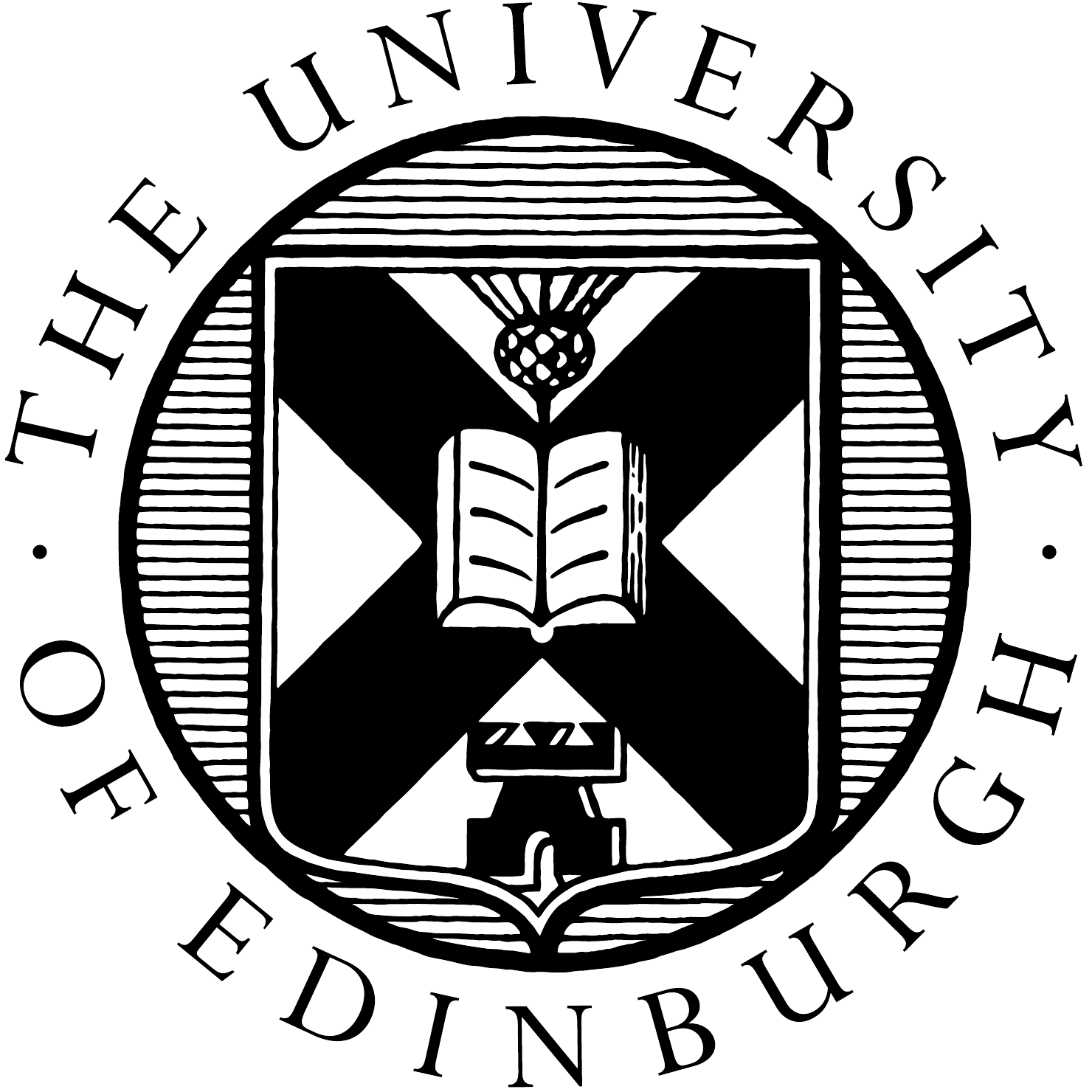}
   \end{tabular}
   \par
   \hrule
   \par
   \begin{center}
   \vspace*{1.5in}
   {\LARGE\@title}
   \par
   \vspace{0.5in}
   {\Large\authors}
   \par
   \vspace{0.5in}
   \large\@date
   \par
   \vfill
   \begin{parbox}{10cm}%
   \large Submitted in satisfaction of the requirements\\
    for the degree of Doctor of Philosophy\\
    in the University of Edinburgh\\
    2008.
   \end{parbox}
   \par
\end{center}
}
\makeatother

%% file: abstract.tex
%
% This is a LaTeX file
%
In the very near future the first data from LHC will be available.
The searches for the Higgs boson and for new physics will require
precise predictions both for the signal and the background processes.
Tree level calculations typically suffer from large renormalization
scale uncertainties.
I present an efficient implementation of an algorithm
for the automated, Feynman diagram based calculation
of one-loop corrections to processes with many external particles.
This algorithm has been successfully applied to compute
the virtual corrections of the process
$u\bar{u}\rightarrow b\bar{b}b\bar{b}$ in massless QCD
and can easily be adapted for other processes which are required
for the LHC.

%% file: preface.tex
\chapter*{Introduction}
Awaiting the first results from the \acf{lhc},
the current problems in particle physics,
through the eyes of a broader public, are very often reduced to a single
particle that is missing for the \acf{sm} to be consistent: the
\person{Higgs} boson. Although the physics programme of the \ac{lhc}
is much richer, the discovery of the \person{Higgs} boson --- or its
exclusion --- is one of the main physics motivations
having lead to the construction of the \ac{lhc}. In~1997, when the
predecessor experiment \acs{lep} was still running\footnote{%
In~1997 there were two particles missing in the \ac{sm}, the
\person{Higgs} boson and the $\nu_\tau$. The latter one was found
in 2000~\cite{Kodama:2000mp}.}, the main goals of the \ac{lhc} were
described as follows~\cite{Womersley:1997qp}:
\begin{quotation}\itshape
The fundamental goal is to uncover and explore the physics behind
electroweak symmetry breaking. This involves the following specific
challenges:
\begin{itemize}
\item Discover or exclude the Standard Modell Higgs and/or
the multiple Higgses of supersymmetry.
\item Discover or exclude supersymmetry over the entire
theoretically allowed mass range.
\item Discover or exclude new dynamics at the electroweak scale.
\end{itemize}
\end{quotation}
\index{standard model|(}
The \person{Higgs} boson gives mass to the fermions and is responsible
for electroweak symmetry breaking in the \acf{sm},
which to our current understanding describes
the physics of the smallest constituents of matter
in terms of a \person{Lorentz} invariant quantum field theory.
Interactions are mediated through gauge fields of the
group structure $\sun[3]_C\times\sun[2]_L\times\mathrm{U}(1)$,
which describe the strong, the weak and the electro-magnetic
force\footnote{Gravitation, the fourth interaction
is not included in the Standard Model. Compared to the coupling
strengths of the other three interactions gravitation is very
weak and therefore can be neglected for the concerns of
collider physics at an energy scale
of~$\mathcal{O}(\unit{1}{\tera\electronvolt})$}.

\index{Higgs mechanism@\person{Higgs} mechanism|(}
The interactions resulting from the 
$\sun[2]_L\times\mathrm{U}(1)$ gauge symmetry describe
the \emph{Electro-Weak Standard
Model}~\cite{Glashow:1961tr,Salam:1968rm,PhysRevLett.19.1264}.
Since left-handed and right-handed fermions couple differently
under the $\sun[2]_L$ interaction, a mass-term for the fermions
is forbidden. Furthermore, the observation
of massive gauge bosons in the \ac{sm}, i.e. the $W^\pm$ and $Z$
bosons\footnote{see e.g.~\cite{PDBook}},
requires that this symmetry group is broken. In~1964
\person{Higgs}~\cite{Higgs:1964ia,Higgs:1964pj,Higgs:1966ev}
and independently \person{Brout} and
\person{Englert}~\cite{Englert:1964et} introduced
a mechanism for mass generation through spontaneous symmetry breaking.
The model starts from a \Lagrangeian{} density without fermion
and gauge boson masses
and introduces a complex scalar $\sun[2]$ doublet~$\Phi$,
\begin{equation}
\Phi=\left(\begin{array}{c}\Phi^+\\\Phi^0\end{array}\right)\text{.}
\end{equation}
In addition to the \Lagrangeian{} density of the pure gauge theory 
one has interactions between the \person{Higgs} doublet and the
$\sun[2]$ gauge fields through a covariant derivative, \person{Yukawa}
type interactions between the fermions and the scalar doublet, and
the \person{Higgs} potential
\begin{equation}
\mathcal{V}(\Phi)=-\mu^2\vert\Phi\vert^2
+\lambda\left(\vert\Phi\vert^2\right)^2\text{.}
\end{equation}
The crucial point here is the negative quadratic term $-\mu^2\vert\Phi\vert^2$:
the potential develops a minimum away from $\vert\Phi\vert=0$ but for
a non-vanishing vacuum expectation value
\begin{equation}
\vert\Phi\vert=\frac{v}{\sqrt{2}}=\sqrt{\frac{\mu^2}{2\lambda}}\text{.}
\end{equation}
Rotational symmetry allows one to write the scalar doublet as
\begin{equation}
\Phi=\frac{1}{\sqrt{2}}%
e^{i\vec{T}\cdot\vec{\xi}/v}%
\left(\begin{matrix}0\\H+v\end{matrix}\right)\text{,}
\end{equation}
requiring a perturbative expansion around the true minimum for the
fields $H$ and $\vec{\xi}$ to obtain the physical degrees of freedom.
It turns out that the scalars
in $\vec{\xi}$ are absorbed by the longitudinal modes of the (former massless,
now massive) gauge bosons, and the \person{Higgs} boson acquires a mass
$M_H=\sqrt{2\lambda}v$. The fermion masses on the other hand are generated
through the \person{Yukawa} interactions and are of the form
$m_f=g_fv/\sqrt{2}$, where $g_f$ is the \person{Yukawa} coupling constant
of the considered fermion.\footnote{%
See for example~\cite{Ellis:QCD,Boehm:2001,Donoghue:1994,Bardin:1999}.}
This implies that within the \ac{sm} one expects the \person{Higgs} boson
to couple predominantly to heavy particles.
\index{Higgs mechanism@\person{Higgs} mechanism|)}

The $\sun[3]_C$ gauge group gives rise to the strong interaction
that binds the partons inside the nucleons; because the charge
of the strong force is called \emph{colour}~\cite{Fritzsch:1973pi}
the theory usually is referred to as \acf{qcd}.

\begin{figure}[hbtp]
\subfloat[]{
\includegraphics[width=0.4\textwidth]{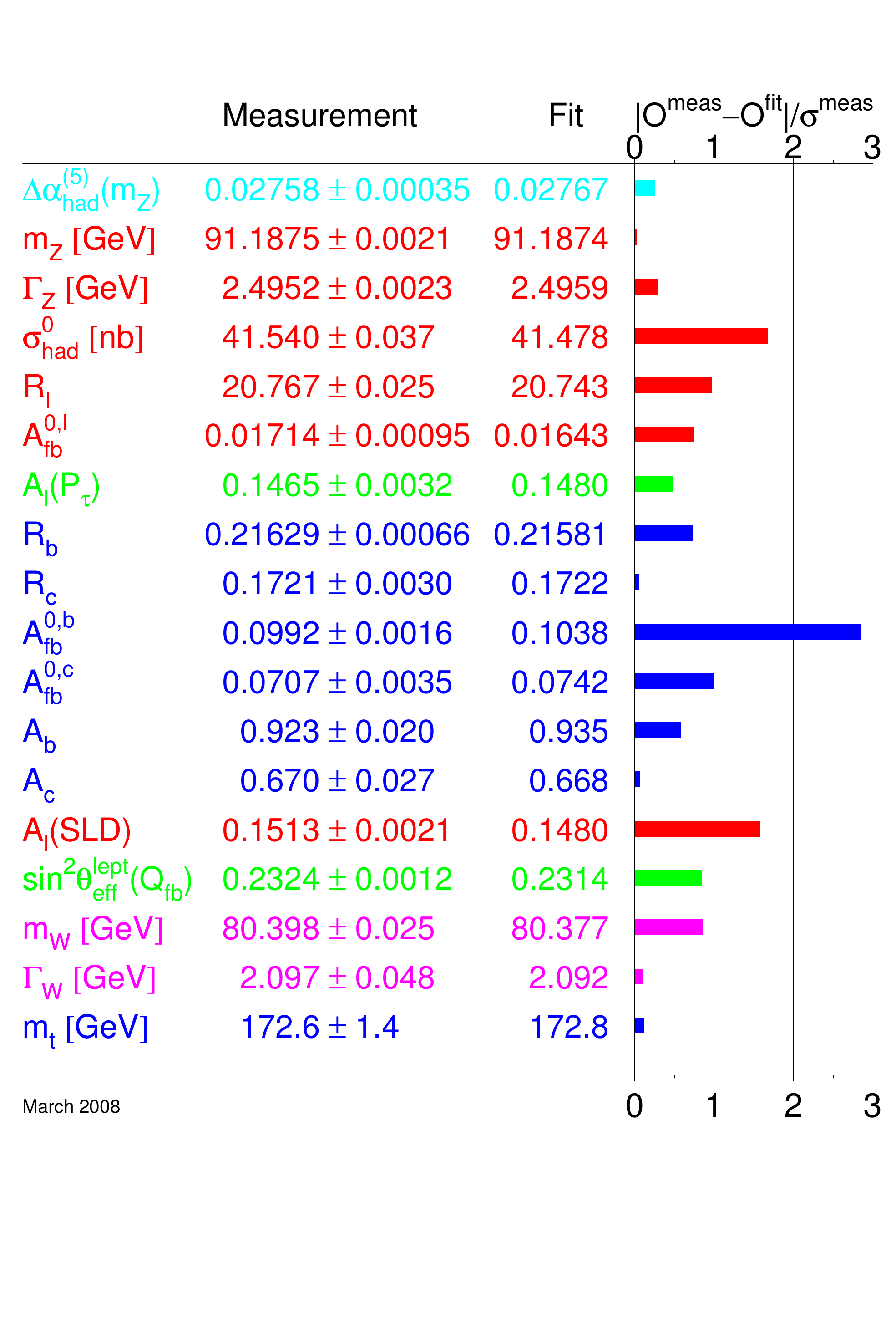}
\label{subfig:preface:pull}
}
\subfloat[]{
\includegraphics[width=0.5\textwidth]{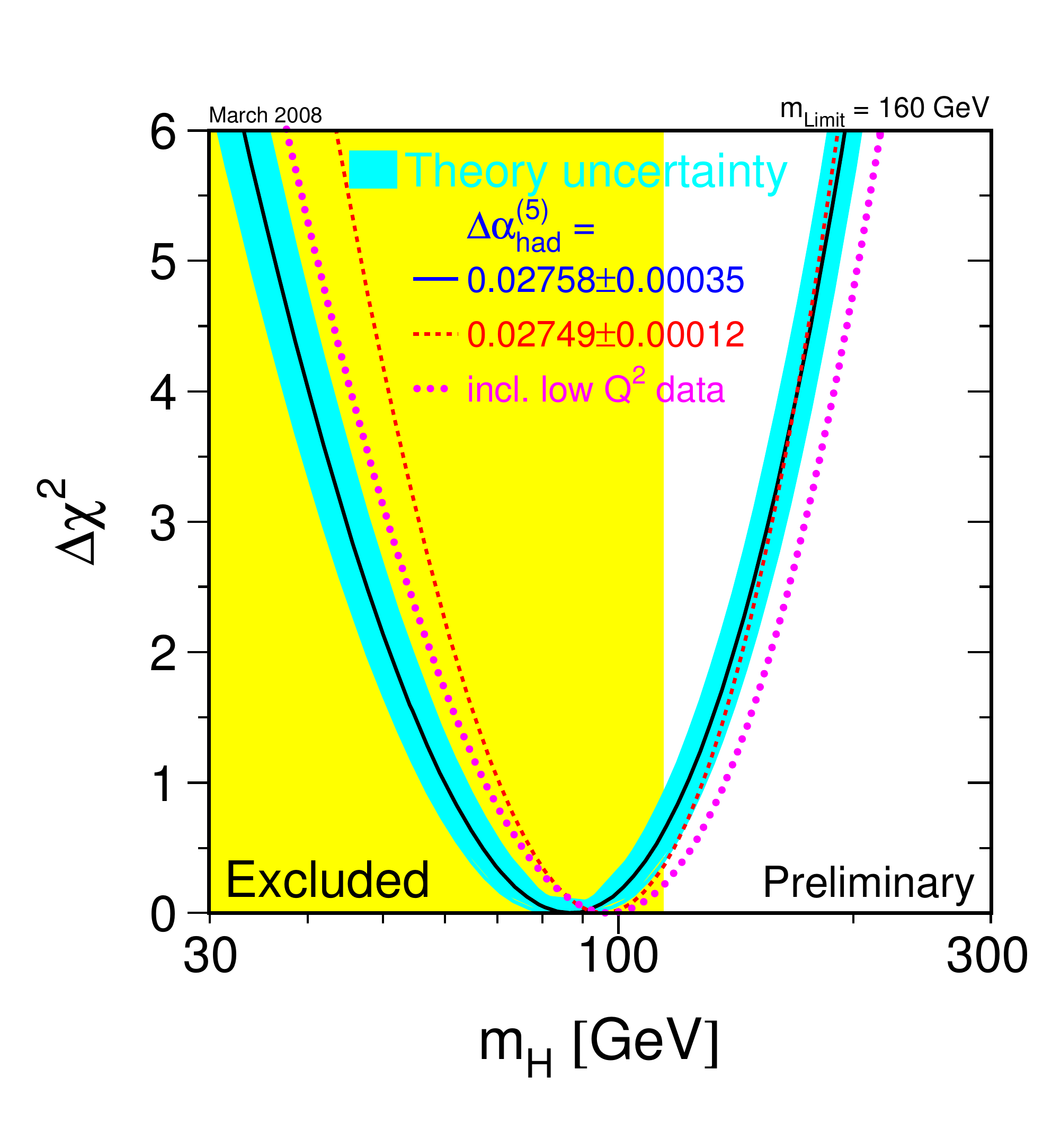}
\label{subfig:preface:blueband}
}
\caption{Results from the analysis of electro-weak precision data
as of March~2008~\cite{Barate:2003sz,:2005ema,showpull}.
Left: global fit of 18~\ac{sm} observables.
Right: the current best fit for the mass of a \ac{sm} \person{Higgs}
boson. The yellow area is experimentally excluded. Combined analysis
of precision measurements predict a light \ac{sm}\ \person{Higgs} boson.}
\label{fig:preface:ewwg}
\end{figure}
The \ac{sm} to date has endured all experimental tests
without showing significant deviations between the \ac{sm}
predictions and the experimental
data~\cite{Amsler:2008zz,Barate:2003sz,:2005ema}
as can be seen from Figure~\ref{fig:preface:ewwg}\,(a).
It shows the results of a simultaneous fit of 18
observables to the \ac{sm} predictions. The bar charts
indicate the deviation of the fitted value from the
measurement of the observable, weighted by the experimental
uncertainty~$\sigma^{\textrm{meas}}$; all values
are within a $3\sigma$ interval demonstrating the
consistency of the \ac{sm}.

Although direct evidence is still missing for a \ac{sm}
\person{Higgs} boson, the precision of the \acs{lep} experiments
allows not only to constrain the range for the \person{Higgs}
mass from below\footnote{through lack of observation} but also
from above, one reason being the influence of
the presence of \person{Higgs} particles on
the decay width of the $Z$ boson.
Figure~\ref{fig:preface:ewwg}\,(b) shows the so-called
\emph{blue-band plot}, a fit of the \person{Higgs} boson mass
obtained from 18 input parameters. The \acs{lep}\,II experiment
excluded the mass range of $M_H<\unit{114{.}4}{\giga\electronvolt}$
for a \ac{sm} \person{Higgs} boson at 
$95\%\,\mathrm{CL}$~\cite{Barate:2003sz}.
The combined fit in~\cite{:2005ema} yields an upper bound
on the \person{Higgs} boson mass of $M_H<\unit{285}{\giga\electronvolt}$
at $95\%\,\mathrm{CL}$ on $\log_{10}(M_H/\unit{1}{\giga\electronvolt})$.

\begin{figure}[hbtp]
\includegraphics{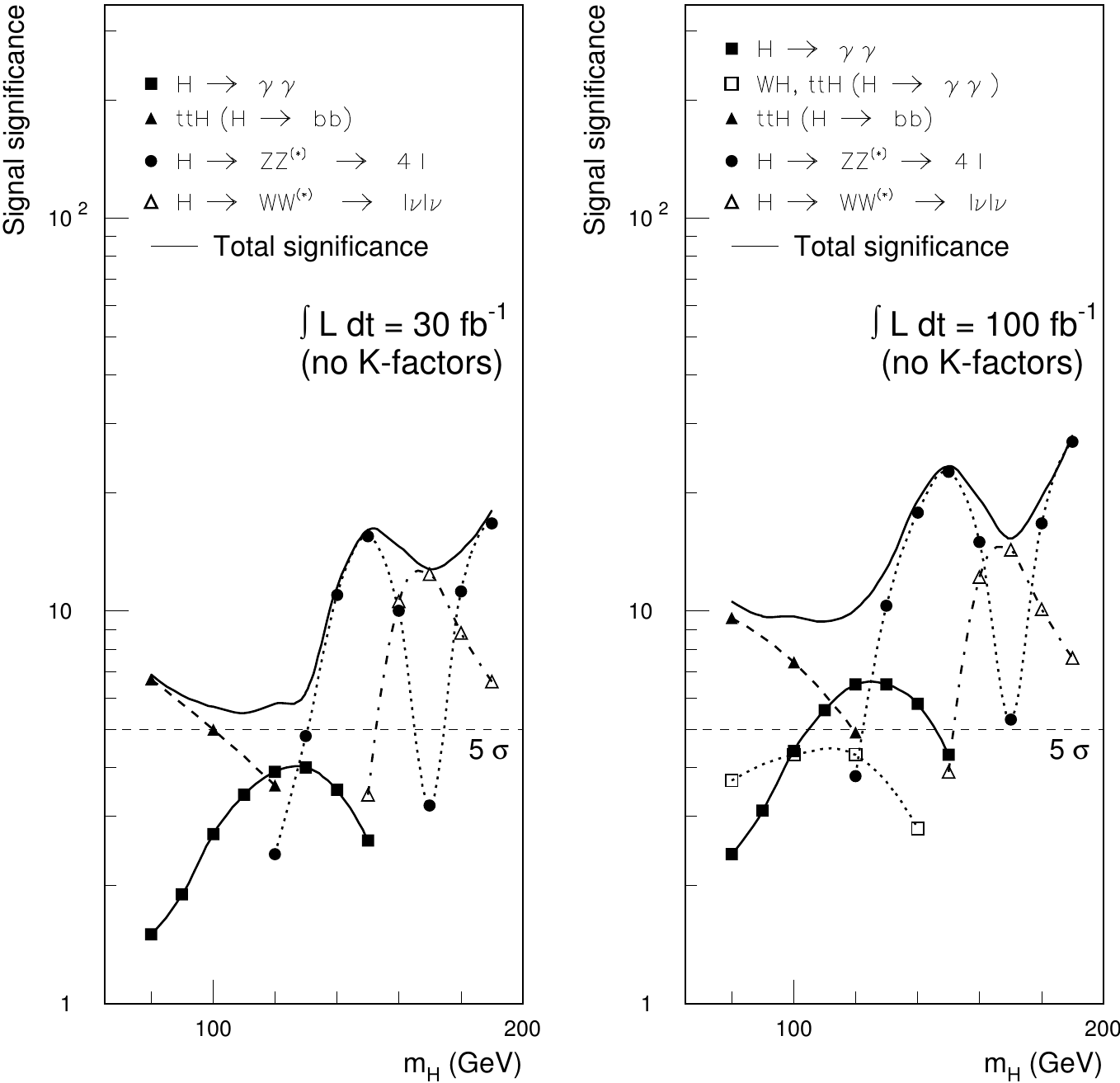}
\caption{Sensitivity of the ATLAS experiment to the discovery of
a \ac{sm} \person{Higgs} boson for an intermediate mass range
for integrated luminosities of \unit{30}{\femto\reciprocal\barn} (left)
and \unit{100}{\femto\reciprocal\barn}. The plots show $S/\sqrt{B}$ where
$S$ is the number of signal events and $B$ is the number of
background events.~\cite{ATLAS:1999fr}}
\label{fig:preface:higgs-significance}
\end{figure}
Figure~\ref{fig:preface:higgs-significance} demonstrates that
the at the \ac{lhc} one will be able to claim a \person{Higgs}
discovery over the entire mass range that is allowed in the
\ac{sm} or to rule out the \ac{sm} if no \person{Higgs} particle
is found. The \ac{lhc} therefore will probe if the \ac{sm}
describes elementary particle physics
at the energy scale of electro-weak symmetry breaking.
However, one of the design goals of the \ac{lhc} is also to
be sensitive to new physics --- \acf{bsm} physics --- if it
leaves signatures in the energy range below
$\mathrm{O}(\unit{10}{\tera\electronvolt})$.
\index{standard model|)}

Despite its big success the \ac{sm} can only be the low energy
effective theory of another, more fundamental theory.
The most obvious reason
why the \ac{sm} cannot be a fundamental theory is the fact
that it does not incorporate gravity. Currently no renormalisable
description of gravity as a gauge theory is known and the most promising
approaches are based on local supersymmetry as an extension of the
\person{Lorentz} symmetry. Another issue which is not addressed by
the \ac{sm} is dark matter: the amount of matter in the universe
predicted by cosmological observations cannot be explained by
the amount of baryonic matter, neither can any of the lighter \ac{sm} 
particles account for the matter content of the universe. Therefore
the existence of new particles beyond the \ac{sm} is well motivated
and indicated experimentally~\cite{Bertone:2004pz,Spergel:2003cb}.

Other indications for the incompleteness of the \ac{sm}
are concerned with a certain lack of explanation rather than direct
experimental motivation. The \ac{sm} does not explain the hierarchy
of masses and mixing angles, nor the presence of three generations,
nor the protection of the \person{Higgs} mass from large radiative
corrections, to name only a few.~\cite{Mohapatra}

The \ac{lhc} will be able to explore energies of
order~\unit{10}{\tera\electronvolt}, which is
the energy range new physics is expected to set in for the above reasons.
Therefore the investigation of different new physics models,
their experimental signatures and their \ac{sm} backgrounds is well motivated.
One of the \ac{bsm} candidates is a \emph{supersymmetric} extension
of the \ac{sm}.
\index{standard model!minimal supersymmetric|(}
The simplest of these models consistent with the current experimental
data is the \acf{mssm}\footnote{See for example~\cite{Mohapatra}}.
For each particle of the \ac{sm} it introduces
a corresponding super-partner, thus promoting each field to a super-field.
To maintain holomorphy of the super-potential and in order to guarantee
that the theory is free of anomalies one introduces a second
\person{Higgs} doublet, one coupling to the up-type quarks and the second
doublet coupling to the down-type quarks. Since pure supersymmetry predicts
equal masses for \ac{sm} particles and their supersymmetric partners
it cannot be an exact symmetry in nature but has to be broken by some
mechanism, introducing a mass hierarchy between the new particles.
In order to conform with the current
bounds on the proton life time, usually only $R$-parity conserving models
are considered, where $R$-parity of a particle
can be defined as $R=(-1)^{2j+L+3B}$,
$j$ being its spin, $L$ its lepton number and $B$ its baryon number.
The conservation of $R$-parity ensures that supersymmetric partners
of the \ac{sm} particles are always pair-produced and that the \ac{lsp}
is~stable.

Although in the \ac{mssm} the \person{Higgs} sector is richer than in the
\ac{sm} one usually obtains stronger bounds on the lightest \person{Higgs}
mass, which will be shown in this paragraph.
The two \person{Higgs} doublets, $\Phi_u$ and $\Phi_d$, manifest
themselves after symmetry breaking in five physical states: two
$\mathcal{CP}$-even, neutral scalar bosons $H^0$ and $h^0$ with
masses $M_{H^0}>M_{h^0}$, a $\mathcal{CP}$-odd scalar field $A^0$,
and two charged states $H^\pm$. The ratio of the vacuum expectation
values of the two doublets is called~$\tan\beta$,
\begin{equation}
\tan\beta=\frac{\vert\langle\Phi_d\rangle\vert}{\vert\langle\Phi_u\rangle\vert}
\text{.}
\end{equation}
At tree-level the mass spectrum of the \person{Higgs} sector
is described completely by the gauge boson masses $m_Z$ and $m_{W}$,
$\tan\beta$ and the mass of the $\mathcal{CP}$-odd scalar, $m_{A^0}$,
\begin{subequations}
\begin{align}
m_{H^\pm}^2&=m_{A^0}^2+m_W^2,\\
2m_{H^0,h^0}^2&=m_{A^0}^2+m_Z^2\pm\sqrt{(m_{A^0}^2+m_Z^2)^2
-4m_Z^2m_{A^0}^2\cos^22\beta}\text{.}
\end{align}
\end{subequations}
The tree-level results imply that $M_{h^0}<M_Z$, a constraint which
would have ruled out the \ac{mssm} already by the \acs{lep} experiment.
However, radiative corrections to the masses are significant and
have to be taken into account~\cite{Haber:Higgsbook}
and in the one-loop leading
logarithmic approximation one obtains the weaker
bound~\cite{Haber:1990aw,Okada:1990vk}
\begin{equation}
m_{h^0}^2\lesssim m_Z^2\cos^2\beta+\frac{3g^2m_t^4}{8\pi^2m_W^2}%
\ln\left(\frac{M_{\tilde{t}_1}M_{\tilde{t}_2}}{m_t^2}\right)
\end{equation}
with $M_{\tilde{t}_{1/2}}$ being the masses of the top-squark
mass eigenstates. This implies that for a supersymmetry breaking
scale at around \unit{1}{\tera\electronvolt} one expects to 
find the lightest \person{Higgs} boson to have a mass
below~\unit{150}{\giga\electronvolt}~\cite{Allanach:2004rh,Djouadi:2005gj}.

For large values of $\tan\beta$ the $H^0b\bar{b}$ coupling
%for $H\in\{H^0,h^0\}$
is enhanced and one expects \person{Higgs}
signals predominantly in $b$-associated
channels~\cite{RichterWas:1997gi,Dai:1994vu,Dai:1996rn}; therefore
it is not surprising that ``\emph{at large $\tan\beta$ the
$b\bar{b}\tau^+\tau^-$ and $b\bar{b}b\bar{b}$ final states may
provide the \emph{only} access to two of the three neutral
\acs{mssm} \person{Higgs} bosons.}''~\cite{Dai:1994vu}.
\begin{figure}[hbtp]
\includegraphics[scale=0.7]{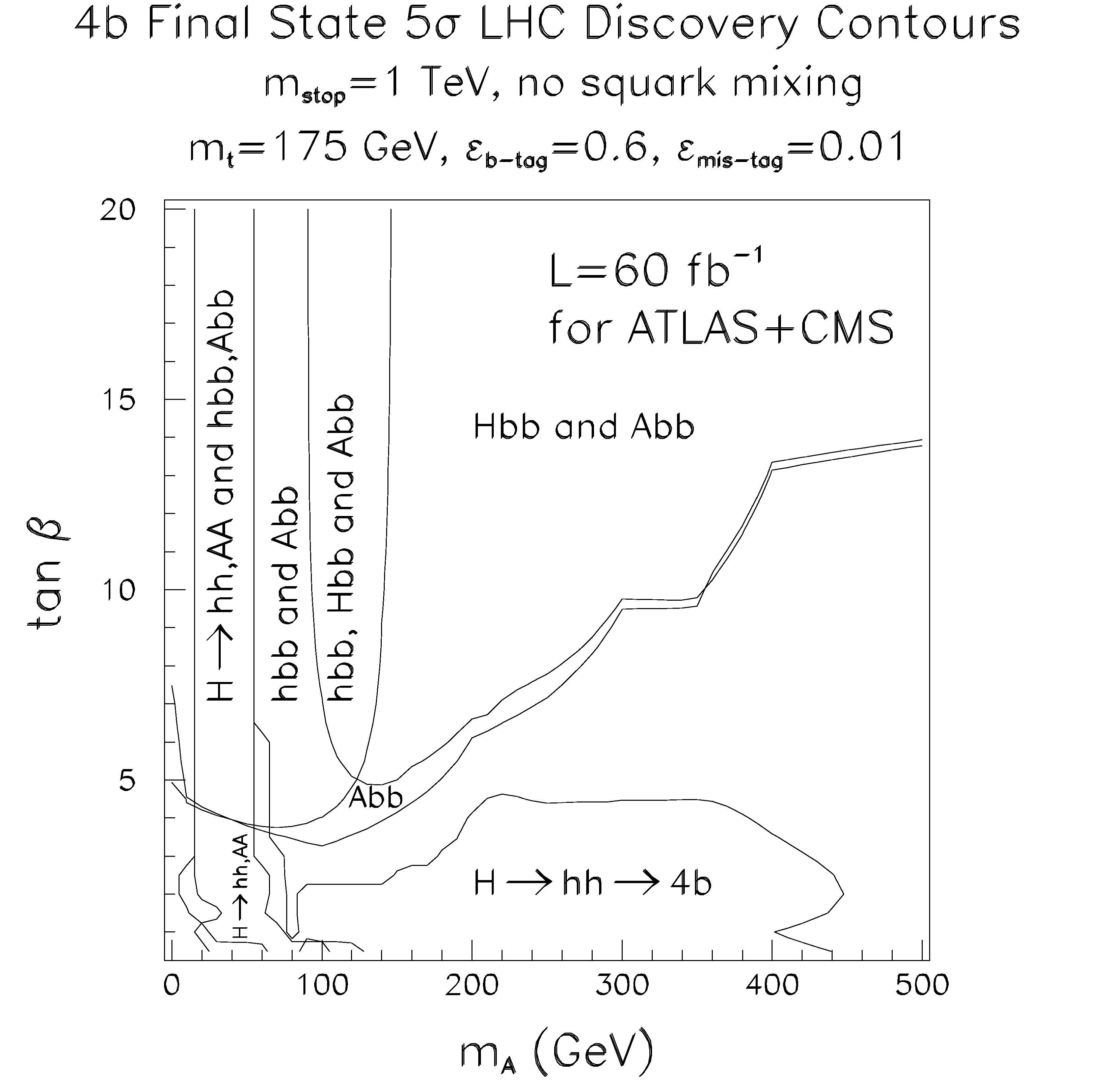}
\caption{$5\sigma$ discovery contours
in the ($M_{A^0}$, $\tan\beta$) parameter plane
for the channels
$gg\rightarrow b\bar{b}H\rightarrow b\bar{b}b\bar{b}$
($H\in\{h^0,H^0,A^0\}$)
and $gg\rightarrow H^0\rightarrow (h^0h^0)/(A^0A^0)\rightarrow
b\bar{b}b\bar{b}$ for an integrated luminosity of
\unit{30}{\femto\reciprocal\barn}
for ATLAS and \acs{cms} individually,
combining their statistic.~\cite{Dai:1996rn}} 
\label{fig:preface:4b-discovery}
\end{figure}

Figure~\ref{fig:preface:4b-discovery} shows the parameter regions
in the ($M_{A^0}$, $\tan\beta$) plane for which a $5\sigma$
discovery is possible at the \ac{lhc} after
both the experiments ATLAS and CMS have collected 
\unit{30}{\femto\reciprocal\barn}
integrated luminosity, analysing events with three and four tagged
$b$-jets in the final state~\cite{Dai:1996rn}\footnote{
Further detailed
studies of discovering \ac{mssm} \person{Higgs} bosons in the
$b\bar{b}b\bar{b}$ final state can for example be found
in~\cite{Djouadi:2000gu,Mahboubi:01}.
}. In~\cite{Dai:1994vu}
the lack of a full background study at \ac{nlo} is accounted for
by the use of a global $K$-factor of~$2$ which compromises the
significance of the study, and hence the authors argue that
``\emph{explicit calculations of the actual $K$ factors are needed}''.
Also in \ac{sm} studies for the \ac{lhc}, missing higher order corrections
to background processes are often limiting the precision with which
the measurements can be evaluated. The ATLAS sensitivity study
for the discovery of a \person{Higgs}
boson (see Figure~\ref{fig:preface:higgs-significance}), for example,
does not include higher order corrections for ``\emph{these $K$-factors are
generally not known for most background processes.}''~\cite{ATLAS:1999fr}
\index{standard model!minimal supersymmetric|)}

At hadron-hadron colliders such as the \ac{lhc} providing
a purely partonic initial state one expects most of the dynamics
to be due to the strong interaction.
Renormalisation introduces an unphysical scale~$\mu_R$
which would drop out if all orders of the perturbative
expansion were summed up. Realistically one computes
cross-sections only to a fixed order in perturbation theory,
which leaves a residual dependence on $\mu_R$.
The prediction only stabilises as higher-order corrections
are added to the \ac{lo} result~\cite{Ellis:QCD}.
The precision with which
the \ac{lhc} will measure the events requires the prediction
of both signal and background processes to at least \acf{nlo}
in the strong coupling constant $\alpha_s$
for processes with up to four particles in the final
state~\cite{Bern:2008ef,Ellis:QCD,ATLAS:1999fr,Draggiotis:2002hm}
--- a limit dictated by what is computable
with current techniques and technology rather than desired
from a experimental point of view~\cite{Buttar:2006zd}.
A list of processes which are well motivated
for \ac{lhc} phenomenology and seem
computationally within reach has been compiled
at the \emph{Les Houches} conferences on
\emph{Physics at TeV Colliders} in recent
years~\cite{Buttar:2006zd,Bern:2008ef}.

Although many important calculations have been accomplished
in recent years\footnote{See~\cite{Dittmaier:2007th,Campbell:2007ev,%
Sanguinetti:2008xt,Campbell:2006xx,Bredenstein:2008ia,Ciccolini:2007ec,%
Ciccolini:2007jr,Lazopoulos:2007ix,Hankele:2007sb,Jager:2006zc,Jager:2006cp,%
Bozzi:2007ur}},
there are still few
$2\rightarrow 4$ processes to be calculated which are crucial
for \person{Higgs} and \ac{bsm} studies at the \ac{lhc}~\cite{Bern:2008ef},
and the first years of the \ac{lhc}'s run time will probably
lead to new requirements of precision predictions.
The high demand for \ac{nlo} calculations for the \ac{lhc}
induces the need for automated tools for the calculation cross-sections
and other observables at the one-loop order.

This thesis presents an algorithm that automates the generation of
the virtual corrections to \ac{nlo} matrix elements. The current
implementation is capable of generating \fortranXC{} code for
the numerical evaluation of massless \ac{qcd} amplitudes\footnote{%
The extension of the implementation to massive amplitudes is in
preparation but beyond the scope of this thesis.}.

As an application the
$q\bar{q}\rightarrow b\bar{b}b\bar{b}$ amplitude at \ac{nlo} in $\alpha_s$
has been calculated, which is part of the \ac{sm} background
to the $b\bar{b}b\bar{b}$ channels in
\ac{mssm} \person{Higgs} searches. 
Although mainly motivated by supersymmetry,
the four-$b$ final state also allows
the study of other interesting \ac{bsm} physics models such
as hidden valley models, where decays of hadrons of an additional
confining gauge group can produce high multiplicities
of $b\bar{b}$ pairs~\cite{Krolikowski:2008qa,Bern:2008ef}.

In Chapter~\ref{chp:qcd} I introduce~\ac{qcd} with its \Lagrangeian{}
density and \person{Feynman} rules.
Chapter~\ref{chp:repqcdamp} describes the structure of
\acs{qcd} amplitudes and techniques for treating the
colour algebra.
Methods for calculating amplitudes
at one-loop precision are discussed in Chapter~\ref{chp:qcdnlo}.
Results for the computation of the virtual corrections
of the process $u\bar{u}\rightarrow b\bar{b}b\bar{b}$
are presented in Chapter~\ref{chp:results}.
A more technical discussion of the underlying implementation
that has been used for this calculation is attached
in Appendix~\ref{chp:implementation}.

%% file: qcd.tex
%%%%%%%%%%%%%%%%%%%%%%%%%%%%%%%%%%%%%%%%%%%%%%%%%%%%%%%%%%%%%%%
\chapter{Principles of \acs{qcd}}
\label{chp:qcd}
\begin{headquote}{Murray Gell-Mann}
The beauty of the basic laws of natural science,
as revealed in the study of particles and of the cosmos,
is allied to the litheness of a merganser diving in a pure Swedish lake,
or the grace of a dolphin leaving shining trails at night in the Gulf of California,
or the loveliness of the ladies assembled at this banquet.
\end{headquote}
\section*{Introduction}

\input{qcd-intro}

%\section{\acs{qcd}: A Gauge Theory of the Strong Interaction}
%\label{sec:stronginteraction}
\input{qcd-gauge}

%%%%%%%%%%%%%%%%%%%%%%%%%%%%%%%%%%%%%%%%%%%%%%%%%%%%%%%%%%%%%%%
\chapter{Representations of \acs{qcd} Amplitudes}
\label{chp:repqcdamp}
\begin{headquote}{Jules Henri Poincar\'e}
It is the harmony of the diverse parts, their symmetry, 
their happy balance; in a word it is all that introduces order, 
all that gives unity, that permits us to see clearly and
to comprehend at once both the ensemble and the details.
\end{headquote}
\section*{Introduction}
\label{seq:amprepintro}
\input{rep-intro}

\section{Colour-Flow Decomposition}
\label{sec:colorordering}
\input{qcd-color}
\clearpage
\section{Irreducible Representations of \acs{sun}}
\label{sec:irreps}
\input{qcd-irreps}

\section{The Spinor Helicity Projection Method}
\label{sec:shelpmethod}
\input{qcd-shelpmethod.tex}

%%%%%%%%%%%%%%%%%%%%%%%%%%%%%%%%%%%%%%%%%%%%%%%%%%%%%%%%%%%%%%%
\chapter{\acs{qcd} at One-Loop Precision}
\begin{headquote}{Werner Heisenberg}
An expert is someone who knows some of the worst mistakes
that can be made in his subject and who manages to avoid them.
\end{headquote}
\label{chp:qcdnlo}

\section*{Introduction}
\label{sec:nlointro}
\input{nlo-intro}

\section{\acs{qcd} in Dimensional Regularisation}
\label{sec:qcddimreg}
\input{qcd-dimreg}

\section{Reduction of the Scalar Integrals}
\label{sec:scalarred}
\input{qcd-scalarred}

\section{Tensor Reduction by Subtraction}
\label{sec:tensorred}
\input{qcd-tensorred}
\input{qcd-tensorred-ibp}

\section{Representation of the Virtual Corrections}
\label{sec:virtrep}
\input{qcd-diagramrep}

\section{Renormalisation of \ac{qcd}}
\label{sec:qcd-renorm}
\input{qcd-renormalisation}

\section{Real Emission Contribution}
\label{sec:realem}
\input{qcd-real}

\section{Dipole Subtraction}
\label{sec:dipoles}
\input{qcd-dipoles}

\section{Phase Space Integration and Monte-Carlo Techniques}
\label{sec:psintegral}
\input{qcd-psintegral}

%% file: qcd-intro.tex
\acf{qcd} is the non-\person{Abel}ian gauge theory of the \sun[3]{} group.
Within the \ac{sm} it describes the strong interaction. The aim of this
thesis is the development of methods to calculate \ac{qcd} amplitudes and
to present results of one example calculation.
The \Lagrangeian{} density of \ac{qcd} is introduced in
Section~\ref{sec:qcd-lagrange} setting up the framework for the rest of
this thesis. The \person{Feynman} rules induced by the \Lagrangeian{} density
build one of the fundamental tools in our calculation.
Section~\ref{sec:qcd-running} describes in short \emph{assymptotic freedom}:
\ac{qcd}, at low energies, is strongly coupled. The coupling strength 
only decreases for energy scales larger than about $1\,\textrm{GeV}$
which justifies a perturbative expansion of~\ac{qcd} to be used as a
description of a hard collision of partons in a hadron-hadron collider.

As a last general aspect the \emph{flavour symmetry} of \ac{qcd} is
discussed in Section~\ref{ssec:qcd-gauge:flavoursym}. The interactions
of \ac{qcd} do not distinguish between quark flavours which induces
an additional, discrete symmetry to this theory. We exploit flavour
symmetry in our calculation as it allows to reduce the number of
different \person{Feynman} diagrams to be calculated.

%% file: qcd-gauge.tex
\section{The \Lagrangeian{} Density of \acs{qcd}}
\label{sec:qcd-lagrange}
Throughout this document I follow the convention for the metric
tensor\footnote{The extension to $n\neq4$ dimensions is described
separately in Section~\ref{sec:qcddimreg}, Chapter~\ref{chp:qcdnlo}.}
\begin{displaymath}
g^{\mu\nu}=\diag(1,-1,-1,-1)\text{,}
\end{displaymath}
the \person{Levi-Civita} tensor
with~$\epsilon_{0123}=+1$ and the
\person{Dirac} $\gamma$-matrices obey the anticommutation relation
\begin{equation}
\{\gamma^\mu,\gamma^\nu\}\equiv\gamma^\mu\gamma^\nu+\gamma^\nu\gamma^\mu=
2g^{\mu\nu}\text{.}
\end{equation}
Contractions of \person{Dirac} matrices with \person{Lorentz} vectors are
denoted by the \person{Feynman}-slash,
$\pslash\equiv\gamma^\mu p_\mu$. Summation over repeated indices is
understood unless explicitly noted differently. This
summation convention is applied not only to \person{Lorentz} indices but
also to \ac{sun} colour indices.
The indices of the representation of the \text{Dirac} algebra are
omitted where no ambiguities are possible.
I work in natural
units\footnote{This leads to the usual conversion factors
$\hbar/\unit{1}{\giga\electronvolt}\approx6{.}58212\cdot\unit{\(10^{-25}\)}{\second}$
and
$\hbar c/\unit{1}{\giga\electronvolt}\approx1{.}97327\cdot\unit{\(10^{-16}\)}{\metre}$;
the units for cross-sections therefore are
$\unit{1}{\milli\barn}\approx2{.}56819\usk(\hbar c)^2/\unit{1}{\giga\squaren\electronvolt}$.
}, $\hbar=c=1$.

\index{QCD|(}
Historically, the concept of colour was introduced in the quark
model to satisfy \person{Dirac} statistics for hadrons with three identical
quarks; the colour symmetry was a global~\sun[3]{} gauge symmetry.
It was a real breakthrough in the success of the quark model when two main
observations, confinement and asymptotic freedom, could be described by
the gauge theory of a local~\sun[3]{} colour symmetry,
i.e. \acf{qcd}\footnote{See for example~\cite{Ellis:QCD}}. 
\ac{qcd} is a strongly coupled theory which in principle requires
non-perturbative methods. Here, lattice~\ac{qcd} plays the
most important role allowing the precise determination of
the properties of \ac{qcd} bound states.

Due to the running of the coupling constant for very high energies 
the coupling constant of \ac{qcd} becomes small and a perturbative
expansion becomes meaningful.
Perturbative \ac{qcd} therefore can be used as a predictive tool
for collider experiments and is the main focus of this work.

\index{QCD!Lagrangian density@\Lagrangeian{} density}
The \Lagrangeian{} density~$\Ld_\text{QCD}$ of \ac{qcd} can be
split into three parts: the classical density~$\Ld_\text{cl}$, the gauge fixing term~$\Ld_\text{gf}$ and the ghost
term~$\Ld_\text{gh}$,
\begin{equation} \label{eq:qcd-gauge:lagrange}
\Ld_\text{QCD}=\Ld_\text{cl}+\Ld_\text{gf}+\Ld_\text{gh}\text{.}
\end{equation}

The classical \Lagrangeian{} density of a non-\person{Abel}ian
gauge theory coupled to fermionic matter reads
\begin{equation}
\Ld_\text{cl}=-\frac14 F_{\mu\nu}^A F^{\mu\nu,A}+\sum_{q\in\text{flav.}}\bar{q}_a(i\Dslash_{ab}-m_q\delta_{ab})q_{b}\text{.}
\end{equation}
The fermion fields are denoted by~$q_a$, 
where the sum over the quark fields $q$ runs over all different flavours
($u$,~$d$, $s$, $c$, $b$ and~$t$) and~$m_q$
stands for the mass of a quark of the respective flavour.
The field strength tensor $F^A_{\mu\nu}$ for the gluon field~$\Glue^A_\mu$ is
\begin{equation}
F_{\mu\nu}^A=\partial_\mu\Glue_\nu^A-\partial_\nu\Glue_\mu^A-g_sf^{ABC}%
\Glue^B_\mu\Glue^C_\nu\text{,}
\end{equation}
where~$g_s$ is the strong coupling constant and~$f^{ABC}$ is the structure constant of the gauge group.
Capital Latin letters denote indices over the adjoint representation of the gauge group and lower case letters stand for
indices in the fundamental representation. The properties of the colour algebra are described in detail in Chapter~\ref{chp:repqcdamp},
Sections \ref{sec:colorordering} and~\ref{sec:irreps}.
I also use the symbol
\begin{equation}
\alpha_s=\frac{g_s^2}{4\pi}\text{.}
\end{equation}
The covariant derivative has the form
\begin{subequations}
\begin{align}
\Diff^\mu_{ab}&=\partial^\mu\delta_{ab}+ig_s\Glue^{\mu,C}t_{ab}^C\quad\text{in the fundamental, and}\\
\Diff^\mu_{AB}&=\partial^\mu\delta_{AB}+ig_s\Glue^{\mu,C}T_{AB}^C\quad\text{in the adjoint representation,}
\end{align}
\end{subequations}
where $t^A_{ab}$ ($T^C_{AB}$) are the generators of the fundamental (adjoint) representation of the gauge group.

\index{gauge!covariant gauge fixing term}
Again following~\cite{Ellis:QCD}, I describe two different families of gauge fixing terms. A covariant gauge fixing term is
provided by
\begin{equation}
\Ld_\text{gf}=-\frac{1}{2\lambda}\left(\partial^\mu\Glue_\mu^A\right)\left(\partial^\nu\Glue_\nu^A\right)
\end{equation}
with the gauge parameter~$\lambda$. This approach requires the introduction of a ghost field via
\begin{equation}
\Ld_\text{gh}=\left(\partial_\mu\eta^A\right)^\dagger\left(\Diff^\mu_{AB}\eta^B\right)
\end{equation}
to remove the remaining unphysical degree of freedom.
The ghost field is a complex scalar field obeying fermionic statistics.
Diagrams with external ghost fields can be avoided by an appropriate choice
of the gluon polarisation vectors.

\index{gauge!axial gauge fixing term}
A second class of gauge fixing terms are the axial gauges, which involve an arbitrary four-vector~$q$,
\begin{equation}
\Ld_\text{gf}=-\frac{1}{2\lambda}\left(q^\mu\Glue_\mu^A\right)\left(q^\nu\Glue_\nu^A\right)\text{.}
\end{equation}
This gauge fixing term does not require the ghost sector but leads to a more complicated gluon propagator.

\index{gauge!Feynman gauge@\person{Feynman} gauge}
The two most prominent gauge choices of covariant gauges are~$\lambda=1$ (i.e. the \person{Feynman} gauge) and
$\lambda\rightarrow0$ (i.e the \person{Landau} gauge). For practical calculations very often \person{Feynman} gauge is chosen for it leads to
a simpler numerator structure than an arbitrary choice of~$\lambda$.

\index{QCD!Feynman rules@\person{Feynman} rules}
\index{Feynman@\person{Feynman}!rules|see{QCD, \person{Feynman} rules}}
The \Lagrangeian{} density~\eqref{eq:qcd-gauge:lagrange} in covariant gauge
leads to the following set of
\person{Feynman} rules, given in~\eqref{eqs:feynmanrules}.
Straight lines represent quarks, gluons are drawn as
curly lines and ghosts as dotted lines. 
These \person{Feynman} rules correspond to the ones given
in~\cite{Boehm:Gauge}; the corresponding rules in~\cite{Ellis:QCD} are
obtained by the transformation $g_s\rightarrow -g_s$.
\begin{fmffile}{qcd-gauge-fmf}
\begin{subequations}\label{eqs:feynmanrules}
\begin{align}
% qq propagator
\parbox[c][\height+1.5\baselineskip][c]{25mm}{%nl
\begin{fmfchar*}(20,20)
\fmfleft{q}\fmfright{qbar}
\fmflabel{$a$}{q}
\fmflabel{$b$}{qbar}
\fmf{fermion,label=$p$}{q,qbar}
\end{fmfchar*}}&\quad=i\delta_{ab}\frac{\pslash+m}{p^2-m^2+i\delta}
\displaybreak[1]
\\
% gg propagator
\parbox[c][\height+1.5\baselineskip][c]{25mm}{%nl
\begin{fmfchar*}(20,20)
\fmfleft{q}\fmfright{qbar}
\fmflabel{$A\comma\mu$}{q}
\fmflabel{$B\comma\nu$}{qbar}
\fmf{gluon,label=$p$,label.dist=5mm}{q,qbar}
\end{fmfchar*}}&\quad=\delta^{AB}\frac{i}{p^2+i\delta}\left(-g^{\mu\nu}+(1-\lambda)\frac{p^\mu p^\nu}{p^2+i\delta}\right)
\displaybreak[1]
\\
% GG propagator
\parbox[c][\height+1.5\baselineskip][c]{25mm}{%nl
\begin{fmfchar*}(20,20)
\fmfleft{q}\fmfright{qbar}
\fmflabel{$A$}{q}
\fmflabel{$B$}{qbar}
\fmf{ghost,label=$p$}{q,qbar}
\end{fmfchar*}}&\quad=\delta^{AB}\frac{i}{p^2+i\delta}
\displaybreak[1]
\\
% ggg vertex
\parbox[c][\height+1.5\baselineskip][c]{25mm}{%nl
\begin{fmfchar*}(20,20)
\fmfleft{g1,g2}\fmfright{g3}\fmfdot{vertex}
\fmflabel{$A$}{g1}
\fmflabel{$B$}{g2}
\fmflabel{$C$}{g3}
\fmf{gluon,label=$p_{1\comma\mu}$}{g1,vertex}
\fmf{gluon,label=$p_{2\comma\nu}$}{g2,vertex}
\fmf{gluon,label=$p_{3\comma\rho}$}{g3,vertex}
\end{fmfchar*}}&\quad%nl
%\begin{minipage}[c]{60mm}
%\begin{displaymath}
\begin{array}{rl}
=g_sf^{ABC}\left[\right.&g^{\mu\nu}(p_1-p_2)^\rho\\
+&g^{\nu\rho}(p_2-p_3)^\mu\\
+&\left.g^{\mu\rho}(p_3-p_1)^\nu\right]
\end{array}
%\end{displaymath}
%\end{minipage}
\displaybreak[1]
\\
% gggg vertex
\parbox[c][\height+1.5\baselineskip][c]{25mm}{%nl
\begin{fmfchar*}(20,20)
\fmfleft{g1,g2}\fmfright{g3,g4}\fmfdot{vertex}
\fmflabel{$A$}{g1}
\fmflabel{$B$}{g2}
\fmflabel{$C$}{g3}
\fmflabel{$D$}{g4}
\fmf{gluon,label=$p_{1\comma\mu}$}{g1,vertex}
\fmf{gluon,label=$p_{2\comma\nu}$}{g2,vertex}
\fmf{gluon,label=$p_{3\comma\rho}$}{g3,vertex}
\fmf{gluon,label=$p_{4\comma\sigma}$}{g4,vertex}
\end{fmfchar*}}&\quad
%\begin{minipage}[c]{60mm}
%\begin{displaymath}
\begin{array}{rl}
=-ig_s^2\left[\right.&f^{ABE}f^{CDE}\left(g^{\mu\rho}g^{\nu\sigma}-g^{\mu\sigma}g^{\nu\rho}\right)\\
+&f^{ACE}f^{BDE}\left(g^{\mu\nu}g^{\rho\sigma}-g^{\mu\sigma}g^{\nu\rho}\right)\\
+&\left.f^{ADE}f^{BCE}\left(g^{\mu\nu}g^{\rho\sigma}-g^{\mu\rho}g^{\nu\sigma}\right)\right]
\end{array}
%\end{displaymath}
%\end{minipage}
\displaybreak[1]
\\
% qqg vertex
\parbox[c][\height+1.5\baselineskip][c]{25mm}{%nl
\begin{fmfchar*}(20,20)
\fmfleft{q,qbar}\fmfright{g}\fmfdot{vertex}
\fmflabel{$a$}{q}
\fmflabel{$b$}{qbar}
\fmflabel{$C$}{g}
\fmf{fermion}{q,vertex,qbar}
\fmf{gluon,label=$p^\mu$,label.dist=5mm}{vertex,g}
\end{fmfchar*}}&\quad=ig_st^C_{ba}\gamma^\mu\displaybreak[1]\\
% GGg vertex
\parbox[c][\height+1.5\baselineskip][c]{25mm}{%nl
\begin{fmfchar*}(20,20)
\fmfleft{q,qbar}\fmfright{g}\fmfdot{vertex}
\fmflabel{$A$}{q}
\fmflabel{$B$}{qbar}
\fmflabel{$C$}{g}
\fmf{ghost}{q,vertex}
\fmf{ghost,label=$p^\mu$}{vertex,qbar}
\fmf{gluon}{vertex,g}
\end{fmfchar*}}&\quad=-g_sf^{ABC}p^\mu
\end{align}
\end{subequations}
These rules are valid in the covariant gauge, all momenta are ingoing at the vertices and following
the arrow along propagators. In axial gauge the gluon propagator has to be replaced by
\begin{displaymath}
\delta^{AB}\frac{i}{p^2+i\delta}\left(-g^{\mu\nu}+\frac{p^\mu q^\nu+q^\mu p^\nu}{q\cdot p}
+\frac{(q^2+\lambda p^2)p^\mu p^\nu}{(q\cdot p)^2}\right)\text{.}
\end{displaymath}
\end{fmffile}
\medskip

\section{The Effective Coupling and Asymptotic Freedom}
\label{sec:qcd-running}
\index{QCD!running coupling constant}
Calculating terms of higher order of $\alpha_s=g_s^2/(4\pi)$ in the
perturbative expansion usually introduces ultraviolet divergences which
have to be cured by renormalisation.
One generic property of regularisation is the appearance of a new mass scale,
which in dimensional regularisation usually is called~$\mu$.
A physical observable~$R$ therefore not only depends on the energy scale~$Q$
of the process but also on the parameter~$\mu$.
Since the second dependency is unphysical
--- every choice of $\mu$ should lead to the same result ---
one may postulate this independence by the
renormalisation group equation~\cite{Ellis:QCD}
\begin{equation}
\mu^2\frac{\diff}{\diff\mu^2}R(Q^2/\mu^2, \alpha_s)=\left[\mu^2\frac{\partial}{\partial\mu^2}+
\mu^2\frac{\partial\alpha_s}{\partial\mu^2}\frac{\partial}{\partial\alpha_s}\right]R(Q^2/\mu^2,\alpha_s)=0\text{.}
\end{equation}
The coefficient of the second term is called the $\beta$-function,
\begin{equation}\label{eq:qcd-gauge:beta}
\beta(\alpha_s)\equiv\mu^2\frac{\partial\alpha_s(\mu^2)}{\partial\mu^2}=Q^2\frac{\partial\alpha_s(Q^2)}{\partial Q^2}\text{,}
\end{equation}
which implicitly defines a scale dependent coupling constant,
the so called \emph{running coupling constant} $\alpha_s(Q^2)$.

In the perturbative regime and for $n_f$ flavours of massless quarks only we can express the $\beta$-function as
\begin{equation}
\beta(\alpha_s)=-b\alpha_s(Q^2)\left[1+b^\prime\alpha_s(Q^2)+{\mathcal{O}}\left(\alpha_s(Q^2)\right)\right]\text{,}
\end{equation}
with the coefficients
\begin{subequations}
\begin{align}
b&=\frac{33-2n_f}{12\pi}\quad\text{and}\\
b^\prime&=\frac{153-19n_f}{24\pi^2b}\text{.}
\end{align}
\end{subequations}
Neglecting~$b^\prime$ and all higher order terms, the partial differential equation~\eqref{eq:qcd-gauge:beta}
can be solved leading to a relation between $\alpha_s(Q^2)$ and $\alpha_s(\mu^2)$,
\begin{equation} \label{eq:qcd-gauge:runningalpha}
\alpha_s(Q^2)=\frac{\alpha_s(\mu^2)}{1+\alpha_s(\mu^2)b\ln(Q^2/\mu^2)}
\end{equation}
and the behaviour of~$\alpha_s$ depends on the number of flavours: 
$b$ is positive as long as $n_f<33/2$. 

The scale at which the denominator
of~\eqref{eq:qcd-gauge:runningalpha} vanishes is called
the \person{Landau} pole~$Q=\Lambda_\text{QCD}$.
\index{Landau@\person{Landau}!pole}
Setting~$\mu$ to the Z-mass $m_z$ in this approximation
and for~$n_f=5$ one obtains using~$\alpha_s(m_Z^2)$~\cite{PDBook}
\begin{equation}
\Lambda_\text{QCD}=m_Z\exp\left[-\frac{1}{b\alpha_s(m_Z^2)}\right]%
\approx\unit{91}{\mega\electronvolt}.
\end{equation}
Better approximations yield values around~\unit{200}{\mega\electronvolt}.
For large energies $\alpha_s(Q^2)$ decreases since the
\acl{sm} contains $n_f=6$ flavours,
a fact that is known as \emph{asymptotic freedom}.
As low energies close to the \person{Landau} pole
are reached,
the coupling constant becomes large and a perturbative expansion
is no longer valid. This strongly coupled regime leads to
quark \emph{confinement} and ensures that in nature no
free coloured particles appear.
\index{QCD|)}

\section{Flavour Symmetry}
\label{ssec:qcd-gauge:flavoursym}
Under the interactions of \ac{qcd} all quark flavours interact in the
same way with the gluons. If in addition one considers the approximation
of massless quarks the \Lagrangeian{} density is invariant under
the exchange of flavour and hence any amplitude calculated in massless
\ac{qcd} is invariant under the exchange of flavours. This is true
only if the combinatorics of the configuration does not change.

In the following section I derive the relation between the two amplitudes
$\mathcal{A}_{u\bar{u}\rightarrow b\bar{b}b\bar{b}}$ and
$\mathcal{A}_{u\bar{u}\rightarrow b\bar{b}s\bar{s}}$.
This relation is useful for the calculation of the first amplitude
from the second one and has been exploited in our calculation of
the former amplitude.
Because of its higher symmetry 
$\mathcal{A}_{u\bar{u}\rightarrow b\bar{b}b\bar{b}}$ consists of
more \person{Feynman} diagrams than 
$\mathcal{A}_{u\bar{u}\rightarrow b\bar{b}s\bar{s}}$. Hence the
second amplitude is easier to calculate and implement.

For the derivation of the relation between the amplitudes
 the following abstractions from \ac{qcd}
can be made: the only requirement of flavour symmetry is that the
quarks couple equally to the other fields in the theory, which
are represented by a single field $\Phi$. We consider a
theory with two quark fields $\Psi$ and $\Xi$. If no contact
interaction of two quark pairs is allowed the \Lagrangeian{}
density can be written as
\begin{equation}
\Ld=\bar\Psi Q\Psi+\bar\Xi Q\Xi
+\Ld_B(\Phi,\partial\Phi)
+\bar\Psi V(\Phi)\Psi
+\bar\Xi V(\Phi)\Xi
\text{.}
\end{equation}
The operator $Q$ stands for the kinetic part of the \Lagrangeian{}
density; usually $Q=(\fmslash{\partial}+m)$, but the exact form is
irrelevant for the discussion. The interaction between the quarks
and $\Phi$ is denoted by $V(\Phi)$.
$\Ld_B$ contains the self interactions
and the kinetic part of the \Lagrangeian{} density for the
field~$\Phi$.

The partition function $Z$ is defined as
\begin{equation}
Z[\bar\eta, \eta, \bar\xi, \xi]=
\int\Diff\bar\Psi\Diff\Psi\Diff\bar\Xi\Diff\Xi\Diff\Phi
\,e^{i\int\diff[d]x\,\Ld
+\bar\eta\Psi-\bar\Psi\eta
+\bar\xi\Xi-\bar\Xi\xi+\Phi J_\Phi
}\text{.}
\end{equation}
The functional integral can be separated from the sources by
completing the square and introducing the \person{Green}'s
function $S$ that fulfils
\begin{equation}
QS(x,x^\prime)=\delta^{(d)}(x-x^\prime)
\end{equation}
as explained in standard textbooks about quantum field
theory~\cite{Peskin:QFT,Boehm:2001}. Using the notation
\begin{align}
\Ld_0[\bar\Psi,\Psi,\bar\Xi,\Xi]&=\bar\Psi Q\Psi+\bar\Xi Q\Xi\quad\text{and}\\
\Ld_I[\bar\Psi,\Psi,\bar\Xi,\Xi,\Phi]&=
\bar\Psi V(\Phi)\Psi
+\bar\Xi V(\Phi)\Xi
+\Ld_B(\Phi,\partial\Phi)
\end{align}
one obtains
\begin{multline}
Z[\bar\eta, \eta, \bar\xi, \xi]=
\exp\left\{i\int\!\!\diff[d]x\,\Ld_I\left[
\frac{\delta}{i\delta\bar\eta(x)},
\frac{\delta}{i\delta\eta(x)},
\frac{\delta}{i\delta\bar\xi(x)},
\frac{\delta}{i\delta\xi(x)},
\frac{\delta}{i\delta J_\Phi(x)}
\right]
\right\}\times\\
\exp\left\{i\int\!\!\diff[d]x\diff[d]y\,
\bar\eta(x)S(x,y)\eta(y)
+\bar\xi(x)S(x,y)\xi(y)
\right\}
\left(\int\Diff\bar\Psi\Diff\Psi\Diff\bar\Xi\Diff\Xi\Diff\Phi
\,e^{i\int\diff[d]x\,\Ld_0
}\right)\text{.}
\end{multline}
The remaining functional integral is called $Z_0$
and is only a constant which does not carry any
functional dependence. 

Now, we can compare the correlation functions
\begin{align}
\braket{T\bar\Psi(x_1)\Psi(x_2)\bar\Xi(x_3)\Xi(x_4)}&=
\left.\frac{1}{Z_0}\frac{\delta^4Z[\bar\eta,\eta,\bar\xi,\xi]}{
\delta\bar\eta(x_1)\delta\eta(x_2)\delta\bar\xi(x_3)\delta\xi(x_4)}
\right\vert_{\begin{array}{l}\scriptstyle\eta=\bar\eta=0\\
\scriptstyle\xi=\bar\xi=J_\Phi=0
\end{array}}
\quad\text{and}\\
\braket{T\bar\Psi(x_1)\Psi(x_2)\bar\Psi(x_3)\Psi(x_4)}&=
\left.\frac{1}{Z_0}\frac{\delta^4Z[\bar\eta,\eta,\bar\xi,\xi]}{
\delta\bar\eta(x_1)\delta\eta(x_2)\delta\bar\eta(x_3)\delta\eta(x_4)}
\right\vert_{\begin{array}{l}\scriptstyle\eta=\bar\eta=0\\
\scriptstyle\xi=\bar\xi=J_\Phi=0\end{array}}
\end{align}
by direct calculation. It is sufficient to compare both correlation
functions for the term
\begin{multline*}
-\frac12\int\!\!\diff[d]x\,V\left(\frac{\delta}{i\delta\Phi(x)}\right)\left(
\frac{\delta}{i\delta\bar\eta(x)}
\frac{\delta}{i\delta\eta(x)}
+\frac{\delta}{i\delta\bar\xi(x)}
\frac{\delta}{i\delta\xi(x)}
\right)\times\\
\int\!\!\diff[d]y\,V\left(\frac{\delta}{i\delta\Phi(y)}\right)\left(
\frac{\delta}{i\delta\bar\eta(y)}
\frac{\delta}{i\delta\eta(y)}
+\frac{\delta}{i\delta\bar\xi(y)}
\frac{\delta}{i\delta\xi(y)}
\right)
\end{multline*}
from the expansion of $\exp(\int{i\Ld_I})$.
This expansion has been done
with the computer algebra system
\form~\cite{Vermaseren:2000nd,Vermaseren:2002}
and the result, written in terms of quarks and gluons is summarised
below:
\begin{align}
\mathcal{A}_{u\bar{u}\rightarrow b\bar{b}b\bar{b}}(a,b;1,2,3,4)&=
\mathcal{A}_{u\bar{u}\rightarrow b\bar{b}s\bar{s}}(a,b;1,2,3,4)
-\mathcal{A}_{u\bar{u}\rightarrow b\bar{b}s\bar{s}}(a,b;1,4,3,2)\text{,}\\
\mathcal{A}_{gg\rightarrow b\bar{b}b\bar{b}}(a,b;1,2,3,4)&=
\mathcal{A}_{gg\rightarrow b\bar{b}s\bar{s}}(a,b;1,2,3,4)
-\mathcal{A}_{gg\rightarrow b\bar{b}s\bar{s}}(a,b;1,4,3,2)\text{.}
\end{align}
The external fields are denoted simply by their
indices ($a$, $b$, $1$, $2$, \dots); this notation implies
that momenta, colour and helicity labels need to be swapped
accordingly.

%% file: rep-intro.tex
In order to make predictions for colliders at high energies
one needs to relate observables with the underlying theory.
In quantum mechanics this relation is given through
the \emph{scattering matrix}\index{Scattering matrix}%
\footnote{In Section~\ref{ssec:qcd-dimreg:loop-integrals}
of Chapter~\ref{chp:qcdnlo} I introduce a matrix $S_{ij}$
that encodes the kinetic information of a
\person{Feynman} diagram at one-loop. Although the two
matrices are unrelated objects, in the literature
for both matrices~$S$ is the commonly used symbol.%
}\index{$S$-matrix@Smatrix}~$S$; the $S$-matrix element
$\braket{f\vert S\vert i}$ describes the transition
from an initial state $\ket{i}$ to a final state $\bra{f}$,
where $\ket{i}$ is taken at time $t\rightarrow-\infty$ and
$\ket{f}$ is a state at $t\rightarrow+\infty$.
The operator~$S$ can be related to the interaction part
of the \Lagrangeian{} density\footnote{See for example~\cite{Boehm:2001}},
\begin{equation}
S=\mathbf{T}e^{i\int\!\!\diff[4]x\mathcal{L}_I}\text{,}
\end{equation}
where $\mathbf{T}$ denotes the time-ordered product.
For momentum eigenstates of momenta $p_i$ and $p_f$ respectively
the $S$-matrix elements can be written as
\begin{equation}
\braket{f\vert S\vert i}=\braket{f\vert i}+i(2\pi)^4\delta^{(4)}(p_i-p_f)\ME_{fi}\text{.}
\end{equation}

The \person{Feynman} rules give a prescription how to obtain 
an analytical expression for \(i\ME\) from a sum of \person{Feynman}
diagrams. In \ac{qcd} each \person{Feynman} diagram can be written
as a product of a colour vector $\ket{c}$ and a kinematical coefficient.
With a common choice of a colour basis~$\mathcal{B}$ for all diagrams, 
the invariant matrix element has the form
\begin{equation}
\ME_{fi}=\sum_{\ket{c}\in\mathcal{B}}
\mathcal{A}_c(p_a, p_b; p_1, \ldots, p_N)\ket{c}\text{,}
\end{equation}
where $p_a$ and $p_b$ denote the momenta of the incoming
particles and $p_1, \ldots, p_N$ those of the final state
particles.
Different choices of bases~$\mathcal{B}$ are discussed in
Sections \ref{sec:colorordering} and~\ref{sec:irreps}.
The coefficient function~\(\mathcal{A}_c\) contains
all dependencies on the momenta. Its calculation
can be simplified through projections on the
physical degrees of freedom for spinors and
polarisation vectors, which leads to the
formalism of spinor helicity projections
described in Section~\ref{sec:shelpmethod}.

Observables measured at colliders can usually be expressed
through a differential cross-section\footnote{As \ac{qcd} partons
cannot be observed as free particles in nature,
one distinguishes between the partonic and the hadronic
cross-section. The latter is obtained by the former
through a convolution with parton distribution functions.
Here, I describe the partonic cross-section.
For a discussion of its hadronic
equivalent see Section~\ref{sec:dipoles} in Chapter~\ref{chp:qcdnlo}},
\begin{equation}
\diff{\sigma}=\frac{1}{2\vert p_a+p_b\vert^2}\frac{1}{n_an_b}\phasespace{N}(p_a+p_b)
\sum_{c,c^\prime\in\mathcal{B}}\bra{c^\prime}\mathcal{A}_{c^\prime}^\ast\mathcal{A}_c\ket{c}
\measureF{N}(p_1,\ldots,p_N)\text{.}
\end{equation}
The measurement function \(\measureF{N}\) defines the observable and usually
contains $\Theta$-functions, defining the experimental cuts, the definition of
jets in the case of jet-observables and, in the case of exclusive observables,
the quantities of which distributions are to be obtained. The $N$-particle
phase space can be parametrised as follows,
\begin{equation}
\phasespace{N}(Q;p_1,\ldots,p_N)=\prod_{j=1}^N\frac{\diff[4]{p_j}}{(2\pi)^3}\Theta(p^0_j)
\delta({p_j}^2-m_j^2)
\cdot(2\pi)^4\delta^{(4)}\left(Q-\sum_{i=1}^Np_i\right)
\end{equation}
and is discussed in more detail in Chapter~\ref{chp:qcdnlo} in Section~\ref{sec:psintegral}.
The factors $n_a$ and $n_b$ denote normalisations induced by spin and colour averages.
A factor of $1/n!$ for has to be included for each set of $n$ final state particles 
that are not distinguished in the observable.

%% file: qcd-color.tex
\subsection{\acs{sun} Diagrammatics}
\begin{fmffile}{birdtracks}
One part of amplitude calculations in non-\person{Abel}ian
gauge theories is the simplification and evaluation of the colour
structure. In this section I will present a diagrammatic approach
which is mainly motivated and introduced in~\cite{Cvitanovic}.
The basic idea is to represent
all indices by external lines and all tensors by vertices;
\person{Kronecker} delta symbols therefore appear
as internal lines. I use dashed lines for the adjoint representation,
fermion lines for the fundamental
representation and dotted lines for the trivial representation.
The dotted lines could be left out in most cases
since they only represent a one; they are drawn here anyway 
to clarify the origin of the formul\ae.
\begin{subequations}
\begin{align}
\parbox[c][\height+1.5\baselineskip][c]{25mm}{%nl
\begin{fmfchar*}(20,2)
\fmfleft{q}\fmfright{qbar}
\fmflabel{$A$}{q}
\fmflabel{$B$}{qbar}
\fmf{dashes}{q,qbar}
\end{fmfchar*}}&\quad=\delta^{AB}
\displaybreak[1]\\
\parbox[c][\height+1.5\baselineskip][c]{25mm}{%nl
\begin{fmfchar*}(20,2)
\fmfleft{q}\fmfright{qbar}
\fmflabel{$a$}{q}
\fmflabel{$b$}{qbar}
\fmf{fermion}{q,qbar}
\end{fmfchar*}}&\quad=\delta^a_b
\displaybreak[1]\\
\parbox[c][\height+1.5\baselineskip][c]{25mm}{%nl
\begin{fmfchar*}(20,2)
\fmfleft{q}\fmfright{qbar}
\fmf{dots}{q,qbar}
\end{fmfchar*}}&\quad=1
\displaybreak[1]\\
\parbox[c][\height+1.5\baselineskip][c]{25mm}{%nl
\begin{fmfchar*}(20,20)
\fmfleft{q}\fmfright{qbar}\fmftop{g}
\fmflabel{$a$}{q}
\fmflabel{$b$}{qbar}
\fmflabel{$C$}{g}
\fmfdot{T}
\fmf{fermion}{q,T,qbar}
\fmffreeze
\fmf{dashes}{g,T}
\end{fmfchar*}}&\quad=t_{ba}^C
\displaybreak[1]\\
\parbox[c][\height+1.5\baselineskip][c]{25mm}{%nl
\begin{fmfchar*}(20,20)
\fmfsurround{gA,gC,gB}
\fmflabel{$A$}{gA}
\fmflabel{$B$}{gB}
\fmflabel{$C$}{gC}
\fmfdot{f}
\fmf{dashes}{gA,f}
\fmf{dashes}{gB,f}
\fmf{dashes}{gC,f}
\end{fmfchar*}}&\quad=if^{ABC}
\end{align}
\end{subequations}
In diagrammatic form the defining equation of the \person{Lie} algebra reads
\begin{equation} \label{eq:qcd-color:LIEalgebra}
\parbox[c][\height+1.5\baselineskip][c]{25mm}{%nl
\begin{fmfchar*}(25,20)
\fmfright{q}\fmfleft{qbar}\fmftop{n1,g2,g1,n2}
\fmfdot{T1}\fmfdot{T2}
\fmf{fermion}{q,T1,T2,qbar}
\fmffreeze
\fmf{dashes}{g1,T1}\fmf{dashes}{g2,T2}
\end{fmfchar*}}
\quad-\quad
\parbox[c][\height+1.5\baselineskip][c]{25mm}{%nl
\begin{fmfchar*}(25,20)
\fmfright{q}\fmfleft{qbar}\fmftop{n1,g2,g1,n2}
\fmfdot{T1}\fmfdot{T2}
\fmf{fermion}{q,T1,T2,qbar}
\fmffreeze
\fmf{dashes}{g2,T1}\fmf{dashes}{g1,T2}
\end{fmfchar*}}
\quad=\quad
\parbox[c][\height+1.5\baselineskip][c]{25mm}{%nl
\begin{fmfchar*}(25,20)
\fmfright{q}\fmfleft{qbar}\fmftop{n1,g1,g2,n2}
\fmfdot{f}\fmfdot{T}
\fmf{fermion}{q,T,qbar}
\fmffreeze
\fmf{dashes}{g1,f,g2}\fmf{dashes}{f,T}
\end{fmfchar*}}
\end{equation}
Up to now no specific gauge group has been chosen. The only ingredient that depends on
the gauge group is the completeness relation stating that the identity can be written as
a sum of projections into all different irreducible representations.
For a tensor product of the fundamental representation with its conjugate
of \ac{sun} this sum simplifies
to two terms, a projection on the adjoint and a projection onto the trivial representation:
\begin{equation}\label{eq:qcd-color:completeness}
\parbox[c][\height+1.5\baselineskip][c]{20mm}{%nl
\begin{fmfchar*}(20,15)
\fmftop{p1,q1}\fmfbottom{q2,p2}
\fmf{fermion}{p1,q1}
\fmf{fermion}{p2,q2}
\end{fmfchar*}}
\quad=
\frac{
\parbox[c][\height+0.3\baselineskip][c]{10mm}{%nl
\begin{fmfchar*}(10,10)
\fmfleft{p}\fmfright{q}
\fmf{dashes,left}{p,q,p}
\end{fmfchar*}}
}{
\parbox[c][\height+0.3\baselineskip][c]{10mm}{%nl
\begin{fmfchar*}(10,10)
\fmfleft{p}\fmfright{q}
\fmfdot{p,q}
\fmf{fermion,left}{p,q,p}
\fmf{dashes}{p,q}
\end{fmfchar*}}
}\quad
\parbox[c][\height+1.5\baselineskip][c]{25mm}{%nl
\begin{fmfchar*}(20,15)
\fmftop{p1,q1}\fmfbottom{q2,p2}
\fmfdot{T1,T2}
\fmf{fermion}{p1,T1,q2}
\fmf{fermion}{p2,T2,q1}
\fmf{dashes}{T1,T2}
\end{fmfchar*}}
+
\frac{
\parbox[c][\height+0.3\baselineskip][c]{10mm}{%nl
\begin{fmfchar*}(10,10)
\fmfleft{p}\fmfright{q}
\fmf{dots,left}{p,q,p}
\end{fmfchar*}}
}{
\parbox[c][\height+0.3\baselineskip][c]{10mm}{%nl
\begin{fmfchar*}(10,10)
\fmfleft{p}\fmfright{q}
\fmfdot{p,q}
\fmf{fermion,left}{p,q,p}
\fmf{dots}{p,q}
\end{fmfchar*}}
}\quad
\parbox[c][\height+1.5\baselineskip][c]{25mm}{%nl
\begin{fmfchar*}(20,15)
\fmftop{p1,q1}\fmfbottom{q2,p2}
\fmfdot{T1,T2}
\fmf{fermion}{p1,T1,q2}
\fmf{fermion}{p2,T2,q1}
\fmf{dots}{T1,T2}
\end{fmfchar*}}
\end{equation}
Diagrams with no external lines always represent scalar.
A circle without a vertex is just the trace over the corresponding
identity matrix and therefore the dimension of that representation.
Lines of the trivial representation can be omitted and therefore
the only unknown symbol is
\begin{equation}
\parbox[c][\height+0.3\baselineskip][c]{10mm}{%nl
\begin{fmfchar*}(10,10)
\fmfleft{p}\fmfright{q}
\fmfdot{p,q}
\fmf{fermion,left}{p,q,p}
\fmf{dashes}{p,q}
\end{fmfchar*}}\quad=\tr{T^AT^A}\equiv T_R\delta^{AA}\text{.}
\end{equation}

The quadratic \person{Casimir} operator~$T_R$ can be chosen as the normalisation
of the generators; following common conventions, I use~$T_R=1/2$.
In~$\sun[N_C]$, the dimensions of the representations
are~$\delta^{AA}=N_C^2-1$ and $\delta^a_a=N_C$.
The completeness relation~\eqref{eq:qcd-color:completeness}
can now be rearranged to
\begin{equation}\label{eq:qcd-color:reduce}
\parbox[c][\height+1.5\baselineskip][c]{20mm}{%nl
\begin{fmfchar*}(20,15)
\fmftop{p1,q1}\fmfbottom{q2,p2}
\fmfdot{T1,T2}
\fmf{fermion}{p1,T1,q2}
\fmf{fermion}{p2,T2,q1}
\fmf{dashes}{T1,T2}
\end{fmfchar*}}=T_R\left(
\parbox[c][\height+1.5\baselineskip][c]{25mm}{%nl
\begin{fmfchar*}(20,15)
\fmftop{p1,q1}\fmfbottom{q2,p2}
\fmf{fermion}{p1,q1}
\fmf{fermion}{p2,q2}
\end{fmfchar*}}
\quad
\quad-\quad
\frac{1}{N_C}\parbox[c][\height+1.5\baselineskip][c]{25mm}{%nl
\begin{fmfchar*}(20,15)
\fmftop{p1,q1}\fmfbottom{q2,p2}
\fmfdot{T1,T2}
\fmf{fermion}{p1,T1,q2}
\fmf{fermion}{p2,T2,q1}
\fmf{dots}{T1,T2}
\end{fmfchar*}}
\right)\text{.}
\end{equation}

With the above relations one already can reduce simple two-point functions
by using \person{Schur}'s lemma,
which allows the \person{Casimir} operators to be written 
in terms of multiples of the identity matrix:
\begin{align}
\label{eq:qcd-color:CFbubble}
\parbox[c][\height+1.5\baselineskip][c]{30mm}{%nl
\begin{fmfchar*}(30,15)
\fmfleft{p1}\fmfright{q1}
\fmfdot{T1,T2}
\fmf{fermion,tension=2}{p1,T1}
\fmf{fermion,tension=2}{T2,q1}
\fmf{fermion,left}{T1,T2}
\fmf{dashes,left}{T2,T1}
\end{fmfchar*}}&=
\frac{
\parbox[c][\height+0.3\baselineskip][c]{10mm}{%nl
\begin{fmfchar*}(10,10)
\fmfleft{p}\fmfright{q}
\fmfdot{p,q}
\fmf{fermion,left}{p,q,p}
\fmf{dashes}{p,q}
\end{fmfchar*}}
}{
\parbox[c][\height+0.3\baselineskip][c]{10mm}{%nl
\begin{fmfchar*}(10,10)
\fmfleft{p}\fmfright{q}
\fmf{fermion,left}{p,q,p}
\end{fmfchar*}}
}
\,\,\parbox[c][\height+1.5\baselineskip][c]{20mm}{%nl
\begin{fmfchar*}(20,2)
\fmfleft{q}\fmfright{qbar}
\fmf{fermion}{q,qbar}
\end{fmfchar*}}
\quad=
C_F\,\,
\parbox[c][\height+1.5\baselineskip][c]{20mm}{%nl
\begin{fmfchar*}(20,2)
\fmfleft{q}\fmfright{qbar}
\fmf{fermion}{q,qbar}
\end{fmfchar*}}
\\
\parbox[c][\height+1.5\baselineskip][c]{30mm}{%nl
\begin{fmfchar*}(30,15)
\fmfleft{p1}\fmfright{q1}
\fmfdot{T1,T2}
\fmf{dashes,tension=2}{p1,T1}
\fmf{dashes,tension=2}{T2,q1}
\fmf{fermion,left}{T1,T2}
\fmf{fermion,left}{T2,T1}
\end{fmfchar*}}&=
\frac{
\parbox[c][\height+0.3\baselineskip][c]{10mm}{%nl
\begin{fmfchar*}(10,10)
\fmfleft{p}\fmfright{q}
\fmfdot{p,q}
\fmf{fermion,left}{p,q,p}
\fmf{dashes}{p,q}
\end{fmfchar*}}
}{
\parbox[c][\height+0.3\baselineskip][c]{10mm}{%nl
\begin{fmfchar*}(10,10)
\fmfleft{p}\fmfright{q}
\fmf{dashes,left}{p,q,p}
\end{fmfchar*}}
}
\,\,\parbox[c][\height+1.5\baselineskip][c]{20mm}{%nl
\begin{fmfchar*}(20,2)
\fmfleft{q}\fmfright{qbar}
\fmf{dashes}{q,qbar}
\end{fmfchar*}}
\quad=
T_R\,\,
\parbox[c][\height+1.5\baselineskip][c]{20mm}{%nl
\begin{fmfchar*}(20,2)
\fmfleft{q}\fmfright{qbar}
\fmf{dashes}{q,qbar}
\end{fmfchar*}}
\\
\label{eq:qcd-color:ggbubble}
\parbox[c][\height+1.5\baselineskip][c]{30mm}{%nl
\begin{fmfchar*}(30,15)
\fmfleft{p1}\fmfright{q1}
\fmfdot{T1,T2}
\fmf{dashes,tension=2}{p1,T1}
\fmf{dashes,tension=2}{T2,q1}
\fmf{dashes,left}{T1,T2}
\fmf{dashes,left}{T2,T1}
\end{fmfchar*}}&=
\frac{
\parbox[c][\height+0.3\baselineskip][c]{10mm}{%nl
\begin{fmfchar*}(10,10)
\fmfleft{p}\fmfright{q}
\fmfdot{p,q}
\fmf{dashes,left}{p,q,p}
\fmf{dashes}{p,q}
\end{fmfchar*}}
}{
\parbox[c][\height+0.3\baselineskip][c]{10mm}{%nl
\begin{fmfchar*}(10,10)
\fmfleft{p}\fmfright{q}
\fmf{dashes,left}{p,q,p}
\end{fmfchar*}}
}
\,\,\parbox[c][\height+1.5\baselineskip][c]{20mm}{%nl
\begin{fmfchar*}(20,2)
\fmfleft{q}\fmfright{qbar}
\fmf{dashes}{q,qbar}
\end{fmfchar*}}
\quad=
C_A\,\,
\parbox[c][\height+1.5\baselineskip][c]{20mm}{%nl
\begin{fmfchar*}(20,2)
\fmfleft{q}\fmfright{qbar}
\fmf{dashes}{q,qbar}
\end{fmfchar*}}
\end{align}
The constant~$C_F$ can be read off directly, 
\begin{equation}
C_F=\frac{T_R\delta^{AA}}{\delta^a_a}=\frac{N_C^2-1}{2N_C}\text{,}
\end{equation}
whereas for $C_A$ another trick is needed.

One can obtain a relation to express the structure constants $f^{ABC}$ in terms of generators
in the fundamental representation by multiplying~\eqref{eq:qcd-color:LIEalgebra} with another generator,
\begin{equation} \label{eq:qcd-color:ftoTTT}
\parbox[c][\height+1.5\baselineskip][c]{20mm}{%nl
\begin{fmfchar*}(20,20)
\fmfsurround{A,C,B}
\fmfdot{x,y,z}
\fmf{fermion,right=0.5}{x,z,y,x}
\fmf{dashes,tension=3}{A,x}
\fmf{dashes,tension=3}{B,y}
\fmf{dashes,tension=3}{C,z}
\end{fmfchar*}}
\,\,-
\parbox[c][\height+1.5\baselineskip][c]{20mm}{%nl
\begin{fmfchar*}(20,20)
\fmfsurround{A,C,B}
\fmfdot{x,y,z}
\fmf{fermion,left=0.5}{x,y,z,x}
\fmf{dashes,tension=3}{A,x}
\fmf{dashes,tension=3}{B,y}
\fmf{dashes,tension=3}{C,z}
\end{fmfchar*}}
\,\,=
\parbox[c][\height+1.5\baselineskip][c]{30mm}{%nl
\begin{fmfchar*}(30,20)
\fmfsurround{A,C,B}
\fmfdot{x,y,z}
\fmf{fermion,left}{x,y,x}
\fmf{dashes,tension=3}{A,x}
\fmf{dashes,tension=3}{y,z}
\fmf{dashes,tension=3}{B,z}
\fmf{dashes,tension=3}{C,z}
\end{fmfchar*}}
\,\,=T_R
\parbox[c][\height+1.5\baselineskip][c]{20mm}{%nl
\begin{fmfchar*}(20,20)
\fmfsurround{A,C,B}
\fmfdot{x}
\fmf{dashes,tension=3}{A,x}
\fmf{dashes,tension=3}{B,x}
\fmf{dashes,tension=3}{C,x}
\end{fmfchar*}}
\end{equation}
Together with~\eqref{eq:qcd-color:reduce} and~\eqref{eq:qcd-color:CFbubble} we obtain one of the
so called \emph{star-triangle relations}:
\index{star-triangle relation}
\begin{equation} \label{eq:qcd-color:startriangle1}
\parbox[c][\height+1.5\baselineskip][c]{25mm}{%nl
\begin{fmfchar*}(25,20)
\fmftop{p,q}
\fmfbottom{g}
\fmfdot{T1,T2,f}
\fmf{fermion}{p,T1}
\fmf{fermion}{T1,T2}
\fmf{fermion}{T2,q}
\fmffreeze
\fmf{dashes}{T1,f,T2}
\fmf{dashes,tension=1.5}{g,f}
\end{fmfchar*}}
\,\,=T_RN_C
\parbox[c][\height+1.5\baselineskip][c]{20mm}{%nl
\begin{fmfchar*}(20,20)
\fmftop{p,q}
\fmfbottom{g}
\fmfdot{T}
\fmf{fermion}{p,T,q}
%\fmffreeze
\fmf{dashes,tension=1.5}{g,T}
\end{fmfchar*}}
\end{equation}
This result can be used to evaluate~$C_A$
replacing one structure constant using~\eqref{eq:qcd-color:ftoTTT}
and then applying the previous star triangle relation;
using that $f^{ABC}$ is antisymmetric one then finds
\begin{equation}
T_R\,\,\parbox[c][\height+0.3\baselineskip][c]{10mm}{%nl
\begin{fmfchar*}(10,10)
\fmfleft{p}\fmfright{q}
\fmfdot{p,q}
\fmf{dashes,left}{p,q,p}
\fmf{dashes}{p,q}
\end{fmfchar*}}
\,\,=
2T_RN_C\quad\parbox[c][\height+0.3\baselineskip][c]{15mm}{%nl
\begin{fmfchar*}(10,10)
\fmfsurround{T1,T2}
\fmfdot{T1,T2}
\fmf{fermion,left}{T1,T2,T1}
\fmf{dashes}{T1,T2}
\end{fmfchar*}}
\,\,=2N_C^2C_FT_R\text{,}
\end{equation}
and hence~$C_A=N_C$.
A second star triangle relation can be obtained in a similar way to this leading to
\begin{equation} \label{eq:qcd-color:startriangle2}
\parbox[c][\height+1.5\baselineskip][c]{25mm}{%nl
\begin{fmfchar*}(25,20)
\fmftop{p,q}
\fmfbottom{g}
\fmfdot{T1,T2,f}
\fmf{dashes}{p,T1}
\fmf{dashes}{T1,T2}
\fmf{dashes}{T2,q}
\fmffreeze
\fmf{dashes}{T1,f,T2}
\fmf{dashes,tension=1.5}{g,f}
\end{fmfchar*}}
\,\,=T_RN_C
\parbox[c][\height+1.5\baselineskip][c]{20mm}{%nl
\begin{fmfchar*}(20,20)
\fmftop{p,q}
\fmfbottom{g}
\fmfdot{T}
\fmf{dashes}{p,T,q}
%\fmffreeze
\fmf{dashes,tension=1.5}{g,T}
\end{fmfchar*}}
\end{equation}

\subsection{Colour Decomposition}
One way of organising \ac{qcd} amplitudes is to project onto a colour basis which separates
the amplitude into gauge invariant subamplitudes\footnote{See for
example~\cite{Dixon:1996wi}}. 
It is therefore convenient to choose a common set of indices
both for the quarks and the gluons.
To achieve this common form every external gluon with
the adjoint index~$A$ is multiplied by $(1/\sqrt{T_R})t^A_{ij}$ to satisfy and eliminate
all external adjoint indices. When the amplitude
is squared instead of carrying out a colour sum over 
the adjoint index~$A$ one has to run two colour sums
over~$i$ and~$j$. The advantage of this procedure
is that the colour structure of the amplitude now is formed
in terms of \person{Kronecker} deltas and hence the
amplitude splits into subamplitudes as follows
\begin{equation}
\mathcal{A}(q_{i_1}, \bar{q}^{j_1}, q_{i_2}, \bar{q}^{j_2}, \ldots, g_{i_m}^{j_m},\ldots,g_{i_n}^{j_n})=
\sum_{\sigma\in \Sym[n]}\delta^{j_1}_{i_{\sigma(1)}}\cdots\delta^{j_N}_{i_{\sigma(N)}}\mathcal{A}_\sigma\text{,}
\end{equation}
where~$q$ ($\bar{q}$) and~$g$ represent the colour structure introduced by
quarks and gluons, \Sym[n] is
the symmetric group\footnote{\Sym[n] is the group of permutations of~$n$ elements.}
and $\mathcal{A}_\sigma$ is the respective
subamplitude. Since all external adjoint indices have
been replaced by a pair of fundamental ones,
Algorithm~\ref{alg:qcd-color:reduction} ensures that all colour
structure is reduced to \person{Kronecker} deltas
and no contracted nor external adjoint indices are left over.
Statements {1--3} in Algorithm~\ref{alg:qcd-color:reduction}
are optional, but they
improve the performance for diagrams containing many gluon self couplings.
An algorithm similar to Algorithm~\ref{alg:qcd-color:reduction}
has been described in~\cite{Hakkinen:1996bb}.

\begin{algorithm}
\caption{Evaluation of the colour structure}\label{alg:qcd-color:reduction}
\begin{algorithmic}[1]
\STATE{Simplify repeating~\eqref{eq:qcd-color:startriangle2}}
\STATE{Simplify repeating~\eqref{eq:qcd-color:startriangle1}}
\STATE{Simplify repeating~\eqref{eq:qcd-color:ggbubble}}
\REPEAT
\STATE{Eliminate~$f^{ABC}$ using~\eqref{eq:qcd-color:ftoTTT}}
\STATE{Eliminate~$t^C_{ab}$ using~\eqref{eq:qcd-color:reduce}}
\UNTIL{no more replacements possible}
\end{algorithmic}
\end{algorithm}

\subsection{Other Colour Bases}
\label{ssec:qcd-color:other-bases}
In the approach I presented above, for an amplitude calculation the traditional
\ac{qcd} \person{Feynman} rules, as given in
Section~\ref{sec:qcd-lagrange} of Chapter~\ref{chp:qcd},
are used. Later the colour related
objects are translated into a graphical notation. External gluons are
multiplied by a generator of the fundamental $\sun[N_C]$ representation
to allow a unified treatment of quarks and gluons in the calculation of
colour factors. Going on step further, one can rewrite the \Lagrangeian{}
density of \ac{qcd} such that
\begin{equation}\label{eq:qcd-color:fundamental-basis1}
\delta^{AB}=\frac{1}{T_R}\tr{t^At^B}=
\left(\frac{1}{\sqrt{T_R}}(t^A)_i^j\right)%
\left(\frac{1}{\sqrt{T_R}}(t^B)_j^i\right)%
\end{equation}
and thus replacing all appearances of the gluon field by
\begin{equation}
\left(\frac{1}{\sqrt{T_R}}(t^A)_i^j\right)%
{\mathcal{A}}_\mu^A
\rightarrow(A_\mu)_i^j
\end{equation}
and treating the \Lagrangeian{} density with respect to
the new variable. This leads to a different representation of the
\person{Feynman} rules which are known as \emph{double-line
notation}~\cite{'tHooft:1973jz,Maltoni:2002mq}. 
\index{double-line notation}
Similarly, one could
also introduce a double-line notation for the \person{Lorentz} part
of the amplitude by replacing
\begin{equation}
g^{\mu\nu}=\frac12\sigma^{\mu}_{\alpha\dot\alpha}%
{\bar\sigma}^{\nu,\dot\alpha\alpha}\text{,}
\end{equation}
which replaces all \person{Lorentz} indices by a pair of
\person{Weyl} spinor indices~\cite{Weinzierl:2005dd}.
\index{Weyl-van der Waerden representation@\person{Weyl}-\person{van der Waerden} representation}
This representation of the spinorial indices is known as the
\person{Weyl}-\person{van der Waerden} representation and is
discussed in detail in Section~\ref{ssec:qcd-shelpmethod:WvdW}.

One of the disadvantages of this approach is the fact that one does not
generate a true basis of the colour space but introduces spurious vectors:
permutations containing a $\delta_j^i$ acting on a gluonic leg $(t^A)_i^j$
will project out the trace $\tr{t^A}=0$ and hence should be removed
explicitly. From counting the number of possibilities of tracing single
generators $t^A$ one finds that such a basis has
\begin{equation}\label{eq:qcd-color:fundamental-basis3}
\sum_{i=0}^G(-1)^i{G\choose i}(2Q+G-i)!
\end{equation}
elements, when $G$ is the number of external gluons and $Q$ is the number
of external quarks.
For amplitudes with many external gluons this removal of zero-vectors
can be cumbersome.
This is one of the reasons why for these amplitudes usually
colour ordering is considered a better solution;
it is described, for example, in~\cite{Dixon:1996wi}.
Here the fact is used that purely gluonic amplitudes
can be decomposed into the form
\begin{multline}
{\mathcal A}(g_{A_1},g_{A_2},\ldots,g_{A_n})=\\
\sum_{\scriptsize\begin{array}{c}{\sigma}
\vec{\lambda}\vdash n\\
\lambda_i\ge2\end{array}}%
\tr{t^{A_{\sigma(1)}}t^{A_{\sigma(2)}}\cdots t^{A_{\sigma(\lambda_1)}}}
\cdots
\tr{t^{A_{\sigma(n-\lambda_p+1)}}\cdots t^{A_{\sigma(n-1)}}t^{A_{\sigma(n)}}}
{\mathcal A}_{\sigma,\vec\lambda}\text{,}
\end{multline}
where only permutations~$\sigma\in \Sym[n]$
leading to distinguishable terms
for the traces\footnote{i.e. all cyclic permutations of the elements
of one trace and all permutations that reorder traces of the same length
are factored out}
are summed over.
The condition $\vec\lambda\vdash n$ denotes that the sum over $\vec\lambda$
traverses all integer partitions of~$n$.

This construction of a basis can easily be extended to the mixed case of
quarks and gluons and can be understood diagrammatically through
a theory defined by the following \person{Feynman} rules
\begin{displaymath}
\parbox[c][\height+1.5\baselineskip][c]{20mm}{%nl
\begin{fmfchar*}(15,15)
\fmfleft{p}\fmfright{q}
\fmftop{g}\fmfdot{T1}
\fmf{fermion}{p,T1,q}
\fmffreeze
\fmf{gluon}{T1,g}
\fmflabel{i}{q}
\fmflabel{j}{p}
\fmflabel{A}{g}
\end{fmfchar*}}
=(t^A)_i^j\quad\text{and}\qquad
\parbox[c][\height+1.5\baselineskip][c]{20mm}{%nl
\begin{fmfchar*}(15,15)
\fmfleft{p}\fmfright{q}
\fmf{fermion}{p,q}
\fmflabel{i}{q}
\fmflabel{j}{p}
\end{fmfchar*}}
=\delta_i^j\text{.}
\end{displaymath}
One then has to create all possibly disconnected diagrams removing tadpoles
and empty traces \parbox[c]{6mm}{
\begin{fmfchar*}(5,5)
\fmfleft{p}\fmfright{q}
\fmf{fermion,right=1}{p,q}
\fmf{plain,right=1}{q,p}
\fmflabel{\phantom{x}}{q}
\end{fmfchar*}}, because they lead to additional linearly dependent vectors.

As long as one works with a general number of colours $N_C$ the
dimension $d_{g;f}$ of such a basis for $g$ gluons and $f$ quark pairs
can be derived from considering the number of possibilities of inserting
an additional gluon; this is equivalent to inserting an additional quark
pair and removing the singlet combination,
\begin{equation}
d_{g+1;f}=d_{g;f+1}-d_{g;f}\text{.}
\end{equation}
In the purely fermionic case one has the $d_{0;f}=f!$ permutations of
the fermion lines. One can prove by direct calculation that
\begin{equation}\label{eq:qcd-color:d-basis}
d_{g;f}=\sum_{i=0}^g(-1)^i{g\choose i}(g+f-i)!\text{.}
\end{equation}
fulfils the two properties.
The closed form, however, suggests the equivalence with the basis
described in Equations \eqref{eq:qcd-color:fundamental-basis1}
to~\eqref{eq:qcd-color:fundamental-basis3}.

Further reductions of this basis can be achieved by considering
the irreducible representations of the symmetric group permuting
the \(g+f\) lines in the fundamental representation. If one fixes
$N_C$ it is clear that no antisymmetrisation over more than $N_C$
lines is possible and hence those combinations of permutations
have to vanish. A systematic treatment of the symmetric group is
given in Section~\ref{sec:irreps}. Table~\ref{tab:qcd-color:basecomp}
shows the number of basis\footnote{The term \emph{basis} should not
be taken in the literal mathematical sense; the spirit of this section
is to show that depending on different assumptions there are additional
relations that render some of the vectors linearly dependent.}
elements in the different representations for a number of different
processes. The colour-basis of the amplitude
is usually smaller than the ones presented because the \person{Bose}
symmetry of the amplitude allows the application of
further symmetrisation of the colour vectors.
However, helicity projections destroy some of the symmetries
of the amplitude and hence one would have to work out a different basis for
each helicity projection. This is important if one tries to achieve a compact
analytical result; the trade-off in a numerical calculation is debatable.

\begin{table}[hp]
\begin{tabular}{l|rrr}
process & $(g+f)!$ & $d_{g;f}$ & $\bar{d}_{g;f}(3)$ \\
\hline % 2 -> 1
$q\bar{q}\rightarrow g$ & 2 & 1 & 1 \\
$gg\rightarrow g$ & 6 & 2 & 2 \\
\hline % 2 -> 2
$q\bar{q}\rightarrow q\bar{q}$ & 2 & 2 & 2 \\
$gg\rightarrow q\bar{q}$ & 6 & 3 & 3 \\
$gg\rightarrow gg$ & 24 & 9 & 8 \\
\hline % 2 -> 3
$q\bar{q}\rightarrow gq\bar{q}$ & 6 & 4 & 4 \\
$gg\rightarrow gq\bar{q}$ & 24 & 11 & 10 \\
$gg\rightarrow ggg$ & 120 & 44 & 32 \\
\hline % 2 -> 4
$q\bar{q}\rightarrow q\bar{q}q\bar{q}$ & 6 & 6 & 6 \\
$gg\rightarrow q\bar{q}q\bar{q}$ & 24 & 14 & 13 \\
$gg\rightarrow ggq\bar{q}$ & 120 & 53 & 40 \\
$gg\rightarrow gggg$ & 720 & 265 & 145 \\
\hline % 2 -> 5
$q\bar{q}\rightarrow gq\bar{q}q\bar{q}$ & 24 & 18 & 17 \\
$gg\rightarrow gq\bar{q}q\bar{q}$ & 120 & 64 & 50 \\
$gg\rightarrow gggq\bar{q}$ & 720 & 309 & 177 \\
$gg\rightarrow ggggg$ & 5040 & 1854 & 702 \\
\hline % 2 -> 6
$q\bar{q}\rightarrow q\bar{q}q\bar{q}q\bar{q}$ & 24 & 24 & 23 \\
$gg\rightarrow q\bar{q}q\bar{q}q\bar{q}$ & 120 & 78 & 63 \\
$gg\rightarrow ggq\bar{q}q\bar{q}$ & 720 & 362 & 217 \\
$gg\rightarrow ggggq\bar{q}$ & 5040 & 2119 & 847 \\
$gg\rightarrow gggggg$ & 40320 & 14833 & 3598 \\
\end{tabular}
\caption[Comparison of different colour representations]{For the different
processes the number of basis elements in different representations of the
colour structure is compared: \((g+f)!\) is the number of vectors obtained
when all gluons are projected onto a quark-antiquark pair. $d_{g;f}$
is the number of vectors after subtracting all spurious vectors.
$\bar{d}_{g;f}(3)$ counts the vectors if the $\Sym[g+f]$ structure
considering $N_C=3$ is taken into account and one works with irreducible
representations only.}
\label{tab:qcd-color:basecomp}

\end{table}
\input{qcd-color-recoupling}

%%%%%%%%%%%%%%%%%%%%%%%%% KEEP THIS LAST %%%%%%%%%%%%%%%%%%%%%%%%%%%%%
\end{fmffile}

%% file: qcd-color-recoupling.tex
\subsection{Recoupling Relations}
\label{ssec:qcd-color-recoupling:recoupling}
In this section I describe recoupling relations for \ac{qcd},
which are useful for a quick reduction of large loops.
The reader be reminded of two basic relations which have been introduced
earlier for the special case of the fundamental and the adjoint representation.
As a direct consequence of \person{Schur}'s lemma any two-point birdtrack
diagram must connect two lines of the same irreducible representation
and hence be proportional to a \person{Kronecker} delta,
\begin{equation}
\label{eq:qcd-color-recoupling:CFbubble}
\parbox[c][\height+1.5\baselineskip][c]{30mm}{%nl
\begin{fmfchar*}(30,15)
\fmfleft{p1}\fmfright{q1}
\fmfdot{T1,T2}
\fmf{fermion,tension=2,label=$\lambda$}{p1,T1}
\fmf{fermion,tension=2,label=$\lambda^\prime$}{T2,q1}
\fmf{fermion,left,label=$\mu$}{T1,T2}
\fmf{fermion,left,label=$\nu$}{T2,T1}
\end{fmfchar*}}=
\frac{
\parbox[c][\height+1.5\baselineskip][c]{10mm}{%nl
\begin{fmfchar*}(10,10)
\fmfleft{p}\fmfright{q}
\fmfdot{p,q}
\fmf{fermion,left,label=$\mu;\nu;\lambda$}{p,q}
\fmf{fermion,left}{q,p}
\fmf{fermion}{q,p}
\end{fmfchar*}}
}{
\parbox[c][\height+0.3\baselineskip][c]{7mm}{%nl
\begin{fmfchar*}(7,7)
\fmfleft{p}\fmfright{q}
\fmf{fermion,left}{q,p}
\fmf{plain,left,label=$\lambda$,label.side=right}{p,q}
\end{fmfchar*}}
}
\,\,\parbox[c][\height+1.5\baselineskip][c]{20mm}{%nl
\begin{fmfchar*}(20,2)
\fmfleft{q}\fmfright{qbar}
\fmf{fermion,label=$\lambda$}{q,qbar}
\end{fmfchar*}}\,\,\delta_{\lambda\lambda^\prime}\text{.}
\end{equation}

A second identity is the completeness relation,
\index{completeness relation}
\begin{equation}\label{eq:qcd-color-recoupling:completeness}
\parbox[c][\height+1.5\baselineskip][c]{20mm}{%nl
\begin{fmfchar*}(20,15)
\fmftop{p1,q1}\fmfbottom{p2,q2}
\fmf{fermion,label=$\mu$}{p1,q1}
\fmf{fermion,label=$\nu$}{p2,q2}
\end{fmfchar*}}
\quad=\sum_{\lambda}
\frac{
\parbox[c][\height+0.3\baselineskip][c]{7mm}{%nl
\begin{fmfchar*}(7,7)
\fmfleft{p}\fmfright{q}
\fmf{fermion,left}{p,q}
\fmf{plain,left,label.side=right,label=$\lambda$}{q,p}
\end{fmfchar*}}
}{
\parbox[c][\height+0.7\baselineskip][c]{10mm}{%nl
\begin{fmfchar*}(10,10)
\fmfleft{p}\fmfright{q}
\fmfdot{p,q}
\fmf{fermion,left}{p,q}
\fmf{fermion,left,label=$\mu;\nu;\lambda$,label.dist=0.4}{q,p}
\fmf{fermion}{p,q}
\end{fmfchar*}}
}\quad
\parbox[c][\height+1.5\baselineskip][c]{25mm}{%nl
\begin{fmfchar*}(20,15)
\fmftop{p1,q1}\fmfbottom{q2,p2}
\fmfdot{T1,T2}
\fmf{fermion,label=$\mu$,label.side=left}{p1,T1}
\fmf{fermion,label=$\nu$,label.side=right}{q2,T1}
\fmf{fermion,label=$\nu$}{T2,p2}
\fmf{fermion,label=$\mu$}{T2,q1}
\fmf{fermion,label=$\lambda$}{T1,T2}
\end{fmfchar*}}\text{,}
\end{equation}
where the sum runs over all irreducible representations of the underlying
\person{Lie} algebra with non-vanishing coupling to the representations
$\mu$ and~$\nu$.

The starting point for the derivation of the recoupling relation
is a four point tree graph with arbitrary, irreducible
representations~\cite{Cvitanovic}, as shown
in~\eqref{eq:qcd-colour-recoupling:tchannel};
Equation~\eqref{eq:qcd-color-recoupling:completeness} is used twice
and the sum over $\lambda^\prime$ is evaluated using
Equation~\eqref{eq:qcd-color-recoupling:CFbubble}.
\begin{equation}\label{eq:qcd-colour-recoupling:tchannel}
\parbox[c][\height+1.5\baselineskip][c]{20mm}{%nl
\begin{fmfchar*}(20,20)
\fmftop{p1,q1}\fmfbottom{p2,q2}
\fmfdot{v1,v2}
\fmf{fermion,label=$\mu$}{p1,v1}
\fmf{fermion,label=$\nu$}{v1,q1}
\fmf{fermion,label=$\rho$}{p2,v2}
\fmf{fermion,label=$\sigma$}{v2,q2}
\fmf{fermion,label=$\omega$}{v1,v2}
\end{fmfchar*}}
\end{equation}
\index{recoupling relation}
Both sides can then be multiplied by a~$\One$
expressed through the completeness relation. To obtain the
final formula \person{Schur}'s lemma is applied to the propagator:
\begin{multline}\label{eq:qcd-color-recoupling:recoupling}
\parbox[c][\height+1.5\baselineskip][c]{20mm}{%nl
\begin{fmfchar*}(20,20)
\fmftop{p1,q1}\fmfbottom{p2,q2}
\fmfdot{v1,v2}
\fmf{fermion,label=$\mu$}{p1,v1}
\fmf{fermion,label=$\nu$}{v1,q1}
\fmf{fermion,label=$\rho$}{p2,v2}
\fmf{fermion,label=$\sigma$}{v2,q2}
\fmf{fermion,label=$\omega$}{v1,v2}
\end{fmfchar*}}
\,\,=\sum_{\lambda}\sum_{\lambda^\prime}
\frac{
\parbox[c][\height+0.3\baselineskip][c]{7mm}{%nl
\begin{fmfchar*}(7,7)
\fmfleft{p}\fmfright{q}
\fmf{fermion,left,label=$\lambda$}{p,q}
\fmf{plain,left}{q,p}
\end{fmfchar*}}
}{
\parbox[c][\height+0.3\baselineskip][c]{10mm}{%nl
\begin{fmfchar*}(10,10)
\fmfleft{p}\fmfright{q}
\fmfdot{p,q}
\fmf{fermion,left}{p,q}
\fmf{fermion,left,label=$\nu;\sigma;\lambda$}{q,p}
\fmf{fermion}{p,q}
\end{fmfchar*}}
}\,\,
\frac{
\parbox[c][\height+0.3\baselineskip][c]{7mm}{%nl
\begin{fmfchar*}(7,7)
\fmfleft{p}\fmfright{q}
\fmf{fermion,left,label=$\lambda^\prime$}{p,q}
\fmf{plain,left}{q,p}
\end{fmfchar*}}
}{
\parbox[c][\height+0.3\baselineskip][c]{10mm}{%nl
\begin{fmfchar*}(10,10)
\fmfleft{p}\fmfright{q}
\fmfdot{p,q}
\fmf{fermion,left}{p,q}
\fmf{fermion,left,label=$\mu;\rho;\lambda^\prime$}{q,p}
\fmf{fermion}{p,q}
\end{fmfchar*}}
}\quad
\parbox[c][\height+1.5\baselineskip][c]{40mm}{%nl
\begin{fmfchar*}(40,20)
\fmftop{p1,q1}\fmfbottom{p2,q2}
\fmftop{top}\fmfbottom{bottom}
\fmfdot{v1,v2,v3,v4,v5,v6}
\fmf{fermion,label=$\mu$}{p1,v3}
\fmf{fermion,label=$\nu$}{v6,q1}
\fmf{fermion,label=$\rho$}{p2,v3}
\fmf{fermion,label=$\sigma$}{v6,q2}
\fmf{fermion,label=$\lambda$,tension=3}{v3,v4}
\fmf{fermion,label=$\lambda^\prime$,tension=3}{v5,v6}
\fmf{fermion,label=$\nu$,left=0.4}{v1,v5}
\fmf{fermion,label=$\sigma$,left=0.4}{v5,v2}
\fmf{fermion,label=$\rho$,left=0.4}{v2,v4}
\fmf{fermion,label=$\mu$,left=0.4}{v4,v1}
\fmf{phantom,tension=3}{top,v1}
\fmf{phantom,tension=3}{bottom,v2}
\fmf{fermion,label=$\omega$}{v1,v2}
\end{fmfchar*}}\displaybreak[1]\\[10pt]
\Longrightarrow\quad
\parbox[c][\height+1.5\baselineskip][c]{20mm}{%nl
\begin{fmfchar*}(20,20)
\fmftop{p1,q1}\fmfbottom{p2,q2}
\fmfdot{v1,v2}
\fmf{fermion,label=$\mu$}{p1,v1}
\fmf{fermion,label=$\nu$}{v1,q1}
\fmf{fermion,label=$\rho$}{p2,v2}
\fmf{fermion,label=$\sigma$}{v2,q2}
\fmf{fermion,label=$\omega$}{v1,v2}
\end{fmfchar*}}
=\,
\sum_\lambda
\frac{
\parbox[c][\height+1.5\baselineskip][c]{15mm}{%nl
\begin{fmfchar*}(15,15)
\fmfsurround{v1,v2,v3}
\fmfdot{v1,v2,v3,z}
\fmf{fermion,right=0.5,label=$\lambda$}{v1,v2}
\fmf{fermion,right=0.5,label=$\mu$}{v2,v3}
\fmf{fermion,right=0.5,label=$\nu$}{v3,v1}
\fmf{fermion,label=$\sigma$,label.dist=1.3}{z,v1}
\fmf{fermion,label=$\rho$,label.dist=0.9}{v2,z}
\fmf{fermion,label=$\omega$,label.dist=0.9}{v3,z}
\end{fmfchar*}}\,\,\,
\parbox[c][\height+1.5\baselineskip][c]{7mm}{%nl
\begin{fmfchar*}(7,7)
\fmfsurround{p,q}
\fmf{fermion,right,label=$\lambda$}{p,q}
\fmf{plain,right}{q,p}
\end{fmfchar*}}
}{
\parbox[c][\height+0.3\baselineskip][c]{10mm}{%nl
\begin{fmfchar*}(10,10)
\fmfleft{p}\fmfright{q}
\fmfdot{p,q}
\fmf{fermion,left}{p,q}
\fmf{fermion,left,label=$\nu;\sigma;\lambda$}{q,p}
\fmf{fermion}{p,q}
\end{fmfchar*}}\quad
\parbox[c][\height+0.3\baselineskip][c]{10mm}{%nl
\begin{fmfchar*}(10,10)
\fmfleft{p}\fmfright{q}
\fmfdot{p,q}
\fmf{fermion,left}{p,q}
\fmf{fermion,left,label=$\mu;\rho;\lambda$}{q,p}
\fmf{fermion}{p,q}
\end{fmfchar*}}
}\quad
\parbox[c][\height+1.5\baselineskip][c]{30mm}{%nl
\begin{fmfchar*}(30,20)
\fmftop{p1,q1}\fmfbottom{p2,q2}
\fmfdot{v3,v6}
\fmf{fermion,label=$\mu$}{p1,v3}
\fmf{fermion,label=$\nu$}{v6,q1}
\fmf{fermion,label=$\rho$}{p2,v3}
\fmf{fermion,label=$\sigma$}{v6,q2}
\fmf{fermion,label=$\lambda$}{v3,v6}
\end{fmfchar*}}
\end{multline}
In literature the coefficients of this relation are referred as \nj{3} and \nj{6} symbols or Wigner coefficients.
\index{six-j symbol@\nj{\text{six}} symbol}
\index{three-j symbol@\nj{\text{three}} symbol}

The \nj{3} symbols already appeared earlier in the \ac{qcd} self-energy graphs,
and together with the \nj{6} symbols one can derive the general version of the
star-triangle relations which were introduced in the special case of quarks
and gluons. Starting from a general triangle \acs{sun}-graph we can
recouple one of its propagators, which reduces the triangle to a
self-energy graph; the latter can be eliminated by \person{Schur}'s lemma
and hence one~gets
\index{star-triangle relation|main}
\begin{equation}\label{eq:qcd-color-recoupling:startriangle}
\parbox[c][\height+1.5\baselineskip][c]{25mm}{%nl
%the triangle:
\begin{fmfchar*}(25,20)
\fmftop{p1,p2}
\fmfbottom{p3}
\fmfdot{T1,T2,T3}
\fmf{fermion,label=$\mu$,label.side=left}{p1,T1}
\fmf{fermion,label=$\nu$}{p2,T2}
\fmf{fermion,label=$\gamma$,label.side=left}{T1,T2}
\fmffreeze
\fmf{fermion,label=$\beta$}{T1,T3}
\fmf{fermion,label=$\alpha$,label.side=left}{T2,T3}
\fmf{fermion,tension=1.5,label=$\rho$}{T3,p3}
\end{fmfchar*}}=
\frac{%the \nj{6}
\parbox[c][\height+1.5\baselineskip][c]{15mm}{%nl
\begin{fmfchar*}(15,15)
\fmfsurround{v1,v2,v3}
\fmfdot{v1,v2,v3,z}
\fmf{fermion,right=0.5,label=$\mu$}{v1,v2}
\fmf{fermion,right=0.5,label=$\beta$}{v2,v3}
\fmf{fermion,right=0.5,label=$\rho$}{v3,v1}
\fmf{fermion,label=$\nu$,label.dist=0.4}{v1,z}
\fmf{fermion,label=$\gamma$,label.dist=0.4,label.side=right}{v2,z}
\fmf{fermion,label=$\alpha$,label.dist=0.4}{z,v3}
\end{fmfchar*}}
}{%the \nj{3}
\parbox[c][\height+0.3\baselineskip][c]{10mm}{%nl
\begin{fmfchar*}(10,10)
\fmfleft{p}\fmfright{q}
\fmfdot{p,q}
\fmf{fermion,left}{p,q}
\fmf{fermion,left,label=$\mu;\nu;\rho$}{q,p}
\fmf{fermion}{p,q}
\end{fmfchar*}}\quad
}\quad
%the tree:
\parbox[c][\height+1.5\baselineskip][c]{25mm}{%nl
\begin{fmfchar*}(25,20)
\fmftop{p1,p2}
\fmfbottom{p3}
\fmfdot{T}
\fmf{fermion,label=$\mu$,label.side=left}{p1,T}
\fmf{fermion,label=$\nu$}{p2,T}
%\fmffreeze
\fmf{fermion,label=$\rho$}{T,p3}
\end{fmfchar*}}
\end{equation}

The recoupling relation provides a useful check for the calculation of
\nj{6} coefficients when applied to two non-adjacent lines of a \nj{6} symbol:
\begin{equation}
% The sixj on the left
\parbox[c][\height+1.5\baselineskip][c]{15mm}{%nl
\begin{fmfchar*}(15,15)
\fmfsurround{v1,v2,v3}
\fmfdot{v1,v2,v3,z}
\fmf{fermion,right=0.5,label=$\mu$}{v1,v2}
\fmf{fermion,right=0.5,label=$\beta$}{v2,v3}
\fmf{fermion,right=0.5,label=$\rho$}{v3,v1}
\fmf{fermion,label=$\nu$,label.dist=1.0}{v1,z}
\fmf{fermion,label=$\gamma$,label.dist=1.0,label.side=right}{v2,z}
\fmf{fermion,label=$\alpha$,label.dist=1.0}{z,v3}
\end{fmfchar*}}
=\sum_\lambda
\frac{
% the dimension
\parbox[c][\height+0.3\baselineskip][c]{7mm}{%nl
\begin{fmfchar*}(7,7)
\fmfleft{p}\fmfright{q}
\fmfdot{p,q}
\fmf{fermion,right,label=$\lambda$}{q,p}
\fmf{plain,right}{p,q}
\end{fmfchar*}}
}{
%the first \nj{3}
\parbox[c][\height+0.3\baselineskip][c]{10mm}{%nl
\begin{fmfchar*}(10,10)
\fmfleft{p}\fmfright{q}
\fmfdot{p,q}
\fmf{fermion,left}{p,q}
\fmf{fermion,left,label=$\gamma;\rho;\lambda$}{q,p}
\fmf{fermion}{p,q}
\end{fmfchar*}}
\quad
%the second \nj{3}
\parbox[c][\height+0.3\baselineskip][c]{10mm}{%nl
\begin{fmfchar*}(10,10)
\fmfleft{p}\fmfright{q}
\fmfdot{p,q}
\fmf{fermion,left}{p,q}
\fmf{fermion,left,label=$\beta;\nu;\lambda$}{q,p}
\fmf{fermion}{p,q}
\end{fmfchar*}}
}
\quad
%the first \nj{6}
\parbox[c][\height+1.5\baselineskip][c]{15mm}{%nl
\begin{fmfchar*}(15,15)
\fmfsurround{v1,v2,v3}
\fmfdot{v1,v2,v3,z}
\fmf{fermion,right=0.5,label=$\lambda$}{v1,v2}
\fmf{fermion,right=0.5,label=$\beta$}{v2,v3}
\fmf{fermion,right=0.5,label=$\rho$}{v3,v1}
\fmf{fermion,label=$\gamma$,label.dist=1.0}{v1,z}
\fmf{fermion,label=$\nu$,label.dist=1.0,label.side=right}{v2,z}
\fmf{fermion,label=$\alpha$,label.dist=1.0}{z,v3}
\end{fmfchar*}}
\qquad
%the second \nj{6}
\parbox[c][\height+1.5\baselineskip][c]{15mm}{%nl
\begin{fmfchar*}(15,15)
\fmfsurround{v1,v2,v3}
\fmfdot{v1,v2,v3,z}
\fmf{fermion,right=0.5,label=$\mu$}{v1,v2}
\fmf{fermion,right=0.5,label.dist=1.0,label=$\gamma$}{v2,v3}
\fmf{fermion,right=0.5,label=$\rho$}{v3,v1}
\fmf{fermion,label=$\nu$,label.dist=1.0}{v1,z}
\fmf{fermion,label=$\beta$,label.dist=1.0,label.side=right}{v2,z}
\fmf{fermion,label=$\lambda$,label.dist=1.0}{z,v3}
\end{fmfchar*}}
\end{equation}

Another 
useful application of the recoupling relations is the systematic reduction
of large loops. The first step of such a reduction would be the projection on
a basis of tree graphs, which in the simplest case could just be formed by
\person{Kronecker} delta symbols if the external legs are in the same
representation. As a result one obtains tree graphs which have
graphs without external legs, so-called
bubble graphs, as coefficients. The bubbles evaluate to scalars
and therefore represent the group-theoretical factor of the underlying diagram.
The knowledge of all \nj{6} and \nj{3} symbols
is enough to evaluate any given bubble diagram\footnote{In fact the 
dimensions of the representations and the \nj{3} symbols are just special
cases of \nj{6} coefficients containing the fundamental representation
in two (resp. one) of the lines.}: Any loop of size two can be reduced
by \person{Schur}'s lemma~\eqref{eq:qcd-color-recoupling:CFbubble}
and cycles of size three are eliminated by the star-triangle relation. Larger
loops can always be split in half using the completeness
relation~\eqref{eq:qcd-color-recoupling:completeness}.
\index{completeness relation}

The statistical run time for the 
reduction of a loop of size~$n$ is ${\mathcal O}(n^{\log_2m})$,
where~$m$ is the average number of irreducible representation that
appear under the sum of the completeness relation\footnote{For~$m=1$ the
run time reduces to~${\mathcal O}(\log_2n)$.}. This is no improvement over
algorithm~\ref{alg:qcd-color:reduction} for the~\sun[3]\ colour factors
of~\ac{qcd}. However, there is no restriction to a special \person{Lie}
group and therefore this approach also works for the orthogonal group.
With few modifications one can include spinor representations and use the
same algorithm for the evaluation of spinor traces, which I will discuss
in Section~\ref{sec:qcddimreg} in Chapter~\ref{chp:qcdnlo}. 
Another advantage is that one is free
in the choice of the basis which the diagram is projected on;
algorithm~\ref{alg:qcd-color:reduction}, however, restricts itself to
the choice of a basis formed by \person{Kronecker} symbols.
As already mentioned above, this approach requires knowledge of
all \nj{3} and \nj{6} symbols that might appear during the computation.
Therefore
efficient methods to calculate these factors should be investigated.

%% file: qcd-irreps.tex
\begin{fmffile}{irreps}
\subsection*{Introduction}
In the previous section I have presented an algorithm for the reduction of
colour-tensors into an irreducible basis that relies on the knowledge of
the \nj1 (i.e. dimensions), \nj3 and \nj6 symbols of all appearing
irreducible representations of the \ac{sun}. In this section I discuss,
in a slightly more general context, how to obtain all irreducible
representations of the \ac{gln} and an algorithm for
the calculation of all relevant coefficients. One can then apply
restrictions to the representations such that the irreducible representations
of the \ac{gln} reduce further down to the ones of its subgroups. For \acs{qcd}
one is interested in the \ac{sun} only but the presented algorithm applies
to other
\person{Lie} groups as well once one knows how to construct the irreducible
representation of those groups.

It is well known that the irreducible representations of the \ac{gln}
are described by the irreducible representations of the symmetric
group~\acs{sym}, which are labelled by partitions or equivalently by Young
diagrams~\cite{Weyl:1939,Murnaghan38,Fulton97,Sagan91}.
In section~\ref{ssec:qcd-irreps:sym} I briefly sketch the
relevant properties of the symmetric group and establish a diagrammatic
notation for permutations and vectors of the module~$\Cset\Sym$ which allows
us to construct irreducible matrix representations of~\acs{sym}.
A matrix representation called \person{Young}'s Natural Representation
is introduced in~\ref{ssec:qcd-irreps:youngrep}.
In the following
section, \ref{ssec:qcd-irreps:nj} these representations are used
to calculate the \nj{n} symbols for the~\ac{sun}.
I conclude this section with the discussion of a complete
diagram-based algorithm for the reduction of \acs{gln}-tensors
in~\ref{ssec:qcd-irreps:reduce}.

\subsection{Diagrammatics for the Symmetric Group}
\label{ssec:qcd-irreps:sym}
\index{symmetric group}\index{.Sk@$\Sym$|see{symmetric group}}
The \acf{sym} describes the set of bijective maps
on a set with $k$ elements. The concatenation of two
elements $\sigma_1,\sigma_2\in\Sym$ with
$\sigma_{1,2}: x_j\mapsto x_{j^\prime}=\sigma_{1,2}(x_j)$
defines a multiplication
$\sigma_1\cdot\sigma_2: x_j\mapsto \sigma_1(\sigma_2(x_j))$.
Several notations for the elements of
\Sym[k] are used in the literature:
a permutation $\sigma: x_j\mapsto x_{j^\prime}=\sigma(x_j)$ can be
written in two rows as follows
\begin{equation}
\sigma=\begin{pmatrix}
x_1&x_2&\ldots&x_k\\
\sigma(x_1)&\sigma(x_2)&\ldots&\sigma(x_k)
\end{pmatrix}
\end{equation}
Let, for instance, $\sigma_1,\sigma_2\in\Sym[4]$ be
\begin{equation}
\sigma_1=\begin{pmatrix}
a_1&a_2&a_3&a_4\\
a_3&a_1&a_2&a_4
\end{pmatrix},\quad\sigma_2=
\begin{pmatrix}
a_1&a_2&a_3&a_4\\
a_4&a_2&a_3&a_1
\end{pmatrix}
\text{;}
\end{equation}
one can turn the rows into columns and connect the position of $a_j$
with that of its image $a_{j^\prime}=\sigma(a_j)$, i.e. if the elements
on the left are ordered according to the indices $j$, connect the
$j$-th row on the left with the $j^\prime$-th row on the right.
Multiplication is then carried out as denoted below;
where unambiguous labels can be omitted:
\begin{equation}
\sigma_1\cdot\sigma_2
= \plabel{{a_1}{a_2}{a_3}{a_4}}\perm{3124}\plabel{{a_2}{a_3}{a_1}{a_4}}
  \cdot
  \plabel{{a_2}{a_3}{a_1}{a_4}}\perm{1243}\plabel{{a_2}{a_3}{a_4}{a_1}}
= \plabel{{a_1}{a_2}{a_3}{a_4}}\perm{3124}\perm{1243}
  \plabel{{a_2}{a_3}{a_4}{a_1}}
=\perm{4123}
=\begin{pmatrix}
a_1&a_2&a_3&a_4\\
a_4&a_1&a_2&a_3
\end{pmatrix}
\end{equation}

Another notation is the so called cycle notation
\index{cycle notation (permutations)|main}%
\index{permutation!cycle notation|main}. This notation is useful
to exhibit the cycle structure of a permutation and hence for finding
the conjugacy classes. Since \Sym[n]\ is a finite group, for any permutation
$\sigma\in\Sym[n]$ and every $a_j$ there must be a positive integer $p\leq n$
with $\sigma^p(a_j)=a_j$. The cycle notation is obtained by writing down
all disjoint cycles $(a_j, \sigma(a_j), \ldots, \sigma^{p-1}(a_j))$; cycles
of length one can be omitted. For the above example one gets
\begin{subequations}
\begin{align}
\sigma_1&=(a_1,a_3,a_2)(a_4)=(a_1,a_3,a_4)\\
\sigma_2&=(a_1,a_4)(a_2)(a_3)=(a_1,a_4)\\
\sigma_1\sigma_2&=(a_1,a_2,a_3,a_4)
\end{align}
\end{subequations}
Graphically one can obtain this notation by closing the loops around the
permutation and reading of the labels as one follows every loop.
For example for $\sigma_1$ one has
\begin{displaymath}
\permtrace4{%
\plabel{{a_1}{a_2}{a_3}{a_4}}\perm{3124}}=
\permtrace1{\plabel{{a_1}}\pid1\plabel{{a_3}}\pid1\plabel{{a_2}}}
\,\permtrace1{\plabel{{a_4}}}\text{.}
\end{displaymath}
As a corollary of the character orthogonality (cf.~\cite{Sagan91})
the conjugacy classes of the symmetric group label its irreducible
representations. This fact is used extensively to obtain a complete
list of all irreducible representations, not only of $\Sym$ but also
of~\acs{gln}.

Whenever acting on a tensor, permutations denote a product of Kronecker
deltas acting on the indices of a tensor:
\begin{align}
(a_1,a_2)(a_3,a_4)T^{a_1a_2a_3a_4}=
\delta_{a_2^\prime}^{a_1}
\delta_{a_1^\prime}^{a_2}
\delta_{a_4^\prime}^{a_3}
\delta_{a_3^\prime}^{a_4}
T^{a_1^\prime a_2^\prime a_3^\prime a_4^\prime}
=T^{a_2a_1a_4a_3}
\end{align}

The conjugacy class\index{conjugacy class} of a group element
$\sigma\in\Sym$
is defined as the set
\begin{equation}
\left\{\pi\sigma\pi^{-1}\vert \pi\in\Sym\right\}
\end{equation}
One can see that the conjugacy classes of the symmetric group are
determined by the cycle structure of each group element, since
the extra pair of permutations $\pi$ and~$\pi^{-1}$ can always
be absorbed in a permutation of the labels $a_1,\ldots,a_k$
without changing the cycle structure of the permutation.
Using $\sigma_1$ and $\sigma_2$ defined above one obtains:
\begin{displaymath}
\permtrace4{\plabel{{a_1}{a_2}{a_3}{a_4}}\pid4%
\xperm{3124}\permid4\perm{1243}\permid4\perm{3124}}=
\permtrace4{\perm{3124}\pid4\plabel{{a_1}{a_2}{a_3}{a_4}}\pid4%
\xperm{3124}\permid4\perm{1243}}=
\permtrace4{\plabel{{a_3}{a_1}{a_2}{a_4}}\pid4%
\xperm{1243}}\,\text{,}
\end{displaymath}
and hence we have $\sigma_1\cdot(a_1,a_4)\cdot\sigma_1^{-1}=(a_2,a_4)$.
In the diagrammatic notation the decomposition into transpositions can
be found by moving the crossings of lines such that all crossings are
horizontally separated and therefore the signum of a permutation is
just determined by the number line crossings. The signum is a group
homomorphism and obeys multiplicativity,
$\sgn(\sigma\pi)=\sgn(\sigma)\sgn(\pi)$.

Each conjugacy class is determined by the cycle structure of its elements
which in turn can be described by an integer partition
\index{integer partition}
denoting the lengths of the cycles.
$\sigma_1$ contains a three-cycle and a one cycle, $\sigma_2$
contains one two-cycle and two one-cycles. One writes $(1^1,2^0,3^1)$ and
$(1^2, 2^1)$ respectively where the exponents denote the multiplicities of
the different cycle lengths~\cite{Sagan91}. Alternatively, \person{Young}
diagrams can be used for the same purpose where the length of each row
corresponds to each cycle length,
\begin{align*}
(1^1, 2^0, 3^1)&\equiv\partition{31}\quad\text{and}\\
(1^2, 2^1)&\equiv\partition{211}\text{.}
\end{align*}

As one can see from the examples, a \person{Young} diagram%
\index{Young@\person{Young}!diagram|main} of shape $\lambda$
is a system of square unit-boxes with $\lambda_j$ boxes in each row
aligned at the left with descending row lengths. The transposed
diagram $\lambda^\prime$ is the diagram with rows and columns exchanged,
e.g.\label{page:qcd-irreps:transposed}
\index{Young@\person{Young}!diagram, transposed|main}
\begin{displaymath}
\left(\partition{5332}\right)^\prime=\partition{44311}\text{.}
\end{displaymath}

Cycles of length two are called transpositions\index{transposition}.
It can be shown that every permutation has a decomposition into
transpositions and one can assign an invariant
\begin{equation}
\sgn(\pi)\equiv(-1)^{\text{number of transpositions in $\pi$}}
\end{equation}
which does not depend on the way $\pi$ is decomposed into transpositions.
The permutation $(1234)=(34)(23)(12)$ and its signum 
is~$\sgn((1234))=(-1)^3=-1$. From this one can construct two one-dimensional
and therefore irreducible representations:
the trivial representation $\rho(\sigma)=1$ and the
alternating representation~$\rho(\sigma)=\sgn(\sigma)$.

For a complete treatment of all irreducible representations, however,
one must extend the discussion to group modules. A module $V$
is called a group module of the group $G$ if there is a multiplication
such that\index{group module}
\begin{subequations}
\begin{align}
gv&\in V,\\
g(cv+dw)&=c(gv)+d(gw),\\
(gh)v&=g(hv)\quad\text{and}\\
\One v&=v,\quad\text{where $\One$ is the identity in $G$,}
\end{align}
\end{subequations}
for all $g, h\in G$ and $v, w\in V$ together with\footnote{It is 
sufficient to restrict the discussion to modules over the field $\Cset$.}
$c, d\in\Cset$. Every group module defines a matrix representation of the
group. Let $\mathcal{B}$ be a basis of $V$ and
$\langle\cdot,\cdot\rangle$
be the canonical inner product which on two basis vectors
$b_i, b_j\in\mathcal{B}$ is $\langle b_i, b_j\rangle=\delta_{ij}$
then the matrices $\rho(g)$,
\begin{equation}
\rho_{ij}(g)\equiv\langle b_i, g b_j\rangle
\end{equation}
form a representation of the group because\footnote{%
Note: $\langle g^{-1}b_i,hb_j\rangle=\delta_{g^{-1}b_i,hb_j}=
\langle b_i, ghb_j\rangle$ because of the equivalence
$g^{-1}b_i=hb_j\Leftrightarrow b_i=ghb_j$.}
\begin{multline}
\left[\rho(g)\rho(h)\right]_{ij}=
\sum_{k}\langle b_i,gb_k\rangle\langle b_k,hb_j\rangle=
\sum_{k^\prime, b_{k^\prime}=gb_k}\langle b_i,b_{k^\prime}\rangle
\langle g^{-1}b_{k^\prime},hb_j\rangle=\\
\langle g^{-1}b_i,hb_j\rangle=
\langle b_i,ghb_j\rangle=\rho(gh)_{ij}\text{.}
\end{multline}

Two \Sym-modules play a special role: when \Sym\ acts on the set
$\mathbf{S}=\{\mathbf{1}, \mathbf{2}, \ldots, \mathbf{k}\}$ of
$k$ formal symbols the space $\Cset\mathbf{S}$ forms a group
module with the action
\begin{displaymath}
\sigma\sum_{\mathbf{s}\in\mathbf{S}} c_{\mathbf{s}}\mathbf{s}=
\sum_{\mathbf{s}\in\mathbf{S}} c_{\mathbf{s}}\sigma(\mathbf{s}),
\quad\sigma\in\Sym\text{.}
\end{displaymath}
This module induces the \emph{defining representation} of the group.%
\index{defining representation|see{representation}}%
\index{representation!defining}

The module $\Cset\left[\Sym\right]$, i.e. the space of all formal linear
combinations $\sum_{\sigma\in\Sym}c_\sigma\cdot\sigma$ is called
the \emph{group algebra}\index{group algebra}, and the multiplication
is defined by the usual group multiplication. The group algebra
defines the \emph{regular representation}%
\index{representation!regular}.

Two particularly useful elements of $\Cset [\Sym]$
are the so-called symmetrisers and antisymmetrisers on $k$ lines,
as introduced in~\cite{Cvitanovic}.
Symmetrisers are represented by empty boxes,
\begin{align}
\plabel{1{\vdots}k}\pid3\symrow{3}\pid3&=
\frac1{k!}\sum_{\sigma\in\Sym}\sigma\\
\intertext{and antisymmetrisers by filled boxes}
\plabel{1{\vdots}k}\pid3\antisymrow{3}\pid3&=
\frac1{k!}\sum_{\sigma\in\Sym}\sgn(\sigma)\cdot\sigma
\end{align}
\index{symmetriser|main}
\index{antisymmetriser|main}

As an example those two symbols are given below for~\Sym[3]:
\begin{align*}
  \pid3\symrow{3}\pid3&=\frac{1}{3!}\left(%
  \perm{123}+\perm{132}+\perm{312}+\perm{321}+\perm{231}+\perm{213}
  \right)\\
  \intertext{and the corresponding antisymmetriser is}
  \pid3\antisymrow{3}\pid3&=\frac{1}{3!}\left(%
  \perm{123}-\perm{132}+\perm{312}-\perm{321}+\perm{231}-\perm{213}
  \right)\text{.}
\end{align*}

These elements have the obvious properties: they are idempotent and
they are eigenvectors of the permutations acting exclusively on the
lines of the (anti-)symmetriser,
\begin{equation}
  \perm{213}\symrow3\pid3=\perm{312}\symrow3\pid3=\pid3\symrow3\pid3\,\,,\qquad
  \perm{213}\antisymrow3\pid3=(-1)\perm{312}\antisymrow3\pid3=%
  (-1)\pid3\antisymrow3\pid3\,\,\text{.}
\end{equation}
Hence symmetrisers and antisymmetrisers that overlap in more than one
line annihilate each other because one can always find a permutation
which has different eigenvalues under each of the two algebra elements,
\begin{equation}
  \label{eq:symmetrizers:annihilate}
  (+1)\,\pid4\symrow{31}\pid4\antisymrow{13}\pid4=
  \left(\,\pid4\symrow{31}\perm{1324}\,\right)\pid4\antisymrow{13}\pid4=
  \pid4\symrow{31}\pid4\left(\,\perm{1324}\antisymrow{13}\pid4\,\right)=
  (-1)\,\pid4\symrow{31}\pid4\antisymrow{13}\pid4=0\text{.}
\end{equation}

The idempotence of the (anti-)symmetrisers is in fact only a special case
of a more general absorption property of smaller (anti-)symmetrisers by
bigger one if all lines of the smaller one are connected to the other one:
\begin{equation}
\pid6\symrow{132}\permid6\symrow{6}\pid6=\pid6\symrow{6}\pid6\,\text{,}\quad
\pid6\antisymrow{132}\permid6\antisymrow{6}\pid6=
\pid6\antisymrow{6}\pid6\,\text{.}
\end{equation}

For actual calculations an expansion of the (anti-)symmetrisers would
lead to a spurious proliferation of terms and is therefore to avoid.
Instead a recursive definition given in~\cite{Cvitanovic} can be used,
\begin{align}\label{eq:qcd-irreps:symrec}
\plabel{12\vdots{p}}\pid4\symrow{4}\pid4&=\frac1p\left(%
\plabel{12\vdots{p}}\pid4\symrow{13}\pid4+(p-1)
\plabel{12\vdots{p}}\pid4\symrow{13}\perm{2134}\symrow{13}\pid4\right)\\
\intertext{and}\label{eq:qcd-irreps:asymrec}
\plabel{12\vdots{p}}\pid4\antisymrow{4}\pid4&=\frac1p\left(%
\plabel{12\vdots{p}}\pid4\antisymrow{13}\pid4-(p-1)
\plabel{12\vdots{p}}\pid4\antisymrow{13}\perm{2134}\antisymrow{13}\pid4\right)%
\text{.}
\end{align}
In this representation of the (anti-)symmetrisers the smaller elements can be
reused in a computation which reduces the run time for the computation of a
symmetriser to polynomial rather than factorial.

\subsection[\person{Young}'s Natural Representation]{%
\person{Garnir} Relations and \person{Young}'s Natural Representation}
\label{ssec:qcd-irreps:youngrep}
\index{Young@\person{Young}!natural representation|main}
The goal of this section is
to define a set of projectors~$P_\lambda$ into invariant 
subspaces~$V^\lambda\subseteq\Cset\Sym$ 
for integer partitions (i.e. \person{Young} diagrams) $\lambda\vdash k$ of $k$.
The \person{Garnir} relations then allow to construct irreducible
matrix representations $\rho_\lambda$ for each group element.

Before the construction of the projectors can be addressed,
a couple of well known
combinatorial facts about \person{Young} diagrams have to be reviewed
(see for example~\cite{Fulton97,Sagan91}). Fist of all, the notion of a
\emph{tableau}\index{Young@\person{Young}!tableau|main}
\index{Tableau|see{\person{Young} tableau}} has to be introduced.

\begin{definition}[\person{Young} Tableau]
Let $\lambda=(\lambda_1,\lambda_2,\ldots,\lambda_p)\vdash n$ be
an integer partition, $\lambda_1\geq\lambda_2\geq\ldots\geq\lambda_p$
and~$A$ be an alphabet ($\epsilon\in A$).
The pair $Y=(\lambda, \rho)$ where $\rho$ is a mapping
$\rho: \Nset\times\Nset\mapsto A$
with $\rho(i, j)=\epsilon$ iff $i>p\vee j>\lambda_i$
defines a \person{Young} tableau of shape $\lambda$ over the alphabet $A$.
If no alphabet is specified the positive integer numbers are understood.
Tableaux are denoted by a \person{Young} diagram with the
values $\rho(i, j)$ filled into its
boxes at the $i$-th row and $j$-th column.

The filling of a tableau $Y=(\lambda, \rho)$ over the positive integer numbers
is the integer partition built from all
entries $\rho(i, j)\neq\epsilon$.

The shape of a diagram $Y=(\lambda,\rho)$
is also denoted as $\shape(Y)=\lambda$.\index{.shY@$\shape(Y)$}
\end{definition}

Examples for \person{Young} tableaux with the filling $(1^2, 2, 3^2)$ are
\begin{displaymath}
\tableau{{11}{23}{3}}, \tableau{{12}{31}{3}}, \tableau{{113}{23}}\text{.}
\end{displaymath}
The first two diagrams have the same shape. Two properties of tableaux
over alphabets with an ordering\footnote{For the ease of notation one can
assume $a<\epsilon \forall a \in A-\{\epsilon\}$.} are the following:

\begin{definition}[(Semi)-Standard Tableau]
\index{Young@\person{Young}!tableau!semi-standard}
A \person{Young} tableau is called a \emph{standard} tableau if for all
elements inside the shape of the tableau have $\rho(i,j)<\rho(i,j+1)$
and $\rho(i,j)<\rho(i+1,j)$, i.e. the entries are strictly increasing along
both the rows and the columns. A tableau is called \emph{semi-standard} if
the elements inside the shape of the tableau have $\rho(i,j)\leq\rho(i,j+1)$
and $\rho(i,j)<\rho(i+1,j)$, i.e. the entries are strictly increasing along
the columns and non-decreasing along the rows.
\end{definition}

In the above example the first and the last tableau are semi-standard.
None of the three tableaux is a standard tableau because no filling
with repeated elements gives a standard tableau.

\index{hook number}\index{.hij@$h_{ij}$|see{hook number}}
One combinatorial quantity which appears in many of the later formul\ae{}
is the hook number $h_\lambda=\prod_{i,j}h_{ij}$:
it is constructed by filling the entries $\rho(i,j)$
of a tableau by the hook length $h_{ij}$ of each box,
where the hook length is the number of boxes below and to the right of
the box including the box itself. For example in the diagram below
$h_{22}=4$, and $h_\lambda=537600$ because
\begin{displaymath}
h_{22}=\tableau{{{}{}{}{}{}}{{}{\bullet}{\bullet}}%
{{}{\bullet}{}}{{}{\bullet}}}=4,\quad
h_\lambda=\prod\tableau{{87521}{542}{431}{21}}=
8\cdot7\cdot5^2\cdot4^2\cdot3\cdot2^3
=537600\text{.}
\end{displaymath}

The number of standard tableaux (with the number $1\ldots k$ as filling)
of a given shape $\lambda\vdash k$ is\index{.fl@$f^\lambda$|main}
\begin{equation}
f^\lambda=\frac{k!}{h_\lambda}\text{.}
\end{equation}
A proof for this formula can be found in~\cite{Sagan91}.

The so-called \person{Young} projectors can be defined as follows:
\begin{definition}[\person{Young} Projector]
\index{Young@\person{Young}!projector|main}
For a given standard tableau $Y$ of shape $\lambda\vdash k$
where the entries of the boxes label a set of $k$ lines,
the product of symmetrisers on all lines according to
the rows of the \person{Young} diagram and antisymmetrisers
(multiplied from the right) according to the columns of the same diagram
together with the normalisation factor\footnote{$\lambda^\prime$ denotes
the transposed \person{Young} diagram, as defined on
Page~\pageref{page:qcd-irreps:transposed}.}
\begin{equation}
\alpha=\frac{\prod_j\lambda_j!\prod_i\lambda^\prime_i!}{h_\lambda}
\end{equation}
and an appropriate permutation between the symmetrisers and the
antisymmetrisers is called the \person{Young} projector $P_Y$
of the standard tableau $Y$.
\end{definition}

As an example $\lambda=\scriptstyle\partition{322}$ is chosen.
The normalisation factor is
\begin{displaymath}
h_\partition{322}=\frac{(3!2!2!)(3!3!1!)}{\prod\tableau{{541}{32}{21}}}=\frac{9}{5}
\end{displaymath}
and hence the projector is
\begin{displaymath}
P_{\tableau{{123}{45}{67}}}=\frac95\cdot
\plabel{1234567}\pid7\projector{322}\pid7\,\text{.}
\end{displaymath}

The above definition seems somewhat arbitrary and it is neither clear that
$P_Y$ is in fact a projector nor if the definition specifies $P_Y$ uniquely.
The following theorems fix that shortcoming and their proofs
follow mainly~\cite{Elvang:2003ue}.

\begin{theorem}[Uniqueness of $P_Y$]\label{thm:qcd-irreps:unique}
There is only one independent, non-vanishing choice of the permutation
that connects the symmetrisers of a projector $P_Y$ on their right
to the antisymmetrisers on their left. Rearranging any non-vanishing
choice of the internal lines
of the projector to any other non-vanishing choice of that permutation
leads to a global factor of $\pm1$.
\end{theorem}

The proof is done by induction over the number of columns~\cite{Elvang:2003ue}.
For only one column of length $k$ the projector consists of $k$ symmetrisers
of one line each and one antisymmetriser of length $k$. In this case swapping
two lines at the left leads to a factor $(-1)$ by the definition of the
antisymmetriser.
Adding a column of length $k$ at the left to a projector corresponding to a
standard tableau $Y$ of shape $\lambda$
with $k^\prime\leq k$ rows corresponds to adding another
line to each existing symmetriser, adding an antisymmetriser of length $k$
and possibly adding $(k-k^\prime)$ new symmetrisers of length $1$ each.
Now there are $k$ symmetrisers of various lengths and at least one
antisymmetriser of length $k$.
Since one cannot connect two lines of that antisymmetriser to
the same symmetriser there is essentially one way of connecting those $k$
lines, and the different possibilities can be canonicalised by allowing
a factor $\pm1$ and swapping legs at the single symmetrisers and
antisymmetrisers. Connecting the remaining lines now reduces to connecting
the lines of the tableau of the previous induction step.

\begin{theorem}[Properties of $P_\lambda$]\label{thm:qcd-irrep:projector}
The \person{Young} projectors have the following properties:
\begin{enumerate}
\item[(a)] $P_Y P_Z = \delta_{YZ} P_Y$,
	where $\shape(Y), \shape(Z) \vdash k$
\item[(b)] $\sum_{Y} P_Y = \One_k$,
	where the sum runs over all standard tableaux $Y$ with
	$\shape(Y)\vdash k$
\item[(c)] $P_Y\sigma P_Y=m_Y(\sigma)P_Y$ and $m_Y(\sigma)\in\{0,\pm1\}$
	for any permutation $\sigma\in\Sym$, $\shape(Y)\vdash k$.
\end{enumerate}
\end{theorem}

Property (c) is the easiest to prove since the antisymmetrisers
of the projector on the left and the symmetrisers of the projector
on the right with $\sigma$ in between can be read as a \person{Young}
projector from right to left.
The uniqueness argument~\ref{thm:qcd-irreps:unique} holds and
proves $m_Y(\sigma)\in\{0,\pm1\}$.

If in property (a) the shapes of $Y$ and $Z$ are the same,
the permutation between the antisymmetrisers of $Y$ and
the symmetrisers of $Z$ is either equal to
the inverse of the internal permutation in $Z$ in which case $Y=Z$,
or it is different (for $Y\neq Z$) in which case $P_YP_Z=0$
due to (c).
For $\shape(Y)\neq\shape(Z)$ we consider the case where
some\footnote{If necessary the same recursive argument
as in Theorem~\ref{thm:qcd-irreps:unique} can be used.}
column of $Y$ is longer than in $Z$. Then it is easy to
see that at least two lines coming from the same antisymmetriser
have to be connected to the same symmetriser at the second
projector and hence the product vanishes. For the reverse
case, i.e. one column in $Z$ is longer than in $Y$,
one can expand out the antisymmetrisers in $Y$ and the
symmetrisers in $Z$; for every term in the sum one
can apply the above argument and hence the product must vanish.

Only the completeness (b) and the idempotence $P_Y P_Y=P_Y$
remain to be shown. Using the uniqueness argument again one can
again expand out the inner symmetrisers and antisymmetrisers of
$P_YP_Y$ and obtains idempotence up to a normalisation:
\begin{equation}
P_YP_Y=P_Y\cdot \mathrm{const}
\end{equation}
and the sum runs over all permutations with their appropriate signs
that appear in the expansion of the internal (anti-)symmetrisers.

In order to complete the proof one has to establish that the
$P_Y$ with $Y$ running over all standard tableaux $\shape(Y)\vdash k$
form a complete basis of $\Cset\Sym$. The integer partitions of
$\lambda\vdash k$ label the conjugacy classes,
and the the standard tableaux of shape $\lambda$ form a basis
of each subspace $V^\lambda\equiv P_\lambda\cdot\Cset\Sym$.
The latter statement is proved\footnote{For a proof the reader
is referred to~\cite{Fulton97,Sagan91}.} by the

\begin{theorem}[\person{Garnir} Relations]\label{thm:qcd-irreps:garnir}
\index{Garnir@\person{Garnir}!relation}
The action of a permutation $\sigma$ acting on a \person{Young}
projector~$P_Y$ in a way that $\sigma Y$ corresponds to
non-standard tableau
can always be expressed in terms of a linear combination of
\person{Young} projectors
\begin{displaymath}
\sigma P_Y=\sum_Z c_{\sigma,Z} P_Z
\end{displaymath}
with complex numbers $c_{\sigma,Z}$ and the sum running
over all standard tableaux $\shape(Z)=\shape(Y)$.
The coefficients are generated by repeatedly applying the
following algorithm, which terminates:

For two adjacent columns $j$ and $(j+1)$ in $\pi Y$ with the elements
$(a_1<a_2<\ldots<a_p)$ and $(b_1<b_2<\ldots<b_q)$ respectively
find the smallest index $r$ such that $a_r>b_r$. Let
$A=\{a_r, a_{r+1},\ldots, a_p\}$ and $B=\{b_1, b_2,\ldots,b_r\}$
and form the \person{Garnir} element
\begin{equation}
g_{A,B}=\sum_{\pi}\sgn(\pi)\pi
\end{equation}
where the sum runs over all permutations $\pi\in\Sym[A\cup B]$
which leave the ordering of each column-strip $\pi(a_r,\ldots,a_p)$
and $\pi(b_1,\ldots,b_r)$ intact.
The \person{Garnir} element annihilates $\sigma P_Y$, and since
the identity is always in the sum, one can replace
\begin{equation}
\sigma P_Y=\sum_{\pi\neq\mathrm{id}}\sgn(\pi)\pi\sigma P_Y\text{.}
\end{equation}
\end{theorem}

To finish the proof of Theorem~\ref{thm:qcd-irrep:projector}
it now is sufficient to show that $\alpha_Y$ is chosen such that
$\sum P_Y=\One$. One can use the fact that in the expansion of
$P_Y$ the identity appears only once (see~\cite{Elvang:2003ue})
and hence can write the completeness relation as
\begin{equation}
\One=\sum_Y\alpha_Y\frac{1}{
\prod_{\lambda_i: \lambda=\shape(Y)}\lambda_i!
\prod_{\lambda^\prime_j: \lambda=\shape(Y)}\lambda^\prime_j!}\One
\end{equation}
we can now use that $\alpha_Y\equiv \alpha_\lambda$ in fact only
depends on the shape
$\lambda=\shape(Y)$ and rewrite the sum as a sum over partitions:
\begin{equation}
1=\sum_{\lambda\vdash k}\frac{\alpha_\lambda f^\lambda}{
\prod_{\lambda_i}\lambda_i!
\prod_{\lambda^\prime_j}\lambda^\prime_j!}
=k!\sum_{\lambda\vdash k}\frac{\alpha_\lambda/h_\lambda}{
\prod_{\lambda_i}\lambda_i!
\prod_{\lambda^\prime_j}\lambda^\prime_j!}\text{.}
\end{equation}
Plugging in the expression for $\alpha_\lambda$ and the
relation for the dimensions of the conjugacy classes
$k!=\sum_\lambda (f^\lambda)^2$ one immediately proves
everything because the idempotence can be obtained from
multiplying $\One=\sum_Y P_Y$ by $P_Z$ and using orthogonality.

Along the way we also proved that those representations are in fact
the irreducible representations which is due to the fact that they are
constructed from the conjugacy classes of the group.

\paragraph{Construction of the Representation Matrices} Before going on
to the actual construction of the representation matrices the results
of the first part of this section are summarised below: the projectors
$\langle P_{Y_1}, P_{Y_2}, \ldots P_{Y_{f^\lambda}}\rangle$,
$Y_i$~being the standard tableaux of shape~$\shape(Y_i)=\lambda\vdash k$,
form an orthogonal basis of the invariant
subspaces~$V^\lambda\subset\Cset\Sym$. The \person{Garnir} relations
establish a constructive proof and allow the calculation of the coefficients
$\rho_{ij}(\sigma)$ as defined below:
\begin{equation}
\sigma P_{Y_i}=\sum_{j=1}^{f^\lambda} \rho_{ij}(\sigma) P_{Y_j}\text{.}
\end{equation}

This notation suggests that the matrices
\begin{equation}
\rho(\sigma)=\left(\rho_{ij}(\sigma)\right)_{i,j=1}^{f^\lambda}
\end{equation}
form a $f^\lambda$-dimensional representation of the symmetric group,
the caveat being that the matrices multiply from the right as
one multiplies permutations to the left, as can be seen below,
\begin{multline}
\sigma_1\sigma_2 P_{Y_i}=
\sigma_1\sum_{j=1}^{f^\lambda} \rho_{ij}(\sigma_2) P_{Y_j}=\\
\sum_{j=1}^{f^\lambda} \rho_{ij}(\sigma_2)\sum_{k=1}^{f^\lambda}
\rho_{jk}(\sigma_1) P_{Y_k}=
\sum_{k=1}^{f^\lambda}
\left(\rho(\sigma_2)\rho(\sigma_1)\right)_{ik} P_{Y_k}\text{.}
\end{multline}

The matrices satisfy $\rho(\sigma_1\sigma_2)=\rho(\sigma_2)\rho(\sigma_1)$,
so in fact $\rho(\sigma)^\transposed$ would make a representation matrix
in the ordinary sense. It is also clear from its definition
that~$\rho({\mathrm{id}})=\One$ as required.

An explicit example for the representation $\lambda=\scriptstyle\partition{221}$
is given below to clarify the construction of the representation matrices.
The basis of this subspace consists of the five standard tableaux
\begin{displaymath}
Y_1=\tableau{{12}{34}{5}},\quad
Y_2=\tableau{{13}{24}{5}},\quad
Y_3=\tableau{{12}{35}{4}},\quad
Y_4=\tableau{{13}{25}{4}},\quad
Y_5=\tableau{{14}{25}{3}}\text{.}
\end{displaymath}
The corresponding projectors are, up to a common normalisation factor,
\begin{displaymath}
\ysymwidth=1mm
P_{Y_1}={\tiny\permid5\projector{221}},
P_{Y_2}={\tiny\xperm{13245}\projector{221}},
P_{Y_3}={\tiny\xperm{12354}\projector{221}},
P_{Y_4}={\tiny\xperm{13254}\projector{221}},
P_{Y_5}={\tiny\xperm{14253}\projector{221}}\text{.}
\end{displaymath}
Applying the permutation $\sigma=(354)$ to $P_{Y_1}$ leads to
a non-standard ordering of the lines,
\begin{equation}\label{eq:qcd-irreps:garnirex1}
\sigma P_{Y_1}=\perm{12534}\pid5\projector{221}
\end{equation}
whereas on $P_{Y_3}$ the action of $\sigma$ is trivial,
\begin{equation}
\sigma P_{Y_3}=\perm{12534}\perm{12354}\projector{221}=
\perm{12435}\projector{221}=\permid5\projector{221}=P_{Y_1}\text{.}
\end{equation}
To reduce~\eqref{eq:qcd-irreps:garnirex1} one has to apply the
\person{Garnir} relations, which as a diagrammatic analogue can be
represented by applying a symmetriser on the left to the legs
according to the set $A\cup B$ in Theorem~\ref{thm:qcd-irreps:garnir}:
\begin{equation}
0=\symrow{113}\perm{12534}\pid5\projector{221}=
\frac{2}{3!}\left(\permid5+\xperm{12453}+\perm{12354}\right)
\projector{221}=\frac{2}{3!}\left(P_{Y_1}+\sigma P_{Y_1}+P_{Y_3}\right)
\end{equation}
The action of $\sigma$ on $V^\lambda$ can summarise in a system of equations,
\begin{equation}
\sigma\begin{pmatrix}
P_{Y_1}\\
P_{Y_2}\\
P_{Y_3}\\
P_{Y_4}\\
P_{Y_5}
\end{pmatrix}^\transposed=\begin{pmatrix}
-P_{Y_1}-P_{Y_3}\\
P_{Y_5}\\
P_{Y_1}\\
-P_{Y_1}-P_{Y_2}\\
-P_{Y_3}-P_{Y_4}
\end{pmatrix}^\transposed=\begin{pmatrix}
P_{Y_1}\\
P_{Y_2}\\
P_{Y_3}\\
P_{Y_4}\\
P_{Y_5}
\end{pmatrix}^\transposed\cdot\begin{pmatrix}
-1& 0& 1&-1& 0\\
 0& 0& 0&-1& 0\\
-1& 0& 0& 0&-1\\
 0& 0& 0& 0&-1\\
 0& 1& 0& 0& 0
\end{pmatrix}
\end{equation}

\paragraph{\acs{gln}-Dimensions} So far the only
\index{representation!gln@$\gln$|(}
representations of the symmetric group have been considered.
It is well known that by acting with the appropriate \Sym-projectors
on the direct product of $k$ copies of vectors
in the fundamental representation of \acs{gln}, one generates the
irreducible representations of \acs{gln}.
The dimensions of these \acs{gln}-representations is
calculated as follows:
Let $Y$ be one of the standard tableaux\footnote{
The representations are labelled by the integer partitions, not by
individual tableaux.
I have chosen an arbitrary basis-element $Y$ only in order to remain
consistent in the notation $P_Y$, where $Y$ has to be a standard tableau.}
of shape $\lambda$.
The \acs{gln}-dimension of the irreducible representation
constructed from the projectors $P_Y$ is
\begin{equation}\label{eq:qcd-irreps:glndim}
\dim_{\gln} P_Y =\tr{P_Y}=\frac{h_\lambda(n)}{h_\lambda}\text{,}
\end{equation}
where $h_\lambda(n)$ is the modified hook number constructed as
follows:
the top-left box of the diagram $\lambda$ is filled with the
symbol $n$, all other boxes are filled increasing by one along
the rows and decreasing along the columns. For
$\lambda=\scriptsize\partition{321}$ one gets
\begin{displaymath}
\dim_{\gln} P_\partition{321}=\frac{\yboxsize=1.7em
\prod\tableau{{{n}{n+1}{n+2}}{{n-1}{n}}{{n-2}}}}{%
\prod\tableau{{531}{31}{1}}}=\frac{n^2(n^2-1)(n^2-4)}{45}\text{.}
\end{displaymath}
Amongst other proofs a diagrammatic version of the proof based on
a colouring algorithm has been presented by the authors
of~\cite{Elvang:2003ue}.
Since every index of the tensor runs
over the value $1,\ldots,n$ and the indices are symmetrised along
the rows and antisymmetrised along the columns of the \person{Young}
diagram $\lambda$, the semi-standard tableaux of shape $\lambda$ over
the alphabet $\{1,\ldots,n\}$ enumerate the independent components
of the tensor which is the same as the number in~\eqref{eq:qcd-irreps:glndim}.

With respect of the irreducible representations,
the only difference between \acs{gln} and its subgroups, such as \acs{sun}
and \acs{son}, are the appearance of additional invariant tensors that can
reduce irreducible representations of \acs{gln} further.
For \acs{sun} the new tensor is the totally antisymmetric
\person{Levi-Civita} tensor which diagrammatically can be represented
by half an antisymmetriser over $n$ lines~\cite{Cvitanovic}:
\begin{subequations}
\begin{align}
\ysymwidth=1.5mm
\epsilon^{a_1a_2\ldots a_n}&=
i^{-n(n-1)/2}\sqrt{n!}\,
\antisymrow{4}\pid4\plabel{{a_1}{a_2}{\vdots}{a_n}}\\
\epsilon_{a_1a_2\ldots a_n}&=
i^{n(n-1)/2}\sqrt{n!}\,
\plabel{{a_1}{a_2}{\vdots}{a_n}}\pid4\antisymrow{4}
\end{align}
\end{subequations}
The normalisation in front of the tensor has been chosen such that
\begin{equation}
\pid4\antisymrow{4}\,\antisymrow{4}\pid4=\pid4\antisymrow{4}\pid4\quad\text{and}\quad
\antisymrow{4}\permid4\antisymrow{4}=1\text{.}
\end{equation}

For the representations the existence of this invariant tensor implies
that any projector that contains a column of length $n$ can be decomposed
and is no longer irreducible. Hence, in \acs{sun} representations with
extra columns of length $n$ have to be regarded equivalent to those with
the columns of length $n$ being stripped off, for $n=3$ for example one has
\begin{displaymath}
\tableau{{{\bullet}{\bullet}{}{}}{{\bullet}{\bullet}{}}{{\bullet}{\bullet}}}
\equiv
\tableau{{{\bullet}{}{}}{{\bullet}{}}{{\bullet}}}
\equiv
\tableau{{{}{}}{{}}}\text{.}
\end{displaymath}

Similarly, one obtains the representations of the orthogonal group
by removing all traces, since in \acs{son} the metric
$g^{\mu}_{\nu}$ forms another invariant tensor
As a consequence the treatment of \acs{son}
is more complicated since the number of boxes is not conserved anymore
and therefore this work is restricted to the treatment of the unitary
group.
\index{representation!gln@$\gln$|)}

\subsection{Calculation of \nj{n} Symbols in \acs{sun}}
\label{ssec:qcd-irreps:nj}

In Section~\ref{ssec:qcd-color-recoupling:recoupling} it has been
shown that only three kinds of numeric invariants are enough to calculate
any vacuum-bubble diagram, i.e. a diagram with no external indices and therefore
corresponds to a scalar. They are the \acs{gln}-dimension (\nj1-symbols),
\begin{align*}
\parbox[c][\height+0.3\baselineskip][c]{10mm}{%nl
\begin{fmfchar*}(10,10)
\fmfleft{p}\fmfright{q}
\fmfdot{p,q}
\fmf{fermion,label={},left}{p,q}
\fmf{fermion,left}{q,p}
\fmf{fermion}{p,q}
\end{fmfchar*}}&\quad\text{(\nj3-symbol) and}\\
\parbox[c][\height+1.5\baselineskip][c]{15mm}{%nl
\begin{fmfchar*}(15,15)
\fmfsurround{v1,v2,v3}
\fmfdot{v1,v2,v3,z}
\fmf{fermion,right=0.5}{v1,v2}
\fmf{fermion,right=0.5}{v2,v3}
\fmf{fermion,right=0.5}{v3,v1}
\fmf{fermion,label={}}{v1,z}
\fmf{fermion}{v2,z}
\fmf{fermion}{z,v3}
\end{fmfchar*}}&\quad\text{(\nj6-symbol).}
\end{align*}

In terms of projectors, the \person{Clebsh}-\person{Gordan}
vertices are up to a phase,
\begin{equation}
\parbox[c][\height+1.5\baselineskip][c]{15mm}{%nl
\begin{fmfchar*}(15,15)
\fmfsurround{v3,v1,v2}
\fmfdot{z}
\fmf{fermion,label=$X$}{v1,z}
\fmf{fermion,label=$Y$}{v2,z}
\fmf{fermion,label=$Z$}{z,v3}
\end{fmfchar*}}=
\parbox[c]{25mm}{\includegraphics[width=25mm]{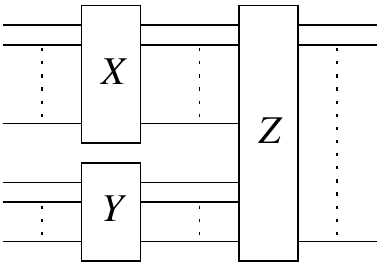}}
\end{equation}
where the boxes represent the appropriate projectors.

Therefore the \nj{n}-symbols are
\begin{align*}
\parbox[c][\height+0.3\baselineskip][c]{10mm}{%nl
\begin{fmfchar*}(10,10)
\fmfleft{p}\fmfright{q}
\fmfdot{p,q}
\fmf{fermion,label=$X$,left}{p,q}
\fmf{fermion,label=$Z$,left}{q,p}
\fmf{fermion,label=$Y$}{p,q}
\end{fmfchar*}}&=\tr{P_X P_Y P_Z}=
\tr{\parbox[c]{30mm}{\includegraphics[width=30mm]{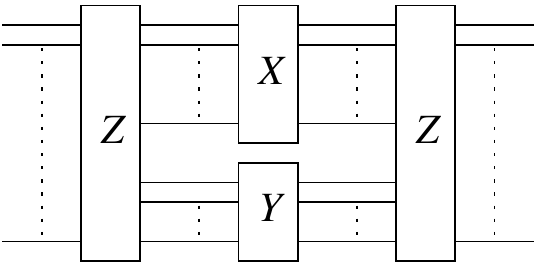}}}\\
\parbox[c][\height+1.5\baselineskip][c]{15mm}{%nl
\begin{fmfchar*}(15,15)
\fmfsurround{v1,v2,v3}
\fmfdot{v1,v2,v3,z}
\fmf{fermion,right=0.5,label=$\mu$}{v1,v2}
\fmf{fermion,right=0.5,label=$\beta$}{v2,v3}
\fmf{fermion,right=0.5,label=$\xi$}{v3,v1}
\fmf{fermion,label=$\nu$,label.dist=1.0}{v1,z}
\fmf{fermion,label=$\gamma$,label.dist=1.0,label.side=right}{v2,z}
\fmf{fermion,label=$\chi$,label.dist=1.0}{z,v3}
\end{fmfchar*}%
}&=\tr{P_\xi P_\nu P_\mu P_\beta P_\gamma P_\chi}=
\tr{\parbox[c]{50mm}{\includegraphics[width=50mm]{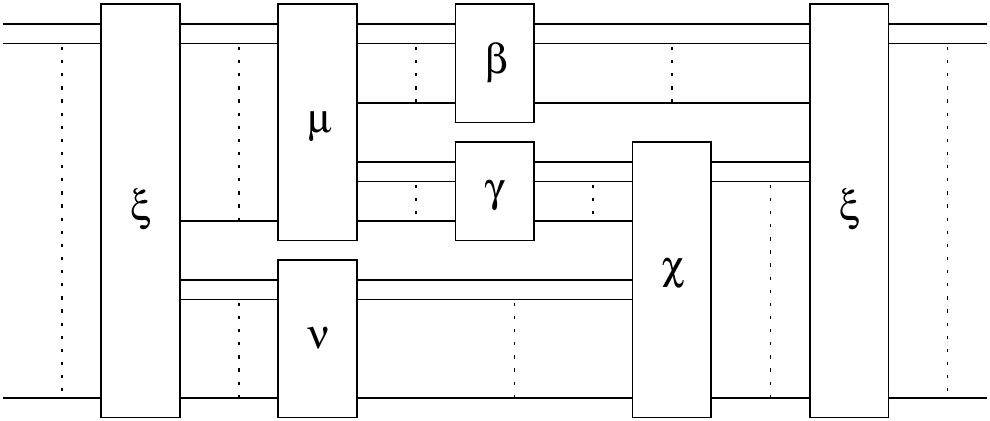}}}
\end{align*}
In the second step the idempotence of the largest projector and the
cyclicity of the trace have been used. It is clear that
the $n$-dependence only comes in through the dimension
$\dim_{\gln}(P_{Z})$ ($\dim_{\gln}(P_\xi)$ resp.) of the highest
weight representation occurring in the diagram:
the smaller projectors sandwiched in between the largest
can be expanded leading only to a sum of permutations $\sigma$
of which the contribution to the result
is $m_Z(\sigma)$ ($m_\xi(\sigma)$ resp.).
Hence, the \nj3- and \nj6-symbols can be evaluated as a trace of
\Sym[k]-representation matrices times the \acs{gln}-dimension of
the highest weight representation,
\begin{align*}
\parbox[c][\height+1.5\baselineskip][c]{10mm}{%nl
\begin{fmfchar*}(10,10)
\fmfleft{p}\fmfright{q}
\fmfdot{p,q}
\fmf{fermion,label=$X$,left}{p,q}
\fmf{fermion,label=$Z$,left}{q,p}
\fmf{fermion,label=$Y$}{p,q}
\end{fmfchar*}}&=\dim_{\gln}(P_Z)\alpha_X\alpha_Y\alpha_Z%
\tr{\rho_Z(P_X)\rho_Z(P_Y)\rho_Z(Z)}
\quad\text{and}\\
\parbox[c][\height+1.5\baselineskip][c]{15mm}{%nl
\begin{fmfchar*}(15,15)
\fmfsurround{v1,v2,v3}
\fmfdot{v1,v2,v3,z}
\fmf{fermion,right=0.5,label=$\mu$}{v1,v2}
\fmf{fermion,right=0.5,label=$\beta$}{v2,v3}
\fmf{fermion,right=0.5,label=$\xi$}{v3,v1}
\fmf{fermion,label=$\nu$,label.dist=1.0}{v1,z}
\fmf{fermion,label=$\gamma$,label.dist=1.0,label.side=right}{v2,z}
\fmf{fermion,label=$\chi$,label.dist=1.0}{z,v3}
\end{fmfchar*}%
}&=\dim_{\gln}(P_\xi)\alpha_\xi\alpha_\mu\alpha_\nu%
\alpha_\beta\alpha_\gamma\alpha_\chi
\tr{\rho_\xi(P_\xi)\rho_\xi(P_\nu)%
\rho_\xi(P_\mu)\rho_\xi(P_\beta)\rho_\xi(P_\gamma)\rho_\xi(P_\chi)}
\text{.}
\end{align*}
In the calculation one can make use of the fact that
$\rho_Z(Z)=vw^\transposed$, where $v_j=\delta_{ZY_j}$,
$w_i=m_Z(\sigma_{Y_i})$, and $\sigma_{Y_i}$ is the permutation that
has to be applied to the canonical projector\footnote{i.e. the projector
corresponding to the tableau which is filled by $1,2,\ldots,k$ when
the rows are read line-wise left to right and top to bottom}
to obtain $P_{Y_i}$. The representation matrices of the permutations
are constructed in \person{Young}'s natural representation as explained
in Section~\ref{ssec:qcd-irreps:youngrep}, the matrices for
the antisymmetrisers and symmetrisers are built recursively using
Equations~\eqref{eq:qcd-irreps:symrec} and~\eqref{eq:qcd-irreps:asymrec}.

\subsection{The \person{Littlewood}-\person{Richardson} Rule}
\label{ssec:qcd-irreps:LR}
\index{Littlewood-Richardson@\person{Littlewood}-\person{Richardson}!rule}
\index{Littlewood-Richardson@\person{Littlewood}-\person{Richardson}!coefficient}
\index{.c@$c_{\mu\nu}^\lambda$|see{\person{Littlewood}-\person{Richardson} coefficient}}
One of the remaining problems for an actual implementation of an algorithm,
using a reduction of the \acs{sun}-group structure of a \person{Feynman}
diagram into irreducible pieces, is the determination of the
representations $\lambda$ occurring in the
completeness relation~\eqref{eq:qcd-color-recoupling:completeness} together
with their multiplicities. The second problem to be solved is fixing the
freedom of having additional permutations between $\mu$, $\nu$ and $\lambda$
in the equation above; especially in the case where the
multiplicity $c_{\mu\nu}^\lambda$
of a representation~$\lambda$ is larger than one, one has to find
$c_{\mu\nu}^\lambda$
independent permutations which span the subspace concerned.

In 1934 \person{Littlewood} and \person{Richardson}~\cite{LR:1934}
formulated a rule to
calculate the multiplicities, that is the
\ac{lr} coefficients
$c_{\mu\nu}^\lambda$ in the decomposition
\begin{equation}
V^\mu\otimes V^\nu=\bigoplus_{\lambda}c_{\mu\nu}^\lambda V^\lambda
\end{equation}
The formulation of the \ac{lr} rule requires us to introduce
the concept of skew-tableaux:
\begin{definition}[Skew Tableau]\index{Young@\person{Young}!tableau!skew}
\index{.l@$\lambda/\mu$|see{\person{Young} tableau, skew}}
The skew tableau $\lambda/\mu$ ($\lambda_i\geq\mu_i,\forall i$)
is obtained from a \person{Young} tableau of shape $\lambda$ by removing
all boxes of the partition $\mu$ from the top left corner. The shape of
the remaining boxes is preserved, i.e. the boxes are not aligned at the
left. Skew tableaux are denoted as follows:
\begin{displaymath}\yboxsize=9pt
\lambda=\partition{431},\mu=\partition{21}\Rightarrow\lambda/\mu=%
\tableau{{{\rule{10pt}{10pt}}{\rule{10pt}{10pt}}{}{}}{{\rule{10pt}{10pt}}{}{}}{{}}}
\end{displaymath}
\end{definition}

The \ac{lr} rule counts skew tableaux of which the row words are \emph{reverse
lattice words}.
\begin{definition}[Lattice word]
A \emph{lattice word}
\index{lattice word} (also lattice permutation, ballot sequence
\index{ballot sequence|see{lattice word}} or
\person{Yamanouchi} word\index{Yamanouchi word@\person{Yamanouchi} word|see{lattice word}})
is a word $x_1x_2\ldots x_p$ of positive integers such that at each position $i\leq p$
for each $x, 1<x\leq\max_{j\leq i}(x_j)$ the number $x-1$ occurs at least as often as $x$.

A \emph{reverse lattice word} is a word $x_1x_2\ldots x_p$ such that
the reverse word $x_px_{p-1}\ldots x_1$ is a lattice word.
\end{definition}

The \ac{lr} rule then states the following.
\begin{theorem}[\acl{lr} rule]
The value of the coefficient~$c_{\mu\nu}^\lambda$ in the decomposition
\begin{displaymath}
V^\mu\otimes V^\nu=\bigoplus_{\lambda}c_{\mu\nu}^\lambda V^\lambda
\end{displaymath}
is the number of skew semi-standard tableaux of shape $\lambda/\mu$
filled with the numbers of the partition $(1^{\nu_1}, 2^{\nu_2}\ldots)$
such that its row word is a reverse lattice word.
The row word is obtained by reading the rows of the skew tableau
from left-to right, bottom to top.
\end{theorem}

As an example we consider the multiplication of
\begin{displaymath}
\mu=\partition{21}\quad\text{and}\quad\nu=\partition{21}
\end{displaymath}
The only non-zero coefficients stem from those partitions $\lambda$
where the number of boxes is the same on both sides of the
equation\footnote{This is an obvious corollary of the \ac{lr} rule.}.

As a first step we have to generate all integer partitions $\lambda\vdash k$,
where~$k$ is the sum of the number of boxes in $\mu$ and $\nu$. Only those
partitions contribute to the result where $\lambda_i\geq\mu_i$ for all rows;
the authors of~\cite{Zoghbi1998} present an algorithm which generates
integer partitions in lexicographic order such that one can stop the iteration
when the first partition is lexicographically smaller than $\mu$.\footnote{%
If a partition $\lambda$ is lexicographically smaller than $\mu$ it
implies that $\lambda_i\geq\mu_i$ fails.} For generating the ballot sequences
we can use the algorithm given in~\cite{Nijenhuis:1978}. For the example we
get the skew tableaux
% xx11 xx11 xx1 xx1 xx1 xx1 xx xx
% x2   x    x12 x1  x2  x   x1 x1
%      2        2   1   1   12 1
%                       2      2
%
%
\begin{displaymath}\yboxsize=9pt\newcommand{\NUL}{\rule{10pt}{10pt}}
\tableau{{{\NUL}{\NUL}11}{{\NUL}2}}\quad
\tableau{{{\NUL}{\NUL}11}{{\NUL}}{2}}\quad
\tableau{{{\NUL}{\NUL}1}{{\NUL}12}}\quad
\tableau{{{\NUL}{\NUL}1}{{\NUL}1}{2}}\quad
\tableau{{{\NUL}{\NUL}1}{{\NUL}2}{1}}\quad
\tableau{{{\NUL}{\NUL}1}{{\NUL}}{1}{2}}\quad
\tableau{{{\NUL}{\NUL}}{{\NUL}1}{12}}\quad
\tableau{{{\NUL}{\NUL}}{{\NUL}1}{1}{2}}.
\end{displaymath}
Hence the partition $\yboxsize=4pt\partition{321}$ has a multiplicity
$c_{\mu\nu}^{\partition{321}}=2$
whereas the other partitions appear only once in the decomposition.

One can check that no representations are missing and all
\ac{lr} coefficients are
correct by ensuring that the dimensions on both sides add up
\begin{equation}
d_{\gln}(\mu)\cdot d_{\gln}(\nu)=
\sum_\lambda c_{\mu\nu}^\lambda d_{\gln}(\lambda)\text{.}
\end{equation}

As mentioned at the beginning of this section, for our algorithm
in addition to the multiplicities one also needs according
permutations that span the subspace $V^\lambda$. This can be achieved
by the observation~\cite{Fulton97} that one can construct a standard
tableau $Y_w$ of shape $\nu$ from the skew tableau $\lambda/\nu$
as constructed above:
for each lattice word $w=x_1x_2\ldots x_p$ occurring
in the \ac{lr} rule, put the number $i$ in the $x_i$-th row, filling
the rows from left to right. The permutation between the tableau
$Y_w$ and the canonical tableau, i.e. where the rows are filled
by the numbers $1,\ldots p$ left-to-right starting from the top row,
is a basis element for this subspace.

In the previous example, the partition $\lambda=(3,2,1)$ appears twice,
associated with the lattice words $112$ and $121$. The smaller partitions
are $\mu=\nu=(2,1)$.
We can reconstruct the two tableaux of shape $\nu$,
\begin{displaymath}
\tableau{{45}{6}}\leadsto\perm{123}\quad\text{and}\quad%
\tableau{{46}{5}}\leadsto\perm{132}\text{.}
\end{displaymath}
From this, two permutations spanning the subspace for the
projector~$P_\lambda$ can be constructed,
\begin{displaymath}
\perm{124356}\quad\text{and}\quad\perm{124365}\,\text{.}
\end{displaymath}

The construction of the tableaux
ensures irreducibility under the \person{Garnir} relations and hence
independence of the associated permutations.
As the multiplicity in \ac{lr} rule corresponds to the dimension of
the subspace this also proves the completeness of this basis.

\subsection{Diagrammatic Tensor Reduction for \acs{gln}}
\label{ssec:qcd-irreps:reduce}
This section describes an algorithm that allows the reduction of
arbitrary \acs{gln} tensors to a basis constructed from irreducible
representations; as scalar coefficients only \nj6, \nj3 and \nj1
coefficients arise during the reduction. Tensors are represented
as directed graphs with edges labelled by partitions and vertices
associated with permutations. The discussion is restricted
to bubble diagrams with no external legs because in a first step
one can always project onto a tree basis with the help of the
\person{Wigner}-\person{Eckhard} theorem, which diagrammatically
reads~\cite{Cvitanovic}
\begin{equation}
\phantom{xxxx}
\parbox[c][\height+1.5\baselineskip][c]{17mm}{%nl
\begin{fmfchar*}(20,40)
\fmfleftn{p}{5}
\fmfrightn{q}{5}
\fmf{fermion,tension=0.9}{p1,s1}
\fmf{fermion,tension=0.7}{p2,s2}
\fmf{fermion,tension=0.5}{p3,s3}
\fmf{fermion,tension=0.7}{p4,s4}
\fmf{fermion,tension=0.9}{p5,s5}
\fmf{phantom}{s1,t1,q1}
\fmf{phantom}{s2,t2,q2}
\fmf{phantom}{s3,t3,q3}
\fmf{phantom}{s4,t4,q4}
\fmf{phantom}{s5,t5,q5}
\fmfpoly{hatched}{t1,t2,t3,t4,t5,s5,s4,s3,s2,s1}
\fmflabel{$\mu_1$}{p1}
\fmflabel{$\mu_2$}{p2}
\fmflabel{$\mu_3$}{p3}
\fmflabel{$\mu_4$}{p4}
\fmflabel{$\mu_5$}{p5}
\end{fmfchar*}}
=\sum_{\nu,\sigma,\lambda}\kappa_{\mu_1\mu_2}^\sigma
\kappa_{\mu_4\mu_5}^\nu \kappa_{\mu_3\nu}^\lambda
\kappa_{\lambda\sigma}^\One\cdot\phantom{xxx}
\parbox[c][\height+0.3\baselineskip][c]{50mm}{%nl
\begin{fmfchar*}(50,30)
\fmfleftn{p}{5}
\fmfrightn{q}{5}
\fmf{fermion,tension=0.7}{p1,v12}
\fmf{fermion,tension=0.7}{p5,v45}
\fmf{fermion,tension=1.0,label=$\nu$}{v45,v345}
\fmf{fermion,tension=1.0,label=$\lambda$}{v345,vl}
\fmf{fermion,tension=0.5,label=$\sigma$}{v12,vl}
\fmfdot{v45,v345,v12,vl}
\fmf{phantom,tension=2.0}{vl,wl}
\fmfdot{w45,w345,w12,wl}
\fmf{phantom,tension=0.5}{wl,w345}
\fmf{phantom,tension=0.2}{w345,w45}
\fmf{phantom,tension=1.0}{wl,w12}
\fmf{phantom,tension=0.4}{w45,q5}
\fmf{phantom,tension=0.5}{w12,q1}
\fmf{phantom,tension=0.6}{s1,t1}
\fmf{phantom,tension=0.6}{s2,t2}
\fmf{phantom,tension=0.6}{s3,t3}
\fmf{phantom,tension=0.6}{s4,t4}
\fmf{phantom,tension=0.6}{s5,t5}
\fmf{phantom}{t1,q1}
\fmf{phantom}{t2,q2}
\fmf{phantom}{t3,q3}
\fmf{phantom}{t4,q4}
\fmf{phantom}{t5,q5}
\fmf{phantom,tension=0.1}{w345,p5}
\fmfpoly{hatched,tension=0.1}{t1,t2,t3,t4,t5,s5,s4,s3,s2,s1}
\fmffreeze
\fmf{fermion}{p2,v12}
\fmf{fermion}{p3,v345}
\fmf{fermion}{p4,v45}
\fmf{fermion,left=0.4,label=$\nu$}{w345,w45}
\fmf{fermion,left=0.2}{wl,w345}
\fmf{fermion,right=0.2,label=$\sigma$}{wl,w12}
\fmf{fermion,right=0.3}{w12,s1}
\fmf{fermion,left=0.3}{w12,s2}
\fmf{fermion}{w345,s3}
\fmf{fermion,right=0.3}{w45,s4}
\fmf{fermion,left=0.3}{w45,s5}
\fmflabel{$\mu_1$}{p1}
\fmflabel{$\mu_2$}{p2}
\fmflabel{$\mu_3$}{p3}
\fmflabel{$\mu_4$}{p4}
\fmflabel{$\mu_5$}{p5}
\end{fmfchar*}}
\end{equation}
The tree is constructed by repeatedly applying the
completeness relation~\eqref{eq:qcd-color-recoupling:completeness},
and \person{Schur}'s lemma ensures that the single representation
in the middle\footnote{represented by the gap rather than a line} is
the trivial representation.

The bubble diagram can be
reduced by repeatedly eliminating the smallest loop. If the smallest
loop is of length 3, or smaller, one can directly apply \person{Schur}'s
lemma or the star-triangle
relation~\eqref{eq:qcd-color-recoupling:startriangle}; larger loops
can be split into half by pinching two opposite edges of that cycle
inserting the completeness
relation~\eqref{eq:qcd-color-recoupling:completeness}.

The remainder of this section describes graph algorithms to determine
the topological properties of the diagram.
Algorithm~\ref{alg:qcd-irreps:graph-reduce}
summarises the steps that are required for the tensor reduction, after one
has projected on the irreducible basis of trees.
\begin{algorithm}
\caption{\textsc{Graph Reduction}(\(G\)).  Returns the value associated with a
bubble graph as a function in~\(n\).}\label{alg:qcd-irreps:graph-reduce}
\begin{algorithmic}
\REQUIRE a bubble graph \(G\) as an array of tuples \((i, j, \lambda)\).
\STATE \(L\) \(\leftarrow\) \textsc{Find Loops}(\(G\)).
\STATE \(S\) \(\leftarrow\) \textsc{Shortest Loop}(\(L\)).
\IF{\textsc{Length}(\(S\)) \(\ge 4\)}
	\STATE Select a pair of edges \(X,Y\) from \(S\).
	\RETURN \(\sum_Z c_Z^{X,Y}\cdot\)\textsc{Graph Reduction}(\(G^\prime\)),
	where \(G^\prime\) is the Graph that is generated by inserting \(Z\)
   	according to Equation~\eqref{eq:qcd-color-recoupling:completeness}.
\ELSIF{\textsc{Length}(\(S\))\(=3\)}
	\RETURN \(c\cdot\)\textsc{Graph Reduction}(\(G^\prime\)),
	where \(G^\prime\) is generated by
	Equation~\eqref{eq:qcd-color-recoupling:startriangle}
   and \(c\) is the appropriate coefficient.
\ELSIF{\textsc{Length}(\(S\))\(=2\)}
	\RETURN \(c\cdot\)\textsc{Graph Reduction}(\(G^\prime\)),
	where \(G^\prime\) is generated by
	Equation~\eqref{eq:qcd-color-recoupling:CFbubble}
   and \(c\) is the appropriate coefficient.
\ELSE
	\RETURN the dimension $\dim(\lambda)$ of the representation~$\lambda$
   corresponding to this line.
\ENDIF
\end{algorithmic}
\end{algorithm}

If one allows for disconnected graphs
Algorithm~\ref{alg:qcd-irreps:graph-reduce}
has to be run on every connected component of the graph.  Efficient
algorithms for finding connected components as well as finding cycles
can be, found for example, in~\cite{Galler:1964,Knuth:TAOCP1,Kennedy:2007sj}:
Algorithm~\ref{alg:qcd-irreps:find-cc} enumerates the connected
components of a graph and Algorithm~\ref{alg:qcd-irreps:find-cycles}
returns a list of all cycles.

All array indices are assumed to be \(1\)-based, and the symbols
\(V_G\) denotes the number of vertices in a graph~\(G\) and \(E_G\) the
number of edges.
The ancestor function \(A\) is defined as~\cite{Kennedy:2007sj},
\begin{equation}
A(i, F)=\left\{\begin{array}{rl}
i,&\;F[i]=0\;\text{or}\;F[i]=i\text{,}\\
A(F[i], F),&\;\text{else.}
\end{array}\right.
\end{equation}

\begin{algorithm}
\caption{\textsc{Find Connected Components}(\(G\)).  Returns a partition
of the indices of the edges into their connected components.}
\label{alg:qcd-irreps:find-cc}
\begin{algorithmic}
\REQUIRE a bubble graph \(G\) as an array of tuples \((i, j, \lambda)\).
\STATE \(P\leftarrow\emptyset\);
	\(F[i]\leftarrow i\quad\forall i\in1\ldots V_G\).
\FORALL{\((i, j, \lambda)\in G\)}
	\IF{\(A(i, F)\neq A(j, F)\)}
		\STATE \(F[A(i, F)]\leftarrow A(j, F)\).
	\ENDIF
\ENDFOR
\STATE \(R\leftarrow\{A(i, F)\vert i=1\ldots V_G\}\).
\FORALL{\(r\in R\)}
	\STATE \(P\leftarrow P\cup\{\{i\in 1\ldots V_G\vert A(i, F) = r\}\}\).
\ENDFOR
\RETURN \(P\)
\end{algorithmic}
\end{algorithm}

A vector of the space \((\mathbb{Z}_3)^{E_G}\) is called
a \(\mathbb{Z}_3\)-chain.
Let \(z=\langle z_1,\ldots,z_{E_G}\rangle\in(\Zset_3)^{E_G}\)
describe a cycle in the graph~\(G\), and
\(G[k]=(i,j,\lambda)\).  \(z_k\) is therefore zero
if the edge of index \(k\) is not an element
of the cycle, \(z_k=+1\) if \(G[k]\) is an element and \(z_k=-1\) if the cycle
contains \(G[k]\) against its orientation, i.e.  the cycle contains
\((j, i, \lambda)\).  Similarly to \(A\) one can define a function \(A_R\)
that keeps also track of the path between two vertices,
\begin{equation}
A_R(i, F)=\left\{\begin{array}{rl}
(i, \langle0\ldots0\rangle),&\;F[i]=0\;\text{or}\;F[i]=i\text{,}\\
(i^\prime, r^\prime+R[i]),&\;\text{else, where \((i^\prime, r^\prime)=A_R(F[i], F)\).}
\end{array}\right.
\end{equation}

\begin{algorithm}
\caption{\textsc{Find Loops}(\(G\)).  Returns a set of \(\Zset_3\)-chains
constituting the cycles of the graph \(G\).}
\label{alg:qcd-irreps:find-cycles}
\begin{algorithmic}
\REQUIRE a bubble graph \(G\) as an array of tuples \((i, j, \lambda)\).
\STATE \(L\leftarrow\emptyset\);
	\(F[i]\leftarrow i\quad\forall i\in1\ldots V_G\);
	\(R[i]\leftarrow\langle0,\ldots,0\rangle\quad\forall i\in1\ldots V_G\).

\FORALL{\(g_k=(i, j, \lambda)\in G\)}
	\STATE \((a_i, p_i)\leftarrow A_R(i, F)\);
		\((a_j, p_j)\leftarrow A_R(j, F)\).
	\STATE \(p_{ij}\leftarrow -p_i + p_j +
		\langle \delta_{kl}\vert l=1..E_G\rangle\).
	\IF{\(a_i\neq a_j\)}
		\STATE \(F[a_i]\leftarrow a_j\);
			\(R[a_i]\leftarrow p_{ij}\).
	\ELSE
		\STATE \(L\leftarrow L\cup\{p_{ij}\}\).
	\ENDIF
\ENDFOR
\RETURN \(L\)
\end{algorithmic}
\end{algorithm}

For the reduction algorithm the selection
of the shortest cycle is not essential: for the the algorithm
to terminate any random selection of a edge pairs \((X, Y)\)
that reduces the length of a cycle\footnote{and does not increase
the length of all other cycles} is sufficient. On the other hand the order
in which the reduction is done can well influence the performance of
the algorithm which has not been studied in detail.
\clearpage
%%%% KEEP THIS LAST %%%%%%%%%%%%%%%%%%%%%%%%%%%%%%%%%%%%
\end{fmffile}

%% file: qcd-shelpmethod.tex
\subsection{Introduction}\index{spinor!helicity projection|(}
The \emph{Spinor Helicity Formalism} has proved to be very convenient
for calculation within the framework of massless \ac{qcd}.
Massless fermions and massless gauge bosons have only two
physical degrees of freedom but are represented by objects
with four components, \person{Dirac} spinors and polarisation vectors,
respectively. This mismatch is cured by projecting on helicity
states leading to more compact expressions than traditional
approaches.

Intrinsically this formalism is designed for the four dimensional
case. Therefore, a prescription has to be defined how to extend
helicity amplitudes to $n=4-2\eps$ dimension for to embed them
into the \ac{dreg} scheme; this will be an issue of
Section~\ref{sec:qcddimreg} in Chapter~\ref{chp:qcdnlo}.
The extension to massive particles is described at the end
of this Chapter in Sections~\ref{ssec:qcd-shelpmethod:massivef}
and~\ref{ssec:qcd-shelpmethod:massiveb}.

The equations of motion for a spin-$\frac12$ field with mass~$m$ are
given by the \person{Dirac} equations
\begin{align}\label{eq:shelp:DiracEquation1}
(\pslash-m)u(p)=(\pslash+m)v(p)&=0\quad\text{and}\\
\label{eq:shelp:DiracEquation2}
\bar{u}(p)(\pslash-m)=\bar{v}(p)(\pslash+m)&=0\text{.}
\end{align}
In the massless case, solutions of positive and negative energy are
degenerate\footnote{See for example \cite{Dixon:1996wi}.} which can be seen from the
operators~$(\pslash-m)$ and~$(\pslash+m)$ becoming the same,
i.e.~$(\pslash)$; hence the solutions
\begin{subequations}
\begin{align}
u_\pm(p)&=\Pi_\pm u(p)\quad\text{and}\\
v_\mp(p)&=\Pi_\mp v(p)
\end{align}
\end{subequations}
can be identified, where I use the helicity projection operators
\index{.Pi@$\Pi_\pm$|main}
\begin{equation}
\Pi_\pm\equiv\frac12\left(\One\pm\gamma_5\right)\text{.}
\end{equation}
I use the common bracket notation~\cite{Xu:1987}
\index{.p@$\ket{p_\pm}$, $\bra{p_\pm}$, $\braket{pq}$|main}%
\index{.qp@$\cbraket{qp}$}
\begin{subequations}
\begin{align}
\ket{p_\pm}&\equiv u_\pm(p)=v_\mp(p)\quad\text{and}\\
\bra{p_\pm}&\equiv \bar{u}_\pm(p)=\bar{v}_\mp(p)\quad\text{together with}\\
\braket{pq}&\equiv\braket{p_-\vert q_+}\quad\text{and}\\
\label{eq:qcd-shelpmethod:conj}%
\cbraket{pq}&\equiv\braket{p_+\vert q_-}=
\mathrm{sgn}(p\cd q)\,\braket{qp}^\ast
\end{align}
\end{subequations}
In the literature the extra sign in the last equation is usually omitted
since it becomes essential only for non-physical kinematics.

The orthogonality of the projectors~$\Pi_\pm$ leads to
the annihilation of all other products
\begin{equation}
\braket{p_+\vert q_+}=\braket{p_-\vert q_-}=0\text{;}
\end{equation}
the completeness relation reads in this notation as
\begin{equation}
\label{eq:qcd-shelpmethod:completeness}%
\ket{p_+}\bra{p_+}+\ket{p_-}\bra{p_-}=\pslash\text{,}
\end{equation}
and hence
\begin{equation}\label{eq:shelp:CompletenessProjection}
\ket{p_\pm}\bra{p_\pm}=\Pi_\pm\pslash\text{.}
\end{equation}
One can conclude that
\begin{multline}
\label{eq:qcd-shelpmethod:sqr}
\braket{pq}\cbraket{qp}=\braket{p_-\vert q_+}\braket{q_+\vert p_-}=
\bar{u}(p)\Pi_+\Pi_+\qslash\Pi_-u(p)=\\
\tr{\pslash\Pi_+\Pi_+\qslash\Pi_-}
=\tr[+]{\pslash\qslash}=2\,p\cd q\text{.}
\end{multline}
Here I make use of the notation\index{.trPM@$\tr[\pm]{\ldots}$}
\begin{equation}
\tr[\pm]{\Gamma}=\tr{\Pi_\pm\Gamma}\text{.}
\end{equation}
Equations \eqref{eq:qcd-shelpmethod:conj} and \eqref{eq:qcd-shelpmethod:sqr}
determine the spinor product $\braket{pq}$ up to a phase; hence, after a certain phase
choice these products, which evaluate to a complex number,
can be computed numerically.
\index{Spinor helicity projections|)}
%%%%%%%%%%%%%%%%%%%%%%%%%%%%%%%%%%%%%%%%%%%%%%%%%%%%%%%%%%%%%%%%%%%%%%%%%
\subsection{Choosing a Representation}
%%%%%%%%%%%%%%%%%%%%%%%%%%%%%%%%%%%%%%%%%%%%%%%%%%%%%%%%%%%%%%%%%%%%%%%%%
Not only for numerical calculations but also to find a
proper fixing of the phase~$\phi_{pq}$ in the defining equation
\begin{equation}
\braket{pq}=\sqrt{\vert (pq)\vert}e^{i\phi_{pq}}
\end{equation}
a certain basis choice is very convenient.

Following~\cite{Kleiss:1985yh,Jadach:1998wp},
a pair of \emph{basic spinors} is introduced:
two four-vectors $\zeta$ and $\eta$
are chosen such that\footnote{In the original paper $\zeta$ and $\eta$
are called $k^0$ and $k^1$ respectively.}
\begin{equation}\label{eq:shelp:basisvectorconditions}
\zeta^2=0,\qquad\eta^2=-1\quad\text{and}\, \zeta\cd\eta=0\text{.}
\end{equation}
Once the spinor $\ket{\zeta_-}$ is defined and obeys
\begin{equation}
\ket{\zeta_-}\bra{\zeta_-}=\Pi_-\fmslash{\zeta}
\end{equation}
it is easy to show that for the positive-helicity state $\ket{\zeta_+}$
the definition
\begin{equation}
\ket{\zeta_+}=\fmslash{\eta}\ket{\zeta_-}
\end{equation}
suffices the \person{Dirac} equation
and~\eqref{eq:shelp:CompletenessProjection}.

Hence, for all lightlike four-vectors~$p$ with~$p\cd\zeta\neq0$
one can define the corresponding spinors as
\begin{equation}
\ket{p_\pm}=\frac{\fmslash{p}}{\sqrt{2p\cd\zeta}}\ket{\zeta_\mp}
\quad\text{and}\quad
\bra{p_\pm}=\bra{\zeta_\mp}\frac{\fmslash{p}}{\sqrt{2p\cd\zeta}}
\text{.}
\end{equation}
The important detail about the definition is that the bra- and ket-spinors
are not exactly conjugates of each other because for negative~$p\cd\zeta$
the denominator becomes imaginary; this feature, however, is important in
order to preserve~\eqref{eq:qcd-shelpmethod:sqr}.

The representation in terms of the reference vectors $\zeta$ and $\eta$
now can be used to give an expression for the spinor products
$\cbraket{pq}$ and $\braket{pq}$ of two lightlike vectors $p$ and~$q$.
\begin{multline}\label{eq:shelp:cbracketsign}
\cbraket{pq}=\braket{p_+\vert q_-}=
\frac{\braket{\zeta_-\vert\fmslash{p}\fmslash{q}\vert\zeta_+}}{%
\sqrt{2p\cd\zeta}\sqrt{2q\cd\zeta}}=
\frac{\tr[-]{\fmslash{\zeta}\fmslash{p}\fmslash{q}\fmslash{\eta}}}{%
2\sqrt{p\cd\zeta}\sqrt{q\cd\zeta}}=\\
\frac{1}{\sqrt{p\cd\zeta}\sqrt{q\cd\zeta}}
\left((p\cd\zeta)(q\cd\eta)-(p\cd\eta)(q\cd\zeta)
-i\epsilon_{\mu\nu\rho\sigma}\zeta^\mu p^\nu q^\rho\eta^\sigma\right)
=-\cbraket{qp}\text{,}
\end{multline}
and in analogy we obtain
\begin{multline}\label{eq:shelp:bracketsign}
\braket{qp}=
\frac{1}{\sqrt{q\cd\zeta}\sqrt{p\cd\zeta}}
\left((p\cd\zeta)(q\cd\eta)-(p\cd\eta)(q\cd\zeta)
+i\epsilon_{\mu\nu\rho\sigma}\zeta^\mu p^\nu q^\rho\eta^\sigma\right)=\\
=-\braket{pq}=\mathrm{sgn}(p\cd\zeta)\mathrm{sgn}(q\cd\zeta)\,\cbraket{pq}^\ast
\text{.}
\end{multline}
The signum functions arise from the ratio
\begin{equation}
\frac{\sqrt{p\cd\zeta}\,^\ast}{\sqrt{p\cd\zeta}}=\left\{
\begin{array}{lcrl}
\sqrt{\vert p\cd\zeta\vert}/\sqrt{\vert p\cd\zeta\vert}&=&
1&\text{if } p\cd\zeta>0\\
-i\sqrt{\vert p\cd\zeta\vert}/i\sqrt{\vert p\cd\zeta\vert}
&=&-1&\text{if } p\cd\zeta<0
\end{array}\right\}=\mathrm{sgn}(p\cd\zeta)
\end{equation}
The product of signs in the last line of~\eqref{eq:shelp:bracketsign}
can be rewritten\footnote{
It is assumed that none of the products vanish.} as
\begin{equation}\label{eq:shp:cauchyschwarz}
\mathrm{sgn}(p\cd\zeta)\mathrm{sgn}(q\cd\zeta)=\mathrm{sgn}(p\cd q)
\end{equation}
since we can represent
\begin{displaymath}
\zeta=\frac{\zeta^0}{\vert\vec{\zeta}\vert}(\vert\vec{\zeta}\vert,\vec{\zeta})
\end{displaymath}
and analogously $p$ and~$q$. One can work out all dot products and apply
the \person{Cauchy}-\person{Schwarz} inequality on the three-dimensional
scalar products to obtain Equation~\eqref{eq:shp:cauchyschwarz} above
in terms of the zero components only.
Similarly, it follows from Equations \eqref{eq:shelp:cbracketsign}
and~\eqref{eq:shelp:bracketsign}~that
\begin{equation}
\braket{-p,q}=\braket{p,-q}=i\braket{pq}\quad\text{and }
\cbraket{-p,q}=\cbraket{p,-q}=i\cbraket{pq}\text{.}
\end{equation}

For a numerical evaluation it is helpful to relate these expressions
directly to the components of the vectors $p$ and~$q$. To achieve this
any choice of $\zeta$ and $\eta$ subjected
to~\eqref{eq:shelp:basisvectorconditions} can be~used:
\begin{subequations}
\begin{align}
\zeta^\mu&=(1,0,0,1)\\
\eta^\mu&=(0,1,0,0)
\end{align}
\end{subequations}
To allow for a compact notation, the abbreviations
$p_\pm=p^0\pm p^3$ and $p_\perp=p^1+ip^2$ are used,
leading to
\begin{equation}
\braket{pq}=\sqrt{\frac{q_-}{p_-}}p_\perp^\ast
-\sqrt{\frac{p_-}{q_-}}q_\perp^\ast
=\sqrt{2\vert p\cd q\vert}e^{i\phi_{pq}}
\end{equation}
with a phase~$e^{i\phi_{pq}}$ which is characterised by
\begin{subequations}
\begin{align}
\cos\phi_{pq}&=\frac{q_-p^1-p_-q^1}{\sqrt{2\vert p\cd q\vert p_-q_-}}
\quad\text{and}\\
\sin\phi_{pq}&=\frac{q_-p^2-p_-q^2}{\sqrt{2\vert p\cd q\vert p_-q_-}}
\text{.}
\end{align}
\end{subequations}
It should be noted that this phase is not \person{Lorentz} invariant.
This can be seen, for example, from a rotation round the $z$-axis by
an angle~$\alpha$, which leads to an additional phase
\begin{equation}
\braket{p^\prime q^\prime}=\sqrt{\vert(pq)\vert}e^{i\phi_{pq}}e^{-i\alpha}
=e^{-i\alpha}\braket{pq}\text{.}
\end{equation}

\subsection{External Massless Gauge Bosons}
%%%%%%%%%%%%%%%%%%%%%%%%%%%%%%%%%%%%%%%%%%%%%%%%%%%%%%%%%%%%%%%%%%%%%%%%%%
\label{ssec:qcd-shelpmethod:extgluons}
\index{polarisation vector!for massless gauge bosons}
The method of \acp{shp} has been introduced to provide a compact
representation of amplitudes. So far we have only regarded external fermions.
In general however, amplitudes containing external gauge bosons are 
needed as well.
This requires an appropriate representation of the polarisation
vectors~$\varepsilon_\pm^\mu(k)$ belonging to a gauge boson of momentum~$k$,
which is introduced according to~\cite{Xu:1987}.
The~definition
\begin{subequations}
\label{eq:qcd-shelpmethod:defepsilon}
\begin{align}
\varepsilon_+^\mu(q,k)&=\frac{\braket{q_-\vert\gamma^\mu\vert k_-}}{\sqrt{2}\braket{qk}}\text{,}\\
\varepsilon_-^\mu(q,k)&=\frac{\braket{q_+\vert\gamma^\mu\vert k_+}}{\sqrt{2}\cbraket{kq}}
\end{align}
\end{subequations}
with $q$ being lightlike represents a polarisation vector in an axial gauge
and hence the completeness relation
\begin{equation}
\label{eq:qcd-shelpmethod:gaugerel}
\varepsilon_+^\mu(q,k)\left(\varepsilon_+^\nu(q,k)\right)^\ast+
\varepsilon_-^\mu(q,k)\left(\varepsilon_-^\nu(q,k)\right)^\ast= 
-g^{\mu\nu}+\frac{k^\mu q^\nu+q^\mu k^\nu}{k\cd q}
\end{equation}
must hold\footnote{See for example~\cite{Dixon:1996wi}}.

To show that~\eqref{eq:qcd-shelpmethod:defepsilon} is a valid definition of a polarisation vector a few
auxiliary relations are needed. The \person{Gordon} identity
\index{Gordon identity@\person{Gordon} identity}
\begin{equation}
\label{eq:qcd-shelpmethod:gordon}
\braket{p_\pm\vert\gamma^\mu\vert p_\pm}=\tr[\pm]{\gamma^\mu\pslash}=2\,p^\mu
\end{equation}
can be used to express lightlike vectors in terms of spinor strings.

The second required ingredient is a \person{Fierz} rearrangement formula,
\index{Fierz identity@\person{Fierz} identity}
\begin{equation}
\label{eq:qcd-shelpmethod:fierz}
\braket{p_+\vert\gamma^\mu\vert q_+}
\braket{r_+\vert\gamma_\mu\vert s_+}=2\cbraket{pr}\braket{sq}\text{,}
\end{equation}
where $p$, $q$, $r$ and $s$ are lightlike vectors.
To show this identity the left hand side can be completed to a single trace by
insertion of
\begin{equation}\label{eq:shelpmethod:provefierz1}
1=% 
\frac{\braket{q_+\vert\fmslash{m}\vert r_+}}{\cbraket{qm}\braket{mr}}
\frac{\braket{s_+\vert\fmslash{n}\vert p_+}}{\cbraket{sn}\braket{np}}\text{,}
\end{equation}
where again $m$ and $n$ are arbitrary, lightlike momenta. Here I use
the fact that
\begin{multline}\label{eq:shelpmethod:spintotrace1}
\braket{q_+\vert\fmslash{m}\vert r_+}=
\bra{q_+}\left(\ket{m_+}\bra{m_+}+\ket{m_-}\bra{m_-}\right)\ket{r_+}=\\
\braket{q_+\vert m_-}\braket{m_-\vert r_+}=
\cbraket{qm}\braket{mr}
\end{multline}
and for any string~$\Gamma$ of \person{Dirac}~matrices
\begin{equation}\label{eq:shelpmethod:spintotrace2}
\braket{q_\pm\vert\Gamma\vert q_\pm}=
\tr{\braket{q_\pm\vert\Gamma\vert q_\pm}}=
\tr{\Gamma\ket{q_\pm}\bra{q_\pm}}=
\tr{\Gamma\Pi_\pm\fmslash{q}}=
\tr[\pm]{\fmslash{q}\Gamma}\text{.}
\end{equation}

The left hand side of~\eqref{eq:shelpmethod:provefierz1} now reads
\begin{multline}\label{eq:qcd-shelpmethod:provefierzefinal}
\frac{\braket{p_+\vert\gamma^\mu\vert q_+}% 
\braket{q_+\vert\fmslash{m}\vert r_+}
\braket{r_+\vert\gamma_\mu\vert s_+}
\braket{s_+\vert\fmslash{n}\vert p_+}}{\cbraket{qm}\braket{mr}\cbraket{sn}\braket{np}}
=
\frac{\tr[+]{\pslash\gamma^\mu\qslash\fmslash{m}\fmslash{r}\gamma_\mu\fmslash{s}
\fmslash{n}}}{\cbraket{qm}\braket{mr}\cbraket{sn}\braket{np}}=\\
\frac{-2\tr[+]{\pslash\fmslash{r}\fmslash{m}\qslash\fmslash{s}
\fmslash{n}}}{\cbraket{qm}\braket{mr}\cbraket{sn}\braket{np}}=
%\frac{-2\tr[+]{\fmslash{n}\fmslash{s}\qslash\fmslash{m}\fmslash{r}\pslash 
%}}{\cbraket{qm}\braket{mr}\cbraket{sn}\braket{np}}=\\
\frac{-2\cbraket{sn}\braket{np}\cbraket{pr}\braket{rm}\cbraket{mq}\braket{qs}}{
\cbraket{qm}\braket{mr}\cbraket{sn}\braket{np}}=\\
-2\cbraket{pr}\braket{qs}=2\cbraket{pr}\braket{sq}\text{.}
\end{multline}
In this derivation I made use of the relation
$\gamma^\mu\gamma^\nu\gamma^\rho\gamma^\sigma\gamma_\mu=
-2\gamma^\sigma\gamma^\rho\gamma^\nu$,
which is only valid in four dimensions. In $n$ dimensions the additional term
$(4-n) \gamma^\nu\gamma^\rho\gamma^\sigma$ arises, which 
changes the right hand side of
Equation~\eqref{eq:qcd-shelpmethod:provefierzefinal}
to $(4-n)\cbraket{rs}\braket{pq}$.
Similarly\footnote{In fact the prove is simpler since an insertion of
$\cbraket{qr}\braket{sp}$ already completes the trace},
the analog of~\eqref{eq:qcd-shelpmethod:fierz} with different signs
can be proved,
\begin{equation}\label{eq:shp:FierzPM}
\braket{p_+\vert\gamma^\mu\vert q_+}
\braket{r_-\vert\gamma_\mu\vert s_-}=2\cbraket{ps}\braket{rq}\text{.}
\end{equation}

Equations~\eqref{eq:qcd-shelpmethod:gordon} and \eqref{eq:qcd-shelpmethod:fierz}
together prove the orthogonality condition
\begin{equation}
\varepsilon_\pm^\mu(q,k) k_\mu\propto
\braket{q_\mp\vert\gamma^\mu\vert k_\mp}
\braket{k_\mp\vert\gamma_\mu\vert k_\mp}=-2\cbraket{qk}\braket{kk}=0
\end{equation}
and by similar arguments $(\varepsilon_\pm)^2=\varepsilon_\pm\cd q=0$ can be shown.

Finally, equation \eqref{eq:qcd-shelpmethod:gaugerel} has to be proved. 
From the definition of $\varepsilon_\pm^\mu(q,k)$ one can read off immediately that
$(\varepsilon_\pm^\mu)^\ast=\varepsilon_\mp^\mu$, and hence the left hand side
of \eqref{eq:qcd-shelpmethod:gaugerel} reads
\begin{displaymath}
\varepsilon_+^\mu(q,k)\varepsilon_-^\nu(q,k)+
\varepsilon_-^\mu(q,k)\varepsilon_+^\nu(q,k)\text{.}
\end{displaymath}
Substituting the definition of~$\varepsilon$
into Equation~\eqref{eq:qcd-shelpmethod:defepsilon}
and using the charge conjugation relation for vector currents,
$\braket{q_\mp\vert\gamma^\mu\vert k_\mp}=\braket{k_\pm\vert\gamma^\mu\vert q_\pm}$,
both terms can be rewritten as traces, which evaluate to
\begin{equation}
\varepsilon_+^\mu(q,k)\varepsilon_-^\nu(q,k)=
\frac12\left(-g^{\mu\nu}+\frac{k^\mu q^\nu+q^\mu k^\nu}{k\cd q}\right)
-\frac{i}2\varepsilon^{\rho\mu\sigma\nu}k_\rho q_\sigma
\end{equation}
and the respective term with $\mu$ and $\nu$ exchanged.
Adding both terms up one reproduces
Equation~\eqref{eq:qcd-shelpmethod:gaugerel}.

As a further result one can show that for any \person{Lorentz} contractions 
of polarisation vectors of same helicity
one can always find a gauge choice such that the dot product reduces to
a pure phase. Using
a suitable choice for the auxiliary vectors together with the \person{Fierz}
identity~\eqref{eq:qcd-shelpmethod:fierz}, one achieves the form
\begin{equation}
\varepsilon_\pm(k_i, q_i=k_j)\cd\varepsilon_\pm(k_j, q_j=k_i)=
\left(\frac{\braket{k_ik_j}}{\cbraket{k_ik_j}}\right)^{\pm1}=
e^{2i\phi_{ij}}\text{.}
\end{equation}
For the case of mixed helicities in this gauge choice the product
vanishes, that is
\begin{equation}
\varepsilon_\pm(k_i, q_i=k_j)\cd\varepsilon_\mp(k_j, q_j=k_i)=
0\text{.}
\end{equation}

%%%%%%%%%%%%%%%%%%%%%%%%%%%%%%%%%%%%%%%%%%%%%%%%%%%%%%%%%%%%%%%%%%%%%%%%%%
\subsection{Massive Fermions}
\label{ssec:qcd-shelpmethod:massivef}
%%%%%%%%%%%%%%%%%%%%%%%%%%%%%%%%%%%%%%%%%%%%%%%%%%%%%%%%%%%%%%%%%%%%%%%%%%
\index{spinor!massive}
For massive fermions the fields obey the massive \person{Dirac} equation
and the spinors $u(p)$ and $v(p)$ have to be distinguished.
A common notation for both can be achieved, following~\cite{Tanaka:1989gu},
by introducing an additional index~$\rho=\pm1$
to distinguish the solutions of the equations:
\begin{equation}\label{eq:shp:DiracWithRho}
(\fmslash{p}-\rho m)\ket{p_\lambda^\rho}=0\quad\text{and}\quad
\bra{p_\lambda^\rho}(\fmslash{p}-\rho m)=0\text{.}
\end{equation}

The construction
\begin{subequations}
\begin{align}\label{eq:shp:SolutionWithRho}
\ket{p_\lambda^\rho}&=\frac{1}{\sqrt{2p\cd\zeta}}
(\fmslash{p}+\rho m)\ket{\zeta_{-\lambda}}\qquad\text{and}\\
\label{eq:shp:SolutionWithRho2}
\bra{p_\lambda^\rho}&=\frac{1}{\sqrt{2p\cd\zeta}}\bra{\zeta_{-\lambda}}
(\fmslash{p}+\rho m)
\end{align}
\end{subequations}
%%%%%% Maybe write also about:
%\nocite{Ballestrero:1994jn}%
clearly obeys~\eqref{eq:shp:DiracWithRho} and hence it remains to be
shown that the solutions form a complete set. From direct calculation
one gets
\begin{equation}
\ket{p_\lambda^\rho}\bra{p_\lambda^\rho}=
\frac12(\One+\lambda\rho\gamma_5\fmslash{s})(\fmslash{p}+\rho m)\text{,}
\end{equation}
where
\begin{equation}
s^\mu=\frac{1}{m}p^\mu-\frac{m}{p\cd\zeta}\zeta^\mu,\quad s^2=-1\text{,}
\end{equation}
and hence summing up the helicities~$\lambda=\pm1$ leads to the required
result.

Another way of constructing massive spinors is by decomposing $p$ into a sum
of two lightlike vectors $p=p_1+p_2$, $p_1^2=p_2^2=0$. Given an arbitrary
lightlike vector~$\zeta$ this can always be achieved by choosing
\begin{equation}
p_1=\frac{m^2}{2p\cd\zeta}\zeta\qquad\text{and}\qquad p_2=p-p_1\text{.}
\end{equation}
Then $p_1\cd p_2=p\cd p_1=p\cd p_2=m^2/2$ and one can
rewrite~\eqref{eq:shp:SolutionWithRho} as
\begin{multline}
\ket{p_\lambda^\rho}=\frac{1}{\sqrt{2p\cd p_1}}(\fmslash{p}+\rho m)
\ket{{p_1}_{-\lambda}}=
\frac{1}{\sqrt{2p\cd p_1}}(\fmslash{p_1}+\fmslash{p_2}+\rho m)
\ket{{p_1}_{-\lambda}}=\\
\frac{1}{\sqrt{2p\cd p_1}}\fmslash{p_1}\ket{{p_1}_{-\lambda}}
+\frac{1}{\sqrt{2p\cd p_2}}\fmslash{p_2}\ket{{p_1}_{-\lambda}}
+\rho\ket{{p_1}_{-\lambda}}=
\ket{{p_2}_{\lambda}}
+\rho\ket{{p_1}_{-\lambda}}\text{.}
\end{multline}

%%%%%%%%%%%%%%%%%%%%%%%%%%%%%%%%%%%%%%%%%%%%%%%%%%%%%%%%%%%%%%%%%%%%%%%%%%
\subsection{Massive Gauge Bosons}
\label{ssec:qcd-shelpmethod:massiveb}
%%%%%%%%%%%%%%%%%%%%%%%%%%%%%%%%%%%%%%%%%%%%%%%%%%%%%%%%%%%%%%%%%%%%%%%%%%
\index{polarisation vector!for massive gauge bosons}
Although the treatment of massive gauge bosons is not part of my thesis
I want to describe two extensions of the massless spinor helicity formalism
for these cases.

The authors of~\cite{Kleiss:1985yh} also describe a formalism that deals
with massive vector bosons by turning the helicity sum into a Monte-Carlo
integral. The idea is, similar to the case of fermions, to split the
massive momentum~$q$ into a sum of lightlike four-vectors~$q=q_1+q_2$
and to introduce the polarisation vector
\begin{equation}
a^\mu=\frac{\braket{{q_1}_-\vert\gamma^\mu\vert{q_2}_-}}{\sqrt{2}m}
\end{equation}
without further constraints. The cross section is still reproduced correctly
if one replaces the spin sum by the integral
\begin{equation}
\sum\epsilon^\mu(\epsilon^\nu)^\ast\rightarrow
\int\!\!\frac{\diff[2]\Omega}{4\pi/3}a^\mu (a^\nu)^\ast=
-g^{\mu\nu}+\frac{q^\mu q^\nu}{m^2}\text{.}
\end{equation}
While this approach is very well suited for a direct numerical evaluation
of unpolarised amplitudes, for an analytical result another technique
should be used where the polarisations become accessible directly.
Formula for such an approach are given in~\cite{Nanava:2003xh} but should
also be proved in what follows.

The candidates for the polarisation vectors for a massive gauge boson 
of momentum $q^\mu=q_1^\mu+q_2^\mu$, $q^2=m^2$ and $q_1^2=q_2^2=0$ are
as follows
\begin{subequations}
\begin{align}
\varepsilon_\pm^\mu(q,m)&=
\frac{\braket{{q_1}_\pm\vert\gamma^\mu\vert{q_2}_\pm}}{\sqrt{2}m}\text{,}\\
\varepsilon_0^\mu(q,m)&=
\frac{\braket{{q_1}_+\vert\gamma^\mu\vert{q_1}_+}}{2m}
-\frac{\braket{{q_2}_+\vert\gamma^\mu\vert{q_2}_+}}{2m}
=\frac{q_1^\mu-q_2^\mu}{m}\text{.}
\end{align}
\end{subequations}
The structure of $\varepsilon_\pm$ is very similar to the case of massless
bosons, and hence the proves for 
$\varepsilon_\pm\cd q=\varepsilon_\pm\cd\varepsilon_pm=0$ and
$(\varepsilon_\pm)^\ast=\varepsilon_\mp$ are just as above.
Hence we need to show that 
\begin{subequations}
\begin{align}
\label{eq:shp:massiveGaugeCond1}
\varepsilon_0\cd q&=0\text{,}\\
\label{eq:shp:massiveGaugeCond2}
\varepsilon_\pm\cd\varepsilon_\mp&=-1\text{,}\\
\label{eq:shp:massiveGaugeCond3}
\varepsilon_0\cd\varepsilon_0&=-1\quad\text{and}\\
\label{eq:shp:massiveGaugeCond4}
\varepsilon_\pm\cd\varepsilon_0&=0\text{.}
\end{align}
\end{subequations}

For the first equation it suffices to split $q$ into $q_1$ and $q_2$; half
of the terms vanish due to $\fmslash{q_i}\ket{{q_i}_+}=0$. For the remaining
terms we get
\begin{multline}
2m\varepsilon_0\cd q=
\braket{{q_1}_+\vert\fmslash{q_2}\vert{q_1}_+}
-\braket{{q_2}_+\vert\fmslash{q_1}\vert{q_2}_+}=\\
\cbraket{q_2q_1}\braket{q_1q_2}-\cbraket{q_2q_1}\braket{q_1q_2}
=2q_1\cd q_2-2q_1\cd q_2=0\text{.}
\end{multline}

To prove~\eqref{eq:shp:massiveGaugeCond4} one can use the \person{Fierz}
identity, which in this case for each term yields one factor of the form
$\braket{q_iq_i}$ or $\cbraket{q_iq_i}$. Similarly,
in~\eqref{eq:shp:massiveGaugeCond3} after applying the \person{Fierz}
identity all but two terms cancel, and one finds
\begin{equation}
\varepsilon_0\cd\varepsilon_0=-2\frac{\cbraket{q_1q_2}\braket{q_1q_2}}{4m^2}
=-1
\end{equation}
since $2q_1\cd q_2=m^2$. In the same spirit one can
evaluate~\eqref{eq:shp:massiveGaugeCond2} by applying~\eqref{eq:shp:FierzPM}.

Finally we have to establish that the spin sum
is complete. We can expand out
\begin{multline}
\varepsilon_+^\mu(\varepsilon_+^\nu)^\ast=\frac1{2m^2}
\braket{{q_1}_+\vert\gamma^\mu\vert{q_2}_+}
\braket{{q_2}_+\vert\gamma^\mu\vert{q_1}_+}=
\frac1{2m^2}\tr[+]{\fmslash{q}_1\gamma^\mu\fmslash{q}_2\gamma^\nu}=\\
\frac2{2m^2}\left(q_1^\mu q_2^\nu+q_1^\nu q_2^\mu-\frac{m^2}2g^{\mu\nu}\right)
=\varepsilon_-^\mu(\varepsilon_-^\nu)^\ast
\end{multline}
and do the same for
\begin{multline}
\varepsilon_0^\mu(\varepsilon_0^\nu)^\ast=
\frac1{4m^2}\left(
\braket{{q_1}_+\vert\gamma^\mu\vert{q_1}_+}
\braket{{q_1}_+\vert\gamma^\nu\vert{q_1}_+}\right.\\
\left.
+\braket{{q_2}_+\vert\gamma^\mu\vert{q_2}_+}
\braket{{q_2}_+\vert\gamma^\nu\vert{q_2}_+}
-\braket{{q_1}_+\vert\gamma^\mu\vert{q_1}_+}
\braket{{q_2}_+\vert\gamma^\nu\vert{q_2}_+}\right.\\
\left.
-\braket{{q_2}_+\vert\gamma^\mu\vert{q_2}_+}
\braket{{q_1}_+\vert\gamma^\nu\vert{q_1}_+}
\right)\\
\shoveleft=
\frac1{4m^2}\left(
\tr[+]{\fmslash{q}_1\gamma^\mu\fmslash{q}_1\gamma^\nu}
+\tr[+]{\fmslash{q}_2\gamma^\mu\fmslash{q}_2\gamma^\nu}
-\tr[+]{\fmslash{q}_1\gamma^\mu}\tr[+]{\fmslash{q}_2\gamma^\nu}\right.\\
\left.
-\tr[+]{\fmslash{q}_2\gamma^\mu}\tr[+]{\fmslash{q}_1\gamma^\nu}
\right)=
\frac1{m^2}\left(
q_1^\mu q_1^\nu + q_2^\mu q_2^\nu - q_1^\mu q_2^\nu - q_2^\mu q_1^\nu
\right)\text{.}
\end{multline}
Gathering all terms the completeness comes out as expected
\begin{equation}
\varepsilon_+^\mu(\varepsilon^\nu_+)^\ast
+\varepsilon_-^\mu(\varepsilon^\nu_-)^\ast
+\varepsilon_0^\mu(\varepsilon^\nu_0)^\ast=
-g^{\mu\nu}+\frac{q^\mu q^\nu}{m^2}\text{.}
\end{equation}

%%%%%%%%%%%%%%%%%%%%%%%%%%%%%%%%%%%%%%%%%%%%%%%%%%%%%%%%%%%%%%%%%%%%%%%%%%
\subsection{Amplitude Representation}
\begin{figure}[hbtp]
\begin{center}
\includegraphics[scale=0.5,bb=0 0 189 148]{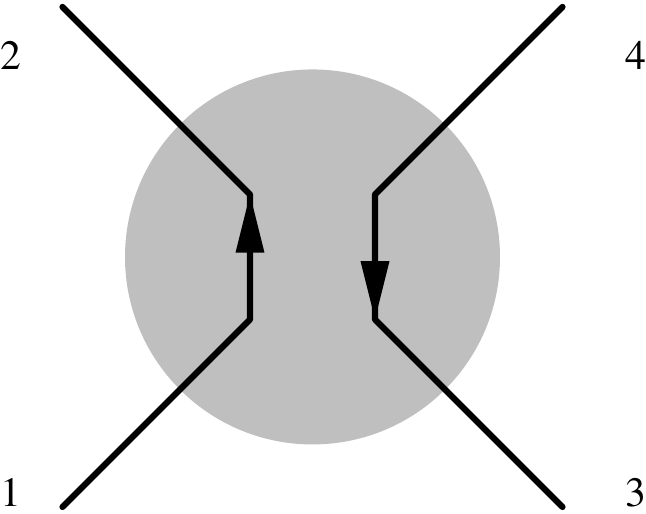}
\end{center}
\caption{Schematic picture of the amplitude $q\bar{q}\rightarrow q^\prime\bar{q}^\prime$.}
\label{fig:qcd-shelpmethod:qqqq-schematic}
\end{figure}

%%%%%%%%%%%%%%%%%%%%%%%%%%%%%%%%%%%%%%%%%%%%%%%%%%%%%%%%%%%%%%%%%%%%%%%%%%
The formalism of \acp{shp} is a very powerful technique for the calculation of
matrix elements because it splits every amplitude in a natural way into gauge invariant
pieces. In principle these subamplitudes are observable in an experiment where all helicities
of the initial particles can be prepared and those of the final state particles are measured.
One generates the subamplitudes by inserting helicity projection operators in
every spinor line. As example we shall have a look at the
$q\bar{q}\rightarrow q^\prime\bar{q}^\prime$ amplitude, of which a sketch is shown in
figure \ref{fig:qcd-shelpmethod:qqqq-schematic}.
The expression for the amplitude~$\mathcal{A}(1,2,3,4)$ has the generic form\footnote{
The complete tensor structure is represented by the direct product~$\otimes$. $\Gamma_{12}$ and $\Gamma_{34}$
are two strings of \person{Dirac} gamma matrices,
and~$\tau$ is the remaining, momentum dependent part of the amplitude.}
\begin{equation}
\mathcal{A}(1,2,3,4)=\bar{v}(p_2)\Gamma_{12}u(p_1)\otimes\tau\otimes\bar{u}(p_3)\Gamma_{34}v(p_4)\text{.}
\end{equation}
I abbreviate the momenta by their indices where this is unambiguous.
We split this expression into subamplitudes by the help of $\One=\Pi_++\Pi_-$,
\begin{multline}
\mathcal{A}(1,2,3,4)=
 \bar{v}(p_2)\Gamma_{12}\Pi_+u(p_1)\otimes\tau\otimes\bar{u}(p_3)\Gamma_{34}\Pi_+v(p_4)\\
+\bar{v}(p_2)\Gamma_{12}\Pi_-u(p_1)\otimes\tau\otimes\bar{u}(p_3)\Gamma_{34}\Pi_+v(p_4)\\
+\bar{v}(p_2)\Gamma_{12}\Pi_+u(p_1)\otimes\tau\otimes\bar{u}(p_3)\Gamma_{34}\Pi_-v(p_4)\\
+\bar{v}(p_2)\Gamma_{12}\Pi_-u(p_1)\otimes\tau\otimes\bar{u}(p_3)\Gamma_{34}\Pi_-v(p_4)\\
\shoveleft{
=\braket{p_2^+\vert\Gamma_{12}\vert p_1^+}\otimes\tau\otimes\braket{p_3^+\vert\Gamma_{34}\vert p_4^+}
+\braket{p_2^-\vert\Gamma_{12}\vert p_1^-}\otimes\tau\otimes\braket{p_3^+\vert\Gamma_{34}\vert p_4^+}}\\
+\braket{p_2^+\vert\Gamma_{12}\vert p_1^+}\otimes\tau\otimes\braket{p_3^-\vert\Gamma_{34}\vert p_4^-}
+\braket{p_2^-\vert\Gamma_{12}\vert p_1^-}\otimes\tau\otimes\braket{p_3^-\vert\Gamma_{34}\vert p_4^-}\\
\shoveleft{\equiv\mathcal{A}^{++++}(1,2,3,4)+\mathcal{A}^{--++}(1,2,3,4)
+\mathcal{A}^{++--}(1,2,3,4)+\mathcal{A}^{----}(1,2,3,4)}\text{.}
\end{multline}
The signs in this notation correspond to the signs of the helicity projectors;
for antifermions the sign is different from the physical helicity.
Further simplifications can be achieved by parity invariance:
for any helicity subamplitude in \ac{qcd},
the action of the parity operator $\mathcal{P}$ is
\begin{equation}
\mathcal{A}^{\lambda_1,\lambda_2,\ldots}(1,2,\ldots)=
\mathcal{P}\mathcal{A}^{\lambda_1,\lambda_2,\ldots}(1,2,\ldots)=
\mathcal{A}^{-\lambda_1,-\lambda_2,\ldots}(\mathcal{P}1,\mathcal{P}2,\ldots)\text{,}
\end{equation}
where $\mathcal{P}i=\mathcal{P}p_i=(p_i^0, -p_i^1, -p_i^2, -p_i^3)$
is the parity conjugated of the momentum vector~$p_i$. 
This allows the computation of the whole amplitude from the knowledge of
just two helicity amplitudes, $\mathcal{A}^{++++}$ and~$\mathcal{A}^{++--}$:
\begin{multline}
\mathcal{A}(1,2,3,4)=\mathcal{A}^{++++}(1,2,3,4)+
\mathcal{A}^{++--}(\mathcal{P}1,\mathcal{P}2,\mathcal{P}3,\mathcal{P}4)\\
+\mathcal{A}^{++--}(1,2,3,4)+\mathcal{A}^{++++}(\mathcal{P}1,\mathcal{P}2,\mathcal{P}3,\mathcal{P}4)\text{.}
\end{multline}

For an algebraic treatment of the amplitude one might want to work
with traces, i.e. with polynomials in the \person{Mandelstam} variables
rather than spinor products; this can be achieved using
a lightlike auxiliary vector~$m$
when applying~\eqref{eq:shelpmethod:spintotrace1}
and~\eqref{eq:shelpmethod:spintotrace2} to
\begin{equation}
\braket{p_2^+\vert\Gamma_{12}\vert p_1^+}=
\frac{\braket{p_1^+\vert\fmslash{m}\vert p_2^+}}{\cbraket{p_1m}\braket{mp_2}}
\braket{p_2^+\vert\Gamma_{12}\vert p_1^+}=
\frac{\tr[+]{\fmslash{p_1}\fmslash{m}\fmslash{p_2}\Gamma_{12}}}{
2\sqrt{\vert p_1\cd m\vert}\sqrt{\vert p_2\cd m\vert}
e^{i\phi_{mp_2}}e^{-i\phi_{mp_1}}}\text{.}
\end{equation}
The advantage of this method is the simpler treatment of the resulting
expressions where the coefficients of the integrals are expressed in
terms of \person{Mandelstam} variables only.
An alternative approach is based on the use of spinor products instead
of \person{Mandelstam} variables~\cite{Pittau:1996ez}: every appearance
of an integration momentum $\fmslash{q}_a$ in the numerator can be extracted
from the spinor line by the use of
\begin{equation}
\label{eq:qcd-shelpmethod:extractq1}
\qslash[a]=\frac{1}{2k_i\cd k_j}\left(
2q_a\cd k_i\,\kslash[j]
+2q_a\cd k_j\,\kslash[i]
-\kslash[i]\qslash[a]\kslash[j]
-\kslash[j]\qslash[a]\kslash[i]
\right)
\end{equation}
with the two lightlike momenta $k_i$ and~$k_j$.\footnote{The application
of this equation allows to separate $\qslash[a]$ from an adjacent
$\gamma^\mu$ if one wants to apply \person{Fierz} identities before
carrying out the integrals because one can split the spinor lines
using $\kslash[i]\qslash[a]\kslash[j]=
\ket{k_{i,\lambda}}\braket{k_{i,\lambda}\vert\qslash[a]\vert k_{j,\lambda}}
\bra{k_{j,\lambda}}$.}
If there are not enough
lightlike vectors in the amplitude there is an easy way of constructing
a pair of massless momenta from two massive ones, by starting from the
ansatz $l_1=k_1+\eta_1k_2$ and $l_2=k_2+\eta_2k_1$ and
imposing~$l_1^2=l_2^2=0$.
Equation~\eqref{eq:qcd-shelpmethod:extractq1} allows
all appearances of $\fmslash{q}_a$ to be rewritten
in terms of dot-products $q_a\cd k_i$,
$q_a\cd k_j$ and simple spinor
products~$\braket{k_{i,\pm}\vert \fmslash{q}_a\vert k_{j,\pm}}$. In the end
one can always achieve having only one such spinor product left
because\footnote{$k_i$ and $k_j$ are assumed to be four-dimensional vectors.}
\begin{equation}
\braket{k_{i,\pm}\vert \fmslash{p}\vert k_{j,\pm}}
\braket{k_{j,\pm}\vert \fmslash{p}\vert k_{i,\pm}}=
\tr{\Pi^\pm\fmslash{k}_i\fmslash{p}\Pi^\pm\fmslash{k}_j\fmslash{p}}=
4(k_i\cd\hat{p})(k_j\cd\hat{p})-2(k_i\cd k_j)\hat{p}^2\text{.}
\end{equation}

%%%%%%%%%%%%%%%%%%%%%%%%%%%%%%%%%%%%%%%%%%%%%%%%%%%%%%%%%%%%%%%%%%%%%%%
\subsection{The \person{Weyl}-\person{van der Waerden} Representation}
%%%%%%%%%%%%%%%%%%%%%%%%%%%%%%%%%%%%%%%%%%%%%%%%%%%%%%%%%%%%%%%%%%%%%%%
\label{ssec:qcd-shelpmethod:WvdW}
\index{Weyl-van der Waerden representation@\person{Weyl}-\person{van der Waerden} representation|main}
The spinor helicity projections as described above on the one hand
are a tool for structuring an amplitude into orthogonal pieces
on a symbolical level, so-called helicity amplitudes. These are
easier to handle than the full amplitude and allow for much more compact
expressions. However, the fact that one projects from the 4 by~4
\person{Dirac} matrices onto two-dimensional subspaces also allows for
a different notation of the \person{Clifford} algebra on these subspaces
and also a more efficient, numerical computation of the traces and spinor
products.

\person{Weyl}~\cite{Weyl:1931} and
\person{van der Waerden}~\cite{Waerden:1932} originally developed the
representation theory of the \person{Lorentz} group in a way that later
lead into spinorial methods in quantum field theory. The basic idea behind
the \ac{wvdw} representation is the fact that the matrices
\begin{equation}
\gamma^\mu=\left(\begin{array}{cc}
0_{2\times2} & \sigma^\mu \\
\bar{\sigma}^\mu & 0_{2\times2}
\end{array}\right)\quad\text{with}\,\sigma^\mu=(\sigma_0, \vec{\sigma})
\quad\text{and}\,\bar{\sigma}^\mu=(\sigma_0,-\vec{\sigma})\text{,}
\end{equation}
where $\sigma_0=\One_{2\times2}$ and $\vec{\sigma}$ are the \person{Pauli}
matrices
\begin{equation}
{\sigma}^1=\begin{pmatrix}0&1\\1&0\end{pmatrix},\quad
{\sigma}^2=\begin{pmatrix}0&-i\\i&0\end{pmatrix},\quad
{\sigma}^3=\begin{pmatrix}1&0\\0&-1\end{pmatrix}\text{,}
\end{equation}
form a representation of the \person{Clifford} algebra
\begin{equation}\label{eq:qcd-shelpmethod:clifford-wvdw1}
\{\gamma^\mu,\gamma^\nu\}=2g^{\mu\nu}\otimes\One_{4\times4}%
\quad\text{because}\quad
\sigma^\mu_{\alpha\dot\beta}\bar{\sigma}^{\nu\,\dot\beta\gamma}
+\sigma^\nu_{\alpha\dot\beta}\bar{\sigma}^{\mu\,\dot\beta\gamma}=
2g^{\mu\nu}\delta_\alpha^\gamma%
\text{.}
\end{equation}
In this representation
the helicity projectors are
\begin{equation}
\Pi_+=\begin{pmatrix}\One&0\\0&0\end{pmatrix}\quad\text{and}\,
\Pi_-=\begin{pmatrix}0&0\\0&\One\end{pmatrix}\text{.}
\end{equation}
Tracing both sides of Equation~\eqref{eq:qcd-shelpmethod:clifford-wvdw1}
and using the \person{Hermit}icity of the \person{Pauli} matrices
yields a decomposition of the metric tensor $g^{\mu\nu}$ into
$\sigma$ and $\bar\sigma$:
\begin{equation}
g^{\mu\nu}=\frac12%
\sigma^{\mu}_{\alpha\dot\beta}%
\bar{\sigma}^{\nu\,\dot\beta\alpha}
\end{equation}
It should be noted that $\alpha$ and $\dot\alpha$ denote
distinct indices; the dot distinguishes the spin-$\frac12$ representation
and its conjugate.

The above relation allows to define a bijective mapping between
four-vectors $k^\mu$ and matrices $k_{\alpha\dot\alpha}$.
\begin{subequations}
\begin{align}
k_{\alpha\dot\alpha}&=k_\mu\sigma^\mu_{\alpha\dot\alpha}
\Leftrightarrow
k^\mu=\frac{1}{2}k_{\alpha\dot\alpha}\bar{\sigma}^{\mu\,\dot\alpha\alpha}\\
\intertext{and similarly}
\bar{k}^{\dot\alpha\alpha}&=k_\mu\bar{\sigma}^{\mu\,\dot\alpha\alpha}
\Leftrightarrow
k^\mu=\frac{1}{2}\bar{k}^{\dot\alpha\alpha}{\sigma}^\mu_{\alpha\dot\alpha}
\text{.}
\end{align}
\end{subequations}
With the abbreviations $p_\pm=p^0\pm p^3$ and $p_\perp=p^1+ip^2$ one
obtains in components
\begin{equation}\label{eq:qcd-shelpmethod:mat-components}
p_{\alpha\dot\alpha}=\begin{pmatrix}
p_-&-p_\perp^\ast\\
-p_\perp&p_+
\end{pmatrix}\quad\text{and}\,
\bar{p}^{\dot\alpha\alpha}=\begin{pmatrix}
p_+&p_\perp^\ast\\
p_\perp&p_-
\end{pmatrix}\text{.}
\end{equation}

Since $k_{\alpha\dot\alpha}$ is a \person{Hermit}ian two by two matrix one can
use the spectral theorem to decompose it into its eigenvectors,
\begin{equation}
k_{\alpha\dot\alpha}=\lambda_+\ket{k_\alpha^+}\bra{k_{\dot\alpha}^+}
+\lambda_-\ket{k_\alpha^-}\bra{k_{\dot\alpha}^-}
\end{equation}
where the inner product is defined by the antisymmetric
spinor-metrics $\varepsilon^{\alpha\beta}$
with $\varepsilon^{12}=\varepsilon_{21}=1$ 
and $\varepsilon^{\dot1\dot2}=\varepsilon_{\dot2\dot1}=1$ 
following the convention of~\cite{Wess:SUSY},
raising and lowering is done by
$\ket{p^\alpha}=\epsilon^{\alpha\beta}\ket{p_\beta}$ and
$\ket{p_\alpha}=\epsilon_{\alpha\beta}\ket{p^\beta}$.
The dotted and undotted spinors are related
by complex conjugation:
$\bra{p_{\dot\alpha}}=\delta^\alpha_{\dot\alpha}(\ket{p^{\alpha}})^\ast$ 
and
$\bra{p^{\alpha}}=\delta^\alpha_{\dot\alpha}(\ket{p_{\dot\alpha}})^\ast$.
Hence
$\braket{p_\alpha\vert q^\alpha}=\ket{p_1}\ket{q_2}-\ket{p_2}\ket{q_1}$.

The eigenvalues can be derived easily from the
components of~\eqref{eq:qcd-shelpmethod:mat-components} as
$\lambda_\pm=p^0\pm\vert\vec{p}\vert$; the eigenvectors have a
compact representation if one chooses spherical 
coordinates for the space like components of
$k^\mu=(k^0,\vert\vec{k}\vert\cos\phi\sin\vartheta,
\vert\vec{k}\vert\sin\phi\sin\vartheta,
\vert\vec{k}\vert\cos\vartheta)$ as for example
in~\cite{Dittmaier:1998nn}
\begin{equation}
\ket{k_\alpha^+}=\begin{pmatrix}e^{-i\phi}\cos\frac{\vartheta}{2}\\
\sin\frac{\vartheta}{2}
\end{pmatrix}\quad\text{and}\quad
\ket{k_\alpha^-}=\begin{pmatrix}
\sin\frac{\vartheta}{2}\\
-e^{+i\phi}\cos\frac{\vartheta}{2}
\end{pmatrix}\text{.}
\end{equation}
Complex conjugation relates therefore the eigenvectors as
$\bra{k_\alpha^+}=\delta_\alpha^{\dot\alpha}%
\varepsilon^{\dot\alpha\dot\beta}\ket{k^-_{\dot\beta}}$.

For lightlike vectors $\lambda_-=0$ and hence with the redefinition
\begin{equation}
\ket{k_\alpha}=\sqrt{2k^0}\ket{k_\alpha^+}
\end{equation}
one is back to the usual spinors as defined in the section before when
$k_{\alpha\dot\alpha}$ and $\bar{k}^{\dot\alpha\alpha}$ are identified
with the projections $\Pi_\pm\kslash$.

Another useful identity on the \person{Pauli}
matrices which can be used for symbolic calculations can be simply
derived by inspection:
\begin{equation}
\sigma^{\mu}_{\alpha\dot\alpha}\bar{\sigma}_\mu^{\dot\beta\beta}=
2\delta_\alpha^\beta\delta_{\dot\alpha}^{\dot\beta}.
\end{equation}
This allows to either cut or sew together traces with contracted
\person{Lorentz} indices similar to the \person{Chisholm} identities
for \person{Dirac} matrices.

An interesting numerical application of the \ac{wvdw} representation
is the efficient evaluation of traces. As has been shown in the
previous section any trace of contracted \person{Dirac} matrices
$\Gamma=\pslash[1]\pslash[2]\cdots\pslash[2n]$ can be
written as a sum of traces of the form $\tr[+]{\Gamma^\prime}$, where
$\Gamma^\prime$ is a cyclic permutation of~$\Gamma$. We can now introduce
projectors, according to the commutation
relation~$\Pi_\pm\pslash=\pslash\Pi_\mp$,
\begin{equation}
\tr[+]{\Gamma}=\tr{(\Pi_+\pslash[1]\Pi_-)(\Pi_-\pslash[2]\Pi_+)%
\cdots(\Pi_-\pslash[2n]\Pi_+)}=
p_{1\,\alpha\dot\beta}\bar{p}_2^{\dot\beta\gamma}\cdots
\bar{p}_{2n}^{\dot\omega\alpha}
\end{equation}
because
\begin{align*}
\Pi_+\pslash\Pi_-&=
\begin{pmatrix}\One&0\\0&0\end{pmatrix}%
\begin{pmatrix}0&p_{\alpha\dot\alpha}\\\bar{p}^{\dot\alpha\alpha}&0%
\end{pmatrix}%
\begin{pmatrix}0&0\\0&\One\end{pmatrix}%
=
\begin{pmatrix}0&p_{\alpha\dot\alpha}\\0&0%
\end{pmatrix}%
\\
\intertext{and}
\Pi_-\pslash\Pi_+&=
\begin{pmatrix}0&0\\0&\One\end{pmatrix}%
\begin{pmatrix}0&p_{\beta\dot\beta}\\\bar{p}^{\dot\beta\beta}&0%
\end{pmatrix}%
\begin{pmatrix}\One&0\\0&0\end{pmatrix}%
=
\begin{pmatrix}0&0\\\bar{p}^{\dot\beta\beta}&0%
\end{pmatrix}%
\text{.}
\end{align*}
This shows that once one uses helicity projections one can evaluate
traces of two by two matrices instead of four by four, which reduces
the number of multiplications involved significantly.

The possible applications of the \ac{wvdw} representation go beyond
what has been described here, and for a more complete treatment the
reader is referred to~\cite{Dittmaier:1998nn,Weinzierl:2005dd}.

%% file: nlo-intro.tex
At the lowest order in the perturbative expansion of a
scattering amplitude one obtains only tree-like diagrams and
various automated tools exist to generate and evaluate
\ac{lo} matrix elements
numerically~\cite{Krauss:2001iv,Moretti:2001zz,Maltoni:2002qb,Yuasa:1999rg,%
Hahn:2000kx,Hahn:2005vh,Boos:2004kh,Mangano:2002ea}.
At \ac{nlo} a cross-section for a $2\rightarrow N$ process
consits of three different terms:
\begin{equation}
\sigma=\born{\sigma}+\left(\real{\sigma}+\virt{\sigma}\right)
   +\mathcal{O}(\alpha_s^{N+2})\text{.}
\end{equation}
The first term, $\born\sigma$, corresponds to the \ac{lo} result
which in \ac{qcd} is of order $\mathcal{O}(\alpha_s^{N})$.
The corrections of order $\mathcal{O}(\alpha_s^{N+1})$ are
given in parenthesis: the real emission corrections $\real\sigma$,
that is the emission of an additional parton, correspond to
a $2\rightarrow(N+1)$ tree-level process whereas the virtual
corrections $\virt\sigma$ are described by a $2\rightarrow N$ process
containing one loop in the \person{Feynman} diagrams.

In four dimensions, both $\virt\sigma$ and $\real\sigma$ are divergent
and need to be regularised in order to lead to a meaningful physical result.
As a regularisation method we choose \acf{dreg}, where one replaces
the number of dimensions by $n=4-2\varepsilon$. Section~\ref{sec:qcd-dimreg}
reviews the most important implications of this regularisation scheme.
In the virtual correction, working in an $d$-dimensional space requires
to solve integrals of the type
\begin{displaymath}
\int\!\!\frac{\diff[d]k}{i\pi^{d/2}}\frac{
k^{\mu_1}\cdots k^{\mu_r}}{%nl
\prod_{j=1}^N((k+r_j)^2-m_j^2+i\delta)}\text{,}
\end{displaymath}
where the divergences show up in poles
of $1/\varepsilon$ and $1/\varepsilon^2$.
For our calculation we use a systematic reduction of these integrals
in order to express them in terms of simpler building blocks.
This reduction method is presented in Sections~\ref{sec:scalarred}
and~\ref{sec:tensorred}. A more fundamental introduction to one-loop
integrals can be found in Appendix~\ref{app:loops}.

The divergencies in the real emission part $\real\sigma$ of the cross-section
are discussed in~\ref{sec:realem}. Applying \ac{dreg} to the real emission
contributions requires to integrate the unobserved particle over
an $n$-dimensional phase space. In practise, however, one carries out the
phase space integration by Monte Carlo techniques, as described in
Section~\ref{sec:psintegral} and hence working in fractional-dimensional
spaces appears to be impractical. Instead, we use a subtraction method
as described in Section~\ref{sec:dipoles}, that subtracts the terms leading
to singularities from the integrand of $\real\sigma$, rendering it finite.
The subtracted terms can be integrated analytically over the one-particle
subspace, leading to poles in $1/\varepsilon$ and $1/\varepsilon^2$; this
integrated subtraction terms are added back to the virtual amplitude,
cancelling the so-called \acf{ir} poles of the amplitude.
The remaining singularities are due to \acf{uv} poles and are cured by
renormalisation as shown in Section~\ref{sec:qcd-renorm}.

For the calculation of \ac{nlo} corrections to cross-sections the level
of automisation in current computer programs is far more limited as for
\ac{lo} calculations. The computation of the real corrections has
recently been automated by different
groups~\cite{Gleisberg:2007md,Frederix:2008hu,Seymour:2008mu}.
The automated computation of the virtual corrections,
although having received much effort, have not reached the same degree
of automisation and current implementations are limited to
$2\rightarrow2$~\cite{Kurihara:2006kt}
or $2\rightarrow3$~\cite{Hahn:2006qw} processes.
Therefore, \ac{nlo} corrections for \ac{qcd} processes with more
than two partons in the final state still remain a computational
challenge, mainly due to the combinatorial complexity of the problem.

%% file: qcd-dimreg.tex
\subsection{Introduction}
\label{sec:qcd-dimreg}
Higher order calculation in four dimensional, continuous field theories
lead to singularities which have to be systematically removed by a
renormalisation procedure. In order to handle these singularities in a
consistent way a regularisation of the loop integrals is needed. 
\index{dimensional regularisation}
The so called \acf{dreg} scheme as proposed by
\person{'t~Hooft} and \person{Veltman}~\cite{'tHooft:1972fi} is one of the
most widely used regularisation schemes and has led to many successful
\ac{sm} calculations over last decades. According to the
\ac{msbar}\index{ms@\acs{msbar}}\index{.gammaE@$\gamma_E$}
scheme I use the subtraction
term\footnote{$\gamma_E=-\Gamma^\prime(1)$ is the \person{Euler} constant.}
\begin{equation}
\Delta=\frac{1}{\varepsilon}-\gamma_E+\ln(4\pi)
\end{equation}

\index{.gamma5@$\gamma_5$|(}
It is well known\footnote{See for example~\cite{Jegerlehner:2000dz}.}
that in presence of gauge anomalies, a consistent continuation
of $\gamma_5$ to~$D\neq4$ dimensions
is not possible while preserving gauge invariance.

As the \ac{qcd} is invariant under space reflections it is free of
those anomalies~\cite{Boehm:Gauge,Bardeen:1969md}.
Hence calculations in this work can be treated using
the \person{'t~Hooft}-\person{Veltman} algebra,
\index{Hooft-Veltman@\person{'t Hooft}-\person{Veltman}!algebra}
which extends the four dimensional \person{Dirac} algebra to general
\son[1, D-1] vectors using the relation
\begin{equation}
\label{eq:qcd-dimreg:tHV-algebra}
\left\{\gamma^\mu,\gamma_5\right\}=\begin{cases}
0,\quad \mu\in\{0,1,2,3\},\cr
2\bar{\gamma}^\mu\gamma_5,\quad \text{otherwise.}
\end{cases}
\end{equation}
Here the symbol $\bar{\gamma}^\mu=\bar{g}^\mu_\nu\gamma^\nu$ has been used,
where the metric is split up into
\begin{equation}
g^{\mu\nu}=\hat{g}^{\mu\nu}+\bar{g}^{\mu\nu}\text{,}
\end{equation}
and $\hat{g}^\mu_\nu$ is a projector on the physical subspace,
\begin{equation}
\hat{g}^\mu_\nu\equiv\begin{cases}
\delta^\mu_\nu,\quad \mu\in\{0,1,2,3\},\cr
0,\quad \text{otherwise.}
\end{cases}
\end{equation}
This notation is applied to all vectors, i.e. $\bar{p}^\mu=\bar{g}^\mu_\nu p^\nu$ and
$\hat{p}^\mu=\hat{g}^\mu_\nu p^\nu$, and since $\hat{g}^\mu_\rho\bar{g}^\rho_\nu=0$
one obtains for the modulus of an arbitrary vector $k^2=\bar{k}^2+\hat{k}^2$.

Equation \eqref{eq:qcd-dimreg:tHV-algebra} can also be read as
\begin{displaymath}
[\gamma_5,\bar{\gamma}^\mu]=0\text{,}
\end{displaymath}
which can be interpreted as $\gamma_5$ acting trivial in the non-physical dimensions.
This behaviour becomes manifest through the definition
\begin{equation}
\label{eq:qcd-dimreg:gamma5}
\gamma_5\equiv\frac{i}{4!}\epsilon_{\mu\nu\rho\sigma}% 
\hat{\gamma}^\mu\hat{\gamma}^\nu\hat{\gamma}^\rho\hat{\gamma}^\sigma\text{.}
\end{equation}
\index{.gamma5@$\gamma_5$|)}

\subsection{Spinor Traces in $D$ Dimensions}
An important issue for practical calculations is how to calculate spinor traces within
\ac{dreg}. In this section I show an algorithm (Algorithm~\ref{alg:qcd-dimreg:traces}) that separates $\hat{\gamma}^\mu$
and $\gamma_5$ matrices from $\bar{\gamma}^\mu$ and allows the separate
evaluation of a purely four dimensional trace and a trace consisting of 
$\bar{\gamma}^\mu$-objects only. We will see that the latter trace leads to
$\mathcal{O}(\varepsilon)$-terms in $n=4-2\varepsilon$ dimensions.

\begin{algorithm}
\begin{algorithmic}[1]
\STATE{$\tr{\ldots\gamma^\mu\ldots}\rightarrow%nl
\tr{\ldots\bar{\gamma}^\mu\ldots}+\tr{\ldots\hat{\gamma}^\mu\ldots}$\label{cmd:qcd-dimreg:traces:1}}
\WHILE{replacements left}
\STATE{$\tr{\ldots \bar{\gamma}^\mu\hat{\gamma}^\nu\ldots}\rightarrow%nl
-\tr{\ldots \hat{\gamma}^\nu\bar{\gamma}^\mu\ldots}$\label{cmd:qcd-dimreg:traces:2}}
\STATE{$\tr{\ldots \bar{\gamma}^\mu\gamma_5\ldots}\rightarrow%nl
+\tr{\ldots \gamma_5\bar{\gamma}^\mu\ldots}$\label{cmd:qcd-dimreg:traces:3}}
\STATE{$\tr{\ldots \hat{\gamma}^\mu\gamma_5\ldots}\rightarrow%nl
-\tr{\ldots \gamma_5\hat{\gamma}^\mu\ldots}$\label{cmd:qcd-dimreg:traces:4}}
\STATE{$\tr{\ldots\gamma_5\gamma_5\ldots}\rightarrow%nl
+\tr{\ldots\One\ldots}$\label{cmd:qcd-dimreg:traces:5}}
\ENDWHILE
\STATE{}
\COMMENT{All traces now have the form $\Trace\{\hat{\Gamma}\bar{\Gamma}\}$ or $\Trace\{\gamma_5\hat{\Gamma}\bar{\Gamma}\}$%nl
\label{cmd:qcd-dimreg:traces:6}}
\STATE{$\tr{\gamma_5\bar{\gamma}^\mu\ldots}\rightarrow%nl
0$\label{cmd:qcd-dimreg:traces:7}}\quad\COMMENT{see Eq.~\eqref{eq:qcd-dimreg:tracesplit} and~\eqref{eq:qcd-dimreg:gamma5}}
\STATE{$\tr{\ldots\hat{\gamma}^\mu\bar{\gamma}^\nu\ldots}\rightarrow%nl
\tr{\ldots\hat{\gamma}^\mu}\tr{\bar{\gamma}^\nu\ldots}/\tr{\One}$\label{cmd:qcd-dimreg:traces:8}}
\STATE{Evaluate traces separately.\label{cmd:qcd-dimreg:traces:9}}
\end{algorithmic}
\caption{Carry Out Traces}
\label{alg:qcd-dimreg:traces}
\end{algorithm}

It is obvious that Algorithm~\ref{alg:qcd-dimreg:traces} is confluent and terminating for it shuffles all $\bar{\gamma}^\mu$
to the right and all $\gamma_5$ to the left%
\footnote{A rigorous proof is easily done by introducing
the lexicographic ordering $\One<\gamma_5<\hat{\gamma}^0<%
\ldots<\hat{\gamma}^3<\bar{\gamma}^0<\ldots<\bar{\gamma}^3$;
for there exists a least element ($\One$) and the algorithm produces
a (lexicographic) descending chain of expression termination is guaranteed
by the generalised induction principle.
The validity of comment~\ref{cmd:qcd-dimreg:traces:7} is guaranteed
since, including the possibilities for either
$\hat{\Gamma}$ and $\bar{\Gamma}$ being $\One$,
not being in the given form would include at least one of the replacements
\ref{cmd:qcd-dimreg:traces:2}--\ref{cmd:qcd-dimreg:traces:5} being applicable. 
To show the confluence of the algorithm one has to prove that all rewriting rules commute, which can be easily
done.}.
Rule~\ref{cmd:qcd-dimreg:traces:2} is valid as the anticommutator
\begin{equation}
\label{eq:qcd-dimreg:4epsanti}
\{\hat{\gamma}^\mu,\bar{\gamma}^\nu\}=\hat{g}^\mu_\rho\bar{g}^\nu_\sigma\{\gamma^\rho,\gamma^\sigma\}=
2\hat{g}^\mu_\rho\bar{g}^\nu_\sigma g^{\rho\sigma}=2\hat{g}^{\mu\rho}\bar{g}^\nu_\rho=0
\end{equation}
vanishes. The rewriting rules \ref{cmd:qcd-dimreg:traces:3}--\ref{cmd:qcd-dimreg:traces:5} follow directly from the
definition of the algebra~\eqref{eq:qcd-dimreg:tHV-algebra}.
Step~\ref{cmd:qcd-dimreg:traces:7} is just a special case of step~\ref{cmd:qcd-dimreg:traces:8}.

The missing bit in the proof of algorithm~\ref{alg:qcd-dimreg:traces} is the equation
\begin{equation}
\label{eq:qcd-dimreg:tracesplit}
\tr{\One}\tr{\hat{\Gamma}\bar{\Gamma}}=\tr{\hat{\Gamma}}\tr{\bar{\Gamma}}\text{,}
\end{equation}
where $\hat{\Gamma}=\hat{\gamma}^{\mu_1}\cdots\hat{\gamma}^{\mu_m}$ and
$\bar{\Gamma}=\bar{\gamma}^{\nu_1}\cdots\bar{\gamma}^{\nu_n}$ for all
non-negative  integers $m$ and $n$.
This also covers the cases involving~$\gamma_5$ being a product
of four-dimensional \person{Dirac} matrices, which can be seen
from~Equation~\eqref{eq:qcd-dimreg:gamma5}.

The proof is done by complete induction over $n$, where $n=0$ can be read off
from~\eqref{eq:qcd-dimreg:tracesplit}. Since the trace is cyclic and due
to~\eqref{eq:qcd-dimreg:4epsanti} one obtains
\begin{multline}
\tr{\One}\tr{\hat{\Gamma}\bar{\Gamma}}=
(-1)^m\tr{\One}\tr{\hat{\Gamma}\bar{\gamma}^{\nu_n}\bar{\gamma}^{\nu_1}\cdots\bar{\gamma}^{\nu_{n-1}}}=\\
(-1)^m\sum_{i=1}^{n-1}(-1)^{i+1}\cdot2\bar{g}^{\nu_i\nu_n}%nl
\tr{\One}\tr{\hat{\Gamma}\bar{\gamma}^{\nu_1}\cdots\bar{\gamma}^{\nu_{i-1}}%nl
\bar{\gamma}^{\nu_{i+1}}\cdots\bar{\gamma}^{\nu_{n-1}}}\\
+(-1)^{m+n-1}%nl
\tr{\One}\tr{\hat{\Gamma}\bar{\Gamma}}\text{.}
\end{multline}

For the traces inside the sum we can substitute the induction
step which leads to
\begin{multline}
\label{eq:qcd-dimreg:intermediate1}
(-1)^m\cdot(1+(-1)^{m+n})\tr{\One}\tr{\hat{\Gamma}\bar{\Gamma}}=\\
\tr{\hat{\Gamma}}\sum_{i=1}^{n-1}(-1)^{i+1}\cdot2\bar{g}^{\nu_i\nu_n}%nl
\tr{\bar{\gamma}^{\nu_1}\cdots\bar{\gamma}^{\nu_{i-1}}%nl
\bar{\gamma}^{\nu_{i+1}}\cdots\bar{\gamma}^{\nu_{n-1}}}=
2\tr{\hat{\Gamma}}\tr{\bar{\Gamma}}\text{.}
\end{multline}
It remains to investigate the cases for the numbers $m$ and~$n$ being even
or odd. If $m$ is odd, one obtains
\begin{multline}
\tr{\hat{\Gamma}\One\bar{\Gamma}}=
\tr{\hat{\Gamma}\gamma_5\gamma_5\bar{\Gamma}}=
-\tr{\gamma_5\hat{\Gamma}\gamma_5\bar{\Gamma}}=\\
-\tr{\hat{\Gamma}\gamma_5\bar{\Gamma}\gamma_5}=
-\tr{\hat{\Gamma}\gamma_5\gamma_5\bar{\Gamma}}=
-\tr{\hat{\Gamma}\bar{\Gamma}}=0\text{.}
\end{multline}
On the other hand, both sides vanish if $n$ is odd. For the left, non-vanishing case
the result agrees with the conjecture, and everything is proved.

It should be noted that the intermediate step in
\eqref{eq:qcd-dimreg:intermediate1} already shows
that the evaluation of the trace $\Trace\{\bar{\Gamma}\}$ yields products
of $\bar{g}^{\mu\nu}$ tensors.
In amplitude calculations these terms vanish unless
they are traced or contracted with integration momenta;
as external momenta in the
\acf{tho} scheme\index{Hooft-Veltman@\person{'t Hooft}-\person{Veltman}!scheme}
are kept in four dimensions
their projection on the $(D-4)$ dimensional subspace vanishes. The trace
$\bar{g}^\mu_\mu=(D-4)$ as well as the terms proportional to $\bar{k}^2$ 
lead to $\mathcal{O}(\varepsilon)$-terms,
giving rise to polynomial terms in the amplitude
if they are multiplied to a $1/\varepsilon$ pole from the loop integrals.
The evaluation of $\tr{\bar{\Gamma}}$ can be implemented straightforward:
we have already seen that only two kinds of non-vanishing terms can
arise from these traces, i.e. $(D-4)$ and $\bar{k}^2$
and therefore no additional care needs to be take
in order to keep the number of terms low.
The most na\"ive reduction formula
\begin{equation}
\label{eq:qcd-dimreg:naive-trace}
\tr{\bar{\gamma}^{\nu_1}\cdots\bar{\gamma}^{\nu_n}}=
\sum_{i=2}^n(-1)^i\bar{g}^{\nu_1\nu_i}\tr{\bar{\gamma}^{\nu_2}\cdots\bar{\gamma}^{\nu_{i-1}}%nl
\bar{\gamma}^{\nu_{i+1}}\bar{\gamma}^{\nu_n}}
\end{equation}
is sufficient.

For the four dimensional trace~$\Trace\{\hat{\Gamma}\}$
we can use another tool: the
\person{Chisholm} identity, which is valid in four dimensions only, allows the evaluation
of traces regardless if they include a~$\gamma_5$ or not. For an algebraic reduction
shorter results are achieved by algorithms that include other relations as
well~\cite{Vermaseren:2002}; on the other hand one can use the \person{Chisholm}
identity to write a numerical evaluation of traces having all \person{Lorentz} indices contracted
with external momenta and avoiding explicit summation over indices.

\index{Chisholm identity@\person{Chisholm} identity|(}
The \person{Chisholm} identity can be derived following the proof in~\cite{Kleiss:1985yh}.
The initial point of the proof is the fact that in the four dimensional \person{Minkowski} space
for the generators of the \person{Clifford} algebra a finite basis exists,
and every product~$S$ consisting of an odd number of \person{Dirac} matrices therefore can
be expressed as
\begin{equation}
S=V_\mu\hat{\gamma}^\mu+A_\mu\gamma_5\hat{\gamma}^\mu\text{,}
\end{equation}
where $V^\mu$ and~$A^\mu$ are the two coefficient vectors. Now one considers the expression
\begin{equation}
\label{eq:qcd-dimreg:chisholm01}
\tr{S\hat{\gamma}^\mu}\hat{\gamma}_\mu=
\tr{V_\nu\hat{\gamma}^\nu\hat{\gamma}^\mu+A_\nu\gamma_5\hat{\gamma}^\nu\hat{\gamma}^\mu}\hat{\gamma}_\mu
=4V^\mu\hat{\gamma}_\mu\text{.}
\end{equation}
The right hand side can also stem from
\begin{equation}
\label{eq:qcd-dimreg:chisholm02}
2(S + S^\mathrm{R})=2(2V_\mu\hat{\gamma}^\mu+A_\mu\{\gamma_5,\hat{\gamma}^\mu\})=4V_\mu\hat{\gamma}^\mu\text{,}
\end{equation}
where $S^\mathrm{R}$ denotes the reverse of the spinor line.
Equating \eqref{eq:qcd-dimreg:chisholm01} and \eqref{eq:qcd-dimreg:chisholm02} leads to the desired identity,
\begin{equation}
\label{eq:qcd-dimreg:chisholm}
\tr{S\hat{\gamma}^\mu}\hat{\gamma}_\mu=2(S + S^\mathrm{R})\text{.}
\end{equation}

The analogous formula for a string $T$ of an even number of $\gamma$-matrices\footnote{%nl
If a~$\gamma_5$ appears in a string of \person{Dirac} matrices it counts as an even number of matrices.}
comes from the representation
\begin{equation}
T=T_{\mu\nu}[\hat{\gamma}^\mu,\hat{\gamma}^\nu]+P\gamma_5+S\One\text{;}
\end{equation}
the same logic as above applies and leads to
\begin{equation}
\label{eq:qcd-dimreg:chisholm-even}
\tr{T\gamma_5}\gamma_5+\tr{T}\One=2(T+T^\mathrm{R})
\end{equation}
\index{Chisholm identity@\person{Chisholm} identity|)}

The identity~\eqref{eq:qcd-dimreg:chisholm} can be made twofold use of.
It leads to a reduction formula for \person{Dirac} traces on one hand,
on the other hand it can be used to eliminate \person{Lorentz} indices that are contracted between two traces
since
\begin{equation}
\label{eq:qcd-dimreg:chisholm-twotraces}
\tr{S\hat{\gamma}^\mu}\tr{S^\prime\hat{\gamma}_\mu}=
\tr{S^\prime\tr{S\hat{\gamma}^\mu}\hat{\gamma}_\mu}=
2\tr{S^\prime S}+2\tr{S^\prime S^\mathrm{R}}\text{.}
\end{equation}
To get rid of pairs of contracted indices inside one trace we can apply~\eqref{eq:qcd-dimreg:chisholm-even}
to a string~$\hat{\gamma}^\mu S\hat{\gamma}_\mu$ of odd length, 
revealing
\begin{equation}
\label{eq:qcd-dimreg:chisholm-paired-odd}
\hat{\gamma}^\mu S\hat{\gamma}_\mu=-2S^\mathrm{R}\text{.}
\end{equation}
Without loss of generality strings of even length can be treated through the case
$\hat{\gamma}^\mu S\hat{\gamma}^\nu\hat{\gamma}_\mu$, where $S$ has odd length and one finds
\begin{equation}
\label{eq:qcd-dimreg:chisholm-paired-even}
\hat{\gamma}^\mu S\hat{\gamma}^\nu\hat{\gamma}_\mu=
2(\hat{\gamma}^\nu S+S^\mathrm{R}\hat{\gamma}^\nu)\text{.}
\end{equation}

For obtaining a reduction formula for traces 
one starts from the specific choice $S=-(i/4)\gamma_5\gamma^\mu\gamma^\nu\gamma^\rho$.
This implies for \eqref{eq:qcd-dimreg:chisholm}
\begin{equation}
\label{eq:qcd-dimreg:trred01}
\tr{S\hat{\gamma}^\sigma}\hat{\gamma}_\sigma=\epsilon^{\mu\nu\rho\sigma}\hat{\gamma}_\sigma
\end{equation}
Here I used, as an implication of \eqref{eq:qcd-dimreg:gamma5}, that
\begin{equation}
\label{eq:qcd-dimreg:def-tr4gamma5}
-\frac{i}{4}\tr{\gamma_5\hat{\gamma}^\mu\hat{\gamma}^\nu\hat{\gamma}^\rho\hat{\gamma}^\sigma}\hat{\gamma}_\sigma= 
\epsilon^{\mu\nu\rho\sigma}\hat{\gamma}_\sigma=\\
-\frac{i}2(\gamma_5\hat{\gamma}^\mu\hat{\gamma}^\nu\hat{\gamma}^\rho+
\hat{\gamma}^\rho\hat{\gamma}^\nu\hat{\gamma}^\mu\gamma_5)\text{.}
\end{equation}
Rearranging the terms of \eqref{eq:qcd-dimreg:trred01} this leads to
\begin{multline}
\label{eq:qcd-dimreg:numeric-trace01}
\tr{\gamma_5\hat{\pslash}_1\hat{\pslash}_2\hat{\pslash}_3\hat{\pslash}_4\cdots\hat{\pslash}_n}=\\
\hat{p}_1\cdot\hat{p}_2 \tr{\gamma_5\hat{\pslash}_3\hat{\pslash}_4\cdots\hat{\pslash}_n}
-\hat{p}_1\cdot\hat{p}_3 \tr{\gamma_5\hat{\pslash}_2\hat{\pslash}_4\cdots\hat{\pslash}_n}\\
+\hat{p}_2\cdot\hat{p}_3 \tr{\gamma_5\hat{\pslash}_1\hat{\pslash}_4\cdots\hat{\pslash}_n}
+i\epsilon^{p_1p_2p_3\mu}\tr{\gamma_\mu\hat{\pslash}_4\cdots\hat{\pslash}_n}\text{.}
\end{multline}
The notation $\epsilon^{p_1\bul\bul\bul}$
is a shorthand notation for $\epsilon^{\mu\bul\bul\bul}p_{1,\mu}$.
Using the pendant to~\eqref{eq:qcd-dimreg:naive-trace},
\begin{equation}
\label{eq:qcd-dimreg:naive-trace4}
\tr{\hat{\gamma}^{\nu_1}\cdots\hat{\gamma}^{\nu_n}}=
\sum_{i=2}^n(-1)^i\hat{g}^{\nu_1\nu_i}\tr{\hat{\gamma}^{\nu_2}\cdots\hat{\gamma}^{\nu_{i-1}}%nl
\hat{\gamma}^{\nu_{i+1}}\hat{\gamma}^{\nu_n}}
\end{equation}
one can eliminate also the explicit index~$\mu$ in the last term of~\eqref{eq:qcd-dimreg:numeric-trace01},
and hence ends up in a formula suitable for numerical evaluation, provided numerical implementations of
$p_i\cdot p_j$ and $\epsilon^{p_ip_jp_mp_n}$ exist,
\begin{multline}
\label{eq:qcd-dimreg:numeric-trace}
\tr{\gamma_5\hat{\pslash}_1\hat{\pslash}_2\hat{\pslash}_3\hat{\pslash}_4\cdots\hat{\pslash}_n}=
\hat{p}_1\cdot\hat{p}_2 \tr{\gamma_5\hat{\pslash}_3\hat{\pslash}_4\cdots\hat{\pslash}_n}\\
-\hat{p}_1\cdot\hat{p}_3 \tr{\gamma_5\hat{\pslash}_2\hat{\pslash}_4\cdots\hat{\pslash}_n}
+\hat{p}_2\cdot\hat{p}_3 \tr{\gamma_5\hat{\pslash}_1\hat{\pslash}_4\cdots\hat{\pslash}_n}\\
+i\sum_{j=4}^n(-1)^j\epsilon^{p_1p_2p_3p_j}\tr{\hat{\pslash}_4\cdots\hat{\pslash}_{j-1}\hat{\pslash}_{j+1}\cdots\hat{\pslash}_n}\text{.}
\end{multline}

We can complete algorithm~\ref{alg:qcd-dimreg:traces}
by specifying its last step, which
leads to algorithm~\ref{alg:qcd-dimreg:finaltracing}.
\begin{algorithm}
\begin{algorithmic}[5]
\STATE{Evaluate $\Trace\{\bar{\Gamma}\}$ using %
\eqref{eq:qcd-dimreg:naive-trace}}
\IF{numeric evaluation}
\WHILE{\person{Lorentz} indices left}
\STATE{Apply \eqref{eq:qcd-dimreg:chisholm-paired-even} or \eqref{eq:qcd-dimreg:chisholm-paired-odd}}
\STATE{Apply \eqref{eq:qcd-dimreg:chisholm-twotraces}}
\WHILE{applicable}
\STATE{$\tr{\ldots \hat{\gamma}^\mu\gamma_5\ldots}\rightarrow%nl
-\tr{\ldots \gamma_5\hat{\gamma}^\mu\ldots}$}
\ENDWHILE
\ENDWHILE
\STATE{}\COMMENT{All $\hat{\gamma}$ matrices are contracted with momenta now.\label{alg:qcd-dimreg:finaltracing:assert}}
\STATE{Evaluate traces via \eqref{eq:qcd-dimreg:numeric-trace} and~\eqref{eq:qcd-dimreg:naive-trace4}
numerically.\label{alg:qcd-dimreg:finaltracing:numerical}}
\ELSE[algebraic evaluation]
\STATE{Use \texttt{FORM} to evaluate $\Trace\{\hat{\Gamma}\}$ algebraically. \label{alg:qcd-dimreg:finaltracing:FORM}}
\ENDIF
\end{algorithmic}
\caption{Evaluate traces separately}
\label{alg:qcd-dimreg:finaltracing}
\end{algorithm}

The assertion~\ref{alg:qcd-dimreg:finaltracing:assert} can be made for \ac{qcd}
where all epsilon tensors stem from the evaluation of traces and therefore no contractions of the
form $\epsilon^{\mu\bul\bul\bul}\tr{\hat{\gamma}_\mu\ldots}$ are possible here. However,
this case would be easy to handle as well, since all epsilon tensors can be eliminated earlier
via reverse application of~\eqref{eq:qcd-dimreg:def-tr4gamma5}.
The steps up to line~\ref{alg:qcd-dimreg:finaltracing:FORM}
are required for a numerical evaluation of the traces in order to
remove all explicit \person{Lorentz} indices; the matrices $\pslash$
can be calculated, multiplied and traced numerically\footnote{%
Line~\ref{alg:qcd-dimreg:finaltracing:FORM}
suggests an alternative implementation.}.
%The algebraic reduction
%in line~\ref{alg:qcd-dimreg:finaltracing:FORM}, of course,
%can also be carried out using the
%formula given in this section or by some other computer algebra program.
%The lines~\ref{alg:qcd-dimreg:finaltracing:numerical}
%in the actual implementation have been replaced by an algorithm,
%which is taking the trace after direct calculation of the
%$\pslash$-matrices in a certain representation; a direct comparison
%of both algorithms has shown that the second leads to a much
%faster evaluation of the traces where the internal \fortran{} function
%\texttt{matmul} is used for the multiplication of the \person{Dirac} matrices.

\subsection{Gluons in \acs{dreg}}
In Section~\ref{sec:shelpmethod} of Chapter~\ref{chp:repqcdamp}
we have already seen that in $D=4$ dimensions
both the quarks and the gluons have two degrees of freedom
and therefore introduced the spinors $\ket{p_\pm}$ for the quarks
and the polarisation vectors $\epsilon_\pm^\mu$ for the gluons.
In the \emph{\acf{ndr}} scheme one finds
the number of polarisations to be $(n-2)$~\cite{Smith:2004ck}; in the \ac{tho} scheme
scheme, which I use, all external particles are strictly kept four dimensional and therefore the number of
polarisations is~$2$. Table~\ref{tbl:qcd-dimreg:prescriptions} summarises the comparison between~\ac{ndr},
the \ac{tho} scheme and \acf{dred}~\cite{Smith:2004ck,Jegerlehner:2000dz}.
%It has already been seen in
%Section~\ref{ssec:qcd-shelpmethod:extgluons} that the orthogonality relations
%are not affected by the choice of the number of dimensions.
\index{regularisation schemes, comparison}
\begin{table}[htbp]
\begin{center}
\begin{minipage}{\textwidth}
\begin{center}
\begin{tabular}{|l|c|c|c|}
\hline
&\acs{ndr}&\acs{tho}&\acs{dred}\\
\hline
&$\gamma^\mu=\hat{\gamma}^\mu+\bar{\gamma}^\mu$
&$\gamma^\mu=\hat{\gamma}^\mu+\bar{\gamma}^\mu$
&$\gamma^\mu=\hat{\gamma}^\mu,\quad\bar{\gamma}^\mu\equiv0$\\
%nl
$\{\gamma_5,\gamma^\mu\}$
&$0$
&eq.~\eqref{eq:qcd-dimreg:tHV-algebra}
&$0$\\
internal momenta
&$k=\hat{k}+\bar{k}$
&$k=\hat{k}+\bar{k}$
&$k=\hat{k}+\bar{k}$\\
external momenta
&$p_i=\hat{p}_i+\bar{p}_i$
&$p_i=\hat{p}_i,\quad\bar{p}_i=0$
&$p_i=\hat{p}_i,\quad\bar{p}_i=0$\\
int. gluon pol.\footnote{number of polarisations of internal gluons}
&$n-2$
&$n-2$
&$2$\\
ext. gluon pol.\footnote{number of polarisations of external gluons}
&$n-2$
&$2$
&$2$\\
\hline
\end{tabular}
\end{center}
\end{minipage}
\end{center}
\caption{Comparison of different regularisation prescriptions}\label{tbl:qcd-dimreg:prescriptions}
\end{table}

\subsection{Loop Integrals}
\label{ssec:qcd-dimreg:loop-integrals}
%\ac{nlo} calculations, in general, can be split in three major parts:
%the \person{Born} amplitude, the real correction
%and the virtual corrections.
%While the \person{Born} amplitude for most processes is a textbook
%exercise, things start to become complicated as soon as the momentum
%integration of a virtual particle enters the game.
%The integration over loop momenta usually introduces \ac{uv}
%and \ac{ir} divergences.
%The first class is cured through renormalisation where the poles
%cancel against counterterm diagrams.
%The second class of divergences is much more subtle since certain infrared
%divergences only cancel between virtual and real corrections.
As discussed in the introduction of this chapter, integration over
an unobserved, virtual particle in the amplitude gives rise to
tensor integrals like
\begin{displaymath}
\int\!\!\frac{\diff[n]k}{i\pi^{n/2}}\frac{k^\mu}{[(k+r_1)^2+i\delta][(k+r_2)^2+i\delta][k^2+i\delta]}\text{.}
\end{displaymath}
These integrals can be evaluated by
the traditional approach by projections of the tensor integral onto the
vectors $r_1^\mu$ and~$r_2^\mu$ and by rewriting
it in terms of scalar integrals.
The form factors, i.e. the coefficients of $r_1^\mu$ and $r_2^\mu$,
then contain the inverse of a \person{Gram}
determinant\footnote{The related \person{Gram} matrix would be defined
via~$G_{ij}=2\,r_i\cdot r_j$.
Later I will use a more convenient definition for the \person{Gram} matrix, 
i.e.~\eqref{eq:qcd-scalarred:defineG}; both definitions are equal up to a
common shift of the momenta~$r_i$.}
\index{Gram@\person{Gram}!determinant}
\begin{displaymath}
\det{G}=4(r_1^2r_2^2-(r_1\cdot r_2)^2)\text{.}
\end{displaymath}
This determinant vanishes for certain kinematic configurations;
this kind of divergences is unphysical and must
cancel in the full calculation.
If not treated algebraically, these inverse \person{Gram} determinants
can spoil the numerical stability of an amplitude calculation and
therefore have to be avoided.

\begin{figure}[hbtp]
\begin{center}
\begin{fmffile}{loopintegral}
\begin{fmfchar*}(50,50)
\fmfsurroundn{p}{7}
\fmf{fermion,label=$p_1$,label.side=left}{p1,v1}
\fmf{fermion,label=$p_2$,label.side=left}{p2,v2}
\fmf{fermion,label=$p_3$,label.side=left}{p3,v3}
\fmf{fermion,label=$p_4$,label.side=left}{p4,v4}
\fmf{fermion,label=$p_5$,label.side=left}{p5,v5}
\fmf{fermion,label=$p_6$,label.side=left}{p6,v6}
\fmf{fermion,label=$p_N$,label.side=left}{p7,vN}
\fmf{dots}{v6,vN}
\fmf{fermion,label=$q_1$,label.side=left}{v1,v2}
\fmf{fermion,label=$q_2$,label.side=left}{v2,v3}
\fmf{fermion,label=$q_3$,label.side=left}{v3,v4}
\fmf{fermion,label=$q_4$,label.side=left}{v4,v5}
\fmf{fermion,label=$q_5$,label.side=left}{v5,v6}
\fmf{fermion,label=$q_N$,label.side=left}{vN,v1}
\end{fmfchar*}
\end{fmffile}
\end{center}
\caption{Definition of the momenta at an arbitrary  $N$-point integral}
\label{fig:qcd-dimreg:integralmomenta}
\end{figure}
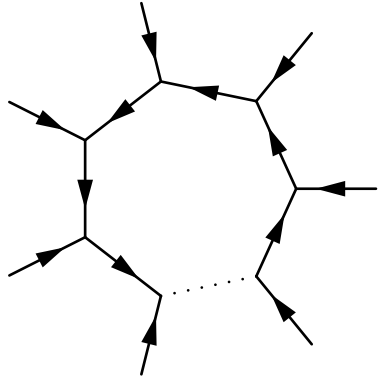

Given an arbitrary loop amplitude we now concentrate on the one-particle
irreducible  part of the diagram,
which always can be sketched as shown in
figure~\ref{fig:qcd-dimreg:integralmomenta}.
The notation is always chosen such that the external momentum flow
of~$p_i, i\in\{1,\ldots,N\}$ is ingoing, and the indices are understood
modulo~$N$, i.e. $p_{N+1}\equiv p_1$. The momenta through the propagators are
\index{.qi@$q_i^\mu$}\index{.ri@$r_i^\mu$}
\begin{equation}
q_i=k+r_i,\qquad\text{where}\quad r_i-r_{i-1}=p_i\text{,}
\end{equation}
and the masses of the particles in the propagators are $m_i$.
The definition of the vectors~$r_i$ resembles the invariance of the loop
integral under shifts of the integration momentum.
In the most general case one has to deal with integrals of the form
\begin{equation}
\label{eq:qcd-dimreg:defTI}
I_N^{d,\alpha;\mu_1\ldots\mu_r}(a_1,\ldots a_r; S)\equiv%nl
\int\!\!\frac{\diff[d]k}{i\pi^{d/2}}\frac{\left(\bar{k}^2\right)^\alpha
\hat{q}_{a_1}^{\mu_1}\cdots \hat{q}_{a_r}^{\mu_r}}{%nl
\prod_{j=1}^N(q_j^2-m_j^2+i\delta)}\text{.}
\end{equation}
It should be noted that in the following discussion I will always
suppress a factor of $(4\pi\mu^2)^{2-n/2}$, which would naturally arise
from the fact that the \Lagrangeian{} density
has to be kept dimensionless and
therefore is multiplied by powers of an arbitrary mass scale~$\mu$.
Furthermore, all logarithms of dimensionful quantities
(e.g.~$\ln(s)$) are understood to be regulated by powers of the same
mass scale~$\mu$ (i.e.~$\ln(s/\mu^2)$); hence in these expressions the
scale dependence on~$\mu$ becomes explicit.

Integrals with $\alpha\neq0$ stem from the evaluation of
$\tr{\bar{\Gamma}}$; only a limited number of them appear
to be non-vanishing, and those left result in very simple expressions which are presented in 
Chapter~\ref{chp:qcdnlo}, Section~\ref{ssec:qcd-dimreg:polyintegrals}.
For $\alpha=0$, I will omit the superscript~$\alpha$.

Contrary to most existing approaches this definition of the tensor integrals
has the propagator momenta $q_i$ in the numerator instead
of the integration momentum $k$.
This generalisation takes into account the origin of the tensor integrals
in the appropriate \person{Feynman} rules.
The way back to standard form is easy enough:
one has to break shift invariance by choosing one
$r_{a_\lambda}=0,a_\lambda\in\{1,\ldots,N\}$
and only considering the integral
$I_N^{d,\alpha;\mu_1\ldots\mu_r}(a_\lambda,\ldots,a_\lambda;S)$.
Another generalisation is to also take into account
integrals with dimensions $d\neq n$, where $n=4-2\varepsilon$.
In the reduction formalism described in
Sections~\ref{sec:scalarred} and~\ref{sec:tensorred}, 
the basis integrals include the use of dimensions $n$, $n+2$ and~$n+4$.
It should also be noted that the momenta in the numerator
are the four dimensional parts in this definition.
Usually the definition with $\hat{q}^\mu_a$ cannot be distinguished
from a definition with $q^\mu_a$ as long as the tensor integrals are
contracted with external momenta which project onto the four-dimensional
subspace.
Care has to be taken if the contraction
$k\cdot k=\hat{g}_{\mu\nu}\hat{k}^\mu\hat{k}^\nu+(\bar{k})^2$
appears in the calculation.

\index{.Sij@$S_{ij}$|main}\index{.Deltaij@$\Delta_{ij}^\mu$}
The matrix~$S$ contains the kinematic invariants in the following form
\begin{equation}
\label{eq:qcd-dimreg:defineSMatrix}
S_{ij}=(\Delta_{ij})^2-m_i^2-m_j^2,\qquad\text{with}\quad\Delta_{ij}^\mu=r_i^\mu-r_j^\mu=q_i^\mu-q_j^\mu
\end{equation}
The importance of the kinematic matrix $S$ becomes clear
if one writes down the loop integrals in \person{Feynman} parameter
space,
\begin{subequations}
\begin{equation}
\label{eq:qcd-dimreg:defSI}
I_N^d(l_1,\ldots,l_r;S)=(-1)^N\Gamma(N-d/2)\int_0\!\!\diff[N]z\delta_z\frac{z_{l_1}\cdots z_{l_r}}{%nl
\left(-\frac12z^\transposed Sz-i\delta\right)^{N-d/2}}\text{.}
\end{equation}
The abbreviations I introduced are\index{.deltaZ@$\delta_z$|main}
\begin{align}
\label{eq:qcd-dimreg:def-deltaz}
\delta_z&=\delta\left(1-\sum\nolimits_{j=1}^Nz_j\right)\text{,}\\
\label{eq:qcd-dimreg:def-int0Theta}
\int_0\!\!\diff[N]z&=\int_{-\infty}^\infty\prod_{j=1}^N\left(\diff{z_j}\Theta(z_j)\right)\quad\text{and}\\
z^\transposed Sz&=\sum_{i,j=1}^Nz_iS_{ij}z_j\text{.}
\end{align}
\end{subequations}

The relation between \eqref{eq:qcd-dimreg:defTI} and~\eqref{eq:qcd-dimreg:defTI} can be found by the usual procedure,
introducing \person{Feynman} parameters and substituting $k\rightarrow k-\sum_{j=1}^Nz_jr_j$
one obtains\footnote{It must be considered that $\bar{k}^2=-\vert\bar{k}^2\vert$.}
\begin{multline}
\label{eq:qcd-dimreg:relTISI01}
I_N^{d,\alpha;\mu_1\ldots\mu_r}(a_1,\ldots a_r; S)=\\%nl
\Gamma(N)\int_0\!\!\diff[N]z\delta_z%nl
\int\!\!\frac{\diff[4]\hat{k}}{i\pi^2}\frac{\diff[d-4]\bar{k}}{\pi^{d/2-2}}%nl
\frac{\left(\bar{k}^2\right)^\alpha \prod_{\nu=1}^r(\hat{k}^{\mu_\nu}-\sum_{i=1}^Nz_i\Delta^{\mu_\nu}_{ia_\nu})}{%nl
\left[\hat{k}^2+\bar{k}^2+\frac12z^\transposed Sz+i\delta\right]^N}\text{.}
\end{multline}

The momentum integration in the $(d-4)$-dimensional subspace can be carried out immediately. Before one can treat the
four-dimensional subspace as well the numerator needs some further investigation. First one observes that when
the numerator is expanded all terms with an odd number of $\hat{k}$ vectors vanishes under symmetric integration.
Any even number~$2l$ of $\hat{k}$ can be reduced to
\begin{equation}
\hat{k}^{\nu_1}\cdots\hat{k}^{\nu_{2l}}=\frac{%nl
[\overbrace{\hat{g}^{\bul\bul}\cdots\hat{g}^{\bul\bul}}^{l}]^{\nu_1\ldots\nu_{2l}}}{%nl
\prod_{j=0}^{l-1}(4+2j)}(\hat{k}^2)^l=
\frac{%nl
[\hat{g}^{\bul\bul}\cdots\hat{g}^{\bul\bul}]^{\nu_1\ldots\nu_{2l}}}{%nl
2^l\Gamma(l+2)}(\hat{k}^2)^l\text{.}
\end{equation}
The square brackets with trailing indices denote the distribution of the indices over the elements inside the brackets.
The combinatorial factor can be obtained from considering all possibilities of connecting the endpoints of $l$~lines where every closed
line counts as factor $4=\hat{g}^\mu_\mu$. The $(j+1)$-th line is added by either connecting its two endpoints to each other and building
an extra circle or by stitching itself to one of the $2j$ existing endpoints, which explains the factor $(4+2j)$. The second relation is then
proved by induction.

Now the four-dimensional momentum integration can be carried out as well and the whole formula reads
\begin{multline}
\label{eq:qcd-dimreg:TItoSI}
I_N^{d,\alpha;\mu_1\ldots\mu_r}(a_1,\ldots,a_r; S)=
(-1)^{r+\alpha}\frac{\Gamma(\alpha+d/2-2)}{\Gamma(d/2-2)}\sum_{l=0}^{\lfloor r/2\rfloor}%nl
\left(-\frac12\right)^l\times\\\sum_{j_1,\ldots,j_{r-2l}=1}^N [\underbrace{%nl
\hat{g}^{\bul\bul}\cdots \hat{g}^{\bul\bul}}_l%nl
\Delta_{j_1\bul}^{\bul}\cdots\Delta_{j_{r-2l}\bul}^{\bul}%nl
]^{\mu_1\ldots\mu_r}_{a_1\ldots a_r}%nl
I_N^{d+2\alpha+2l}({j_1},\ldots,{j_{r-2l}};S)\text{.}
\end{multline}

\subsection{Polynomial Loop Integrals}
\label{ssec:qcd-dimreg:polyintegrals}
As mentioned earlier only a limited number of integrals with $\alpha>0$
are non-zero.
The reason therefore is the factor
\begin{equation}
c_\alpha^d=(-1)^\alpha\frac{\Gamma(\alpha+d/2-2)}{\Gamma(d/2-2)}\text{.}
\end{equation}
For $\alpha=0$ this factor is $c^d_0=1$.
In the case $\alpha>0$ we only need to consider $d=n$ and get the result
\begin{equation}
c_\alpha^n
=(-1)^{\alpha-1}\varepsilon\frac{\Gamma(\alpha-\varepsilon)}{%
\Gamma(1-\varepsilon)}
=(-1)^{\alpha-1}(\alpha-1)!\varepsilon + {\mathcal{O}}(\varepsilon^2).
\end{equation}
This coefficient has to be combined with the integral $I_N^{n+2\alpha+2l}$
from Equation~\eqref{eq:qcd-dimreg:TItoSI}. For phenomenological applications
terms of order $\mathcal{O}(\varepsilon)$ are irrelevant, therefore
one needs to consider only integrals that contain divergences.
The dimension of the integral is always strictly larger than\footnote{%
We only consider the case $\alpha\neq0$.}~$n$ and hence the integrals
are free of \ac{ir} singularities, which will be proved in
Section~\ref{ssec:qcd-scalarred:IRdivergences}.
Hence the integrals leading to a finite contribution
in the final result must contain a \ac{uv} divergence 
coming from the $\Gamma$ function,
\begin{equation}
\Gamma\left(N-\frac{n+2\alpha+2l}{2}\right)=\Gamma(N-2-\alpha-l+\varepsilon)
\equiv\Gamma(\varepsilon-\eta)\text{.}
\end{equation}
In order to produce a \ac{uv} divergence, the integer part of the argument
needs to fulfil $\eta\geq0$. Taking into account the $\varepsilon$ stemming
from $c_\alpha^n$ one obtains
\begin{equation}
\label{eq:qcd-dimreg:polynomial-loop01}
\varepsilon I_N^{n+2l+2\alpha}(l_1,\ldots,l_r;S)=\left\{\begin{array}{ll}
{\mathcal{O}}(\varepsilon),&-\eta>0\text{,}\\
(-1)^N\frac{1}{2^\eta\eta!}\int\limits_0\!\!\diff[N]z\,%
\delta_z\prod\limits_{j=1}^rz_{l_j}[z^\transposed Sz]^\eta
+{\mathcal{O}}(\varepsilon),\;&%nl
\phantom{-}\eta\ge0\text{.}
\end{array}\right.
\end{equation}
Finally, the \person{Feynman} parameter integral
\begin{equation}
\label{eq:qcd-dimreg:fppoly01}
P_{\alpha_1,\alpha_2,\ldots,\alpha_N}=\int\!\diff[N]{z}\delta_z\prod_{j=1}^Nz_j^{\alpha_j},\quad\text{with}\,\forall j\in\{1,\ldots,N\}:\alpha_j\ge0
\end{equation}
can be solved and the solution is given below.

We start from the equation
\begin{equation}
f_p(r,s)=\int_0^p\!\!\diff{x}x^r(p-x)^s=\frac{r!s!}{(r+s+1)!}p^{r+s+1}\text{,}
\end{equation}
which is shown by induction and using integration by parts.
Now we go back to~\eqref{eq:qcd-dimreg:fppoly01}, where the integration over~$z_N$ is
carried out over the $\delta$-function. What remains is the integral
\begin{equation}
P_{\alpha_1,\alpha_2,\ldots,\alpha_N}=
\prod_{j=1}^{N-1}\left(\int_0^{p_{N-j}}\diff{z_{N-j}}z_{N-j}^{\alpha_{N-j}}%
\right)(p_1-z_1)^{\alpha_N}
\end{equation}
with the upper bounds~$p_{i-1}=p_{i}-z_{i}$ and~$p_{N-1}=1$.
The rightmost integral is recognised as
\begin{equation}
\int_0^{p_1}\!\!\diff{z_1}z_1^{\alpha_1}(p_1-z_1)^{\alpha_N}=%nl
f_{p_1}(\alpha_1,\alpha_N)=%nl
\frac{\alpha_1!\alpha_N!}{(\alpha_1+\alpha_N+1)!}(p_2-z_2)^{%nl
\alpha_1+\alpha_N+1}
\end{equation}
and analogously one iterates through all the integrals to the left.
After the left integration the result is
\begin{equation}\label{eq:qcd-dimreg:solvePolyloop}
P_{\alpha_1,\alpha_2,\ldots,\alpha_N}=
\frac{\prod_{j=1}^N(\alpha_j!)}{(N-1+\sum_{i=1}^N\alpha_i)!}\text{.}
\end{equation}
Finally the symbol $P_N(j_1,\ldots,j_s)$ is introduced,
which counts the indices in a expression:
\begin{equation}
P_N(j_1,\ldots,j_s)=P_{\left(\sum_{i=1}^s\delta_{1,j_i}\right),\ldots,\left(\sum_{i=1}^s\delta_{N,j_i}\right)}\text{.}
\end{equation}
Substituted back into~\eqref{eq:qcd-dimreg:polynomial-loop01},
for the case $\eta\ge0$ one obtains
\begin{multline}\label{eq:qcd-dimreg:common-poly-solution}
\varepsilon I_N^{n-4+2(N+\eta)}(l_1,\ldots,l_r;S)=\\
\frac{(-1)^N}{2^\eta\eta!}\sum_{j_1,\ldots j_{2\eta}=1}^N S_{j_1j_2}\cdots S_{j_{2\eta-1},j_{2\eta}}
P_N(l_1,\ldots,l_r,j_1,\ldots,j_{2\eta})\text{.}
\end{multline}

Working in the \person{Feynman} gauge in \ac{qcd} one can put
a limit on the degree of the numerator, as the \person{Feynman} rules
in this case ensure that the tensorial rank of the integral never
exceeds the number~$N$ of loop propagators,
\begin{equation}
2l+2\alpha\le N\text{.}
\end{equation}
Using the definition of~$\eta=l+\alpha+2-N$ one obtains
\begin{equation}
0\le 2\eta\le 4-N\quad\Rightarrow\quad N\le4\quad\Rightarrow\quad l+\alpha\le2\text{.}
\end{equation}
Taking all these formul\ae\ together reveals that one does not need to know any integral with $\eta\ge2$.
On the other hand for~$0\le\eta\le1$ the formul\ae\ become very simple: for $\eta=0$ one has~$n+2l+2\alpha=n-4+N$
and hence
\begin{align}
\varepsilon I_N^{n-4+2N}(l_1,\ldots,l_r;S)&=(-1)^NP_N(l_1,\ldots,l_r)\text{.}
\intertext{For~$\eta=1$ one can replace~$n+2l+2\alpha=n+2(N+1)$ which yields}
\varepsilon I_N^{n-4+2(N+1)}(l_1,\ldots,l_r;S)&=\frac{(-1)^N}{2}\sum_{j_1,j_2=1}^NS_{j_1j_2}P_N(j_1,j_2,l_1,\ldots,l_r)\text{.}
\end{align}
Explicit formul\ae\ for the required cases are given in Appendix~\ref{app:appendix-integrals},
Section~\ref{app:appendix-integrals:polynomial}.

In this context a caveat of \ac{dreg} should be addressed:
the order of different limits can have important consequences and therefore
has to be treated with special care. Let us therefore refer back to the
definition of the tensor integrals~\eqref{eq:qcd-dimreg:defTI} and compare it
to an equally valid definition, where the~$\hat{q}^\mu$ in the numerator are
replaced by~$q^\mu$:
\begin{equation}
\label{eq:qcd-dimreg:defTIdifferently}
\check{I}_N^{d,\alpha;\mu_1\ldots\mu_r}(a_1,\ldots, a_r; S)\equiv%nl
\int\!\!\frac{\diff[d]k}{i\pi^{n/2}}\frac{\left(\bar{k}^2\right)^\alpha q_{a_1}^{\mu_1}\cdots q_{a_r}^{\mu_r}}{%nl
\prod_{j=1}^N(q_j^2-m_j^2+i\delta)}
\end{equation}
After a short calculation one finds for example
\begin{subequations}
\begin{align}
I_4^{n,1;\mu\nu}(a_1, a_2; S)&=
   -\left(\frac{1}{12}+{\mathcal O}(\varepsilon)\right)\hat{g}^{\mu\nu}%nl
\quad\text{versus}\\
\check{I}_4^{n,1;\mu\nu}(a_1, a_2; S)&=
   -\left(\frac{1}{8}+{\mathcal O}(\varepsilon)\right)g^{\mu\nu}\quad\text{.}
\end{align}
\end{subequations}

Since external vectors were assumed as four-dimensional
from na\"\i{}ve treatment of the limits one would expect that for a
vector~$p$ with~$p^2\neq0$ the expressions~$p_\mu p_\nu I^{n,1;\mu\nu}_4$
and~$p_\mu p_\nu \check{I}^{n,1;\mu\nu}_4$ should be the same but they
apparently are not. The reason can be seen if one splits the
second expression into 
\begin{equation}
\label{eq:qcd-dimreg:splitcoefficientinpolyint}
\check{I}_4^{n,1;\mu\nu}(a_1, a_2; S)=
Ag^{\mu\nu} = \hat{A}\hat{g}^{\mu\nu} + \bar{A}\bar{g}^{\mu\nu}\text{,}
\end{equation}
where one obtains the values
\begin{subequations}
\begin{align}
A&=-\frac{1}{8} + {\mathcal O}(\varepsilon)\text{,}\\
\hat{A}&=-\frac{1}{12} + {\mathcal O}(\varepsilon)\text{,}\\
\bar{A}&=\frac{1}{12\varepsilon} + {\mathcal O}(1)\text{.}
\end{align}
\end{subequations}
Contracting both sides of~\eqref{eq:qcd-dimreg:splitcoefficientinpolyint}
with $g_{\mu\nu}$ leads to the correct result but the projections on the
subspaces by contracting with $\hat{g}_{\mu\nu}$ and $\bar{g}_{\mu\nu}$
are different.
Multiplying the equation with $p_\mu p_\nu$ reveals that
\begin{equation}
\left(-\frac18 + {\mathcal O}(\varepsilon)\right)p^2\neq
-\left(\frac{1}{12} + {\mathcal O}(\varepsilon)\right)\hat{p}^2
+\left(\frac{1}{12\varepsilon} + {\mathcal O}(1)\right)\bar{p}^2\text{,}
\end{equation}
%and one clearly sees that the physical limit $\bar{p}\rightarrow0$ has to be
%taken before $\varepsilon\rightarrow0$ can be carried out. This example
%shows that the limits in our regularisation scheme do not commute.
%Furthermore, 
and hence, when using dimension splitting\footnote{Conversely, in \ac{ndr}
this type of integrals with powers of $\bar{k}^2$ in the numerator do
not appear.}
the above integral cannot be
decomposed into a $g^{\mu\nu}$ component only but must be treated as a
linear combination of $\hat{g}^{\mu\nu}$ and $\bar{g}^{\mu\nu}$ instead.

%% file: qcd-scalarred.tex
\subsection{Introduction}
In the previous section one-loop integrals have been introduced and
discussed in their general form, and a translation into \person{Feynman}
parameter integrals has been given. In the following section relations
between scalar integrals are established that lead to a reduction
algorithm which allows any one-loop tensor integral to be expressed
in terms of a limited set of standard scalar integrals.

The observation of relations between scalar integrals~\cite{Melrose:1965kb}
has been a key development for the first computations of
higher order corrections, e.g.
for $e^++e^-\rightarrow \mu^++\mu^-$~\cite{Passarino:1978jh} and
later $e^++e^-\rightarrow e^++e^-+X$ with a pseudoscalar $X$ in
the final state~\cite{vanNeerven:1984ak}, and
the techniques used
for these early computations led into the systematic development
of reduction techniques for one-loop
integrals~\cite{Davydychev:1991va,Fleischer:1999hq}. A systematic
treatment of critical phase-space regions has been addressed
by different approaches~\cite{Binoth:1999sp,%
Denner:2002ii,Binoth:2005ff,Denner:2005nn}

In my present calculation the reduction method
proposed in~\cite{Binoth:2005ff} is used.
This chapter describes the foundation of this method;
for a full list of form factors one should refer to the original paper.
A complete list of the scalar one-loop integrals that are required
as a basis has been compiled in~\cite{Ellis:2007qk} and is provided
as a numerical implementation by the same authors~\cite{qcdloop}.
 
\subsection{Infrared Divergences}\label{ssec:qcd-scalarred:IRdivergences}
It has been mentioned already that the reduction formalisms,
both for scalar integrals and for tensor integrals
separates infrared poles in the integrals and groups them
such that cancellations can be carried out easily.
While ultraviolet poles in dimensional regularisation
stem from singularities of the $\Gamma$-function
and are removed systematically by renormalisation,
infrared divergences appear when massless particles
propagate through the loop. 
These singularities usually are kept during the calculation
and have to cancel in the end.

Two classes of infrared divergences have to be distinguished: A \emph{soft divergence} arises
when the integration momentum~$k$ becomes soft, i.e. $k^\mu\rightarrow0$.
Let all masses~$m_i$ be zero and all external particles be lightlike.
Using shift invariance the integral~$I_N^{d;\mu_1\ldots\mu_r}(S)$ under~$k\rightarrow \lambda k-r_a$,
where~$\lambda$ denotes an arbitrary real variable,
becomes\footnote{I do not distinguish between four and $n$-dimensional vectors here. It should be clear from former
definitions where to use~$\hat{q}$ instead of~$q$. For the discussion of infrared singularities this difference, however,
is irrelevant.}
\begin{equation}
I_N^{d;\mu_1\ldots\mu_r}(a_1,\ldots a_r; S)=%nl
\int\!\!\frac{\lambda^d\diff[d]k}{i\pi^{n/2}}\frac{(\lambda k^{\mu_1}+\Delta_{a_1a}^{\mu_1})\cdots(\lambda k^{\mu_r}+\Delta_{a_r}^{\mu_r})}{%nl
\lambda^2k^2\prod_{j\in S^{\{a\}}}\left((\lambda k+\Delta_{ja})^2+i\delta\right)}\text{.}
\end{equation}
Now the soft limit is taken by~$\lambda\rightarrow0$. Therefore the soft infrared behaviour of the integral is determined by the overall power of
lambda. Hence a non-trivial numerator can improve the infrared behaviour, i.e. increase power of~$\lambda$, but never generate singularities.
In this respect the scalar functions are the worst case to be studied. In a scalar function there are three sources for~$\lambda^{-l}$ 
with negative exponent~$(-l)$. Obviously the~$k^2$ term in the denominator contributes~$\lambda^{-1}$ but also two more propagators,
\begin{equation}
(\lambda k+\Delta_{(a\pm1)a})^2=
\lambda^2k^2+2\lambda k\cdot\Delta_{(a\pm1)a}+\Delta_{(a\pm1)a}^2=
2\lambda k\cdot\Delta_{(a\pm1)a}+\mathcal{O}(\lambda^2)
\end{equation}
cause problems. Here I used the fact that~$\Delta_{(a\pm1)a}^2=p_{a\pm1}^2=0$.
Including also the differential which yields a~$\lambda^d$ the overall degree of divergence is~$(d-4)$. This is negative
for~$d=n=4-2\varepsilon$ but always positive for any $d>4$. Therefore all higher dimensional scalar integrals are free of
soft divergences.

The second class of divergences is called \emph{soft collinear infrared singularities}. They arise when two partons become
collinear.
The singularities can be exposed by the following procedure: the integration momentum is replaced
by~$k\rightarrow\lambda_\parallel k_\parallel+\lambda_\perp k_\perp-r_a$,
where~$k_\perp\cdot p_a=k_\perp\cdot k_\parallel=0$
and~$k_\parallel$ lies in the one dimensional subspace of the $d$-dimensional
\person{Minkowski} space that is spanned by~$\langle p_a\rangle$.

Now one can examine the collinear behaviour of scalar an arbitrary scalar integral,
\begin{equation}
I_N^{d}(S)=
\int\!\!\frac{\lambda_\perp^{d-1}\lambda_\parallel\diff[d-1]k_\perp\diff k_\parallel}{i\pi^{n/2}}
\frac{1}{%nl
\prod_{j\in S}\left((\lambda k+\Delta_{ja})^2+i\delta\right)}\text{.}
\end{equation}
The dangerous propagators in that case are
\begin{subequations}
\begin{align}
(k+\Delta_{aa})^2&=\lambda_\perp^2k_\perp^2\text{,}\\
(k+\Delta_{(a-1)a})^2&=\lambda_\perp^2k_\perp^2
%+2\lambda_\parallel k_\parallel\cdot p_a
\quad\text{and}\\
(k+\Delta_{(a+1)a})^2&=\lambda_\perp^2k_\perp^2+2\lambda_\parallel k_\parallel\cdot p_{a+1}+
\lambda_\perp k_\perp\cdot p_{a+1}\text{.}
\end{align}
\end{subequations}
Now first the collinear limit is taken, i.e.~$\lambda_\perp\rightarrow0$; this limit causes no poles for $d>5$.
Performing~$\lambda_\parallel\rightarrow0$ in addition is safe as well since the~$\lambda_\parallel$ in numerator
and denominator cancel exactly. This proves that all integrals in~$d\ge6-2\epsilon$ are infrared safe.
Similar techniques can be applied to reveal infrared divergences in the real emission part of the amplitude.

\subsection{Subtraction Method for Scalar Integrals}
In this section I consider only integrals of the type
\begin{multline}
\label{eq:qcd-scalarred:SI01}
I_N^{d}(S)=%nl
\int\!\!\frac{\diff[n]k}{i\pi^{n/2}}\frac{1}{%nl
\prod_{j\in S_\#}(q_j^2-m_j^2+i\delta)}=\\
(-1)^N\Gamma(N-d/2)\int_0\!\!\diff[N]z\delta_z\frac{1}{%nl
\left(-\frac12z^\transposed Sz-i\delta\right)^{N-d/2}}\text{.}
\end{multline}
With $S_\#$ I denote the support of the matrix~$S\in\Rset^{N\times N}$, 
\begin{equation}
S_\#\equiv\left\{i\in\{1,\ldots,N\}\,\vert\,\exists j\in\{1,\ldots,N\}:\,S_{ij}\neq0\right\}\text{.}
\end{equation}
This notation will become important because pinches of the 
matrix $S$, which are defined by
\begin{equation}
S_{ij}^{\{l_1,l_2,\ldots,l_{m}\}}=\left\{\begin{array}{ll}
S_{ij},&\{i,j\}\cap\{l_1,l_2,\ldots,l_m\}=\emptyset,\\
0,&\text{otherwise}\text{,}
\end{array}\right.
\end{equation}
play a central role in the formulation of the reduction algorithm\footnote{
Alternatively, the algorithm could be written down in terms of
matrices of different sizes.}.
For example, starting from a non-singular matrix~$S$,
$S^{\{j_1,j_2\}}_\#=\{1,\ldots,N\}-\{j_1, j_2\}$.
All pinched matrices,
that is all matrices where $\{l_1, l_2,\ldots\}\neq\emptyset$,
are singular by definition. As a proper way to treat them
 will use the \person{Moore}-\person{Penrose}
pseudoinverse\footnote{A detailed definition
of the \person{Moore}-\person{Penrose} pseudoinverse
can be found in Appendix~\ref{app:MPinverse}.}
$\tilde{S}^{\{l_1,\ldots,l_m\}}$
of the pinched matrix~$S^{\{l_1,\ldots,l_m\}}$.
In the presence of pinched matrices some of the
previous definitions need to be slightly changed,\index{.deltaZ@$\delta_z$}
\begin{align*}
\delta_z&=\delta\left(1-\sum\nolimits_{j\in S_\#}z_j\right)\text{,}\\
\quad z^\transposed Sz&=\sum_{i,j\in S_\#}z_iS_{ij}z_j\\
\intertext{and}
\int_0\!\!\diff[N]z&=\int_{-\infty}^\infty\prod_{j\in S_\#}^N\left(\diff{z_j}\Theta(z_j)\right)\text{.}
\end{align*}

The aim of the reduction is to split the integral into an infrared safe part and a remainder that contains all
possible sources for infrared singularities,
\begin{equation}\label{eq:qcd-scalarred:splitting}
I_N^n(S)=I_\text{div}+I_\text{fin}\text{.}
\end{equation}
We will see that at the end of each reduction chain the infrared
poles are always contained in the three-point functions.

To obtain a form like~\eqref{eq:qcd-scalarred:splitting} the numerator of the integral is rewritten as
a linear combination of the propagators,
\begin{multline}\label{eq:qcd-scalarred:split01}
I_N^n(S)=\sum_{j\in S_\#}b_j(S)\int\!\!\frac{\diff[n]k}{i\pi^{n/2}}\frac{(q_j^2-m_j^2)}{%nl
\prod_{j\in S_\#}(q_j^2-m_j^2+i\delta)}\\+\int\!\!\frac{\diff[n]k}{i\pi^{n/2}}\frac{1-\sum_{j\in S_\#}b_j(S)(q_j^2-m_j^2)}{%nl
\prod_{j\in S_\#}(q_j^2-m_j^2+i\delta)}\text{.}
\end{multline}

The first term of the expression \eqref{eq:qcd-scalarred:split01} is the sum of
pinched integrals
\begin{equation}
I_\text{div}=\sum_{j\in S_\#}b_j(S)I_{N-1}^n(S^{\{j\}})
\end{equation}
In the second term, analogous to the procedure in section~\ref{ssec:qcd-dimreg:loop-integrals}, we introduce \person{Feynman}
parameters and shift the origin of the integration momentum by $k\rightarrow k-\sum_{i\in S_\#}z_ir_i$. The denominator can be
written, as usual, in quadratic form; the numerator becomes
\begin{multline}
1-\sum_{j\in S_\#}b_j(S)\left((k-\sum_{i\in S_\#}z_i\Delta_{ij})^2-m_j^2\right)=\\
-\left(\sum_{j\in S_\#}b_j\right)\left(k^2-\frac12z^\transposed Sz\right)+\left[1-\sum_{j,k\in S_\#}z_k(b_jS_{jk}-2\Delta_{kj}\cdot k)\right]\text{.}
\end{multline}
To obtain that result one needs the relation $2\Delta_{ij}\cdot\Delta_{kl}=S_{il}+S_{jk}-S_{ik}-S_{jl}$. The term linear in the integration
momentum vanishes under symmetric integration. Now we can choose the~$b_j(S)$, which are still undetermined, such that the square
bracket vanishes. Once more we use the fact that~$\sum_{k\in S_\#}z_k=1$ and hence find the condition
\begin{equation}
\label{eq:qcd-scalarred:bj-condition}
\sum_{j\in S_\#}S_{ij}b_j(S)=1
\end{equation}
for the bracket to vanish. Before finding the solution of the above equation the remaining integral shall be brought back into
standard form. Therefore I introduce the symbol~$B(S)=\sum_{j\in S_\#}b_j(S)$, and hence
\begin{equation}
\label{qcd-scalarred:d+2:1}
I_\text{fin}=-B(S)\Gamma(N)\int_0\!\!\diff[N]z\delta_z\int\!\!\frac{\diff[n]k}{i\pi^{n/2}}\frac{(k^2-\frac12 z^\transposed Sz)}{%nl
[k^2+\frac12 z^\transposed Sz+i\delta]^N}
\end{equation}
Carrying out the momentum integration leaves us with
\begin{equation}
I_\text{fin}=-B(S)\Gamma(N)\int_0\!\!\diff[N]z\delta_z\frac{(-1)^{N+1}\Gamma(N)}{%nl
\Gamma(\frac{n}{2})[-\frac12z^\transposed Sz-i\delta]^{N-\frac{n+2}{2}}}%nl
\int_0^\infty\!\!\diff{x}\frac{x-1}{(x+1)^N}x^{\frac{n-2}{2}}\text{.}
\end{equation}
The integral over~$x$ is a difference of beta-functions,
\begin{multline}
\int_0^\infty\!\!\diff{x}\frac{x-1}{(x+1)^N}x^{\frac{n-2}{2}}=%nl
\frac{\Gamma(\frac{n+2}{2})\Gamma(N-\frac{n+2}{2})}{\Gamma(N)}-\frac{\Gamma(\frac{n}{2})\Gamma(N-\frac{n}{2})}{\Gamma(N)}=\\
\frac{\Gamma(\frac{n}{2})\Gamma(N-\frac{n+2}{2})}{\Gamma(N)}\left(\frac{n+2}{2}-N+\frac{n}{2}+1\right)\text{,}
\end{multline}
and therefore the whole integral is
\begin{equation}
\label{qcd-scalarred:d+2:2}
I_\text{fin}=-B(S)(N-n-1)I_N^{n+2}(S)\text{.}
\end{equation}

In the remaining part of this section I will review the result of~\cite{Binoth:2005ff}, that~\eqref{eq:qcd-scalarred:bj-condition} for any case
can be solved using the pseudoinverse~$\tilde{S}$,
\begin{equation}
b_j(S)=\sum_{i\in S_\#}\tilde{S}_{ij}\text{.}
\end{equation}

One always can rewrite~$S$ in terms of the \person{Gram} matrix~$G^{(a)}$
\index{Gram@\person{Gram}!matrix|main}
and a remainder,
\begin{subequations}
\label{eq:qcd-scalarred:SComponents}
\begin{align}\label{eq:qcd-scalarred:splitS}
S&=-G^{(a)}+v^{(a)}\eta^\transposed+\eta v^{(a)\transposed},\quad\text{with}\\
\label{eq:qcd-scalarred:defineG}G^{(a)}_{ij}&=2\Delta_{ia}\cdot\Delta_{ja}\text{,}\\
v^{(a)}_i&=\Delta_{ia}^2-m_i^2\quad\text{and}\\
\eta_i&=1,\quad\forall i\in S_\#\text{.}
\end{align}
\end{subequations}
The vectors~$\eta$ and~$v^{(a)}$ are column vectors; all (explicit and implicit) sums are understood over the support~$S_\#$ of~$S$.
In what follows the index~$a$ is kept constant and therefore the superscript~$\phantom{i}^{(a)}$ is omitted. The definition of~$G^{(a)}$
exposes that $G^{(a)}_{ia}=G^{(a)}_{ai}=0,\forall i\in S_\#$.

In this notation equation~\eqref{eq:qcd-scalarred:bj-condition} reads
\begin{equation}
-Gb+v\eta^\transposed b+\eta v^\transposed b=\eta\text{,}
\end{equation}
and can equally be written as a set of two equations,
\begin{subequations} \label{eq:qcd-scalarred:splitLES}
\begin{align}
\label{eq:qcd-scalarred:splitLES01}
Gb&=(\eta^\transposed b)v-Bv_a\eta=B(v-v_a\eta)\text{,}\\
\label{eq:qcd-scalarred:splitLES02}
(v^\transposed b)\eta&=(1-Bv_a)\eta\quad\Leftrightarrow v^\transposed b=1-Bv_a\text{.}
\end{align}
\end{subequations}

In the case~$\det{S}\neq0$, the matrix~$S$ is
regular and the pseudoinverse is equal to the normal inverse~$S^{-1}$.
The required condition\footnote{$S=S^\transposed$ is used.}
\begin{equation}
\sum_{j\in S_\#}S_{ij}b_j(S)=\sum_{j\in S_\#}\sum_{k\in S_\#}S_{ij}S_{ik}^{-1}=\sum_{k\in S_\#}\delta_{ik}=1
\end{equation}
trivially holds. If, however, $\det{S}=0$ one has to ensure that
\begin{equation} \label{eq:qcd-scalarred:consistency-eta}
S\tilde{S}\eta=\eta
\end{equation}
is fulfilled. According to theorem~\ref{thrm:penrose} constructing a solution to the
linear system~\eqref{eq:qcd-scalarred:splitLES}
implicitly proves equation~\eqref{eq:qcd-scalarred:consistency-eta}.

\index{Gram@\person{Gram}|(}
First the \person{Gram} matrix~$G$ is represented in an or orthonormal basis~$e^\mu_1,\ldots,e^\mu_r$ of the
subspace~$\langle\Delta_{ia}^\mu\vert i\in S_\#\rangle$, where $r$~is the rank of~$G$, $r=\rank{G}$. 
Since the physical \person{Minkowski} space has dimension~$4$ there are never
more than~4 linear independent external momenta, or equally one always has~$\rank{G}\leq\min(N-1,4)$.
Now one can write
\begin{equation}\label{eq:qcd-scalarred:DeltaAndR}
\Delta_{ia}^\mu=\sum_{m=1}^rR^{(a)}_{mi}e_m^\mu,\quad \hat{G}^{(a)}_{ij}=2e_i\cdot e_j=2\delta_{ij},\quad R^{(a)}_{ij}=e_i\cdot\Delta_{ja}\text{.}
\end{equation}
Hence one has~$R\in\Rset^{r\times N}$, $\hat{G}, RR^\transposed\in\Rset^{r\times r}$ and $G,R^\transposed R\in\Rset^{N\times N}$,
and $G$ and~$\hat{G}$ are related by $G=R^\transposed\hat{G}R$. The matrix~$R$ has full line rank and therefore $RR^\transposed$ is invertible. This
allows to construct the pseudoinverse of~$G$ explicitly,
\begin{equation}
\tilde{G}=R^\transposed(RR^\transposed)^{-1}\hat{G}^{-1}(RR^\transposed)^{-1}R
\end{equation}
\index{Gram@\person{Gram}|)}%

To invert \eqref{eq:qcd-scalarred:splitLES01} the consistency condition from theorem~\ref{thrm:penrose} has to be fulfilled,
\begin{equation}\label{eq:qcd-scalarred:consistencyG}
B (\One-G\tilde{G})(v-v_a\eta)=0\text{.}
\end{equation}
Clearly, for $\rank{G}<N-1$ and general~$v$ this equation can only be satisfied\footnote{Still a special $v$ could be such that
$(\One-G\tilde{G})(v-v_a\eta)=0$. In that case theorem~\ref{thrm:penrose} also holds for $B\neq0$ and the solution is given by
the theorem through the pseudoinverse, i.e.~$b=B\tilde{G}(v-v_a\eta)+(\One-\tilde{G}G)u$, where one always can choose
$u$ such that the two remaining constraints~$\eta^\transposed b=B$ and~$v^\transposed b=1+Bv_a$ are met.}
when~$B=0$, since
$(\One-G\tilde{G})$ projects on $\Kern{G}$, and $\dim(\Kern{G})\ge2$.
For the case $\rank{G}=N-1$ and~$\det{S}\neq0$ the solution has been given before, and hence only the case~$B=0$ is given further investigation.
The problematic regions where $\rank{G}=N-1$ but at the same time~$\det{S}=0$
are discussed separately in section~\ref{ssec:scalarred:smatrixzero}.

%the matrix~$R^{\{a\}}$, which similar to the definition of~$S^{\{a\}}$ is obtained by deleting the $a$-th
%row from~$R$ is regular\footnote{$R^{\{a\}}\in\Rset^{(N-1)\times(N-1)}$ describes an orthogonal
%transformation on the~$\Delta^\mu_{ia}$.}. This allows to single out the $a$-th row from
%equation~\eqref{eq:qcd-scalarred:consistencyG}, which is always fulfilled. For the remaining rows I use
%the fact that
%\begin{multline}
%G\tilde{G}=R^\transposed(RR^\transposed)^{-1}R
%\Rightarrow
%R^{\{a\}\transposed}(RR^\transposed)^{-1}R^{\{a\}}=\\
%R^{\{a\}\transposed}(R^{\{a\}}R^{\{a\}\transposed})^{-1}R^{\{a\}}=
%R^{\{a\}\transposed}R^{\{a\}\transposed\,-1}R^{\{a\}\,-1}R^{\{a\}}=\One_{N-1}\text{.}
%\end{multline}

 The system~\eqref{eq:qcd-scalarred:splitLES} now simplifies to
\begin{equation}
Gb=0,\quad v^\transposed b=1,\quad B=\eta^\transposed b=0\text{.}
\end{equation}
With the abbreviations~$\delta v=v-v_a\eta$ and~$K_G=\One-\tilde{G}G$, which is the projector on~$\Kern{G}$,
the solution is easily constructed:
\begin{subequations}
\begin{align}
b_i&=\frac{(K_G\delta{v})_i}{\delta{v}^\transposed K_G\delta{v}}+\sum_{j=1}^{N-\rank{G}-2}\beta_ju^{(j)}_i, \quad\text{if}\,i\in S^{\{a\}}_\#\text{,}\\
b_a&=-\sum_{j\in S^{\{a\}}_\#}b_j\text{,}
\end{align}
\end{subequations}
where $u^{(j)}$ form a basis of~$\Kern{G}\cap\langle\delta{v},\delta_a\rangle^\perp$, and
$\langle\delta{v},\delta_a\rangle^\perp$ is the space orthogonal to~$\delta{v}$ and~$\delta_a$, where~$(\delta_a)_i=\delta_{ai}$.
The vector~$\delta_a$ has to be projected out because in the solution~$b(S)$ the remnant of~$\delta_a$ is the null-vector.
The upper bound in the sum is the dimension of the solution space,
\begin{equation}
\dim(\Kern{G}\cap\langle\delta{v},\delta_a\rangle^\perp)=\dim(\Kern{G})-\dim(\langle\delta{v},\delta_a\rangle)=
N-\rank{G}-2\text{.}
\end{equation}
The real constants~$\beta_j$ parametrise the solution~$b(S)$. This allows to determine the rank of~$S$ for the
dimension of the solution space of $Sb=\eta$ is~$N-\rank{G}-2$; hence
\begin{equation}\label{eq:qcd-scalarred:rkS=rkG+2}
\rank{S}=\rank{G}+2,\quad\text{for}\,\rank{G}\leq N-2\text{.}
\end{equation}

Though the solvability of $Sb=\eta$ has already been shown, the result can be exploited further. We have seen that
the rank of~$G$ distinguishes two kinematic situations: the case~$\rank{G}=\min(4,N-1)$ is called \emph{non-exceptional kinematics},
conversely~$\rank{G}<\min(4,N-1)$ is called~\emph{exceptional kinematic}.
For non-exceptional kinematics
relation~\eqref{eq:qcd-scalarred:rkS=rkG+2} can be expressed as
\begin{equation}
\rank{S}=\min(N,6)\text{.}
\end{equation}
Since~$S$ can be split through~\eqref{eq:qcd-scalarred:splitS} one finds another interpretation of~\eqref{eq:qcd-scalarred:rkS=rkG+2}.
The matrix
\begin{equation}
M\equiv v\eta^\transposed+\eta v^\transposed
\end{equation}
is of rank~2 as I will show now.

The eigenvectors of~$M$ are
\begin{equation}
x_\pm=\alpha_\pm\left(v\pm\sqrt{\frac{v^\transposed v}{N}}\eta\right)
\end{equation}
with the eigenvalues~$\lambda_\pm=v^\transposed\eta\pm\sqrt{Nv^\transposed v}$ and
the normalisation constants~$\alpha_\pm$ such that $x_\pm^\transposed x_\pm=1$, and
an orthonormal basis~$x_i,\,i=1,\ldots,N-2$ of the eigenspace for the eigenvalue~$0$,
\begin{equation}
\langle x_1,\ldots,x_{N-2}\rangle=\langle v,\eta\rangle^\perp\text{.}
\end{equation}
The orthogonal transformation matrix $O=(x_1,\ldots,x_{N-2},x_+,x_-)$ diagonalises~$M$,
\begin{equation}
M=O^\transposed\diag(0,\ldots,0,\lambda_+,\lambda_-)O
\end{equation}
and hence the rank of M is~$2$. The pseudoinverse of~M is given by
\begin{equation}
\tilde{M}=O^\transposed\diag(0,\ldots,0,1/\lambda_+,1/\lambda_-)O.
\end{equation}
This proves the rank formula
\begin{equation}\label{eq:qcd-scalarred:rankformulaSGM}
\rank{S}=\rank{(-G+M)}=\rank{G}+\rank{M}
\end{equation}
for exceptional kinematics.
The fact that~\eqref{eq:qcd-scalarred:rankformulaSGM} is fulfilled
allows to use explicit formul\ae\ for the pseudoinverse~$\tilde{S}$ 
which are given in~\cite{Henderson81,fill00moorepenrose}.
However, for a numerical implementation an appropriate choice is the
\person{Greville} algorithm~\cite{263213}.

To conclude this section I briefly review what one has gained so far. The reduction algorithm for scalar integrals
splits the integrals into a infrared finite, higher dimensional part and a sum of pinched integrals containing
the infrared divergences:
\begin{equation}\label{eq:qcd-scalarred:final}
I^n_N(S)=\sum_{k\in S_\#}b_k(S) I_{N-1}^n(S^{\{k\}})-B(S)(N-n-1)I_N^{n+2}(S)
\end{equation}
The solutions~$b_k(S)$ are constructed such that~$B(S)$ generally vanishes for~$N\geq6$;
in exceptional kinematics~$B(S)$ vanishes for arbitrary~$N$. In the case~$N=5$ the coefficient
$(N-n-1)=\mathcal{O}(\varepsilon)$ and can be dropped in phenomenological applications
since~$I_5^{n+2}(S)$ is both, ultraviolet and infrared finite.

%%%%%%%%%%%%%%%%%%%%%%%%%%%%%%%%%%%%%%%%%%%%%%%%%%%%%%%%%%%%%%%%%%%
\subsection{A Determinant Relation}
In~\cite{Binoth:1999sp,Bern:1993kr} a relation between the
\person{Gram} determinant and~$\det{S}$ is given,
\begin{equation}\label{eq:qcd-scalarred:detrelation}
\det{G^{\fmslash{a}}}=(-1)^{N-1}B\det{S}\text{,}
\end{equation}
where $G^{\fmslash{a}}$ refers to the matrix constructed from~$G^{(a)}$ by
eliminating the $a$-th row and column; analogously I call
$\delta v^{\fmslash{a}}$ the result of leaving out the $a$-th row
from~$\delta v^{(a)}$.

This equation reveals that the term proportional to~$I_N^{d+2}$ must vanish
in~\eqref{eq:qcd-scalarred:final} for~$N\geq6$ where $\det G=0$ is generally
true. On the other hand this relation proves, that
in~\ref{ssec:qcd-tensorred-ibp:TensorRed} the application
of~\eqref{eq:qcd-scalarred:final} introduces inverse \person{Gram} determinants
through the $1/B$~terms.

The proof is given for invertible~$G^{\fmslash{a}}$.
We have seen in the previous section that for $\det{G^{\fmslash{a}}}=0$
the $b_j$ can always be chosen such that~$B=0$, which
satisfies~\eqref{eq:qcd-scalarred:detrelation} trivially.
It is
\begin{equation}
\label{eq:qcd-scalarred:detrel01}
B\det{S}=B\det(-G^{(a)}+v\eta^\transposed+\eta v^\transposed)\text{,}
\end{equation}
and by standard determinant manipulations one can rewrite this
determinant in block matrix form,
\begin{equation}
B\det{S}=B\det\left(\begin{array}{cc}
-G^{\fmslash{a}}&\delta v^{\fmslash{a}}\\
(\delta v^{\fmslash{a}})^\transposed & 2v^{(a)}_a
\end{array}\right)\text{.}
\end{equation}
For~$\det{G^{\fmslash{a}}}\neq0$ a proper inverse of~$G^{\fmslash{a}}$
exists, and therefore one can write
\begin{multline}
B\det{S}=
B\det(-G^{\fmslash{a}})\cdot\det(2v^{(a)}_a+
(\delta v^{\fmslash{a}})^\transposed (G^{\fmslash{a}})^{-1}
\delta v^{\fmslash{a}})=\\
(-1)^{N-1}\det(G^{\fmslash{a}})\cdot
(2Bv^{(a)}_a+
(\delta v^{\fmslash{a}})^\transposed (G^{\fmslash{a}})^{-1}
B\delta v^{\fmslash{a}})\text{.}
\end{multline}

The bracket simplifies to one because
\eqref{eq:qcd-scalarred:splitLES} can be expressed as
\begin{subequations}
\begin{align}
B\delta v^{\fmslash{a}}&=G^{\fmslash{a}}b\quad\text{and}\\
(\delta v^{\fmslash{a}})^\transposed b &= 1 - 2Bv^{(a)}_a\text{,}
\end{align}
\end{subequations}
and hence leads to
\begin{multline}
2Bv^{(a)}_a+
(\delta v^{\fmslash{a}})^\transposed (G^{\fmslash{a}})^{-1}
B\delta v^{\fmslash{a}}=\\
2Bv^{(a)}_a+
(\delta v^{\fmslash{a}})^\transposed (G^{\fmslash{a}})^{-1}
G^{\fmslash{a}}b=
2Bv^{(a)}_a+1-2Bv^{a}_a=1\text{.}
\end{multline}

\subsection{Problematic Phase Space Regions}
\label{ssec:scalarred:smatrixzero}
\begin{fmffile}{phasespace}
The previous sections reveal clearly that the reduction for scalar integrals
as described above is unproblematic in all except one cases: as can be seen
from the determinant relation~\eqref{eq:qcd-scalarred:detrelation}, no solution
for~$B$ can be found when~$\det{S}$ vanishes while~$\det{G^{\fmslash{a}}}$
remains non-zero.

Due to the complexity of the full problem\footnote{
In general there are $N(N-1)/2$ \person{Mandelstam} variables plus
$N$ propagator masses which have to considered when finding the roots
of~$\det{S}$} I restrict the discussion to the case of
$2\rightarrow(N-2)$~scattering, with $N\leq6$, where all the external and
internal particles are massless.

The conjecture to be shown in this section is that all problematic phase
space regions, i.e.~$\det{S}=0\wedge\det{G^{\fmslash{a}}}\neq0$, in fully
massless $2\rightarrow(N-2)$-processes for $N\leq6$ lie only on the soft
and collinear phase space boundaries. On the other hand, $N$ is also bounded
from below: since triangles are the endpoint of the reduction only 
(sub-)determinants down to size four are considered.\footnote{The tensor
reduction of three-point functions is described in
section~\ref{ssec:qcd-tensorred-ibp:threepoint}
and does not involve~$\det{S}$.}

\begin{figure}[hbtp]
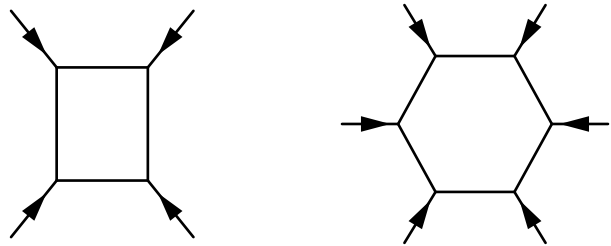

\centering
\subfloat[][A box topology]{\label{fig:qcd-scalarred:top:box}
\parbox[c][45mm][c]{40mm}{
\centering
\begin{fmfchar*}(30,30)
\fmfleft{s1,s2}\fmfright{s4,s3}
\fmflabel{$k_1$}{s1}
\fmflabel{$k_2$}{s2}
\fmflabel{$k_3$}{s3}
\fmflabel{$k_4$}{s4}
\fmf{fermion}{s1,q1}\fmf{fermion}{s2,q2}
\fmf{fermion}{s3,q3}\fmf{fermion}{s4,q4}
\fmf{plain,tension=0.5}{q1,q2,q3,q4,q1}
\end{fmfchar*}}
}
\qquad
\subfloat[][A hexagon topology]{\label{fig:qcd-scalarred:top:hex}
\parbox[c][45mm][c]{40mm}{
\begin{fmfchar*}(35,35)
\fmfsurroundn{s}{6}
\begin{fmffor}{n}{1}{1}{6}
   \fmf{fermion}{s[n],q[n]}
\end{fmffor}
\fmfcyclen{plain,tension=6/8}{q}{6}
\fmflabel{$k_1$}{s1}
\fmflabel{$k_2$}{s2}
\fmflabel{$k_3$}{s3}
\fmflabel{$k_4$}{s4}
\fmflabel{$k_5$}{s5}
\fmflabel{$k_6$}{s6}
\end{fmfchar*}}
}
\caption{The topologies used in the discussion.}
\label{fig:qcd-scalarred:top}
\end{figure}

In the case $N=4$ according to \figref{fig:qcd-scalarred:top:box} the
determinant of~$S$ is just the product
\begin{equation}
\det{S}=s_{12}^2s_{23}^2\text{.}
\end{equation}
The \person{Mandelstam} variables are defined as
\begin{subequations}
\begin{align}
s_i&=k_i^2\text{,}\\
s_{ij}&=(k_i+k_j)^2\text{,}\\
s_{ijl}&=(k_i+k_j+k_j)^2\text{,}\\
\ldots\nonumber
\end{align}
\end{subequations}
where the $k_i$ are the external momenta,
defined as all ingoing. Where appropriate
I identify~$s_{12}\equiv s$, the centre
of mass energy of the colliding partons.
%%%%%%%%%%%%%%%%%%

Hence the determinant only is zero when either~$s_{12}$ or~$s_{23}$ vanishes.
By choosing an appropriate explicit representation of the kinematics the
vanishing of a \person{Mandelstam} variable can be related to an infrared
situation. Therefore I consider the centre of mass system of the ingoing
particles, in which the four-vectors read
\begin{subequations}
\begin{align}
k_1&=\sqrt{s}/2(1,0,0,1)\text{,}\\
k_2&=\sqrt{s}/2(1,0,0,-1)\text{,}\\
k_3&=E_3(-1,0,\sin{\varphi},\cos{\varphi})\text{,}\\
k_4&=-(k_1+k_2+k_3)\text{.}
\end{align}
\end{subequations}
In this parametrisation we have~$s_{12}=s$
and~$s_{23}=-\sqrt{s}E_3(1+\cos\varphi)$. The roots
of $\det{S}$ lie where one of the energies vanishes and
where the outgoing particles become collinear with the
beam axis, both of which are infrared situations. On the other hand
the point~$\cos\varphi=-1$ can be related directly to vanishing
transverse momentum $p_{3,T}=E_3\sin\varphi=0$. 

Analogous treatment to the four-particle case is sufficient for 
$N=5$. The determinant is~$\det{S}=2s_{12}s_{23}s_{34}s_{45}s_{51}$,
and hence the roots of the determinant clearly lie on the phase space
boundaries.

\begin{figure}[hbtp]
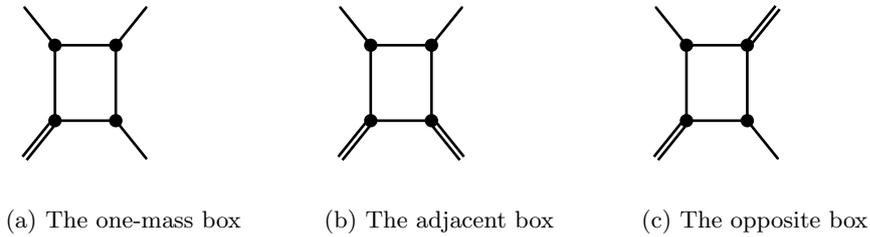

\centering
\subfloat[][The one-mass box]{\label{fig:qcd-scalarred:boxes:1m}
\parbox[c][30mm][c]{30mm}{
\begin{fmfchar*}(20,20)
\fmfleft{s1,s2}\fmfright{s4,s3}
\fmf{double}{s1,q1}
\fmf{plain}{s2,q2}
\fmf{plain}{s3,q3}
\fmf{plain}{s4,q4}
\fmfcyclen{plain,tension=1/2}{q}{4}
\fmflabel{$1,2,3$}{s1}
\fmflabel{$4$}{s2}
\fmflabel{$5$}{s3}
\fmflabel{$6$}{s4}
\fmfdotn{q}{4}
\end{fmfchar*}}
}
\qquad
\subfloat[][The adjacent box]{\label{fig:qcd-scalarred:boxes:ad}
\parbox[c][30mm][c]{30mm}{
\begin{fmfchar*}(20,20)
\fmfleft{s1,s2}\fmfright{s4,s3}
\fmf{double}{s1,q1}
\fmf{plain}{s2,q2}
\fmf{plain}{s3,q3}
\fmf{double}{s4,q4}
\fmfcyclen{plain,tension=1/2}{q}{4}
\fmflabel{$6,1$}{s1}
\fmflabel{$2$}{s2}
\fmflabel{$3$}{s3}
\fmflabel{$4,5$}{s4}
\fmfdotn{q}{4}
\end{fmfchar*}}
}
\qquad
\subfloat[][The opposite box]{\label{fig:qcd-scalarred:boxes:op}
\parbox[c][30mm][c]{30mm}{
\begin{fmfchar*}(20,20)
\fmfleft{s1,s2}\fmfright{s4,s3}
\fmf{double}{s1,q1}
\fmf{plain}{s2,q2}
\fmf{double}{s3,q3}
\fmf{plain}{s4,q4}
\fmfcyclen{plain,tension=1/2}{q}{4}
\fmflabel{$6,1$}{s1}
\fmflabel{$2$}{s2}
\fmflabel{$3,4$}{s3}
\fmflabel{$5$}{s4}
\fmfdotn{q}{4}
\end{fmfchar*}}
}
%\label{fig:qcd-scalarred:top:h}
\caption{The three different types of pinched boxes that can arise from
the reduction of a six-point diagram. Only one representative permutation
of external legs is considered.}
\label{fig:qcd-scalarred:boxes}
\end{figure}

Next, I consider the situation for $N=6$ kinematics
(s.~\figref{fig:qcd-scalarred:top:hex}). The corresponding $S$-matrix
is
\begin{equation}
S=\begin{pmatrix}
0&0&s_{23}&s_{234}&s_{61}&0\\
0&0&0&s_{34}&s_{345}&s_{12}\\
s_{23}&0&0&0&s_{45}&s_{123}\\
s_{234}&s_{34}&0&0&0&s_{56}\\
s_{61}&s_{345}&s_{45}&0&0&0\\
0&s_{12}&s_{123}&s_{56}&0&0
\end{pmatrix}
\end{equation}
The interest is only to the
subdeterminants of sizes~4 and~5. There are only three types of pinches
from a hexagon down to boxes, which are shown
in~\figref{fig:qcd-scalarred:boxes}: the one-mass box, the adjacent box and
the opposite box.
The one-mass box and the adjacent box lead to $S$-matrices of which the
determinant is simply a product of \person{Mandelstam} variables. However, the
opposite boxes is accompanied by the determinant
\begin{equation}
\det{}S^{\{3,6\}}=\left(\left\vert\begin{array}{cc}
s_{234}&s_{61}\\
s_{34}&s_{345}
\end{array}\right\vert\right)^2\equiv D_3^2
\end{equation}
and cyclic permutations of the indices of the \person{Mandelstam} variables
respectively. The determinants~$D_1$, $D_2$ and $D_3$ are semi-definite
and vanish only at the phase space boundary. This, again, can be shown by
introducing an appropriate parametrisation of the kinematics. To do the
prove for~$E_3$ one can choose the reference frame where $k_2$ and $(-k_5)$
are back to back\footnote{Consider that~$(-k_5)$ is the physical, outgoing
momentum, whereas $k_5$ itself is defined as the ingoing vector.}:
\begin{subequations}
\begin{align}
\label{subeq:qcd-scalarred:k61}
k_{61}&=k_6+k_1=-k_2-k_{34}-k_5\text{,}\\
k_2&=\sqrt{s}/2(1,0,0,1)\text{,}\\
k_{34}&=k_3+k_4=(-E_{45},0,p\sin\varphi,p\cos\varphi)\quad\text{and}\\
k_5&=E_5(-1,0,0,1)\text{.}
\end{align}
\end{subequations}
With the additional definition of the transverse momentum $p_T=p\sin\varphi$
one can write~$D_3$ as
\begin{equation}\label{eq:qcd-scalarred:D3}
D_3=-2\sqrt{s}E_5p_T^2\leq0\text{,}
\end{equation}
which shows the proposed conjecture. Similar expressions one can derive for
$D_1$ and $D_2$, and in fact for all corresponding determinants for
non-trivial permutations of the external legs, which also shows the
semi-definiteness of these expressions. In particular, one finds~$D_1\geq0$
and~$D_2\leq0$ for any physical $2\rightarrow4$~kinematics.
The same arguments apply to determinants with one pinched propagator because
one obtains a very similar structure as one can see from
\begin{equation}
\det{S}^{\{3\}}=s_{23}s_{34}s_{45}D_3\geq0\text{.}
\end{equation}
%In four-dimensional space-time~$\det{G^{\fmslash{a}}}$
%vanishes for all~$N\geq6$. Therefore only $S$-matrices have to be considered
%that belong to pentagon functions and below.

The case of $D_3\rightarrow0$ can be understood as a \person{Landau}
singularity and is discussed for the six-photon amplitude
in~\cite{Bernicot:2007hs}. Equation~\eqref{eq:qcd-scalarred:D3}
shows that the only non-trivial limit for $D_3\rightarrow0$ is
caused by the collinear situation~$p_T\rightarrow0$, in which
case one can write $k_{34}=-xk_5-yk_2$ for $x$ and $y$ being
uniquely defined by $E_{45}$ and~$p$.
Equation~\eqref{subeq:qcd-scalarred:k61} implies
$k_{61}=-(1-x)k_5-(1-y)k_2$. In other words, the propagators
of the loop carry momenta
\begin{subequations}
\begin{align}
q_1&=-(1-y)k_2\text{,}\\
q_2&=yk_2\text{,}\\
q_4&=-xk_5\quad\text{and}\\
q_5&=(1-x)k_5\text{.}
\end{align}
\end{subequations}
This kinematical situation corresponds to a \emph{double parton scattering}%
\index{double parton scattering} situation,
where each of the partons~2 and~5 split into pairs of partons and
each of the internal particles is collinear with either $k_2$
or~$k_5$, $q_1\sim q_2\sim k_2$ and $q_4\sim q_5\sim k_5$.
Therefore all four internal propagators are on-shell and
obey the \person{Landau} equations~\cite{Landau:1959fi},
\begin{equation}
q_1^2=q_2^2=q_4^2=q_5^2=0\text{.}
\end{equation}

\end{fmffile}

%% file: qcd-tensorred.tex
\subsection{Form Factor Representation for Tensor Integrals}
Before discussing the reduction of tensor integrals I introduce a
form factor representation of the tensor integrals
according to~\cite{Binoth:2005ff}. Equation~\eqref{eq:qcd-dimreg:TItoSI}
already suggests to write tensor integrals as a tensor product of a
structure carrying the \person{Lorentz} structure with a 
\person{Lorentz} invariant form factor,
\begin{equation}\label{eq:qcd-tensorred:formfactorrep}
\begin{split}I_N^{n;\mu_1\ldots\mu_r}&(a_1,\ldots,a_r;S)=\\
&\phantom{+}\sum_{j_1,\ldots,j_r\in S_{\#}}%
\left[\,\Delta_{j_1\bul}^{\bul}\cdots\Delta_{j_r\bul}^{\bul}%
\,\right]^{\mu_1\ldots\mu_r}_{a_1\ldots a_r}A_{j_1\ldots j_r}^{N,r}(S)\\
&+\sum_{j_1,\ldots,j_{r-2}\in S_{\#}}%
\left[\,g^{\bul\bul}\Delta_{j_1\bul}^{\bul}\cdots%
\Delta_{j_{r-2}\bul}^{\bul}\,\right]^{\mu_1\ldots\mu_r}_{a_1\ldots a_r}%
B_{j_1\ldots j_{r-2}}^{N,r}(S)\\
&+\sum_{j_1,\ldots,j_{r-4}\in S_{\#}}
\left[\,g^{\bul\bul}g^{\bul\bul}%
\Delta_{j_1\bul}^{\bul}\cdots\Delta_{j_{r-4}\bul}^{\bul}%
\,\right]^{\mu_1\ldots\mu_r}_{a_1\ldots a_r}%
C_{j_1\ldots j_{r-4}}^{N,r}(S)\text{.}
\end{split}
\end{equation}
The square brackets are interpreted as follows:
\begin{subequations}
\begin{equation}
\left[\,\Delta_{j_1\bul}^{\bul}\cdots\Delta_{j_r\bul}^{\bul}\,%
\right]^{\mu_1\ldots\mu_r}_{a_1\ldots a_r}=
\Delta_{j_1a_1}^{\mu_1}\cdots \Delta_{j_ra_r}^{\mu_r}\text{,}
\end{equation}
\begin{equation}
[\,\underbrace{g^{\bul\bul}\cdots g^{\bul\bul}}_{l}%
\Delta_{j_1\bul}^{\bul}\cdots\Delta_{j_{r-2l}\bul}^{\bul}%
\,]^{\mu_1\ldots\mu_r}_{a_1\ldots a_r}=
\sum_{\rlap{\def\arraystretch{.7}
\begin{array}[b]{r@{}l}
\scriptstyle A\uplus B=&\scriptstyle\{1,\ldots,r\}\\
\scriptstyle \vert A\vert=&\scriptstyle 2l
\end{array}}}%
\left[\,g^{\bul\bul}\cdots g^{\bul\bul}%
\,\right]^{(\mu_i)_{i\in A}}\cdot
\prod_{k\in B}\Delta_{j_ka_k}^{\mu_k}
\end{equation}
\end{subequations}
and $[g^{\bul\bul}\cdots g^{\bul\bul}]^{\mu_1\ldots\mu_{2l}}$
the sum of all distinguishable distributions of the indices,
as explained earlier; the notation $A\uplus B$ denotes the
union of two sets $A$ and $B$ where $A\cap B=\emptyset$.
No more than two metric tensors arise in
the tensors on the right hand side of the form factor representation
for calculations in the \person{Feynman} gauge where the rank of an
integral is never greater than~$N$; this is a result
of~\eqref{eq:qcd-tensorred:Ngt6reduction} as shown in the following
section.

\subsection{Tensor Reduction by Subtraction}
The same logic of the section about the reduction of scalar integrals
can be applied to tensor integrals~\cite{Binoth:2005ff}.
Starting from definition~\eqref{eq:qcd-dimreg:defTI} one can split the
tensor integral into\footnote{The derivation is written down for~$\alpha=0$
since it is more convenient to write. No changes have to be made for~$\alpha\neq0$.}
\begin{multline}
I_N^{d;\mu_1\ldots\mu_r}(a_1,\ldots a_r; S)=%nl
\int\!\!\frac{\diff[d]k}{i\pi^{n/2}}\frac{\left[q_{a_1}^{\mu_1}+\sum_{j\in S_\#}C^{\mu_1}_{ja_1}(q_j^2-m_j^2)\right]%nl
q_{a_2}^{\mu_2}\cdots q_{a_r}^{\mu_r}}{\prod_{j=1}^N(q_j^2-m_j^2+i\delta)}\\
-\sum_{j\in S_\#}C^{\mu_1}_{ja_1}\int\!\!\frac{\diff[d]k}{i\pi^{n/2}}\frac{(q_j^2-m_j^2)q_{a_2}^{\mu_2}\cdots q_{a_r}^{\mu_r}}{%nl
\prod_{j=1}^N(q_j^2-m_j^2+i\delta)}
\text{.}
\end{multline}
This corresponds to the splitting in the scalar case, and one can write
\begin{align}
I_N^{d;\mu_1\ldots\mu_r}(a_1,\ldots a_r; S)&=I_\text{div}+I_\text{fin}\text{, with}\\
I_\text{div}&=-\sum_{j\in S_\#}C^{\mu_1}_{ja_1}I_{N-1}^{n;\mu_2\ldots\mu_r}(a_2,\ldots,a_r;S^{\{j\}})\quad\text{and}\\
I_\text{fin}&=\int\!\!\frac{\diff[d]k}{i\pi^{n/2}}\frac{A_{a_1}^{\mu_1}q_{a_2}^{\mu_2}\cdots q_{a_r}^{\mu_r}}{%nl
\prod_{j=1}^N(q_j^2-m_j^2+i\delta)}\text{.}
\end{align}
I introduced the vector
\begin{equation}
A_{a_1}^{\mu_1}\equiv q_{a_1}^{\mu_1}+\sum_{j\in S_\#}C^{\mu_1}_{ja_1}(q_j^2-m_j^2)
\end{equation}
which must be brought in a form that ensures that~$I_\text{fin}$ is infrared safe. Therefore one proceeds exactly as in the scalar case
by introducing \person{Feynman} parameters and shifting the integration momentum~$k\rightarrow k-\sum_{i\in S_\#}z_ir_i$. From
\begin{multline}
A_{b}^{\mu}=k^\mu+\left(\sum_{j\in S_\#}C_{jb}^\mu\right)\left(k^2-\frac12z^\transposed Sz\right)\\
+\sum_{k\in S_\#}z_k\left[\sum_{j\in S_\#}C_{jb}^\mu(S_{jk}+2k\cdot\Delta_{jk})-\Delta_{kb}^\mu\right]
\end{multline}
one obtains a infrared safe integral if
\begin{equation}\label{eq:qcd-tensorred:LES}
\sum_{j\in S_\#}S_{kj}C_{jb}^\mu=\Delta_{kb}^\mu
\end{equation}
in analogy to~\eqref{eq:qcd-scalarred:bj-condition}; this numerator makes the integral infrared safe because
all terms are either proportional to~$k$ or to~$(k^2-1/2z^\transposed Sz)$, the first yielding an additional~$\lambda$
when referring to Section~\ref{ssec:qcd-scalarred:IRdivergences}, the second leading to a higher dimensional integral.

It remains to show the solubility of~\eqref{eq:qcd-tensorred:LES}. For invertible~$S$ the solution is
\begin{equation}
C_{ib}^\mu=\sum_{j\in S_\#}(S^{-1})_{ij}\Delta_{jb}\text{.}
\end{equation}
For singular~$S$ the result can be given in terms of the pseudoinverse, if and only if the consistency condition
\begin{equation} \label{eq:qcd-tensorred:consistencyS}
(\One-S\tilde{S})\cdot C_b^\mu=0
\end{equation}
holds. Again the reverse way is chosen and the solution is constructed explicitly to prove~\eqref{eq:qcd-tensorred:consistencyS}.

The matrix~$S$ is split into its components~$-G+\eta v^\transposed+v\eta^\transposed$, introducing a natural
splitting on~\eqref{eq:qcd-tensorred:LES}, where the abbreviation
\begin{equation}
\mathcal{V}_b^\mu\equiv\sum_{j\in S_\#}C_{jb}^\mu
\end{equation}
becomes helpful:
\begin{subequations}
\begin{align}
\sum_{j\in S_\#}G^{(a)}_{ij}C_{jb}^\mu=\delta v_i\mathcal{V}_b^\mu-\Delta_{ib}^\mu\text{,}
\label{eq:qcd-tensorred:LES01}\\
\sum_{j\in S_\#}\delta v_jC_{jb}^\mu=\Delta_{ab}^\mu-2v_a\mathcal{V}_b^\mu.
\end{align}
\end{subequations}

The system~\eqref{eq:qcd-tensorred:LES01} admits solutions if and only if
\begin{equation}
\sum_{j\in S_\#}K_{G,ij}(\Delta_{jb}^\mu-\delta v_j\mathcal{V}_b^\mu)=0\text{.}
\end{equation}
we have seen earlier that for general~$\delta v$, $K_G\delta v\neq v$.
On the other hand using~\eqref{eq:qcd-scalarred:DeltaAndR} for~$\Delta_{jb}^\mu$ and
$K_G=\One-R^\transposed(RR^\transposed)^{-1}R$ proves that
\begin{equation}
\sum_{j\in S_\#}K_{G,ij}\Delta_{jb}^\mu=0\text{,}
\end{equation}
and hence for~$\mathcal{V}_b^\mu=0$ solutions are constructable. Up to the choice of
the parametrisation of~$\Kern{G}$ the solution is defined by theorem~\ref{thrm:penrose},
\begin{subequations}
\begin{align}\label{eq:qcd-tensorred:solution01}
C_{ib}^\mu&=-\sum_{j\in S_\#}\tilde{G}_{ij}\Delta_{jb}^\mu+W^\mu_i,\quad i\in S^{\{a\}}_\#\text{,}\\
C_{ab}^\mu&=-\sum_{j\in S^{\{a\}}_\#}C_{jb}^\mu,\quad\text{where}\\
W_i^\mu&=\frac{(K_G\delta v)_i}{\delta v^\transposed K_G\delta v}\left(\Delta_{ab}^\mu+
\sum_{j,k\in S^{\{a\}}_\#}\delta v_j\tilde{G}_{jk}\Delta_{kb}^\mu\right)+\sum_{j=1}^{N-\rank{G}-2}\beta_j^\mu u_i^{(j)}
\end{align}
\end{subequations}
As already in Section~\ref{sec:scalarred}, $u^{(j)}$
form a basis of~$\Kern{G}\cap\langle\delta{v},\delta_a\rangle^\perp$.

This solution is now regarded only for~$N\ge6$; in this case one always ha~$\mathcal{V}_b~\mu=0$.
Equation~\eqref{eq:qcd-tensorred:solution01} can now be plugged in back into the expression for~$A_b^\mu$,
\begin{equation}
A_b^\mu=(\hat{g}^{\mu\nu}+2\sum_{j,k\in S_\#}z_kC^\mu_{jb}\Delta^\nu_{jk})\hat{k}_\nu\text{.}
\end{equation}
Using~\eqref{eq:qcd-scalarred:DeltaAndR} and the fact~$W^\mu\in\Kern{G}$ --- which implies
\begin{displaymath}
\sum_{j\in S_\#}\Delta_{jb}^\nu W^\mu_j=\sum_{j\in S_\#}\Delta_{jb}^\nu (K_GW^\mu)_j=
\sum_{j,k\in S_\#}(K_{G,jk}\Delta_{kb}^\nu) W^\mu_j=0\quad\text{---}
\end{displaymath}
it follows that
\begin{equation}
\sum_{j\in S_\#}C^\mu_{jb}\Delta^\nu_{jk}=-\sum_{ij\in S_\#}\Delta^\mu_{ib}\tilde{G}_{ij}\Delta^\nu_{jk}=
-\frac12\sum_{m=1}^{\rank{G}}e_m^\mu e_m^\nu\text{.}
\end{equation}
Since for non-exceptional kinematics\footnote{Still the constraint~$N\ge6$ is assumed.} we have~$\rank{G}=4$ and
therefore the right hand side of the above equation fulfils the completeness relation,
\begin{equation}\label{eq:qcd-tensorred:completeness}
\sum_{j\in S_\#}C^\mu_{jb}\Delta^\nu_{jk}=-\frac12g^{\mu\nu}\text{.}
\end{equation}
and hence~$A_b^\mu$ vanishes, which implies that for~$N\ge6$ only the pinched integrals survive. In phenomenological
applications the remnant of~$A_b^\mu$ can only be contracted
with~$\Delta_{ij}^\mu=\Delta_{ia}^\mu+\Delta_{ja}^\mu$, and
one is left with
\begin{equation}
\Delta_{ia}^\mu A_{b,\mu}=\hat{k}_\nu\left(\Delta_{ia}^\nu-\sum_{m=1}^{\rank{G}}e_m^\nu(e_m\cdot\Delta_{ia})\right)\text{,}
\end{equation}
which simplifies by the help of
\begin{equation}
\sum_{m=1}^{\rank{G}}e_m^\nu(e_m\cdot\Delta_{ia})=\sum_{m=1}^{\rank{G}}e_m^\nu R^{(a)}_{mi}=\Delta_{ia}^\nu
\end{equation}
to~$\Delta_{ij}^\mu A_{b,\mu}=0$.
Therefore, for phenomenology it is safe to conclude that
\begin{equation}\label{eq:qcd-tensorred:Ngt6reduction}
I_N^{n;\mu_1\ldots\mu_r}(a_1,\ldots,a_r;S)=-\sum_{j\in S_\#}C_{ja_1}^{\mu_1} I_{N-1}^{n;\mu_2\ldots\mu_r}(a_2,\ldots,a_r;S^{\{j\}}),
\quad\text{for $N\geq6$}\text{.}
\end{equation}

%% file: qcd-tensorred-ibp.tex
\subsection{Reduction of Non-Trivial Numerators in Integrals}
\label{ssec:fpnumerators}
The previous approach to tensor reduction leads to an integral basis that
contains integrals with non-trivial polynomials of~$\{z_i\}_{i=1}^N$
in their numerators. For a full reduction to simple scalar integrals I follow
the approach of~\cite{Binoth:1999sp}. This approach is based on the fact
that the \person{Feynman} parameter integrals are of the common form
\index{Feynman@\person{Feynman}!parameter}
\begin{equation}
I_N^d(l_1,\ldots,l_r;S)=%nl
\int_{-\infty}^{\infty}\!\!\diff{z_i}\int_{-\infty}^{\infty}\!\!\diff{z_j}\,%nl
\delta(C-z_i-z_j)g(z_i,z_j)\text{.}
\end{equation}
On the other hand it is clear that
\begin{equation}
\int_{-\infty}^{\infty}\!\!\diff{z_i}\int_{-\infty}^{\infty}\!\!\diff{z_j}\frac{\partial}{\partial z_i}\left[\delta(C-z_i-z_j)g(z_i,z_j)\right]=0\text{.}
\end{equation}

Now one can integrate out the $\delta$-function through the $z_j$-integration and apply the chain rule to the differentiation
and reintroduce the $\delta$-function after:
\begin{multline}
\label{eq:qcd-tensorred-ibp:ibp-eq}
\int_{-\infty}^{\infty}\!\!\diff{z_i}\int_{-\infty}^{\infty}\!\!\diff{z_j}\frac{\partial}{\partial z_i}\left[\delta(C-z_i-z_j)g(z_i,z_j)\right]=\\
\int_{-\infty}^{\infty}\!\!\diff{z_i}\frac{\partial}{\partial z_i}g(z_i,C-z_i)=
\int_{-\infty}^{\infty}\!\!\diff{z_i}\left.\left(\frac{\partial g(z_i,z_j)}{\partial z_i}-\frac{\partial g(z_i,z_j)}{\partial z_j}\right)\right\vert_{z_j=C-z_i}
%\\=\int_{-\infty}^{\infty}\!\!\diff{z_i}\int_{-\infty}^{\infty}\!\!\diff{z_j}\,
%\delta(C-z_i-z_j)%nl
%  \left(\frac{\partial g(z_i,z_j)}{\partial z_i}
%-\frac{\partial g(z_i,z_j)}{\partial z_j}\right)
\text{.}
\end{multline}
%Hence, for any pair of indices~$i,j$ one can write
%\begin{multline} \label{eq:qcd-tensorred-ibp:ibp-eq}
%\int_{-\infty}^{\infty}\!\!\diff{z_i}\int_{-\infty}^{\infty}\!\!\diff{z_j}\,
%\delta(C-z_i-z_j)%nl
%  \frac{\partial g(z_i,z_j)}{\partial z_i}\\
%=\int_{-\infty}^{\infty}\!\!\diff{z_i}\int_{-\infty}^{\infty}\!\!\diff{z_j}\,
%\delta(C-z_i-z_j)%nl
%  \frac{\partial g(z_i,z_j)}{\partial z_j}\text{.}
%\end{multline}

Up to the integration over all $z_l$ with $l\neq i, l\neq j$ and the respective $\Theta$-functions 
for~$I_N^d(l_1,\ldots,l_r;S)$ the actual form of~$g(z_i, z_j)$ is
\begin{equation}
g(z_i, z_j)=c_{ij}\,(-1)^N\Gamma(N-d/2)\Theta(z_i)\Theta(z_j)%
\left(\prod_{k=1}^rz_{l_k}\right)\left[-\frac12z^\transposed Sz\right]^{-(N-d/2)}
\end{equation}

Based on~\eqref{eq:qcd-scalarred:splitS} one can write
\begin{equation}
z^\transposed Sz=-z^\transposed G^{(a)}z+2\,(z\cd\delta v^{(a)})\,(z\cd \eta)
+2\,v^{(a)}_a\,(z\cd \eta)^2
\end{equation}
where~$z\cd\eta=1$ under the integral due to the $\delta$-function. 
The required differentiation for the denominator then is
\begin{equation}
\frac{\partial}{\partial z_j}\left[-\frac12z^\transposed Sz\right]=\left(G^{(a)}\cd z-\delta v^{(a)}\right)_j\text{.}
\end{equation}
We will also use that pinched integrals can always be rewritten as
integrals with an additional $\delta$-function,
\begin{multline}
I_{N-1}^d(l_1,\ldots,l_r;S^{\{a\}})=\\(-1)^{N-1}\Gamma(N-1-d/2)%
\int_0\!\!\diff[N]z\delta_z\frac{\delta(z_a)\prod_{k=1}^r z_{l_k}}{%nl
\left[-\frac12z^\transposed Sz-i\delta\right]^{N-1-d/2}}\text{.}
\end{multline}

Now one index can be chosen to be~$a$, such that~$G^{(a)}_{ak}$ and~$\delta v^{(a)}_a$ vanish;
by the help of~\eqref{eq:qcd-tensorred-ibp:ibp-eq} one can relate integrals with a different number
of \person{Feynman} parameters in the numerator,
\begin{multline}
-I_{N-1}^{d-2}(l_1,\ldots,l_r;S^{\{i\}})+\sum_{k=1}^r\delta_{il_k}I_{N}^{d}(l_1,\ldots,l_{k-1},l_{k+1},l_r;S)\\
-\sum_{l_0\in S_\#}G^{(a)}_{il_0}I_N^{d-2}(l_0,\ldots,l_r;S)+\delta v_i^{(a)}I_N^{d-2}(l_1,\ldots,l_r;S)\\
=-I_{N-1}^{d-2}(l_1,\ldots,l_r;S^{\{a\}})+\sum_{k=1}^r\delta_{al_k}I_{N}^{d}(l_1,\ldots,l_{k-1},l_{k+1},l_r;S)\text{.}
\end{multline}

%%%%%%%%%%%%%%%%%%%%%%%%%%%%%%%%%%%%%%%%%
$S$ can be introduced back again while shifting 
the dimension~$d\rightarrow d+2$. The term containing $\delta_{al_0}$ since
the $z_a$-integration is used to eliminate the $\delta$-function and therefore
no~$z_a$ appears in the numerator\footnote{However, the second term
on the right hand side survives, since the corresponding
$\theta$-function appears either on the left or on the right hand side of
the expression.}
\begin{multline}\label{eq:qcd-tensorred-ibp:intermediate01}
\sum_{l_0\in S_\#}S_{il_0}I_N^{d}(l_0,\ldots,l_r;S)
+\sum_{k=1}^r\delta_{il_k}I_{N}^{d+2}(l_1,\ldots,l_{k-1},l_{k+1},\ldots,l_r;S)\\
-I_{N-1}^{d}(l_1,\ldots,l_r;S^{\{i\}})
-2v^{(a)}_aI_N^d(l_1,\ldots,l_r;S)\\
-\sum_{l_0\in S_{\#}}\delta v^{(a)}_{l_0}I_N^d(l_0,\ldots,l_r;S)=
-I_{N-1}^{d}(l_1,\ldots,l_r;S^{\{a\}})
\text{.}
\end{multline}
A crucial simplification can be achieved 
by eliminating~$I_{N-1}^{d}(l_1,\ldots,l_r;S^{\{a\}})$ through an
auxiliary relation:
\begin{multline}
I_{N-1}^d(l_1,\ldots,l_r;S^{\{a\}})=(N-d-r-1)I^{d+2}_N(l_1,\ldots,l_r;S)\\
+\sum_{l_0\in S_{\#}}\delta v^{(a)}_{l_0}I_N^d(l_0,\ldots,l_r;S)
+2v^{(a)}_aI_N^d(l_1,\ldots,l_r;S)
\end{multline}

The prove is done via induction over~$r$. For~$p=0$ one finds, applying
Formul\ae~\eqref{qcd-scalarred:d+2:1} and~\eqref{qcd-scalarred:d+2:2},
\begin{multline}
I_{N-1}^d(S^{\{a\}})=\int\frac{\diff^dk}{i\pi^{d/2}}\frac{%
q_a^2-m_a^2}{\prod_{j\in S_{\#}}(q_j^2-m_j^2+i\delta)}=\\
(N-d-1)I_N^{d+2}(S)+\sum_{l\in S_{\#}}\delta v^{(a)}_lI_N^d(l;S)
+2v^{(a)}_aI^d_N(S)\text{.}
\end{multline}
This result is achieved by manipulating the numerator after introducing
\person{Feynman} parameter and completing the square in the denominator,
\begin{multline}
\left(k-\sum_{j\in S_{\#}}z_j\Delta_{ja}\right)^2-m_a^2=
k^2-2\sum_{j\in S_{\#}}z_jk\cdot\Delta_{ja}+
\sum_{j_1,j_2\in S_{\#}}z_iz_j\Delta_{j_1a}\Delta_{j_2a}-m_a^2\text{.}
\end{multline}
The linear term vanishes under symmetric integration. Furthermore
relations~\eqref{eq:qcd-scalarred:SComponents}, \eqref{qcd-scalarred:d+2:1}
and~\eqref{qcd-scalarred:d+2:2}
are used to achieve the above result.

%%%%% Induction step
The induction step can be carried out,
regarding~$\delta v^{(a)}_j$ as independent variables,
by partial differentiation
with respect to~$\delta v_j^{(a)}$, because
\begin{equation}
\frac{\partial}{\partial\delta v_j^{(a)}}\left[-\frac12z^\transposed Sz\right]=-z_j
\end{equation}
allows the introduction of additional \person{Feynman} parameters in the
numerator,
\begin{equation}
\frac{\partial}{\partial\delta v_{l_{p+1}}^{(a)}}I_N^d(l_1,\ldots,l_r;S)=
I_N^{d-2}(l_1,\ldots,l_{p+1};S)\text{.}
\end{equation}

This result, substituted into
equation~\eqref{eq:qcd-tensorred-ibp:intermediate01} reads
\begin{multline}\label{eq:qcd-tensorred-ibp:intermediate02}
\sum_{l_0\in S_\#}S_{il_0}I_N^{d}(l_0,\ldots,l_r;S)=
+I_{N-1}^{d}(l_1,\ldots,l_r;S^{\{i\}})\\
-\sum_{k=1}^r\delta_{il_k}I_{N}^{d+2}(l_1,\ldots,l_{k-1},l_{k+1},\ldots,l_r;S)
-(N-d-r-1)I_N^{d+2}(l_1,\ldots,l_r;S)
\text{.}
\end{multline}
For $N\leq6$ one can invert this equation, obtaining the desired result,
\begin{multline}\label{eq:qcd-tensorred-ibp:result}
I_N^{d}(l_0,\ldots,l_r;S)=
-\sum_{k=1}^rS_{l_0l_k}^{-1}%
I_{N}^{d+2}(l_1,\ldots,l_{k-1},l_{k+1},\ldots,l_r;S)\\
+\sum_{i\in S_{\#}}S_{l_0i}^{-1}I_{N-1}^{d}(l_1,\ldots,l_r;S^{\{i\}})
-b_{l_0}(N-d-r-1)I_N^{d+2}(l_1,\ldots,l_r;S)\text{.}
\end{multline}

For the case~$N=1$ the result can be obtained by explicit 
calculation\footnote{The result is given only up %
 to~${\mathcal{O}}(\varepsilon)$}:
\begin{multline}
I_1^{n+2\alpha}(m^2)=
I_1^{n+2\alpha}(l;m^2)=\\
\frac{(-1)^\alpha(m^2)^{\alpha+1}}{2^\alpha(\alpha+1)!}
\left[\Delta-\ln\left(\frac{m^2-i\delta}{\mu^2}\right)
+1+\sum_{\nu=1}^\alpha\left(\frac{1}{1+\nu}\right)\right]=\\
\frac{(-1)^\alpha(m^2)^{\alpha+1}}{2^\alpha(\alpha+1)!}\left[
I_2^n(0;0,m^2)+\sum_{\nu=1}^\alpha\left(\frac{1}{1+\nu}\right)
\right]
\end{multline}
This result is valid for integer values $\alpha\geq0$.

\subsection{Tensor Reduction through Integration by Parts}
\label{ssec:qcd-tensorred-ibp:TensorRed}
Together with~\eqref{eq:qcd-dimreg:TItoSI} and~\eqref{eq:qcd-scalarred:final},
the recurrence relation~\eqref{eq:qcd-tensorred-ibp:result} provides
a self-contained scheme for tensor reduction. By recursively
applying~\eqref{eq:qcd-tensorred-ibp:result} the highest dimension
that appears in an integral is increased by two in each reduction step.
These integrals
then can be brought back to the standard basis by reverse application 
of~\eqref{eq:qcd-scalarred:final}. The only exception is the
integral~$I_5^{n+2}$ where the inversion of~\eqref{eq:qcd-scalarred:final}
would lead to terms~$\propto\varepsilon^{-1}$; however, terms containing
that integral have been found to cancel in all practical cases. It should be
noticed that the reduction involving the 
inverse of~\eqref{eq:qcd-scalarred:final} is the only source for
$1/B$~terms, i.e. for inverse \person{Gram} determinants.

%%%%%%%%%%%%%%%%%%%%%%%%%%%%%%%%%%%%%%%%%%%%%%%%%%%%%%%%%%%%%%%%%%%%%%%
\subsection{Reduction of Three-Point Tensor Integrals}
\label{ssec:qcd-tensorred-ibp:threepoint}
In certain kinematical cases the $S$-matrix of the three-point
functions can become singular while $\det{G^{\fmslash{a}}}$ does not.
While it was shown earlier in section~\ref{ssec:scalarred:smatrixzero}
that these cases for~$N\geq4$ in the massless limit arise only on
the infrared phase-space boundaries, for~$N=3$ the situation is worse due
to a number of pinched matrices~$S^{\{i,j,k\}}$ stemming from
six-particle cases, where~$\det{S^{\{i,j,k\}}}$ vanishes identically.

It turns out, however, that the \person{Gram} matrix is the better
choice for a tensor reduction in the three-particle case.
The \person{Gram} determinant in three-point kinematics is the
\person{K\"allen} function
\begin{equation}
\det{G^{\fmslash{a}}}=-\lambda(s_1,s_2,s_3)=
-\left(s_1^2+s_2^2+s_3^2-2s_1s_2-2s_1s_3-2s_2s_3\right)
\end{equation}
Solving the equation $\det{G^{\fmslash{a}}}=0$ only allows for the
solutions
\begin{subequations}
\begin{align}
\label{eq:qcd-tensorred-ibp:lambdaIsZero-1}
s_1^{\pm}&=(\sqrt{s_2}\pm\sqrt{s_3})^2,\quad s_2,s_3\geq0\text{,}\\
s_1^{\pm}&=-(\sqrt{\vert s_2\vert}\pm\sqrt{\vert s_3\vert})^2,\quad s_2,s_3\leq0\text{.}
\end{align}
\end{subequations}
The second solution cannot appear at one-loop for any physical~$2\rightarrow N$
kinematics. For the solutions~\eqref{eq:qcd-tensorred-ibp:lambdaIsZero-1} also~$s_1$
must be non-negative and hence one can derive the kinematical constraint
\begin{equation}
s_1\geq(\sqrt{s_2}+\sqrt{s_3})^2=s_1^{+}\text{.}
\end{equation}
This means that singularities in~$1/\det{G^{\fmslash{a}}}$
lie only on the border
of the phase space and correspond to physical thresholds.

The three-point functions therefore cannot be treated by the
standard approach as introduced in~\ref{ssec:fpnumerators}.
Explicit formul\ae{} for the required integrals can be obtained
from~\eqref{eq:qcd-dimreg:relTISI01} by applying a
\person{Passarino}-\person{Veltman} like tensor-reduction.
By multiplying the equation by products of $2\Delta^{\mu_i}_{la_i}$
one generates products of~$G^{(a)}_{ij}$ on the
right hand side, while the numerators in the tensor integrals can be completed
to propagators which partly cancel the denominators. One finds
\begin{align}
2\Delta_{ia}\cdot q_a&=S_{aa}-S_{ia}+[q_i^2-m_i^2]-[q_a^2-m_a^2]
\quad\text{and}\\
2q_a\cdot q_b&=-S_{ab}+[q_a^2-m_a^2]+[q_b^2-m_b^2]\text{.}
\end{align}
Since~$G^{\fmslash{a}}$ is invertible, the integrals $I_3^d(l_0,\ldots,l_r;S)$
can be extracted by solving a linear system of equations; however the
inversion is only possible for $l_j\neq a$. No formula can be derived
directly for the integral $I_3^d(l_1,l_2,l_3;S)$ for $l_1,l_2,l_3$
being mutually different since $a$ cannot be chosen different from each
of the~$l_i$ simultaneously. This problem can be circumvented applying
the fact that the \person{Feynman} parameters have to sum up to one and
therefore the problematic integral can be replaced via
\begin{equation}
\sum_{l_r\in S_{\#}}I_N^d(l_1,\ldots,l_{r-1},l_r;S)=
   I_3^d(l_1,\ldots,l_{r-1};S)
\end{equation}
The reduction formul\ae{} are given below.
It should be noted that the reduction of the higher dimensional scalar
integrals down to four dimensions can be done using the standard approach,
because of
\begin{equation}
I_3^{d+2}(S)=\frac{1}{B(d-2)}\left[I_3^d(S)
-\sum_{j\in S_{\#}}b_jI_2^d(S^{\{j\}})\right]\text{,}
\end{equation}
where only~$1/B$ and~$b_j/B$ appear, which contain no factors
of~$1/\det{S}$.

In~\eqref{eq:qcd-tensorred-ibp:three-point-reductions} the constraint
$a\not\in\{l_1,l_2,l_3\}$ is assumed. The inverse of~$G^{(a)}$ is
understood in terms of the pseudoinverse.
%\vspace{32pt}
%\hrule width 1.0\linewidth height .5pt
\begin{subequations}
\label{eq:qcd-tensorred-ibp:three-point-reductions}
\begin{multline}
I_3^d(l_1;S)=\left[\sum_{i\in S_{\#}}G_{l_1i}^{(a)\,-1}\left(
S_{ia}-S_{aa}\right)\right]I_3^d(S)\\
-\sum_{i\in S_{\#}}G_{l_1i}^{(a)\,-1}\left[
I_2^d\left(S^{\{i\}}\right)-I_2^d\left(S^{\{a\}}\right)
\right]
\end{multline}
\begin{multline}
I_3^d(l_1,l_2;S)=G^{(a)\,-1}_{l_1l_2}I_3^{d+2}(S)
-\left[\sum_{i\in S_{\#}}G_{l_1i}^{(a)\,-1}\left(
S_{aa}-S_{ia}\right)\right]I_3^d(l_2;S)\\
-\sum_{i\in S_{\#}}G_{l_1i}^{(a)\,-1}\left[
I_2^d\left(l_2;S^{\{i\}}\right)-I_2^d\left(l_2;S^{\{a\}}\right)
\right]
\end{multline}
\begin{multline}
I_3^d(l_1,l_2,l_3;S)=\\
G^{(a)\,-1}_{l_1l_2}I_3^{d+2}(l_3;S)
+G^{(a)\,-1}_{l_1l_3}I_3^{d+2}(l_2;S)
+G^{(a)\,-1}_{l_2l_3}I_3^{d+2}(l_1;S)\\
-\left[\sum_{i\in S_{\#}}G_{l_1i}^{(a)\,-1}\left(
S_{ia}-S_{aa}\right)\right]\left(
I_3^d(l_2,l_3;S)-G^{(a)\,-1}_{l_2l_3}I_3^{d+2}(S)\right)\\
-\sum_{i\in S_{\#}}G_{l_1i}^{(a)\,-1}
\left[I_2^d(l_2,l_3;S^{\{i\}})-G^{(a)\,-1}_{l_2l_3}I_2^{d+2}(S^{\{i\}})\right]\\
+\left(\sum_{i\in S_{\#}}G_{l_1i}^{(a)\,-1}\right)
\left[I_2^d(l_2,l_3;S^{\{a\}})-G^{(a)\,-1}_{l_2l_3}I_2^{d+2}(S^{\{a\}})\right]
\end{multline}
\end{subequations}
%\end{minipage}
%
%First I give the formula for the massless case with one on-shell leg,
%where $s_1=m_1^2=m_2^2=m_3^2=0$.
%I introduce the abbreviation~$I_2^d(s)\equiv I_2^d(s;0,0)$ for the massless
%two-point integral.
%
%\begin{subequations}
%\begin{align}
%I_3^d(1;S)&=-\frac{s_3}{s_2-s_3}I_3^n(S)
%+\frac{2s_3I_2^d(s_3)-(s_2+s_3)I_2^d(s_2)}{(s_2-s_3)^2}\\
%I_3^d(3;S)&=\frac{s_2}{s_2-s_3}I_3^n(S)
%+\frac{2s_2I_2^d(s_2)-(s_2+s_3)I_2^d(s_3)}{(s_2-s_3)^2}\\
%I_3^d(2;S)&=-\frac{I^d_2(s_2)-I^d_s(s_3)}{s_2-s_3}
%\end{align}
%\end{subequations}
%
%The integrals for two on-shell legs are given for the
%partly massive case, since the limits~$m_i\rightarrow0$
%are taken trivially.
%
%First I give the results for the case $s_2=s_3=m_1^2=m_2^2=0$.
%\begin{subequations}
%\begin{align}
%I_3^d(1;S)&=\frac{1}{s_1}\left[m_3^2I_3^d(S)-I_2^d(s_1;m_3^2,0)\right]\\
%I_3^d(2;S)&=\frac{1}{s_1}\left[(s_1-m_3^2)I_3^d(S)
%+2I_2^d(s_1;m_3^2,0)-I_2^d(0;m_3^2,0)\right]\\
%I_3^d(3;S)&=\frac{1}{s_1}\left[I_2^d(0;m_3^2,0)-I_2^d(s_1;m_3^2,0)\right]
%\end{align}
%\end{subequations}
%A very similar result yields the kinematics $s_2=s_3=m_2^2=m_3^2=0$.
%\begin{subequations}
%\begin{align}
%I_3^d(1;S)&=\frac{1}{s_1}\left[I_2^d(0;m_1^2,0)-I_2^d(s_1;m_1^2,0)\right]\\
%I_3^d(2;S)&=\frac{1}{s_1}\left[(s_1-m_1^2)I_3^d(S)
%+2I_2^d(s_1;m_1^2,0)-I_2^d(0;m_1^2,0)\right]\\
%I_3^d(3;S)&=\frac{1}{s_1}\left[m_1^2I_3^d(S)-I_2^d(s_1;m_1^2,0)\right]
%\end{align}
%\end{subequations}
%
%The remaining cases can be obtained by cyclical permutation of all the
%indices of $z_j$ and $s_j$.

%% file: qcd-diagramrep.tex
\subsection{Basis Integrals}
In Sections~\ref{sec:scalarred} and~\ref{sec:tensorred} it is shown that
all one-loop integrals can be mapped onto a set of of basis integrals
using a set of reduction relations. One set of functions to represent
an arbitrary amplitude~\cite{Binoth:1999sp} consists of scalar
integrals with \person{Feynman} parameters in the numerators,
\index{Feynman@\person{Feynman}!parameter}
\begin{subequations}\label{eq:qcd-diagramrep:Integralsets}
\begin{multline}\label{eq:qcd-diagramrep:IN}
{\mathcal{I}}_{\mathcal{N}}=\{
I_2^n(S), I_2^n(l_1;S), I_2^n(l_1,l_2;S),\\
I_3^n(S), I_3^n(l_1;S), I_3^n(l_1,l_2;S),I_3^n(l_1,l_2,l_3;S),
I_3^{n+2}(S),\\
I_4^{n+2}(S), I_4^{n+2}(l_1;S), I_4^{n+2}(l_1,l_2;S),I_4^{n+2}(l_1,l_2,l_3;S),
I_4^{n+4}(S),I_4^{n+4}(l_1;S)
\}\text{.}
\end{multline}

In another step these integrals can be reduced further to scalar integrals
with trivial numerators,
\begin{equation}\label{eq:qcd-diagramrep:IS}
{\mathcal{I}}_{\mathcal{S}}=\{
I_2^n(S), I_3^n(S), I_4^{6}(S)
\}\text{.}
\end{equation}
\end{subequations}
However, this reduction step introduces inverse \person{Gram} determinants,
whereas while working with the set~${\mathcal{I}}_{\mathcal{N}}$ only
determinants of~$S$ can arise in the denominator.
Although higher-dimensional pentagons ($I_5^{n+2}$, $I_5^{n+4}$, \dots)
formally appear in the reduction, too, it has been shown
in~\cite{Binoth:2005ff} that the coefficients of these integrals
are always of order $\mathcal{O}(\varepsilon)$ and therefore drop out
in phenomenological calculations.

Hence we can write each diagram as
\begin{equation}
{\mathcal{D}}(S)=
\sum_{I(S^\prime)\in{\mathcal{I}}_{\mathcal{N}}}
\frac{{\mathcal{P}}_I}{{\mathcal{Q}}_I}I(S^\prime)\text{.}
\end{equation}
${\mathcal{P}}_I$ is a polynomial in the \person{Mandelstam} variables
and contractions of the external momenta with the
\person{Levi-Civita} tensor; ${\mathcal{Q}}_I$ is a product of
determinants~$\det{S^{\{\ldots\}}}$, which arise from the reduction
of the integrals. The~$S^\prime$ represents the pinched submatrices
of~$S$. 

Similarly, one could use the second set of integrals,
\begin{equation}
{\mathcal{D}}(S)=
\sum_{I(S^\prime)\in{\mathcal{I}}_{\mathcal{S}}}
\frac{{\mathcal{P}}_I}{{\mathcal{Q}}_I}I(S^\prime)\text{.}
\end{equation}
The only principal difference is, that now the polynomials~${\mathcal{Q}}_I$
can also contain \person{Gram} determinants, which should cancel against
the numerator~${\mathcal{P}}_I$.

\subsection{\person{Mandelstam} Variables}
\label{ssec:qcd-diagramrep:mandelstam}
\begin{figure}[hbtp]
\centering
\includegraphics[scale=0.9]{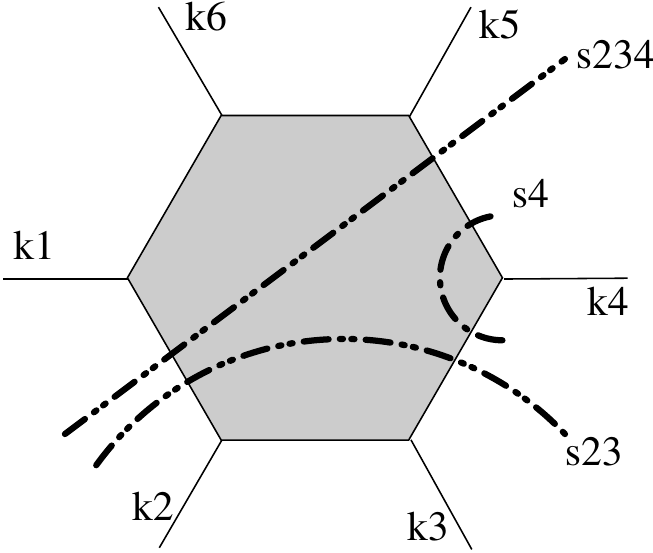}
\caption{Three out of the fifteen possible cuts
of a six-particle amplitude.}
\label{fig:qcd-diagrep:cuts}
\end{figure}

\index{Mandelstam variable@\person{Mandelstam} variable|(}
In this short section I want to present a way of finding an independent
set of \person{Mandelstam} variables for an arbitrary number~$N$ of external
legs. It can be easily implemented as an algorithm that carries out the
substitution of dot products~$k_i\cdot k_j$ by \person{Mandelstam} variables.
The number of dot-products in an $N$-particle scattering problem 
is~$N(N-1)/2$ as one of the momenta can be eliminated by momentum
conservation. This number of independent variables can be achieved by
considering all possible cuts through a generic $N$-particle diagram,
as shown in \figref{fig:qcd-diagrep:cuts}.
Every cut through the diagram corresponds to a partition of the set
of external momenta and hence to a \person{Mandelstam} variable, which
is obtained from the square of the sum of all momenta in one subset of
the partition. The cuts in the diagram of \figref{fig:qcd-diagrep:cuts}
are
\begin{subequations}
\begin{align}
s_4&=(k_4)^2=(k_5+k_6+k_1+k_2+k_3)^2\text{,}\\
s_{23}&=(k_2+k_3)^2=(k_4+k_5+k_6+k_1)^2\quad\text{and}\\
s_{234}&=(k_2+k_3+k_4)^2=(k_5+k_6+k_1)^2\text{.}
\end{align}
\end{subequations}
To get a unique naming scheme I always choose the smaller of both
subsets and, if both are of equal size, the one starting, in a cyclical sense,
with the smaller index. This means for the case~$N=6$
that I always use $s_{23}$ instead of~$s_{4561}$ and $s_{234}$ instead
of~$s_{561}$. Indices are always understood modulo $N$ in generic expressions
like~$s_{i,i+1}$.

To show that all dot-products can be mapped onto this set of
\person{Mandelstam} variables, 
I calculate~$k_i\cdot k_{i+d}$ from $s_{i,i+1,\ldots,i+d}$,
\begin{multline}
s_{i,\ldots,i+d}=(k_i+\ldots+k_{i+d})^2=
k_i^2+k_{i+d}^2+
2 k_i\cdot k_{i+d}+\\
(k_{i+1}+\ldots+k_{i+d-1})^2 + 
2(k_i+k_{i+d})\cdot(k_{i+1}+\ldots+k_{i+d-1})=\\
2k_i\cdot k_{i+d}+s_{i,\ldots,i+d-1}+s_{i+1,\ldots,i+d}
-s_{i+1,\ldots,i+d-1}\text{,}
\end{multline}
and hence one finds
\begin{equation}\label{eq:qcd-diagramrep:dot-to-mandelstam}
2k_i\cdot k_{i+d}=
s_{i,\ldots,i+d}+s_{i+1,\ldots,i+d-1}
-s_{i,\ldots,i+d-1}-s_{i+1,\ldots,i+d}
\text{.}
\end{equation}
\index{Mandelstam variable@\person{Mandelstam} variable|)}

The other kind of \person{Lorentz}-invariant contractions of the
external vectors are the contractions with the
\person{Levi-Civita} symbol. These, however, always appear linear,
since products of two $\epsilon$-tensors can be expressed in a determinant
of \person{Kronecker} $\delta$-symbols, and therefore in contraction with
external momenta one ends up in a polynomial of \person{Mandelstam} variables again.
The number of distinctive contractions is $(N-1)!/[4!(N-1-4)!]$ due to the
antisymmetric character of the $\epsilon$-tensor.

%% file: qcd-renormalisation.tex
\subsection{Introduction}
After having introduced the tools to evaluate integrals and 
\person{Dirac} traces in a dimensionally regularised theory,
we are in the position to calculate the counterterms
which are necessary in order to obtain finite answers from
the calculation of any observables in \ac{qcd}.

Equation~\eqref{eq:qcd-gauge:lagrange} gives the \Lagrangeian{}
density of~\ac{qcd}. In order to renormalise the theory one replaces
the quantities in the original \Lagrangeian{} by their bare
counterparts which are related to each other by a multiplicative
factor
\begin{equation}
\Ld_0=
\Ld_\text{QCD}(
g\rightarrow g_0,
q_a\rightarrow q_{a,0},
{\mathcal A}\rightarrow{\mathcal A_0},\ldots) =
\Ld_\text{QCD}(Z_1g,\sqrt{Z_2}q_a,\sqrt{Z_3}{\mathcal A},\ldots)\text{.}
\end{equation}
As described in Section~\ref{sec:qcd-running} of Chapter~\ref{chp:qcd},
in dimensional regularisation one introduces an arbitrary energy
scale~$\mu$ in order to continue the dimension of the \Lagrangeian{} density
consistently to $n\neq4$ dimensions. This leads to a redefinition
of the coupling constant $g\rightarrow\mu^\varepsilon g$, which is not
discussed in the following. For a more detailed discussion
the reader is referred to the conventions
in Appendix~\ref{app:appendix-integrals} and, for example, to~\cite{Muta:qcd}.

The renormalisation constants $Z_1$, $Z_2$, \dots can be expanded in
a power series in $\alpha_s$,
\begin{equation}
Z_i=1+\delta_i+{\mathcal O}(\alpha_s^2)\text{,}
\end{equation}
where one assumes that $\delta_i={\mathcal O}(\alpha_s)$.
Usually these terms in the \Lagrangeian{} are split into a
renormalised \Lagrangeian{}~$\Ld_\text{ren}$ and a
counterterm \Lagrangeian{}~$\Ld_\text{ct}$ that contains
all terms proportional to~$\delta_i$,
\begin{equation}
\Ld_\text{QCD}=\Ld_\text{ren} + \Ld_\text{ct}\text{.}
\end{equation}

This procedure allows for some freedom as to which terms,
in addition to the poles,
are taken into account in the calculation of~$\delta_i$. In order to
fix this ambiguity I work in the \ac{msbar}~scheme. This
is a prescription according to which the counterterms contain only terms
proportional to~$\Delta=1/\varepsilon-\gamma_E+\ln(4\pi)$. 

\subsection{Renormalisation of the Gluon Propagator}
\begin{fmffile}{qcd-rengluon-fmf}
The gluon propagator at \ac{nlo} receives corrections from quark,
gluon and ghost loops:
\begin{multline}
\Pi^{AB,\mu\nu}(p^2)=
% PI
\parbox[c][\height+1.5\baselineskip][c]{22mm}{
\begin{fmfchar}(20,5)
\fmfleft{v1}\fmfright{v2}
\fmf{gluon}{v3,v1}
\fmf{gluon}{v2,v3}
\fmfv{decor.shape=circle,decor.size=.3w,decor.filled=30}{v3}
\end{fmfchar}}=
% = G1
\parbox[c][\height+1.5\baselineskip][c]{22mm}{
\begin{fmfchar}(20,10)
\fmfleft{v1}\fmfright{v2}
\fmfdot{v3,v4}
\fmf{gluon}{v3,v1}
\fmf{gluon}{v2,v4}
\fmf{fermion,left,tension=0.3}{v3,v4,v3}
\end{fmfchar}}+
% + G2 ( tadpole )
\parbox[c][\height+1.5\baselineskip][c]{22mm}{
\begin{fmfchar}(20,10)
\fmfleft{v1}\fmfright{v2}
\fmfdot{v3}
\fmf{gluon}{v3,v1}
\fmf{gluon}{v2,v3}
\fmf{gluon,right=1.5,tension=1,curly_len=1}{v3,v3}
\end{fmfchar}}\\+
% + G3 ( gluon )
\parbox[c][\height+1.5\baselineskip][c]{22mm}{
\begin{fmfchar}(20,10)
\fmfleft{v1}\fmfright{v2}
\fmfdot{v3,v4}
\fmf{gluon}{v3,v1}
\fmf{gluon}{v2,v4}
\fmf{gluon,left=1.3,tension=0.3}{v3,v4,v3}
\end{fmfchar}}+
% + G4 ( ghost )
\parbox[c][\height+1.5\baselineskip][c]{22mm}{
\begin{fmfchar}(20,5)
\fmfleft{v1}\fmfright{v2}
\fmfdot{v3,v4}
\fmf{gluon}{v3,v1}
\fmf{gluon}{v2,v4}
\fmf{ghost,left,tension=0.3}{v3,v4,v3}
\end{fmfchar}}+
% + G5 ( counter term )
\parbox[c][\height+1.5\baselineskip][c]{20mm}{
\begin{fmfchar}(20,5)
\fmfleft{v1}\fmfright{v2}
\fmf{gluon}{v3,v1}
\fmf{gluon}{v2,v3}
\fmfv{decor.shape=cross,decor.size=9thick}{v3}
\end{fmfchar}}
=\\
G_1+G_2+G_2+G_3+G_4+G_5\text{.}
\end{multline}
Gauge invariance requires\footnote{I calculate the graphs for
a four-dimensional gluon, which is sufficient for one-loop
renormalisation.}
\begin{equation}
p_\mu\Pi^{AB,\mu\nu}(p^2)=0
\end{equation}
and hence imposes the tensor structure
\begin{equation}
\Pi^{AB,\mu\nu}=\left[p^\mu p^\nu-p^2\hat{g}^{\mu\nu}\right]
\delta^{AB}\Pi(p^2)\text{.}
\end{equation}

The diagram~$G_1$ is the only contribution that involves a fermion
loop and therefore, in contrast to all other contributions, depends
on the number of flavours~$n_F$. Since gauge invariance must not
depend on the flavours, this contribution must be gauge invariant
on its own, whereas it turns out that only the sum of the diagrams
$G_2+G_3+G_4$ is gauge invariant. Direct calculation of the diagrams
shows
\begin{subequations}
\begin{align}
G_1&=n_FT_R\frac{\alpha_s}{4\pi}%
\frac{1-\varepsilon}{1-\frac23\varepsilon}I_2^n(p^2)%
\cdot\frac43%
\left[p^\mu p^\nu-p^2\hat{g}^{\mu\nu}\right]%
\delta^{AB}\text{,}\\
G_2&=0\text{,}\\
G_3&=\frac12C_A\frac{\alpha_s}{4\pi}%
\frac{1}{1-\frac23\varepsilon}I_2^n(p^2)\\
&\nonumber\times\left[%
\left(-\frac{11}3+\frac{7}3\varepsilon\right)p^\mu p^\nu
+\left(\frac{19}{6}-2\varepsilon\right)p^2\hat{g}^{\mu\nu}\right]%
\delta^{AB}\text{,}\\
G_4&=C_A\frac{\alpha_s}{4\pi}%
\frac{1}{1-\frac23\varepsilon}I_2^n(p^2)\\
&\nonumber\times\left[\left(\frac16-\frac16\varepsilon\right)p^\mu p^\nu%
+\frac1{12}p^2\hat{g}^{\mu\nu}\right]%
\delta^{AB}\text{,}\\
G_5&=(Z_3-1)\left[p^\mu p^\nu-p^2\hat{g}^{\mu\nu}\right]\text{.}
\end{align}
\end{subequations}
One can expand the diagrams in epsilon to extract the pole part,
which determines $Z_3$ as
\begin{equation}
\label{eq:renormalisation:Z3}
Z_3=1+\frac{\alpha_s}{4\pi}\left(\frac53C_A-\frac43n_FT_R\right)\Delta
\end{equation}
This result is gauge dependent and only true for~$\lambda=1$.
In general gauge the coefficient of $C_A$
in~\eqref{eq:renormalisation:Z3} is~\cite{Peskin:QFT}
\begin{equation}
\frac{13}{6}-\frac{\lambda}{2}%
\stackrel{\lambda\rightarrow1}{\longrightarrow}\frac53
\end{equation}

Finally, the full expression for~$\Pi(p^2)$ in \person{Feynman} gauge and 
\ac{msbar} is
\begin{multline}
\Pi(p^2)=\frac{\alpha_s}{4\pi}\left[\frac43n_FT_R(1-\varepsilon)
-\left(\frac53-\varepsilon\right)C_A\right]
\frac{1}{1-\frac23\varepsilon}\left(I_2^n(p^2)-\Delta\right)\\
+\frac{\alpha_s}{4\pi}\left[\frac53C_A-\frac43n_FT_R\right]\Delta\text{.}
\end{multline}
\end{fmffile}

\subsection{Renormalisation of the Fermion Propagator}
\begin{fmffile}{qcd-renfermion-fmf}
The renormalisation of the fermion propagator consists of a field strength
renormalisation and a mass renormalisation~\cite{Grozin:QED}. The term of interest from
the \Lagrangeian{} density is
\begin{equation}\label{eq:qcd-renorm:lagrange}
\bar{q}_0(i\fmslash{\partial}-m_0)q_0=
Z_2\bar{q}(i\fmslash{\partial}-Z_mm)q\text{.}
\end{equation}
The bare propagator therefore reads
\begin{equation}
S_0(p)=\frac{\fmslash{p}+m_0}{p^2-m_0^2}\text{,}
\end{equation}
and summing up all 1PI corrections $\Sigma(p)$ one obtains the full
propagator~$S(p)$ which can be written recursively as
\begin{equation}
S(p) = S_0(p) + S_0(p)\Sigma(p)S(p)
\end{equation}
or by solving the above equation as
\begin{equation}
S(p) = (S_0(p)^{-1}-\Sigma(p))^{-1}\text{.}
\end{equation}
Since the two-point function only involves one external vector~$p$
the tensor structure of~$\Sigma(p)$ has only two components
\begin{equation}
\Sigma(p)=\fmslash{p}\Sigma_V(p)+m_0\Sigma_S(p)\text{.}
\end{equation}
Therefore the full propagator can be written as
\begin{equation}
S(p)=\frac{1}{1-\Sigma_V(p)}\left(\fmslash{p}
-\frac{1+\Sigma_S(p)}{1-\Sigma_V(p)}m_0\right)^{-1}
\end{equation}
Comparing with equation~\eqref{eq:qcd-renorm:lagrange} one can read
off the two renormalisation conditions
\begin{equation}\label{eq:qcd-renormalisation:fermcond}
(1-\Sigma_V(p))Z_2=\text{finite}\quad\text{and}\quad
\frac{1+\Sigma_S(p)}{1-\Sigma_V(p)}Z_m=\text{finite.}
\end{equation}
The only diagram that contributes to~$\Sigma(p)$ is
\begin{equation}
i\Sigma(p)=
\parbox[c][\height+1.5\baselineskip][c]{22mm}{
\begin{fmfchar}(20,5)
\fmfleft{v1}\fmfright{v4}
\fmf{fermion}{v1,v2}
\fmf{fermion, tension=0.75}{v2,v3}
\fmf{fermion}{v3,v4}
\fmffreeze
\fmf{gluon,left=1.3,tension=0.75}{v2,v3}
\fmfdot{v2,v3}
\end{fmfchar}}=\frac{\alpha_s}{4\pi}C_F\left\{
\frac32\fmslash{p}-m\right\}\Delta+{\mathcal O}(\varepsilon^0)
\text{.}
\end{equation}
From this we can read off the conditions in order to
satisfy~\eqref{eq:qcd-renormalisation:fermcond},
\begin{subequations}
\begin{align}
Z_2&=1+\frac32\frac{\alpha_s}{4\pi}C_F\Delta+{\mathcal O}(\varepsilon)
\quad\text{and}\\
Z_m&=1-\frac{\alpha_s}{4\pi}C_F\Delta+{\mathcal O}(\varepsilon)\text{.}
\end{align}
\end{subequations}

\subsection{Renormalisation of the Coupling Constant}
Similarly, one can calculate the renormalisation constant~$Z_1$ from the
one-loop corrections to the gluon-quark vertex. The requirement of
finiteness of the renormalised vertex 
in the \ac{msbar} scheme leads to the equation\footnote{%
See for example~\cite{Boehm:2001}}
\begin{equation}
\delta_1+\delta_2+\frac12\delta_3=-(C_A+C_F)\frac{\alpha_s}{4\pi}\Delta\text{,}
\end{equation}
which can be solved for $Z_1$,
\begin{equation}
Z_1=1+\left(\frac23T_RN_F-\frac{11}6C_A\right)%
\frac{\alpha_s}{4\pi}\Delta\text{.}
\end{equation}
It should be noted that the counter terms of the gluon self-interaction
vertices and the ghost-sector are fixed by $Z_1$, $Z_2$ and $Z_3$ through
gauge invariance.
\end{fmffile}

%% file: qcd-real.tex
\begin{fmffile}{eeqq}
\subsection{Introduction}
For a systematic expansion of the scattering amplitude in $\alpha_s$ one
has to consider not only loop diagrams but also diagrams belonging to the
process which includes the emission of one extra, unresolved parton. The
cross section up to \ac{nlo} in $\alpha_s$ for a process with $N$~final
state partons and $I$~partons in the initial state
therefore has the form\cite{Catani:1996vz}
\begin{equation}
\label{eq:qcd-real:sigma-naive}
\sigma=\born{\sigma}+\left(\real{\sigma}+\virt{\sigma}\right)
   +\mathcal{O}(\alpha_s^{N+I})\text{.}
\end{equation}
The leading term $\born{\sigma}$ is the tree level contribution to the
process and can be written as
\begin{equation}
\born{\sigma}=\int\!\!\phasespace{N}\vert\born{\ME}\vert^2\measureF{N}\text{,}
\end{equation}
where $\phasespace{N}$ denotes a $N$-particle phase space,
$\born{\ME}$ is the \person{Born} matrix element and
$\measureF{N}$ is the jet measurement function.
\index{.Fj@$\measureF{N}$|see{measurement function}}
\index{measurement function}
In order to result in a
meaningful matching with the experiment the measurement function has to
be \ac{ir} safe, i.e. the function has to be defined such that
in the collinear and soft limits one obtains
\begin{equation}
\label{eq:qcd-real:measure-ir}
\measureF{N+1}\rightarrow\measureF{N}\text{.}
\end{equation}

The terms $\virt\sigma$ and $\real\sigma$ are the NLO corrections to the
process; $\virt\sigma$ contains the one-loop diagrams whereas $\real\sigma$
is the tree level process with $N+1$ final state particles:
\begin{align}
\virt{\sigma}&=\int\!\!\phasespace{N}(\virt{\ME}{\born{\ME}}^\ast
+{\virt{\ME}}^\ast\born{\ME})\measureF{N}\\
\real{\sigma}&=\int\!\!\phasespace{N+1}\vert\real{\ME}\vert^2\measureF{N+1}
\end{align}

Although this looks as if one could simply do a phase space integration over
$\phasespace{N+1}$ for the real corrections and separately integrate
the other contributions over~$\phasespace{N}$ this naive approach derails
due to \ac{ir} singularities. These singularities cancel in inclusive
cross-sections between
$\virt\sigma$ and~$\real\sigma$~\cite{PhysRev.52.54,Kinoshita:1962ur,%
Lee:1964is,Cutkosky:1960sp}.

In the virtual correction the \ac{ir} singularities
can be extracted explicitly and usually are regularised and expressed
as poles in~$\varepsilon$.
For the real corrections one has to choose a different
approach: in order to obtain an amplitude that is finite at every point in
phase space one subtracts a term that contains the collinear and soft
approximation of the matrix element $\real\ME$. In order to account for the
subtraction a corresponding term has to be added to the virtual correction.
Equation~\eqref{eq:qcd-real:sigma-naive} then turns into
\begin{multline}
\sigma=\born\sigma
+\int\!\!\phasespace{N+1}\left(\vert\born\ME\vert^2\measureF{N+1}
-\diff\sigma^{\mathcal A}\right)_{\varepsilon=0}\\
+\int\!\!\phasespace{N}\left(
(\virt{\ME}{\born{\ME}}^\ast+{\virt{\ME}}^\ast\born{\ME})\measureF{N}
+\int\!\!\phasespace{1}\diff\sigma^{\mathcal A}\right)_{\varepsilon=0}
   +\mathcal{O}(\alpha_s^{N+I})\text{.}
\end{multline}

The calculation of the \ac{ir} counterterm $\diff\sigma^{\mathcal A}$
relies on the fact that in the collinear limit the amplitude factorises
into a hard subprocess and a soft part.
Explicit expressions of counterterms have been derived
in~\cite{Ellis:1980wv}; in this work the fully process independent approach
of Reference~\cite{Catani:1996vz} is used.

\subsection{A Complete Example: $e^+e^-\rightarrow q\bar{q}$}
In this chapter the calculation of $e^+e^-\rightarrow q\bar{q}$ is
reviewed in great detail~\cite{Ellis:1980wv} to explicitly show the
cancellation of \ac{ir} poles and to motivate the use of the dipole
subtraction method~\cite{Catani:1996vz}.

In the following example the amplitude of the process
\begin{equation}
e^+(p_1) + e^-(p_2) \rightarrow \gamma^\ast(p_1+p_2)
\rightarrow q(k_1) + \bar{q}(k_2)
\end{equation}
is considered at \ac{nlo} in $\alpha_s$. In the real correction
an additional gluon $g(k_3)$ is radiated off the quarks in the final state.

Since the initial state is not partonic this example avoids the discussion
of initial state radiation which has been devoted a later section to.

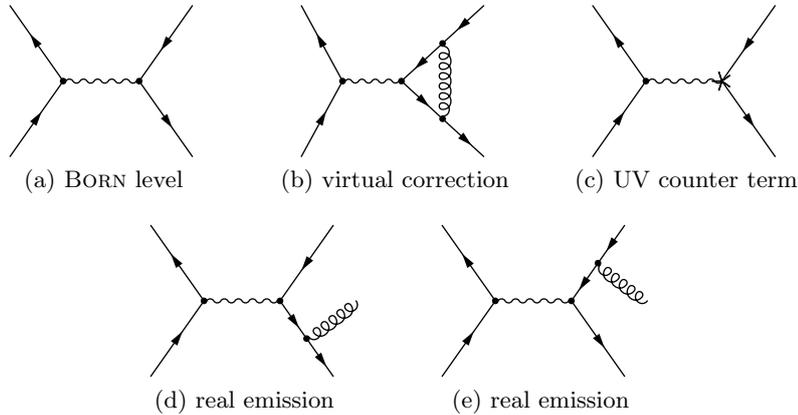
\begin{figure}[hbtp]
\begin{center}
\subfloat[\person{Born} level]{\scalebox{0.5}{\input{eeqq/born}}%
\label{fig:qcd-real:eeqq-born}%
}%
\qquad%
\subfloat[virtual correction]{\scalebox{0.5}{\input{eeqq/virt}}%
\label{fig:qcd-real:eeqq-virt}%
}%
\qquad%
\subfloat[\ac{uv} counter term]{\scalebox{0.5}{\input{eeqq/ct}}%
\label{fig:qcd-real:eeqq-ct}%
}%
\qquad%
\subfloat[real emission]{\scalebox{0.5}{\input{eeqq/real1}}%
\label{fig:qcd-real:eeqq-real1}%
}%
\qquad%
\subfloat[real emission]{\scalebox{0.5}{\input{eeqq/real2}}
\label{fig:qcd-real:eeqq-real2}%
}%
\end{center}
\caption{Diagrams contributing to the process $e^+e^-\rightarrow q\bar{q}$.
\label{fig:qcd-real:eeqq-diagrams}%
}
\end{figure}

\subsubsection{The Tree Level Contribution}
The tree level contribution consists of only one \person{Feynman} diagram,
an $s$-channel photon exchange between the electrons and the quarks.
I use physical, i.e. $p_1+p_2=k_1+k_3$ resp.
$p_1+p_2=k_1+k_2+k_3$ in the real emission.
I use the \person{Mandelstam} variables
$t_{ij}=(p_i-k_j)^2$, $s_{ij}=(k_i+k_j)^2$ and $s=(p_1+p_2)^2$. Since all
particles are assumed to be on-shell and massless these definitions boil
down to $t_{ij}=-2p_i\cd k_j$, $s_{ij}=2k_i\cd k_j$ and $s=2p_1\cd p_2$.

The squared matrix element averaged over initial state spins
and summed over final state spins and colours is
\begin{equation}
\overline{\vert\born\ME\vert^2}=2(4\pi\alpha)^2Q_f^2N_C%
\left(\frac{t_{11}^2+t_{21}^2}{s^2}-\varepsilon\right).
\end{equation}
Here $Q_f$ is the electrical charge of the quark flavour and $N_C$ the
number of \sun[\ensuremath{N_C}] colours.

\subsubsection{The Virtual Corrections}
Next we determine the virtual correction to this process.
The squared amplitude that stems from
\figref{fig:qcd-real:eeqq-diagrams}~\subref{fig:qcd-real:eeqq-virt}
is
\begin{multline}
\overline{\vert\virt\ME\vert^2}=
\parbox[c][\height+1.5\baselineskip][c]{40mm}{%
\scalebox{0.66}{\input{eeqq/virt2a}}}+
\parbox[c][\height+1.5\baselineskip][c]{40mm}{%
\scalebox{0.66}{\input{eeqq/virt2b}}}=\\
\frac{(4\pi\alpha)^2Q_f^2(4\pi\alpha_s)\tr{t^At^A}}{%
4s^2\cdot2^n\pi^{\frac{n}2}}
%\times\\
\int\!\!\frac{\diff[n]k}{i\pi^{\frac{n}2}}
\frac{\tr{\pslash[1]\gamma^\mu\pslash[2]\gamma^\nu}%
\tr{\kslash[2]\gamma^\rho(\kslash-\kslash[2])%
\gamma_\mu(\kslash+\kslash[1])\gamma_\rho\kslash[1]\gamma_\nu}}{%
[k^2+i\delta][(k+k_1)^2+i\delta][(k-k_2)^2+i\delta]}\\
+ \text{\acs{hc}}\text{.}
\end{multline}

which can be simplified to the form given in~\cite{Catani:1996vz}:
\begin{equation}\label{eq:qcd-real:eeqq-virt-final}
\overline{\vert\virt\ME\vert^2}=
\overline{\vert\born\ME\vert^2}\,\cdot\frac{C_F\alpha_s}{2\pi}
\left(\frac{4\pi\mu^2}{s}\right)^\varepsilon
\frac{1}{\Gamma(1-\varepsilon)}\left(%
-\frac{2}{\varepsilon^2}-\frac{3}{\varepsilon}-8+\pi^2
+\mathcal{O}(\varepsilon)\right)\text{.}
\end{equation}
It should be noted that the factor $s^{-\varepsilon}/\Gamma(1-\varepsilon)$
must be considered for a full expansion in $\varepsilon$; however, for the
purpose of the subtraction the given form is more convenient. Furthermore
one should be aware that the expression contains \ac{ir} poles only;
\ac{uv} renormalisation is trivial in this example since the counterterm
from the vertex correction 
\figref{fig:qcd-real:eeqq-diagrams}~\subref{fig:qcd-real:eeqq-ct}
is cancelled by the wave function renormalisation.

\subsubsection{The Real Emission}
We now consider the the real emission diagrams~\cite{Binoth:QCDLecture}
in \figref{fig:qcd-real:eeqq-diagrams}
\subref{fig:qcd-real:eeqq-real1} and~\subref{fig:qcd-real:eeqq-real2},
\begin{multline}
\label{eq:qcd-real:ME-real}
\overline{\vert\real\ME\vert^2}=
\parbox[c][20mm][c]{40mm}{%
\scalebox{0.66}{\input{eeqq/real2a}}}+
\parbox[c][20mm][c]{40mm}{%
\scalebox{0.66}{\input{eeqq/real2b}}}\\
\shoveright{+\text{\acs{ir} finite diagrams} =}\displaybreak[1]\\
\shoveleft{} \frac{1}{4}(4\pi\alpha)^2(4\pi\alpha_s)Q_f^2\tr{t^at^a}%
\frac{1}{s^2}\tr{\pslash[2]\gamma_\mu\pslash[1]\gamma_\nu}\times\\
\left\{
+\frac{1}{s_{13}^2}\tr{\kslash[2]\gamma^\nu(\kslash[1]+\kslash[3])%
   \gamma^\rho\kslash[1]\gamma^\sigma(\kslash[1]+\kslash[3])\gamma^\mu}
\right.\\
\qquad-\frac{1}{s_{13}s_{23}}
	\tr{\kslash[2]\gamma^\nu(\kslash[1]+\kslash[3])%
	\gamma^\rho\kslash[1]\gamma^\mu%
	(\kslash[2]+\kslash[3])\gamma^\sigma}
\\
\qquad-\frac{1}{s_{13}s_{23}}
	\tr{\kslash[2]\gamma^\rho(\kslash[2]+\kslash[3])%
	\gamma^\nu\kslash[1]\gamma^\sigma%
	(\kslash[1]+\kslash[3])\gamma^\mu}
\\\left.\shoveright{
+\frac{1}{s_{23}^2}\tr{\kslash[2]\gamma^\rho(\kslash[2]+\kslash[3])%
   \gamma^\nu\kslash[1]\gamma^\mu(\kslash[2]+\kslash[3])\gamma^\sigma}
\right\}d_{\rho\sigma}(k_3) =}\\
\shoveleft{8\pi\alpha_sC_F\frac{1}{s_{13}s_{23}s}\left\{
2(4\pi\alpha)^2Q_f^2N_C
\left[ %%%%%%%%%%%%%%%%%%%
   t_{11}^2+t_{21}^2+t_{12}^2+t_{22}^2 %
 \right.\right.}\\
	-\varepsilon(t_{11}-t_{22})^2-\varepsilon(t_{12}-t_{21})^2 %
\left.\left. %
	-\varepsilon s^2 - \varepsilon s_{12}^2 %
	+ \varepsilon^2 (s - s_{12})^2\right]\right\}
\end{multline}
The tensor $d_{\mu\nu}(k_3)$ stems from the polarisation sum
\begin{equation}
d_{\mu\nu}(k)=\sum_{\text{pol.}}\epsilon_\mu(k)\epsilon_\nu^\ast(k)=
-g_{\mu\nu}+\frac{k_\mu r_\nu + k_\nu r_\mu}{k\cd r}\text{,}
\end{equation}
using axial gauge\index{Gauge!axial} with an arbitrary lightlike vector $r$.

By integrating over the phase space one runs into singularities whenever
$k_3$ becomes soft or if it is in the collinear region with either $k_1$
or $k_2$, which induces $s_{13}$ and (or) $s_{23}$ to tend to zero.

\subsubsection{The Collinear Limit}
In the collinear limit the amplitude factorises into a splitting function,
a \ac{ir} divergence and the tree level amplitude. To carry out the limit
in a controlled manner one introduces a
\person{Sudakov} parametrisation~\cite{Catani:1996vz}:
\index{Sudakov parametrisation@\person{Sudakov} parametrisation}
\begin{subequations}
\begin{align}
k_i^\mu &= z p^\mu + k_\perp^\mu - \frac{k_\perp^2}{2(p\cd n)z}n^\mu%
\qquad\text{and}\\
k_3^\mu &= (1-z) p^\mu - k_\perp^\mu - \frac{k_\perp^2}{2(p\cd n)(1-z)}n^\mu
\end{align}
\end{subequations}
where $p$ is the common collinear direction of $k_3$ and $k_i$ ($i=1,2$),
$k_\perp$ is the transverse direction and $n$ is an auxiliary lightlike vector.
The vectors $p$, $k_\perp$ and $n$ are obeying $p^2=n^2=0$, $k_\perp^2<0$
and $k_\perp\cd n=k_\perp\cd p=0$.
The limit $k_\perp^2\rightarrow0$ represents the collinear case and
$z\rightarrow1$ leads to the soft case.
It is easy to show that this parametrisation preserves $k_i^2=k_3^2=0$
and with $j=3-i$ in the limit $k_\perp^2\rightarrow0$
the \person{Mandelstam} variables read
\begin{subequations}
\begin{align}
s_{i3}&= -\frac{k_\perp^2}{z(1-z)}\rightarrow0\text{,}\\
s_{j3}&=(1-z)s\text{,}\\
s_{12}&=zs\text{,}\\
t_{1i}&=zt_{2j}\qquad\text{and}\\
t_{2i}&=zt_{1j}\text{.}
\end{align}
\end{subequations}
The real emission part~\eqref{eq:qcd-real:ME-real} in the collinear
limit hence becomes
\begin{equation}
\overline{\vert\real\ME\vert^2}\xrightarrow{k_\perp^2\rightarrow0}
8\pi\alpha_s\frac{1}{s_{i3}}%
C_F\left(\frac{1+z^2}{1-z}-\varepsilon(1-z)\right)%
\overline{\vert\born\ME\vert^2}=
8\pi\alpha_s\frac{1}{s_{i3}}P_{qq}(z)%
\overline{\vert\born\ME\vert^2}\text{.}
\end{equation}
Here I introduced the \person{Altarelli}-\person{Parisi} splitting function
$P_{qq}(z)$ which is a process independent function that only depends on
the two particles involved in the soft subprocess\footnote{Here the soft
subprocess is a gluon splitting off a quark.}.

In order to use the achieved result as a \ac{ir} counter term one needs
a momentum mapping from the $(N+1)$ to the $N$-particle final state, i.e.
one needs to express the quantities $p$, $k_\perp$, $n$ and $z$ in terms of
the momenta $k_i$, $k_j$ and $k_3$ such that the following properties are
preserved:
\begin{subequations}
\begin{align}
\label{eq:qcd-real:tilde-cond1}
k_i+k_j+k_3=\tilde{k}_i+\tilde{k}_j&\qquad\text{(momentum conservation)}\\
\label{eq:qcd-real:tilde-cond2}
\tilde{k}_i^2=\tilde{k}_j^2=0&\qquad\text{(on-shell condition),}
\end{align}
\end{subequations}
where the momenta of the $N$-particle final state are denoted with a tilde.
The mapping
\begin{subequations}
\begin{align}
\tilde{k}_i&=k_i+k_3-\frac{y}{1-y}k_j\\
\tilde{k}_j&=\frac{1}{1-y}k_j\\
\intertext{with the parameter}
y&=\frac{s_{i3}}{s_{12}+s_{i3}+s_{j3}}%\xrightarrow{k_\perp^2\rightarrow0}0
\quad\Leftrightarrow\;\frac{1}{1-y}=1+\frac{y}{1-y}=1+\frac{s_{i3}}{s_{12}+s_{j3}}
\end{align}
\end{subequations}
clearly obeys the conditions \eqref{eq:qcd-real:tilde-cond1}
and~\eqref{eq:qcd-real:tilde-cond2}. Furthermore one introduces
\begin{equation}
\tilde{z}_3=\frac{s_{j3}}{s_{12}+s_{j3}}%\xrightarrow{k_\perp^2\rightarrow0}z%
\text{.}
\end{equation}
Since in the collinear limit $\tilde{z}_3\rightarrow z$ and $y\rightarrow0$
the expression
\begin{equation}
\label{eq:qcd-real:ct-eeqq}
\diff\sigma^{\mathcal A}=\sum_{i=1}^2%
8\pi\alpha_s\frac{1}{s_{i3}}P_{qq}(\tilde{z}_3)
\overline{\vert\born\ME(\tilde{k}_1,\tilde{k}_2)\vert^2}%
\measureF{N}(\tilde{k}_1,\tilde{k}_2)\phasespace{3}(k_i,k_j,k_3)
\end{equation}
has the same \ac{ir} behaviour as the real correction 
Equation~\eqref{eq:qcd-real:ct-eeqq} defines a valid counter term.
It should be noted that one has the freedom to add any \ac{ir} safe
terms to the counter term. Here one makes use of the \ac{ir} property
\eqref{eq:qcd-real:measure-ir}
of the measurement function which is crucial for the definition of the
counter term.

\subsubsection{Phase Space Factorisation}
In order to add the subtraction term back to the virtual contribution
one must integrate over the one particle phase space of the extra parton.
This requires to factorise the phase space $\phasespace{3}(k_i,k_j,k_3)$
into $\phasespace{2}(\tilde{k}_i,\tilde{k}_j)\phasespace{1}(k_3)$,
\begin{equation}
\phasespace{3}(k_i,k_j,k_3)=\frac{\diff[n]{k_3}}{(2\pi)^{n-1}}%
\delta(k_3^2)\Theta({k_3}^0)\times%
\phasespace{2}(\tilde{k}_i, \tilde{k}_j)%
\left(\frac{\partial(\tilde{k}_i,\tilde{k}_j)}{%
	\partial(k_i,k_j)}\right)_{k_3=\mathrm{const}}^{-1}
\end{equation}
with the \person{Jacobi}an factor
\begin{equation}
\mathcal{J}=\left(\frac{\partial(\tilde{k}_i,\tilde{k}_j)}{%
	\partial(k_i,k_j)}\right)_{k_3=\mathrm{const}}
=\det\begin{pmatrix}
\frac{\partial\tilde{k}_i^\mu}{\partial k_i^\nu} &
\frac{\partial\tilde{k}_j^\mu}{\partial k_i^\nu} \\
\frac{\partial\tilde{k}_i^\mu}{\partial k_j^\nu} &
\frac{\partial\tilde{k}_j^\mu}{\partial k_j^\nu}
\end{pmatrix}\text{.}
\end{equation}
The matrix in the determinant has dimension $(2n)\times(2n)$.
For convenience the scale $Q=k_ik_j + k_3k_j$ is introduced,
and one gets
\begin{multline}
\mathcal{J}=\det\begin{pmatrix}
\delta^\mu_\nu-\frac{1}{Q}k_j^\mu\left(k_{3\,\nu}-\frac{y}{1-y}k_{j\,\nu}\right)&
\frac{1}{Q}k_j^\mu\left(k_{3\,\nu}-\frac{y}{1-y}k_{j\,\nu}\right)\\
-\frac{y}{1-y}\left(\delta^\mu_\nu-\frac{1}{Q}k_j^\mu(k_i^\nu+k_3^\nu)\right)&
\frac{1}{1-y}\delta^\mu_\nu-\frac{y}{1-y}\frac{1}{Q}k_j^\mu(k_{i\,\nu}+k_{3\,\nu})
\end{pmatrix}=\\
%\det\begin{pmatrix}
%\delta^\mu_\nu&
%\frac{1}{Q}k_j^\mu\left(k_{3\,\nu}-\frac{y}{1-y}k_{j\,\nu}\right)\\
%\delta^\mu_\nu&
%\frac{1}{1-y}\delta^\mu_\nu-\frac{y}{1-y}\frac{1}{Q}k_j^\mu(k_{i\,\nu}+k_{3\,\nu})
%\end{pmatrix}=
\det\begin{pmatrix}
\delta^\mu_\nu&
\frac{1}{Q}k_j^\mu\left(k_{3\,\nu}-\frac{y}{1-y}k_{j\,\nu}\right)\\
0&\frac{1}{1-y}\left(\delta^\mu_\nu-\frac{1}{Q}k_j^\mu(yk_{i\,\nu}-yk_{j\,\nu}+k_{3\,\nu})\right)
\end{pmatrix}=\\
(1-y)^{-n}\det\left(%
\delta^\mu_\nu-\frac{1}{Q}k_j^\mu(yk_{i\,\nu}-yk_{j\,\nu}+k_{3\,\nu})\right)=
(1-y)^{1-n}(1-\tilde{z}_3)\text{.}
\end{multline}
In the last line the identity $\det(\One+uv^\transposed{})=1+v^\transposed{}u$ has been used.
Another factor is introduced by the replacement $\delta(k_j^2)=(1-y)^{2}\delta(\tilde{k}_j^2)$,
and hence the factorisation reads
\begin{align}
\label{eq:qcd-real:factorized01}
\phasespace{3}(k_i,k_j,k_3)&=\phasespace{2}(\tilde{k}_i,\tilde{k}_j)\frac{\diff[n]{k_3}}{(2\pi)^{n-1}}%
\delta(k_3^2)\Theta({k_3}^0)\frac{(1-y)^{n-3}}{(1-\tilde{z}_3)}\nonumber\\
&=\phasespace{2}(\tilde{k}_i,\tilde{k}_j)%
\frac{\diff[n-1]{\mathbf{k}_3}}{(2\pi)^{n-1}(2E_3)}%
\frac{(1-y)^{n-3}}{(1-\tilde{z}_3)}\text{.}
\end{align}
In the centre of mass system of $\tilde{k}_i$ and~$\tilde{k}_j$,
\begin{subequations}
\begin{align}
\tilde{k}_i&=\sqrt{\frac{\tilde{k}_i\cd\tilde{k}_j}{2}}%
(1,\mathbf{0}^{(n-2)},1)\text{,}\\
\tilde{k}_j&=\sqrt{\frac{\tilde{k}_i\cd\tilde{k}_j}{2}}%
(1,\mathbf{0}^{(n-2)},-1)\text{,}\\
k_3&=E_3(1,\mathbf{k}_\perp^{(n-2)}(\theta),\cos(\theta))
\end{align}
\end{subequations}
one can express $\diff[n-1]{k_3}$ in terms of spherical coordinates
relative to the axis defined by $\tilde{\mathbf{k}}_i$, and
Equation~\eqref{eq:qcd-real:factorized01} becomes
\begin{equation}
\phasespace{3}(k_i,k_j,k_3)=
\phasespace{2}(\tilde{k}_i,\tilde{k}_j)%
\frac{\diff{E_3}\,E_3^{n-2}}{2E_3(2\pi)^{n-1}}%
\diff{(\cos\theta)}(\sin\theta)^{n-4}\diff{\Omega_{n-2}}
\frac{(1-y)^{n-3}}{(1-\tilde{z}_3)}\text{,}
\end{equation}
where $\diff{\Omega_{n-2}}$\index{.Omega@$\Omega_d$}
describes the subspace orthogonal to
the other integration directions as in~\eqref{eq:app-intdetails:nsphere}.
In terms of these coordinates, the variables $\tilde{z}_3$ and~$y$ have
the form
\begin{align}
\tilde{z}_3&=\frac{\tilde{k}_j\cd k_3}{\tilde{k}_i\tilde{k}_j}=
\frac{E_3}{\sqrt{2\tilde{k}_i\tilde{k}_j}}(1+\cos\theta)\text{,}\\
y&=\frac{\tilde{k}_i\cd k_3}{%
	\tilde{k}_i\tilde{k}_j-\tilde{k}_jk_3}=
-\frac{E_3(1-\cos\theta)}{E_3(1+\cos\theta)%
	-\sqrt{2\tilde{k}_i\tilde{k}_j}}\text{.}
\end{align}
Another variable transform allows to replace $E_3$ and $\cos\theta$
by $y$ and~$\tilde{z}_3$,
\begin{equation}
E_3=\frac{\sqrt{(2\tilde{k}_i\cd\tilde{k}_j)yz(1-\tilde{z}_3)}}{%
	\sin\theta}\quad%
\text{and}\quad\frac{\partial(E_3,\cos\theta)}{\partial(y,z)}=
\frac{(2\tilde{k}_i\cd\tilde{k}_j)(1-\tilde{z}_3)}{2E_3}\text{,}
\end{equation}
and hence
\begin{multline}
\phasespace{3}(k_i,k_j,k_3)=
\phasespace{2}(\tilde{k}_i,\tilde{k}_j)%
\diff{y}\diff{\tilde{z}_3}\diff{\Omega_{n-2}}%
\frac{(2\tilde{k}_i\cd\tilde{k}_j)}{16\pi^2(2\pi)^{1-2\varepsilon}}%
(E_3\sin\theta)^{n-4}
(1-y)^{n-3}\\
=\phasespace{2}(\tilde{k}_i,\tilde{k}_j)%
   (2\tilde{k}_i\cd\tilde{k}_j)^{1-\varepsilon}%
\diff{\Omega_{n-2}}%
\frac{\diff{y}y^{-\varepsilon}(1-y)^{1-2\varepsilon}}{%
16\pi^2(2\pi)^{1-2\varepsilon}}%
\diff{\tilde{z}_3}(\tilde{z}_3(1-\tilde{z}_3))^{-\varepsilon}
\text{.}
\end{multline}
Together with the limits on the integration variables,
$\Theta(y(1-y))\Theta(\tilde{z}_3(1-\tilde{z}_3))$ one obtains
the form given in~\cite{Catani:1996vz}.

We can now carry out the one-particle subspace~$\phasespace{1}(k_3)$
in Equation~\eqref{eq:qcd-real:ct-eeqq} by
using~$s_{i3}=y(2\tilde{k}_i\tilde{k}_j)$:
\begin{multline}\label{eq:qcd-real:intdipole1}
\int_{k_3}\!\!\diff\sigma^{\mathcal A}=%
\frac{\alpha_s\mu^\varepsilon}{2\pi}\sum_{i=1}^2%
\phasespace{2}(\tilde{k}_i,\tilde{k}_j)
	\overline{\vert\born\ME(\tilde{k}_1,\tilde{k}_2)\vert^2}%
	\measureF{N}(\tilde{k}_1,\tilde{k}_2)%
	(2\tilde{k}_i\cd\tilde{k}_j)^{-\varepsilon}\\
\times\frac{\Omega_{n-2}}{(2\pi)^{1-2\varepsilon}}%
\int_0^1\!\!\diff{y}y^{-1-\varepsilon}(1-y)^{1-2\varepsilon}%
\int_0^1\!\!\diff{\tilde{z}_3}(\tilde{z}_3(1-\tilde{z}_3))^{-\varepsilon}%
P_{qq}(\tilde{z}_3;y)\text{.}
\end{multline}

Naively one might want to use $P_{qq}(\tilde{z})$ instead of
\begin{equation}
P_{qq}(z;y)=C_F\left[\frac{2}{1-z(1-y)}-(1+z)-\varepsilon(1-z)\right]%
\text{.}
\end{equation}
However, only the latter form takes care of overlapping singularities,
and using $P_{qq}(z)$ would not reproduce the correct poles to cancel
the ones in the virtual part of the amplitude. $P_{qq}(z;y)$ maps
smoothly on the original function $P_{qq}(z)$ in the soft and
collinear limit.

The second line of~\eqref{eq:qcd-real:intdipole1} can be evaluated
using the trick in Equation~\eqref{eq:app-intdetails:GammaSum01},
\begin{multline}
C_F\frac{2\pi^{1-\varepsilon}}{%
	\Gamma(1-\varepsilon)(2\pi)^{1-2\varepsilon}}
\int_0^1\!\!\diff{\tilde{z}_3}\int_0^1\!\!\diff{y}
	\tilde{z}_3^{-\varepsilon}(1-\tilde{z}_3)^{-\varepsilon}
	y^{-1-\varepsilon}(1-y)^{1-2\varepsilon}\\
\times\left(
\frac{2}{1-\tilde{z}_3(1-y)}-(1+\tilde{z}_3)-\varepsilon(1-\tilde{z}_3)
\right)\\
=\frac{C_F(4\pi)^\varepsilon}{\Gamma(1-\varepsilon)}%
	\left(\frac{2}{\varepsilon^2}+\frac{3}{\varepsilon}+10-\pi^2
	+ \mathcal{O}(\varepsilon)\right)\text{.}
\end{multline}

The poles cancel with those in the virtual part of the amplitude, and
one can write down the modified differential cross-section in
the limit $\varepsilon\rightarrow0$,
which is suitable for direct numerical integration:
\begin{equation}
\diff{\virt{\sigma}}+\int_1\diff{\sigma^{\mathcal A}}=
\frac{\alpha_sC_F}{2\pi}
%\frac{1}{\Gamma(1-\varepsilon)}
%\left(\frac{4\pi\mu^2}{s}\right)^\varepsilon
\cdot2\overline{\vert\born\ME\vert^2}\measureF{N}
\end{equation}

\end{fmffile}

%% file: eeqq/born.tex
\begin{fmfchar*}(60,40)
\fmfleft{em,ep}\fmfright{q,qbar}
\fmfdot{v1,v2}
\fmf{electron,label=\large$e^-$}{em,v1}
\fmf{electron,label=\large$e^+$}{v1,ep}
\fmf{quark,label=\large$q$}{qbar,v2}
\fmf{quark,label=\large$\bar{q}$}{v2,q}
\fmf{photon,tension=1.4}{v1,v2}
\end{fmfchar*}

%% file: eeqq/virt.tex
\begin{fmfchar*}(60,40)
\fmfleft{em,ep}\fmfright{q,qbar}
\fmfdot{v1,v2,g1,g2}
\fmf{electron}{em,v1,ep}
\fmf{quark,tension=1}{qbar,g1,v2,g2,q}
\fmf{photon,tension=1.4}{v1,v2}
\fmffreeze
\fmf{gluon}{g2,g1}
\end{fmfchar*}

%% file: eeqq/ct.tex
\begin{fmfchar*}(60,40)
\fmfleft{em,ep}\fmfright{q,qbar}
\fmfdot{v1}
\fmfv{decor.shape=pentacross}{v2}
\fmf{electron}{em,v1,ep}
\fmf{quark}{qbar,v2,q}
\fmf{photon,tension=1.4}{v1,v2}
\end{fmfchar*}

%% file: eeqq/real1.tex
\begin{fmfchar*}(60,40)
\fmfleft{em,ep}\fmfright{q,g,qbar}
\fmfdot{v1,v2,g1}
\fmf{electron}{em,v1,ep}
\fmf{quark}{qbar,v2}
\fmf{quark,tension=2}{v2,g1,q}
\fmf{photon,tension=1.4}{v1,v2}
\fmffreeze
\fmf{gluon}{g1,g}
\end{fmfchar*}

%% file: eeqq/real2.tex
\begin{fmfchar*}(60,40)
\fmfleft{em,ep}\fmfright{q,g,qbar}
\fmfdot{v1,v2,g1}
\fmf{electron}{em,v1,ep}
\fmf{quark,tension=2}{qbar,g1,v2}
\fmf{quark,tension=1}{v2,q}
\fmf{photon,tension=1.4}{v1,v2}
\fmffreeze
\fmf{gluon}{g1,g}
\end{fmfchar*}

%% file: eeqq/virt2a.tex
\begin{fmfchar*}(60,30)
\fmftop{cut1}\fmfbottom{cut2}
\fmfleft{em1,ep1}\fmfright{em2,ep2}
\fmf{dashes}{cut1,q1,q0,q2,cut2}
\fmffreeze
\fmfdot{v1l,v2l,v1r,v2r}
\fmf{electron}{em1,v1l,ep1}
\fmf{electron}{em2,v1r,ep2}
\fmf{photon,tension=1.5}{v1l,v2l}
\fmf{photon,tension=1.5}{v1r,v2r}
\fmf{quark,tension=1.5}{q1,v1g}\fmf{quark}{v1g,v2l}
\fmf{quark}{v2l,v2g}\fmf{quark,tension=1.5}{v2g,q2}
\fmf{quark,tension=0.5}{q2,v2r}
\fmf{quark,tension=0.5}{v2r,q1}
\fmffreeze
\fmf{gluon}{v2g,v1g}
\end{fmfchar*}

%% file: eeqq/virt2b.tex
\begin{fmfchar*}(60,30)
\fmftop{cut1}\fmfbottom{cut2}
\fmfleft{em1,ep1}\fmfright{em2,ep2}
\fmf{dashes}{cut1,q1,q0,q2,cut2}
\fmffreeze
\fmfdot{v1l,v2l,v1r,v2r}
\fmf{electron}{em1,v1l,ep1}
\fmf{electron}{em2,v1r,ep2}
\fmf{photon,tension=1.5}{v1l,v2l}
\fmf{photon,tension=1.5}{v1r,v2r}
\fmf{quark,tension=0.5}{q1,v2l}
\fmf{quark,tension=0.5}{v2l,q2}
\fmf{quark,tension=1.5}{q2,v1g}\fmf{quark}{v1g,v2r}
\fmf{quark}{v2r,v2g}\fmf{quark,tension=1.5}{v2g,q1}
\fmffreeze
\fmf{gluon}{v2g,v1g}
\end{fmfchar*}

%% file: eeqq/real2a.tex
\begin{fmfchar*}(60,30)
\fmftop{cut1}\fmfbottom{cut2}
\fmfleft{em1,ep1}\fmfright{em2,ep2}
\fmf{dashes}{cut1,q1,q0,q2,cut2}
\fmffreeze
\fmfdot{v1l,v2l,v1r,v2r}
\fmf{electron}{em1,v1l,ep1}
\fmf{electron}{em2,v1r,ep2}
\fmf{photon,tension=1.5}{v1l,v2l}
\fmf{photon,tension=1.5}{v1r,v2r}
\fmf{quark}{q1,v1g,v2l}
\fmf{quark,tension=0.5}{v2l,q2}
\fmf{quark}{q2,v2g,v2r}
\fmf{quark,tension=0.5}{v2r,q1}
\fmffreeze
\fmf{gluon}{v1g,v2g}
\end{fmfchar*}

%% file: eeqq/real2b.tex
\begin{fmfchar*}(60,30)
\fmftop{cut1}\fmfbottom{cut2}
\fmfleft{em1,ep1}\fmfright{em2,ep2}
\fmf{dashes}{cut1,q1,q0,q2,cut2}
\fmffreeze
\fmfdot{v1l,v2l,v1r,v2r}
\fmf{electron}{em1,v1l,ep1}
\fmf{electron}{em2,v1r,ep2}
\fmf{photon,tension=1.5}{v1l,v2l}
\fmf{photon,tension=1.5}{v1r,v2r}
\fmf{quark,tension=0.5}{q1,v2l}
\fmf{quark}{v2l,v1g,q2}
\fmf{quark,tension=0.5}{q2,v2r}
\fmf{quark}{v2r,v2g,q1}
\fmffreeze
\fmf{gluon}{v1g,v2g}
\end{fmfchar*}

%% file: qcd-dipoles.tex
\subsection{Introduction}
In the previous section it has been shown that for a complete
\ac{nlo} calculation, both the virtual one-loop corrections
and the real emission of an additional particle have to be
calculated to achieve a physically meaningful result. Moreover,
both parts of the calculation are divergent on their own but the
divergences cancel when the parts are put together.

Although other methods are available as well~\cite{Harris:2001sx,Eynck:2001en}
I will focus only on a method developed by \person{Catani} and
\person{Seymour}~\cite{Catani:1996vz} that works with local counter
terms, so-called \emph{dipoles},
that cancel the poles of both parts independently. This
method has been extended also for the massive case~\cite{Catani:2002hc}
and proves to be suitable for
automation~\cite{Gleisberg:2007md,Frederix:2008hu,Seymour:2008mu}.

In the previous section some of the subtleties of such a subtraction
procedure have been neglected since they are discussed in detail
in the original literature~\cite{Catani:1996vz}; instead, I will give
two examples of how to apply the original formul\ae{} to actual processes,
the amplitudes $u\bar{u}\rightarrow b\bar{b}s\bar{s}$ and
$gg\rightarrow b\bar{b}s\bar{s}$. These two amplitudes are the
basic ingredients for the calculation of the physical process
$pp\rightarrow b\bar{b}b\bar{b}$. Although for the real emission
besides an additional gluon one also has to consider the processes
with one gluon in the initial state like $gq\rightarrow b\bar{b}s\bar{s}q$
these subprocesses do not contribute to the cancellation of the poles
in the virtual amplitude
and therefore are not needed to be included in the dipole subtraction.

In the following the notation of~\cite{Catani:1996vz} is adapted and only
massless partons are considered. For a $2\rightarrow N$ process at a
proton-proton collider an inclusive cross section can be calculated
as a convolution of a partonic cross section $\sigma_{ab}$
and the parton distribution functions $f_a(x)$ and~$f_b(x)$,
\begin{equation}
\sigma_{\text{tot.}}=
\int_0^1\!\!\diff{x_1}\int_0^1\!\!\diff{x_2}
\sum_{a,b}f_a(x_1)f_b(x_2)\sigma_{ab}(x_1P_1, x_2P_2)
\end{equation}
where the sum over $a$ and $b$ runs over all contributing parton
flavours\footnote{\emph{Flavour} includes quark flavours
as well as the possibility of an initial state gluon.}. The proton
momenta are $P_1$ and~$P_2$. Here only jet cross sections are discussed
and therefore the topic of fragmentation is neglected.

For the partonic cross section one has
\begin{multline}\label{eq:qcd-dipoles:outline1}
\sigma_{ab}(p_a, p_b) =
\born\sigma_{ab}(p_a, p_b)+\sigma^{\text{NLO}}_{ab}(p_a,p_b)=\\
\born\sigma_{ab}(p_a, p_b)
+\sigma^{\text{NLO}\,\{N+1\}}_{ab}(p_a, p_b)
+\sigma^{\text{NLO}\,\{N\}}_{ab}(p_a, p_b)
+\sigma^{\text{C}\,\{N\}}_{ab}(p_a, p_b)
\text{.}
\end{multline}
As before, $\born\sigma$ is the leading order cross-section. It can
be written as
\begin{multline}
\born\sigma_{ab}(p_a, p_b)=
\frac{1}{2\vert p_a+p_b\vert^2}
\frac{1}{n_an_b}\sum_{\lambda_i}\int\!\!\phasespace{N}(Q)
\measureF{N}(p_1, \ldots, p_N)\\
\times\bra{c^\prime}
\born{\mathcal{A}}_{c^\prime}(p_a^{\lambda_a}, p_b^{\lambda_b};
	p_1^{\lambda_1},\ldots, p_N^{\lambda_N})^\ast
\born{\mathcal{A}}_{c}(p_a^{\lambda_a}, p_b^{\lambda_b};
	p_1^{\lambda_1},\ldots, p_N^{\lambda_N})
\ket{c}\text{.}
\end{multline}
In an appropriate colour basis there is one helicity amplitude
$\mathcal{A}=\born{\mathcal{A}}$
for each colour vector $\ket{c}$; the basis is not necessarily orthogonal
and hence one has to take the correlation matrix $\braket{c^\prime\vert c}$
into account. The normalisation constants $n_a$ and $n_b$ denote the
additional constants for spin and colour average and only depend on the
types of the incoming particles. The total incoming parton momentum
is $Q=p_a+p_b$

The second term in Equation~\eqref{eq:qcd-dipoles:outline1}
contains the real emission and an appropriate counter term such that
the term is finite:
\begin{equation}
\sigma^{\text{NLO}\,\{N+1\}}_{ab}(p_a,p_b)=\int\!\!\phasespace{N+1}(Q)
\left[\left(\frac{\diff{\real\sigma}}{\phasespace{N+1}}\right)_{\varepsilon=0}
-\left(\frac{\diff{\sigma^{\mathcal{A}}}}{%
\phasespace{N+1}}\right)_{\varepsilon=0}
\right]\text{.}
\end{equation}

Similar to the \person{Born} cross section the real emission is
\begin{multline}
\frac{\diff{\real\sigma}}{\phasespace{N+1}}=
\frac{1}{n_an_b}\sum_{\lambda_i}
\measureF{N+1}(p_1,\ldots,p_{N+1})\\
\times\bra{c^\prime}
{\real{\mathcal{A}}_{c^\prime}}(p_a^{\lambda_a}, p_b^{\lambda_b};
	p_1^{\lambda_1},\ldots, p_{N+1}^{\lambda_{N+1}})^\ast
\real{\mathcal{A}}_{c}(p_a^{\lambda_a}, p_b^{\lambda_b};
	p_1^{\lambda_1},\ldots, p_{N+1}^{\lambda_{N+1}})
\ket{c}
\end{multline}
with a corresponding colour basis that includes the extra parton.

The subtraction term is a sum over all dipoles,
\begin{multline}
\frac{\diff{\sigma^{\mathcal{A}}}}{\phasespace{N+1}}=\frac{1}{n_an_b}
\frac{1}{S}\times\\
\left[
\sum_{\{i,j\}=1}^{N+1}\sum_{\rlap{\def\arraystretch{.5}
\begin{array}[b]{r@{}l}
\scriptstyle k&\scriptstyle =1\\
\scriptstyle k&\scriptstyle\neq i,j
\end{array}}}^{N+1}\Dipole_{ij,k}(p_a,p_b;p_1,\ldots, p_{N+1})
\measureF{N}(p_a,p_b;p_1,\ldots,\tilde{p}_{ij},\tilde{p}_k\ldots,p_{N+1})
\right.
\displaybreak[1]\\
+\sum_{\{i,j\}=1}^{N+1}
\Dipole_{ij}^{a}(p_a,p_b;p_1,\ldots, p_{N+1})
\measureF{N}(\tilde{p}_a,p_b;p_1,\ldots,\tilde{p}_{ij},\ldots,p_{N+1})
\displaybreak[1]\\
+\sum_{\{i,j\}=1}^{N+1}
\Dipole_{ij}^{b}(p_a,p_b;p_1,\ldots, p_{N+1})
\measureF{N}(p_a,\tilde{p}_{b};p_1,\ldots,\tilde{p}_{ij},\ldots,p_{N+1})
\displaybreak[1]\\
+\sum_{i=1}^{N+1}
\sum_{\rlap{\def\arraystretch{.5}
\begin{array}[b]{r@{}l}
\scriptstyle k&\scriptstyle =1\\
\scriptstyle k&\scriptstyle\neq i
\end{array}}}^{N+1}
\Dipole_{k}^{ia}(p_a,p_b;p_1,\ldots, p_{N+1})
\measureF{N}(\tilde{p}_{ai},p_b;p_1,\ldots,\tilde{p}_k\ldots,p_{N+1})
\displaybreak[1]\\
+\sum_{i=1}^{N+1}
\sum_{\rlap{\def\arraystretch{.5}
\begin{array}[b]{r@{}l}
\scriptstyle k&\scriptstyle =1\\
\scriptstyle k&\scriptstyle\neq i
\end{array}}}^{N+1}
\Dipole_{k}^{ib}(p_a,p_b;p_1,\ldots, p_{N+1})
\measureF{N}(p_a,\tilde{p}_{bi};p_1,\ldots,\tilde{p}_k\ldots,p_{N+1})
\displaybreak[1]\\
+\sum_{i=1}^{N+1}
\Dipole^{ai,b}(p_a,p_b;p_1,\ldots, p_{N+1})
\measureF{N}(\tilde{p}_{ai},\tilde{p}_{b};p_1,\ldots,p_{N+1})
\\
+\left.
\sum_{i=1}^{N+1}
\Dipole^{bi,a}(p_a,p_b;p_1,\ldots, p_{N+1})
\measureF{N}(\tilde{p}_{a},\tilde{p}_{bi};p_1,\ldots,p_{N+1})
\right]
\end{multline}
Formally there
would be an additional sum over all $(N+1)$ particle configurations
that contribute to the process~\cite{Catani:1996vz};
instead, here it is assumed that every subprocess is treated as a 
separate calculation. The factor $1/S$ is the \person{Bose} symmetry
factor which has to be included if the additional parton changes
the multiplicity of identical particles; it is discussed in detail
in~\cite{Catani:1996vz}.

The summation $\sum_{\{i,j\}=1}^{N+1}$ is over pairs of final state
particles; for example the first sum in the case $N=2$ would read
\begin{multline}
\sum_{\{i,j\}=1}^{3}\sum_{\rlap{\def\arraystretch{.5}
\begin{array}[b]{r@{}l}
\scriptstyle k&\scriptstyle =1\\
\scriptstyle k&\scriptstyle\neq i,j
\end{array}}}^3\Dipole_{ij,k}(p_a,p_b;p_1,\ldots, p_{3})
\measureF{N}(p_a,p_b;p_1,\ldots,\tilde{p}_{ij},\tilde{p}_k\ldots,p_{3})
=\\
\Dipole_{12,3}(p_a,p_b;p_1,p_2,p_{3})
\measureF{2}(p_a,p_b;\tilde{p}_{12},\tilde{p}_{3})\\
+\Dipole_{13,2}(p_a,p_b;p_1,p_2,p_{3})
\measureF{2}(p_a,p_b;\tilde{p}_{13},\tilde{p}_{2})\\
+\Dipole_{23,1}(p_a,p_b;p_1,p_2,p_{3})
\measureF{2}(p_a,p_b;\tilde{p}_{1},\tilde{p}_{23})\\
\end{multline}

The index structure of the dipoles denotes incoming particles with
upper indices and final state particles with lower indices. The double
index denotes the pair of partons that can become collinear, i.e. the
\emph{emitter}, the third index stands for the \emph{spectator}.
The momentum mapping from the $(N+1)$ to the $N$ particle kinematics
obeys momentum conservation and depends on the configuration, i.e.
if the emitter and spectator are initial or final state partons.

Before the structure of the dipoles is discussed in detail for each
case in the following sections the discussion of the partonic
amplitude is continued. The last two terms in
Equation~\eqref{eq:qcd-dipoles:outline1} both have a $2\rightarrow N$
kinematics. The virtual corrections plus the integrated subtraction
terms are infrared finite in $n=4$ dimensions,
\begin{equation}
\sigma^{\text{NLO}\,\{N\}}_{ab}(p_a,p_b)=\int\!\!\phasespace{N}(Q)
\left[\frac{\diff{\virt\sigma}}{\phasespace{N}}
+\int\!\!\phasespace{1}\frac{\diff{\sigma^{\mathcal{A}}}}{%
\phasespace{N+1}}
\right]_{\varepsilon=0}\text{.}
\end{equation}

The virtual correction is the interference term between the one-loop
and the tree-level amplitude,
\begin{multline}
\frac{\diff{\virt\sigma}}{\phasespace{N}}=
\frac{1}{n_an_b}\sum_{\lambda_i}
\measureF{N}(p_1,\ldots,p_{N})\\
\times\bra{c^\prime}
{\virt{\mathcal{A}}_{c^\prime}}(p_a^{\lambda_a}, p_b^{\lambda_b};
	p_1^{\lambda_1},\ldots, p_{N}^{\lambda_{N}})^\ast
\born{\mathcal{A}}_{c}(p_a^{\lambda_a}, p_b^{\lambda_b};
	p_1^{\lambda_1},\ldots, p_{N}^{\lambda_{N}})
\ket{c} + \text{h.c.}
\end{multline}

The integrated subtraction term is a tensor product of an insertion
operator and the \person{Born} level amplitude,
\begin{multline}
\int\!\!\phasespace{1}\frac{\diff{\sigma^{\mathcal{A}}}}{%
\phasespace{N+1}}=
-\frac{\alpha_s}{2\pi}\frac{1}{\Gamma(1-\varepsilon)}
\frac{1}{n_an_b}
\measureF{N}(p_1,\ldots,p_{N})\\
\times
\sum_{\alpha\in\{a,b,1,\ldots N\}}\frac{1}{\mathbf{T}_\alpha^2}
\mathcal{V}_\alpha(\varepsilon)
\sum_{\rlap\rlap{\def\arraystretch{.5}
\begin{array}[b]{c}
\scriptstyle \beta\scriptstyle \in\{a,b,1,\ldots,N\}\\
\scriptstyle \beta\scriptstyle\neq \alpha
\end{array}}}
\!\!\!\bra{c^\prime}\mathbf{T}_\alpha\cdot\mathbf{T}_\beta\ket{c}
\left(\frac{4\pi\mu^2}{2p_\alpha\cdot p_\beta}\right)^\varepsilon\\
\times\sum_{\lambda_i}
{\born{\mathcal{A}}_{c^\prime}}(p_a^{\lambda_a}, p_b^{\lambda_b};
	p_1^{\lambda_1},\ldots, p_{N}^{\lambda_{N}})^\ast
\born{\mathcal{A}}_{c}(p_a^{\lambda_a}, p_b^{\lambda_b};
	p_1^{\lambda_1},\ldots, p_{N}^{\lambda_{N}})
\end{multline}

The colour operators $\mathbf{T}_\alpha$ are a $t_{\alpha^\prime\alpha}^A$
in the case of a quark, $-t_{\alpha^\prime\alpha}^A$ in the case of
an antiquark and $f^{A\alpha^\prime\alpha}$ in the case of a gluon and
hence $\mathbf{T}_\alpha^2$ is $C_F$ for quarks and antiquarks and $C_A$
for gluons. The singular function $\mathcal{V}_\alpha(\varepsilon)$ is
\begin{equation}
\mathcal{V}_\alpha(\varepsilon)=
\mathbf{T}_\alpha^2\left(\frac1{\varepsilon^2}-\frac{\pi^2}{3}\right)
+\gamma_\alpha\frac1{\varepsilon}+\gamma_\alpha+K_\alpha
+\mathcal{O}(\varepsilon)
\end{equation}
The constants $\gamma_\alpha$ and $K_\alpha$ depend on the flavour
of the parton and are
\begin{align}
\gamma_q=\gamma_{\bar{q}}=\frac32C_F,\qquad
&\gamma_g=\frac{11}6C_A-\frac23T_RN_f\text{,}\\
\intertext{and}
K_q=K_{\bar{q}}=\left(\frac72-\frac{\pi^2}{6}\right)C_F,\qquad
&K_g=\left(\frac{67}{18}-\frac{\pi^2}{6}\right)C_A-\frac{10}9T_RN_f\text{.}
\end{align}

The last piece of Equation~\eqref{eq:qcd-dipoles:outline1} contains the
renormalisation scale dependent part of the subtraction terms
and is free of singularities.
It can be written as a convolution
of the leading order matrix element with two colour-charge operators
$\mathbf{K}$ and $\mathbf{P}$
\begin{multline}
\sigma^{\text{C}\,\{N\}}_{ab}(p_a, p_b)=
\int_0^1\!\!\diff{x}
\int\!\!\phasespace{N}(xp_a+p_b)
\measureF{N}(xp_a, p_b;p_1, \ldots, p_N)\\
\times
\frac{1}{n_an_b}\sum_{\lambda_i}
\born{\mathcal{A}}_{c^\prime}(xp_a^{\lambda_a}, p_b^{\lambda_b};
	p_1^{\lambda_1},\ldots, p_N^{\lambda_N})^\ast
\born{\mathcal{A}}_{c}(xp_a^{\lambda_a}, p_b^{\lambda_b};
	p_1^{\lambda_1},\ldots, p_N^{\lambda_N})\\
\times\sum_{a^\prime}\bra{c^\prime}\left(\mathbf{K}^{a,a^\prime}(x)+
\mathbf{P}^{a,a^\prime}(xp_a,x;\mu_F^2\right)\ket{c}\\
\shoveleft+\int_0^1\!\!\diff{x}
\int\!\!\phasespace{N}(p_a+xp_b)
\measureF{N}(p_a, xp_b;p_1, \ldots, p_N)\\
\times
\frac{1}{n_an_b}\sum_{\lambda_i}
\born{\mathcal{A}}_{c^\prime}(p_a^{\lambda_a}, xp_b^{\lambda_b};
	p_1^{\lambda_1},\ldots, p_N^{\lambda_N})^\ast
\born{\mathcal{A}}_{c}(p_a^{\lambda_a}, xp_b^{\lambda_b};
	p_1^{\lambda_1},\ldots, p_N^{\lambda_N})\\
\times\sum_{b^\prime}\bra{c^\prime}\left(\mathbf{K}^{b,b^\prime}(x)+
\mathbf{P}^{b,b^\prime}(xp_b,x;\mu_F^2\right)\ket{c}
\text{.}
\end{multline}
The sums over $a^\prime$ and $b^\prime$ run over all possible splittings
$aa^\prime$ ($bb^\prime$ respectively) for the given initial state
parton $a$ ($b$ resp.). The two operators $\mathbf{K}^{a,a^\prime}$ and
$\mathbf{P}^{a,a^\prime}$ have the following structure
in~\ac{msbar}~\cite{Catani:1996vz}:
\begin{multline}
\mathbf{K}^{a,a^\prime}(x)=\frac{\alpha_s}{2\pi}
\left\{\overline{K}^{aa^\prime}(x)
+\delta^{aa^\prime}\sum_{i=1}^N \mathbf{T}_i\cdot\mathbf{T}_a\frac{\gamma_i}{%
\mathbf{T}_i^2}\left[\left(\frac{1}{1-x}\right)_++\delta(1-x)\right]
\right.\\\left.\nonumber
-\mathbf{T}_b\cdot\mathbf{T}_{a^\prime}\frac1{\mathbf{T}_{a^\prime}^2}
\tilde{K}^{aa^\prime}(x)
\right\}
\end{multline}
with the plus distribution as defined in
Equation~\eqref{eq:appendix-distributions:plusdef},
and
\begin{multline}\label{eq:qcd-dipoles:K-operator}
\mathbf{P}^{a,a^\prime}(xp_a, p_b, p_1,\ldots,p_N; x, \mu_F^2)=
\frac{\alpha_s}{2\pi}P^{aa^\prime}(x)\\
\times\frac1{\mathbf{T}_{a^\prime}^2}\left[
\sum_{i=1}^N\mathbf{T}_i\cdot\mathbf{T}_{a^\prime}\ln\frac{\mu_F^2}{2xp_ap_i}
+\mathbf{T}_b\cdot\mathbf{T}_{a^\prime}\ln\frac{\mu_F^2}{2xp_ap_b}
\right]\text{.}
\end{multline}
In the latter equation one has to insert the regularised
\person{Altarelli}-\person{Parisi} splitting functions
\begin{subequations}
\begin{align}
P^{qg}(x)&=P^{\bar{q}g}(x)=C_F\frac{1+(1-x)^2}{x}\text{,}\\
P^{gq}(x)&=P^{g\bar{q}}(x)=T_R\left(x^2+(1-x)^2\right)\text{,}\\
P^{qq}(x)&=P^{\bar{q}\bar{q}}(x)=T_F\left(\frac{1+x^2}{1-x}\right)_+
\end{align}
and
\begin{multline}
P^{gg}(x)=2C_A\left[\left(\frac{1}{1-x}\right)_++\frac{1-x}{x}-1+x(1-x)\right]
\\+\delta(1-x)\left[\frac{11}6C_A-\frac23N_fT_R\right]\text{.}
\end{multline}
\end{subequations}

Equation~\eqref{eq:qcd-dipoles:K-operator} requires the integration
kernels
\begin{subequations}
\begin{align}
\overline{K}^{qg}(x)&=\overline{K}^{\bar{q}g}(x)=
P^{qg}(x)\ln\frac{1-x}{x}
+C_F\,x\text{,}\\
\overline{K}^{gq}(x)&=\overline{K}^{g\bar{q}}(x)=
P^{gq}(x)\ln\frac{1-x}{x}
+T_R\cdot2x(1-x)\text{,}\\
\overline{K}^{qq}(x)&=\overline{K}^{\bar{q}\bar{q}}(x)=
C_F\left[2\left(\frac{\ln\left(\frac{1-x}{x}\right)}{1-x}\right)_+
-(1+x)\ln\frac{1-x}{x}+(1-x)\right]\\
\nonumber&\qquad-\delta(1-x)(5-\pi^2)C_F\text{,}\\
\overline{K}^{q\bar{q}}(x)&=\overline{K}^{\bar{q}q}(x)=0\quad\text{and}\\
\overline{K}^{gg}(x)&=
2C_A\left[
\left(\frac{\ln\left(\frac{1-x}{x}\right)}{1-x}\right)_+
+\left(\frac{1-x}{x}-1+x(1-x)\right)\ln\frac{1-x}{x}
\right]
\\&\nonumber\qquad-\delta(1-x)\left[\left(\frac{50}9-\pi^2\right)C_A-
\frac{16}9T_RN_f
\right]\text{.}
\end{align}
\end{subequations}

Finally the terms $\tilde{K}^{aa^\prime}$ are required,
\begin{subequations}
\begin{align}
\tilde{K}^{qg}(x)&=\tilde{K}^{\bar{q}g}(x)=P^{qg}(x)\ln(1-x)\text{,}\\
\tilde{K}^{gq}(x)&=\tilde{K}^{g\bar{q}}(x)=P^{gq}(x)\ln(1-x)\text{,}\\
\tilde{K}^{qq}(x)&=\tilde{K}^{\bar{q}\bar{q}}(x)=
-C_F(1+x)\ln(1-x)\\
&\nonumber\qquad+C_F\left[
2\left(\frac{\ln(1-x)}{1-x}\right)_+-\frac{\pi^2}3\delta(1-x)\right]\text{,}\\
\tilde{K}^{gg}&=2C_A\left[\frac{1-x}{x}-1+x(1-x)\right]\ln(1-x)\\
&\nonumber\qquad+C_A\left[
2\left(\frac{\ln(1-x)}{1-x}\right)_+-\frac{\pi^2}3\delta(1-x)\right]
\text{.}
\end{align}
\end{subequations}

%%%%%%%%%%%%%%%%%%%%%%%%%%%%%%%%%%%%%%%%%%%%%%%%%%%%%%%%%%%%%%%%%%%%%
\subsection{Final state emitter, final state spectator:
$\Dipole_{ij,k}$}\label{ssec:qcd-dipoles:FFdipole}
%%%%%%%%%%%%%%%%%%%%%%%%%%%%%%%%%%%%%%%%%%%%%%%%%%%%%%%%%%%%%%%%%%%%%
In the case where both emitter and spectator are final state
partons the following momentum mapping has to be used:
\begin{subequations}
\begin{align}
\tilde{p}_{ij}&=p_i+p_j-\frac{y_{ij,k}}{1-y_{ij,k}}p_k\text{,}\\
\tilde{p}_{k}&=\frac{1}{1-y_{ij,k}}p_k\text{,}\\
\intertext{with}
y_{ij,k}&=\frac{p_i p_j}{p_i p_j+(p_i+p_j) p_k}\quad\text{and}\\
\tilde{z}_i&=\frac{p_i p_k}{(p_i+p_j) p_k}\text{.}
\end{align}
\end{subequations}
All other momenta remain unchanged, $\tilde{p}_n=p_n$. Along
with $\tilde{z}_i$ one defines $\tilde{z}_j$ such that
$\tilde{z}_j=1-\tilde{z}_i$.

The dipole term is
\begin{multline}
\Dipole_{ij,k}(p_a, p_b; p_1,\ldots,p_{N+1})=
-\frac1{2p_i\cdot p_j}
\bra{c^\prime}\frac{\mathbf{T}_{k}\cdot\mathbf{T}_{(ij)}}{%
\mathbf{T}_{(ij)}^2}\ket{c}\\
\times\sum_{\lambda_i,\lambda_j^\prime}
{\born{\mathcal{A}}_{c^\prime}}(p_a^{\lambda_a}, p_b^{\lambda_b};
	\tilde{p}_1^{\lambda_1},\ldots, \tilde{p}_{N}^{\lambda_{N}})^\ast
\born{\mathcal{A}}_{c}(p_a^{\lambda_a}, p_b^{\lambda_b};
	\tilde{p}_1^{\lambda_1^\prime},\ldots,
	\tilde{p}_{N}^{\lambda_{N}^\prime})V_{ij,k}^{\lambda,\lambda^\prime}
\text{,}
\end{multline}
and the operator for the three different splittings are
\begin{align}
\left.V_{ij,k}\right\vert_{i=q}^{j=g}&=8\pi\mu^{2\varepsilon}\alpha_s
C_F\left[\frac2{1-\tilde{z}_i(1-y_{ij,k})}
-(1+\tilde{z}_i)
-\varepsilon(1-\tilde{z}_i)
\right]\delta^{\lambda_{(ij)}\lambda_{(ij)}^\prime}\\
\left.V_{ij,k}\right\vert_{i=q}^{j=\bar{q}}&=8\pi\mu^{2\varepsilon}\alpha_s
T_R\left[
-g^{\mu\nu}-\frac2{p_i\cdot p_j}
(\tilde{z}_ip_i^\mu-\tilde{z}_jp_j^\mu)
(\tilde{z}_ip_i^\nu-\tilde{z}_jp_j^\nu)
\right]
\end{align}
and
\begin{multline}
\left.V_{ij,k}\right\vert_{i=g}^{j=g}=16\pi\mu^{2\varepsilon}\alpha_s
C_A
\left[
-g^{\mu\nu}\left(\frac{1}{1-\tilde{z}_i(1-y_{ij,k})}
+\frac{1}{1-\tilde{z}_j(1-y_{ij,k})}-2\right)\right.\\
\left.+(1-\varepsilon)\frac1{p_i\cdot p_j}
(\tilde{z}_ip_i^\mu-\tilde{z}_jp_j^\mu)
(\tilde{z}_ip_i^\nu-\tilde{z}_jp_j^\nu)
\right]\text{.}
\end{multline}

Here, $\mu$ and $\nu$ denote the spin indices of the gluon $(ij)$
in $\mathcal{A}$ and $\mathcal{A}^\ast$ respectively. The operators
are orthogonal in all other spin indices, i.e. include implicit
factors $\delta^{\lambda_n\lambda_n^\prime}$ for $n\neq(ij)$.

%%%%%%%%%%%%%%%%%%%%%%%%%%%%%%%%%%%%%%%%%%%%%%%%%%%%%%%%%%%%%%%%%%%%%
\subsection{Final state emitter, initial state spectator:
$\Dipole_{ij}^a$}
%%%%%%%%%%%%%%%%%%%%%%%%%%%%%%%%%%%%%%%%%%%%%%%%%%%%%%%%%%%%%%%%%%%%%
The second case one has to consider is when the collinear singularity
stems from final state partons but the spectator is in the initial
state. The appropriate momentum mapping is given below:
\begin{subequations}
\begin{align}
\tilde{p}_{ij}&=p_i+p_j-(1-x_{ij,a})p_a\text{,}\\
\tilde{p}_{a}&=x_{ij,a}p_a\text{,}\\
\intertext{with}
x_{ij,a}&=\frac{(p_i+p_j)p_a-p_ip_j}{(p_i+p_j) p_a}\quad\text{and}\\
\tilde{z}_i&=\frac{p_i p_a}{(p_i+p_j) p_a}\text{.}
\end{align}
\end{subequations}
All other momenta remain unchanged, $\tilde{p}_n=p_n$. Along
with $\tilde{z}_i$ one defines $\tilde{z}_j$ such that
$\tilde{z}_j=1-\tilde{z}_i$.

The dipole term is
\begin{multline}
\Dipole_{ij}^a(p_a, p_b; p_1,\ldots,p_{N+1})=
-\frac1{2p_i\cdot p_j}\frac1{x_{ij,a}}
\bra{c^\prime}\frac{\mathbf{T}_{a}\cdot\mathbf{T}_{(ij)}}{%
\mathbf{T}_{(ij)}^2}\ket{c}\\
\times\sum_{\lambda_i,\lambda_j^\prime}
{\born{\mathcal{A}}_{c^\prime}}(\tilde{p}_a^{\lambda_a}, p_b^{\lambda_b};
	\tilde{p}_1^{\lambda_1},\ldots, \tilde{p}_{N}^{\lambda_{N}})^\ast
\born{\mathcal{A}}_{c}(\tilde{p}_a^{\lambda_a}, p_b^{\lambda_b};
	\tilde{p}_1^{\lambda_1^\prime},\ldots,
	\tilde{p}_{N}^{\lambda_{N}^\prime})V_{ij}^{a\,\lambda,\lambda^\prime}
\text{,}
\end{multline}
and one obtains the operator
\begin{align}
\left.V_{ij}^a\right\vert_{i=q}^{j=g}&=8\pi\mu^{2\varepsilon}\alpha_s
C_F\left[
\frac2{1-\tilde{z}_i+(1-x_{ij,a})}
-(1+\tilde{z}_i)
-\varepsilon(1-\tilde{z}_i)
\right]\delta^{\lambda_{(ij)}\lambda_{(ij)}^\prime}\\
\intertext{for the quark-gluon splitting,}
\left.V_{ij}^a\right\vert_{i=q}^{j=\bar{q}}&=8\pi\mu^{2\varepsilon}\alpha_s
T_R\left[
-g^{\mu\nu}
-\frac2{p_i\cdot p_j}
(\tilde{z}_ip_i^\mu-\tilde{z}_jp_j^\mu)
(\tilde{z}_ip_i^\nu-\tilde{z}_jp_j^\nu)
\right]
\end{align}
in the case of gluon-quark splitting and
\begin{multline}
\left.V_{ij}^a\right\vert_{i=g}^{j=g}=16\pi\mu^{2\varepsilon}\alpha_s
C_A
\left[
-g^{\mu\nu}\left(\frac{1}{1-\tilde{z}_i(1-y_{ij,k})}
+\frac{1}{1-\tilde{z}_j(1-y_{ij,k})}-2\right)\right.\\
\left.+(1-\varepsilon)\frac1{p_i\cdot p_j}
(\tilde{z}_ip_i^\mu-\tilde{z}_jp_j^\mu)
(\tilde{z}_ip_i^\nu-\tilde{z}_jp_j^\nu)
\right]
\end{multline}
for gluon-gluon splitting.

The spin correlations and the indices $\mu$ and $\nu$ are the same
as in Section~\ref{ssec:qcd-dipoles:FFdipole}.

%%%%%%%%%%%%%%%%%%%%%%%%%%%%%%%%%%%%%%%%%%%%%%%%%%%%%%%%%%%%%%%%%%%%%
\subsection{Initial state emitter, final state spectator:
$\Dipole^{ai}_k$}
%%%%%%%%%%%%%%%%%%%%%%%%%%%%%%%%%%%%%%%%%%%%%%%%%%%%%%%%%%%%%%%%%%%%%
The reverse case of the previous one is the situation where the
singularity is in the initial state but the spectator is a final
state parton. Here, one uses the momentum mapping
\begin{subequations}
\begin{align}
\tilde{p}_{ai}&=x_{ik,a}p_a\text{,}\\
\tilde{p}_{k}&=p_k+p_i-(1-x_{ik,a})p_a\text{,}\\
\intertext{with}
x_{ik,a}&=\frac{(p_i+p_k)p_a-p_ip_k}{(p_i+p_k) p_a}\quad\text{and}\\
u_i&=\frac{p_i p_a}{(p_i+p_k) p_a}\text{.}
\end{align}
\end{subequations}
All other momenta remain unchanged, $\tilde{p}_n=p_n$. 

The dipole term is
\begin{multline}
\Dipole^{ai}_k(p_a, p_b; p_1,\ldots,p_{N+1})=
-\frac1{2p_a\cdot p_i}\frac1{x_{ik,a}}
\bra{c^\prime}\frac{\mathbf{T}_{k}\cdot\mathbf{T}_{(ai)}}{%
\mathbf{T}_{(ai)}^2}\ket{c}\\
\times\sum_{\lambda_i,\lambda_j^\prime}
{\born{\mathcal{A}}_{c^\prime}}(\tilde{p}_{ai}^{\lambda_a}, p_b^{\lambda_b};
	\tilde{p}_1^{\lambda_1},\ldots, \tilde{p}_{N}^{\lambda_{N}})^\ast
\born{\mathcal{A}}_{c}(\tilde{p}_{ai}^{\lambda_a}, p_b^{\lambda_b};
	\tilde{p}_1^{\lambda_1^\prime},\ldots,
	\tilde{p}_{N}^{\lambda_{N}^\prime})V_{k}^{ai\,\lambda,\lambda^\prime}
\text{,}
\end{multline}
and one obtains the operator $V^{ai}_k$ for the three
different splittings,
\begin{align}
\left.V^{ai}_k\right\vert_{a=q}^{i=g}&=8\pi\mu^{2\varepsilon}\alpha_s
C_F\left[
\frac2{1-x_{ik,a}+u_i}
-(1+\tilde{x}_{ik,a})
-\varepsilon(1-\tilde{x}_{ik,a})
\right]\delta^{\lambda_{(ai)}\lambda_{(ai)}^\prime}\text{,}\\
%%%%%%%%
\left.V^{ai}_k\right\vert_{a=g}^{i=\bar{q}}&=8\pi\mu^{2\varepsilon}\alpha_s
T_R\left[
(1-\varepsilon)-2x_{ik,a}(1-x_{ik,a})
\right]\delta^{\lambda_{(ai)}\lambda_{(ai)}^\prime}\text{,}
\end{align}
\begin{multline}
\left.V^{ai}_k\right\vert_{a=q}^{i=q}=8\pi\mu^{2\varepsilon}\alpha_s
C_F\\\times\left[
-g^{\mu\nu}x_{ik,a}
+\frac{1-x_{ik,a}}{x_{ik,a}}\frac{2u_i(1-u_i)}{p_i\cdot p_k}
\left(\frac{p_i^\mu}{u_i}-\frac{p_k^\mu}{1-u_i}\right)
\left(\frac{p_i^\nu}{u_i}-\frac{p_k^\nu}{1-u_i}\right)
\right]
\end{multline}
and
\begin{multline}
\left.V^{ai}_k\right\vert_{a=g}^{i=g}=16\pi\mu^{2\varepsilon}\alpha_s
C_A \left[
-g^{\mu\nu}
	\left(\frac{1}{1-x_{ik,a}+u_i}-1+x_{ik,a}(1-x_{ik,a})\right)
\right.\\\left.
+(1-\varepsilon)
\frac{1-x_{ik,a}}{x_{ik,a}}\frac{u_i(1-u_i)}{p_i\cdot p_k}
\left(\frac{p_i^\mu}{u_i}-\frac{p_k^\mu}{1-u_i}\right)
\left(\frac{p_i^\nu}{u_i}-\frac{p_k^\nu}{1-u_i}\right)
\right]\text{.}
\end{multline}

The spin correlations and the indices $\mu$ and $\nu$ are the same
as in Section~\ref{ssec:qcd-dipoles:FFdipole}.

%%%%%%%%%%%%%%%%%%%%%%%%%%%%%%%%%%%%%%%%%%%%%%%%%%%%%%%%%%%%%%%%%%%%%
\subsection{Initial state emitter, initial state spectator:
$\Dipole^{ai,b}$}
%%%%%%%%%%%%%%%%%%%%%%%%%%%%%%%%%%%%%%%%%%%%%%%%%%%%%%%%%%%%%%%%%%%%%
The last case to be considered for our purpose is the situation
with an initial state singularity and an initial state spectator.
In this case the momentum mapping involves all final state
particles, not only the \ac{qcd} partons; it is
\begin{subequations}
\begin{align}
\tilde{p}_{ai}&=x_{i,ab}p_a\text{,}\\
\tilde{p}_{b}&=p_b\text{,}\\
\tilde{p}_j&=p_j-\frac{2p_j\cdot(K+\tilde{K})}{(K+\tilde{K})^2}(K+\tilde{K})
	+\frac{2p_jK}{K^2}\tilde{K}\\
\intertext{with}
x_{i,ab}&=\frac{p_ap_b-p_i(p_a+p_b)}{p_ap_b}\text{,}\\
K&=p_a+p_b-p_i\quad\text{and}\\
\tilde{K}&=\tilde{p}_{ai}+p_b\text{.}
\end{align}
\end{subequations}

The dipole for this case is
\begin{multline}
\Dipole^{ai,b}(p_a, p_b; p_1,\ldots,p_{N+1})=
-\frac1{2p_a\cdot p_i}\frac1{x_{i,ab}}
\bra{c^\prime}\frac{\mathbf{T}_{b}\cdot\mathbf{T}_{(ai)}}{%
\mathbf{T}_{(ai)}^2}\ket{c}\\
\times\sum_{\lambda_i,\lambda_j^\prime}
{\born{\mathcal{A}}_{c^\prime}}(\tilde{p}_{ai}^{\lambda_a}, p_b^{\lambda_b};
	\tilde{p}_1^{\lambda_1},\ldots, \tilde{p}_{N}^{\lambda_{N}})^\ast
\born{\mathcal{A}}_{c}(\tilde{p}_{ai}^{\lambda_a}, p_b^{\lambda_b};
	\tilde{p}_1^{\lambda_1^\prime},\ldots,
	\tilde{p}_{N}^{\lambda_{N}^\prime})V^{ai,b\,\lambda,\lambda^\prime}
\end{multline}

The splitting operators are
\begin{align}
\left.V^{ai,b}\right\vert_{a=q}^{i=g}&=8\pi\mu^{2\varepsilon}\alpha_s
C_F\left[
\frac2{1-x_{i,ab}}
-(1+x_{i,ab})
-\varepsilon(1-x_{i,ab})
\right]\delta^{\lambda_{(ai)}\lambda_{(ai)}^\prime}\text{,}\\
%%%%%%%%
\left.V^{ai,b}\right\vert_{a=g}^{i=\bar{q}}&=8\pi\mu^{2\varepsilon}\alpha_s
T_R\left[
(1-\varepsilon)-2x_{i,ab}(1-x_{i,ab})
\right]\delta^{\lambda_{(ai)}\lambda_{(ai)}^\prime}\text{,}
\end{align}
and for the gluon induced cases
\begin{multline}
\left.V^{ai,b}\right\vert_{a=q}^{i=q}=8\pi\mu^{2\varepsilon}\alpha_s
C_F\\\times\left[
-g^{\mu\nu}x_{i,ab}
+\frac{1-x_{i,ab}}{x_{i,ab}}\frac{2p_a\cdot p_b}{(p_i\cdot p_a)(p_i\cdot p_b)}
\left(p_i^\mu-\frac{p_ip_a}{p_ap_b}p_b^\mu\right)
\left(p_i^\nu-\frac{p_ip_a}{p_ap_b}p_b^\nu\right)
\right]
\end{multline}
and
\begin{multline}
\left.V^{ai,b}\right\vert_{a=g}^{i=g}=16\pi\mu^{2\varepsilon}\alpha_s
C_A
\left[
-g^{\mu\nu}
	\left(\frac{x_{i,ab}}{1-x_{i,ab}}+x_{i,ab}(1-x_{i,ab})\right)
\right.\\\left.
+(1-\varepsilon)
\frac{1-x_{i,ab}}{x_{i,ab}}\frac{p_a\cdot p_b}{(p_i\cdot p_a)(p_i\cdot p_b)}
\left(p_i^\mu-\frac{p_ip_a}{p_ap_b}p_b^\mu\right)
\left(p_i^\nu-\frac{p_ip_a}{p_ap_b}p_b^\nu\right)
\right]\text{.}
\end{multline}

The spin correlations and the indices $\mu$ and $\nu$ are the same
as in Section~\ref{ssec:qcd-dipoles:FFdipole}.

%%%%%%%%%%%%%%%%%%%%%%%%%%%%%%%%%%%%%%%%%%%%%%%%%%%%%%%%%%%%%%%%%%%%%%%%%%%%%%
\subsection{Counterterms for the Six-Quark Amplitude}
%%%%%%%%%%%%%%%%%%%%%%%%%%%%%%%%%%%%%%%%%%%%%%%%%%%%%%%%%%%%%%%%%%%%%%%%%%%%%%
For the amplitude
$u(p_a)+\bar{u}(p_b)\rightarrow d(p_1)+\bar{d}(p_2) +b(p_3)+\bar{b}(p_4)$
a convenient colour basis is expressed in terms of \person{Kronecker} deltas,
because the colour structure of any diagram can be reduced to this basis
using Equation~\eqref{eq:qcd-color:reduce}. Explicitly, one basis choice is
\begin{equation}
\begin{array}{rcl@{\qquad}rcl@{\qquad}rcl}
\ket{1}&=&\delta_{i_a}^{j_b}\delta_{i_2}^{j_1}\delta_{i_4}^{j_3}&
\ket{2}&=&\delta_{i_a}^{j_b}\delta_{i_4}^{j_1}\delta_{i_2}^{j_3}&
\ket{3}&=&\delta_{i_2}^{j_b}\delta_{i_4}^{j_1}\delta_{i_a}^{j_3}\\
\ket{4}&=&\delta_{i_4}^{j_b}\delta_{i_2}^{j_1}\delta_{i_a}^{j_3}&
\ket{5}&=&\delta_{i_4}^{j_b}\delta_{i_a}^{j_1}\delta_{i_2}^{j_3}&
\ket{6}&=&\delta_{i_2}^{j_b}\delta_{i_a}^{j_1}\delta_{i_4}^{j_3}
\end{array}
\end{equation}
In this basis the colour correlation matrix has the following form
\begin{equation}
\braket{c\vert c^\prime}=\begin{pmatrix}
      N_C^3&N_C^2&N_C&N_C^2&N_C&N_C^2\cr
      N_C^2&N_C^3&N_C^2&N_C&N_C^2&N_C\cr
      N_C&N_C^2&N_C^3&N_C^2&N_C&N_C^2\cr
      N_C^2&N_C&N_C^2&N_C^3&N_C^2&N_C\cr
      N_C&N_C^2&N_C&N_C^2&N_C^3&N_C^2\cr
      N_C^2&N_C&N_C^2&N_C&N_C^2&N_C^3
\end{pmatrix}
\end{equation}

Similarly, one obtains the colour correlations for the
dipoles; as an example $\mathbf{T}_{(ai)}\cdot\mathbf{T}_3$ has been
chosen\footnote{Note that the extra minus comes from the convention
for an initial state quark $\mathbf{T}_{ai}=-t^A_{i_aj_a}$},
\begin{equation}
\bra{c}\frac{\mathbf{T}_{(ai)}\cd\mathbf{T}_3}{\mathbf{T}_{(ai)}^2}
\ket{c^\prime}=
\frac1{C_F}\bra{c}(-t^{A}_{c_ac^\prime_a})t^{A}_{c_3c^\prime_3}\ket{c^\prime}=
-\frac{T_R}{C_F}\left(
\bra{c}\delta_{c_a}^{c_3}\delta^{c_a^\prime}_{c_3^\prime}\ket{c^\prime}
-\frac{\braket{c\vert c^\prime}}{N_C}\right)
\end{equation}
where Equation~\eqref{eq:qcd-color:reduce} has been used to reduce the
product of generators; the explicit matrix in this case is
\begin{equation}
\bra{c}\frac{\mathbf{T}_{(ai)}\cdot\mathbf{T}_3}{\mathbf{T}_{(ai)}^2}
\ket{c^\prime}=
-\begin{pmatrix}
	   0&    0& N_C    &   N_C^2&   N_C&   0\cr
	   0&    0& N_C^2  &     N_C&    0&  N_C\cr
	  N_C& N_C^2& N_C^3&   N_C^2&   N_C&N_C^2\cr
	N_C^2&   N_C&   N_C^2& N_C^3& N_C^2&  N_C\cr
	  N_C&    0&     N_C&   N_C^2&    0&   0\cr
	   0&   N_C&   N_C^2&     N_C&    0&   0
\end{pmatrix}
\end{equation}

For the real emission one has to consider the process with an additional
gluon in the final state,
$u(p_a)+\bar{u}(p_b)
\rightarrow d(p_1)+\bar{d}(p_2) +b(p_3)+\bar{b}(p_4)+g(p_5)$.
The only dipoles that can be produced with the \person{Born} level
matrix element are those where a (anti-)quark splits into a gluon
plus (anti-)quark. The subtraction term therefore is as follows:
\begin{equation}
\frac{\diff{\sigma^{\mathcal{A}}}}{\diff{\phasespace{5}}}=
\frac{1}{2N_C}\frac{1}{2N_C}\sum_{\{\Dipole\}}
\measureF{4}(\{\tilde{p}\})\Dipole
\end{equation}
where the sum over all dipole runs over
\begin{multline}
\{\Dipole\}=\{\Dipole_{15,2},\Dipole_{15,3},\Dipole_{15,4},
\Dipole_{15}^a,\Dipole_{15}^b,
\Dipole_{25,1},\Dipole_{25,3},\Dipole_{25,4},
\Dipole_{25}^a,\Dipole_{25}^b,\\
\Dipole_{35,1},\Dipole_{35,2},\Dipole_{35,4},
\Dipole_{35}^a,\Dipole_{35}^b,
\Dipole_{45,1},\Dipole_{45,2},\Dipole_{45,3},
\Dipole_{45}^a,\Dipole_{45}^b,\\
\Dipole^{a5}_1,\Dipole^{a5}_2,\Dipole^{a5}_3,\Dipole^{a5}_4,
\Dipole^{b5}_1,\Dipole^{b5}_2,\Dipole^{b5}_3,\Dipole^{b5}_4,
\Dipole^{a5,b},\Dipole^{b5,a}
\}\text{.}
\end{multline}

The term $\sigma^{\text{C}\,\{N\}}_{ab}$ for this amplitude is
\begin{multline}
\sigma^{\text{C}\,\{N\}}_{ab}(p_a, p_b)=
\int_0^1\!\!\diff{x}
\int\!\!\phasespace{4}(xp_a+p_b)
\measureF{4}(xp_a, p_b;p_1, \ldots, p_4)\\
\times
\frac{1}{(2N_C)^2}\sum_{\lambda_i}
\born{\mathcal{A}}_{c^\prime}(xp_a^{\lambda_a}, p_b^{\lambda_b};
	p_1^{\lambda_1},\ldots, p_4^{\lambda_4})^\ast
\born{\mathcal{A}}_{c}(xp_a^{\lambda_a}, p_b^{\lambda_b};
	p_1^{\lambda_1},\ldots, p_4^{\lambda_4})\\
\times\bra{c^\prime}\left(\mathbf{K}^{q,q}(x)+
\mathbf{P}^{q,q}(xp_a,x;\mu_F^2\right)\ket{c}\\
\shoveleft+\int_0^1\!\!\diff{x}
\int\!\!\phasespace{4}(p_a+xp_b)
\measureF{4}(p_a, xp_b;p_1, \ldots, p_4)\\
\times
\frac{1}{n_an_b}\sum_{\lambda_i}
\born{\mathcal{A}}_{c^\prime}(p_a^{\lambda_a}, xp_b^{\lambda_b};
	p_1^{\lambda_1},\ldots, p_4^{\lambda_4})^\ast
\born{\mathcal{A}}_{c}(p_a^{\lambda_a}, xp_b^{\lambda_b};
	p_1^{\lambda_1},\ldots, p_4^{\lambda_4})\\
\times\bra{c^\prime}\left(\mathbf{K}^{\bar{q},\bar{q}}(x)+
\mathbf{P}^{\bar{q},\bar{q}}(xp_b,x;\mu_F^2\right)\ket{c}
\text{.}
\end{multline}

%%%%%%%%%%%%%%%%%%%%%%%%%%%%%%%%%%%%%%%%%%%%%%%%%%%%%%%%%%%%%%%%%%%%%%%%%%%%%%
\subsection{Counterterms for the Two-Gluon plus Four-Quark Amplitude}
%%%%%%%%%%%%%%%%%%%%%%%%%%%%%%%%%%%%%%%%%%%%%%%%%%%%%%%%%%%%%%%%%%%%%%%%%%%%%%
The process $g(p_a)+g(p_b)\rightarrow d(p_1)+\bar{d}(p_2)+b(p_3)+\bar{b}(p_4)$
can be projected onto the colour basis
\begin{equation}
\begin{array}{rcl@{\qquad}rcl@{\qquad}rcl}
\ket{1}&=&t_{i_2j_1}^At_{i_4j_3}^B&
\ket{2}&=&t_{i_2j_1}^Bt_{i_4j_3}^A&
\ket{3}&=&t_{i_4j_1}^At_{i_2j_3}^B\\
\ket{4}&=&t_{i_4j_1}^Bt_{i_2j_3}^A&
\ket{5}&=&t_{i_2j^\prime}^At_{j^\prime j_1}^B\delta_{i_4}^{j_3}&
\ket{6}&=&t_{i_2j^\prime}^Bt_{j^\prime j_1}^A\delta_{i_4}^{j_3}\\
\ket{7}&=&t_{i_4j^\prime}^At_{j^\prime j_1}^B\delta_{i_2}^{j_3}&
\ket{8}&=&t_{i_4j^\prime}^Bt_{j^\prime j_1}^A\delta_{i_2}^{j_3}&
\ket{9}&=&t_{i_2j^\prime}^At_{j^\prime j_3}^B\delta_{i_4}^{j_1}\\
\ket{10}&=&t_{i_2j^\prime}^Bt_{j^\prime j_3}^A\delta_{i_4}^{j_1}&
\ket{11}&=&t_{i_4j^\prime}^At_{j^\prime j_3}^B\delta_{i_2}^{j_1}&
\ket{12}&=&t_{i_4j^\prime}^Bt_{j^\prime j_3}^A\delta_{i_2}^{j_1}\\
\ket{13}&=&t_{i^\prime j^\prime}^At_{j^\prime i^\prime}^B\delta_{i_2}^{j_1}\delta_{i_4}^{j_3}&
\ket{14}&=&t_{i^\prime j^\prime}^At_{j^\prime i^\prime}^B\delta_{i_4}^{j_1}\delta_{i_2}^{j_3}
\end{array}
\end{equation}
where for $N_C=3$ one vector can be eliminated by the relation
\begin{equation}
  \ket{14} - \ket{13} + \ket{12} + \ket{11} - \ket{10} 
- \ket{9} - \ket{8} - \ket{7} + \ket{6} + \ket{5}
- \ket{4} - \ket{3} + \ket{2} + \ket{1}=0
\end{equation}
which reflects the fact that an antisymmetrisation over more
than $N_C$  fundamental indices of an \sun-tensor is zero.

The counterterms are as in the previous section, with the
only difference that the flavours $a=g$, $b=g$ have to be
replaced in all formul\ae.

\subsection{Outlook and Improvements}
Implementations of the dipole subtraction show that the time which is
spent on the computation of the subtraction terms is comparable to the
computational cost of the real emission matrix element itself.
This drawback has been improved by \person{Nagy}~\cite{Nagy:2003tz} in the
following way: the subtraction term $\sigma^{\mathcal{A}}$ is only relevant
at the border of the phase-space, where collinear and soft divergences can
appear. A new parameter $0<\alpha\leq1$ separates the possibly divergent
region from the unproblematic bulk of the phase space; $\alpha=1$ corresponds
to the full dipole subtraction as presented in~\cite{Catani:1996vz}.
The dipoles are restricted to the critical phase space by the modification
\begin{subequations}
\begin{align}
\Dipole_{ij,k}&\rightarrow\Dipole_{ij,k}\cdot\Theta(y_{ij,k}<\alpha)\text{,}\\
\Dipole^{ai}_{k}&\rightarrow\Dipole^{ai}_{k}\cdot\Theta(u_i<\alpha)\text{,}\\
\Dipole_{ij}^a&\rightarrow\Dipole_{ij}^a\cdot\Theta(1-x_{ij,a}<\alpha)\text{,}\\
\Dipole^{ai,b}&\rightarrow\Dipole^{ai,b}\cdot\Theta(\tilde{v}_i<\alpha)\text{.}
\end{align}
\end{subequations}
A new parameter has been introduced for the dipole $\Dipole^{ai,b}$,
$\tilde{v}_i=p_a\cdot p_i/p_a\cdot p_b$. The modification of the dipoles
induces a change also in the integrated dipoles. Here only the constant
$K_i$ has to be replaced by
\begin{equation}
K_i\rightarrow K_i(\alpha)=K_i-\mathbf{T}_i^2\ln^2(\alpha)+
\gamma_i\cdot(\alpha-1-\ln\alpha)\text{.}
\end{equation}
Further modifications have to be taken into account for the $\mathbf{K}$
and $\mathbf{P}$ operators; the reader is referred to Equations~(13)--(17)
in the original work~\cite{Nagy:2003tz}.

%% file: qcd-psintegral.tex
\subsection{Introduction}
So far the discussion was focused on the computation
of matrix elements. However, for a complete calculation
one also needs to perform the phase-space integral
for the final state particles. The $N$-particle
phase-space is defined as~\cite{Weinzierl:2000wd}
\begin{multline}
\phasespace{N}(Q;p_1,\ldots,p_N)=\prod_{j=1}^N\frac{\diff[4]{p_j}}{(2\pi)^3}\Theta(p^0_j)
\delta({p_j}^2-m_j^2)
\cdot(2\pi)^4\delta^{(4)}\left(Q-\sum_{i=1}^Np_i\right)=
\\
\prod_{j=1}^N\frac{\diff[3]{\vec{p}_j}}{(2\pi)^3\cdot 2\omega_j}
\cdot(2\pi)^4\delta^{(4)}\left(Q-\sum_{i=1}^Np_i\right)
\end{multline}
where $\omega_j=\sqrt{{p_j}^2-m_j^2}$. This phase space is $(3N-4)$
dimensional; for non-trivial physical processes these integrals
are difficult to be done analytically
and can classical adaptive methods cannot solve the
integral in an efficient manner. Furthermore, the constraints
on the phase-space imposed by experimental cuts render a
analytic treatment of the integrals impossible.
Instead, Monte-Carlo methods are
used~\cite{Hammersley:1965,Weinzierl:2000wd}.
For classical methods,
e.g. the \person{Gauss}ian quadrature the error bound drops like
$\mathcal{O}(n^{-2/d})$ where $n$ denotes the number of evaluations
of the integrand. In Monte Carlo methods the behaviour of the
error~$\mathcal{O}(1/\sqrt{n})$ is constant in the number of
dimensions and therefore outperforms the classical methods for
higher-dimensional integrals.

\subsection{Monte-Carlo Integration}
The basic idea behind Monte Carlo integration is to evaluate the
integrand function $f$ in $n$ randomly chosen points $\vec{x}^{(j)}$ to obtain
an estimator for the exact integral
\begin{equation}\label{eq:qcd-psintegral:def01}
I=\int_0^1\!\!\diff[d]{\vec{x}}f(\vec{x})=
\int_0^1\!\!\diff{x_1}\cdots\int_0^1\!\!\diff{x_d}f(x_1,\ldots,x_d)
\approx E=\frac{1}{n}\sum_{j=1}^nf(x^{(j)}_1,\ldots, x_d^{(j)})\text{.}
\end{equation}
The $\vec{x}^{(j)}$ are chosen according to a uniform distribution
in the hypercube $0\leq x^{(j)}_i\leq1$. The restriction to the
interval~$[0;1]$ can always be overcome by a variable transformation.

The statistical error can
be obtained by integrating out all random variables,
\begin{equation}
\int_0^1\!\!\diff[d]{\vec{x}^{(1)}}
\int_0^1\!\!\diff[d]{\vec{x}^{(2)}}
\cdots\int_0^1\!\!\diff[d]{\vec{x}^{(n)}}
\left(\frac{1}{n}\sum_{j=1}^nf(\vec{x}^{(j)})-I
\right)^2=\frac{\sigma^2(f)}{n}\text{,}
\end{equation}
where the variance of the the function $f$ has been introduced,
\begin{equation}
\sigma^2(f)=\int_0^1\!\!\diff[d]{\vec{x}}\left(f(\vec{x})-I\right)^2%
\text{.}
\end{equation}
From the central limit theorem one can conclude that the estimate
of the integral lies in the interval~\cite{Weinzierl:2000wd}
\begin{equation}
\lim_{n\rightarrow\infty}\mathrm{Prob}\left(-a\frac{\sigma(f)}{\sqrt{n}}
\leq\frac{1}{n}\sum_{j=1}^nf(\vec{x}^{(j)})-I\leq
b\frac{\sigma(f)}{\sqrt{n}}\right)=
\frac{1}{\sqrt{2\pi}}\int_{-a}^b\!\!\diff{t}\,e^{-\frac{t^2}{2}}\text{.}
\end{equation}
The exact knowledge of $\sigma(f)$ would render a Monte Carlo integration
unnecessary and is usually not available. Therefore, in practical applications
one estimates the error by Monte Carlo techniques, too,
\begin{equation}\label{eq:qcd-psintegral:mcerror}
\sigma^2(f)\approx S\equiv\frac{1}{n-1}\sum_{j=1}^n\left(f(\vec{x}^{(j)})-E\right)^2%
\text{.}
\end{equation}
Here, $E$ is the estimator as defined in Equation~\eqref{eq:qcd-psintegral:def01}.

In the following two methods are discussed which can be used to
reduce the variance $\sigma^2(f)$ and therefore for a reduction
of the error, \emph{stratified sampling} and \emph{importance
sampling}. A combination of these two ideas leads to an adaptive
algorithm for Monte Carlo integration, \emph{VEGAS}, which closes
the discussion of Monte Carlo techniques.
In Section~\ref{ssec:qcd-psintegral:nlo} I introduce a method
that uses importance sampling for the integration of 
the virtual contribution of \ac{nlo} cross-sections. This
method has been successfully applied in the calculation of
the process~$u\bar{u}\rightarrow b\bar{b}b\bar{b}$, results
of which are presented in Chapter~\ref{chp:results}.

\emph{Stratified sampling}\index{Stratified sampling|main}
uses the simple fact that the integration
region can be split into disjoint subsets
\begin{equation}
[0;1]^d=\biguplus_{\nu=1}^k M_\nu\text{.}
\end{equation}
The volume of each subset is
\begin{equation}
\mathrm{vol}(M_\nu)=\int_0^1\!\!\diff[d]{\vec{x}}\,%
\Theta\left(\vec{x}\in M_\nu\right)
\end{equation}
and therefore $\sum_{\nu=1}^k\mathrm{vol}(M_\nu)=1$.
If in each region the Monte Carlo estimator $E=\sum_{\nu=1}^kE_\nu$ is evaluated
with $N_\nu$ points, $\sum_{\nu=1}^k N_\nu=N$,
\begin{equation}
E_\nu=\frac{\mathrm{vol}(M_\nu)}{N_\nu}\sum_{j=1}^{N_\nu}f(\vec{x}^{(\nu;j)}),
\quad \vec{x}^{(\nu;j)}\in M_\nu
\end{equation}
the new error estimate is
\begin{equation}
\sum_{\nu=1}^k\mathrm{vol}(M_\nu)^2\frac{\sigma^2(f)\vert_{M_\nu}}{N_\nu}
\end{equation}
with the variance restricted on the subspace $M_\nu$
\begin{align}
\sigma^2(f)\vert_{M_\nu}&=
\frac{1}{\mathrm{vol}(M_\nu)}\int_0^1\diff[d]{\vec{x}}\left(
f(\vec{x})\Theta(\vec{x}\in M_\nu)-I\vert_{M_\nu}\right)^2
\intertext{and}
I\vert_{M_\nu}&=\frac{1}{\mathrm{vol}(M_\nu)}\int_0^1\!\!\diff[d]{\vec{x}}\,%
f(\vec{x})\text{.}
\end{align}
For a given partition $\{M_\nu\}$ of the hypercube $[0;1]^d$ the error
is minimised for the choice
\begin{equation}
\frac{N_\mu}{N}=\frac{\mathrm{vol}(M_\mu)\sigma^2(f)\vert_{M_\mu}}{%
\sum_{\nu=1}^k\mathrm{vol}(M_\nu)\sigma^2(f)\vert_{M_\nu}}\text{.}
\end{equation}

A second useful technique is \emph{importance sampling}
\index{Importance sampling|main}, which corresponds to
a variable transformation in usual integration,
\begin{equation}
\int_0^1\!\!\diff[d]{\vec{x}}\,f(\vec{x})=
\int_0^1\!\!\diff[d]{\vec{x}}\frac{f(\vec{x})}{p(\vec{x})}p(\vec{x})\text{.}
\end{equation}
If $p(\vec{x})$ is positive and normalised such that
\begin{equation}
\int_0^1\!\!\diff[d]{\vec{x}}\,p(\vec{x})=1
\end{equation}
then $p(\vec{x})$ can be interpreted as the density of the
probability distribution $P(\vec{x})$,
\begin{equation}
p(\vec{x})=\frac{\partial^d}{\partial{x_1}\cdots\partial{x_d}}P(\vec{x})\text{.}
\end{equation}
The Monte Carlo estimator can be replaced by
\begin{equation}
E=\frac{1}{N}\sum_{j=1}^n\frac{f(\vec{x}^{(j)})}{p(\vec{x}^{(j)})}
\end{equation}
and the points $\vec{x}^{(j)}$ are chosen according to the probability
distribution $P(\vec{x})$. The statistical error is $\sigma(f/p)/\sqrt{n}$;
if $f$ and $p$ are very similar the ratio $f/p$ becomes flat and hence
the error decreases.

In the VEGAS\index{VEGAS|main} algorithm~\cite{Lepage:1977sw}, one combines
the above ideas in the following way: the probability density is
approximated by a grid
\begin{equation}
p(\vec{x})=\prod_{j=1}^d p_j(x_j)\quad\text{with}\,%
p_j(x_j)=\frac{1}{k_j}\sum_{i=1}^{k_j}
\frac{\Theta(x_{i-1,j}\leq x_{j}<x_{i,j})}{x_{i,j}-x_{i-1,j}}
\end{equation}
The grid is separated at points $0=x_{0,j}<x_{1,j}<\ldots<x_{k_j,j}=1$
and is adjusted step-wise such that each bin contributes
\begin{equation}
\frac{1}{\prod_{j=1}^d k_j}\int_0^1\!\!\diff[d]{\vec{x}}\vert f(\vec{x})\vert
\text{.}
\end{equation}
A sequence of $m$ adaption steps leads to the estimates
$E_{(1)},\ldots,E_{(m)}$ and estimates for the variance
$S_{(1)},\ldots,S_{(m)}$. If the number of points in step~$j$
is~$N_{(j)}$ one obtains a combined result
\begin{equation}
E_{\text{combined}}=\left.{\sum_{j=1}^m\frac{N_{(j)}E_{(j)}}{S_{(j)}^2}}
\right/\sum_{i=1}^m\frac{N_{(i)}}{S_{(i)}^2}
\end{equation}
One also can check the consistency of the estimators using
$\chi^2$ per degree of freedom,
\begin{equation}
\chi^2/\text{d.o.f.}=\frac{1}{m-1}\sum_{j=1}^m\frac{(E_{(j)}-E)^2}{S_{(j)}^2}%
\text{.}
\end{equation}

\subsection{Phase Space mapping: RAMBO}
\index{RAMBO|(}
Although nowadays many different and optimised methods are
available to map the
hypercube $[0;1]^{4N}$ to the phase space $\phasespace{N}$,
here I will only focus on the RAMBO~\cite{Kleiss:1985gy};
a broader overview over different phase space mappings is given
for example in~\cite{Weinzierl:2000wd}.

In the first step of the RAMBO algorithm, one generates a set of
$N$ lightlike vectors $q_i$ with isotropic angular distribution and
energy distribution $Ee^{-E}\diff{E}$. The vector
$q_i^\mu=E_i(1,\cos\varphi_i\sin\theta_i,\sin\varphi_i\sin\theta_i,\cos\theta_i)$
is generated from four uniformly distributed random numbers
$u_{i1}\ldots u_{i4}$ in the hypercube $(0;1]^4$ with the
transformation
\begin{subequations}
\begin{align}
\cos\theta_i&=2u_{i1}-1\text{,}\\
\varphi_i&=2\pi u_{i2}\text{,}\\
E_i&=-\ln(u_{i3}u_{i4})\text{.}
\end{align}
\end{subequations}

To obtain physical momenta that obey momentum conservation
$\tilde{Q}=\sum_{j=1}^n p_j^\mu$ for a time like vector
$\tilde{Q}=(Q^0,0,0,0)$
in a second step, one has to apply
the \person{Lorentz} transformation~$p_i^\mu=\Lambda^\mu_\nu q_i^\nu$,
\begin{equation}
\Lambda^\mu_\nu=x\left(\begin{array}{c|c}
\gamma & \vec{b}^\transposed\\
\hline
\vec{b} & \One+a\,\vec{b}\,\vec{b}^\transposed
\end{array}\right)
\end{equation}
parametrised by
\begin{align}
\vec{b}&=-\frac{1}{\sqrt{v^2}}\vec{v},
\quad\text{with}\,v^\mu=\sum_{j=1}^N q_j^\mu\text{,}\\
\gamma&=\sqrt{1+\vert\vec{b}\vert^2}\text{,}\\
a&=\frac{1}{1+\gamma}\quad\text{and}\\
x&=\sqrt{\frac{\tilde{Q}^2}{v^2}}\text{.}
\end{align}

One can show~\cite{Kleiss:1985gy,Weinzierl:2000wd}
that from the transformation each event obtains the weight
\begin{equation}
w=w_0=(2\pi)^{4-3N}\left(\frac{\pi}2\right)^{N-1}\frac{(Q^2)^{N-2}}{%
\Gamma(N)\Gamma(N-1)}\text{.}
\end{equation}

In order to obtain momenta $k_i$ for massive particles from the
momenta $p_i$ as defined above, one applies a second transformation
\begin{equation}
k_i^0=\sqrt{m_i^2+\xi^2(p_i^0)^2},\qquad\vec{k}_i=\xi\vec{p}_i\text{,}
\end{equation}
where the parameter $\xi$ is a solution of the equation
\begin{equation}
\sqrt{\tilde{Q}^2}=\sum_{j=1}^N\sqrt{m_j^2+\xi^2(p_i^0)^2}\text{.}
\end{equation}
Each event obtains a new weight $w=w_0w_m$, where $w_m$ is
\begin{equation}
w_m=(\tilde{Q}^2)^{2-N}\left(\sum_{j-1}^N\vert\vec{k}_j\vert^2\right)^{2N-3}
\left.\left(\prod_{j=1}^N\frac{\vert\vec{k}_j\vert}{k_j^0}\right)\right/
\left(\sum_{j=1}^N\frac{\vert\vec{k}_j\vert^2}{k_j^0}\right)
\end{equation}
\index{RAMBO|)}

\subsection{Integration of \acs{nlo} Amplitudes}
\label{ssec:qcd-psintegral:nlo}
In this section I present a method which is well suited for an
efficient implementation of an event generator at one-loop
level.~\cite{Binoth:2008gx} Traditionally, one uses an
adaptive Monte Carlo program such as VEGAS and call it with the
differential cross-section $\diff\virt\sigma$ in order
to obtain the virtual corrections $\virt\sigma$
to the inclusive cross-section. 

Working with the virtual correction already in the initialisation
of the integrator has two disadvantages:
the virtual corrections are computationally much more
expensive than the \ac{lo} matrix element
and hence one adds to the run time of the integration.
Secondly,
the integrator is likely to adapt to integrable singularities
in the phase space
which destabilises the adaption process.

The method presented
below avoids part of the computational cost using the \ac{lo}
matrix element for the initialisation steps
and calling the virtual corrections only on a set of phase-space points
which have been obtained from the integration of the \ac{lo} cross-section,
thus also avoiding destabilisation of the adaption phase of the integrator.

An observable for a collider
can be defined as the integral\footnote{The presence of
real corrections is not important for this discussion because their
contribution can be computed by standard techniques.}
\begin{multline}
\braket{O}=\int_0^1\!\!\diff{x_1}\int_0^1\!\!\diff{x_2}f_a(x_1)f_b(x_2)
\int\!\!\phasespace{N}(x_1P_1+x_2P_2;p_1,\ldots,p_N)
\\\times
\frac{\diff{\sigma_{ab}(x_1P_1,x_2P_2;p_1,\ldots,p_N}}{%
	\phasespace{N}}O(x_1P_1,x_2P_2;p_1,\ldots,p_N)
\\
\approx\frac{1}{n}\sum_{j=1}^nf_a(x_1^{(j)})f_b(x_2^{(j)})
\frac{\diff{\sigma_{ab}(x_1^{(j)}P_1,x_2^{(j)}P_2;p_1^{(j)},\ldots,p_N^{(j)})}
}{\phasespace{N}}
O(x_1^{(j)}P_1,x_2^{(j)}P_2;p_1^{(j)},\ldots,p_N^{(j)})
\end{multline}
At \ac{lo} one has $\diff\sigma_{ab}=\diff\born\sigma_{ab}$ and
$\sigma_{ab}=\braket{1}$. Most \ac{lo} event generators are
capable of producing \emph{unweighted} events\index{unweighted event},
i.e. from all events
$e_{(j)}=(x_1^{(j)}, x_2^{(j)};p^{(j)}_1,\ldots, p^{(j)}_N)$ a subset~$U$
is chosen such that for a large number of events
\begin{equation}
\braket{O}=
\frac{\born\sigma_{a,b}}{\vert U\vert}\sum_{e_{(j)}\in U}
O(x_1^{(j)}P_1,x_2^{(j)}P_2;p_1^{(j)},\ldots,p_N^{(j)})\text{.}
\end{equation}
The right hand corresponds to a Monte Carlo estimator using importance
sampling with the probability density
\begin{equation}
p(e_{(j)})=
\frac{1}{\born\sigma_{a,b}}f_a(x_1^{(j)})f_b(x_2^{(j)})
\frac{\diff{\sigma_{ab}(x_1^{(j)}P_1,x_2^{(j)}P_2;p_1^{(j)},\ldots,p_N^{(j)})}
}{\phasespace{N}}
\end{equation}

One can carry this idea of importance sampling further and evaluate
the one-loop corrections to the \person{Born} process using the estimate
\begin{equation}\label{eq:qcd-psintegral:reweighting01}
\virt{\braket{O}}=\braket{O\cdot K}=
\frac{\born\sigma_{a,b}}{\vert U\vert}\sum_{e_{(j)}\in U}
O(e_{(j)})K(e_{(j)})\text{,}
\end{equation}
where the local $K$-factor is defined as%
\index{K-factor@$K$-factor!local|main}
\begin{equation}\label{eq:results:reweighting02}
K(e_{(j)})=\frac{\diff\virt\sigma(e_{(j)})}{\diff\born\sigma(e_{(j)})}\text{.}
\end{equation}
for a phase-space point $e_{(j)}$.
The advantage is that an adaptive Monte Carlo program is trained ---
i.e. the grid is optimised in the case of VEGAS ---
according to the \person{Born} level
matrix element; therefore the initialisation is very fast and robust,
as explained above. As this method is well defined as an application
of importance sampling, one can calculate $S/\sqrt{N}$ as an estimate
of the error introduced by the integration over the unweighted events.
Our experience with the $u\bar{u}\rightarrow b\bar{b}b\bar{b}$ shows
that this contribution to the total error is comparable with the
error on the \ac{lo} matrix element, which is shown
in~Chapter~\ref{chp:results}.

%% file: results.tex
\chapter{Virtual \acs{nlo} Corrections
for \texorpdfstring{$u\bar{u}\rightarrow b\bar{b}b\bar{b}$}{%
uu-bar to bb-bar bb-bar}}
\label{chp:results}
\begin{headquote}{Winston Churchill}
However beautiful the strategy,
you should occasionally look at the results.
\end{headquote}
\section*{Introduction}
\label{sec:res-intro}
\input{res-intro}

\section{Performance and Accuracy}
The Monte Carlo integration has been carried out on
the \ac{ecdf}\footnote{See Section~\ref{ssec:imp-fortran:ecdf}
of Appendix~\ref{chp:implementation}} using the method described
in Section~\ref{ssec:qcd-psintegral:nlo} of Chapter~\ref{chp:qcdnlo}.
The implementation of the matrix element is available in
double precision and quadruple precision; a performance of
8{.}9\,s (17{.}6\,s) per phase-space point has been
achieved for a single node\footnote{Intel Xeon 5450 quad-core
\unit{3}{\giga\hertz}}
in double (quadruple) precision.

\begin{figure}[hbtp]
\subfloat[local $K$-factor]{
\includegraphics[width=0.31\textwidth]{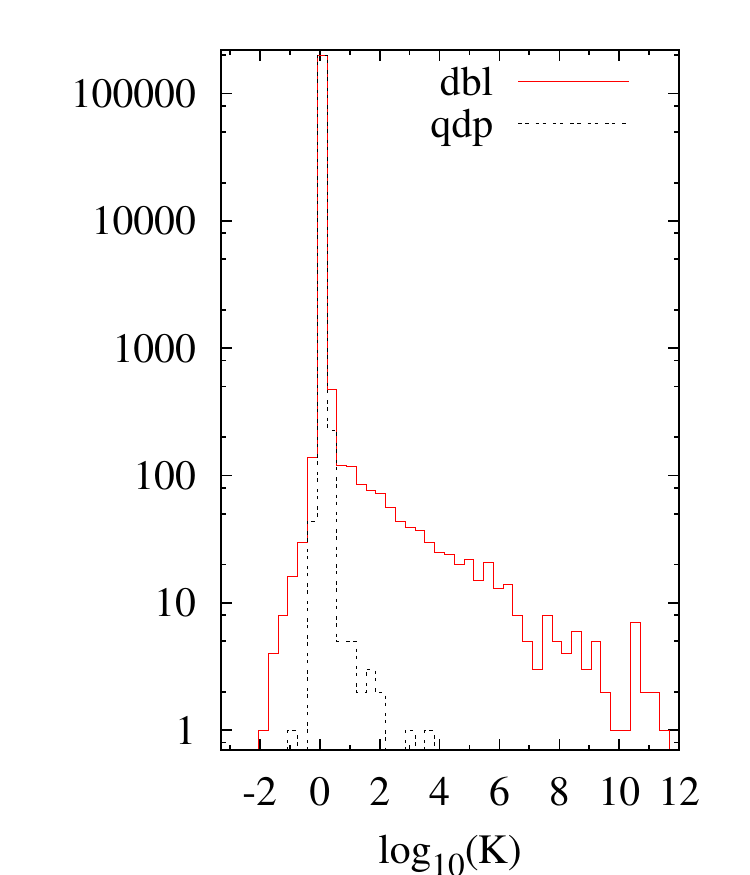}
\label{subfig:results:k-hist}
}
\subfloat[single pole]{
\includegraphics[width=0.31\textwidth]{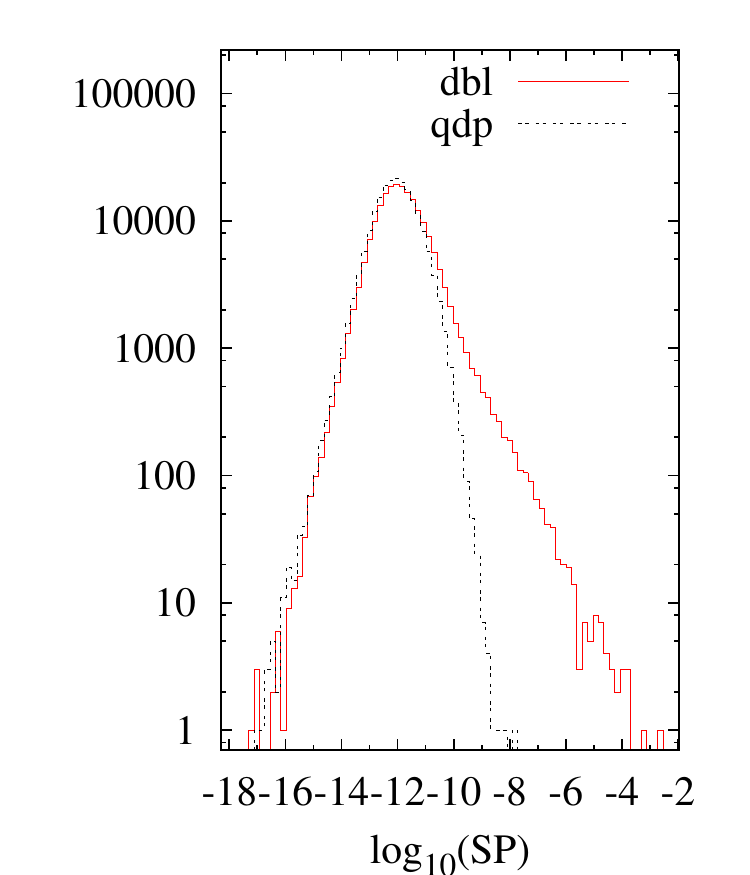}
\label{subfig:results:s-hist}
}
\subfloat[double pole]{
\includegraphics[width=0.31\textwidth]{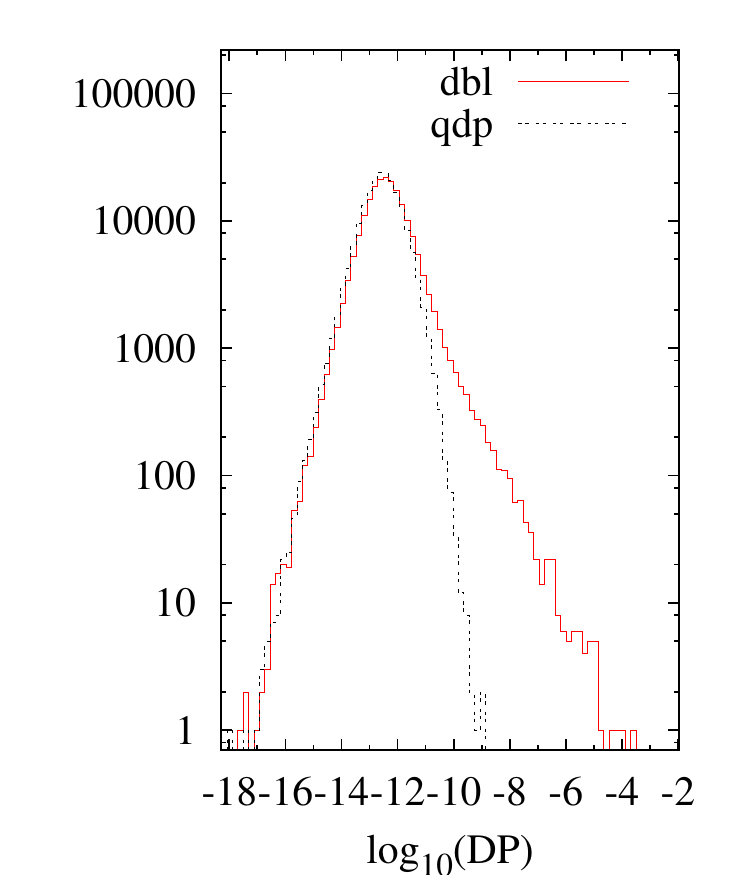}
\label{subfig:results:d-hist}
}
\caption{Distributions for double and quadruple precision obtained
from 200{.}000 randomly generated phase space points. Left: the
local $K$-factor. Middle: the single pole of the squared
\ac{nlo} matrix element normalised by the square of the \ac{lo}
matrix element. Right: as middle, but for the double pole.
} % caption
\label{fig:results:precision-hists}
\end{figure}

The reason to have an implementation in two different
precisions is to reduce and to quantify the error on the
result due to numerical instabilities: in some phase space
regions our amplitude representation induces large terms
cancelling against each other, which in a numerical implementation
leads to a loss of significant digits. Typically, for those points
one observes large ratios between \ac{nlo} and \ac{lo} result,
which is referred to as
\emph{local $K$-factor}\index{K-factor@$K$-factor!local},
or an incomplete cancellation of the pole parts of the amplitude.
The histograms in Figure~\ref{fig:results:precision-hists}
are generated for 200{.}000 phase-space points.
The tail of each distribution
stretching towards large values of the respective quantity is
reduced by an evaluation in higher precision, indicating
that these quantities can be used to single out points which
cause numerical problems. However, these are only indicators but
neither necessary nor sufficient conditions for numerical instabilities.

In order to enjoy the benefits of the improved stability of an
evaluation in quadruple precision and the speed of the calculation
in double precision, we investigated different criteria
applied for each phase-space point, such that
\begin{itemize}
\item the point is calculated in double precision if it passes the test,
\item the point is re-evaluated in higher precision if the test fails,
\item the outcome of the test only depends on the double precision result.
\end{itemize}

The above motivation has lead to the comparison of three different
criteria. The first test uses the local $K$-factor as defined
in Equation~\eqref{eq:results:reweighting02},
\begin{displaymath}
K(e_{(j)})=\frac{\diff\virt\sigma(e_{(j)})}{\diff\born\sigma(e_{(j)})}\text{;}
\end{displaymath}
the other two tests compare the residual value of the single (resp. double)
pole of the local $K$-factor, which in an evaluation with arbitrary precision
would be zero\footnote{The cancellation of the poles is also influenced by the
precision of the kinematics as an input parameter, which in our calculation
is fixed to be double precision.}. If the considered value is larger than
a given, fixed cut-off the test fails and the evaluation is repeated in
higher precision.
\begin{figure}[phbt]
\includegraphics{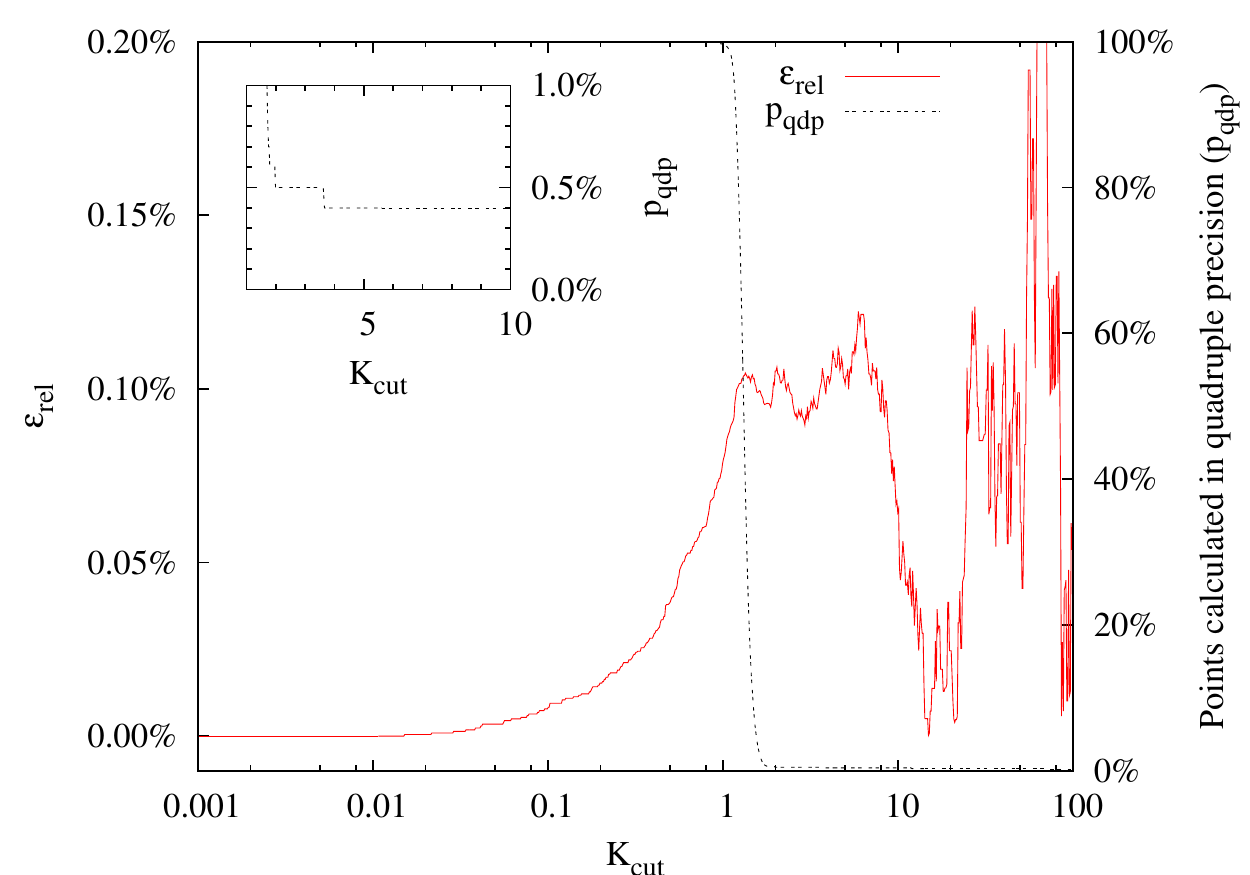}
\caption{Relative error $\varepsilon_{\text{rel}}=
\frac{\vert\sigma(K_\text{cut})-\sigma(0)\vert}{\sigma(0)}$
of the integral over a sample of $200{.}000$ random points.
For each phase-space point the local $K$-factor has been
calculated in double precision ($K_\text{dbl}$); if
$\vert K_\text{dbl}\vert\geq K_\text{cut}$ the
$K$-factor evaluated in quadruple precision ($K_\text{qdp}$)
entered the result, otherwise $K_\text{dbl}$ has been used.
The second curve shows the percentage of points $p_\text{qdp}$
required in quadruple precision to evaluate $\sigma(K_\text{cut})$.
The figure has been produced from 1{.}000 different
values of $K_\text{cut}$,
distributed linearly in $\log(K_\text{cut})$.
The inlay shows $p_\text{qdp}$
for the region between $2\leq K_\text{cut}\leq10$.
} % caption
\label{fig:results:k-acc}
\end{figure}

\begin{figure}[phbt]
\subfloat[single pole]{
\includegraphics[width=0.47\textwidth]{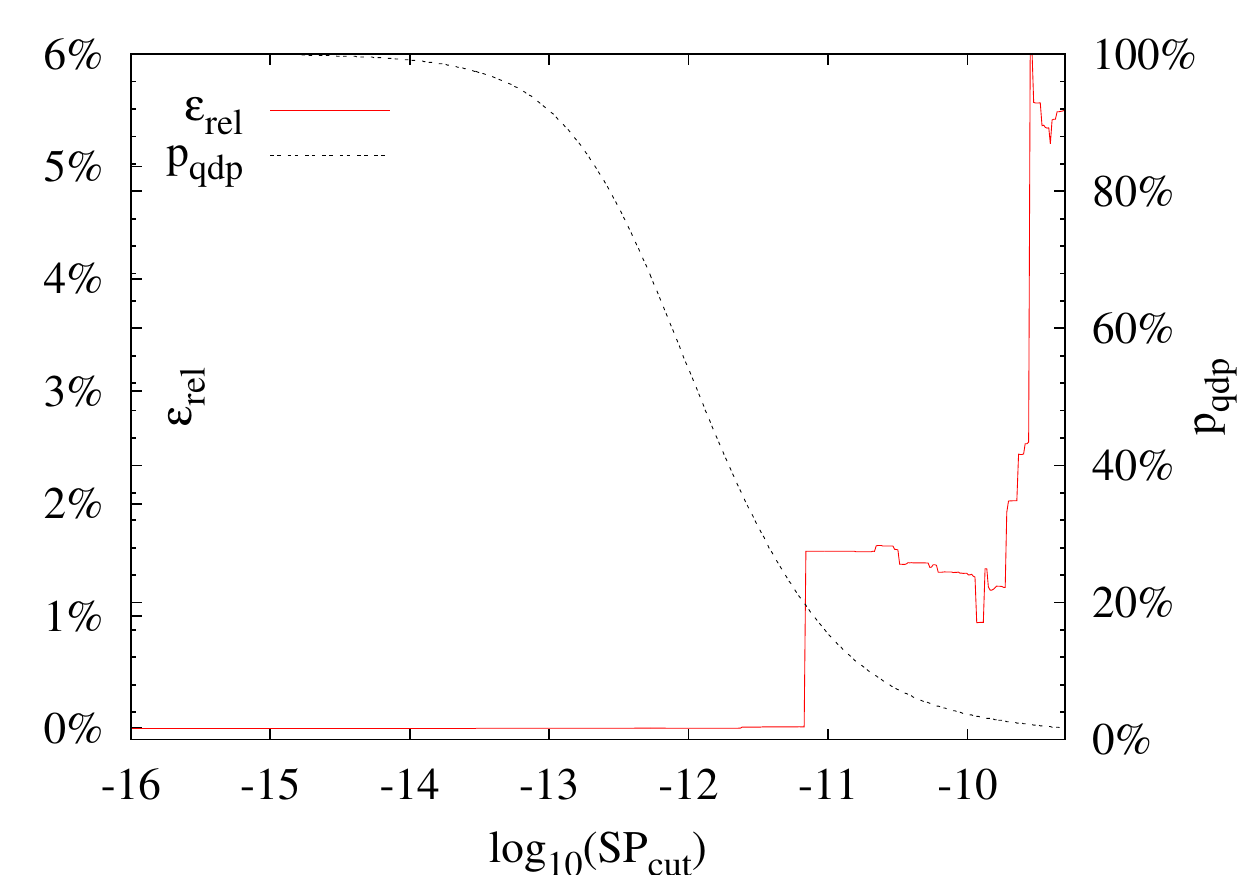}
\label{subfig:results:s-acc}
}
\subfloat[double pole]{
\includegraphics[width=0.47\textwidth]{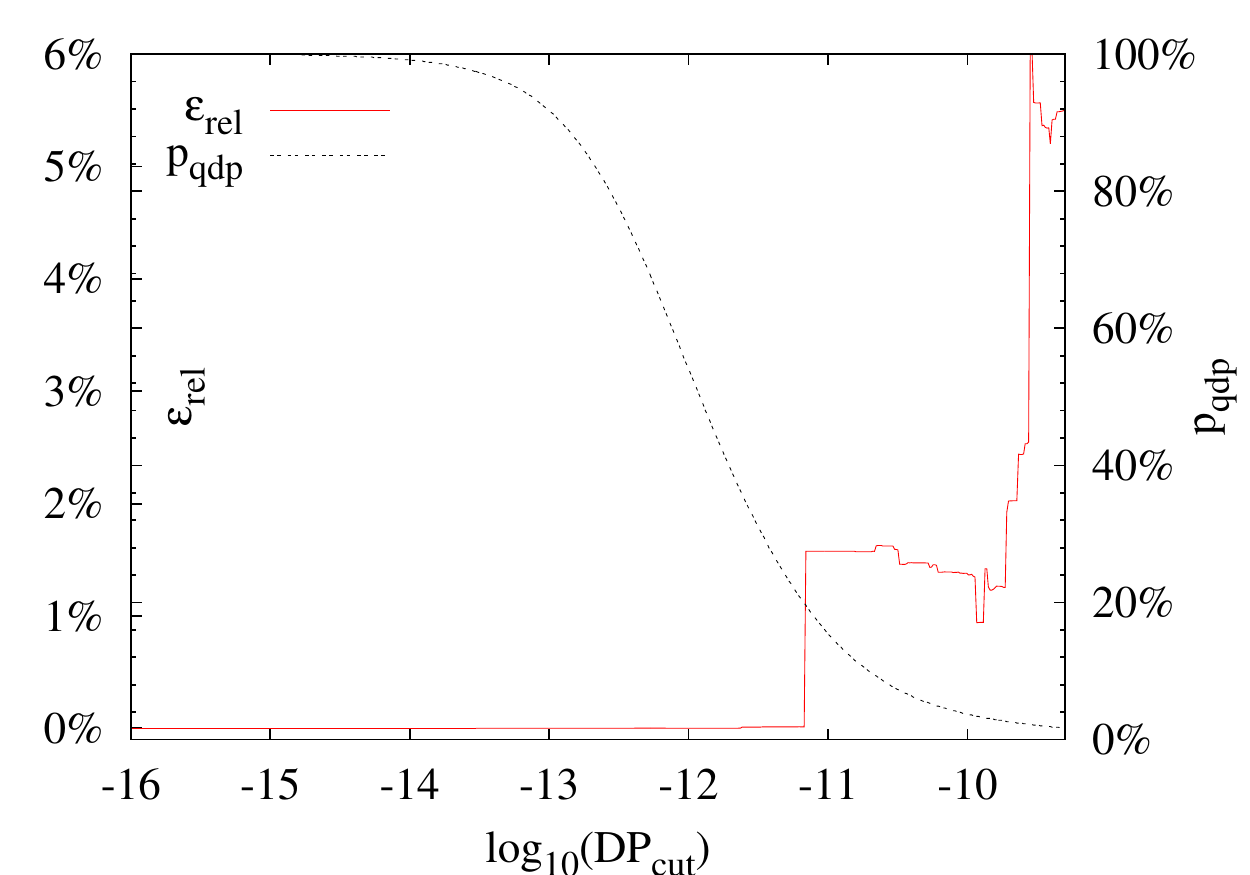}
\label{subfig:results:d-acc}
}
\caption{Relative error $\varepsilon_{\text{rel}}=
\frac{\vert\sigma(P_\text{cut})-\sigma(0)\vert}{\sigma(0)}$
of the integral over a sample of $200{.}000$ random points.
The procedure is the same as in Figure~\ref{fig:results:k-acc}
but instead of $K_\text{cut}$ the single (resp. double) pole
of the $K$-factor,
$P_\text{cut}=\{SP_\text{cut}, DP_\text{cut}\}$ respectively,
have been used as the criterion for an evaluation with higher precision.
} % caption
\label{fig:results:sd-acc}
\end{figure}

Figures \ref{fig:results:k-acc} and~\ref{fig:results:sd-acc} show the
relative errors on the inclusive Monte Carlo integral obtained from
200{.}000 phase space points for different values of the cut-off
parameters. In figure~\ref{fig:results:k-acc} we vary the value of
$K_\text{cut}$ and re-evaluate phase-space points $e_{(j)}$,
where $K(e_{(j)})\geq K_\text{cut}$.
For values of $K_\text{cut}\lesssim10$
the relative error on the Monte Carlo estimate as opposed to a
complete evaluation in quadruple precision
remains small ($\varepsilon\lesssim0{.}1\%$). On the other hand, the number
of points to be calculated in quadruple precision drops below~$0{.}5\%$
around~$K_\text{cut}\approx1.5$, and hence moderate values
of~$K_\text{cut}$ ($2\leq K_\text{cut}\leq10$) yield an estimate that
is within an error of $\mathcal{O}(0{.}1\%)$ compared to the
result obtained in quadruple precision with a performance close to
the speed of a calculation in double precision.

The authors of~\cite{Giele:2008bc} propose to use the correlation
between the accuracy with which the poles in $\varepsilon=(4-n)/2$
cancel in order to obtain an estimate for the stability of 
the numerical evaluation of a phase space point.
In Figure~\ref{fig:results:sd-acc} a similar method as in
Figure~\ref{fig:results:k-acc} has been used to obtain estimates
for the error on a Monte Carlo integral over 200{.}000 points:
instead of the local $K$-factor, the pole-parts
of the squared \ac{nlo} matrix element normalised by 
the square of the \ac{lo} matrix element
have been used as a cut-off parameter to switch between double precision
and quadruple precision. However, for the
$u\bar{u}\rightarrow b\bar{b}b\bar{b}$ amplitude we observe a
identification of unstable points which is worse than in the case
where the local $K$-factor has been used as a discriminant.
In order to obtain a relative error of less than $0{.}1\%$ one
has to evaluate $\mathcal{O}(20\%)$ of the phase space points in
higher precision. The results are very similar for the single
pole (Figure~\ref{fig:results:sd-acc}\,\subref{subfig:results:s-acc}) and the
double pole (Figure~\ref{fig:results:sd-acc}\,\subref{subfig:results:d-acc}).

As an estimate of the error on the \ac{nlo} results we consider the sum of three
contributions:

\renewcommand{\arraystretch}{1.5}
\begin{tabular}[t]{lp{0.7\textwidth}}
$K\cdot\delta\sigma_{\mathrm{LO}}$ &
is the propagation of the error on $\sigma_{\mathrm{LO}}$ introduced by the
Monte Carlo integration of the leading order result. \\
$\frac{S}{\sqrt{N}}$ &
estimates the error resulting from the integration
of the \ac{nlo} matrix element, where $N$ is the number of unweighted
events used for the integration and $S$ is the estimate of the standard
deviation as defined in Equation~\eqref{eq:qcd-psintegral:mcerror}.\\
$\varepsilon_{\mathrm{rel}}$ & as obtained from Figure~\ref{fig:results:k-acc}
is added to the error to account for the limited numerical precision.
In all our calculations we choose $K_{\mathrm{cut}}=4.0$ and
$\varepsilon_{\mathrm{rel}}=0{.}1\%$.
\end{tabular}
\renewcommand{\arraystretch}{1.0}

Table~\ref{tab:results:inclusive} shows the error contributions for
two integrations of the virtual part of the inclusive cross-section.
The errors in both cases are very small; in the larger sample of
$10^6$ phase-space points both the error induced by the integration
of the \ac{lo} part and the error from the integration of the \ac{nlo}
matrix element are comparable.
\begin{table}[hbtp]
\begin{tabular}{lrr}
$N$ & $100{,}000$ & $1{,}000{,}000$\\
\hline
$K\cdot\delta\sigma_{\mathrm{LO}}$ & $0{.}71\%$ & $0{.}58\%$\\
${S}/{\sqrt{N}}$ & $2{.}79\%$ & $0{.}49\%$ \\
$\varepsilon_{\mathrm{rel}}$ & $0{.}10\%$ & $0{.}10\%$ \\
\hline
total error & $3{.}59\%$ & $1{.}17\%$
\end{tabular}
\caption{Relative error contributions to the
virtual corrections of the inclusive cross-section for
$u\bar{u}\rightarrow b\bar{b}b\bar{b}$ at \ac{nlo}.
Each result is based on an independent sample of $N$ events
with a $p_T$ cut of
\unit{25}{\giga\electronvolt},
a rapidity cut of $\vert\eta\vert<2{.}5$ and a separation cut~$\Delta R>0{.}4$
applied. The scales are chosen as $\mu_F=\mu_R=\sum p_T/4$ and
CTEQ\,6{.}5 \acp{pdf}~\cite{Tung:2006tb} have been~used.}
\label{tab:results:inclusive}
\end{table}

We have varied the renormalisation scale~$\mu_R$ and the factorisation
scale~$\mu_F$ in an interval from $1/8$ to $8$ times
of their central value $\mu_0=\sum_{i=1}^4p_T^{(i)}/4$ to study the
influence of the scale choice on the result. The \ac{nlo} corrections
in this calculation consist of the virtual corrections and the insertion
operator~$\langle I(\varepsilon)\rangle$; this subset of terms ensures
that all poles cancel and the results presented are \ac{ir} finite.
The real corrections together with the dipoles and the subtraction
term $\sigma^{C\,\{N\}}$ as described in Section~\ref{sec:dipoles}
of Chapter~\ref{chp:qcdnlo} need to be added to obtain the full
\ac{nlo} corrections.

\begin{figure}[hbtp]
\subfloat[parallel]{
\includegraphics[width=0.47\textwidth]{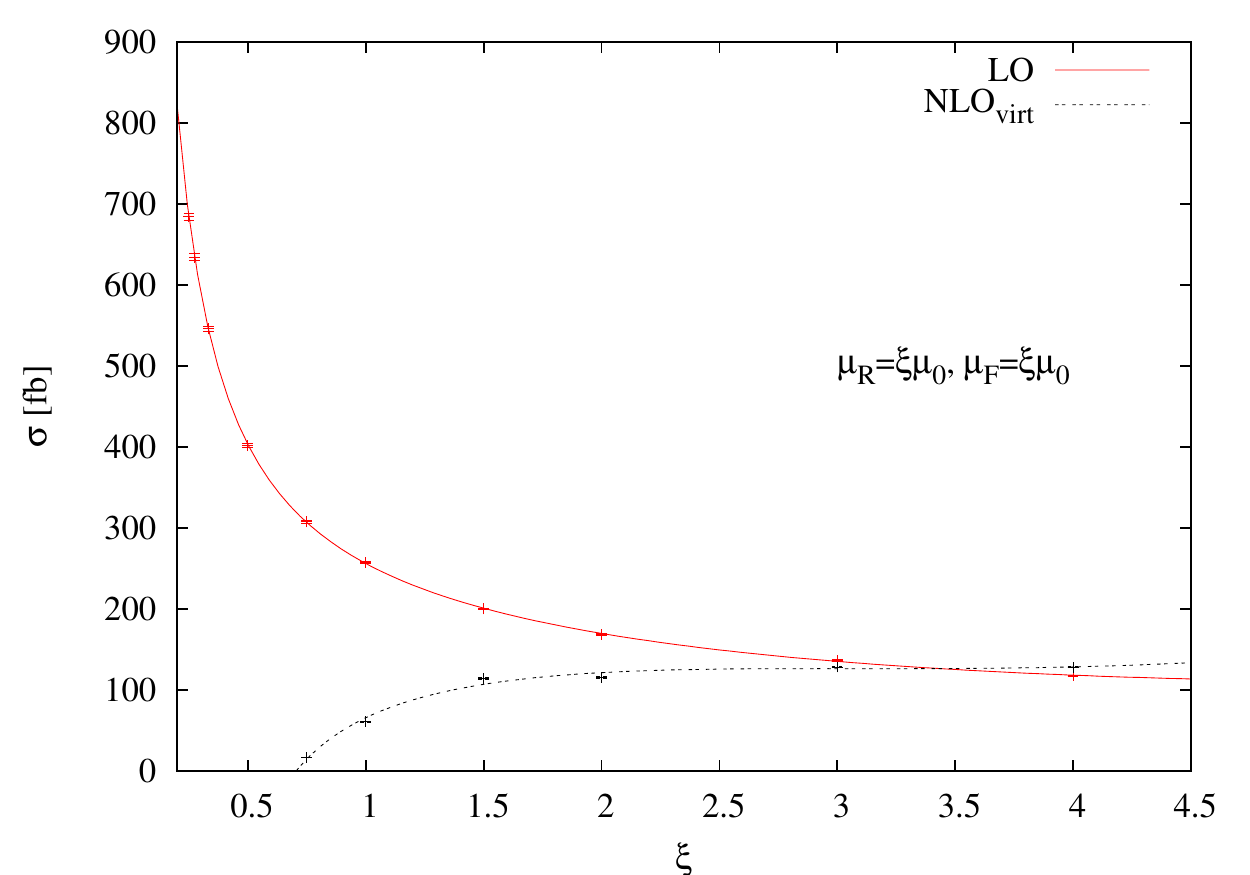}
\label{subfig:results:scalevar-par}
}
\subfloat[antipodal]{
\includegraphics[width=0.47\textwidth]{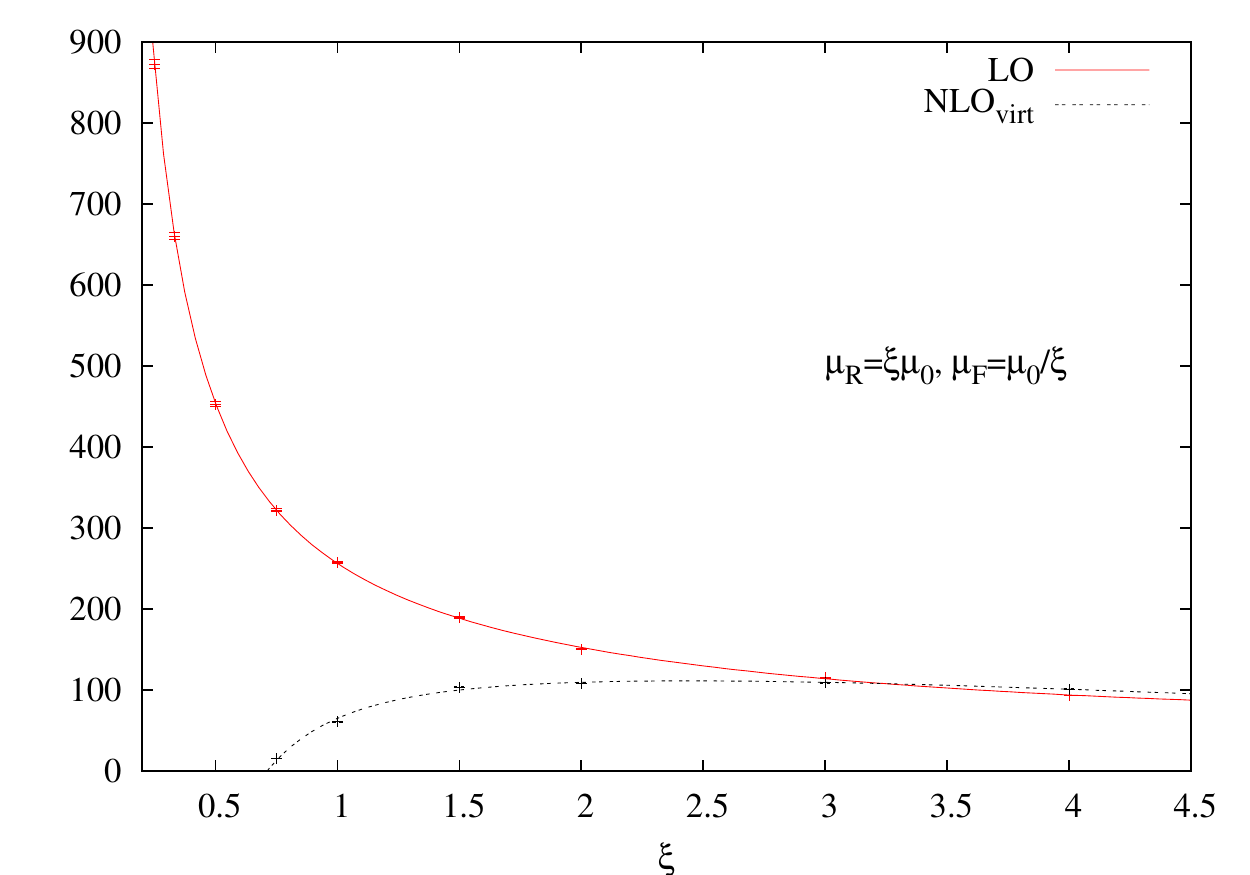}
\label{subfig:results:scalevar-opp}
}
\caption{Scale dependence of the virtual corrections. The two
plots compare the \ac{lo} with the \ac{nlo} result of the
inclusive cross-section for different choices of the scales $\mu_R$
and~$\mu_F$. The \ac{nlo} graph contains the contributions
$\sigma^{\text{NLO}_{\text{virt}}}=\born\sigma+
\virt\sigma+\langle I(\varepsilon)\rangle$,
where $\langle I(\varepsilon)\rangle$ is the sum of the
integrated dipoles. Each data point is generated from a Monte
Carlo integral over $10^5$ phase-space points. In the left
graph the scales have been varied in parallel, $\mu_F=\mu_R=\xi\mu_0$.
The plot on the right shows the antipodal variation, $\mu_F=\mu_0/\xi$,
$\mu_R=\xi\mu_0$, where $\mu_0$ is the average $p_T$ of the four jets.}
\label{fig:results:scalevariation}
\end{figure}

The result for the inclusive cross-section is shown
in Figure~\ref{fig:results:scalevariation} for the case where
both scales are varied in parallel and the case of antiparallel (antipodal)
scale variation. Both graphs show a plateau region for $2\lesssim\xi\lesssim4$.
In the region around $\xi\approx1$ one expects the real emission to
significantly change the result and hence we eschew a interpretation
of the data in this region until the \ac{nlo} corrections are complete.

\section{Exclusive Observables}
In this section we study distributions of exclusive observables,
which are relevant for example in
\ac{mssm} \person{Higgs} boson searches in the $b\bar{b}b\bar{b}$ channel
at the ~\ac{lhc}~\cite{Djouadi:2000gu,Mahboubi:01,Dai:1994vu}.

\begin{figure}[hbtp]
\subfloat[1st (hardest) jet]{
\includegraphics[width=0.47\textwidth]{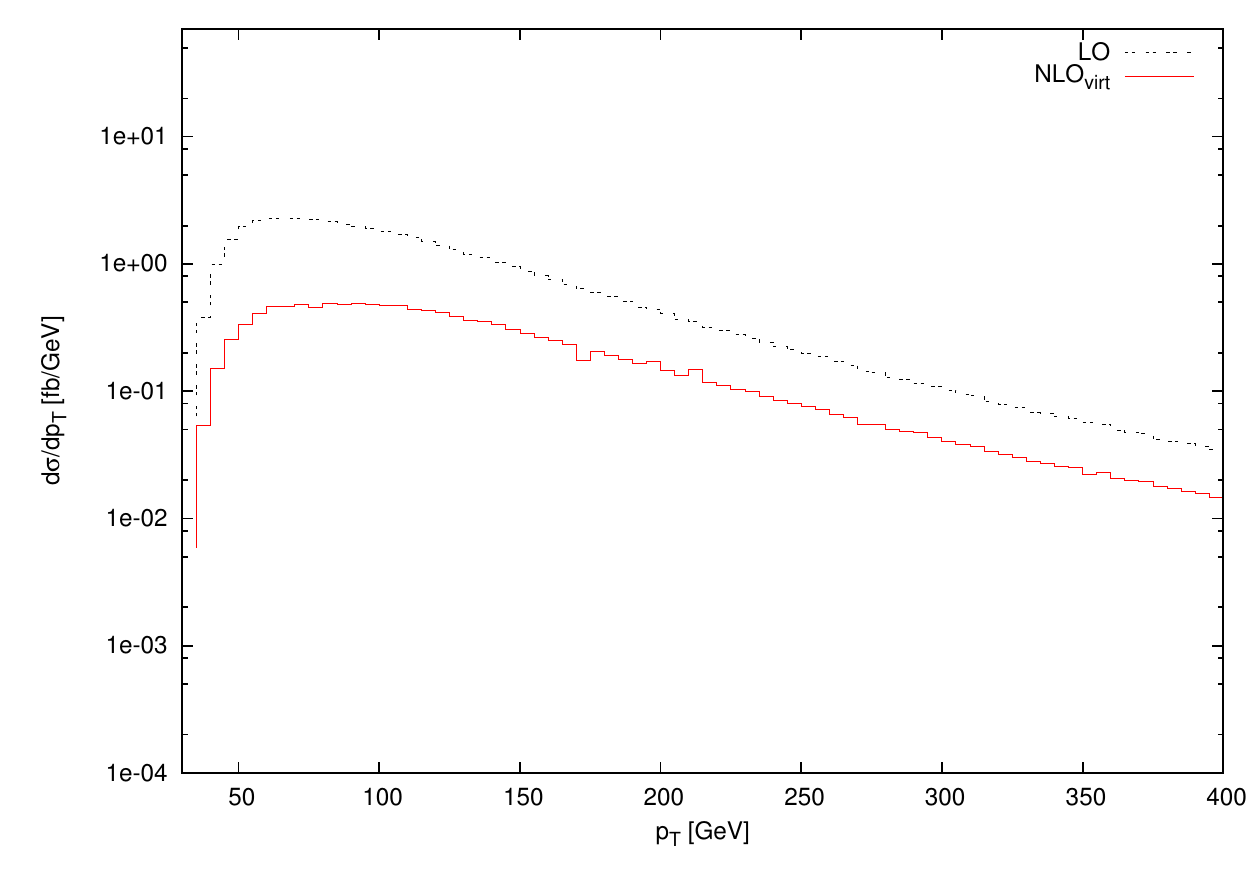}
\label{subfig:results:jet_pt-1st}
}
\subfloat[2nd jet]{
\includegraphics[width=0.47\textwidth]{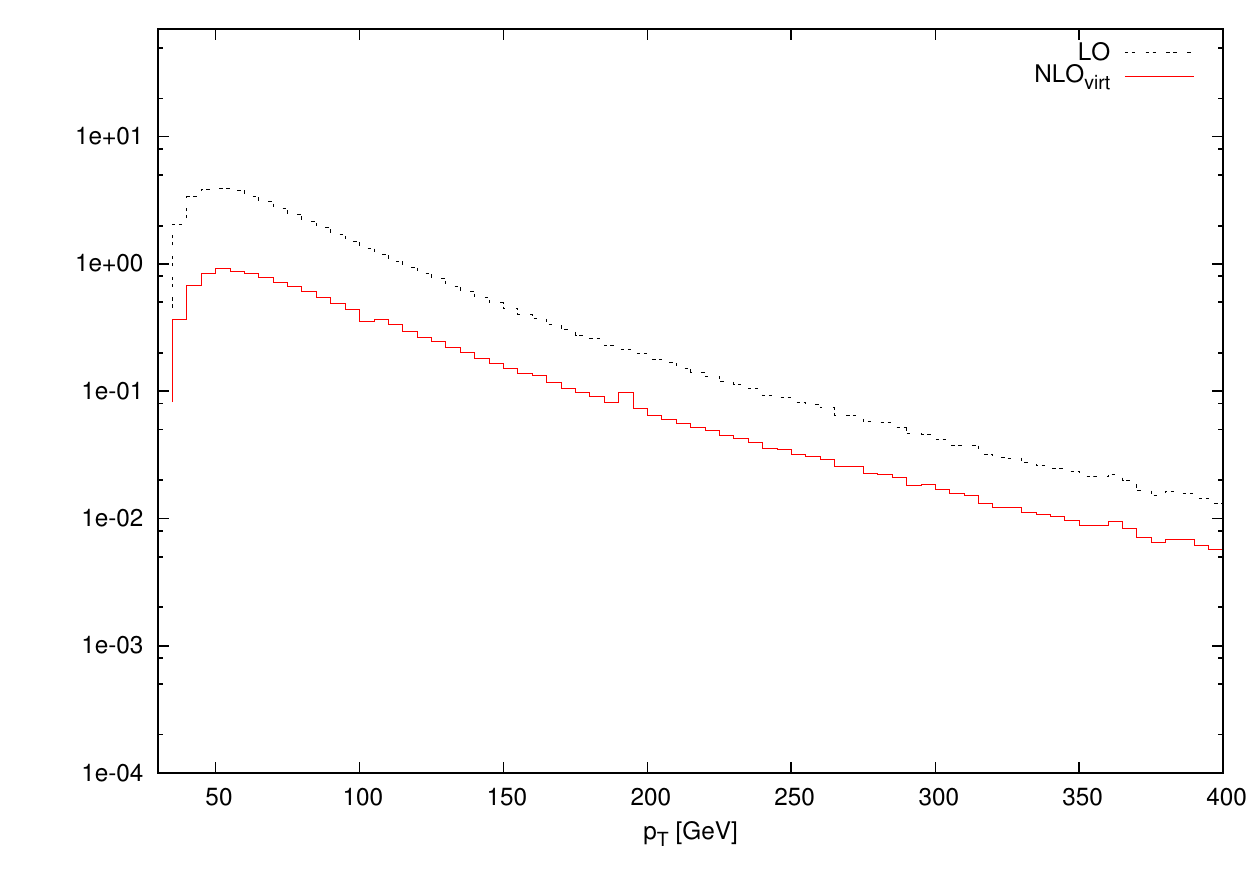}
\label{subfig:results:jet_pt-2nd}
}\\
\subfloat[3rd jet]{
\includegraphics[width=0.47\textwidth]{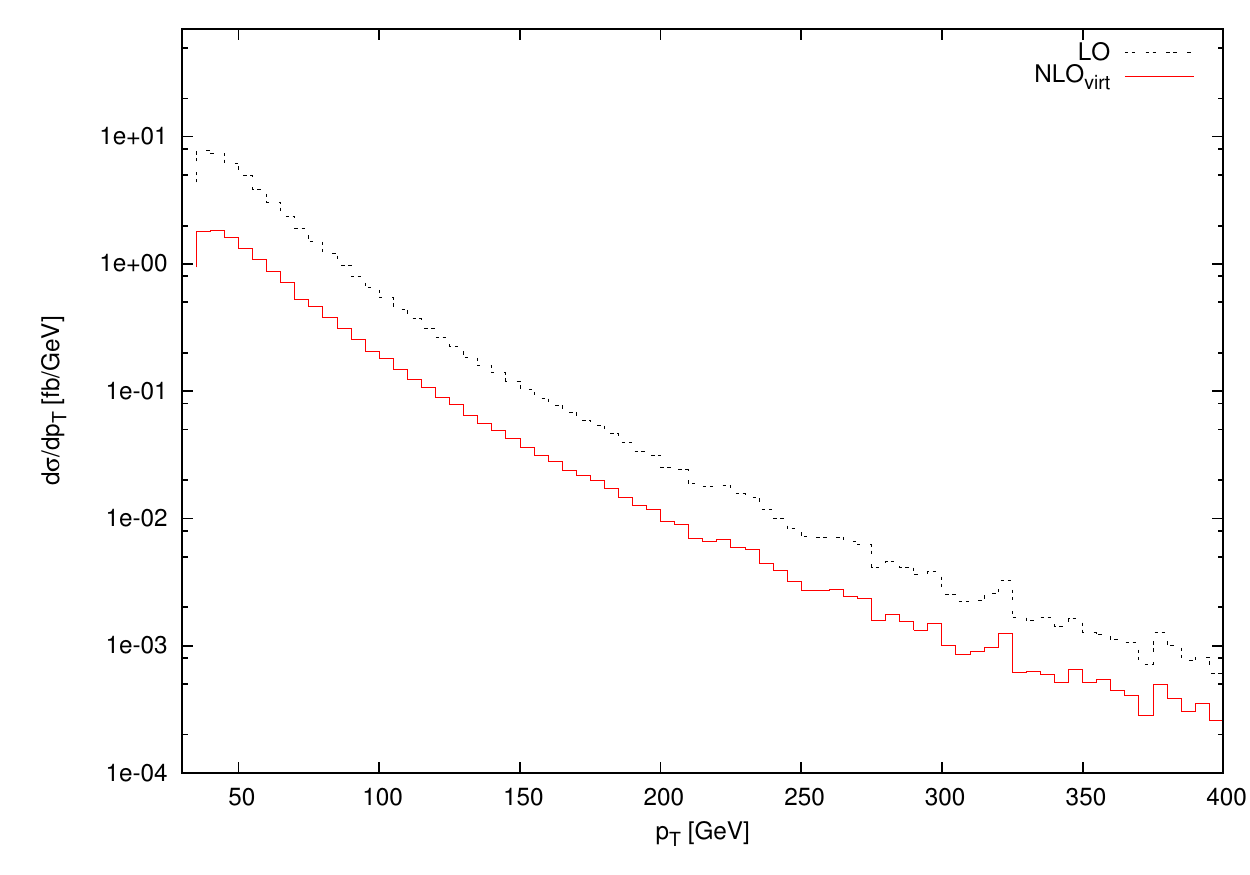}
\label{subfig:results:jet_pt-3rd}
}
\subfloat[4th (softest) jet]{
\includegraphics[width=0.47\textwidth]{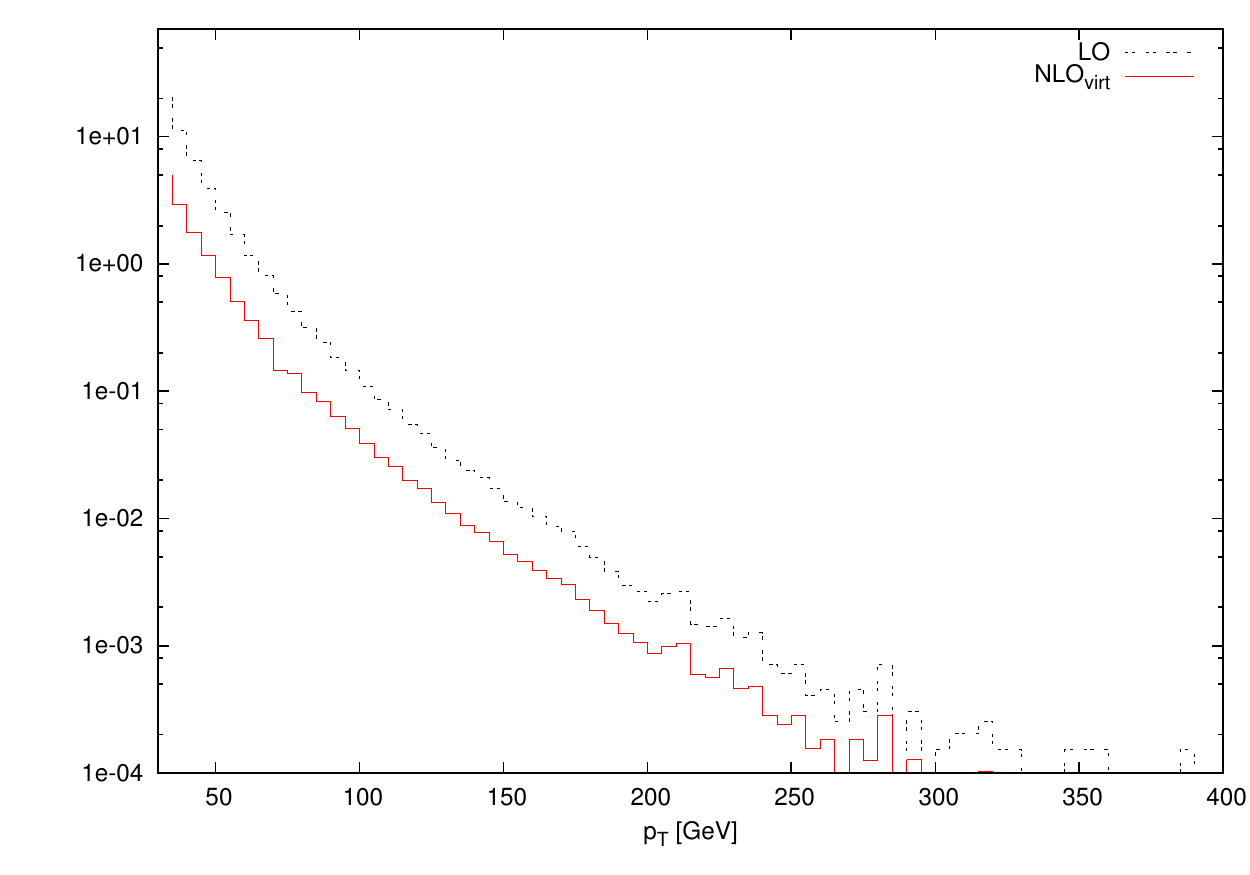}
\label{subfig:results:jet_pt-4th}
}
\caption{Transverse momentum ($p_T$) distributions of the
four $b$-jets. After the jets are ordered by their $p_T$,
the $p_T$ values of the hardest jets, second hardest etc
are distributed on the individual histograms. 
The distributions are based on a sample of $10^6$ events
and the $p_T$ region separated into slices
of~\unit{5}{\giga\electronvolt}.
The curve for $\text{NLO}_\text{virt}$ consists of the
contributions~$\born\sigma+\virt\sigma+\langle I(\varepsilon)\rangle$.}
\label{fig:results:pt}
\end{figure}

\begin{figure}[hbtp]
\includegraphics[width=0.7\textwidth]{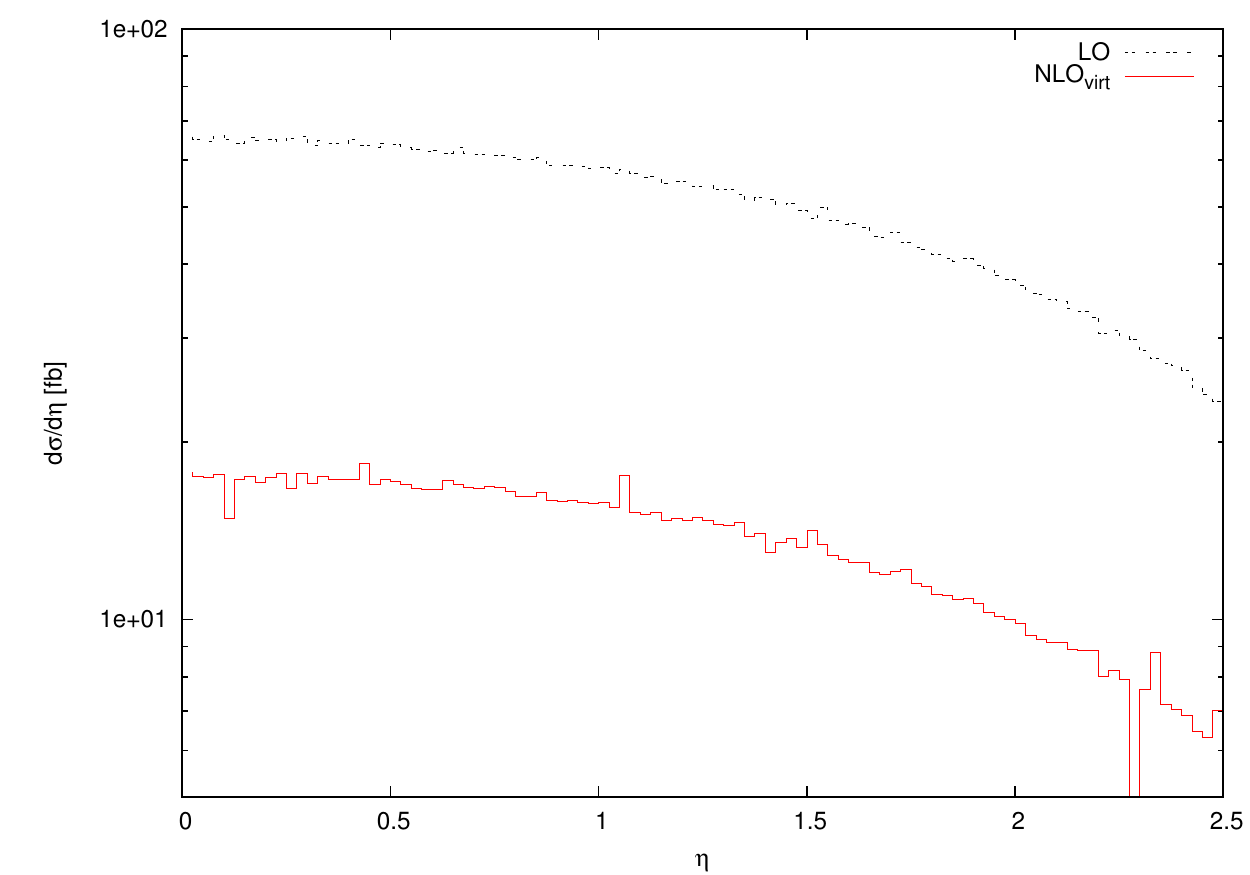}
\caption{Rapidity distribution of the hardest of the four $b$-jets.
This distribution has been produced from $10^6$ phase-space points,
where in every point the pseudorapidity~$\eta$
of the jet of highest $p_T$ has been used for the histogram.
The curve for $\text{NLO}_\text{virt}$ consists of the
contributions~$\born\sigma+\virt\sigma+\langle I(\varepsilon)\rangle$.}
\label{fig:results:eta-1st}
\end{figure}

Figure~\ref{fig:results:pt} shows the distribution of the transverse
momentum $p_T$ of each jet, where the jets are $p_T$-ordered
($p_T^{\text{1st}}\geq p_T^{\text{2nd}}\geq p_T^{\text{3rd}}
\geq p_T^{\text{4th}}$). The reader be reminded that the histograms
only contain a subset of the full \ac{nlo} correction.
The results are obtained from $10^6$
phase space points and only show small fluctuations, confirming
our earlier estimates of the statistical error.
The distributions of the pseudorapidity~$\eta$,
\index{pseudorapidity}
\begin{equation}
\eta(\vec{p})=-\ln\tan(\theta_p/2)\text{,}
\end{equation}
where $\theta_p$ is the angle between the momentum $\vec{p}$ and the beam axis,
look very similar for all four ($p_T$-ordered) jets.
Figure~\ref{fig:results:eta-1st} shows the $\eta$-distribution of the
hardest jet within the applied cuts, $\vert\eta\vert<2{.}5$; the interval
has been divided into 100~bins.

Two important discrimination criteria between signal and background
for \person{Higgs} boson searches
at the \ac{lhc} in the four-$b$ channel are the distributions of the
invariant mass of the four-$b$ and the two-$b$ systems.
\begin{figure}[hbtp]
\subfloat[$m_T(bb)$]{
\includegraphics[width=0.47\textwidth]{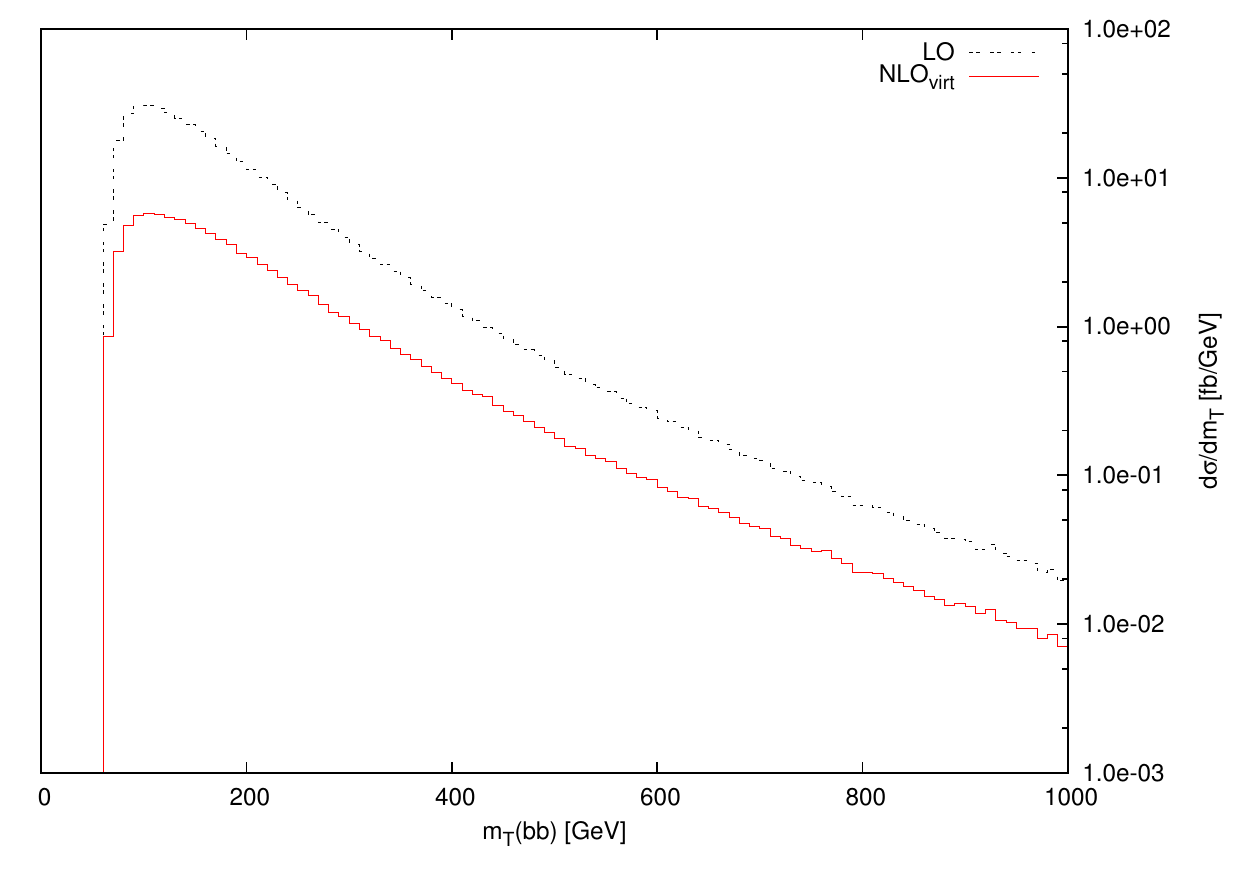}
\label{subfig:results:mt-bb}
}
\subfloat[$m_T(bbbb)$]{
\includegraphics[width=0.47\textwidth]{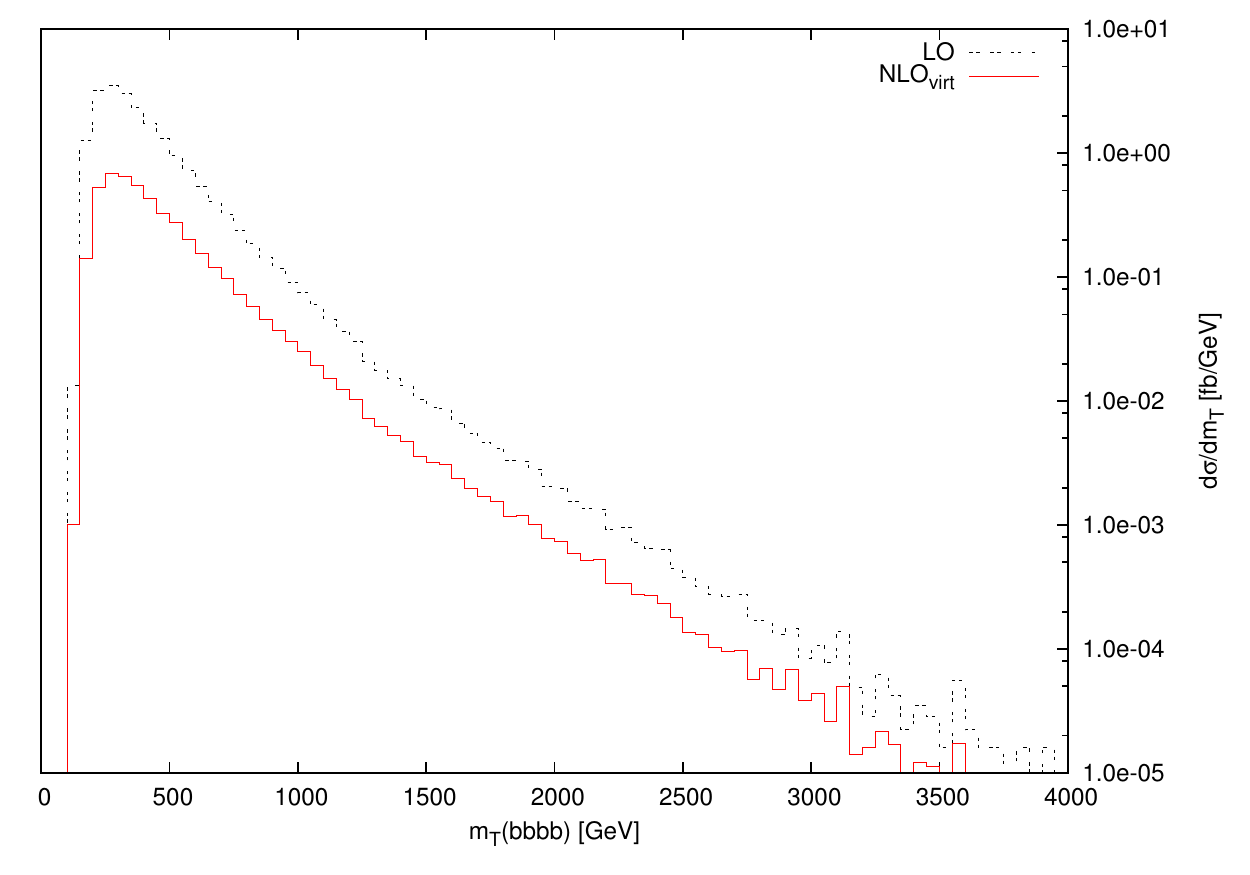}
\label{subfig:results:mt-bbbb}
}
\caption{Invariant transverse mass $m_T$ of the two- and four-$b$ jet
systems. Histogram~(a),
on the left, shows the distribution of $m_T$
for the two-jet systems. The histogram takes into account all six
possibilities of choosing a pair of $b$-jets, not distinguishing between
$b$ and~$\bar{b}$. Figure~(b), on the right, shows the transverse mass
of the four-jet system. In both cases
$\text{NLO}_\text{virt}$ denotes the combination
$\born\sigma+\virt\sigma+\langle I(\varepsilon)\rangle$.}
\label{fig:results:mt}
\end{figure}
The transverse mass of a particle of momentum $p$
is $m_T^2(p)=m^2+p_T^2$ and for a system of particles we define
\begin{align}
m_T(bbbb)&=m_T(p_1)+m_T(p_2)+m_T(p_3)+m_T(p_4)\text{,}\\
m_T(bb)&=m_T(p_i)+m_T(p_j),\quad i\neq j\text{.}
\end{align}
Figure~\ref{subfig:results:mt-bb} shows the distribution of
the transverse mass for the two-jet systems, where all six
possible ways of choosing $i$ and~$j$ have been taken into
account, which explains that the total cross-section is enhanced
by a factor of six.
In Figure~\ref{subfig:results:mt-bbbb} the transverse
mass of the four-jet system, $m_T(bbbb)$ is shown. Both histograms
have been produced from a sample of $10^6$ phase-space~points.

It should be pointed out that other observables can be obtained
easily with our approach as one stores a set of weighted events
both for the \ac{lo} and \ac{nlo} part of the amplitude.
These event files are independent of the observable and have to
be created only once.

%% file: res-intro.tex
In the previous chapter all theoretical foundations necessary
for a computation of \ac{nlo} matrix elements have been presented, many
of which relate directly to algorithms suitable for computer algebra
systems. An overview over the implementation of a
program for such a computation is given in Appendix~\ref{chp:implementation}.
One of the main ideas
behind~\cite{Binoth:2005ff} is the freedom of branching between
numerical evaluation and algebraic reduction of the amplitude for
different sets of basis functions: in the fully numerical approach
the form factors $A^{N,r}_{j_1\ldots j_r}$,
$B^{N,r}_{j_1,\ldots j_{r-2}}$ and~$C^{N,r}_{j_1\ldots j_{r-4}}$
are implemented in a numerical library and only a minimal set of
algebraic simplifications is carried out in order to bring the
expression into a form suitable for compilation with \fortran.
The other possibility of an evaluation is the reduction of the form
factors down to scalar integrals. 
Here one has the choice between the two function
sets $\mathcal{I}_\mathcal{N}$ and~$\mathcal{I}_\mathcal{S}$ as specified
in~\eqref{eq:qcd-diagramrep:Integralsets}. Having chosen the latter one
introduces inverse \person{Gram} determinants which have to cancel
algebraically in order to guarantee a numerically stable evaluation of
the expression. The results below which are given for the virtual
corrections have been obtained with the numerical implementation
of the reduction algorithm. An implementation of the algebraic
reduction has provided an independent check to verify the correctness
of the program.

As an application,
we have calculated the \ac{nlo} corrections to the
process $u\bar{u}\rightarrow b\bar{b}b\bar{b}$ in massless
\ac{qcd}. Continuing from the previous chapter, again
the focus is on the virtual corrections.
The real corrections are obtained using
\texttt{Whizard}~\cite{Whizard1.0,RealCorrections} but are
not included in the results shown below.
For the \ac{ir} subtraction terms we have used
\person{Catani}-\person{Seymour} dipole
subtraction~\cite{Catani:1996vz} with the modifications
as suggested in~\cite{Nagy:2003tz}. In our integration we
have set the cut off for the dipoles to~$\alpha=0{.}1$.

All data sets are generated with the following cuts, which are chosen
to agree with the study of~\cite{Djouadi:2000gu},
to which the process $u\bar{u}\rightarrow b\bar{b}b\bar{b}$ is an
important background. In particular, these cuts are
\begin{itemize}
\item a $p_T$ cut of $p_T>\unit{25}{\giga\electronvolt}$,
\item a rapidity cut of $\vert\eta\vert<2{.}5$ and
\item a separation cut of~$\Delta R>0{.}4$\text{.}
\end{itemize}
Unless stated otherwise, the factorisation scale~$\mu_F$
and the renormalisation scale~$\mu_R$ are chosen as
the average transverse momentum of the final state partons,
\begin{equation}
\mu_F=\mu_R=\sum_{i=1}^4 \frac{p_T}{4}\text{.}
\end{equation}
For the \acp{pdf} we have used the set CTEQ\,6{.}5~\cite{Tung:2006tb}.

%% file: conclusion.tex
\chapter{Conclusion}
\label{chp:conclusion}
\begin{headquote}{Marie Curie}
One never notices what has been done; one can only see what remains to be done
\end{headquote}

\section{Summary}
The \ac{lhc} is of outstanding importance for the next decade of particle
physics. For the first time a collider experiment can access the
energies of electroweak symmetry breaking directly and thus almost
certainly answer the question if the \ac{sm} is an adequate description
of elementary particle physics at this energy scale.
The precision that will be achieved by the experiment, however, must
be met by the phenomenological predictions for the \ac{sm} or any
model describing \ac{bsm} physics and I have motivated earlier that
this precision goal implies that for many processes \ac{qcd} predictions 
must be made with at least one-loop~accuracy~\cite{Bern:2008ef}.

In this work I presented and implemented an algorithm for the calculation of
the virtual corrections of processes with many particles in the final state.
The virtual contributions currently form the bottleneck of most
cross-section calculations at \ac{nlo}; while for the \person{Born} level
and the real emission contributions one can rely on automated tools,
the computation of the virtual corrections very often remains
highly customised and process specific.
The \ac{lhc}'s demand for \ac{nlo} predictions motivates the
automation of one-loop calculations, especially for the case of
\ac{qcd}~corrections.

As a possible solution I presented an algorithm
based on the calculation of \person{Feynman} diagrams
in which the method of spinor helicity projections is used to obtain
compact results. Different approaches for the treatment of the \ac{sun}
colour algebra have been presented and compared.
For the calculation of the tensor integrals,
arising from the momentum integration of a virtual particle,
a reduction algorithm has been discussed which
projects the integrals on to a basis of form factors,
which then can be evaluated
numerically~\cite{Binoth:2005ff,Golem90:2008inprep}.
The integration of the differential \ac{nlo} cross-section
has been improved by a method that avoids the destabilisation
of adaptive Monte Carlo programs in their initialisation~phase. 

The algorithm outlined above has been implemented for massless particles
and has been used to compute the virtual corrections for the
process $u\bar{u}\rightarrow b\bar{b}b\bar{b}$ in the massless limit.
The results indicate that our method is well suited for the
computation of \ac{qcd} processes with up to four final state particles
at~\ac{nlo}.

\section{Outlook}
One of the disadvantages of \person{Feynman} diagram based methods
is the factorial growth of the number of diagams as one increases
the number of external particles in a process.
This drawback is avoided by alternative constructions of
\ac{qcd} tree-level
amplitudes~\cite{Cachazo:2004kj,Parke:1986gb,Britto:2004ap,Britto:2005fq}
combined with the observation that one-loop amplitudes to a large extend
are defined by their analytical properties~\cite{Bern:1994zx,Bern:1994cg},
having triggered the development of \emph{unitarity based methods}
for the calculation of \ac{nlo}
matrix elements~\cite{Berger:2008sj,Giele:2008bc,Ossola:2007ax}.

This calculation demonstrates, however, that processes with
four-particle final states can be dealt with in an efficient
manner by our method
and it can be anticipated that it will provide a tool for
many processes becoming relevant in the near future.

In order to provide a precision prediction for the 
\ac{qcd} corrections to $pp\rightarrow b\bar{b}b\bar{b}$
some work beyond the scope of this thesis remains to be done.
The calculation of the real corrections for the
process~$u\bar{u}\rightarrow b\bar{b}b\bar{b}$ is currently
in progress and the calculation of the amplitude
$gg\rightarrow b\bar{b}b\bar{b}$ is under investigation.
The modifications for the latter case turn out to be
straight forward due to the genericity in
the design of the implementation. An extension to
massive amplitudes and non-\ac{qcd} particles 
is currently under consideration.
The extension to masses in internal propagators is
relevant for many processes with vector bosons or top
quarks in the final state, such as $VVb\bar{b}$,
$VV+2\,\text{jets}$ or $VVV+\text{jet}$, where $V$
represents either a $W^\pm$ or a $Z$ boson, or, for example,
$t\bar{t}b\bar{b}$\footnote{The process $pp\rightarrow t\bar{t}b\bar{b}$
is currently being calculated by the authors of~\cite{Bredenstein:2008ia}}
and $t\bar{t}+2\,\text{jets}$. All of these processes are
motivated either by \ac{sm} \person{Higgs} boson searches or
\ac{bsm} physics~\cite{Bern:2008ef};
in the near future the \ac{lhc} experiment
will release its first results and the above cross-sections
will be required as an input to the analysis of the data.
Along with the \ac{lhc} the particle physics community has
set up a large scale computing infrastructure,
the Grid~\cite{Bird:2005js}. The program described in
Appendix~\ref{chp:implementation} is well suited
and, in fact, has been designed for a distributed computing
environment like the Grid. Hence, the chosen approach has
the capabilities to lead into an automatic tool for the
computation of virtual corrections at one-loop,
complementing existing tools for tree-level calculations.

%% file: appendix-distributions.tex
\chapter{Distributions}
\label{app:distributions}
\begin{headquote}{Paul Adrien Maurice Dirac}
I consider that I understand an equation when I can predict the
properties of its solutions, without actually solving it. 
\end{headquote}
\section*{Introduction}
For a general introduction to generalised functions the reader is referred
to mathematical standard text books such as~\cite{Vladimirov:2002}.
For this work it is sufficient to define distributions by their action
on test functions under integration. Let $G(x)$ be a distribution and
$f(x)$ be a smooth, continuous test function. The distribution $G(x)$
defines an integral transform
\begin{equation}
F(y)=\int_{-\infty}^\infty\!\!\diff{x}G(x-y)f(x)
\end{equation}
for all values of $y$ where the integral converges.
In many practical applications one is only interested in
the integral
\begin{equation}
F(0)=\int_{-\infty}^\infty\!\!\diff{x}G(x)f(x)\text{.}
\end{equation}

Another way of defining distributions is by a sequence of ordinary
functions $G_\epsilon(x)$ with the property
\begin{equation}
F(y)=\lim_{\epsilon\rightarrow0}
\int_{-\infty}^\infty\!\!\diff{x}G_\epsilon(x-y)f(x)
=\int_{-\infty}^\infty\!\!\diff{x}G(x-y)f(x)\text{.}
\end{equation}

\section{The \texorpdfstring{$\delta$--}{Delta} Distribution}
One of the most commonly used distributions is the $\delta$--distribution.
It is defined by
\begin{equation}
f(y)=\int_{-\infty}^\infty\!\!\diff{x}\delta(x-y)f(x)\text{.}
\end{equation}
It can be represented as the limit $\epsilon\rightarrow0$ 
of the sequence of functions
\begin{equation}
\delta_\epsilon(x)=\frac1{\epsilon\sqrt{\pi}}e^{-\frac{x^2}{\epsilon^2}}
\end{equation}
or by its \person{Fourier} transform
\begin{equation}
\delta(x)=\frac1{2\pi}\int_{-\infty}^\infty\!\!\diff{k}e^{ikx}
\end{equation}

A useful identity for the $\delta$--distribution is
\begin{equation}
\delta(g(x))=\sum_{i}\frac1{\vert g^\prime(x_i)\vert}\delta(x-x_i)\text{,}
\end{equation}
where $x_i$ are all roots of $g(x_i)=0$.

\section{The Plus Distribution}
The plus distribution $(\cdot)_+$ is defined as
\begin{equation}\label{eq:appendix-distributions:plusdef}
F=\int_0^1\!\!\diff{x}\left(g(x)\right)_+f(x)=
\int_0^1\!\!\diff{x}\left(f(x)g(x)-f(1)g(x)\right)
\end{equation}
where the test function $f$ is a regular function and
$g$ is singular at $x=1$. Typically these singular
functions are $g(x)=1/(1-x)$ or $g(x)=\ln(1-x)/(1-x)$.

Many authors also use $g(x)=1/x$. Then it is implied
that in Equation~\eqref{eq:appendix-distributions:plusdef}
one replaces $f(1)$ by $f(0)$.

\chapter{The \person{Moore}-\person{Penrose} Inverse}
\label{app:MPinverse}
\begin{headquote}{Whitfield Diffie}
We in science are spoiled by the success of mathematics.
Mathematics is the study of problems so simple that they
have good solutions. 
\end{headquote}

\index{Pseudo-inverse|see{\person{Moore}-\person{Penrose} pseudoinverse}}
\index{Penrose ps@\person{Penrose} pseudoinverse|see{\person{Moore}-\person{Penrose} pseudoinverse}}
\index{Moore Penrose@\person{Moore}-\person{Penrose} pseudoinverse|(main}
The pseudoinverse~$\tilde{M}$ of an arbitrary, real matrix~$M$
has to satisfy\footnote{Since~$S$ is real I will use~$S^\transposed$ instead of~$S^\dagger$.}
\begin{subequations}
\begin{align}
M\tilde{M}M&=M\text{,}\\
\tilde{M}M\tilde{M}&=\tilde{M}\text{,}\\
(\tilde{M}M)^\transposed&=\tilde{M}M\quad\text{and}\\
(M\tilde{M})^\transposed&=M\tilde{M}\text{.}
\end{align}
\end{subequations}
These properties ensure the following
\begin{theorem}\label{thrm:penrose} Let $M\in\Rset^{m\times n}$ be an arbitrary matrix and~$\tilde{M}$
its \person{Moore}-\person{Penrose} pseudoinverse. The linear equation $Mb=v$ has a
solution, if and only if $M\tilde{M}v=v$, and the solution $b(u)=\tilde{M}v+(\One-\tilde{M}M)u$
is the most general solution, where~$u\in\Rset^n$ is an arbitrary vector that parametrises
the homogeneous part of the solution.
\end{theorem}

It is trivially shown that the condition~$M\tilde{M}v=v$ is sufficient for $b(u)$ to be a solution
of the linear system. Let $b^\prime$ another solution to the system, i.e. $Mb^\prime=v$; then
$b^\prime=\tilde{M}Mb^\prime+(\One-\tilde{M}M)b^\prime=\tilde{M}v+(\One-\tilde{M}M)b^\prime=b(b^\prime)$.
This proves the generality of~$b(u)$. To prove the other direction, we assume $b_0$ to be a solution of the linear
equation, and hence $y=Mb_0=M\tilde{M}Mb_0=M\tilde{M}y$, and everything is proved.

For a symmetric squared matrix~$M\in\Rset^{N\times N}$ of rank~$r$ the
 pseudoinverse is
unique, and for~$r=N$ it is identical with the inverse~$M^{-1}$. This is easily shown using that for
any symmetric matrix we find an orthogonal transformation matrix~$U$ such that
\begin{displaymath}
UMU^\transposed=\diag(\lambda_1,\ldots,\lambda_r,\underbrace{0,\ldots,0}_{N-r}),\quad\lambda_i\neq0,i\in\{1,\ldots,r\}\text{.}
\end{displaymath}
The pseudoinverse then is constructed as
\begin{displaymath}
\tilde{M}=U^\transposed\diag(\lambda_1^{-1},\ldots,\lambda_r^{-1},\underbrace{0,\ldots,0}_{N-r})U\text{.}
\end{displaymath}
\index{Moore Penrose@\person{Moore}-\person{Penrose} pseudoinverse|)}

%% file: appendix-intdetails.tex
\chapter{Loop Integrals}
\label{app:loops}
\begin{headquote}{Edsger Dijkstra}
The traditional mathematician recognizes and appreciates
mathematical elegance when he sees it.
I propose to go one step further,
and to consider elegance an essential ingredient of mathematics:
if it is clumsy, it is not mathematics.
\end{headquote}

In this appendix I describe the procedure of transforming loop integrals into
\person{Feynman} parameter in more detail than I did in the chapters before.
This part of the calculation is also described in many textbooks about
particle physics, as for example in~\cite{Peskin:QFT}, but some proofs
that contribute to the understanding of the underlying maths are usually left
out.

In the Section~\ref{sec:app-intdetails:maths} I establish basic facts 
about the $\Gamma$ and $B$-function which are essential for the
introduction of \person{Feynman} parameters. The basic relation is
\index{Feynman@\person{Feynman}!parameter}
\begin{equation}
\label{eq:loops:FP}
\frac{1}{\prod_{k=1}^nA_k^{\alpha_k}}=\int_0\!\!\diff[n]{z}\delta_z
\frac{\prod_{k=1}^nz_k^{\alpha_k-1}}{\left(\sum_{k=1}^nA_kz_k\right)^\alpha}
\frac{\Gamma(\alpha)}{\prod_{k=1}^n\Gamma(\alpha_k)}
\end{equation}
for general complex\footnote{In fact for $\Re(\alpha_k)>0$}~$\alpha_k$
and~$\alpha$, where $\alpha=\sum_{k=1}^n \alpha_k$.
As a direct corollary one can rewrite
equation~\eqref{eq:qcd-dimreg:fppoly01} for non-integer exponents.
Other than most text books,
which, if at all, prove Equation~\eqref{eq:loops:FP}
by induction I derive this
formula by showing its equivalence to \person{Schwinger} parameters,
which are more convenient to make the connection from the axiomatic
introduction of loop integrals in
Section~\ref{sec:int-details:axiomatic}.

At that point one is ready to go through the remaining steps,
the \person{Wick} rotation and the integration of $d$-dimensional spherical
coordinates, the \person{Feynman} parametrisation and finally the expansion
in~$\varepsilon$, which are explained in
Section~\ref{sec:app-intdetails:evalint}.

%%%%%%%%%%%%%%%%%%%%%%%%%%%%%%%%%%%%%%%%%%%%%%%%%%%%%%%%%%%%%%%%%%%%%
\section{Mathematical Prerequisites}\label{sec:app-intdetails:maths}
%%%%%%%%%%%%%%%%%%%%%%%%%%%%%%%%%%%%%%%%%%%%%%%%%%%%%%%%%%%%%%%%%%%%%

%%%%%%%%%%%%%%%%%%%%%%%%%%%%%%%%%%%%%%%%%%%%%%%%%%%%%%%%%%%%%%%%%%%%%
\subsection{The Gamma-Function}
%%%%%%%%%%%%%%%%%%%%%%%%%%%%%%%%%%%%%%%%%%%%%%%%%%%%%%%%%%%%%%%%%%%%%
In this chapter I follow closely chapter~11 of~\cite{Barner:Analysis}.
The authors give the theorems about the $\Gamma$-function in form of
exercises to the reader.

The definition of~$\Gamma(t)$ is given via
\begin{equation}
\label{eq:gamma:definition}
\Gamma(t)=\int_0^\infty x^{t-1}e^{-x}\diff{x},\quad t>0
\end{equation}
By direct calculation one finds
\begin{equation}
\Gamma(1)=1\text{,}
\end{equation}
and we will the later that this normalisation defines the $\Gamma$-function
together with the two properties in the following
\begin{lemma}
The $\Gamma$-function has the following two properties:
\begin{enumerate}
\item $\Gamma(t+1)=t\Gamma(t)$ and
\item $\frac{\diff[2]}{\diff{t^2}}\ln\Gamma(t)>0$ for all $t>0$.
\end{enumerate}
\end{lemma}
The first property is shown by integration by parts
of~\eqref{eq:gamma:definition}. The second property is equivalent to
\begin{equation}
\left\vert\begin{array}{ll}\Gamma(t)&\Gamma^\prime(t)\\
\Gamma^\prime(t)&\Gamma^{\prime\prime}(t)\end{array}\right\vert>0\text{.}
\end{equation}
This can also be interpreted as the condition that the equation
\begin{equation}
\varphi_t(\lambda)\equiv
\lambda^2\Gamma(t)+2\lambda\Gamma^\prime(t)+\Gamma^{\prime\prime}(t)=0
\end{equation}
has no real roots~$\lambda$. We can calculate
\begin{equation}
\frac{\diff[n]}{\diff{t^n}}\Gamma(t)=
\int\left(\ln x\right)^n x^{t-1}e^{-x}\diff{x}
\end{equation}
directly and hence find
\begin{equation}
\varphi_t(\lambda)=
\int\left(\lambda+\ln x\right)^2 x^{t-1}e^{-x}\diff{x}>0
\quad\forall\lambda\in\Rset\text{.}\quad\Box
\end{equation}
These properties allow to extend the definition of~$\Gamma(t)$ to all
non-inter negative values of~$t$ by the recursive definition
\begin{equation}
\Gamma(t)\equiv\frac{\Gamma(t+1)}{t},%
\quad\forall t\in\Rset^{-}-\{0,-1,-2,\ldots\}\text{.}
\end{equation}
From this definition one also finds that~$\Gamma(t)$ has single Poles at
all negative integer numbers where the residue is
\begin{equation}
\mathrm{Res}_{z=-n}\Gamma(z)=\frac{(-1)^n}{n!},\quad n=0,1,2,\ldots
\end{equation}

The following lemma prepares the theorem about the uniqueness of the
$\Gamma$-function.
\begin{lemma}\label{lemma:gamma:differenceeq}
Let~$g(t)$ be a differentiable function defined for~$t>0$ which obeys the
conditions
\begin{enumerate}
\item $g(t+1)-g(t)=\frac{1}{t}$ and\\
\item $g^\prime(t)\geq0$.
\end{enumerate}
There is a $c\in\Rset$ such that
\begin{displaymath}
g(t)=c-\frac{1}{t}
+\sum_{k=1}^\infty\left(\frac{1}{k}-\frac{1}{k+t}\right)\text{.}
\end{displaymath}
\end{lemma}
If one constructs the function
\begin{displaymath}
p(t)\equiv g(t)+\frac{1}{t}
-\sum_{k=1}^\infty\left(\frac{1}{k}-\frac{1}{k+t}\right)
\end{displaymath}
one can use condition~(1) to prove that~$p(t+1)=p(t)$. If $p(t)$ is not
constant then there is a $t_0\in (t;t+1)$ such that $p(t_0)\neq p(t)$, and
hence either $p(t_0)-p(t)$ or $p(t+1)-p(t_0)$ must be negative. According to
the mean value theorem there lies a point~$t_1$ between $t$ and~$t_0$
($t_0$ and $t+1$ respectively) where
$(t_0-t)\cdot p^\prime(t_1)=p(t_0)-p(t)$ 
or $(t+1-t_0)\cdot p^\prime(t_1)=p(t+1)-p(t_0)$ respectively and hence we
can find a positive number~$\epsilon$ such that~$p^\prime(t_1)=-\epsilon$.
Plugging in the definition of~$p(t)$ one obtains
\begin{equation}
\sum_{k=0}^\infty\frac{1}{(k+t_1)^2}-\epsilon=g^\prime(t_1)\text{.}
\end{equation}
The series $\sum(1/k^2)$ converges which means that for any given~$\epsilon$
one can find an~$N=N(\epsilon)$ such that
\begin{equation}
\epsilon>\sum_{k=N}^\infty\frac{1}{k^2}=\sum_{k=0}^\infty\frac{1}{(k+N)^2}
\text{.}
\end{equation}
Choosing~$t>N$ which also means~$t_1>t>N$ one has
\begin{equation}
\sum_{k=0}^\infty\frac{1}{(k+t_1)^2}<\sum_{k=0}^\infty\frac{1}{(k+N)^2}
<\epsilon\quad\Rightarrow\quad g^\prime(t_1)<0\text{,}
\end{equation}
which is in contradiction to the assumptions and one must conclude that
$p^\prime(t)=0$ for all~$t>0$, or equally $p(t)=c$. $\Box$

\begin{corollary}
\label{corollary:gamma:h(t)}
Let~$h(t)$ be a double-differentiable function that is
defined for~$t>0$ and fulfils
\begin{enumerate}
\item $h(t+1)-h(t)=\ln(t)$ and
\item $h^{\prime\prime}(t)\geq0$.
\end{enumerate}
Two functions obeying~(1) and~(2) differ by only a additive constant.
\end{corollary}

Clearly $h^\prime(t)=g(t)$ conforms with the assumptions of 
Lemma~\ref{lemma:gamma:differenceeq} and therefore ~$h(t)$ must be an
antiderivative of~$g(t)$,
\begin{equation}
\label{eq:gamma:h(t)}
h(t)=c-Ct-\ln t -\sum_{k=1}^\infty\left(\ln(1+\frac{t}{k})-\frac{t}{k}\right)%
\text{,}
\end{equation}
however, (1)~requires~$C$ to be fixed as
\begin{equation}
C=\lim_{N\rightarrow\infty}\left(\sum_{k=1}^N\frac{1}{k}-\ln N\right)%
\equiv\gamma_E\text{.}\Box
\end{equation}

\begin{theorem}\label{thm:appendix-intdetails:GammaIsUnique}
Given a double differentiable function~$f(t)$ that is defined
for~$t>0$ and fulfils
\begin{enumerate}
\item $f(t+1)=tf(t)$ and
\item $f(t)f^{\prime\prime}(t)-(f^\prime(t))^2\leq0$.
Then there is a~$c\in\Rset$ such that
\begin{equation}
f(t)=c\Gamma(t)\text{.}
\end{equation}
\end{enumerate}
\end{theorem}
The theorem is a direct consequence of the fact that both,
$\ln f(t)$ and~$\ln\Gamma(t)$ fulfil the assumptions
of Corollary~\ref{corollary:gamma:h(t)}. Therefore
\begin{equation}
\ln f(t) = c + \ln\Gamma(t)\text{,}
\end{equation}
and by exponentiation everything is proved.~$\Box$

Taking the explicit form of~$h(t)$ from~\eqref{eq:gamma:h(t)}, the
normalisation~$\Gamma(1)=1$ leads to~$\Gamma^\prime(1)=-C=-\gamma_E$,
which is usually referred to as \person{Euler}'s constant. This leads to
the expansion of the $\Gamma$-function for small values~$\varepsilon$,
\begin{equation}
\Gamma(\varepsilon)=\frac{1}{\varepsilon}\Gamma(1+\varepsilon)=
\frac{1}{\varepsilon}\left(\Gamma(1)+\varepsilon\Gamma^\prime(1)+
{\mathcal O}(\varepsilon^2)\right)
=\frac{1}{\varepsilon}-\gamma_E+{\mathcal O}(\varepsilon)
\end{equation}

Some integrals, however, require to take also higher order terms into
account. Therefore we need the values of higher derivatives of the
Gamma function, i.e. $\Gamma^\prime(1)$, $\Gamma^{\prime\prime}(1)$
and so on. A convenient notation can be achieved by introducing the
digamma and polygamma functions~\cite{Espinosa:2004},
\begin{align}
\Psi(q) &=\frac{\diff}{\diff{q}}\ln\Gamma(q)\quad\text{and}\\
\Psi^{(m)}(q)&=\frac{\diff[m]}{\diff{q}^m}\Psi(q)
\quad\text{respectively.}
\end{align}
These functions are directly related to the
\person{Hurwitz} zeta function
\begin{equation}
\zeta(z,q)=\sum_{n=0}^\infty\frac{1}{(q+n)^z}
\end{equation}
which is a generalisation of the \person{Riemann} zeta function
\begin{equation}
\zeta(z)=\zeta(z,1)=\sum_{n=1}^\infty\frac{1}{n^z}\text{.}
\end{equation}
By inductions one can show that
\begin{equation}
\left(\frac{\partial}{\partial q}\right)^m \zeta(z,q)=
(-1)^m(z)_m\zeta(z+m,q)
\end{equation}
with the \person{Pochhammer}
symbol\index{Pochhammer@\person{Pochhammer}!symbol|main}
\index{()@$(k)_n$|see{\person{Pochhammer} symbol}}
\begin{equation}
(z)_m\equiv\frac{\Gamma(z+m)}{\Gamma(z)}=(z+m-1)(z+m-2)\cdots(z+1)z\text{.}
\end{equation}
As a direct consequence for $m>0$ one obtains
\begin{equation}
\Psi^{(m)}(q)=(-1)^{m+1}m!\zeta(m+1,q)\text{.}
\end{equation}

On the other hand $\Psi^{(m)}(q)$ is related to the derivatives of the
Gamma~function. For the first two derivatives we get
\begin{align}
\Psi^{(1)}(q)&=\frac{\Gamma^{\prime}(q)}{\Gamma(q)}\quad\text{and}\\
\Psi^{(2)}(q)&=\frac{\Gamma^{\prime\prime}(q)\Gamma(q)-(\Gamma^\prime(q))^2}{(\Gamma(q))^2}\text{.}
\end{align}
Evaluating the second equation at $q=1$ leads to
\begin{equation}
\Psi^{(2)}(1)=\zeta(2)=\frac{\pi^2}{6}=\Gamma^{\prime\prime}(1)-\gamma_E^2\text{.}
\end{equation}
Therefore one can extend the expansion of the Gamma function to the required accuracy,
\begin{equation}
\Gamma(1+\varepsilon)=1-\gamma_E\varepsilon+\left(\frac{\pi^2}{12}+\frac{\gamma_E^2}{2}\right)\varepsilon^2
+{\mathcal O}(\varepsilon^3)
\end{equation}

%%%%%%%%%%%%%%%%%%%%%%%%%%%%%%%%%%%%%%%%%%%%%%%%%%%%%%%%%%%%%%
\subsection{The Beta-Function}
%%%%%%%%%%%%%%%%%%%%%%%%%%%%%%%%%%%%%%%%%%%%%%%%%%%%%%%%%%%%%%
The Beta function is closely related to the previously discussed Gamma
function; in the literature these function often share the common notation
of \person{Euler} integrals of first respective second kind.
In this section I will prove the relation
\begin{equation}\label{eq:app:BetaFunction}
B(s,t)\equiv\int_0^\infty\!\!\diff{x}\frac{x^{s-1}}{(1+x)^{s+t}}
=\int_0^1\!\!\diff{y}y^{s-1}(1-y)^{t-1}=
\frac{\Gamma(s)\Gamma(t)}{\Gamma(s+t)},
\quad\mathrm{Re}(s),\mathrm{Re}(t)>0\text{.}
\end{equation}

We first prove
\begin{equation}
\int_0^\infty\!\!\diff{x}\frac{x^{s-1}}{(1+x)^{s+t}}=
\frac{\Gamma(s)\Gamma(t)}{\Gamma(s+t)}\text{.}
\end{equation}
By the substitution~$y=1/x$ one can show directly that~$B(s,t)=B(t,s)$,
integration by part proves $(s+t)B(s,t+1)=tB(s,t)$. With
\begin{equation}
f_s(t)\equiv B(s,t)\Gamma(s+t)
\end{equation}
one can use Theorem~\ref{thm:appendix-intdetails:GammaIsUnique} to show
that~$f_s(t)=g(s)\Gamma(t)$ for some function~$g(s)$. To determine~$g$
we evaluate the integral
\begin{equation}
g(s)=f_s(1)=\Gamma(s+1)\int_0^\infty\!\!\diff{x}\frac{1}{(1+x)^{s+1}}=
\Gamma(s)\text{.}
\end{equation}
Therefore we find
\begin{equation}
B(s,t)=\frac{\Gamma(s)\Gamma(t)}{\Gamma(s+t)}\text{.}
\end{equation}
To show that the two integrals in~\eqref{eq:app:BetaFunction}
are the same one would carry out the substitution $y=x/(x+1)$.

%%%%%%%%%%%%%%%%%%%%%%%%%%%%%%%%%%%%%%%%%%%%%%%%%%%%%%%%%%%%%%
\subsection{Some Useful Relations}
%%%%%%%%%%%%%%%%%%%%%%%%%%%%%%%%%%%%%%%%%%%%%%%%%%%%%%%%%%%%%%
The different representations of Beta- and Gamma functions are important
to establish relations for products and series of Gamma functions.

The following relations can be used for the integration of the dipoles
in~\cite{Catani:1996vz},
\begin{equation}\label{eq:app-intdetails:GammaSum01}
\sum_{\nu=0}^\infty\frac{\Gamma(a+\nu)}{\Gamma(b+\nu)}=
\frac{\Gamma(b-a-1)\Gamma(a)}{\Gamma(b-a)\Gamma(b-1)}\text{.}
\end{equation}
The proof is as follows:
\begin{multline}
\sum_{\nu=0}^\infty\frac{\Gamma(a+\nu)}{\Gamma(b+\nu)}=
\frac{1}{\Gamma(b-a)}\sum_{\nu=0}^\infty B(a+\nu, b-a)=\\
\frac{1}{\Gamma(b-a)}%
\sum_{\nu=0}^\infty\int_0^1\!\!\diff{t}\,t^{a-1+\nu}(1-t)^{b-a-1}=\\
\frac{1}{\Gamma(b-a)}%
\int_0^1\!\!\diff{t}\,t^{a-1}(1-t)^{b-a-1}\sum_{\nu=0}^\infty t^\nu=\\
\frac{1}{\Gamma(b-a)}%
\int_0^1\!\!\diff{t}\,t^{a-1}\frac{(1-t)^{b-a-1}}{(1-t)}=
\frac{B(a,b-a-1)}{\Gamma(b-a)}\text{.}
\end{multline}

As an example this can be applied to the integral
\begin{multline*}
\int_0^1\!\!\diff{z}z^{-\varepsilon}(1-z)^{-\varepsilon}
\int_0^1\!\!\diff{y}y^{-1-\varepsilon}(1-y)^{1-2\varepsilon}
\frac{2}{1-z(1-y)}
\displaybreak[1]
=\\
2\int_0^1\!\!\diff{z}z^{-\varepsilon}(1-z)^{-\varepsilon}
\int_0^1\!\!\diff{y}y^{-1-\varepsilon}(1-y)^{1-2\varepsilon}
\sum_{\nu=0}^\infty z^\nu(1-y)^\nu=
\displaybreak[1]
\\
2\sum_{\nu=0}^\infty B(1-\varepsilon+\nu,1-\varepsilon)%
B(2-2\varepsilon+\nu,-\varepsilon)=
\displaybreak[1]
\\
2\sum_{\nu=0}^\infty
\frac{\Gamma(1-\varepsilon+\nu)\Gamma(1-\varepsilon)}{%
\Gamma(2-2\varepsilon+\nu)}
\frac{\Gamma(2-2\varepsilon+\nu)\Gamma(-\varepsilon)}{%
\Gamma(2-3\varepsilon+\nu)}=
\displaybreak[1]
\\
2\Gamma(-\varepsilon)\Gamma(1-\varepsilon)\sum_{\nu=0}^\infty
\frac{\Gamma(1-\varepsilon+\nu)}{\Gamma(2-3\varepsilon+\nu)}=
\displaybreak[1]
\\
2\Gamma(-\varepsilon)\Gamma(1-\varepsilon)
\frac{\Gamma(1-\varepsilon)\Gamma(-2\varepsilon)}{%
\Gamma(1-2\varepsilon)\Gamma(1-3\varepsilon)}=\\
2B(1-\varepsilon,-2\varepsilon)B(1-\varepsilon, -\varepsilon)\text{.}
\end{multline*}

%%%%%%%%%%%%%%%%%%%%%%%%%%%%%%%%%%%%%%%%%%%%%%%%%%%%%%%%%%%%%%%%%%%%%%%
\section{An Axiomatic Approach}\label{sec:int-details:axiomatic}
%%%%%%%%%%%%%%%%%%%%%%%%%%%%%%%%%%%%%%%%%%%%%%%%%%%%%%%%%%%%%%%%%%%%%%%
\index{dimensional regularisation!axioms|(}
The introduction of dimensional regularisation arises from the observation that
many of the loop integrals that diverge in four dimensions become convergent in
a space with other than four dimensions. To take a smooth limit to four dimensions,
starting from a dimensionality where the integral is well behaved, involves the
concept of non-integer dimensions: The dimension is defined to be a complex
parameter~$d$, and the divergence of the integral shows up in poles for certain
integer values of~$d$, as we will see later. However, an integral in
non-integer dimensions can only be defined on an infinite dimensional vector space;
It is the definition of the integral that carries the parameter~$d$, not the space
it acts on. To make sure we get the right results back in four dimensions we require
the integration to have certain properties.~\cite{Wilson:1972cf}

Given a linear space~$V$
with a \person{Hermit}ian form~$p\cdot q\in\Rset$ for all~$p,q\in V$ such
that the four dimensional \person{Minkowski} space~$\mathcal M$ is a subspace of~$V$
and the dot-product on~$\mathcal M$ is the restriction of~$\cdot$ on~$\mathcal M$.
For any complex~$d$ we introduce the functional
\begin{equation}
\int_{k\in V}\!\!\diff[d]k f(k)
\end{equation}
on \person{Lorentz} covariant 
functions~$f: V\rightarrow\Cset$,
i.e.~$f(k)=\tilde{f}(k^2, k\cdot q_1, k\cdot q_2,\ldots, k\cdot q_N)$ where
$k,q_1,\ldots,q_N\in V$.
The following properties uniquely define the integration
\begin{subequations}
\begin{description}
\item[Linearity] For any two complex numbers $a$ and $b$ and functions $f$ and $g$
\begin{displaymath}
\int_{k\in V}\!\!\diff[d]k \left(a f(k)+b g(k)\right)
=a\int_{k\in V}\!\!\diff[d]k f(k)+b\int_{k\in V}\!\!\diff[d]k g(k)\text{,}
\end{displaymath}
\item[Translation Invariance] For any vector~$q\in V$
\begin{displaymath}
\int_{k\in V}\!\!\diff[d]k f(k+q)=\int_{k\in V}\!\!\diff[d]k f(k)\text{,}
\end{displaymath}
\item[Scaling] For any complex number~$s$
\begin{displaymath}
\int_{k\in V}\!\!\diff[d]k f(sk)=s^{-d}\int_{k\in V}\!\!\diff[d]k f(k)\text{,}
\end{displaymath}
\item[Normalisation]
\begin{displaymath}
\int_{k\in V}\!\!\diff[d]k e^{-k^2}=\pi^{d/2}\text{.}
\end{displaymath}
\end{description}
\end{subequations}
\index{dimensional regularisation!axioms|)}

To prove this one can use a\index{Generating function}
generating function~$f(k) = e^{-sk^2+k\cdot p}$ for
a complex parameter~$s$ and a vector~$p\in V$. We can solve the integral
over~$f$ by just using the above conditions:
\begin{multline}
\int_{k\in V}\!\!\diff[d]k e^{-sk^2+k\cdot p}=
\int_{k\in V}\!\!\diff[d]k e^{-s\left(k-p/(2s)\right)^2-p^2/(4s^2)}=\\
\int_{k\in V}\!\!\diff[d]k e^{-s\left(k-p/(2s)\right)^2-p^2/(4s)}=
s^{-d}e^{-p^2/(4s)}\int_{k\in V}\!\!\diff[d]k e^{-k^2}=
s^{-d}e^{-p^2/(4s)}\pi^{d/2}\text{.}
\end{multline}
To be formally correct one had to show that this~$f(k)$ indeed generates
all functions one wants to address.
Apparently one can generate all functions that have power series expansions
in~$k^2$ and~$k\cdot p_i$. We can set~$p=\sum_i=1^n s_i p_i$ to find
\begin{align}
(k^2)^j&=\left.\left(-\frac{\partial}{\partial s}\right)^j
e^{-sk^2+s_1k\cdot p_1+s_2k\cdot p_2+\ldots s_nk\cdot p_n}%
\right\vert_{s,s_1,\ldots,s_n=0}\quad\text{and}\\
(k\cdot p_l)^j&=\left.\left(\frac{\partial}{\partial s_l}\right)^j
e^{-sk^2+s_1k\cdot p_1+s_2k\cdot p_2+\ldots s_nk\cdot p_n}%
\right\vert_{s,s_1,\ldots,s_n=0}\text{.}
\end{align}

In loop calculation we have to deal with functions that arise from
products of propagators,
\begin{equation}\label{eq:appendix-intdetails:schwingerProps}
\frac{1}{A_1^{\alpha_1}A_2^{\alpha_2}\cdots A_N^{\alpha_N}}\text{,}
\end{equation}
where $A_j$ is of the form $A_j=[(k+r_j)^2-m_j^2+i\delta]$ and
$\Re(\alpha_j)>0$ for all $j\in\{1\ldots N\}$.
\person{Schwinger} noticed that one can achieve
the above exponential form by introducing an extra parameter for
each propagator,
\index{Schwinger@\person{Schwinger}!parametrisation|main}
\begin{equation}\label{eq:appendix-intdetails:schwingerTrick}
\frac{1}{A_j^{\alpha_j}}=\frac{1}{\Gamma(\alpha_j)}%
\int_0^\infty\!\!\diff{t_j}\;t_j^{\alpha_j-1}e^{-t_jA_j}\text{,}
\end{equation}
where the condition $\Re(A_j)>0$ must hold\footnote{To formally
achieve this for propagators $[(k+p)^2-m^2+i\delta]$ one can carry
out a \person{Wick} rotation\index{Wick rotation@\person{Wick} rotation}
 first to ensure $(k+p)^2\ge0$ and do
the rest of the calculation for $m^2<0$. Analytically continuation
allows to get a result for real masses after the integration has been
carried out.}.
Finally, for expression~\eqref{eq:appendix-intdetails:schwingerProps}
one can write
\begin{equation}\label{eq:appendix-intdetails:schwingerFinal}
\frac{1}{A_1^{\alpha_1}A_2^{\alpha_2}\cdots A_N^{\alpha_N}}=
\frac{1}{\Gamma(\alpha_1)\cdots\Gamma(\alpha_N)}
\int_0^\infty\left(\prod_{j=1}^N\diff{t_j}\;t_j^{\alpha_j-1}\right)%
e^{-\sum_{\nu=1}^N t_\nu A_\nu}
\end{equation}

%%%%%%%%%%%%%%%%%%%%%%%%%%%%%%%%%%%%%%%%%%%%%%%%%%%%%%%%%%%%%%
\subsection{\person{Feynman} Parameters}
%%%%%%%%%%%%%%%%%%%%%%%%%%%%%%%%%%%%%%%%%%%%%%%%%%%%%%%%%%%%%%
\index{Feynman@\person{Feynman}!parameter|main}
Although the \person{Schwinger} parametrisation is well suited
to show the soundness of dimensional regularisation and the existence
of the loop integrals in a mathematical sense, for actual loop calculations
very often another parametrisation is more convenient.

In this section I introduce \person{Feynman} parameters
starting from Equation~\eqref{eq:appendix-intdetails:schwingerFinal}
and hence I show both the equivalence of both parametrisation and
the validity of Equation~\eqref{eq:loops:FP}.

One can introduce a new parameter $t=\sum_{j=1}^Nt_j$ and substitute
$t_j=tz_j$ in~\eqref{eq:appendix-intdetails:schwingerFinal},
\begin{multline}\label{eq:appendix-intdetails:FP001}
\frac{\Gamma(\alpha_1)\Gamma(\alpha_2)\cdots\Gamma(\alpha_N)}%
{A_1^{\alpha_1}A_2^{\alpha_2}\cdots A_N^{\alpha_N}}=\\
\int_0^\infty\!\!\diff{t}\;t^n%
\int_0^\infty\left(\prod_{j=1}^N\diff{z_j}\;z_j^{\alpha_j-1}\right)t^{\alpha-n}%
e^{-t\sum_{\nu=1}^N z_\nu A_\nu}\delta\left(t-t\sum_{\nu=1}^Nz_\nu\right)%
\text{,}
\end{multline}
where~$\alpha=\sum_{j=1}^N\alpha_j$. Using the homogeneity of the
$\delta$-function we can now carry out the $t$-integration by reversing
\person{Schwinger}'s trick~\eqref{eq:appendix-intdetails:schwingerTrick},
\begin{multline}
\int_0^\infty\!\!\diff{t}\;t^{\alpha-1}%
\int_0^\infty\left(\prod_{j=1}^N\diff{z_j}\;z_j^{\alpha_j-1}\right)%
e^{-t\sum_{\nu=1}^N z_\nu A_\nu}\delta\left(1-\sum_{\nu=1}^Nz_\nu\right)=\\
\Gamma(\alpha)\int_0^\infty\diff{z_1}\cdots\diff{z_N}%
\delta\left(1-\sum_{j=1}^Nz_j\right)\frac{\prod_{j=1}^Nz_j^{\alpha_j-1}}%
{\left(\sum_{j=1}^Nz_jA_j\right)^\alpha}\text{.}
\end{multline}

With the earlier definitions \eqref{eq:qcd-dimreg:def-deltaz}
and~\eqref{eq:qcd-dimreg:def-int0Theta} the \person{Feynman} parameters
are\index{.deltaZ@$\delta_z$}
\begin{equation}
\frac{1}{\prod_{j=1}^NA_j^{\alpha_j}}=
\frac{\Gamma(\alpha)}{\prod_{j=1}^N\Gamma(\alpha_j)}
\int_0\!\!\diff[N]{z}\,\delta_z%
\frac{\prod_{j=1}^Nz_j^{\alpha_j-1}}%
{\left(\sum_{j=1}^Nz_jA_j\right)^\alpha}\text{.}
\end{equation}

\section{Evaluation of Loop Integrals}\label{sec:app-intdetails:evalint}
%%%%%%%%%%%%%%%%%%%%%%%%%%%%%%%%%%%%%%%%%%%%%%%%%%%%%%%%%%%%%%
In the following section I show the omitted steps which are necessary
to solve integrals~\eqref{eq:qcd-dimreg:defTI}. As a first step
in Chapter~\ref{chp:qcdnlo}, Section~\ref{ssec:qcd-dimreg:loop-integrals} we introduced
a \person{Feynman} parametrisation by using~\eqref{eq:loops:FP}
with $A_j=(q_j^2-m_j^2+i\delta)$ and their exponents being $1$, which
finally led to~\eqref{eq:qcd-dimreg:relTISI01}. It was shown that the
tensor structure $\hat{k}^\mu\hat{k}^\nu\ldots$ always leads to a factor
$(\hat{k}^2)^l$ in the numerator for some positive, integer~$l$. The integral
we started from hence is split up into a sum of integrals of the form
\begin{equation}
I_N^{d,\alpha,l}(l_1,\ldots,l_N;S)=
\Gamma(N)\int_0\!\!\diff[N]z\delta_z%nl
\int\!\!\frac{\diff[4]\hat{k}}{i\pi^2}\frac{\diff[d-4]\bar{k}}{\pi^{d/2-2}}%nl
\frac{\left(\bar{k}^2\right)^\alpha (\hat{k}^2)^l%
\prod_{\nu=1}^Nz_\nu^{l_\nu}}{%nl
\left[\hat{k}^2+\bar{k}^2+\frac12z^\transposed Sz+i\delta\right]^N}
\end{equation}

\begin{figure}[hbtp]
\begin{center}
\includegraphics[scale=0.3]{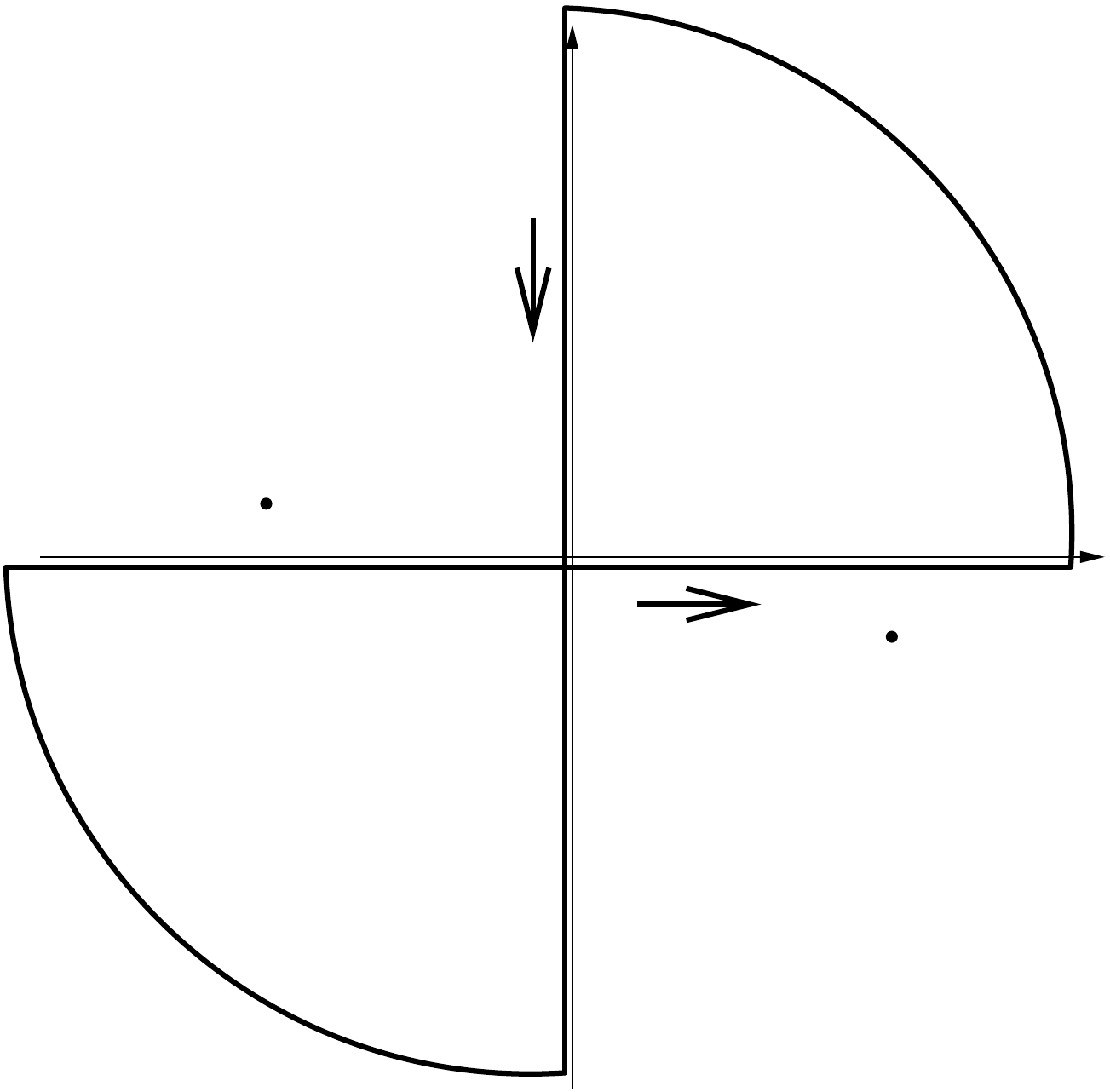}
\end{center}
\caption{The integration contour in the complex $\hat{k}^0$-plane
that is used for the \person{Wick} rotation. The poles of the propagators,
which lie outside the enclosed region, are indicated by dots.}
\label{fig:loops:intcontour}
\end{figure}

\index{Wick rotation@\person{Wick} rotation|main}
To carry out any further integration steps it is much easier to
analytically continue the integral to \person{Euclid}ean space.
This can be achieved by what is known as a \person{Wick} rotation:
The variable $\hat{k}_0$ can be extended from the real axis into the
complex plane, and the integration contour can be closed at infinity
in the first and third quadrant following the imaginary axis in between
(see fig.~\ref{fig:loops:intcontour}). Since the poles of the integrand
lie in the other quadrants, the integral over the closed contour has to
vanish. The curved pieces of the contour do not contribute and hence we
can replace the integration over the real axis by the integration over the
imaginary axis, which after substituting~$ik_0=K_0$
\begin{equation}
\int_{-\infty}^{\infty}\!\!\diff\hat{k}_0\,f(k_0^2-\vec{k}^2)=
-\int_{i\infty}^{-i\infty}\!\!\diff\hat{k}_0\,f(k_0^2-\vec{k}^2)=
i\int_{-\infty}^{\infty}\!\!\diff\hat{K}_0\,f(-K_0^2-\vec{k}^2)
\end{equation}
can be reinterpreted as an integration over one component of a vector
$K=(K_0,\vec{k})$ in a
\person{Euclid}ean vector space with a positive definite inner product.

After this step the integral has the form\footnote{Note that the notation
earlier has been defined such that $\bar{k}^2$ is negative.}
\begin{multline}
I_N^{d,\alpha,l}(l_1,\ldots,l_N;S)=\\
(-1)^{N+\alpha+l}\Gamma(N)\int_0\!\!\diff[N]z\delta_z%nl
\int\!\!\frac{\diff[4]K}{\pi^2}\frac{\diff[d-4]\bar{k}}{\pi^{d/2-2}}%nl
\frac{\left(K^2\right)^\alpha \vert\hat{k}^2\vert^l%
\prod_{\nu=1}^Nz_\nu^{l_\nu}}{%nl
\left[K^2+\vert\bar{k}^2\vert-\frac12z^\transposed Sz-i\delta\right]^N}
\end{multline}
For both parts of the momentum integral the integration can be done in
spherical coordinates. Here we use the axiom that the angular integration
in $d$ dimensions is the surface of the $d$-dimensional unit sphere, %
\index{.Omega@$\Omega_d$|main}
\begin{equation}\label{eq:app-intdetails:nsphere}
\Omega_d=\frac{2\pi^{d/2}}{\Gamma(d/2)}\text{,}
\end{equation}
and hence we get
\begin{multline}
I_N^{d,\alpha,l}(l_1,\ldots,l_N;S)=(-1)^{N+\alpha+l}%
\frac{4\Gamma(N)}{\Gamma(d/2-2)}\\
\int_0\!\!\diff[N]z\delta_z%nl
\int_0^\infty\!\!\int_0^\infty\!\!\diff\varkappa\diff\rho%
\,\,\varkappa^3\rho^{d-5}%nl
\frac{\rho^{2\alpha}\varkappa^{2l}%
\prod_{\nu=1}^Nz_\nu^{l_\nu}}{%nl
\left[\varkappa^2+\rho^2-\frac12z^\transposed Sz-i\delta\right]^N}
\end{multline}

The integrals over $\varkappa$ and~$\rho$ can be identified as
Beta functions (see~\eqref{eq:app:BetaFunction})  and hence
we obtain the desired form of the integral,
\begin{multline}
I_N^{d,\alpha,l}(l_1,\ldots,l_N;S)=%
(-1)^{N+\alpha+l}\Gamma(l+2)%
\frac{\Gamma(\frac{d}2-2+\alpha)}{\Gamma(\frac{d}2-2)}\\
\Gamma(N-\frac{d}2-l-\alpha)\int_0\!\!\diff[N]z\delta_z%nl
\frac{\prod_{\nu=1}^Nz_\nu^{l_\nu}}{%nl
\left[-\frac12z^\transposed Sz-i\delta\right]^{N-d/2-l-\alpha}}\text{,}
\end{multline}
which then can be reinterpreted as
\begin{equation}
I_N^{d,\alpha,l}(S;l_1,\ldots,l_N)=
(-1)^{\alpha+l}\frac{\Gamma(l+2)}{\Gamma(2)}%
\frac{\Gamma(\frac{d}2-2+\alpha)}{\Gamma(\frac{d}2-2)}
I_N^{d+2\alpha+2l}(l_1,\ldots,l_N;S)\text{,}
\end{equation}
where
\begin{equation}
I_N^d(l_1,\ldots,l_N;S)=
(-1)^N\Gamma(N-\frac{d}2)\int_0\!\!\diff[N]z\delta_z%nl
\frac{\prod_{\nu=1}^Nz_\nu^{l_\nu}}{%nl
\left[-\frac12z^\transposed Sz-i\delta\right]^{N-d/2}}\text{.}
\end{equation}

%% file: appendix-integrals.tex
\chapter{Integral Tables}
\label{app:appendix-integrals}
\begin{headquote}{John Billings}
Consider the postage stamp,
its usefulness consists in the ability to stick
to one thing till it gets there.
\end{headquote}

\section{Conventions}
There is a set of conventions in the notation of loop integrals due to
the fact, that many factors are common to most elementary integrals.
One ubiquitous factor is
\begin{equation}
\index{.rGamma@$r_\Gamma$|main}r_\Gamma\equiv%
\frac{\Gamma(1+\varepsilon)\Gamma(1-\varepsilon)^2}{%
\Gamma(1-2\varepsilon)}\text{.}
\end{equation}
The expansion of $r_\Gamma$ is
\begin{equation}
r_\Gamma=1-\gamma_E\varepsilon
+\left(\frac{\gamma_E^2}{2}-\frac{\pi^2}{12}\right)\varepsilon^2
+\mathcal{O}(\varepsilon^3)\text{,}
\end{equation}
where~$\gamma_E=-\Gamma^\prime(1)$.\index{.gammaE@$\gamma_E$|main}
Together with\footnote{See Section~\ref{ssec:qcd-dimreg:loop-integrals}}
\begin{equation}
(4\pi\mu^2)^{2-n/2}=(\mu^2)^\varepsilon\left(1+\varepsilon\ln(4\pi)
+\mathcal{O}(\varepsilon^2)\right)
\end{equation}
this factor constitutes the \ac{uv} subtraction
\index{.Delta@$\Delta$|main}%
$\Delta\equiv 1/\varepsilon-\gamma_E+\ln(4\pi)$ in
the \ac{msbar}~scheme,
\begin{equation}
(4\pi\mu^2)^\varepsilon\frac{r_\Gamma}{\varepsilon}=
(\mu^2)^\varepsilon\left(\Delta+\mathcal{O}(\varepsilon)\right)
\end{equation}
The factor of $(\mu^2)^\varepsilon$ fixes the dimension of expressions
of the form
\begin{equation}
\ln(-s-i\delta)\equiv\ln\left(\frac{-s-i\delta}{\mu^2}\right)\text{.}
\end{equation}

For double poles it is convenient to also pull out a factor of
$e^{-\varepsilon\gamma_E}$ which then allows for the simple
result
\begin{equation}
\frac{r_\Gamma}{\varepsilon^2}=e^{-\varepsilon\gamma_E}
\left(\frac{1}{\varepsilon^2} - \frac{\pi^2}{12}+{\mathcal O}(\varepsilon)
\right)\text{.}
\end{equation}

In many cases it is convenient to express the loop integrals
in terms of the functions (see~\cite{Binoth:2005ff})
\begin{subequations}
\begin{align}
H_0(x,\alpha)&=\frac{(-x-i\delta)^\alpha}{(\mu^2)^{-\varepsilon}x},\\
H_1(x,y,\alpha)&=\frac{1}{(x-y)}\left(%
xH_0(x,\alpha)
-\frac{\alpha}{0+\alpha}yH_0(y,\alpha)
\right)\quad\text{and}\\
H_{N+1}(x,y,\alpha)&=\frac{1}{(x-y)}\left(%
\frac{N}{N+\alpha}xH_N(x,y,\alpha)
-\frac{\alpha}{N+\alpha}yH_0(y,\alpha)
\right),\,N>1\text{.}
\end{align}
\end{subequations}

Using the fact that
\begin{equation}
r_\Gamma=\frac{1}{\Gamma(1-\varepsilon)}
	\left(1+\mathcal{O}(\varepsilon^3)\right)
\end{equation}
the epsilon expansions of simple cross-section in~\ac{qcd}
are very often found in a form like
\begin{equation}
\frac{C_F\alpha_s}{2\pi}
\left(\frac{4\pi\mu^2}{Q^2}\right)^\varepsilon
\frac{1}{\Gamma(1-\varepsilon)}
\left(\frac{A}{\varepsilon^2}+\frac{B}{\varepsilon}
+C+D\pi^2+\mathcal{O}(\varepsilon)\right)\ldots\;\text{,}
\end{equation}
where $A$, $B$, $C$ and $D$ are complex numbers.

\section{Relations for One- and Two-Point Functions}
The scalar tadpole function can be evaluated directly,
\begin{equation}
I_1^n(m^2)=m^2(m^2-i\delta)^{-\varepsilon}\frac{1}{\varepsilon}
\frac{1}{1-\varepsilon}\Gamma(1+\varepsilon)=m^2\left[
\Delta-\ln(m^2-i\delta)-1\right]+{\mathcal O}(\varepsilon)
\end{equation}

The scalar two-point function in~$n$ dimensions can be expressed as
\begin{multline}
I_2^n(S)=I_2^n(s;m_1^2,m_2^2)=\\
\Delta-\int_0^1\!\!\diff{z}\ln\left(
-sz(1-z)+m_1^2z+m_2^2(1-z)-i\delta\right)+{\mathcal{O}}(\varepsilon)\text{.}
\end{multline}
all other one- and two-point functions can be expressed in terms of that
function. The underlying $S$-matrix for $N=2$ is parametrised as
\begin{equation}
S=\left(\begin{array}{cc}
-2m_1^2&s-m_1^2-m_2^2\\
s-m_1^2-m_2^2&-2m_2^2
\end{array}\right)\text{.}
\end{equation}
For the one-point functions $I^d_1(m^2)$ the underlying 
$S$-matrix is~$S=(-2m^2)$.

\begin{subequations}
\begin{align}
&I_2^n(l_0,l_1;S)=\\&\qquad -S_{l_0l_1}^{-1}I_2^{n+2}(S)
-b_{l_0}I_2^{n+2}(l_1;S)
+\sum_{k\in S_{\#}}S_{l_0k}^{-1}I_1^{n+2}(S^{\{k\}})\nonumber\\
&I_2^{n+2}(l_0;S)=\frac{b_{l_0}}{B}I_2^{n+2}(S)
+\sum_{k\in S_{\#}}\left(S^{-1}_{l_0k}-\frac{b_{l_0}b_k}{B}\right)
I_1^{n+2}(S^{\{k\}})\\
&I_2^{n}(l_0;S)=\frac{b_{l_0}}{B}I_2^{n}(S)
+\sum_{k\in S_{\#}}\left(S^{-1}_{l_0k}-\frac{b_{l_0}b_k}{B}\right)
I_1^{n}(S^{\{k\}})\\
&I^{n+2}_2(S)=\frac{1}{B(n-1)}\left[I_2^n(S)
-\sum_{k\in S_{\#}}b_kI_1^n(S^{\{k\}})\right]\\
&I_1^{n+2}(m^2)=\frac{m^4}{4}\left(I_2^n(0;0,m^2)+\frac{1}{2}\right)\\
&I_1^{n}(m^2)=m^2I_2^n(0;0,m^2)
\end{align}
\end{subequations}

\section{Massless Two- and Three-Point Integrals}
The easiest case is given when all propagators are massless because then
the $S$-matrix takes a very simple form. The massless tadpole vanishes
identically in dimensional regularisation. The $S$-matrix is
\begin{equation}
S=\left(\begin{array}{cc}
0&s\\
s&0
\end{array}\right)\text{,}
\end{equation}
and the two-point integral reads as follows,
\begin{multline}
I_2^d(S)=\Gamma(2-d/2)\int_0^1\!\!\diff{z}%
\left[(-s-i\delta)z(1-z)\right]^{d/2-2}=\\
\Gamma(2-d/2)%
\frac{\Gamma(d/2-1)^2}{\Gamma(d-2)}%
sH_0(s,d/2-2)\text{,}
\end{multline}
which in the case $d=n=4-2\varepsilon$ becomes
\begin{equation}
I_2^n(S)=\frac{r_\Gamma}{\varepsilon}\frac{1}{(1-2\varepsilon)}
sH_0(s,-\varepsilon)\text{.}
\end{equation}
Hence, the expansion of~$I_2^n(S)$ is
\begin{equation}
I_2^n(S)=\Delta-\ln(-s-i\delta)+2+{\mathcal O}(\varepsilon)\text{.}
\end{equation}
Similarly the other relevant two-point integrals are calculated,
\begin{align}
%%%%%%%%%%%%%%%%%%%%%%%%%%%%%%%%%%%%%%%%%%%%%%%%%%%%%%%%%%%%%%%%
I_2^{n+2}(S)&=-\frac{r_\Gamma}{\varepsilon}
\frac{1-\varepsilon}{(1-2\varepsilon)_3}sH_0(s,1-\varepsilon)=
\frac{s}{2(3-2\varepsilon)}I_2^n(S)
%\Delta-\ln(-s-i\delta)+1+{\mathcal O}(\varepsilon)
\text{,}\\
%%%%%%%%%%%%%%%%%%%%%%%%%%%%%%%%%%%%%%%%%%%%%%%%%%%%%%%%%%%%%%%%
I_2^n(l;S)&=\frac{r_\Gamma}{\varepsilon}\frac{1}{2(1-2\varepsilon)}
sH_0(s,-\varepsilon)
=\frac12I_2^n(S)\text{,}\\
%%%%%%%%%%%%%%%%%%%%%%%%%%%%%%%%%%%%%%%%%%%%%%%%%%%%%%%%%%%%%%%%
I_2^n(l,l;S)&=-\frac{r_\Gamma}{\varepsilon^2}
\frac{2-\varepsilon}{2(3-2\varepsilon)(1-2\varepsilon)}
sH_0(s,-\varepsilon)
=\frac{2-\varepsilon}{2(3-2\varepsilon)}I_2^n(S)\text{,}\\
%%%%%%%%%%%%%%%%%%%%%%%%%%%%%%%%%%%%%%%%%%%%%%%%%%%%%%%%%%%%%%%%
I_2^n(1,2;S)&=\frac{r_\Gamma}{\varepsilon}
\frac{(1-\varepsilon)^2}{(1-2\varepsilon)_3}
sH_0(s,-\varepsilon)
=\frac{1-\varepsilon}{2(3-2\varepsilon)}I_2^n(S)
%\Delta-\ln(-s-i\delta)+2+{\mathcal O}(\varepsilon)
\text{,}\\
%%%%%%%%%%%%%%%%%%%%%%%%%%%%%%%%%%%%%%%%%%%%%%%%%%%%%%%%%%%%%%%%
I_2^{n,1}(S)&=\varepsilon I_2^{n+2}(S)=
\frac{s}{6}+{\mathcal O}(\varepsilon)
\text{.}
\end{align}

For the case of three-point function one has to distinguish the
cases when $\det{S}$ vanishes, i.e. if one or two scales vanish.
Explicit formul\ae{} are given for the case when for
\begin{equation}
S=\begin{pmatrix}
0&s&u\\
s&0&t\\
u&t&0
\end{pmatrix}
\end{equation}
one or two of the variables vanish. Below I restrict to the cases
$u=0$ and $t=u=0$. For all three variables non-vanishing the usual
reduction formula applies.
\begin{align}
%%%%%%%%%%%%%%%%%%%%%%%%%%%%%%%%%%%%%%%%%%%%%%%%%%%%%%%%%%%%%%%%
I_3^n(S_{t=u=0})&=\frac{r_\Gamma}{\varepsilon^2}H_0(s,-\varepsilon)\\
%%%%%%%%%%%%%%%%%%%%%%%%%%%%%%%%%%%%%%%%%%%%%%%%%%%%%%%%%%%%%%%%
I_3^{n}(1;S_{t=u=0})&=I_3^n(2;S_{t=u=0})=-\frac{r_\Gamma}{\varepsilon}%
\frac{1}{1-2\varepsilon}H_0(s,-\varepsilon)\\
%%%%%%%%%%%%%%%%%%%%%%%%%%%%%%%%%%%%%%%%%%%%%%%%%%%%%%%%%%%%%%%%
I_3^{n}(3;S_{t=u=0})&=\frac{r_\Gamma}{\varepsilon^2}%
\frac{1}{1-2\varepsilon}H_0(s,-\varepsilon)\\
%%%%%%%%%%%%%%%%%%%%%%%%%%%%%%%%%%%%%%%%%%%%%%%%%%%%%%%%%%%%%%%%
I_3^n(1,1;S_{t=u=0})&=I_3^n(2,2;t=u=0)=-\frac{r_\Gamma}{\varepsilon}
\frac{1}{2(1-2\varepsilon)}H_0(s,-\varepsilon)\\
%%%%%%%%%%%%%%%%%%%%%%%%%%%%%%%%%%%%%%%%%%%%%%%%%%%%%%%%%%%%%%%%
I_3^n(1,2;S_{t=u=0})&=I_3^n(2,2;t=u=0)=r_\Gamma
\frac{1}{2(1-\varepsilon)(1-2\varepsilon)}H_0(s,-\varepsilon)\\
%%%%%%%%%%%%%%%%%%%%%%%%%%%%%%%%%%%%%%%%%%%%%%%%%%%%%%%%%%%%%%%%
I_3^{n+2}(S_{t=u=0})&=\frac{r_\Gamma}{\varepsilon}%
\frac{1}{2(1-\varepsilon)(1-2\varepsilon)}H_0(s,1-\varepsilon)
\end{align}
A full list of the three-point functions which are used in our
tensor reduction for massless internal propagators are
in~\cite{Binoth:2005ff}. A more general review on one-loop integrals
with a compilation of the relevant formul\ae{} can be found
in~\cite{Ellis:2007qk}.

\section{Polynomial Loop Integrals}
\label{app:appendix-integrals:polynomial}
In this appendix I present explicit expressions for the integrals of type, 
\begin{eqnarray}
&&\varepsilon I_N^{n-4+2N}(l_1,\ldots,l_r;S)=
(-1)^NP_N(l_1,\ldots,l_r)\text{,}\\
&&\varepsilon I_N^{n-4+2(N+1)}(l_1,\ldots,l_r;S)=
\frac{(-1)^N}{2}\sum_{j_1,j_2=1}^NS_{j_1j_2}P_N(j_1,j_2,l_1,\ldots,l_r)\text{,}
\end{eqnarray}
which are introduced in section~\ref{ssec:qcd-dimreg:polyintegrals}.
The list includes those integrals that can arise in calculations using 
\person{Feynman}~gauge.
Formula for~$\eta>1$ and larger numbers of \person{Feynman} parameters
in the numerator can be derived
using~\eqref{eq:qcd-dimreg:common-poly-solution}. Unless stated differently,
all expressions are given up to order ${\mathcal O}(\varepsilon)$.

\begin{subequations}
\begin{align}
&\varepsilon I_N^{n-4+2N}(S)=\frac{(-1)^N}{(N-1)!}\displaybreak[1]\\
&\varepsilon I_N^{n-4+2N}(l_1;S)=\frac{(-1)^N}{N!}\displaybreak[1]\\
&\varepsilon I_N^{n-4+2N}(l_1, l_2;S)=
\frac{(-1)^N}{(N+1)!}\left(1+\delta_{l_1l_2}\right)\displaybreak[1]\\
&\varepsilon I_N^{n-4+2N}(l_1, l_2,l_3;S)=\frac{(-1)^N}{(N+2)!}\\
&\qquad\times\left(1+\delta_{l_1l_2}+\delta_{l_1l_3}
+\delta_{l_2l_3}+2\delta_{l_1l_2}\delta_{l_2l_3}\right)\nonumber\displaybreak[1]\\
&\varepsilon I_N^{n-4+2N}(l_1, l_2,l_3,l_4;S)=\frac{(-1)^N}{(N+3)!}\\
&\qquad\times(\delta_{l_1l_2}(6\delta_{l_1l_3}\delta_{l_2l_4}+2\delta_{l_1l_3}
+2\delta_{l_2l_4}+\delta_{l_3l_4})\nonumber\\
&\qquad\quad+2\delta_{l_3l_4}(\delta_{l_1l_3}+\delta_{l_2l_4})
+\delta_{l_1l_3}\delta_{l_2l_4}+\delta_{l_1l_4}\delta_{l_2l_3}\nonumber\\
&\qquad\quad+\delta_{l_1l_2}+\delta_{l_1l_3}+\delta_{l_1l_4}
+\delta_{l_2l_3}+\delta_{l_2l_4}+\delta_{l_3l_4}+1)\nonumber\displaybreak[1]\\
&\varepsilon I_N^{n-4+2(N+1)}(S)=\frac{(-1)^N}{2(N+1)!}\left(\sum_{j_1,j_2=1}^NS_{j_1j_2}+\tr{S}\right)\displaybreak[1]\\
&\varepsilon I_N^{n-4+2(N+1)}(l_1;S)=\frac{(-1)^N}{2(N+2)!}\sum_{j_1,j_2=1}^NS_{j_1j_2}\left(1+\delta_{j_1j_2}\right)\\
&\qquad\times\left(1+\delta_{l_1j_1}
+\delta_{l_1j_2}\right)\nonumber\displaybreak[1]\\
&\varepsilon I_N^{n-4+2(N+1)}(l_1,l_2;S)=
\frac{(-1)^N}{2(N+3)!}\sum_{j_1,j_2=1}^NS_{j_1j_2}\\
&\qquad\times(\delta_{j_1j_2}(6\delta_{j_1l_1}\delta_{j_2l_2}
+2\delta_{j_1l_1}+2\delta_{j_2l_2}+\delta_{l_1l_2})\nonumber\\
&\qquad\quad+2\delta_{l_1l_2}(\delta_{j_1l_1}+\delta_{j_2l_2})
+\delta_{j_1l_1}\delta_{j_2l_2}+\delta_{j_1l_2}\delta_{j_2l_1}\nonumber\\
&\qquad\quad+\delta_{j_1j_2}+\delta_{j_1l_1}+\delta_{j_1l_2}
+\delta_{j_2l_1}+\delta_{j_2l_2}+\delta_{l_1l_2}+1)\nonumber
\end{align}
\end{subequations}

The corresponding integrals in $n$ dimensions are,
by applying~\eqref{eq:qcd-dimreg:TItoSI}:
\begin{subequations}
\begin{align}
&I_N^{n,N-2}(S)=-\frac{1}{(N-1)(N-2)}\displaybreak[1]\\
&I_N^{n,N-1}(S)=\frac{1}{2(N+1)N(N-1)}\left(\sum_{j_1,j_2=1}^NS_{j_1j_2}+\tr{S}\right)\displaybreak[1]\\
&I_2^{n,1}(S)=\frac{1}{12}\left(2\Delta_{12}^2-6(m_1^2+m_2^2)\right)\displaybreak[1]\\
&\varepsilon I_2^{n}(S)=1\displaybreak[1]\\
&\varepsilon I_1^{n}(S)=-\frac{1}{2}S_{11}\displaybreak[1]\\
&I_N^{n,N-2;\mu}(a; S)=\frac{1}{N(N-1)(N-2)}\sum_{j=1}^N\Delta_{ja}^\mu\displaybreak[1]\\
&I_N^{n,N-1;\mu}(a; S)=-\frac{(N-2)!}{2(N+2)!}\sum_{j=1}^N\Delta_{ja}^\mu\\
&\qquad\times\left(\sum_{j_1,j_2=1}^NS_{j_1j_2}+\tr{S}
+2\sum_{l=1}^NS_{jl}+S_{jj}
\right)\nonumber\displaybreak[1]\\
&\varepsilon I_2^{n;\mu}(a; S)=-\frac{1}{2}\sum_{j=1}^N\Delta_{ja}^\mu\displaybreak[1]\\
&\varepsilon I_3^{n;\mu\nu}(a_1, a_2; S)=\frac{g^{\mu\nu}}{4}\\
&I_4^{n,1;\mu\nu}(a_1, a_2; S)=-\frac{1}{12}g^{\mu\nu}
\end{align}
\end{subequations}

%% file: implementation.tex
\chapter{Implementation of Amplitude Computations}
\label{chp:implementation}
%\begin{headquote}{Donald Knuth}
%Beware of bugs in the above code;
%I have only proved it correct, not tried~it.
%\end{headquote}
\begin{headquote}{Press et al.\footnote{Numerical Recipies in C, 1992}}
The practical scientist is trying to solve tomorrow’s problem with yesterday’s
computer; the computer scientist, we think, often has it the other way around.
\end{headquote}

\input{imp-sections}

%% file: imp-sections.tex
\section*{Introduction}
\label{sec:imp-intro}
% vim: syntax=nuweb:ts=3:sw=3
This chapter describes the implementation of cross-section calculations
based on the strategy explained in the main part of this thesis.
This code has been successfully tested for the 
$u\bar{u}\rightarrow b\bar{b}b\bar{b}$ amplitude, results of which
are presented in~\ref{chp:results}.

One of the main technical challenges of an amplitude calculation
in \ac{qcd} at \ac{nlo} precision is the computation of the
virtual corrections due to
the number of terms involved during the reduction of the diagrams.
The size of the expressions makes the calculation computational expensive
already prior to the numerical evaluation and gives rise to a high consumption
of resources, not only with respect to computing time but also memory
allocation in order to store intermediate results. This high demand also
addresses the software used for the computation as most of the standard
software like computer algebra systems and compilers are not prepared to
handle huge amounts of data. An implementation of a \ac{nlo} calculation
with many external particles therefore has to address the resource usage
under different viewpoints. From the theoretical, mathematical side one
has to choose a representation of the expression that avoids producing
unnecessarily large amounts of terms. As a more technical issue one has
to provide means to make the required computing resources accessible;
this includes the choice of software capable of dealing with large amounts
of data on the one hand and parallel and distributed computing techniques
on the other hand.

Furthermore, general software design goals must be borne in
mind~\cite{Bugden:SoftwareDesign}.
Two of the major aims for my project are \emph{reusability}
and~\emph{extensibility}; the importance of these design attributes
can be seen from decomposing the process of matrix element evaluations.
The major part of a calculation is process independent, like the evaluation
of colour and \person{Dirac} traces or the reduction and evaluation of
\person{Feynman} parameter integrals. Process and model dependencies only
enter through few parameters like the number of ingoing and outgoing
particles together with their masses, through the \person{Feynman} rules
and in the graph generation. These dependencies can be separated
through \emph{modularity} from an invariant, reusable application core
which can serve as a general purpose tool. Once these criteria are met
one is in the position to release the code to a broader public. 
However, this step involves the necessity of \emph{usability},
\emph{maintainability} and~\emph{portability}. Although from a first look
it seems as if these design metrics have to be considered only on a
very late stage of the development process, it should be clear that their
early disesteem most likely entails rewriting large parts of the code.

\subsection*{Literate Programming}
\index{literate programming|(}
Literate programming\index{Literate programming} has been developed
by \person{Knuth} in the 1980s~\cite{DBLP:journals/cj/Knuth84,Knuth:1992}.
This concept describes the combination of computer programs and
type setting in a way that from a common source both, a compilable
program and a high quality document can be obtained. Rather than
decorating a source code with comments literate programming understands
a program as part of the document that describes the program. A computer
program is organised in little chunks of which the order in the document
does not necessarily correspond to the order in the final program code.

The original version of \texttt{WEB} implemented two tool, \texttt{weave}
translates the \texttt{WEB} document into \TeX{}, \texttt{tangle} extracts
the program fragment from the \texttt{WEB} document and generates the program
code.
\index{literate programming|)}

Here I give an overview overview over the literate programming tool
\texttt{nuweb}~\cite{nuweb}. Its main design goals are simplicity and language
independence. Instead of two separate tools as in the case of
\texttt{weave} and \texttt{tangle}, \texttt{nuweb} consists of a single
command line tool that produces both program and documentation in one go.
In the following, a simple example shall explain the main features and
advantages of literate programming.

The example shows a \form-program which generates a colour basis
for a given partonic process. To calculate the colour basis for
the process $gg\rightarrow q\bar{q}q\bar{q}$ one would specify
\begin{flushleft} \small
\begin{minipage}{\linewidth} \label{scrap1}
$\langle\,$Process Specification\nobreak\ {\footnotesize \NWtarget{nuweb1}{1}}$\,\rangle\equiv$
\vspace{-1ex}
\begin{list}{}{} \item
\mbox{}\verb@@\hbox{\sffamily\bfseries Local}\verb@ colour =@\\
\mbox{}\verb@   #@\hbox{\sffamily\bfseries call}\verb@ insertgluons(2)@\\
\mbox{}\verb@   #@\hbox{\sffamily\bfseries call}\verb@ insertquarks(2)@\\
\mbox{}\verb@   ;@{\NWsep}
\end{list}
\vspace{-1ex}
\footnotesize\addtolength{\baselineskip}{-1ex}
\begin{list}{}{\setlength{\itemsep}{-\parsep}\setlength{\itemindent}{-\leftmargin}}
\item \NWtxtMacroRefIn\ \NWlink{nuweb2}{2}.
\end{list}
\end{minipage}\\[4ex]
\end{flushleft}
The above paragraph has been created with the \texttt{nuweb}
commands
\begin{Verbatim}[numbers=left,frame=lines,framerule=2pt]
The example shows a \form-program ... one would specify 
@d Process Specification
@{@_Local@_ colour =
	#@_call@_ insertgluons(2)
	#@_call@_ insertquarks(2)
	;@|
colour @}
\end{Verbatim}
Line~1 contains ordinary \LaTeX{} text; line~2 introduces a macro
called \emph{Process Specification}.
All \texttt{nuweb} markup start with an at (@) sign.
Lines 3--7 contain a \emph{scrap}, i.e. a short piece of
embedded program code. Scraps usually are delimited by a pair
of \texttt{@\{} and \texttt{@\}}.

Since \texttt{nuweb} does not support automatic syntax highlighting
it gives the user the opportunity to format the code using the
directive \texttt{@\_}, which formats the text in between in bold
font; the formatting does not affect the generated source code.

The above example contains another \texttt{nuweb} instruction:
the character sequence \texttt{@|} introduces a list of identifiers
that are defined in the according scrap\footnote{It is irrelevant if the
identifiers actually appear in this scrap} for which an entry
in the list of identifiers is generated. \texttt{nuweb} automatically
creates a list of user-defined identifiers using the \texttt{@u}
instruction (see Section~\ref{ssec:implementation:idx-identifiers}),
and similarly a list of macros with the \texttt{@m} command (see
Section~\ref{ssec:implementation:idx-macros}) and with the
\texttt{@f} instruction a list of files
(see Section~\ref{ssec:implementation:idx-files}).

The creation of output files for program code is initiated by
the \texttt{@o} command\footnote{Both the commands \texttt{@d}
and \texttt{@o} have a capitalised variant (\texttt{@O} and \texttt{@D}
resp.) which generate long scraps, i.e. the program fragments may
span over several pages. However, for readability it is recommended
that each scrap consists of up to 12 lines.}.
The next definition shows the overall structure of the program
file \texttt{colour.frm}. The declaration section is terminated
by the module separator \texttt{.global} and contains the definition
of all relevant symbols, functions and procedures. In the second
paragraph of the program the expression is transformed and the third
paragraph prints the expression term by term.
\begin{flushleft} \small \label{scrap2}
\verb@"colour.frm"@\nobreak\ {\footnotesize \NWtarget{nuweb2}{2} }$\equiv$
\vspace{-1ex}
\begin{list}{}{} \item
\mbox{}\verb@@\hbox{$\langle\,$Symbol Definitions\nobreak\ {\footnotesize \NWlink{nuweb3}{3}}$\,\rangle$}\verb@@\\
\mbox{}\verb@@\hbox{$\langle\,$Procedure definition \texttt{insertquarks}\nobreak\ {\footnotesize \NWlink{nuweb4}{4}}$\,\rangle$}\verb@@\\
\mbox{}\verb@@\hbox{$\langle\,$Procedure definition \texttt{insertgluons}\nobreak\ {\footnotesize \NWlink{nuweb6}{6}}$\,\rangle$}\verb@@\\
\mbox{}\verb@@\hbox{$\langle\,$Procedure definition \texttt{stripcoeff}\nobreak\ {\footnotesize \NWlink{nuweb7}{7}}$\,\rangle$}\verb@@\\
\mbox{}\verb@.@\hbox{\sffamily\bfseries global}\verb@@\\
\mbox{}\verb@@\\
\mbox{}\verb@@\hbox{$\langle\,$Process Specification\nobreak\ {\footnotesize \NWlink{nuweb1}{1}}$\,\rangle$}\verb@@\\
\mbox{}\verb@@\hbox{$\langle\,$Perform Insertions\nobreak\ {\footnotesize \NWlink{nuweb8}{8}}$\,\rangle$}\verb@@\\
\mbox{}\verb@@\hbox{$\langle\,$Simplify Result\nobreak\ {\footnotesize \NWlink{nuweb12}{12}}$\,\rangle$}\verb@@\\
\mbox{}\verb@@\\
\mbox{}\verb@#$num = 1;@\\
\mbox{}\verb@@\hbox{\sffamily\bfseries Print}\verb@ "color%$=%T", $num;@\\
\mbox{}\verb@$num = $num + 1;@\\
\mbox{}\verb@.@\hbox{\sffamily\bfseries end}\verb@@\\
\mbox{}\verb@@{\NWsep}
\end{list}
\vspace{-2ex}
\end{flushleft}
The above scrap has been generated by the following
piece of \texttt{nuweb} code:
\begin{Verbatim}[numbers=left,frame=lines,framerule=2pt,%
commandchars=\\\{\}]
@o colour.frm
@\{@<Symbol Definitions@>
@<Procedure definition \textbackslash{}texttt\{insertquarks\}@>
\dots
#$num = 1;
@_Print@_ "color%$=%T", $num;
$num = $num + 1;
.@_end@_
@| $num@\}
\end{Verbatim}
Macros are referenced by putting their names inside angle brackets
\texttt{@<}\dots\texttt{@>}. Macros can be referenced before they
are defined, i.e. the order of the macro definitions in the document
is irrelevant. The above example also shows that macro names can contain
virtually any \LaTeX{} commands.

Lines 5--7 show how in \form{} one can enumerate all terms in an expression
by using dollar-variables and the \texttt{Print} command. Line~5 defines
the dollar variable \texttt{\$num} when the preprocessor runs, i.e. before
the module is executed. Then for every term in the expression lines
6 and~7 are invoked. The format sequence \texttt{\%\$} prints out
the first dollar variable from the list after the format string which is
the number of the current term and the sequence \texttt{\%T} prints
the current term. Finally, line~7 increases the counter before the next
term is processed.

This concludes the discussion of the main features of \texttt{nuweb}
which are necessary for the understanding of the code and for the
understanding of the main concepts of Literate Programming. A complete
documentation of \texttt{nuweb} can be found in~\cite{nuweb}. The rest
of this section is concerned with the description of the main part of
the \form-program.

The symbol \texttt{x} is used as a pattern to represent arbitrary
symbolic expressions. The functions \texttt{insertgluon}, \texttt{insertq}
and \texttt{insertt} act on an expression like differential operators as
explained below. The function \texttt{delta(i,j)} stands for a quark
line~$\delta_i^j$ and
\texttt{t(i,j,g)} for a generator $t_{ij}^g$. The result is expressed
in terms of the commuting functions \texttt{line(i,g1,\ldots,gn,j)} 
which represents the product of generators
$t_{ij_1}^{g_1}t_{j_1j_2}^{g_2}\cdots t_{j_nj}^{g_n}$ and \texttt{tr}
for traces of products of generators. Indices $i_1, i_2,\ldots$ are used
for quarks, $j_1,j_2,\ldots$ for antiquarks and $g_1,g_2,\ldots$ for
gluons; for quarks and antiquarks also the two sets \texttt{quarks} and
\texttt{aquarks} are defined.

\begin{flushleft} \small
\begin{minipage}{\linewidth} \label{scrap3}
$\langle\,$Symbol Definitions\nobreak\ {\footnotesize \NWtarget{nuweb3}{3}}$\,\rangle\equiv$
\vspace{-1ex}
\begin{list}{}{} \item
\mbox{}\verb@@\hbox{\sffamily\bfseries Symbol}\verb@ x;@\\
\mbox{}\verb@@\hbox{\sffamily\bfseries Functions}\verb@ insertgluon, delta, t, insertq, insertt;@\\
\mbox{}\verb@@\hbox{\sffamily\bfseries CFunctions}\verb@ tr(@\hbox{\sffamily\bfseries cyclic}\verb@), line;@\\
\mbox{}\verb@@\hbox{\sffamily\bfseries Autodeclare}\verb@ @\hbox{\sffamily\bfseries Indices}\verb@ i, j, g;@\\
\mbox{}\verb@@\\
\mbox{}\verb@@\hbox{\sffamily\bfseries Set}\verb@ quarks: i1, ..., i10;@\\
\mbox{}\verb@@\hbox{\sffamily\bfseries Set}\verb@ aquarks: j1, ..., j10;@{\NWsep}
\end{list}
\vspace{-1ex}
\footnotesize\addtolength{\baselineskip}{-1ex}
\begin{list}{}{\setlength{\itemsep}{-\parsep}\setlength{\itemindent}{-\leftmargin}}
\item \NWtxtMacroRefIn\ \NWlink{nuweb2}{2}.
\end{list}
\end{minipage}\\[4ex]
\end{flushleft}
The procedure \texttt{insertquarks(N)} generates a basis
for \texttt{N} quark-antiquark pairs. If $\mathtt{N}=0$ a
closed quark line \texttt{delta(i1,i1)} is inserted
as a seed for the insertion of the gluons.
Otherwise, by the product of two $\varepsilon$-tensors
an antisymmetriser over all quark lines is generated.
\begin{flushleft} \small
\begin{minipage}{\linewidth} \label{scrap4}
$\langle\,$Procedure definition \texttt{insertquarks}\nobreak\ {\footnotesize \NWtarget{nuweb4}{4}}$\,\rangle\equiv$
\vspace{-1ex}
\begin{list}{}{} \item
\mbox{}\verb@#@\hbox{\sffamily\bfseries procedure}\verb@ insertquarks(N)@\\
\mbox{}\verb@   #@\hbox{\sffamily\bfseries if}\verb@ `N'>0@\\
\mbox{}\verb@      @\hbox{\sffamily\bfseries e}\verb@_(i1, ..., i`N') * @\hbox{\sffamily\bfseries e}\verb@_(j1, ..., j`N')@\\
\mbox{}\verb@   #@\hbox{\sffamily\bfseries else}\verb@@\\
\mbox{}\verb@      delta(i1,i1)@\\
\mbox{}\verb@   #@\hbox{\sffamily\bfseries endif}\verb@@\\
\mbox{}\verb@#@\hbox{\sffamily\bfseries endprocedure}\verb@@{\NWsep}
\end{list}
\vspace{-1ex}
\footnotesize\addtolength{\baselineskip}{-1ex}
\begin{list}{}{\setlength{\itemsep}{-\parsep}\setlength{\itemindent}{-\leftmargin}}
\item \NWtxtMacroRefIn\ \NWlink{nuweb2}{2}.
\end{list}
\end{minipage}\\[4ex]
\end{flushleft}
After contracting the pair of $\varepsilon$-tensors
this antisymmetriser
is turned into a symmetriser by discarding the signs.
\begin{flushleft} \small
\begin{minipage}{\linewidth} \label{scrap5}
$\langle\,$Build Symmetriser\nobreak\ {\footnotesize \NWtarget{nuweb5}{5}}$\,\rangle\equiv$
\vspace{-1ex}
\begin{list}{}{} \item
\mbox{}\verb@@\hbox{\sffamily\bfseries Contract}\verb@;@\\
\mbox{}\verb@#@\hbox{\sffamily\bfseries call}\verb@ stripcoeff(insertgluon,@\hbox{\sffamily\bfseries d}\verb@_,delta)@\\
\mbox{}\verb@.@\hbox{\sffamily\bfseries sort}\verb@@\\
\mbox{}\verb@@{\NWsep}
\end{list}
\vspace{-1ex}
\footnotesize\addtolength{\baselineskip}{-1ex}
\begin{list}{}{\setlength{\itemsep}{-\parsep}\setlength{\itemindent}{-\leftmargin}}
\item \NWtxtMacroRefIn\ \NWlink{nuweb8}{8}.
\end{list}
\end{minipage}\\[4ex]
\end{flushleft}
For each of the \texttt{N} gluon
the procedure \texttt{insertgluons(N)} multiplies
the expression by an \texttt{insertgluon}-operator.
\begin{flushleft} \small
\begin{minipage}{\linewidth} \label{scrap6}
$\langle\,$Procedure definition \texttt{insertgluons}\nobreak\ {\footnotesize \NWtarget{nuweb6}{6}}$\,\rangle\equiv$
\vspace{-1ex}
\begin{list}{}{} \item
\mbox{}\verb@#@\hbox{\sffamily\bfseries procedure}\verb@ insertgluons(N)@\\
\mbox{}\verb@   #@\hbox{\sffamily\bfseries do}\verb@ i=1,`N'@\\
\mbox{}\verb@      insertgluon(g`i') *@\\
\mbox{}\verb@   #@\hbox{\sffamily\bfseries enddo}\verb@@\\
\mbox{}\verb@#@\hbox{\sffamily\bfseries endprocedure}\verb@@{\NWsep}
\end{list}
\vspace{-1ex}
\footnotesize\addtolength{\baselineskip}{-1ex}
\begin{list}{}{\setlength{\itemsep}{-\parsep}\setlength{\itemindent}{-\leftmargin}}
\item \NWtxtMacroRefIn\ \NWlink{nuweb2}{2}.
\end{list}
\end{minipage}\\[4ex]
\end{flushleft}
The program does not attempt to generate each basis vector
in colour space exactly once; it only ensures that each vector
is generated with a positive coefficient. Therefore, the
procedure \texttt{stripcoeff} strips off all coefficients
of the symbols which are given as arguments.

The implementation works as follows: The symbols in the
argument list \texttt{?f} are bracketed off and all
remaining factors are collected as arguments of the
built-in function \texttt{dum\_}. Then all occurrences of
the function \texttt{dum\_} are replaced by~1, i.e.
the factors in the arguments are discarded.
\begin{flushleft} \small
\begin{minipage}{\linewidth} \label{scrap7}
$\langle\,$Procedure definition \texttt{stripcoeff}\nobreak\ {\footnotesize \NWtarget{nuweb7}{7}}$\,\rangle\equiv$
\vspace{-1ex}
\begin{list}{}{} \item
\mbox{}\verb@#@\hbox{\sffamily\bfseries procedure}\verb@ stripcoeff(?f)@\\
\mbox{}\verb@   @\hbox{\sffamily\bfseries Bracket}\verb@ `?f';@\\
\mbox{}\verb@.@\hbox{\sffamily\bfseries sort}\verb@@\\
\mbox{}\verb@   @\hbox{\sffamily\bfseries Collect}\verb@ @\hbox{\sffamily\bfseries dum}\verb@_;@\\
\mbox{}\verb@   @\hbox{\sffamily\bfseries Id}\verb@ @\hbox{\sffamily\bfseries dum}\verb@_(x?) = 1;@\\
\mbox{}\verb@#@\hbox{\sffamily\bfseries endprocedure}\verb@@{\NWsep}
\end{list}
\vspace{-1ex}
\footnotesize\addtolength{\baselineskip}{-1ex}
\begin{list}{}{\setlength{\itemsep}{-\parsep}\setlength{\itemindent}{-\leftmargin}}
\item \NWtxtMacroRefIn\ \NWlink{nuweb2}{2}.
\end{list}
\end{minipage}\\[4ex]
\end{flushleft}
In order to insert the gluons first all quark lines
are cut in all possible ways; then for each generated
pair of cuts a generator $t_{ij}^g$ is inserted in
all possible ways.
\begin{flushleft} \small
\begin{minipage}{\linewidth} \label{scrap8}
$\langle\,$Perform Insertions\nobreak\ {\footnotesize \NWtarget{nuweb8}{8}}$\,\rangle\equiv$
\vspace{-1ex}
\begin{list}{}{} \item
\mbox{}\verb@@\hbox{$\langle\,$Build Symmetriser\nobreak\ {\footnotesize \NWlink{nuweb5}{5}}$\,\rangle$}\verb@@\\
\mbox{}\verb@@\hbox{$\langle\,$Insert a pair of cuts\nobreak\ {\footnotesize \NWlink{nuweb9}{9}}$\,\rangle$}\verb@@\\
\mbox{}\verb@@\hbox{$\langle\,$Cut quark-lines\nobreak\ {\footnotesize \NWlink{nuweb10}{10}}$\,\rangle$}\verb@@\\
\mbox{}\verb@@\hbox{$\langle\,$Insert $t_{ij}^g$\nobreak\ {\footnotesize \NWlink{nuweb11}{11}}$\,\rangle$}\verb@@{\NWsep}
\end{list}
\vspace{-1ex}
\footnotesize\addtolength{\baselineskip}{-1ex}
\begin{list}{}{\setlength{\itemsep}{-\parsep}\setlength{\itemindent}{-\leftmargin}}
\item \NWtxtMacroRefIn\ \NWlink{nuweb2}{2}.
\end{list}
\end{minipage}\\[4ex]
\end{flushleft}
Since the program works with operators the commuting function
\texttt{d\_} has to be replaced by a non-commuting function
\texttt{delta}. Then for each gluon a pair of dummy indices
is introduced.
\begin{flushleft} \small
\begin{minipage}{\linewidth} \label{scrap9}
$\langle\,$Insert a pair of cuts\nobreak\ {\footnotesize \NWtarget{nuweb9}{9}}$\,\rangle\equiv$
\vspace{-1ex}
\begin{list}{}{} \item
\mbox{}\verb@@\hbox{\sffamily\bfseries Id}\verb@ @\hbox{\sffamily\bfseries d}\verb@_(i1?quarks, i2?aquarks) = delta(i1, i2);@\\
\mbox{}\verb@@\hbox{\sffamily\bfseries Repeat}\verb@;@\\
\mbox{}\verb@   @\hbox{\sffamily\bfseries Id}\verb@ @\hbox{\sffamily\bfseries Once}\verb@ insertgluon(g?) = insertq(i0, ia) * insertt(i0, ia, g);@\\
\mbox{}\verb@   @\hbox{\sffamily\bfseries Sum}\verb@ i0, ia;@\\
\mbox{}\verb@@\hbox{\sffamily\bfseries EndRepeat}\verb@;@{\NWsep}
\end{list}
\vspace{-1ex}
\footnotesize\addtolength{\baselineskip}{-1ex}
\begin{list}{}{\setlength{\itemsep}{-\parsep}\setlength{\itemindent}{-\leftmargin}}
\item \NWtxtMacroRefIn\ \NWlink{nuweb8}{8}.
\end{list}
\end{minipage}\\[4ex]
\end{flushleft}
There are three replacements for the insertion of the
cuts: first, all \texttt{inserttt}-operators are permuted
to the left such that the \texttt{insertq}-operators can
act on the quark-lines \texttt{delta}. The second
replacement implements the commutation relation
\begin{equation}
[\mathtt{insertq}(i,j),\delta_{i^\prime}^{j^\prime}]=
\delta_i^{j^\prime}\delta_{i^\prime}^j\text{.}
\end{equation}
If a \texttt{insertq} stands to the right of the terms
the according term is discarded by the third replacement.
\begin{flushleft} \small
\begin{minipage}{\linewidth} \label{scrap10}
$\langle\,$Cut quark-lines\nobreak\ {\footnotesize \NWtarget{nuweb10}{10}}$\,\rangle\equiv$
\vspace{-1ex}
\begin{list}{}{} \item
\mbox{}\verb@@\hbox{\sffamily\bfseries Repeat}\verb@ @\hbox{\sffamily\bfseries Id}\verb@ insertq(?any1) * insertt(?any2) =@\\
\mbox{}\verb@      insertt(?any2) * insertq(?any1);@\\
\mbox{}\verb@@\hbox{\sffamily\bfseries Repeat}\verb@ @\hbox{\sffamily\bfseries Id}\verb@ insertq(i0?, ia?) * delta(i1?, i2?) =@\\
\mbox{}\verb@      + delta(i1, ia) * delta(i0, i2)@\\
\mbox{}\verb@      + delta(i1, i2) * insertq(i0, ia);@\\
\mbox{}\verb@@\hbox{\sffamily\bfseries Id}\verb@ insertq(?any) = 0;@{\NWsep}
\end{list}
\vspace{-1ex}
\footnotesize\addtolength{\baselineskip}{-1ex}
\begin{list}{}{\setlength{\itemsep}{-\parsep}\setlength{\itemindent}{-\leftmargin}}
\item \NWtxtMacroRefIn\ \NWlink{nuweb8}{8}.
\end{list}
\end{minipage}\\[4ex]
\end{flushleft}
Similar to the previous set of rewriting rules the insertion
of the generators again cuts the diagram in all possible ways
and then fills generators into the gaps. It should be noted that
here it is necessary to also consider insertions to the left
and to the right of existing generators. The commutation relations
in this case are
\begin{align}
[\mathtt{insertt}(j,i,g),\delta_{i^\prime}^{j^\prime}]&=
\delta_{i^\prime}^jt_{ij^\prime}^g
+\delta_{i}^{j^\prime}t_{i^\prime j}^g
\quad\text{and}\\
[\mathtt{insertt}(j,i,g),t_{i^\prime j^\prime}^{g^\prime}]&=
t_{i^\prime j}^{g}t_{ij^\prime}^{g^\prime}
+t_{i^\prime j}^{g^\prime}t_{ij^\prime}^{g}
\text{.}
\end{align}
\begin{flushleft} \small
\begin{minipage}{\linewidth} \label{scrap11}
$\langle\,$Insert $t_{ij}^g$\nobreak\ {\footnotesize \NWtarget{nuweb11}{11}}$\,\rangle\equiv$
\vspace{-1ex}
\begin{list}{}{} \item
\mbox{}\verb@@\hbox{\sffamily\bfseries Repeat}\verb@;@\\
\mbox{}\verb@   @\hbox{\sffamily\bfseries Id}\verb@ insertt(i0?, ia?, g?) * delta(i1?, i2?) =@\\
\mbox{}\verb@      + delta(i1, i0) * t(ia, i2, g)@\\
\mbox{}\verb@      + t(i1, i0, g) * delta(ia, i2)@\\
\mbox{}\verb@      + delta(i1, i2) * insertt(i0, ia, g);@\\
\mbox{}\verb@   @\hbox{\sffamily\bfseries Id}\verb@ insertt(i0?, ia?, g) * t(i1?, i2?, g0?) =@\\
\mbox{}\verb@      + t(i1, i0, g) * t(ia, i2, g0)@\\
\mbox{}\verb@      + t(i1, i0, g0) * t(ia, i2, g)@\\
\mbox{}\verb@      + t(i1, i2, g0) * insertt(i0, ia, g);@\\
\mbox{}\verb@@\hbox{\sffamily\bfseries EndRepeat}\verb@;@\\
\mbox{}\verb@@\hbox{\sffamily\bfseries Id}\verb@ insertt(?any) = 0;@{\NWsep}
\end{list}
\vspace{-1ex}
\footnotesize\addtolength{\baselineskip}{-1ex}
\begin{list}{}{\setlength{\itemsep}{-\parsep}\setlength{\itemindent}{-\leftmargin}}
\item \NWtxtMacroRefIn\ \NWlink{nuweb8}{8}.
\end{list}
\end{minipage}\\[4ex]
\end{flushleft}
The last step consists of the simplification of the result:
all dummy indices are contracted, traces of one or zero
generators are replaced. Finally the numerical coefficients are stripped
off.
\begin{flushleft} \small
\begin{minipage}{\linewidth} \label{scrap12}
$\langle\,$Simplify Result\nobreak\ {\footnotesize \NWtarget{nuweb12}{12}}$\,\rangle\equiv$
\vspace{-1ex}
\begin{list}{}{} \item
\mbox{}\verb@@\hbox{\sffamily\bfseries Id}\verb@ delta(i1?, i2?) = line(i1, i2);@\\
\mbox{}\verb@@\hbox{\sffamily\bfseries Id}\verb@ t(i1?, i2?, g?) = line(i1, i2, g);@\\
\mbox{}\verb@@\hbox{\sffamily\bfseries Repeat}\verb@ @\hbox{\sffamily\bfseries Id}\verb@ line(i1?, i2?, ?head) * line(i2?, i3?, ?tail) =@\\
\mbox{}\verb@   line(i1, i3, ?head, ?tail);@\\
\mbox{}\verb@@\hbox{\sffamily\bfseries Id}\verb@ line(i1?, i1?, ?tail) = tr(?tail);@\\
\mbox{}\verb@@\\
\mbox{}\verb@@\hbox{\sffamily\bfseries Id}\verb@ tr(g?) = 0;@\\
\mbox{}\verb@@\hbox{\sffamily\bfseries Id}\verb@ tr() = 1;@\\
\mbox{}\verb@#@\hbox{\sffamily\bfseries call}\verb@ stripcoeff(line,tr)@{\NWsep}
\end{list}
\vspace{-1ex}
\footnotesize\addtolength{\baselineskip}{-1ex}
\begin{list}{}{\setlength{\itemsep}{-\parsep}\setlength{\itemindent}{-\leftmargin}}
\item \NWtxtMacroRefIn\ \NWlink{nuweb2}{2}.
\end{list}
\end{minipage}\\[4ex]
\end{flushleft}
The order of the vectors in the result depends on the internal
term ordering of the \form{} implementation. The output of
the program for the considered process $gg\rightarrow q\bar{q}q\bar{q}$
might look like the following:
\begin{Verbatim}
color1=tr(g1,g2)*line(i1,j1)*line(i2,j2)
color2=tr(g1,g2)*line(i1,j2)*line(i2,j1)
color3=line(i1,j1)*line(i2,j2,g1,g2)
color4=line(i1,j1)*line(i2,j2,g2,g1)
color5=line(i1,j1,g1)*line(i2,j2,g2)
color6=line(i1,j1,g1,g2)*line(i2,j2)
color7=line(i1,j1,g2)*line(i2,j2,g1)
color8=line(i1,j1,g2,g1)*line(i2,j2)
color9=line(i1,j2)*line(i2,j1,g1,g2)
color10=line(i1,j2)*line(i2,j1,g2,g1)
color11=line(i1,j2,g1)*line(i2,j1,g2)
color12=line(i1,j2,g1,g2)*line(i2,j1)
color13=line(i1,j2,g2)*line(i2,j1,g1)
color14=line(i1,j2,g2,g1)*line(i2,j1)
\end{Verbatim}
The occurrence of 14~basis vectors confirms
Equation~\eqref{eq:qcd-color:d-basis}. The program
has been tested for all configurations of
Table~\ref{tab:qcd-color:basecomp} with up to
six gluons.

%%%%%%%%%%%%%%%%%%%%%%%%%%%%%%%%%%%%%%%%%%%%%%%%%%%%%%%%%%%%%%%%%%%%%%%%
\subsection*{Programming with Contracts}
\index{contract, programming by}
The concept of \emph{Programming by Contract} has been proposed
by \person{Bertrand Meyer}~\cite{Meyer:DBC,Meyer:OOP}
and implemented in the programming language
\texttt{Eiffel}~\cite{Meyer:Eiffel}. Since then the concept
has been adapted in other languages either by direct integration
into the language definition or by additional libraries
and tools such as preprocessors. Programming by contract incorporates
three types of contracts between the caller of a method and the class
that implements the method: \emph{preconditions} are checked before a
method is invoked, \emph{postconditions} are checked after a method
returns from execution and \emph{class invariants} are checked before
and after each call to a public method of a class.

In this work I use the programming language
\texttt{Python}~\cite{rossum-python} with the additional package
\texttt{contract}~\cite{pythonpbc}. This combination allows to specify
contracts inside the interface documentation of \texttt{Python} classes.

As an example below are shown parts of the implementation of the
implementation of a class for permutations.
\begin{Verbatim}[commandchars=\\\{\},numbers=left,frame=lines,framerule=2pt]
{\color{blue}import} contract

{\color{blue}class} Permutation:
	"""
	Implements permutations\dots
	
	Internally, the permutations are stored in cycle representation
	with all cycles of length 1 omitted.

	{\color{red}inv:}
		all({\color{blue}len}(c) > 1 {\color{blue}for} c {\color{blue}in} self.cycles)
	"""

	{\color{blue}def} isIdentity(self):
		{\color{blue}return} {\color{blue}len}(self.cycles) == 0

	{\color{blue}def} inverse(self):
		"""
			Computes the inverse of this permutation

			{\color{red}post[]:}
				(self * {\color{red}__return__}).isIdentity()
		"""
		\ldots

contract.checkmod(__name__)
\end{Verbatim}

Lines 10 and~11 define a class invariant which checks that no cycles
of length one are stored. As usual in \texttt{Python}, indentation is
meaningful also within the contracts, i.e. the invariant spans over all
subsequent indented lines following line~10. Since the invariant is
checked before and after all methods, in \texttt{isIdentity} one can
rely on assertion that the identity is the only permutation with no
cycles of length larger than one\footnote{All cycles have to be disjoint
which is not checked here to maintain the simplicity of the example.}.

Lines 21--22 define a postcondition for the calculation of the
inverse element of a permutation. The square brackets after the
keyword \texttt{post} contain a list of variables that may be
modified by the method. The empty list asserts that the method does
not modify its environment at all. The postcondition itself specifies
the defining equation for the inverse element, $gg^{-1}=\textrm{id}$.

The last line \texttt{contract.checkmod(\_\_name\_\_)} is necessary
to activate the module \texttt{contract}, i.e. to parse the comments
for the keywords \texttt{inv}, \texttt{pre} and \texttt{post} and
wrap the methods inside new methods of the following format
\begin{Verbatim}[commandchars=\\\{\}]
{\color{blue}def} wrapper(\ldots):
	check class invariants
	check preconditions
	{\color{red}\_\_old\_\_} = save old values
	{\color{red}\_\_return\_\_} = call original method
	check postconditions
	{\color{blue}return} {\color{red} \_\_return\_\_}
\end{Verbatim}

The object \texttt{\_\_old\_\_} is created to allow the access
to the old values of global variables where methods modify their
environment. An example taken from the documentation of
the package \text{contract} shows its use in a function
that sorts a list in-place:

\begin{Verbatim}[commandchars=\\\{\},numbers=left,frame=lines,framerule=2pt]
{\color{blue}def} sort(a):
    """Sort a list.

    {\color{red}pre:} {\color{blue}isinstance}(a, {\color{blue}type}({\color{blue}list}))
    {\color{red}post[}a{\color{red}]:}
        # array size is unchanged
        {\color{blue}len}(a) == {\color{blue}len}({\color{red}\_\_old\_\_}.a)

        # array is ordered
        {\color{red}forall}([a[i] >= a[i-1] {\color{blue}for} i {\color{blue}in} {\color{blue}range}(1, {\color{blue}len}(a))])

        # all the old elements are still in the array
        {\color{red}forall}({\color{red}\_\_old\_\_}.a, {\color{blue}lambda} e: {\color{red}\_\_old\_\_}.a.count(e) == a.count(e))
    """
    \ldots
\end{Verbatim}

The term \emph{Programming by Contract} for this programming concept
can be explained by having contracts between the caller of a method
and the class as the two different parties of the contract. Both parties
have benefits and obligations. The class, as an obligation, has to ensure
that the postconditions of each method hold; its benefit from the contract
is, that it can rely on the preconditions to be true. The converse is true
for the caller: it can rely on the postconditions to be true and is
obliged to ensure the preconditions of the methods it calls. For the class
invariants all obligations remain within the class. However, both the class
and the caller then can rely on the class invariants. The example of
the permutation showed that this can sometimes lead to more efficient
implementations. If the contracts are kept very tight they can be used
as a tool of software verification and one, in principle could prove
the correctness of a program. In practise, however, very often the
challenge is to find and implement the correct pre- and
postconditions which are appropriate to ensure program correctness
and at the same time are sufficiently fast to test them for non-trivial
examples.

\subsection*{Program Correctness}
\label{sec:imp-correctness}
\input{imp-checks}

\section{Overview}
\label{sec:imp-overview}
\input{imp-overview}

\section{Diagram Generation}
\label{sec:imp-qgraf}
\input{imp-qgraf}

\section{Automatic Code Generation}
\label{sec:imp-autocode}
\input{imp-autocode}

\input{appendix-ffactory}

\section{Algebraic Simplification}
\label{sec:imp-form}
% vim: syntax=nuweb:ts=3:sw=3
%\renewcommand{\NWtarget}[2]{\hypertarget{#1}{#2}}
%\renewcommand{\NWlink}[2]{\hyperlink{#1}{#2}}
\newcommand{\PARAMETER}{\textbf{Parameter}}
\newcommand{\SIDEEFFECTS}{\textbf{Side effects}}

\subsection*{Introduction}
This section describes the \form{} code that is used to
generate \fortranXC{} files from the output of the
diagram generator \qgraf{}. The main goal of this code
is to keep the output as compact as possible. The arguments
in favour of this approach are shorter compilation times and
robustness against failures during the translation when the
requirements of the computer algebra program
exceed the resources provided by the system. On the other
hand, the code which is generated this way is generally
slower than an equivalent output that has been achieved by
more subtle simplification routines that take into account
all possible cancellations. However, the latter approach usually
requires a higher degree of process dependent fine-tuning
and is therefore less suitable for the implementation of a
general purpose tool.

The \ac{cas} \form~\cite{Vermaseren:2000nd,Vermaseren:2002,Vermaseren:2006ag}
in contrast to most general purpose \acp{cas} has originally been developed
mainly as a pure term rewriting system with added capability
to handle \person{Dirac} traces and vectors and, more general,
higher rank tensors. Expressions are represented as lists of terms,
and the canonical form is the fully expanded representation.
\form{} programs are structured as a list of modules; each module is applied
term by term, and only at the end of a module all terms are sorted and
the expression is brought into canonical form again.

The major drawback of this restriction to local replacements
is the incapability of having rewriting rules like
$a+b\rightarrow c$ because that would require the inspection of more
than one term at the same time. However, this is the price to
pay for two of the main advantages of \form: it can handle
arbitrarily big expressions
(limited only by the resources of the computer) and it is very fast
for it avoids the complexity of AC-unification~\cite{Kapur:1992}.

\subsection{The Main Program}
The \form{} program for the algebraic simplification is not called
directly by the user but is invoked by a \texttt{Python} program
that coordinates the translation (see Section~\ref{sec:imp-autocode}).
Command line options are passed to the program through preprocessor
definitions using the \form{} command line parameter \texttt{-D} as
follows:
\begin{center}
	\texttt{form -D DIAG=}\cmdvar{diagram}
		\texttt{-D PREFIX=}\cmdvar{prefix}
		\texttt{-D HELICITY=}\cmdvar{helicity}
		\texttt{preprocess.frm}
\end{center}
The parameter~\cmdvar{diagram} is the index of the diagram to be
processed and corresponds to the labelling assigned by \qgraf.
The prefix usually is one of \texttt{born}, \texttt{virt} or
\texttt{real} and selects which part of the amplitude is calculated;
this, however, is just a naming convention. The diagrams always are
read from the current \texttt{diagrams.h} file and the program relies
on the controlling \texttt{Python} program to set up the environment
correctly. The parameter \cmdvar{helicity} is the binary encoding
of the helicity to be calculated. In the massless case, where the
helicities $\lambda_i=\pm1$ this number is
\begin{equation}
\text{\cmdvar{helicity}}=\sum_{i=1}^{N}\frac{\lambda_{i}+1}{2}\cdot2^{i-1}
\end{equation}
Before the actual program starts it verifies that all three parameters
are present.
\begin{flushleft} \small \label{scrap13}
$\langle\,$check command line arguments\nobreak\ {\footnotesize \NWtarget{nuweb13}{13}}$\,\rangle\equiv$
\vspace{-1ex}
\begin{list}{}{} \item
\mbox{}\verb@#@\hbox{\sffamily\bfseries IfNDef}\verb@ `DIAG'@\\
\mbox{}\verb@   #@\hbox{\sffamily\bfseries Message}\verb@ "Please, run with -D DIAG=<diagram> from command line."@\\
\mbox{}\verb@   #@\hbox{\sffamily\bfseries Terminate}\verb@@\\
\mbox{}\verb@#@\hbox{\sffamily\bfseries EndIf}\verb@@\\
\mbox{}\verb@@\\
\mbox{}\verb@#@\hbox{\sffamily\bfseries IfNDef}\verb@ `PREFIX'@\\
\mbox{}\verb@   #@\hbox{\sffamily\bfseries Message}\verb@ "Please, run with -D PREFIX=<fileprefix> from command line."@\\
\mbox{}\verb@   #@\hbox{\sffamily\bfseries Terminate}\verb@@\\
\mbox{}\verb@#@\hbox{\sffamily\bfseries EndIf}\verb@@\\
\mbox{}\verb@@\\
\mbox{}\verb@#@\hbox{\sffamily\bfseries IfNDef}\verb@ `HELICITY'@\\
\mbox{}\verb@   #@\hbox{\sffamily\bfseries Message}\verb@ "Please, run with -D HELICITY=<helicity> from command line."@\\
\mbox{}\verb@   #@\hbox{\sffamily\bfseries Terminate}\verb@@\\
\mbox{}\verb@#@\hbox{\sffamily\bfseries EndIf}\verb@@{\NWsep}
\end{list}
\vspace{-1ex}
\footnotesize\addtolength{\baselineskip}{-1ex}
\begin{list}{}{\setlength{\itemsep}{-\parsep}\setlength{\itemindent}{-\leftmargin}}
\item \NWtxtMacroRefIn\ \NWlink{nuweb23}{23}.
\end{list}
\end{flushleft}
Before the program starts the actual simplification it checks
for the consistency of the diagram number. The total number of
diagrams for a subprocess is found in a fold called \texttt{\#global}
in the file \texttt{diagrams.h}.
\begin{flushleft} \small \label{scrap14}
$\langle\,$check bounds on diagram number\nobreak\ {\footnotesize \NWtarget{nuweb14}{14}}$\,\rangle\equiv$
\vspace{-1ex}
\begin{list}{}{} \item
\mbox{}\verb@#@\hbox{\sffamily\bfseries Define}\verb@ DIAGRAM "diagram"@\\
\mbox{}\verb@#@\hbox{\sffamily\bfseries Include}\verb@- diagrams.h #global@\\
\mbox{}\verb@#@\hbox{\sffamily\bfseries If}\verb@ `DIAG' > `DIAGRAMCOUNT'@\\
\mbox{}\verb@   #@\hbox{\sffamily\bfseries Message}\verb@ "DIAG (`DIAG') > DIAGRAMCOUNT (`DIAGRAMCOUNT')"@\\
\mbox{}\verb@   #@\hbox{\sffamily\bfseries Terminate}\verb@@\\
\mbox{}\verb@#@\hbox{\sffamily\bfseries EndIf}\verb@@{\NWsep}
\end{list}
\vspace{-1ex}
\footnotesize\addtolength{\baselineskip}{-1ex}
\begin{list}{}{\setlength{\itemsep}{-\parsep}\setlength{\itemindent}{-\leftmargin}}
\item \NWtxtMacroRefIn\ \NWlink{nuweb24}{24}.
\end{list}
\end{flushleft}
The program should also check if it has been called with external
channels set up correctly:
\begin{flushleft} \small \label{scrap15}
$\langle\,$check communication channels\nobreak\ {\footnotesize \NWtarget{nuweb15}{15}}$\,\rangle\equiv$
\vspace{-1ex}
\begin{list}{}{} \item
\mbox{}\verb@#@\hbox{\sffamily\bfseries IfnDef}\verb@ `PIPES_'@\\
\mbox{}\verb@   #@\hbox{\sffamily\bfseries Message}\verb@ "This program must be called from within GOLEM."@\\
\mbox{}\verb@   #@\hbox{\sffamily\bfseries Terminate}\verb@@\\
\mbox{}\verb@#@\hbox{\sffamily\bfseries EndIf}\verb@@\\
\mbox{}\verb@#@\hbox{\sffamily\bfseries SetExternal}\verb@ `PIPE1_'@{\NWsep}
\end{list}
\vspace{-1ex}
\footnotesize\addtolength{\baselineskip}{-1ex}
\begin{list}{}{\setlength{\itemsep}{-\parsep}\setlength{\itemindent}{-\leftmargin}}
\item \NWtxtMacroRefIn\ \NWlink{nuweb23}{23}.
\end{list}
\end{flushleft}
Before the program can read in the \person{Feynman} diagram
all occurring symbols need to be declared. Here, two classes
of symbols are distinguished: symbols that appear only in the
\person{Feynman} rules are defined in the according file,
e.g. \texttt{smqcd.h} for \acl{sm} \ac{qcd}, generic symbols
that appear during the simplification are defined in a
file called \texttt{symbols.h}. At the end of that file
there is a list of automatic declarations mainly for symbols
that are used locally only.
\begin{flushleft} \small
\begin{minipage}{\linewidth} \label{scrap16}
\verb@"symbols.h"@\nobreak\ {\footnotesize \NWtarget{nuweb16}{16} }$\equiv$
\vspace{-1ex}
\begin{list}{}{} \item
\mbox{}\verb@@\hbox{$\langle\,$define symbols for colour algebra\nobreak\ {\footnotesize \NWlink{nuweb17}{17}}$\,\rangle$}\verb@@\\
\mbox{}\verb@@\hbox{$\langle\,$define symbols for \person{Lorentz} and \person{Dirac} algebra\nobreak\ {\footnotesize \NWlink{nuweb18}{18}}$\,\rangle$}\verb@@\\
\mbox{}\verb@@\hbox{$\langle\,$define vectors\nobreak\ {\footnotesize \NWlink{nuweb19}{19}}$\,\rangle$}\verb@@\\
\mbox{}\verb@@\hbox{$\langle\,$define topological functions\nobreak\ {\footnotesize \NWlink{nuweb20}{20}}$\,\rangle$}\verb@@\\
\mbox{}\verb@@\hbox{$\langle\,$define auxiliary functions and symbols\nobreak\ {\footnotesize \NWlink{nuweb21}{21}}$\,\rangle$}\verb@@\\
\mbox{}\verb@@\hbox{$\langle\,$define form factors\nobreak\ {\footnotesize \NWlink{nuweb22}{22}}$\,\rangle$}\verb@@\\
\mbox{}\verb@@\hbox{\sffamily\bfseries Symbol}\verb@ g;@\\
\mbox{}\verb@@\\
\mbox{}\verb@@\hbox{\sffamily\bfseries AutoDeclare}\verb@ @\hbox{\sffamily\bfseries Indices}\verb@ i;@\\
\mbox{}\verb@@\hbox{\sffamily\bfseries AutoDeclare}\verb@ @\hbox{\sffamily\bfseries CFunctions}\verb@ ANY, TEMP;@\\
\mbox{}\verb@@\hbox{\sffamily\bfseries AutoDeclare}\verb@ @\hbox{\sffamily\bfseries Functions}\verb@ NCTEMP;@\\
\mbox{}\verb@@\hbox{\sffamily\bfseries AutoDeclare}\verb@ @\hbox{\sffamily\bfseries Symbols}\verb@ cc, color, ff, tr;@\\
\mbox{}\verb@@\hbox{\sffamily\bfseries AutoDeclare}\verb@ @\hbox{\sffamily\bfseries Vectors}\verb@ vec;@{\NWsep}
\end{list}
\vspace{-2ex}
\end{minipage}\\[4ex]
\end{flushleft}
The symbol \texttt{g} stands for the coupling constant.
Symbols starting with \texttt{ff} and \texttt{tr} represent
form factors and traces respectively. The same symbols are
used by the \texttt{Python} code that maintains a global
list of both.

For the colour algebra the scalar objects
$\mathtt{dF}\equiv N_C$, $\mathtt{dA}\equiv N_C^2-1$,
$\mathtt{TR}\equiv 1/2$ and $\mathtt{CA}\equiv C_A$
are used. The generators are called \texttt{T} in the
fundamental and \texttt{f} in the adjoint representation,
$\mathtt{f(A, B, C)}\equiv f^{ABC}$ and
$\mathtt{T(i, j, A)}\equiv t_{ij}^A$.
\begin{flushleft} \small \label{scrap17}
$\langle\,$define symbols for colour algebra\nobreak\ {\footnotesize \NWtarget{nuweb17}{17}}$\,\rangle\equiv$
\vspace{-1ex}
\begin{list}{}{} \item
\mbox{}\verb@@\hbox{\sffamily\bfseries Symbols}\verb@ dF, dA, TR, CA;@\\
\mbox{}\verb@@\hbox{\sffamily\bfseries CFunctions}\verb@ f, T;@\\
\mbox{}\verb@@\hbox{\sffamily\bfseries CFunction}\verb@ AdjointID(@\hbox{\sffamily\bfseries symmetric}\verb@);@\\
\mbox{}\verb@@\hbox{\sffamily\bfseries CFunction}\verb@ FundamentalID(@\hbox{\sffamily\bfseries symmetric}\verb@);@{\NWsep}
\end{list}
\vspace{-1ex}
\footnotesize\addtolength{\baselineskip}{-1ex}
\begin{list}{}{\setlength{\itemsep}{-\parsep}\setlength{\itemindent}{-\leftmargin}}
\item \NWtxtMacroRefIn\ \NWlink{nuweb16}{16}.
\end{list}
\end{flushleft}
For the \person{Lorentz} algebra and \person{Dirac} algebra
the following conventions are used:
the \person{Dirac} matrices are called \texttt{gg},
$\mathtt{gg(i, j, mu)}\equiv(\gamma^\mu)_{ij}$ and the
corresponding identity matrix is \texttt{gammaID};
$\gamma_5$ is \texttt{gamma5} which defines
$\mathtt{hProjector}(\pm1)\equiv(\One\pm\gamma_5)/2$.

Spinor objects are represented by
$\mathtt{Spinor}(k_i,\pm1)\equiv\ket{k_i^\pm}$,
$\mathtt{AdjSpinor}(k_i,\pm1)\equiv\bra{k_i^\pm}$ and
$\mathtt{SpinorLine}(k_i,\lambda_i,\mu_1, \mu_2, \ldots, \mu_r, k_j, \lambda_j)
\equiv\braket{\left.k_i^{\lambda_i}\right\vert \mu_1, \mu_2, \ldots, \mu_r
\left\vert k_j^{\lambda_j}\right.}$. The metric tensor is denoted
by \texttt{gTensor}. The dimension of the \person{Minkowski} space
is $\mathtt{n}=4-2\cdot\mathtt{eps}=4+\mathtt{[-2eps]}$.
\begin{flushleft} \small \label{scrap18}
$\langle\,$define symbols for \person{Lorentz} and \person{Dirac} algebra\nobreak\ {\footnotesize \NWtarget{nuweb18}{18}}$\,\rangle\equiv$
\vspace{-1ex}
\begin{list}{}{} \item
\mbox{}\verb@@\hbox{\sffamily\bfseries CFunctions}\verb@ gg, gammaID(@\hbox{\sffamily\bfseries symmetric}\verb@), gamma5, hProjector;@\\
\mbox{}\verb@@\hbox{\sffamily\bfseries CFunctions}\verb@ Spinor, AdjSpinor;@\\
\mbox{}\verb@@\hbox{\sffamily\bfseries CFunctions}\verb@ SpinorLine, SpinorTrace;@\\
\mbox{}\verb@@\hbox{\sffamily\bfseries CFunction}\verb@ gTensor(@\hbox{\sffamily\bfseries symmetric}\verb@);@\\
\mbox{}\verb@@\hbox{\sffamily\bfseries Symbols}\verb@ n, eps, [-2eps];@\\
\mbox{}\verb@@{\NWsep}
\end{list}
\vspace{-1ex}
\footnotesize\addtolength{\baselineskip}{-1ex}
\begin{list}{}{\setlength{\itemsep}{-\parsep}\setlength{\itemindent}{-\leftmargin}}
\item \NWtxtMacroRefIn\ \NWlink{nuweb16}{16}.
\end{list}
\end{flushleft}
The four-momenta used in the calculation are
$k_1,k_2,\ldots$ for the external momenta and
$p_1$ for the integration momentum. The translation
of polarisation vectors into spinor helicity
notation (see Equation~\eqref{eq:qcd-shelpmethod:defepsilon})
makes it necessary to have additional gauge vectors
which are called \texttt{qGauge1}, \texttt{qGauge2}
and so on where the index denotes the corresponding particle
(\texttt{k1}, \texttt{k2}, \dots resp.).
\begin{flushleft} \small \label{scrap19}
$\langle\,$define vectors\nobreak\ {\footnotesize \NWtarget{nuweb19}{19}}$\,\rangle\equiv$
\vspace{-1ex}
\begin{list}{}{} \item
\mbox{}\verb@#@\hbox{\sffamily\bfseries define}\verb@ MAXLEGS "7"@\\
\mbox{}\verb@@\hbox{\sffamily\bfseries Vectors}\verb@ k1, ..., k`MAXLEGS'; @\\
\mbox{}\verb@@\hbox{\sffamily\bfseries Vectors}\verb@ p1;@\\
\mbox{}\verb@@\hbox{\sffamily\bfseries Vectors}\verb@ qGauge1, ..., qGauge`MAXLEGS';@{\NWsep}
\end{list}
\vspace{-1ex}
\footnotesize\addtolength{\baselineskip}{-1ex}
\begin{list}{}{\setlength{\itemsep}{-\parsep}\setlength{\itemindent}{-\leftmargin}}
\item \NWtxtMacroRefIn\ \NWlink{nuweb16}{16}.
\end{list}
\end{flushleft}
For the analysis of the topology of each diagram the
function \texttt{edge} is introduced for each propagator,
the function \texttt{node} for every vertex and later
\texttt{circle} to indicate the loop in a one-loop diagram.
\begin{flushleft} \small \label{scrap20}
$\langle\,$define topological functions\nobreak\ {\footnotesize \NWtarget{nuweb20}{20}}$\,\rangle\equiv$
\vspace{-1ex}
\begin{list}{}{} \item
\mbox{}\verb@@\hbox{\sffamily\bfseries CFunctions}\verb@ node, edge(@\hbox{\sffamily\bfseries symmetric}\verb@), circle(@\hbox{\sffamily\bfseries cyclic}\verb@);@{\NWsep}
\end{list}
\vspace{-1ex}
\footnotesize\addtolength{\baselineskip}{-1ex}
\begin{list}{}{\setlength{\itemsep}{-\parsep}\setlength{\itemindent}{-\leftmargin}}
\item \NWtxtMacroRefIn\ \NWlink{nuweb16}{16}.
\end{list}
\end{flushleft}
The function \texttt{POW} denotes powers, $\mathtt{POW}(a, b)=a^b$;
this is especially useful to treat denominators as
$\mathtt{POW}(\dots,-1)$. Spinor products are represented as
$\braket{p_{\lambda_1}\vert q_{\lambda_2}}=
\mathtt{braket}(p, \lambda_1, q, \lambda_2)$. The function
\texttt{MOMENTUM} prevents momenta from automatic contraction
which is desired at several places in the program. Propagators
are translated into $1/(q^2-m^2)=\mathtt{PROP}(q, m)$. For the
translation of the tensor integrals into form factors at an
intermediate step the function \texttt{TI} is introduced to
represent a tensor integral. Its arguments are
\begin{equation}
I_N^{n;\mu_1\mu_2\ldots\mu_r}(a_1,a_2,\ldots,a_r;S)=
\mathtt{TI}(N, r, r_{a_1}, \mu_1, r_{a_2}, \mu_2,\ldots, r_{a_r}, \mu_r)
\text{.}
\end{equation}
The functions \texttt{PREFACTOR} and \texttt{COLORBASIS} are used
to separate parts of the expression into the argument of the according
function. The difference vectors $\Delta_{ij}^\mu=r_i^\mu-r_j^\mu$ are
encoded into the function $\mathtt{DELTA}(i, j, \mu)$. The symbol
\texttt{SNULL} represents an empty list of pinches of the $S$-matrix.

\begin{flushleft} \small \label{scrap21}
$\langle\,$define auxiliary functions and symbols\nobreak\ {\footnotesize \NWtarget{nuweb21}{21}}$\,\rangle\equiv$
\vspace{-1ex}
\begin{list}{}{} \item
\mbox{}\verb@@\hbox{\sffamily\bfseries CFunction}\verb@ POW, braket;@\\
\mbox{}\verb@@\hbox{\sffamily\bfseries CFunctions}\verb@ MOMENTUM, PROP;@\\
\mbox{}\verb@@\hbox{\sffamily\bfseries CFunctions}\verb@ TI;@\\
\mbox{}\verb@@\hbox{\sffamily\bfseries CFunctions}\verb@ PREFACTOR, COLORBASIS;@\\
\mbox{}\verb@@\hbox{\sffamily\bfseries CFunction}\verb@ DELTA;@\\
\mbox{}\verb@@\\
\mbox{}\verb@#@\hbox{\sffamily\bfseries define}\verb@ SNULL "nullarray"@\\
\mbox{}\verb@@\hbox{\sffamily\bfseries Symbol}\verb@ `SNULL';@{\NWsep}
\end{list}
\vspace{-1ex}
\footnotesize\addtolength{\baselineskip}{-1ex}
\begin{list}{}{\setlength{\itemsep}{-\parsep}\setlength{\itemindent}{-\leftmargin}}
\item \NWtxtMacroRefIn\ \NWlink{nuweb16}{16}.
\end{list}
\end{flushleft}
The following list defines all integral form factors
that can appear during the reduction, up to six-point
function and tensor rank six.
\begin{flushleft} \small \label{scrap22}
$\langle\,$define form factors\nobreak\ {\footnotesize \NWtarget{nuweb22}{22}}$\,\rangle\equiv$
\vspace{-1ex}
\begin{list}{}{} \item
\mbox{}\verb@@\hbox{\sffamily\bfseries CFunctions}\verb@@\\
\mbox{}\verb@   a10, a20, a30, a40, a50, a60, a11, a21, a31,@\\
\mbox{}\verb@   a41, a51, a61, a22, a32, a42, a52, a62, a33,@\\
\mbox{}\verb@   a43, a53, a63, a44, a54, a64, a55, a65, a66;@\\
\mbox{}\verb@@\hbox{\sffamily\bfseries CFunctions}\verb@@\\
\mbox{}\verb@   b10, b20, b30, b40, b50, b11, b21, b31, b41, b51,@\\
\mbox{}\verb@   b22, b32, b42, b52, b33, b43, b53, b44, b54, b55;@\\
\mbox{}\verb@@\hbox{\sffamily\bfseries CFunctions}\verb@@\\
\mbox{}\verb@   c10, c20, c30, c40, c50, c11, c21, c31, c41, c51,@\\
\mbox{}\verb@   c22, c32, c42, c52, c33, c43, c53, c44, c54, c55;@\\
\mbox{}\verb@@\hbox{\sffamily\bfseries Set}\verb@ FormFactors:@\\
\mbox{}\verb@   a10, a20, a30, a40, a50, a60, a11, a21, a31, a41,@\\
\mbox{}\verb@   a51, a61, a22, a32, a42, a52, a62, a33, a43, a53,@\\
\mbox{}\verb@   a63, a44, a54, a64, a55, a65, a66, b10, b20, b30,@\\
\mbox{}\verb@   b40, b50, b11, b21, b31, b41, b51, b22, b32, b42,@\\
\mbox{}\verb@   b52, b33, b43, b53, b44, b54, b55, c10, c20, c30,@\\
\mbox{}\verb@   c40, c50, c11, c21, c31, c41, c51, c22, c32, c42,@\\
\mbox{}\verb@   c52, c33, c43, c53, c44, c54, c55;@{\NWsep}
\end{list}
\vspace{-1ex}
\footnotesize\addtolength{\baselineskip}{-1ex}
\begin{list}{}{\setlength{\itemsep}{-\parsep}\setlength{\itemindent}{-\leftmargin}}
\item \NWtxtMacroRefIn\ \NWlink{nuweb16}{16}.
\end{list}
\end{flushleft}
The structure of the main program follows below. After the
program has checked the parameters and defined all symbols
the main part of the program simplifies the expression for
one diagram at the specified helicity and writes out a
\fortranXC{} program.
\begin{flushleft} \small \label{scrap23}
\verb@"preprocess.frm"@\nobreak\ {\footnotesize \NWtarget{nuweb23}{23} }$\equiv$
\vspace{-1ex}
\begin{list}{}{} \item
\mbox{}\verb@#-@\\
\mbox{}\verb@#:@\hbox{\sffamily\bfseries WorkSpace}\verb@ 10M@\\
\mbox{}\verb@@\hbox{\sffamily\bfseries On}\verb@ @\hbox{\sffamily\bfseries ShortStatistics}\verb@;@\\
\mbox{}\verb@@\hbox{\sffamily\bfseries Off}\verb@ @\hbox{\sffamily\bfseries Statistics}\verb@;@\\
\mbox{}\verb@@\\
\mbox{}\verb@@\hbox{$\langle\,$check communication channels\nobreak\ {\footnotesize \NWlink{nuweb15}{15}}$\,\rangle$}\verb@@\\
\mbox{}\verb@@\hbox{$\langle\,$check command line arguments\nobreak\ {\footnotesize \NWlink{nuweb13}{13}}$\,\rangle$}\verb@@\\
\mbox{}\verb@#@\hbox{\sffamily\bfseries Define}\verb@ OUT "`PREFIX'`DIAG'_`HELICITY'.f90"@\\
\mbox{}\verb@@\\
\mbox{}\verb@@\hbox{$\langle\,$read libraries and configuration\nobreak\ {\footnotesize \NWlink{nuweb24}{24}}$\,\rangle$}\verb@@\\
\mbox{}\verb@@\hbox{$\langle\,$determine the helicities of the external particles\nobreak\ {\footnotesize \NWlink{nuweb25}{25}}$\,\rangle$}\verb@@\\
\mbox{}\verb@@\hbox{$\langle\,$define procedures\nobreak\ {\footnotesize \NWlink{nuweb61}{61}}$\,\rangle$}\verb@@\\
\mbox{}\verb@.@\hbox{\sffamily\bfseries sort}\verb@@\\
\mbox{}\verb@@\hbox{$\langle\,$simplification algorithm\nobreak\ {\footnotesize \NWlink{nuweb53}{53}}$\,\rangle$}\verb@@\\
\mbox{}\verb@@\hbox{$\langle\,$output section\nobreak\ {\footnotesize \NWlink{nuweb54}{54}}$\,\rangle$}\verb@@\\
\mbox{}\verb@#@\hbox{\sffamily\bfseries ToExternal}\verb@ "DONE\n"@\\
\mbox{}\verb@.@\hbox{\sffamily\bfseries end}\verb@@{\NWsep}
\end{list}
\vspace{-2ex}
\end{flushleft}
The main program finishes with notifying the \texttt{Python}
program about its termination through sending a line
containing the word ``\texttt{DONE}'' through the external
channels.

The first file that needs to be included is the
file \texttt{symbols.h} which has been explained above.
The file \texttt{`PREFIX'-color.h} is automatically generated
by the \texttt{Python} program \texttt{golem.py} and provides
information about the colour basis. The file \texttt{diagrams.h}
which is generated by \qgraf{} contains a variable \texttt{THEORY}
in its global section; this specifies the file that must be used
to translate the \person{Feynman} rules into a theory-independent
expression. The file \texttt{kin`LEGS'.h} defines the
\person{Mandelstam} variables according to the number of external
particles, which is stored in the variable \texttt{LEGS}.
\begin{flushleft} \small \label{scrap24}
$\langle\,$read libraries and configuration\nobreak\ {\footnotesize \NWtarget{nuweb24}{24}}$\,\rangle\equiv$
\vspace{-1ex}
\begin{list}{}{} \item
\mbox{}\verb@#@\hbox{\sffamily\bfseries Include}\verb@- symbols.h@\\
\mbox{}\verb@@\hbox{$\langle\,$check bounds on diagram number\nobreak\ {\footnotesize \NWlink{nuweb14}{14}}$\,\rangle$}\verb@@\\
\mbox{}\verb@#@\hbox{\sffamily\bfseries Include}\verb@- `PREFIX'-color.h@\\
\mbox{}\verb@#@\hbox{\sffamily\bfseries Include}\verb@- `THEORY'.h@\\
\mbox{}\verb@#@\hbox{\sffamily\bfseries Include}\verb@- diagrams.h #d`DIAG'@\\
\mbox{}\verb@.@\hbox{\sffamily\bfseries sort}\verb@@\\
\mbox{}\verb@#@\hbox{\sffamily\bfseries Include}\verb@- kin`LEGS'.h@\\
\mbox{}\verb@#@\hbox{\sffamily\bfseries Include}\verb@- process.h@{\NWsep}
\end{list}
\vspace{-1ex}
\footnotesize\addtolength{\baselineskip}{-1ex}
\begin{list}{}{\setlength{\itemsep}{-\parsep}\setlength{\itemindent}{-\leftmargin}}
\item \NWtxtMacroRefIn\ \NWlink{nuweb23}{23}.
\end{list}
\end{flushleft}
It should be noted that the order of the include statements
matters in the sense that some files depend on the information
supplied by other files such as the number of external particles.

The helicities of the external particles are encoded in an integer
number in binary. If $\lambda_i=\pm1$ is the helicity of the particle
associated with the momentum $k_i$ then this number is
\begin{displaymath}
\sum_{i=1}^N\left(\frac{\lambda_i+1}{2}\right)\cdot 2^{i-1}\text{.}
\end{displaymath}
In the program the variables $\mathtt{HEL}i=\lambda_i$ are used.
\begin{flushleft} \small \label{scrap25}
$\langle\,$determine the helicities of the external particles\nobreak\ {\footnotesize \NWtarget{nuweb25}{25}}$\,\rangle\equiv$
\vspace{-1ex}
\begin{list}{}{} \item
\mbox{}\verb@#@\hbox{\sffamily\bfseries Define}\verb@ HEL "`HELICITY'"@\\
\mbox{}\verb@#@\hbox{\sffamily\bfseries Do}\verb@ i=1,`LEGS'@\\
\mbox{}\verb@   #@\hbox{\sffamily\bfseries Define}\verb@ HEL`i' "{2*(`HEL'%2)-1}"@\\
\mbox{}\verb@   #@\hbox{\sffamily\bfseries Redefine}\verb@ HEL "{`HEL'/2}"@\\
\mbox{}\verb@#@\hbox{\sffamily\bfseries EndDo}\verb@@\\
\mbox{}\verb@#@\hbox{\sffamily\bfseries UnDefine}\verb@ HEL@{\NWsep}
\end{list}
\vspace{-1ex}
\footnotesize\addtolength{\baselineskip}{-1ex}
\begin{list}{}{\setlength{\itemsep}{-\parsep}\setlength{\itemindent}{-\leftmargin}}
\item \NWtxtMacroRefIn\ \NWlink{nuweb23}{23}.
\end{list}
\end{flushleft}
\subsection{Simplification Algorithm}
\subsubsection{Topological Analysis}
Rather than the process $1+2\rightarrow3+4+\ldots+N$ we consider
$1+2+\ldots+N\rightarrow 0$ by crossing all outgoing legs to the
initial state.
\begin{flushleft} \small \label{scrap26}
$\langle\,$make all momenta ingoing\nobreak\ {\footnotesize \NWtarget{nuweb26}{26}}$\,\rangle\equiv$
\vspace{-1ex}
\begin{list}{}{} \item
\mbox{}\verb@#@\hbox{\sffamily\bfseries Do}\verb@ i=3,`LEGS'@\\
\mbox{}\verb@   @\hbox{\sffamily\bfseries Multiply}\verb@ @\hbox{\sffamily\bfseries replace}\verb@_(k`i', vec`i');@\\
\mbox{}\verb@   @\hbox{\sffamily\bfseries Multiply}\verb@ @\hbox{\sffamily\bfseries replace}\verb@_(vec`i', -k`i');@\\
\mbox{}\verb@#@\hbox{\sffamily\bfseries EndDo}\verb@@{\NWsep}
\end{list}
\vspace{-1ex}
\footnotesize\addtolength{\baselineskip}{-1ex}
\begin{list}{}{\setlength{\itemsep}{-\parsep}\setlength{\itemindent}{-\leftmargin}}
\item \NWtxtMacroRefIn\ \NWlink{nuweb53}{53}.
\end{list}
\end{flushleft}
For one-loop diagrams we introduce $r_i=k_1+\ldots+k_i$,
$q_i=p+r_i$, where $p$ is the integration momentum, and
$\Delta_{ij}=\mathtt{D}i\mathtt{x}j=r_i-r_j$. The procedure
\texttt{IntroduceRMomenta} replaces the sums of momenta by
the abbreviations $r_i$, $q_i$ and $\Delta_{ij}$. The
\texttt{Argument} statement is necessary in order to
replace the momenta inside function arguments, too.
\begin{flushleft} \small \label{scrap27}
$\langle\,$introduce momenta $q_i$ and $r_i$ for one-loop processes\nobreak\ {\footnotesize \NWtarget{nuweb27}{27}}$\,\rangle\equiv$
\vspace{-1ex}
\begin{list}{}{} \item
\mbox{}\verb@#@\hbox{\sffamily\bfseries If}\verb@ `LOOPS' == 1@\\
\mbox{}\verb@.@\hbox{\sffamily\bfseries sort}\verb@@\\
\mbox{}\verb@   @\hbox{\sffamily\bfseries Vectors}\verb@ q1, ..., q`$loopsize';@\\
\mbox{}\verb@   @\hbox{\sffamily\bfseries Vectors}\verb@ r1, ..., r`$loopsize';@\\
\mbox{}\verb@   #@\hbox{\sffamily\bfseries Do}\verb@ i=1,{`$loopsize'-1}@\\
\mbox{}\verb@      #@\hbox{\sffamily\bfseries Do}\verb@ j={`i'+1}, `$loopsize'@\\
\mbox{}\verb@         @\hbox{\sffamily\bfseries Vector}\verb@ D`i'x`j';@\\
\mbox{}\verb@      #@\hbox{\sffamily\bfseries EndDo}\verb@@\\
\mbox{}\verb@   #@\hbox{\sffamily\bfseries EndDo}\verb@@\\
\mbox{}\verb@   @\hbox{\sffamily\bfseries Set}\verb@ qSET: q1, ..., q'$loopsize';@\\
\mbox{}\verb@   @\hbox{\sffamily\bfseries Set}\verb@ rSET: r1, ..., r'$loopsize';@\\
\mbox{}\verb@@\\
\mbox{}\verb@   #@\hbox{\sffamily\bfseries Call}\verb@ IntroduceRMomenta()@\\
\mbox{}\verb@   @\hbox{\sffamily\bfseries Argument}\verb@ VertexFunction;@\\
\mbox{}\verb@      #@\hbox{\sffamily\bfseries Call}\verb@ IntroduceRMomenta()@\\
\mbox{}\verb@   @\hbox{\sffamily\bfseries EndArgument}\verb@;@\\
\mbox{}\verb@#@\hbox{\sffamily\bfseries EndIf}\verb@@{\NWsep}
\end{list}
\vspace{-1ex}
\footnotesize\addtolength{\baselineskip}{-1ex}
\begin{list}{}{\setlength{\itemsep}{-\parsep}\setlength{\itemindent}{-\leftmargin}}
\item \NWtxtMacroRefIn\ \NWlink{nuweb53}{53}.
\end{list}
\end{flushleft}
The procedure \texttt{TopologyInfo} determines the
loop size, the pinched propagators
and the permutation of the external legs.
This information is passed to the \texttt{Python}
program through the external channels.
\begin{flushleft} \small \label{scrap28}
$\langle\,$determine graph topology\nobreak\ {\footnotesize \NWtarget{nuweb28}{28}}$\,\rangle\equiv$
\vspace{-1ex}
\begin{list}{}{} \item
\mbox{}\verb@#@\hbox{\sffamily\bfseries Call}\verb@ TopologyInfo()@\\
\mbox{}\verb@#@\hbox{\sffamily\bfseries ToExternal}\verb@ "LO `LOOPS'\n"@\\
\mbox{}\verb@#@\hbox{\sffamily\bfseries ToExternal}\verb@ "LE `LEGS'\n"@\\
\mbox{}\verb@#@\hbox{\sffamily\bfseries ToExternal}\verb@ "PE `LEGPERMUTATION'\n"@\\
\mbox{}\verb@#@\hbox{\sffamily\bfseries ToExternal}\verb@ "PI `PINCHES'\n"@\\
\mbox{}\verb@#@\hbox{\sffamily\bfseries If}\verb@ (`LOOPS' == 1)@\\
\mbox{}\verb@   #@\hbox{\sffamily\bfseries ToExternal}\verb@ "LS `$loopsize'\n"@\\
\mbox{}\verb@#@\hbox{\sffamily\bfseries Else}\verb@@\\
\mbox{}\verb@   #@\hbox{\sffamily\bfseries ToExternal}\verb@ "LS 0\n"@\\
\mbox{}\verb@#@\hbox{\sffamily\bfseries EndIf}\verb@@{\NWsep}
\end{list}
\vspace{-1ex}
\footnotesize\addtolength{\baselineskip}{-1ex}
\begin{list}{}{\setlength{\itemsep}{-\parsep}\setlength{\itemindent}{-\leftmargin}}
\item \NWtxtMacroRefIn\ \NWlink{nuweb53}{53}.
\end{list}
\end{flushleft}
\subsubsection{Colour Algebra}
For an efficient evaluation of the diagram one has to avoid
multiple reevaluation of the same expressions. Therefore
linear combinations of colour basis elements are grouped
together such that different colour structures only arise
from the four-gluon vertices where colour and spin information
does not factorise. The \person{Feynman} rules are constructed
such that each different colour factor is labelled by a function
\texttt{ANYCS(...)} with a unique argument. If no four-gluon
vertex is in the diagram the factor \texttt{ANYCS(1)} ensures
that the algorithm still works as desired and exactly one
colour structure is built.

In order to label the colour structures by an increasing
index one can make use of \form's capability of interacting
between the preprocessor and the compiled program using
dollar-variables. The preprocessor variable \texttt{cs} is
reset to zero every time the first \texttt{Id} statement
finds the pattern on its left-hand-side; the argument of
the function \texttt{TEMPCS} that matches is written
to the dollar variable \texttt{\$cs}.
\begin{flushleft} \small \label{scrap29}
$\langle\,$find next colour structure\nobreak\ {\footnotesize \NWtarget{nuweb29}{29}}$\,\rangle\equiv$
\vspace{-1ex}
\begin{list}{}{} \item
\mbox{}\verb@@\hbox{\sffamily\bfseries Id}\verb@ @\hbox{\sffamily\bfseries IfMatch}\verb@->cstruelab`$dummy' TEMPCS(cc?$cs) = TEMPCS(cc);@\\
\mbox{}\verb@@\hbox{\sffamily\bfseries Goto}\verb@ csfalselab`$dummy';@\\
\mbox{}\verb@@\hbox{\sffamily\bfseries Label}\verb@ cstruelab`$dummy';@\\
\mbox{}\verb@@\hbox{\sffamily\bfseries ReDefine}\verb@ cs, "0";@\\
\mbox{}\verb@@\hbox{\sffamily\bfseries Label}\verb@ csfalselab`$dummy';@{\NWsep}
\end{list}
\vspace{-1ex}
\footnotesize\addtolength{\baselineskip}{-1ex}
\begin{list}{}{\setlength{\itemsep}{-\parsep}\setlength{\itemindent}{-\leftmargin}}
\item \NWtxtMacroRefIn\ \NWlink{nuweb31}{31}.
\end{list}
\end{flushleft}
After a \texttt{.sort}
the preprocessor comes back into action and increases
the counter in case of a match, i.e. when \texttt{cs} is
zero at this point. The second \texttt{Id} statement uses
the content of \texttt{\$cs} to identify all occurrences of that
colour structure and labels it by the function
\texttt{TEMPKIN} with the counter as an argument.
\begin{flushleft} \small \label{scrap30}
$\langle\,$actual replacement of colour structure\nobreak\ {\footnotesize \NWtarget{nuweb30}{30}}$\,\rangle\equiv$
\vspace{-1ex}
\begin{list}{}{} \item
\mbox{}\verb@#$dummy = {`$dummy'+1};@\\
\mbox{}\verb@#@\hbox{\sffamily\bfseries If}\verb@ `cs' == 0@\\
\mbox{}\verb@   #$counter = {`$counter'+1};@\\
\mbox{}\verb@   @\hbox{\sffamily\bfseries Id}\verb@ TEMPCS($cs) = TEMPKIN(`$counter');@\\
\mbox{}\verb@#@\hbox{\sffamily\bfseries EndIf}\verb@@{\NWsep}
\end{list}
\vspace{-1ex}
\footnotesize\addtolength{\baselineskip}{-1ex}
\begin{list}{}{\setlength{\itemsep}{-\parsep}\setlength{\itemindent}{-\leftmargin}}
\item \NWtxtMacroRefIn\ \NWlink{nuweb31}{31}.
\end{list}
\end{flushleft}
When eventually no more terms match the variable \texttt{cs}
is not reset anymore and the loop terminates.
\begin{flushleft} \small \label{scrap31}
$\langle\,$label colour structures\nobreak\ {\footnotesize \NWtarget{nuweb31}{31}}$\,\rangle\equiv$
\vspace{-1ex}
\begin{list}{}{} \item
\mbox{}\verb@#$dummy = 0;@\\
\mbox{}\verb@#$counter = 0;@\\
\mbox{}\verb@#@\hbox{\sffamily\bfseries Do}\verb@ cs=1,1@\\
\mbox{}\verb@   @\hbox{$\langle\,$find next colour structure\nobreak\ {\footnotesize \NWlink{nuweb29}{29}}$\,\rangle$}\verb@@\\
\mbox{}\verb@.@\hbox{\sffamily\bfseries sort}\verb@@\\
\mbox{}\verb@   @\hbox{$\langle\,$actual replacement of colour structure\nobreak\ {\footnotesize \NWlink{nuweb30}{30}}$\,\rangle$}\verb@@\\
\mbox{}\verb@.@\hbox{\sffamily\bfseries sort}\verb@@\\
\mbox{}\verb@#@\hbox{\sffamily\bfseries EndDo}\verb@@{\NWsep}
\end{list}
\vspace{-1ex}
\footnotesize\addtolength{\baselineskip}{-1ex}
\begin{list}{}{\setlength{\itemsep}{-\parsep}\setlength{\itemindent}{-\leftmargin}}
\item \NWtxtMacroRefIn\ \NWlink{nuweb32}{32}.
\end{list}
\end{flushleft}
Finally, the expression holding the entire
\person{Feynman} diagram is discarded in favour of the expressions
\texttt{struct`i'} that hold the single colour structures.
\begin{flushleft} \small \label{scrap32}
$\langle\,$split into colour structures\nobreak\ {\footnotesize \NWtarget{nuweb32}{32}}$\,\rangle\equiv$
\vspace{-1ex}
\begin{list}{}{} \item
\mbox{}\verb@@\hbox{\sffamily\bfseries Multiply}\verb@ ANYCS(1);@\\
\mbox{}\verb@@\hbox{\sffamily\bfseries Id}\verb@ ANYCS(?all) = TEMPCS(ANYCS(?all));@\\
\mbox{}\verb@@\hbox{\sffamily\bfseries ChainIn}\verb@ TEMPCS;@\\
\mbox{}\verb@@\hbox{\sffamily\bfseries Repeat}\verb@ @\hbox{\sffamily\bfseries Id}\verb@ TEMPCS(cc0?, cc1?, ?tail) = TEMPCS(cc0*cc1, ?tail);@\\
\mbox{}\verb@.@\hbox{\sffamily\bfseries sort}\verb@@\\
\mbox{}\verb@@\hbox{$\langle\,$label colour structures\nobreak\ {\footnotesize \NWlink{nuweb31}{31}}$\,\rangle$}\verb@@\\
\mbox{}\verb@@\hbox{\sffamily\bfseries Bracket}\verb@ TEMPKIN;@\\
\mbox{}\verb@.@\hbox{\sffamily\bfseries sort}\verb@@\\
\mbox{}\verb@#@\hbox{\sffamily\bfseries Do}\verb@ i=1,`$counter'@\\
\mbox{}\verb@   @\hbox{\sffamily\bfseries Local}\verb@ struct`i' = `DIAGRAM'`DIAG'[TEMPKIN(`i')] * TEMPKIN(`i');@\\
\mbox{}\verb@#@\hbox{\sffamily\bfseries EndDo}\verb@@\\
\mbox{}\verb@.@\hbox{\sffamily\bfseries sort}\verb@@\\
\mbox{}\verb@@\hbox{\sffamily\bfseries Drop}\verb@ `DIAGRAM'`DIAG';@{\NWsep}
\end{list}
\vspace{-1ex}
\footnotesize\addtolength{\baselineskip}{-1ex}
\begin{list}{}{\setlength{\itemsep}{-\parsep}\setlength{\itemindent}{-\leftmargin}}
\item \NWtxtMacroRefIn\ \NWlink{nuweb53}{53}.
\end{list}
\end{flushleft}
To project on the colour basis elements first all
products of \person{Kronecker} deltas
(here: \texttt{FundamentalID}) are replaced by
symbolic names (\texttt{color1}, \texttt{color2}, \ldots).

The steps it takes to create a colour vector are the following.
One starts from an expression like
\begin{displaymath}
\sum_{i=1}^{\mathtt{NUMCS}}c_i\cdot\mathtt{color}i\texttt{.}
\end{displaymath}
The \texttt{Collect} statement puts the whole expression
into the argument of a function; the function argument is then
copied \texttt{NUMCS} times, and in each copy one of the
basis elements is set to one. All remaining colour basis
elements are replaced by zero. Finally, the function
\texttt{COLORBASIS} contains the arguments
\(
\mathtt{COLORBASIS}(c_1, c_2, \ldots, c_{\mathtt{NUMCS}})
\).
\begin{flushleft} \small \label{scrap33}
$\langle\,$create colour vector\nobreak\ {\footnotesize \NWtarget{nuweb33}{33}}$\,\rangle\equiv$
\vspace{-1ex}
\begin{list}{}{} \item
\mbox{}\verb@@\hbox{\sffamily\bfseries AntiBracket}\verb@ T, f, dF, dA, TR, color1, ..., color`NUMCS';@\\
\mbox{}\verb@.@\hbox{\sffamily\bfseries sort}\verb@@\\
\mbox{}\verb@@\hbox{\sffamily\bfseries Collect}\verb@ TEMPCOLOR;@\\
\mbox{}\verb@@\hbox{\sffamily\bfseries Normalize}\verb@ TEMPCOLOR;@\\
\mbox{}\verb@@\hbox{\sffamily\bfseries Id}\verb@ TEMPCOLOR(cc0?) = COLORBASIS(cc0 * @\hbox{\sffamily\bfseries replace}\verb@_(color1,1)@\\
\mbox{}\verb@#@\hbox{\sffamily\bfseries Do}\verb@ c=2,`NUMCS'@\\
\mbox{}\verb@   , cc0 * @\hbox{\sffamily\bfseries replace}\verb@_(color`c',1)@\\
\mbox{}\verb@#@\hbox{\sffamily\bfseries EndDo}\verb@@\\
\mbox{}\verb@);@\\
\mbox{}\verb@#@\hbox{\sffamily\bfseries Do}\verb@ c=1,`NUMCS'@\\
\mbox{}\verb@   @\hbox{\sffamily\bfseries Multiply}\verb@ @\hbox{\sffamily\bfseries replace}\verb@_(color`c',0);@\\
\mbox{}\verb@#@\hbox{\sffamily\bfseries EndDo}\verb@@{\NWsep}
\end{list}
\vspace{-1ex}
\footnotesize\addtolength{\baselineskip}{-1ex}
\begin{list}{}{\setlength{\itemsep}{-\parsep}\setlength{\itemindent}{-\leftmargin}}
\item \NWtxtMacroRefIn\ \NWlink{nuweb34}{34}.
\end{list}
\end{flushleft}
The elements of the colour vectors for each colour structure
are read into dollar variables \texttt{\$basis`i'x`c'}
for each colour structure \texttt{i}
and the basis element \texttt{c}.
\begin{flushleft} \small \label{scrap34}
$\langle\,$project on colour basis\nobreak\ {\footnotesize \NWtarget{nuweb34}{34}}$\,\rangle\equiv$
\vspace{-1ex}
\begin{list}{}{} \item
\mbox{}\verb@#@\hbox{\sffamily\bfseries Call}\verb@ colorstructures() @\\
\mbox{}\verb@@\hbox{\sffamily\bfseries Id}\verb@ POW(TR, -1/2)^2 = 1/TR;@\\
\mbox{}\verb@#@\hbox{\sffamily\bfseries Do}\verb@ k=1,10@\\
\mbox{}\verb@   @\hbox{\sffamily\bfseries Sum}\verb@ i`k'r1, i`k'r2, i`k'r3,i`k'r4;@\\
\mbox{}\verb@#@\hbox{\sffamily\bfseries EndDo}\verb@@\\
\mbox{}\verb@@\hbox{$\langle\,$create colour vector\nobreak\ {\footnotesize \NWlink{nuweb33}{33}}$\,\rangle$}\verb@@\\
\mbox{}\verb@#@\hbox{\sffamily\bfseries Do}\verb@ i=1,`$counter'@\\
\mbox{}\verb@   #@\hbox{\sffamily\bfseries Do}\verb@ c=1,`NUMCS'@\\
\mbox{}\verb@      #$basis`i'x`c' = 0;@\\
\mbox{}\verb@      @\hbox{\sffamily\bfseries Id}\verb@ TEMPKIN(`i') * COLORBASIS(cc?$basis`i'x`c', ?tail) =@\\
\mbox{}\verb@         TEMPKIN(`i') * COLORBASIS(?tail);@\\
\mbox{}\verb@   #@\hbox{\sffamily\bfseries EndDo}\verb@@\\
\mbox{}\verb@#@\hbox{\sffamily\bfseries EndDo}\verb@@\\
\mbox{}\verb@@\hbox{\sffamily\bfseries Id}\verb@ TEMPKIN(cc?) * COLORBASIS = 1;@\\
\mbox{}\verb@.@\hbox{\sffamily\bfseries sort}\verb@@{\NWsep}
\end{list}
\vspace{-1ex}
\footnotesize\addtolength{\baselineskip}{-1ex}
\begin{list}{}{\setlength{\itemsep}{-\parsep}\setlength{\itemindent}{-\leftmargin}}
\item \NWtxtMacroRefIn\ \NWlink{nuweb53}{53}.
\end{list}
\end{flushleft}
One of the last steps of the program is to evaluate
the colour vector numerically by plugging in $N_C=3$.
\begin{flushleft} \small
\begin{minipage}{\linewidth} \label{scrap35}
$\langle\,$evaluate colour vector numerically\nobreak\ {\footnotesize \NWtarget{nuweb35}{35}}$\,\rangle\equiv$
\vspace{-1ex}
\begin{list}{}{} \item
\mbox{}\verb@#@\hbox{\sffamily\bfseries Do}\verb@ i=1,`$counter'@\\
\mbox{}\verb@   #@\hbox{\sffamily\bfseries Do}\verb@ c=1,`NUMCS'@\\
\mbox{}\verb@      @\hbox{\sffamily\bfseries Local}\verb@ basis`i'x`c' = $basis`i'x`c';@\\
\mbox{}\verb@   #@\hbox{\sffamily\bfseries EndDo}\verb@@\\
\mbox{}\verb@#@\hbox{\sffamily\bfseries EndDo}\verb@@\\
\mbox{}\verb@@\hbox{\sffamily\bfseries Id}\verb@ dA = 8;@\\
\mbox{}\verb@@\hbox{\sffamily\bfseries Id}\verb@ dF = 3;@\\
\mbox{}\verb@@\hbox{\sffamily\bfseries Id}\verb@ 1/dF = 1/3;@\\
\mbox{}\verb@@\hbox{\sffamily\bfseries Id}\verb@ TR = 1/2;@{\NWsep}
\end{list}
\vspace{-1ex}
\footnotesize\addtolength{\baselineskip}{-1ex}
\begin{list}{}{\setlength{\itemsep}{-\parsep}\setlength{\itemindent}{-\leftmargin}}
\item \NWtxtMacroRefIn\ \NWlink{nuweb53}{53}.
\end{list}
\end{minipage}\\[4ex]
\end{flushleft}
\subsubsection{Integration}
Although in the current implementation all internal masses are
expected to be zero they are already written to dollar variables.
Once a massive implementation is being developed this information
needs to be provided to the \texttt{Golem90} library in order to
set up the correct $S$-matrix.
\begin{flushleft} \small \label{scrap36}
$\langle\,$read propagator masses\nobreak\ {\footnotesize \NWtarget{nuweb36}{36}}$\,\rangle\equiv$
\vspace{-1ex}
\begin{list}{}{} \item
\mbox{}\verb@#@\hbox{\sffamily\bfseries Do}\verb@ i=1,`$loopsize'@\\
\mbox{}\verb@   #$mass`i' = 0;@\\
\mbox{}\verb@   @\hbox{\sffamily\bfseries Id}\verb@ PROP(-q`i', cc?) = PROP(q`i', cc);@\\
\mbox{}\verb@   @\hbox{\sffamily\bfseries Id}\verb@ PROP(-q`i', 0) = PROP(q`i', 0);@\\
\mbox{}\verb@   @\hbox{\sffamily\bfseries Id}\verb@ PROP(q`i', cc?$mass`i') = 1;@\\
\mbox{}\verb@   @\hbox{\sffamily\bfseries Id}\verb@ PROP(q`i', 0) = 1;@\\
\mbox{}\verb@#@\hbox{\sffamily\bfseries EndDo}\verb@@{\NWsep}
\end{list}
\vspace{-1ex}
\footnotesize\addtolength{\baselineskip}{-1ex}
\begin{list}{}{\setlength{\itemsep}{-\parsep}\setlength{\itemindent}{-\leftmargin}}
\item \NWtxtMacroRefIn\ \NWlink{nuweb39}{39}.
\end{list}
\end{flushleft}
The numerator of each tensor integral is written into a
temporary function; the arguments are pairs of momenta
and indices $(r_i, \mu)$ for each $q_i^\mu$ in the numerator.
for each such pair one power of the symbol \texttt{ccCOUNT} is
multiplied to the corresponding term. Since the variable
\texttt{\$rank} is determined for each term, together with the
\texttt{If}-statement one calculates the maximal rank that
occurs in this diagram in the variable \texttt{\$maxrank}.
The imaginary $i$ in front of the function \texttt{TI}
ensures that the definition of the tensor integral is the same
as in Equation~\eqref{eq:qcd-dimreg:defTI}.
\begin{flushleft} \small \label{scrap37}
$\langle\,$construct tensor integral\nobreak\ {\footnotesize \NWtarget{nuweb37}{37}}$\,\rangle\equiv$
\vspace{-1ex}
\begin{list}{}{} \item
\mbox{}\verb@@\hbox{\sffamily\bfseries Multiply}\verb@ TEMP;@\\
\mbox{}\verb@@\hbox{\sffamily\bfseries Id}\verb@ MOMENTUM(n, vec?qSET?rSET, iMU?) = TEMP(vec, iMU) * ccCOUNT;@\\
\mbox{}\verb@@\hbox{\sffamily\bfseries ChainIn}\verb@ TEMP;@\\
\mbox{}\verb@@\hbox{\sffamily\bfseries Id}\verb@ ccCOUNT^n?$rank = 1;@\\
\mbox{}\verb@@\hbox{\sffamily\bfseries If}\verb@($rank > $maxrank);@\\
\mbox{}\verb@   $maxrank = $rank;@\\
\mbox{}\verb@@\hbox{\sffamily\bfseries EndIf}\verb@;@\\
\mbox{}\verb@@\hbox{\sffamily\bfseries Id}\verb@ TEMP(?all) = @\hbox{\sffamily\bfseries i}\verb@_ * TI(`$loopsize', @\hbox{\sffamily\bfseries nargs}\verb@_(?all)/2, ?all);@{\NWsep}
\end{list}
\vspace{-1ex}
\footnotesize\addtolength{\baselineskip}{-1ex}
\begin{list}{}{\setlength{\itemsep}{-\parsep}\setlength{\itemindent}{-\leftmargin}}
\item \NWtxtMacroRefIn\ \NWlink{nuweb39}{39}.
\end{list}
\end{flushleft}
The rewriting rules for the translation of tensor integrals into
form factors are automatically generated by a \texttt{Java} program
which is explained in Section~\ref{sec:FormFactory.java}.
The one-point form factors are
replaced by zero as only massless integrals are considered. This needs
to be changed for a massive calculation and ideally be implemented
in the \texttt{Golem90} library.
\begin{flushleft} \small \label{scrap38}
$\langle\,$introduce form factors\nobreak\ {\footnotesize \NWtarget{nuweb38}{38}}$\,\rangle\equiv$
\vspace{-1ex}
\begin{list}{}{} \item
\mbox{}\verb@#@\hbox{\sffamily\bfseries Do}\verb@ r=0,`$maxrank'@\\
\mbox{}\verb@   #@\hbox{\sffamily\bfseries Include}\verb@- ff-`$loopsize'-`r'.h@\\
\mbox{}\verb@#@\hbox{\sffamily\bfseries EndDo}\verb@@\\
\mbox{}\verb@@\hbox{\sffamily\bfseries Id}\verb@ a10(?all) = 0;@\\
\mbox{}\verb@@\hbox{\sffamily\bfseries Id}\verb@ a11(?all) = 0;@{\NWsep}
\end{list}
\vspace{-1ex}
\footnotesize\addtolength{\baselineskip}{-1ex}
\begin{list}{}{\setlength{\itemsep}{-\parsep}\setlength{\itemindent}{-\leftmargin}}
\item \NWtxtMacroRefIn\ \NWlink{nuweb39}{39}.
\end{list}
\end{flushleft}
In a last step the notation is changed from the function
\texttt{DELTA} to the vectors \texttt{D`i'x`j'}.
\begin{flushleft} \small \label{scrap39}
$\langle\,$perform integration\nobreak\ {\footnotesize \NWtarget{nuweb39}{39}}$\,\rangle\equiv$
\vspace{-1ex}
\begin{list}{}{} \item
\mbox{}\verb@#$maxrank = 0;@\\
\mbox{}\verb@#$rank = 0;@\\
\mbox{}\verb@@\hbox{\sffamily\bfseries Id}\verb@ MOMENTUM(n, -vec?qSET, iMU?) = - MOMENTUM(n, vec, iMU);@\\
\mbox{}\verb@@\hbox{$\langle\,$read propagator masses\nobreak\ {\footnotesize \NWlink{nuweb36}{36}}$\,\rangle$}\verb@@\\
\mbox{}\verb@@\hbox{$\langle\,$construct tensor integral\nobreak\ {\footnotesize \NWlink{nuweb37}{37}}$\,\rangle$}\verb@@\\
\mbox{}\verb@.@\hbox{\sffamily\bfseries sort}\verb@@\\
\mbox{}\verb@@\hbox{$\langle\,$introduce form factors\nobreak\ {\footnotesize \NWlink{nuweb38}{38}}$\,\rangle$}\verb@@\\
\mbox{}\verb@@\hbox{\sffamily\bfseries Id}\verb@ DELTA(vec?, vec?, iMU?) = 0;@\\
\mbox{}\verb@#@\hbox{\sffamily\bfseries Do}\verb@ i=1,{`$loopsize'-1}@\\
\mbox{}\verb@   #@\hbox{\sffamily\bfseries Do}\verb@ j={`i'+1},`$loopsize'@\\
\mbox{}\verb@      @\hbox{\sffamily\bfseries Id}\verb@ DELTA(r`i', r`j', iMU?) = MOMENTUM(4, D`i'x`j', iMU);@\\
\mbox{}\verb@      @\hbox{\sffamily\bfseries Id}\verb@ DELTA(r`j', r`i', iMU?) = -MOMENTUM(4, D`i'x`j', iMU);@\\
\mbox{}\verb@   #@\hbox{\sffamily\bfseries EndDo}\verb@@\\
\mbox{}\verb@#@\hbox{\sffamily\bfseries EndDo}\verb@@{\NWsep}
\end{list}
\vspace{-1ex}
\footnotesize\addtolength{\baselineskip}{-1ex}
\begin{list}{}{\setlength{\itemsep}{-\parsep}\setlength{\itemindent}{-\leftmargin}}
\item \NWtxtMacroRefIn\ \NWlink{nuweb53}{53}.
\end{list}
\end{flushleft}
\subsubsection{\texorpdfstring{\son}{SO(n)} Algebra}
After integration many simplifications can be made to the \person{Lorentz}
algebra since all $n\neq4$ dimensional vectors have been
eliminated during integration. All momenta and metric tensors are
contracted as far as possible.
\begin{flushleft} \small \label{scrap40}
$\langle\,$carry out \person{Lorentz} algebra\nobreak\ {\footnotesize \NWtarget{nuweb40}{40}}$\,\rangle\equiv$
\vspace{-1ex}
\begin{list}{}{} \item
\mbox{}\verb@@\hbox{\sffamily\bfseries Id}\verb@ gTensor(n, i1?, i1?) = 4 - 2 * eps;@\\
\mbox{}\verb@@\hbox{\sffamily\bfseries Argument}\verb@ MOMENTUM;@\\
\mbox{}\verb@   #@\hbox{\sffamily\bfseries Call}\verb@ kinematics()@\\
\mbox{}\verb@@\hbox{\sffamily\bfseries EndArgument}\verb@;@\\
\mbox{}\verb@@\hbox{\sffamily\bfseries Id}\verb@ MOMENTUM(n, iNU?, iMU?) = MOMENTUM(4, iNU, iMU);@\\
\mbox{}\verb@@\hbox{\sffamily\bfseries Repeat}\verb@ @\hbox{\sffamily\bfseries Id}\verb@ gTensor(n, iMU?, iNU?) * MOMENTUM(4, vec?, iMU?) = @\\
\mbox{}\verb@   MOMENTUM(4, vec, iNU);@\\
\mbox{}\verb@@\hbox{\sffamily\bfseries Repeat}\verb@ @\hbox{\sffamily\bfseries Id}\verb@ gTensor(n, iMU?, iNU?) * SpinorTrace(?head, gg(n, iMU?), ?tail) =@\\
\mbox{}\verb@   SpinorTrace(?head, gg(n, iNU), ?tail);@\\
\mbox{}\verb@@\hbox{\sffamily\bfseries Repeat}\verb@ @\hbox{\sffamily\bfseries Id}\verb@ MOMENTUM(4, vec?, iMU?) * SpinorTrace(?head, gg(n, iMU?), ?tail) =@\\
\mbox{}\verb@   SpinorTrace(?head, gg(4, vec), ?tail);@\\
\mbox{}\verb@@\hbox{\sffamily\bfseries Id}\verb@ MOMENTUM(4, vec1?, iMU?) * MOMENTUM(4, vec2?, iMU?) = vec1.vec2;@\\
\mbox{}\verb@#@\hbox{\sffamily\bfseries Call}\verb@ kinematics()@\\
\mbox{}\verb@@\hbox{\sffamily\bfseries Argument}\verb@ SpinorTrace;@\\
\mbox{}\verb@   @\hbox{\sffamily\bfseries Id}\verb@ vec? = gg(4, vec);@\\
\mbox{}\verb@@\hbox{\sffamily\bfseries EndArgument}\verb@;@{\NWsep}
\end{list}
\vspace{-1ex}
\footnotesize\addtolength{\baselineskip}{-1ex}
\begin{list}{}{\setlength{\itemsep}{-\parsep}\setlength{\itemindent}{-\leftmargin}}
\item \NWtxtMacroRefIn\ \NWlink{nuweb53}{53}.
\end{list}
\end{flushleft}
Although no $n$-dimensional vectors remain after integration one
still has to deal with $n$-dimensional \person{Dirac} matrices
unless one decides to work in the \ac{dred} scheme. The code
below implements the commutator according to the \ac{tho}
scheme,
\begin{equation}
(\hat{\gamma}^\mu+\bar{\gamma}^\mu)\gamma_5=
\gamma_5(-\hat{\gamma}^\mu+\bar{\gamma}^\mu)
\end{equation}
and follows largely the Algorithms \ref{alg:qcd-dimreg:traces}
and~\ref{alg:qcd-dimreg:finaltracing}.
\begin{flushleft} \small \label{scrap41}
$\langle\,$$n$-dimensional spinor algebra\nobreak\ {\footnotesize \NWtarget{nuweb41}{41}}$\,\rangle\equiv$
\vspace{-1ex}
\begin{list}{}{} \item
\mbox{}\verb@@\hbox{\sffamily\bfseries Normalize}\verb@ SpinorTrace;@\\
\mbox{}\verb@@\hbox{\sffamily\bfseries Repeat}\verb@ @\hbox{\sffamily\bfseries Id}\verb@ SpinorTrace(?head, gg(n, iMU?), ?tail) =@\\
\mbox{}\verb@   SpinorTrace(?head, gg(4, iMU), ?tail) +@\\
\mbox{}\verb@   SpinorTrace(?head, gg([-2eps], iMU), ?tail);@\\
\mbox{}\verb@@\hbox{$\langle\,$move $\bar\gamma^\mu$ right\nobreak\ {\footnotesize \NWlink{nuweb42}{42}}$\,\rangle$}\verb@@\\
\mbox{}\verb@@\hbox{$\langle\,$split traces\nobreak\ {\footnotesize \NWlink{nuweb43}{43}}$\,\rangle$}\verb@@\\
\mbox{}\verb@@\hbox{$\langle\,$evaluate $(n-4)$-dimensional traces\nobreak\ {\footnotesize \NWlink{nuweb44}{44}}$\,\rangle$}\verb@@\\
\mbox{}\verb@.@\hbox{\sffamily\bfseries sort}\verb@@\\
\mbox{}\verb@@\\
\mbox{}\verb@* Get rid of the gTensors:@\\
\mbox{}\verb@Repeat Id gTensor([-2eps], i1?, i2?) * gTensor([-2eps], i2?, i3?) =@\\
\mbox{}\verb@   gTensor([-2eps], i1, i3);@\\
\mbox{}\verb@@\\
\mbox{}\verb@Id gTensor([-2eps], i1?, i1?) = -2 * eps;@\\
\mbox{}\verb@Id gTensor([-2eps], i1?, i2?) * SpinorTrace(?head, gg(4, i2?), ?tail) = 0;@\\
\mbox{}\verb@@{\NWsep}
\end{list}
\vspace{-1ex}
\footnotesize\addtolength{\baselineskip}{-1ex}
\begin{list}{}{\setlength{\itemsep}{-\parsep}\setlength{\itemindent}{-\leftmargin}}
\item \NWtxtMacroRefIn\ \NWlink{nuweb53}{53}.
\end{list}
\end{flushleft}
In order to split the traces into a $4$-dimensional trace and a
$(n-4)$-dimensional one all $\bar\gamma^\mu$ are shuffled to the
right according to the commutation rules.
\begin{flushleft} \small \label{scrap42}
$\langle\,$move $\bar\gamma^\mu$ right\nobreak\ {\footnotesize \NWtarget{nuweb42}{42}}$\,\rangle\equiv$
\vspace{-1ex}
\begin{list}{}{} \item
\mbox{}\verb@@\hbox{\sffamily\bfseries Repeat}\verb@;@\\
\mbox{}\verb@   @\hbox{\sffamily\bfseries Id}\verb@ SpinorTrace(?head gg([-2eps], iMU?), gg(4, ?any), ?tail) =@\\
\mbox{}\verb@      -SpinorTrace(?head, gg(4, ?any), gg([-2eps], iMU), ?tail);@\\
\mbox{}\verb@   @\hbox{\sffamily\bfseries Id}\verb@ SpinorTrace(?head gg([-2eps], iMU?), hProjector(cc0?), ?tail) =@\\
\mbox{}\verb@      SpinorTrace(?head hProjector(cc0), gg([-2eps], iMU), ?tail);@\\
\mbox{}\verb@   @\hbox{\sffamily\bfseries Id}\verb@ SpinorTrace(?head gg(4, ?any), hProjector(cc0?), ?tail) =@\\
\mbox{}\verb@      SpinorTrace(?head hProjector(-cc0), gg(4, ?any), ?tail);@\\
\mbox{}\verb@   @\hbox{\sffamily\bfseries Id}\verb@ SpinorTrace(hProjector(cc0?), hProjector(cc0?), ?tail) =@\\
\mbox{}\verb@      SpinorTrace(hProjector(cc0), ?tail);@\\
\mbox{}\verb@   @\hbox{\sffamily\bfseries Id}\verb@ SpinorTrace(hProjector(1), hProjector(-1), ?tail) = 0;@\\
\mbox{}\verb@   @\hbox{\sffamily\bfseries Id}\verb@ SpinorTrace(hProjector(-1), hProjector(1), ?tail) = 0;@\\
\mbox{}\verb@@\hbox{\sffamily\bfseries EndRepeat}\verb@;@{\NWsep}
\end{list}
\vspace{-1ex}
\footnotesize\addtolength{\baselineskip}{-1ex}
\begin{list}{}{\setlength{\itemsep}{-\parsep}\setlength{\itemindent}{-\leftmargin}}
\item \NWtxtMacroRefIn\ \NWlink{nuweb41}{41}.
\end{list}
\end{flushleft}
For the splitting three cases have to be considered
for technical reasons: the
first $(n-4)$-dimensional matrix being adjacent to a
$4$-dimensional \person{Dirac} matrix, being
adjacent to a projector and finally being the first matrix
in the trace. The function \texttt{TEMPTRACE} carries the number
of \person{Dirac} matrices as the first argument. The splitting
into two traces is based on Equation~\eqref{eq:qcd-dimreg:tracesplit}.
\begin{flushleft} \small \label{scrap43}
$\langle\,$split traces\nobreak\ {\footnotesize \NWtarget{nuweb43}{43}}$\,\rangle\equiv$
\vspace{-1ex}
\begin{list}{}{} \item
\mbox{}\verb@@\hbox{\sffamily\bfseries Id}\verb@ SpinorTrace(?head, gg(4, ?any), gg([-2eps], iNU?), ?tail) =@\\
\mbox{}\verb@   SpinorTrace(?head, gg(4, ?any)) *@\\
\mbox{}\verb@   TEMPTRACE(@\hbox{\sffamily\bfseries nargs}\verb@_(?tail) + 1, gg([-2eps], iNU), ?tail);@\\
\mbox{}\verb@@\hbox{\sffamily\bfseries Id}\verb@ SpinorTrace(hProjector(cc0?), gg([-2eps], iNU?), ?tail) =@\\
\mbox{}\verb@   4 * (1/2) * TEMPTRACE(@\hbox{\sffamily\bfseries nargs}\verb@_(?tail) + 1, gg([-2eps], iNU), ?tail);@\\
\mbox{}\verb@@\hbox{\sffamily\bfseries Id}\verb@ SpinorTrace(gg([-2eps], iNU?), ?tail) =@\\
\mbox{}\verb@   4 * TEMPTRACE(@\hbox{\sffamily\bfseries nargs}\verb@_(?tail) + 1, gg([-2eps], iNU), ?tail);@{\NWsep}
\end{list}
\vspace{-1ex}
\footnotesize\addtolength{\baselineskip}{-1ex}
\begin{list}{}{\setlength{\itemsep}{-\parsep}\setlength{\itemindent}{-\leftmargin}}
\item \NWtxtMacroRefIn\ \NWlink{nuweb41}{41}.
\end{list}
\end{flushleft}
The $(n-4)$ dimensional trace can be evaluated by the usual rules.
In fact Equation~\eqref{eq:qcd-dimreg:naive-trace} is sufficient for
a reduction of the traces, as has been shown earlier. The algorithm
below implements this equation. The initial replacement
$\mathtt{gg([-2eps], iMU?)}\rightarrow\mathtt{iMU}$ starts the
loop over the sum of all swaps in
Equation~\eqref{eq:qcd-dimreg:naive-trace}. For traces of length~4
and below explicit formul\ae{} are used for efficiency.
\begin{flushleft} \small \label{scrap44}
$\langle\,$evaluate $(n-4)$-dimensional traces\nobreak\ {\footnotesize \NWtarget{nuweb44}{44}}$\,\rangle\equiv$
\vspace{-1ex}
\begin{list}{}{} \item
\mbox{}\verb@@\hbox{\sffamily\bfseries Id}\verb@ TEMPTRACE(cc0?odd_, ?tail) = 0;@\\
\mbox{}\verb@@\hbox{\sffamily\bfseries Repeat}\verb@;@\\
\mbox{}\verb@   @\hbox{\sffamily\bfseries Id}\verb@ TEMPTRACE(cc0?{>4}, gg([-2eps], iMU?), ?tail) =@\\
\mbox{}\verb@      TEMPTRACE(cc0, iMU, ?tail);@\\
\mbox{}\verb@   @\hbox{\sffamily\bfseries Repeat}\verb@ @\hbox{\sffamily\bfseries Id}\verb@ TEMPTRACE(cc0?{>4}, ?head, iMU?,@\\
\mbox{}\verb@         gg([-2eps], iNU?), ?tail) =@\\
\mbox{}\verb@      gTensor([-2eps], iMU, iNU) * TEMPTRACE(cc0 - 2, ?head, ?tail)@\\
\mbox{}\verb@      - TEMPTRACE(cc0, ?head, gg([-2eps], iNU), iMU ?tail);@\\
\mbox{}\verb@   @\hbox{\sffamily\bfseries Id}\verb@ TEMPTRACE(cc0?{>4}, ?head, iMU?) = 0;@\\
\mbox{}\verb@@\hbox{\sffamily\bfseries EndRepeat}\verb@;@\\
\mbox{}\verb@@\hbox{\sffamily\bfseries Id}\verb@ TEMPTRACE(2, gg([-2eps], i1?), gg([-2eps], i2?)) =@\\
\mbox{}\verb@   gTensor([-2eps], i1, i2);@\\
\mbox{}\verb@@\hbox{\sffamily\bfseries Id}\verb@ TEMPTRACE(4, gg([-2eps], i1?), gg([-2eps], i2?),@\\
\mbox{}\verb@      gg([-2eps], i3?), gg([-2eps], i4?)) =@\\
\mbox{}\verb@   + gTensor([-2eps], i1, i2) * gTensor([-2eps], i3, i4)@\\
\mbox{}\verb@   - gTensor([-2eps], i1, i3) * gTensor([-2eps], i2, i4)@\\
\mbox{}\verb@   + gTensor([-2eps], i1, i4) * gTensor([-2eps], i2, i3);@{\NWsep}
\end{list}
\vspace{-1ex}
\footnotesize\addtolength{\baselineskip}{-1ex}
\begin{list}{}{\setlength{\itemsep}{-\parsep}\setlength{\itemindent}{-\leftmargin}}
\item \NWtxtMacroRefIn\ \NWlink{nuweb41}{41}.
\end{list}
\end{flushleft}
The only remaining \person{Lorentz} indices in the expression
are those connecting \person{Dirac} matrices in the traces.
In Section~\ref{sec:qcddimreg} of Chapter~\ref{chp:qcd} it
has been shown that the \person{Chisholm} identities can be
used to achieve an index-free expression. The procedure
\texttt{hProjector\hspace{0pt}Simplify} needs to be invoked repeatedly
to restore the canonical order of the trace, i.e. to
shuffle all helicity projectors to the left after one of
the \person{Chisholm} identities has been applied.
The \texttt{Bracket} statement allows for a more efficient
evaluation since only the relevant factors of each term are
considered in the replacements and the overall number of terms
can be temporarily reduced, since \form{} groups terms with the
same spinor traces into one bracket.

The last four replacements carry out immediate simplifications
and bring the traces into a standard form which is used later:
the first argument of each trace is a number $\{\pm1,0\}$ that
either represents the projector or the identity matrix.
\begin{flushleft} \small \label{scrap45}
$\langle\,$eliminate \person{Lorentz} indices\nobreak\ {\footnotesize \NWtarget{nuweb45}{45}}$\,\rangle\equiv$
\vspace{-1ex}
\begin{list}{}{} \item
\mbox{}\verb@@\hbox{\sffamily\bfseries Repeat}\verb@ @\hbox{\sffamily\bfseries Id}\verb@ SpinorTrace(?head, gg(4, i1?), ?tail) =@\\
\mbox{}\verb@   SpinorTrace(?head, i1, ?tail);@\\
\mbox{}\verb@@\hbox{\sffamily\bfseries Bracket}\verb@ SpinorTrace;@\\
\mbox{}\verb@.@\hbox{\sffamily\bfseries sort}\verb@; @\\
\mbox{}\verb@@\hbox{\sffamily\bfseries Keep}\verb@ @\hbox{\sffamily\bfseries Brackets}\verb@;@\\
\mbox{}\verb@#@\hbox{\sffamily\bfseries Call}\verb@ hProjectorSimplify()@\\
\mbox{}\verb@@\hbox{\sffamily\bfseries Repeat}\verb@;@\\
\mbox{}\verb@   @\hbox{$\langle\,$deal with situation $\tr{\cdots\gamma^\mu\cdots\gamma_\mu\cdots}$\nobreak\ {\footnotesize \NWlink{nuweb46}{46}}$\,\rangle$}\verb@@\\
\mbox{}\verb@   @\hbox{$\langle\,$deal with situation $\tr{\cdots\gamma^\mu\cdots}\tr{\cdots\gamma_\mu\cdots}$\nobreak\ {\footnotesize \NWlink{nuweb47}{47}}$\,\rangle$}\verb@@\\
\mbox{}\verb@   #@\hbox{\sffamily\bfseries Call}\verb@ hProjectorSimplify()@\\
\mbox{}\verb@@\hbox{\sffamily\bfseries Endrepeat}\verb@;@\\
\mbox{}\verb@@\hbox{\sffamily\bfseries Id}\verb@ SpinorTrace() = 4;@\\
\mbox{}\verb@@\hbox{\sffamily\bfseries Id}\verb@ SpinorTrace(hProjector(cc0?)) = 2;@\\
\mbox{}\verb@@\hbox{\sffamily\bfseries Id}\verb@ SpinorTrace(hProjector(cc0?), ?tail) = SpinorTrace(cc0, ?tail);@\\
\mbox{}\verb@@\hbox{\sffamily\bfseries Id}\verb@ SpinorTrace(vec?, ?tail) = SpinorTrace(0, vec, ?tail);@{\NWsep}
\end{list}
\vspace{-1ex}
\footnotesize\addtolength{\baselineskip}{-1ex}
\begin{list}{}{\setlength{\itemsep}{-\parsep}\setlength{\itemindent}{-\leftmargin}}
\item \NWtxtMacroRefIn\ \NWlink{nuweb53}{53}.
\end{list}
\end{flushleft}
One possible contraction of \person{Lorentz} indices is inside
the same trace. In the first rule the part of the trace
between the two contracted matrices is stored in the function
\texttt{ANY0} together with the length of the strip. The second
rule ensures that all projectors have been moved to the left
properly. The remaining rules distinguish between the cases
where the strip is of even or odd length according to
Equations~\eqref{eq:qcd-dimreg:chisholm}
and~\eqref{eq:qcd-dimreg:chisholm-even}.
\begin{flushleft} \small \label{scrap46}
$\langle\,$deal with situation $\tr{\cdots\gamma^\mu\cdots\gamma_\mu\cdots}$\nobreak\ {\footnotesize \NWtarget{nuweb46}{46}}$\,\rangle\equiv$
\vspace{-1ex}
\begin{list}{}{} \item
\mbox{}\verb@@\hbox{\sffamily\bfseries Id}\verb@ @\hbox{\sffamily\bfseries Once}\verb@ SpinorTrace(?head, i1?, ?mid, i1?, ?tail) =@\\
\mbox{}\verb@    ANY0(@\hbox{\sffamily\bfseries nargs}\verb@_(?mid), ?mid) * SpinorTrace(?head, ANY0, ?tail);@\\
\mbox{}\verb@@\hbox{\sffamily\bfseries Id}\verb@ ANY0(?head, hProjector(cc0?), ?tail) = TEMPERRORTOKEN;@\\
\mbox{}\verb@@\hbox{\sffamily\bfseries Id}\verb@ ANY0(cc0?@\hbox{\sffamily\bfseries odd}\verb@_, ?mid) * SpinorTrace(?head, ANY0, ?tail) =@\\
\mbox{}\verb@   - 2 * SpinorTrace(?head, @\hbox{\sffamily\bfseries reverse}\verb@_(?mid), ?tail);@\\
\mbox{}\verb@@\hbox{\sffamily\bfseries Id}\verb@ ANY0(cc0?@\hbox{\sffamily\bfseries even}\verb@_, ?mid, vec?) * SpinorTrace(?head, ANY0, ?tail) =@\\
\mbox{}\verb@   + 2 * SpinorTrace(?head, vec, ?mid, ?tail)@\\
\mbox{}\verb@   + 2 * SpinorTrace(?head, @\hbox{\sffamily\bfseries reverse}\verb@_(?mid), vec, ?tail);@\\
\mbox{}\verb@@\hbox{\sffamily\bfseries Id}\verb@ ANY0(cc0?@\hbox{\sffamily\bfseries even}\verb@_, ?mid, i1?) * SpinorTrace(?head, ANY0, ?tail) =@\\
\mbox{}\verb@   + 2 * SpinorTrace(?head, i1, ?mid, ?tail)@\\
\mbox{}\verb@   + 2 * SpinorTrace(?head, @\hbox{\sffamily\bfseries reverse}\verb@_(?mid), i1, ?tail);@\\
\mbox{}\verb@@\hbox{\sffamily\bfseries Id}\verb@ ANY0(0) * SpinorTrace(?head, ANY0, ?tail) =@\\
\mbox{}\verb@   + 4 * SpinorTrace(?head, ?tail);@{\NWsep}
\end{list}
\vspace{-1ex}
\footnotesize\addtolength{\baselineskip}{-1ex}
\begin{list}{}{\setlength{\itemsep}{-\parsep}\setlength{\itemindent}{-\leftmargin}}
\item \NWtxtMacroRefIn\ \NWlink{nuweb45}{45}.
\end{list}
\end{flushleft}
The remaining case is the situation with a product of two traces
connected by a \person{Lorentz} contraction. Here
Equation~\eqref{eq:qcd-dimreg:chisholm-twotraces} applies.
\begin{flushleft} \small
\begin{minipage}{\linewidth} \label{scrap47}
$\langle\,$deal with situation $\tr{\cdots\gamma^\mu\cdots}\tr{\cdots\gamma_\mu\cdots}$\nobreak\ {\footnotesize \NWtarget{nuweb47}{47}}$\,\rangle\equiv$
\vspace{-1ex}
\begin{list}{}{} \item
\mbox{}\verb@@\hbox{\sffamily\bfseries Id}\verb@ SpinorTrace(?h1, i1?, ?t1) * SpinorTrace(?h2, i1?, ?t2) =@\\
\mbox{}\verb@   + 2 * SpinorTrace(?t2, ?h2, ?t1, ?h1)@\\
\mbox{}\verb@   + 2 * SpinorTrace(?t2, ?h2, @\hbox{\sffamily\bfseries reverse}\verb@_(?t1, ?h1));@{\NWsep}
\end{list}
\vspace{-1ex}
\footnotesize\addtolength{\baselineskip}{-1ex}
\begin{list}{}{\setlength{\itemsep}{-\parsep}\setlength{\itemindent}{-\leftmargin}}
\item \NWtxtMacroRefIn\ \NWlink{nuweb45}{45}.
\end{list}
\end{minipage}\\[4ex]
\end{flushleft}
\subsubsection{Subexpression Elimination}
The following section describes the simplest form of an incomplete
elimination of common subexpressions. In a first step all remaining
propagators that are not part of the loop integral are stripped from
the diagram and stored in a dollar variable.
\begin{flushleft} \small
\begin{minipage}{\linewidth} \label{scrap48}
$\langle\,$strip propagators\nobreak\ {\footnotesize \NWtarget{nuweb48}{48}}$\,\rangle\equiv$
\vspace{-1ex}
\begin{list}{}{} \item
\mbox{}\verb@@\hbox{\sffamily\bfseries Argument}\verb@ POW;@\\
\mbox{}\verb@   #@\hbox{\sffamily\bfseries Call}\verb@ kinematics()@\\
\mbox{}\verb@@\hbox{\sffamily\bfseries EndArgument}\verb@;@\\
\mbox{}\verb@@\hbox{\sffamily\bfseries Id}\verb@ POW(cc0?, -1) = @\hbox{\sffamily\bfseries TEMP}\verb@(cc0);@\\
\mbox{}\verb@@\hbox{\sffamily\bfseries Multiply}\verb@ TEMP(1);@\\
\mbox{}\verb@@\hbox{\sffamily\bfseries Repeat}\verb@ @\hbox{\sffamily\bfseries Id}\verb@ TEMP(cc0?) * TEMP(cc1?) = TEMP(cc0 * cc1);@\\
\mbox{}\verb@@\hbox{\sffamily\bfseries Bracket}\verb@ TEMP;@\\
\mbox{}\verb@.@\hbox{\sffamily\bfseries sort}\verb@@\\
\mbox{}\verb@#$props = 1;@\\
\mbox{}\verb@@\hbox{\sffamily\bfseries Keep}\verb@ @\hbox{\sffamily\bfseries Brackets}\verb@;@\hbox{\sffamily\bfseries Id}\verb@ TEMP(cc0?$props) = 1;@\\
\mbox{}\verb@.@\hbox{\sffamily\bfseries sort}\verb@@{\NWsep}
\end{list}
\vspace{-1ex}
\footnotesize\addtolength{\baselineskip}{-1ex}
\begin{list}{}{\setlength{\itemsep}{-\parsep}\setlength{\itemindent}{-\leftmargin}}
\item \NWtxtMacroRefIn\ \NWlink{nuweb53}{53}.
\end{list}
\end{minipage}\\[4ex]
\end{flushleft}
In a next step all form factors are replaced by symbols.
These symbols are generated by the \texttt{Python}
program \texttt{golem.py} such that equivalent form factors
have the same symbol across all diagrams.

Before the actual replacement all form factors are symmetrised
over their indices which to minimise the number of different
terms.
\begin{flushleft} \small
\begin{minipage}{\linewidth} \label{scrap49}
$\langle\,$symmetrise form factors\nobreak\ {\footnotesize \NWtarget{nuweb49}{49}}$\,\rangle\equiv$
\vspace{-1ex}
\begin{list}{}{} \item
\mbox{}\verb@#@\hbox{\sffamily\bfseries Do}\verb@ r=2,{`$maxrank'}@\\
\mbox{}\verb@   @\hbox{\sffamily\bfseries Symmetrize}\verb@ a`$loopsize'`r', 1, ...,`r';@\\
\mbox{}\verb@   #@\hbox{\sffamily\bfseries If}\verb@ (`r' > 2) && (`$loopsize' < 6)@\\
\mbox{}\verb@      @\hbox{\sffamily\bfseries Symmetrize}\verb@ b`$loopsize'`r', 1, ...,{`r'-2};@\\
\mbox{}\verb@   #@\hbox{\sffamily\bfseries EndIf}\verb@@\\
\mbox{}\verb@   #@\hbox{\sffamily\bfseries If}\verb@ (`r' > 4) && (`$loopsize' < 6)@\\
\mbox{}\verb@      @\hbox{\sffamily\bfseries Symmetrize}\verb@ b`$loopsize'`r', 1, ...,{`r'-4};@\\
\mbox{}\verb@   #@\hbox{\sffamily\bfseries Endif}\verb@@\\
\mbox{}\verb@#@\hbox{\sffamily\bfseries EndDo}\verb@@{\NWsep}
\end{list}
\vspace{-1ex}
\footnotesize\addtolength{\baselineskip}{-1ex}
\begin{list}{}{\setlength{\itemsep}{-\parsep}\setlength{\itemindent}{-\leftmargin}}
\item \NWtxtMacroRefIn\ \NWlink{nuweb50}{50}.
\end{list}
\end{minipage}\\[4ex]
\end{flushleft}
The procedure \texttt{recfind} builds all possible
form factors and carries out the necessary steps
for their replacement. All powers of $\varepsilon$
higher than $\varepsilon^2$ are removed beforehand
to avoid spurious calculations\footnote{The highest
order of
divergence in a form factor is $\varepsilon^-2$
at one-loop.}
\begin{flushleft} \small \label{scrap50}
$\langle\,$replace form factors by symbols\nobreak\ {\footnotesize \NWtarget{nuweb50}{50}}$\,\rangle\equiv$
\vspace{-1ex}
\begin{list}{}{} \item
\mbox{}\verb@@\hbox{$\langle\,$symmetrise form factors\nobreak\ {\footnotesize \NWlink{nuweb49}{49}}$\,\rangle$}\verb@@\\
\mbox{}\verb@@\hbox{\sffamily\bfseries Id}\verb@ eps^n?{>=3} = 0;@\\
\mbox{}\verb@#@\hbox{\sffamily\bfseries Do}\verb@ r=0,'$maxrank'@\\
\mbox{}\verb@   #@\hbox{\sffamily\bfseries Call}\verb@ recfind(a`$loopsize'`r',`$loopsize',`r')@\\
\mbox{}\verb@#@\hbox{\sffamily\bfseries EndDo}\verb@@\\
\mbox{}\verb@#@\hbox{\sffamily\bfseries If}\verb@ `$loopsize' < 6@\\
\mbox{}\verb@   #@\hbox{\sffamily\bfseries Do}\verb@ r=2,`$maxrank'@\\
\mbox{}\verb@      #@\hbox{\sffamily\bfseries Call}\verb@ recfind(b`$loopsize'`r',`$loopsize',{`r'-2})@\\
\mbox{}\verb@   #@\hbox{\sffamily\bfseries EndDo}\verb@@\\
\mbox{}\verb@   #@\hbox{\sffamily\bfseries Do}\verb@ r=4,`$maxrank'@\\
\mbox{}\verb@      #@\hbox{\sffamily\bfseries Call}\verb@ recfind(c`$loopsize'`r',`$loopsize',{`r'-4})@\\
\mbox{}\verb@   #@\hbox{\sffamily\bfseries EndDo}\verb@@\\
\mbox{}\verb@#@\hbox{\sffamily\bfseries EndIf}\verb@@{\NWsep}
\end{list}
\vspace{-1ex}
\footnotesize\addtolength{\baselineskip}{-1ex}
\begin{list}{}{\setlength{\itemsep}{-\parsep}\setlength{\itemindent}{-\leftmargin}}
\item \NWtxtMacroRefIn\ \NWlink{nuweb53}{53}.
\end{list}
\end{flushleft}
A similar strategy as for the form factors is pursued for
the spinor traces. For each trace an expression is obtained
from the \texttt{Python} program \texttt{golem.py} which
consists of a polynomial of symbols representing canonical
traces; therefore the \texttt{FromExternal} statement receives
a replacement rule that replaces the trace which has been
matched by the previous \texttt{Id} statement in the loop.
\begin{flushleft} \small
\begin{minipage}{\linewidth} \label{scrap51}
$\langle\,$replace spinor traces by constants\nobreak\ {\footnotesize \NWtarget{nuweb51}{51}}$\,\rangle\equiv$
\vspace{-1ex}
\begin{list}{}{} \item
\mbox{}\verb@#$dummy = 0;@\\
\mbox{}\verb@#@\hbox{\sffamily\bfseries Define}\verb@ TREXP "ERR"@\\
\mbox{}\verb@#@\hbox{\sffamily\bfseries Do}\verb@ sp=1,1@\\
\mbox{}\verb@   #$dummy = {`$dummy'+1};@\\
\mbox{}\verb@   @\hbox{\sffamily\bfseries Id}\verb@ @\hbox{\sffamily\bfseries IfMatch}\verb@->labsp`$dummy'@\\
\mbox{}\verb@      SpinorTrace(cc?$spsign, ?tail$spmoms) = SpinorTrace(cc, ?tail);@\\
\mbox{}\verb@   @\hbox{\sffamily\bfseries Goto}\verb@ labsp`$dummy'fail;@\\
\mbox{}\verb@   @\hbox{\sffamily\bfseries Label}\verb@ labsp`$dummy';@\\
\mbox{}\verb@      @\hbox{\sffamily\bfseries ReDefine}\verb@ sp, "0";@\\
\mbox{}\verb@   @\hbox{\sffamily\bfseries Label}\verb@ labsp`$dummy'fail;@\\
\mbox{}\verb@.@\hbox{\sffamily\bfseries sort}\verb@@\\
\mbox{}\verb@   #@\hbox{\sffamily\bfseries If}\verb@ `sp' == 0@\\
\mbox{}\verb@      #@\hbox{\sffamily\bfseries ToExternal}\verb@ "TR `$spsign',`$spmoms'\n"@\\
\mbox{}\verb@      #@\hbox{\sffamily\bfseries FromExternal}\verb@@\\
\mbox{}\verb@      @\hbox{\sffamily\bfseries Id}\verb@ SpinorTrace(`$spsign', `$spmoms') = `TREXP';@\\
\mbox{}\verb@   #@\hbox{\sffamily\bfseries EndIf}\verb@@\\
\mbox{}\verb@#@\hbox{\sffamily\bfseries EndDo}\verb@@{\NWsep}
\end{list}
\vspace{-1ex}
\footnotesize\addtolength{\baselineskip}{-1ex}
\begin{list}{}{\setlength{\itemsep}{-\parsep}\setlength{\itemindent}{-\leftmargin}}
\item \NWtxtMacroRefIn\ \NWlink{nuweb53}{53}.
\end{list}
\end{minipage}\\[4ex]
\end{flushleft}
\subsubsection{Summary}
Below the topics which have been discussed above are put
in order. The procedure \texttt{FeynmanRules} is defined
in a file \texttt{`THEORY'.h} and replaces the propagator-
and vertex-functions and the external states by their
actual representations according to the \person{Feynman}
rules. The procedure \texttt{masses} is defined in
\texttt{process.h} and typically contains a statement
to replaces particle masses by zero, e.g. to use a massless
approximation for the $u$, $d$ and $s$ quark one would
define the procedure as follows:
\begin{Verbatim}
#procedure masses()
   Multiply replace_(emu, 0, emd, 0, ems, 0);
#endprocedure
\end{Verbatim}

After splitting the expressions into the distinct colour
structures the \acs{sun} algebra is carried out by
calling the procedure \texttt{sunsimplify}; then each
colour structure is projected on the colour basis.

The spinor lines are completed to traces by insertions
of appropriate ratios of the form
\begin{displaymath}
\braket{k_{i, \pm}\vert\Gamma_{ij}\vert k_{j, \pm}}=
\braket{k_{i, \pm}\vert\Gamma_{ij}\vert k_{j, \pm}}
\frac{\braket{k_{j, \pm}\vert\fmslash{m}\vert k_{i, \pm}}
}{
\braket{k_{j, \pm}\vert m_{\mp}}\braket{m_{\mp}\vert k_{i, \pm}}
}=
\frac{
\tr[\pm]{\kslash[i]\Gamma_{ij}\kslash[j]\fmslash{m}}
}{\braket{k_{j, \pm}\vert m_{\mp}}\braket{m_{\mp}\vert k_{i, \pm}}}\text{.}
\end{displaymath}
The brackets in the denominator are evaluated numerically
and go into a global prefactor:
\begin{flushleft} \small \label{scrap52}
$\langle\,$strip global factor\nobreak\ {\footnotesize \NWtarget{nuweb52}{52}}$\,\rangle\equiv$
\vspace{-1ex}
\begin{list}{}{} \item
\mbox{}\verb@@\hbox{\sffamily\bfseries Id}\verb@ 1/braket(?all) = ANYBRAKET(?all);@\\
\mbox{}\verb@@\hbox{\sffamily\bfseries AntiBracket}\verb@ ANYBRAKET, g;@\\
\mbox{}\verb@.@\hbox{\sffamily\bfseries sort}\verb@@\\
\mbox{}\verb@@\hbox{\sffamily\bfseries Collect}\verb@ PREFACTOR;@\\
\mbox{}\verb@@\hbox{\sffamily\bfseries Normalize}\verb@ PREFACTOR;@\\
\mbox{}\verb@@\hbox{\sffamily\bfseries Id}\verb@ PREFACTOR(cc0?$prefactor) = 1;@\\
\mbox{}\verb@.@\hbox{\sffamily\bfseries sort}\verb@@{\NWsep}
\end{list}
\vspace{-1ex}
\footnotesize\addtolength{\baselineskip}{-1ex}
\begin{list}{}{\setlength{\itemsep}{-\parsep}\setlength{\itemindent}{-\leftmargin}}
\item \NWtxtMacroRefIn\ \NWlink{nuweb53}{53}.
\end{list}
\end{flushleft}
A call to the procedure \texttt{buildspinorlines}
does the contractions of the
form~$\ket{k_\pm}\bra{k_pm}=\Pi_\pm\kslash$.

The integrals are translated into form factors after that step
so that no $n$-dimensional vectors remain in the expression.
All traces containing the $(n-4)$-dimensional \person{Dirac}
matrices $\bar{\gamma}^\mu$ are carried out in the next step
and all explicit appearances of \person{Lorentz} indices
are eliminated. Finally some of the subexpressions are
replaced by symbols before the expression is written into
a \fortran{} file.
\begin{flushleft} \small \label{scrap53}
$\langle\,$simplification algorithm\nobreak\ {\footnotesize \NWtarget{nuweb53}{53}}$\,\rangle\equiv$
\vspace{-1ex}
\begin{list}{}{} \item
\mbox{}\verb@@\hbox{$\langle\,$make all momenta ingoing\nobreak\ {\footnotesize \NWlink{nuweb26}{26}}$\,\rangle$}\verb@@\\
\mbox{}\verb@@\hbox{$\langle\,$determine graph topology\nobreak\ {\footnotesize \NWlink{nuweb28}{28}}$\,\rangle$}\verb@@\\
\mbox{}\verb@@\hbox{$\langle\,$introduce momenta $q_i$ and $r_i$ for one-loop processes\nobreak\ {\footnotesize \NWlink{nuweb27}{27}}$\,\rangle$}\verb@@\\
\mbox{}\verb@#@\hbox{\sffamily\bfseries Call}\verb@ FeynmanRules()@\\
\mbox{}\verb@#@\hbox{\sffamily\bfseries Call}\verb@ masses()@\\
\mbox{}\verb@#@\hbox{\sffamily\bfseries Call}\verb@ RemoveMetricTensors()@\\
\mbox{}\verb@.@\hbox{\sffamily\bfseries sort}\verb@@\\
\mbox{}\verb@@\hbox{$\langle\,$split into colour structures\nobreak\ {\footnotesize \NWlink{nuweb32}{32}}$\,\rangle$}\verb@@\\
\mbox{}\verb@#@\hbox{\sffamily\bfseries Call}\verb@ sunsimplify()@\\
\mbox{}\verb@@\hbox{$\langle\,$project on colour basis\nobreak\ {\footnotesize \NWlink{nuweb34}{34}}$\,\rangle$}\verb@@\\
\mbox{}\verb@#@\hbox{\sffamily\bfseries Call}\verb@ spinorties()@\\
\mbox{}\verb@@\hbox{$\langle\,$strip global factor\nobreak\ {\footnotesize \NWlink{nuweb52}{52}}$\,\rangle$}\verb@@\\
\mbox{}\verb@#@\hbox{\sffamily\bfseries Call}\verb@ buildspinorlines()@\\
\mbox{}\verb@.@\hbox{\sffamily\bfseries sort}\verb@@\\
\mbox{}\verb@#@\hbox{\sffamily\bfseries If}\verb@ `LOOPS' == 1@\\
\mbox{}\verb@   @\hbox{$\langle\,$perform integration\nobreak\ {\footnotesize \NWlink{nuweb39}{39}}$\,\rangle$}\verb@@\\
\mbox{}\verb@#@\hbox{\sffamily\bfseries EndIf}\verb@@\\
\mbox{}\verb@@\hbox{\sffamily\bfseries Id}\verb@ PROP(vec?, cc?) = POW(vec.vec - cc^2, -1);@\\
\mbox{}\verb@@\hbox{\sffamily\bfseries Id}\verb@ PROP(vec?, 0) = POW(vec.vec, -1);@\\
\mbox{}\verb@.@\hbox{\sffamily\bfseries sort}\verb@@\\
\mbox{}\verb@@\hbox{$\langle\,$carry out \person{Lorentz} algebra\nobreak\ {\footnotesize \NWlink{nuweb40}{40}}$\,\rangle$}\verb@@\\
\mbox{}\verb@@\hbox{$\langle\,$$n$-dimensional spinor algebra\nobreak\ {\footnotesize \NWlink{nuweb41}{41}}$\,\rangle$}\verb@@\\
\mbox{}\verb@@\hbox{$\langle\,$eliminate \person{Lorentz} indices\nobreak\ {\footnotesize \NWlink{nuweb45}{45}}$\,\rangle$}\verb@@\\
\mbox{}\verb@.@\hbox{\sffamily\bfseries sort}\verb@@\\
\mbox{}\verb@@\hbox{$\langle\,$strip propagators\nobreak\ {\footnotesize \NWlink{nuweb48}{48}}$\,\rangle$}\verb@@\\
\mbox{}\verb@#@\hbox{\sffamily\bfseries If}\verb@ (`LOOPS' == 1)@\\
\mbox{}\verb@   @\hbox{$\langle\,$replace form factors by symbols\nobreak\ {\footnotesize \NWlink{nuweb50}{50}}$\,\rangle$}\verb@@\\
\mbox{}\verb@#@\hbox{\sffamily\bfseries Else}\verb@@\\
\mbox{}\verb@   @\hbox{\sffamily\bfseries Id}\verb@ eps^n?{>0} = 0;@\\
\mbox{}\verb@#@\hbox{\sffamily\bfseries EndIf}\verb@@\\
\mbox{}\verb@@\hbox{$\langle\,$replace spinor traces by constants\nobreak\ {\footnotesize \NWlink{nuweb51}{51}}$\,\rangle$}\verb@@\\
\mbox{}\verb@.@\hbox{\sffamily\bfseries sort}\verb@@\\
\mbox{}\verb@@\hbox{$\langle\,$evaluate colour vector numerically\nobreak\ {\footnotesize \NWlink{nuweb35}{35}}$\,\rangle$}\verb@@{\NWsep}
\end{list}
\vspace{-1ex}
\footnotesize\addtolength{\baselineskip}{-1ex}
\begin{list}{}{\setlength{\itemsep}{-\parsep}\setlength{\itemindent}{-\leftmargin}}
\item \NWtxtMacroRefIn\ \NWlink{nuweb23}{23}.
\end{list}
\end{flushleft}
\subsection{Generation of the \fortran{} File}
In this section the output generated by the \form{}
program is described. \form{} supports the
programmer in generating files in other languages
than \form{} by providing the \texttt{Format}
statement which sets the format of the output
to the desired format. However, this format only
applies to expressions; therefore the dollar
variables \texttt{\$prefactor} and \texttt{\$props}
are written to local expressions to allow \form{}
to take control over their format.
\begin{flushleft} \small \label{scrap54}
$\langle\,$output section\nobreak\ {\footnotesize \NWtarget{nuweb54}{54}}$\,\rangle\equiv$
\vspace{-1ex}
\begin{list}{}{} \item
\mbox{}\verb@#@\hbox{\sffamily\bfseries Define}\verb@ COUNTER "`$counter'"@\\
\mbox{}\verb@#@\hbox{\sffamily\bfseries Define}\verb@ LOOPSIZE "`$loopsize'"@\\
\mbox{}\verb@@\hbox{\sffamily\bfseries Format}\verb@ Fortran;@\\
\mbox{}\verb@.@\hbox{\sffamily\bfseries sort}\verb@@\\
\mbox{}\verb@@\hbox{\sffamily\bfseries Local}\verb@ prefactor = `$prefactor';@\\
\mbox{}\verb@@\hbox{\sffamily\bfseries Local}\verb@ props = `$props';@\\
\mbox{}\verb@@\hbox{\sffamily\bfseries Id}\verb@ ANYBRAKET(?all) = 1/braket(?all);@\\
\mbox{}\verb@.@\hbox{\sffamily\bfseries sort}\verb@@\\
\mbox{}\verb@@\\
\mbox{}\verb@#@\hbox{\sffamily\bfseries Write}\verb@ <`OUT'> "@\hbox{\sffamily\bfseries module}\verb@     `PREFIX'`DIAG'_`HELICITY'"@\\
\mbox{}\verb@@\hbox{$\langle\,$write header of \fortran{} module\nobreak\ {\footnotesize \NWlink{nuweb55}{55}}$\,\rangle$}\verb@@\\
\mbox{}\verb@#@\hbox{\sffamily\bfseries Write}\verb@ <`OUT'> "@\hbox{\sffamily\bfseries contains}\verb@"@\\
\mbox{}\verb@@\hbox{$\langle\,$write function for diagram\nobreak\ {\footnotesize \NWlink{nuweb56}{56}}$\,\rangle$}\verb@@\\
\mbox{}\verb@#@\hbox{\sffamily\bfseries Write}\verb@ <`OUT'> "@\hbox{\sffamily\bfseries end}\verb@ @\hbox{\sffamily\bfseries module}\verb@ `PREFIX'`DIAG'_`HELICITY'"@{\NWsep}
\end{list}
\vspace{-1ex}
\footnotesize\addtolength{\baselineskip}{-1ex}
\begin{list}{}{\setlength{\itemsep}{-\parsep}\setlength{\itemindent}{-\leftmargin}}
\item \NWtxtMacroRefIn\ \NWlink{nuweb23}{23}.
\end{list}
\end{flushleft}
The header of the \fortran{} file contains also information about
the file creation such as the \form{} version and the creation date;
this can be valuable if one tries to trace back which files are affected
by problems in other components of the code. The rest of the header
is the import of all relevant module files.
\begin{flushleft} \small \label{scrap55}
$\langle\,$write header of \fortran{} module\nobreak\ {\footnotesize \NWtarget{nuweb55}{55}}$\,\rangle\equiv$
\vspace{-1ex}
\begin{list}{}{} \item
\mbox{}\verb@#@\hbox{\sffamily\bfseries Write}\verb@ <`OUT'> "! Created by FORM `@\hbox{\sffamily\bfseries VERSION}\verb@_'.`@\hbox{\sffamily\bfseries SUBVERSION}\verb@_'%"@\\
\mbox{}\verb@#@\hbox{\sffamily\bfseries Write}\verb@ <`OUT'> " `@\hbox{\sffamily\bfseries NAMEVERSION}\verb@_'"@\\
\mbox{}\verb@#@\hbox{\sffamily\bfseries Write}\verb@ <`OUT'> "!         from file `@\hbox{\sffamily\bfseries NAME}\verb@_', `@\hbox{\sffamily\bfseries DATE}\verb@_'"@\\
\mbox{}\verb@#@\hbox{\sffamily\bfseries Write}\verb@ <`OUT'> "!"@\\
\mbox{}\verb@#@\hbox{\sffamily\bfseries Write}\verb@ <`OUT'> ""@\\
\mbox{}\verb@#@\hbox{\sffamily\bfseries Write}\verb@ <`OUT'> "   @\hbox{\sffamily\bfseries use}\verb@ precision"@\\
\mbox{}\verb@#@\hbox{\sffamily\bfseries Write}\verb@ <`OUT'> "   @\hbox{\sffamily\bfseries use}\verb@ form_factor_type"@\\
\mbox{}\verb@#@\hbox{\sffamily\bfseries Write}\verb@ <`OUT'> "   @\hbox{\sffamily\bfseries use}\verb@ algebra"@\\
\mbox{}\verb@#@\hbox{\sffamily\bfseries Write}\verb@ <`OUT'> "   @\hbox{\sffamily\bfseries use}\verb@ param"@\\
\mbox{}\verb@#@\hbox{\sffamily\bfseries Write}\verb@ <`OUT'> "   @\hbox{\sffamily\bfseries use}\verb@ `PREFIX'_tr"@\\
\mbox{}\verb@#@\hbox{\sffamily\bfseries If}\verb@ `LOOPS'==1@\\
\mbox{}\verb@   #@\hbox{\sffamily\bfseries Write}\verb@ <`OUT'> "   @\hbox{\sffamily\bfseries use}\verb@ `PREFIX'_ff"@\\
\mbox{}\verb@#@\hbox{\sffamily\bfseries EndIf}\verb@@\\
\mbox{}\verb@#@\hbox{\sffamily\bfseries Write}\verb@ <`OUT'> "   @\hbox{\sffamily\bfseries use}\verb@ mandelstam`LEGS'"@\\
\mbox{}\verb@#@\hbox{\sffamily\bfseries Write}\verb@ <`OUT'> "   @\hbox{\sffamily\bfseries implicit}\verb@ @\hbox{\sffamily\bfseries none}\verb@"@{\NWsep}
\end{list}
\vspace{-1ex}
\footnotesize\addtolength{\baselineskip}{-1ex}
\begin{list}{}{\setlength{\itemsep}{-\parsep}\setlength{\itemindent}{-\leftmargin}}
\item \NWtxtMacroRefIn\ \NWlink{nuweb54}{54}.
\end{list}
\end{flushleft}
The second part of the output section writes a function that
calculates a single diagram for the given helicity. After all
variables are defined and initialised the colour vector is calculated;
in the last section the expressions for each colour structure
are printed out and all parts are summed up.
\begin{flushleft} \small \label{scrap56}
$\langle\,$write function for diagram\nobreak\ {\footnotesize \NWtarget{nuweb56}{56}}$\,\rangle\equiv$
\vspace{-1ex}
\begin{list}{}{} \item
\mbox{}\verb@#@\hbox{\sffamily\bfseries Write}\verb@ <`OUT'> "@\hbox{\sffamily\bfseries function}\verb@     `PREFIX'`DIAG'h`HELICITY'(vecs) @\hbox{\sffamily\bfseries result}\verb@(res)"@\\
\mbox{}\verb@#@\hbox{\sffamily\bfseries Write}\verb@ <`OUT'> "   @\hbox{\sffamily\bfseries implicit}\verb@ @\hbox{\sffamily\bfseries none}\verb@"@\\
\mbox{}\verb@@\hbox{$\langle\,$write variable declarations\nobreak\ {\footnotesize \NWlink{nuweb57}{57}}$\,\rangle$}\verb@@\\
\mbox{}\verb@@\hbox{$\langle\,$initialise local variables\nobreak\ {\footnotesize \NWlink{nuweb59}{59}}$\,\rangle$}\verb@@\\
\mbox{}\verb@.@\hbox{\sffamily\bfseries sort}\verb@@\\
\mbox{}\verb@@\hbox{$\langle\,$write colour vector\nobreak\ {\footnotesize \NWlink{nuweb58}{58}}$\,\rangle$}\verb@@\\
\mbox{}\verb@@\hbox{$\langle\,$combine result\nobreak\ {\footnotesize \NWlink{nuweb60}{60}}$\,\rangle$}\verb@@\\
\mbox{}\verb@#@\hbox{\sffamily\bfseries Write}\verb@ <`OUT'> "@\hbox{\sffamily\bfseries end}\verb@ @\hbox{\sffamily\bfseries function}\verb@ `PREFIX'`DIAG'h`HELICITY'"@{\NWsep}
\end{list}
\vspace{-1ex}
\footnotesize\addtolength{\baselineskip}{-1ex}
\begin{list}{}{\setlength{\itemsep}{-\parsep}\setlength{\itemindent}{-\leftmargin}}
\item \NWtxtMacroRefIn\ \NWlink{nuweb54}{54}.
\end{list}
\end{flushleft}
The input to the function is a list of the four-vectors $k_1,\ldots, k_N$
in unphysical, i.e. all ingoing kinematics. As result the function
returns a vector containing a form factor type for each colour basis element.
The variables \texttt{props} and \texttt{prefactor} are the quantities
which have been stripped off the expression earlier. The variables
\texttt{result}$i$ contain the coefficients of each colour structure
and the variables \texttt{basis}$i$ contain the associated colour vector.
\begin{flushleft} \small \label{scrap57}
$\langle\,$write variable declarations\nobreak\ {\footnotesize \NWtarget{nuweb57}{57}}$\,\rangle\equiv$
\vspace{-1ex}
\begin{list}{}{} \item
\mbox{}\verb@#@\hbox{\sffamily\bfseries Write}\verb@ <`OUT'> "   @\hbox{\sffamily\bfseries real}\verb@(ki), @\hbox{\sffamily\bfseries dimension}\verb@(`LEGS',4), @\hbox{\sffamily\bfseries intent}\verb@(@\hbox{\sffamily\bfseries in}\verb@) :: vecs"@\\
\mbox{}\verb@#@\hbox{\sffamily\bfseries Write}\verb@ <`OUT'> "   @\hbox{\sffamily\bfseries type}\verb@(form_factor), @\hbox{\sffamily\bfseries dimension}\verb@(1:`NUMCS') :: res"@\\
\mbox{}\verb@#@\hbox{\sffamily\bfseries Write}\verb@ <`OUT'> "   @\hbox{\sffamily\bfseries integer}\verb@ :: i"@\\
\mbox{}\verb@#@\hbox{\sffamily\bfseries Write}\verb@ <`OUT'> "   @\hbox{\sffamily\bfseries complex}\verb@(ki) :: props, prefactor"@\\
\mbox{}\verb@#@\hbox{\sffamily\bfseries Do}\verb@ i=1,`COUNTER'@\\
\mbox{}\verb@   #@\hbox{\sffamily\bfseries Write}\verb@ <`OUT'> "   @\hbox{\sffamily\bfseries type}\verb@(form_factor) :: result`i'"@\\
\mbox{}\verb@   #@\hbox{\sffamily\bfseries Write}\verb@ <`OUT'> "   @\hbox{\sffamily\bfseries real}\verb@(ki), @\hbox{\sffamily\bfseries dimension}\verb@(1:`NUMCS') :: basis`i'"@\\
\mbox{}\verb@#@\hbox{\sffamily\bfseries EndDo}\verb@@\\
\mbox{}\verb@#@\hbox{\sffamily\bfseries Write}\verb@ <`OUT'> "   @\hbox{\sffamily\bfseries real}\verb@(ki), @\hbox{\sffamily\bfseries dimension}\verb@(4) :: k1%"@\\
\mbox{}\verb@#@\hbox{\sffamily\bfseries Do}\verb@ i=2,`LEGS'@\\
\mbox{}\verb@   #@\hbox{\sffamily\bfseries Write}\verb@ <`OUT'> ",k`i'%"@\\
\mbox{}\verb@#@\hbox{\sffamily\bfseries EndDo}\verb@@\\
\mbox{}\verb@#@\hbox{\sffamily\bfseries Write}\verb@ <`OUT'> ""@{\NWsep}
\end{list}
\vspace{-1ex}
\footnotesize\addtolength{\baselineskip}{-1ex}
\begin{list}{}{\setlength{\itemsep}{-\parsep}\setlength{\itemindent}{-\leftmargin}}
\item \NWtxtMacroRefIn\ \NWlink{nuweb56}{56}.
\end{list}
\end{flushleft}
In order to keep the code human-readable all entries of the colour vectors
are denoted both numerically and as a comment symbolically. After each
colour vector the according coefficient is printed. The line
\texttt{! VAR result`i'} is a sentinel for the \texttt{awk} script that
postprocesses the output: if the format is set to produce \fortran{} code,
\form{} breaks each expression into chunks which are separated by the
indicator ``\texttt{\_ = \_}'', where the underscores need to be replaced by
the variable name. The reason for this is that most \fortran{} compilers
only allow a limited number of continuation lines. The comment line above
instructs the \texttt{awk} script to set \texttt{result`i'} as the new
variable name and replaces the sentinel sequence accordingly.\footnote{
At that point \texttt{`i'} is already replaced by a number.}
\begin{flushleft} \small \label{scrap58}
$\langle\,$write colour vector\nobreak\ {\footnotesize \NWtarget{nuweb58}{58}}$\,\rangle\equiv$
\vspace{-1ex}
\begin{list}{}{} \item
\mbox{}\verb@#@\hbox{\sffamily\bfseries Do}\verb@ i=1,`COUNTER'@\\
\mbox{}\verb@   #@\hbox{\sffamily\bfseries Do}\verb@ c=1,`NUMCS'@\\
\mbox{}\verb@      #@\hbox{\sffamily\bfseries Write}\verb@ <`OUT'> "!  basis`i'(`c') = %$", $basis`i'x`c'@\\
\mbox{}\verb@      #@\hbox{\sffamily\bfseries Write}\verb@ <`OUT'> "   basis'i'('c') = %E", basis'i'x'c'@\\
\mbox{}\verb@   #@\hbox{\sffamily\bfseries EndDo}\verb@@\\
\mbox{}\verb@   #@\hbox{\sffamily\bfseries Write}\verb@ <`OUT'> "   ! VAR result`i'"@\\
\mbox{}\verb@   #@\hbox{\sffamily\bfseries Write}\verb@ <`OUT'> "   result`i' = %E", struct`i'@\\
\mbox{}\verb@#@\hbox{\sffamily\bfseries EndDo}\verb@@{\NWsep}
\end{list}
\vspace{-1ex}
\footnotesize\addtolength{\baselineskip}{-1ex}
\begin{list}{}{\setlength{\itemsep}{-\parsep}\setlength{\itemindent}{-\leftmargin}}
\item \NWtxtMacroRefIn\ \NWlink{nuweb56}{56}.
\end{list}
\end{flushleft}
The code sets up variables $k_i$ because the variable
\texttt{LEGPERMUTATION} is given in that format.
The subroutine \texttt{yvariables} globally defines symbols
for the \person{Mandelstam} variables according to the
current permutation of external legs.
\begin{flushleft} \small \label{scrap59}
$\langle\,$initialise local variables\nobreak\ {\footnotesize \NWtarget{nuweb59}{59}}$\,\rangle\equiv$
\vspace{-1ex}
\begin{list}{}{} \item
\mbox{}\verb@#@\hbox{\sffamily\bfseries Do}\verb@ i=1,`LEGS'@\\
\mbox{}\verb@   #@\hbox{\sffamily\bfseries Write}\verb@ <`OUT'> "   k`i'(1:4)=vecs(`i', 1:4)"@\\
\mbox{}\verb@#@\hbox{\sffamily\bfseries EndDo}\verb@@\\
\mbox{}\verb@#@\hbox{\sffamily\bfseries Write}\verb@ <`OUT'> "   @\hbox{\sffamily\bfseries call}\verb@ yvariables(`LEGPERMUTATION')"@\\
\mbox{}\verb@.@\hbox{\sffamily\bfseries sort}\verb@@\\
\mbox{}\verb@#@\hbox{\sffamily\bfseries Write}\verb@ <`OUT'> "   ! VAR props"@\\
\mbox{}\verb@#@\hbox{\sffamily\bfseries Write}\verb@ <`OUT'> "   props = %E", props@\\
\mbox{}\verb@#@\hbox{\sffamily\bfseries Write}\verb@ <`OUT'> "   ! VAR prefactor"@\\
\mbox{}\verb@#@\hbox{\sffamily\bfseries Write}\verb@ <`OUT'> "   prefactor = %E", prefactor@\\
\mbox{}\verb@#@\hbox{\sffamily\bfseries Write}\verb@ <`OUT'> "   prefactor = prefactor / props"@{\NWsep}
\end{list}
\vspace{-1ex}
\footnotesize\addtolength{\baselineskip}{-1ex}
\begin{list}{}{\setlength{\itemsep}{-\parsep}\setlength{\itemindent}{-\leftmargin}}
\item \NWtxtMacroRefIn\ \NWlink{nuweb56}{56}.
\end{list}
\end{flushleft}
The last part of the function runs a loop over all
colour structures to add up the results. In the very
end the prefactor that is global to all terms is multiplied.
\begin{flushleft} \small \label{scrap60}
$\langle\,$combine result\nobreak\ {\footnotesize \NWtarget{nuweb60}{60}}$\,\rangle\equiv$
\vspace{-1ex}
\begin{list}{}{} \item
\mbox{}\verb@#@\hbox{\sffamily\bfseries Write}\verb@ <`OUT'> "   @\hbox{\sffamily\bfseries do}\verb@ i = 1, `NUMCS'"@\\
\mbox{}\verb@#@\hbox{\sffamily\bfseries Write}\verb@ <`OUT'> "      res(i) = result1 * basis1(i)%"@\\
\mbox{}\verb@#@\hbox{\sffamily\bfseries Do}\verb@ i = 2,`COUNTER'@\\
\mbox{}\verb@   #@\hbox{\sffamily\bfseries Write}\verb@ <`OUT'> "&"@\\
\mbox{}\verb@   #@\hbox{\sffamily\bfseries Write}\verb@ <`OUT'> "      &      + result`i' * basis`i'(i)%"@\\
\mbox{}\verb@#@\hbox{\sffamily\bfseries EndDo}\verb@@\\
\mbox{}\verb@#@\hbox{\sffamily\bfseries Write}\verb@ <`OUT'> ""@\\
\mbox{}\verb@#@\hbox{\sffamily\bfseries Write}\verb@ <`OUT'> "      res(i) = res(i) * prefactor"@\\
\mbox{}\verb@#@\hbox{\sffamily\bfseries Write}\verb@ <`OUT'> "   @\hbox{\sffamily\bfseries end}\verb@ @\hbox{\sffamily\bfseries do}\verb@"@{\NWsep}
\end{list}
\vspace{-1ex}
\footnotesize\addtolength{\baselineskip}{-1ex}
\begin{list}{}{\setlength{\itemsep}{-\parsep}\setlength{\itemindent}{-\leftmargin}}
\item \NWtxtMacroRefIn\ \NWlink{nuweb56}{56}.
\end{list}
\end{flushleft}
\subsection{Subroutines}
Subroutines --- or procedures, as they are called in \form{} ---
serve several purposes. One reason for using procedures
is to structure the source code into smaller pieces that can be
understood independently; in \form{} one can write procedures
into separate files which adds a physical structure to the code
on top of the logical structure.
In addition procedures can be parametrised
and hence be reused in different places of the code. 
They also support recursion,
a feature that has been used in the procedure \texttt{findff},
which together with \texttt{recfind} replaces the form factors.
\begin{flushleft} \small \label{scrap61}
$\langle\,$define procedures\nobreak\ {\footnotesize \NWtarget{nuweb61}{61}}$\,\rangle\equiv$
\vspace{-1ex}
\begin{list}{}{} \item
\mbox{}\verb@@\hbox{$\langle\,$define procedure \texttt{TopologyInfo}\nobreak\ {\footnotesize \NWlink{nuweb62}{62}}$\,\rangle$}\verb@@\\
\mbox{}\verb@@\hbox{$\langle\,$define procedure \texttt{IntroduceRMomenta}\nobreak\ {\footnotesize \NWlink{nuweb71}{71}}$\,\rangle$}\verb@@\\
\mbox{}\verb@@\hbox{$\langle\,$define procedure \texttt{RemoveMetricTensors}\nobreak\ {\footnotesize \NWlink{nuweb74}{74}}$\,\rangle$}\verb@@\\
\mbox{}\verb@@\hbox{$\langle\,$define procedure \texttt{recfind}\nobreak\ {\footnotesize \NWlink{nuweb75}{75}}$\,\rangle$}\verb@@\\
\mbox{}\verb@@\hbox{$\langle\,$define procedure \texttt{kinematics}\nobreak\ {\footnotesize \NWlink{nuweb79}{79}}$\,\rangle$}\verb@@\\
\mbox{}\verb@@\hbox{$\langle\,$define procedure \texttt{hProjectorSimplify}\nobreak\ {\footnotesize \NWlink{nuweb80}{80}}$\,\rangle$}\verb@@\\
\mbox{}\verb@@\hbox{$\langle\,$define procedure \texttt{sunsimplify}\nobreak\ {\footnotesize \NWlink{nuweb87}{87}}$\,\rangle$}\verb@@{\NWsep}
\end{list}
\vspace{-1ex}
\footnotesize\addtolength{\baselineskip}{-1ex}
\begin{list}{}{\setlength{\itemsep}{-\parsep}\setlength{\itemindent}{-\leftmargin}}
\item \NWtxtMacroRefIn\ \NWlink{nuweb23}{23}.
\end{list}
\end{flushleft}
\subsubsection{Determination of the Loop Topology}
In the massless case the $S$ matrix is fully determined by two
pieces of information: given any planar embedding of the
\person{Feynman} diagram into the two-dimensional drawing
plane then the order in which the external lines are traversed
when walking around the diagram defines a permutation of the
symbols $\{k_1,\ldots,k_N\}$. In an unpinched graph, i.e.
a graph which contains a loop of size~$N$ this permutation
is unique up to a cyclic permutation of the legs and a $\Zset_2$
symmetry. These two symmetries define the freedom of choosing
a starting point and the direction in which one starts the
traversal. For a pinched graph, i.e. a graph of girth smaller
than~$N$, there are additional symmetries.

If $\pi$ is the permutation denoting the order of the legs
(``\texttt{LEGPERMUTATION}'')
and in the pinched kinematics
$r_{i+1}-r_i=k_{a_1}+\ldots+k_{a_p}$ for some sequence
$a_1,\ldots,a_p$ such that\footnote{As usual the indices
have to be cyclically continued $i+N\equiv i$.} $a_i=\pi(i_0+i)$
then $k_{a_2},\ldots,k_{a_p}$ are part of the pinch list,
and the pinch list (``\texttt{PINCHES}'')
is built by all those sequences for $i$ running over
all loop propagators. If the graph is tree level the
permutation is fixed to be the identity and all legs
are in the pinch list.
\begin{flushleft} \small \label{scrap62}
$\langle\,$define procedure \texttt{TopologyInfo}\nobreak\ {\footnotesize \NWtarget{nuweb62}{62}}$\,\rangle\equiv$
\vspace{-1ex}
\begin{list}{}{} \item
\mbox{}\verb@#@\hbox{\sffamily\bfseries Define}\verb@ LEGPERMUTATION ""@\\
\mbox{}\verb@#@\hbox{\sffamily\bfseries Define}\verb@ PINCHES ""@\\
\mbox{}\verb@#@\hbox{\sffamily\bfseries Procedure}\verb@ TopologyInfo()@\\
\mbox{}\verb@   @\hbox{$\langle\,$introduce \texttt{edge} and \texttt{node} functions\nobreak\ {\footnotesize \NWlink{nuweb63}{63}}$\,\rangle$}\verb@@\\
\mbox{}\verb@   @\hbox{$\langle\,$remove \texttt{node} functions\nobreak\ {\footnotesize \NWlink{nuweb64}{64}}$\,\rangle$}\verb@@\\
\mbox{}\verb@   @\hbox{$\langle\,$collect momenta in the loop\nobreak\ {\footnotesize \NWlink{nuweb65}{65}}$\,\rangle$}\verb@@\\
\mbox{}\verb@   @\hbox{$\langle\,$determine \texttt{\$loopsize} and $r_i$\nobreak\ {\footnotesize \NWlink{nuweb66}{66}}$\,\rangle$}\verb@@\\
\mbox{}\verb@   .@\hbox{\sffamily\bfseries sort}\verb@@\\
\mbox{}\verb@   #@\hbox{\sffamily\bfseries If}\verb@ `$loopsize' > 1@\\
\mbox{}\verb@      @\hbox{$\langle\,$calculate $\Delta_{i,i+1}$\nobreak\ {\footnotesize \NWlink{nuweb69}{69}}$\,\rangle$}\verb@@\\
\mbox{}\verb@      @\hbox{$\langle\,$determine the permutation of the legs and the pinches\nobreak\ {\footnotesize \NWlink{nuweb70}{70}}$\,\rangle$}\verb@@\\
\mbox{}\verb@   #@\hbox{\sffamily\bfseries Else}\verb@@\\
\mbox{}\verb@      #@\hbox{\sffamily\bfseries ReDefine}\verb@ LEGPERMUTATION "k1"@\\
\mbox{}\verb@      #@\hbox{\sffamily\bfseries Do}\verb@ i=2,`LEGS'@\\
\mbox{}\verb@         #@\hbox{\sffamily\bfseries ReDefine}\verb@ LEGPERMUTATION "`LEGPERMUTATION',k`i'"@\\
\mbox{}\verb@      #@\hbox{\sffamily\bfseries EndDo}\verb@@\\
\mbox{}\verb@      #@\hbox{\sffamily\bfseries ReDefine}\verb@ PINCHES "`LEGPERMUTATION'"@\\
\mbox{}\verb@   #@\hbox{\sffamily\bfseries EndIf}\verb@@\\
\mbox{}\verb@#@\hbox{\sffamily\bfseries EndProcedure}\verb@@{\NWsep}
\end{list}
\vspace{-1ex}
\footnotesize\addtolength{\baselineskip}{-1ex}
\begin{list}{}{\setlength{\itemsep}{-\parsep}\setlength{\itemindent}{-\leftmargin}}
\item \NWtxtMacroRefIn\ \NWlink{nuweb61}{61}.
\end{list}
\end{flushleft}
The code decorates any propagator function with a function
called \texttt{edge} that contains the momentum and the
indices denoting both ends of the propagator as parameters;
all vertices, similarly are multiplied by a function called
\texttt{node} that contains a index labelling the vertex and
the indices of the according end-points of all adjacent propagators.

The program restricts the degree of the vertices to three
and four, which can be changed if one needs to examine other
quantum field theories that contain higher degree vertices.
\begin{flushleft} \small \label{scrap63}
$\langle\,$introduce \texttt{edge} and \texttt{node} functions\nobreak\ {\footnotesize \NWtarget{nuweb63}{63}}$\,\rangle\equiv$
\vspace{-1ex}
\begin{list}{}{} \item
\mbox{}\verb@@\hbox{\sffamily\bfseries Id}\verb@ ANY?PropagatorFunction@\\
\mbox{}\verb@      (ccFlavour?, iPROP?, k1?, ccMass?, iFrom?, iTo?) =@\\
\mbox{}\verb@   ANY(ccFlavour, iPROP, k1, ccMass, iFrom, iTo) *@\\
\mbox{}\verb@   edge(k1, iFrom, iTo);@\\
\mbox{}\verb@@\\
\mbox{}\verb@#@\hbox{\sffamily\bfseries Do}\verb@ deg=3,4@\\
\mbox{}\verb@   id ANY?VertexFunction(iVertex?@\\
\mbox{}\verb@   #@\hbox{\sffamily\bfseries Do}\verb@ i=1,`deg'@\\
\mbox{}\verb@      , ANYP`i'?(ccFlavour`i'?, iPROP`i'?, k`i'?, ccMass`i'?@\\
\mbox{}\verb@      , iFrom`i'?, iTo`i'?)@\\
\mbox{}\verb@   #@\hbox{\sffamily\bfseries EndDo}\verb@@\\
\mbox{}\verb@      ) =@\\
\mbox{}\verb@   ANY(iVertex@\\
\mbox{}\verb@   #@\hbox{\sffamily\bfseries Do}\verb@ i=1,`deg'@\\
\mbox{}\verb@      , ANYP`i'(ccFlavour`i', iPROP`i', k`i', ccMass`i'@\\
\mbox{}\verb@      , iFrom`i', iTo`i')@\\
\mbox{}\verb@   #@\hbox{\sffamily\bfseries EndDo}\verb@@\\
\mbox{}\verb@   ) * node(iVertex, iFrom1, ..., iFrom'deg');@\\
\mbox{}\verb@#@\hbox{\sffamily\bfseries EndDo}\verb@@{\NWsep}
\end{list}
\vspace{-1ex}
\footnotesize\addtolength{\baselineskip}{-1ex}
\begin{list}{}{\setlength{\itemsep}{-\parsep}\setlength{\itemindent}{-\leftmargin}}
\item \NWtxtMacroRefIn\ \NWlink{nuweb62}{62}.
\end{list}
\end{flushleft}
The functions \texttt{node} can be eliminated if the indices
denoting the endpoints of each edge are replaced by the indices
denoting the vertices themselves; the full information about
the topology is still preserved.
\begin{flushleft} \small \label{scrap64}
$\langle\,$remove \texttt{node} functions\nobreak\ {\footnotesize \NWtarget{nuweb64}{64}}$\,\rangle\equiv$
\vspace{-1ex}
\begin{list}{}{} \item
\mbox{}\verb@@\hbox{\sffamily\bfseries Repeat}\verb@ @\hbox{\sffamily\bfseries Id}\verb@ node(iVertex?, ?mid, iFrom?, ?end) *@\\
\mbox{}\verb@      edge(p1?, iFrom?, i2?) = @\\
\mbox{}\verb@   node(iVertex, ?mid, iFrom, ?end) * edge(p1, iVertex, i2);@\\
\mbox{}\verb@@\hbox{\sffamily\bfseries Repeat}\verb@ @\hbox{\sffamily\bfseries Id}\verb@ node(iVertex?, ?mid, iFrom?, ?end) *@\\
\mbox{}\verb@      edge(p1?, i1?, iFrom?) = @\\
\mbox{}\verb@   node(iVertex, ?mid, iFrom, ?end) * edge(p1, i1, iVertex);@\\
\mbox{}\verb@@\hbox{\sffamily\bfseries Id}\verb@ node(?all) = 1;@{\NWsep}
\end{list}
\vspace{-1ex}
\footnotesize\addtolength{\baselineskip}{-1ex}
\begin{list}{}{\setlength{\itemsep}{-\parsep}\setlength{\itemindent}{-\leftmargin}}
\item \NWtxtMacroRefIn\ \NWlink{nuweb62}{62}.
\end{list}
\end{flushleft}
The form of the expression which has just been achieved is now
suited for the use with the \texttt{ReplaceLoop} statement.
It scans the arguments of the \texttt{edge} functions for a
closed loop of indices and replaces the list of remaining
arguments, which in this case is a list of the momenta of the
propagators in the loop, by the function \texttt{circle}.
The important feature of the \texttt{ReplaceLoop} statement
is that it preserves the order of the arguments; the function
\texttt{circle} is defined as cycle-symmetric; to fix the starting
point for counting the first argument is chosen to be \texttt{p1}.
The function \texttt{node} is chosen instead of \texttt{circle}
because of the symmetry of the \texttt{circle} function\footnote{
The function \texttt{circle} would ``forget'' about the earlier
choice of a starting point.}. The statement \texttt{SplitArg} allows
to separate the $q_i=p+r_i$ into the pair $(p, r_i)$.
\begin{flushleft} \small \label{scrap65}
$\langle\,$collect momenta in the loop\nobreak\ {\footnotesize \NWtarget{nuweb65}{65}}$\,\rangle\equiv$
\vspace{-1ex}
\begin{list}{}{} \item
\mbox{}\verb@@\hbox{\sffamily\bfseries ReplaceLoop}\verb@ edge, @\hbox{\sffamily\bfseries arguments}\verb@=3, @\hbox{\sffamily\bfseries loopsize}\verb@=all, @\hbox{\sffamily\bfseries outfun}\verb@=circle;@\\
\mbox{}\verb@@\hbox{\sffamily\bfseries Id}\verb@ edge(k1?, iFrom?, iFrom?) = circle(k1);@\\
\mbox{}\verb@@\hbox{\sffamily\bfseries Id}\verb@ edge(k1?, iFrom?, iTo?) = 1;@\\
\mbox{}\verb@@\hbox{\sffamily\bfseries Id}\verb@ circle(p1, ?tail) = node(0, ?tail);@\\
\mbox{}\verb@@\hbox{\sffamily\bfseries Id}\verb@ circle(-p1, ?tail) = node(0, ?tail);@\\
\mbox{}\verb@@\hbox{\sffamily\bfseries Repeat}\verb@;@\\
\mbox{}\verb@   @\hbox{\sffamily\bfseries SplitArg}\verb@ (p1), node;@\\
\mbox{}\verb@   @\hbox{\sffamily\bfseries Id}\verb@ node(?head, k1?, -p1, ?tail) = node(?head, -k1, ?tail);@\\
\mbox{}\verb@   @\hbox{\sffamily\bfseries Id}\verb@ node(?head, k1?, p1, ?tail) = node(?head, k1, ?tail);@\\
\mbox{}\verb@@\hbox{\sffamily\bfseries EndRepeat}\verb@;@{\NWsep}
\end{list}
\vspace{-1ex}
\footnotesize\addtolength{\baselineskip}{-1ex}
\begin{list}{}{\setlength{\itemsep}{-\parsep}\setlength{\itemindent}{-\leftmargin}}
\item \NWtxtMacroRefIn\ \NWlink{nuweb62}{62}.
\end{list}
\end{flushleft}
The arguments of the function \texttt{node} are read into the dollar
variables \texttt{\$r}$i$. The size of the loop can easily be determined
by counting the arguments using the built-in function \texttt{nargs\_}.
It should be noted that the vectors $r_i$ are chosen such
that~$r_{\mathtt{\$loopsize}}=0$.
\begin{flushleft} \small \label{scrap66}
$\langle\,$determine \texttt{\$loopsize} and $r_i$\nobreak\ {\footnotesize \NWtarget{nuweb66}{66}}$\,\rangle\equiv$
\vspace{-1ex}
\begin{list}{}{} \item
\mbox{}\verb@@\hbox{\sffamily\bfseries Id}\verb@ node(?all) = node(@\hbox{\sffamily\bfseries nargs}\verb@_(?all), ?all);@\\
\mbox{}\verb@@\hbox{\sffamily\bfseries Id}\verb@ node(cc0?$loopsize, ?all) = node(?all);@\\
\mbox{}\verb@@\hbox{\sffamily\bfseries .sort}\verb@@\\
\mbox{}\verb@@\hbox{\sffamily\bfseries Id}\verb@ node(cc0?$r`$loopsize'@\\
\mbox{}\verb@   #@\hbox{\sffamily\bfseries Do}\verb@ i=1, {`$loopsize'-1}@\\
\mbox{}\verb@      , k`i'?$r`i'@\\
\mbox{}\verb@   #@\hbox{\sffamily\bfseries EndDo}\verb@@\\
\mbox{}\verb@   ) = 1;@{\NWsep}
\end{list}
\vspace{-1ex}
\footnotesize\addtolength{\baselineskip}{-1ex}
\begin{list}{}{\setlength{\itemsep}{-\parsep}\setlength{\itemindent}{-\leftmargin}}
\item \NWtxtMacroRefIn\ \NWlink{nuweb62}{62}.
\end{list}
\end{flushleft}
In the next scrap the program makes use of the
fact that all external momenta are chosen to be
ingoing. The main difficulty in determining
the permutation of the external vectors and
the pinch list is the ambiguity in the representation
of a sum of external vectors $v=k_{i_1}+\ldots+k_{i_p}$: 
because of momentum conservation $z=k_1+\ldots+k_N=0$
the vectors $v$ and~$(v\pm z)$ are the same. However,
the program has to choose one of the three representations
which ensures that one ends up with a list of all vectors.

The algorithm starts from a list of all $\Delta_{i,i+1}$
which is stored in the argument list of
a function~\texttt{TEMPLegs}.
\begin{flushleft} \small \label{scrap67}
$\langle\,$list all $\Delta_{i,i+1}$\nobreak\ {\footnotesize \NWtarget{nuweb67}{67}}$\,\rangle\equiv$
\vspace{-1ex}
\begin{list}{}{} \item
\mbox{}\verb@@\hbox{\sffamily\bfseries Multiply}\verb@ TEMPLegs(@\\
\mbox{}\verb@      $r1 - $r'$loopsize'@\\
\mbox{}\verb@      #@\hbox{\sffamily\bfseries Do}\verb@ i=2,`$loopsize'@\\
\mbox{}\verb@         , $r`i' - $r{`i'-1}@\\
\mbox{}\verb@      #@\hbox{\sffamily\bfseries EndDo}\verb@@\\
\mbox{}\verb@      );@{\NWsep}
\end{list}
\vspace{-1ex}
\footnotesize\addtolength{\baselineskip}{-1ex}
\begin{list}{}{\setlength{\itemsep}{-\parsep}\setlength{\itemindent}{-\leftmargin}}
\item \NWtxtMacroRefIn\ \NWlink{nuweb69}{69}.
\end{list}
\end{flushleft}
Then each argument $v$ in the list is replaced by
a pair $(v\cdot\eta, v)$; after one substitutes
$k_i\cdot\eta=1$ the first entry in the pair becomes
an integer number which counts the number of external
vectors in the sum and is either positive if $v$ is
the list of vectors to be used in the permutation,
or negative if instead the list is $(z-v)$.
\begin{flushleft} \small \label{scrap68}
$\langle\,$discriminate $v$ and $(v-z)$\nobreak\ {\footnotesize \NWtarget{nuweb68}{68}}$\,\rangle\equiv$
\vspace{-1ex}
\begin{list}{}{} \item
\mbox{}\verb@@\hbox{\sffamily\bfseries Repeat}\verb@ @\hbox{\sffamily\bfseries Id}\verb@ TEMPLegs(?head, p1?, ?tail) =@\\
\mbox{}\verb@   TEMPLegs(?head, node(p1.vec, p1), ?tail);@\\
\mbox{}\verb@@\hbox{\sffamily\bfseries Argument}\verb@ TEMPLegs;@\\
\mbox{}\verb@   @\hbox{\sffamily\bfseries Argument}\verb@ node;@\\
\mbox{}\verb@      @\hbox{\sffamily\bfseries Id}\verb@ p1?{k1,...,k`LEGS'}.vec = 1;@\\
\mbox{}\verb@   @\hbox{\sffamily\bfseries EndArgument}\verb@;@\\
\mbox{}\verb@   @\hbox{\sffamily\bfseries Id}\verb@ node(cc0?@\hbox{\sffamily\bfseries neg}\verb@_, p1?) =@\\
\mbox{}\verb@      node(-p1, 0, (k1 +...+ k`LEGS') + p1);@\\
\mbox{}\verb@   @\hbox{\sffamily\bfseries Id}\verb@ node(cc0?@\hbox{\sffamily\bfseries pos}\verb@_, p1?) =@\\
\mbox{}\verb@      node((k1 +...+ k`LEGS') - p1, 0, p1);@\\
\mbox{}\verb@   @\hbox{\sffamily\bfseries SplitArg}\verb@ node;@\\
\mbox{}\verb@@\hbox{\sffamily\bfseries EndArgument}\verb@;@{\NWsep}
\end{list}
\vspace{-1ex}
\footnotesize\addtolength{\baselineskip}{-1ex}
\begin{list}{}{\setlength{\itemsep}{-\parsep}\setlength{\itemindent}{-\leftmargin}}
\item \NWtxtMacroRefIn\ \NWlink{nuweb69}{69}.
\end{list}
\end{flushleft}
The function \texttt{TEMPLegs} is split into
two lists \texttt{TEMPHeads} and \texttt{TEMPTails},
each containing either all $v$ or all $(z-v)$.
\begin{flushleft} \small \label{scrap69}
$\langle\,$calculate $\Delta_{i,i+1}$\nobreak\ {\footnotesize \NWtarget{nuweb69}{69}}$\,\rangle\equiv$
\vspace{-1ex}
\begin{list}{}{} \item
\mbox{}\verb@@\hbox{$\langle\,$list all $\Delta_{i,i+1}$\nobreak\ {\footnotesize \NWlink{nuweb67}{67}}$\,\rangle$}\verb@@\\
\mbox{}\verb@@\hbox{$\langle\,$discriminate $v$ and $(v-z)$\nobreak\ {\footnotesize \NWlink{nuweb68}{68}}$\,\rangle$}\verb@@\\
\mbox{}\verb@@\hbox{\sffamily\bfseries Id}\verb@ TEMPLegs(?all) = TEMPHeads(?all) * TEMPTails(?all);@\\
\mbox{}\verb@@\hbox{\sffamily\bfseries Repeat}\verb@ @\hbox{\sffamily\bfseries Id}\verb@ TEMPHeads(?a, node(?head, 0, ?tail), ?b) = @\\
\mbox{}\verb@   TEMPHeads(?a, node(?head), ?b);@\\
\mbox{}\verb@@\hbox{\sffamily\bfseries Repeat}\verb@ @\hbox{\sffamily\bfseries Id}\verb@ TEMPTails(?a, node(?head, 0, ?tail), ?b) =@\\
\mbox{}\verb@   TEMPTails(?a, node(?tail), ?b);@{\NWsep}
\end{list}
\vspace{-1ex}
\footnotesize\addtolength{\baselineskip}{-1ex}
\begin{list}{}{\setlength{\itemsep}{-\parsep}\setlength{\itemindent}{-\leftmargin}}
\item \NWtxtMacroRefIn\ \NWlink{nuweb62}{62}.
\end{list}
\end{flushleft}
A priori it is not clear how the diagram generator chooses
to denote the momenta in the graph. Therefore both choices,
$v$ and $(z-v)$ are considered as possibilities. In the end
those lists are selected which have the right number of arguments,
i.e. the list of legs must have \texttt{`LEGS'} elements and
the list of pinches $(\mathtt{`LEGS'}-\mathtt{\$loopsize})$.
\begin{flushleft} \small \label{scrap70}
$\langle\,$determine the permutation of the legs and the pinches\nobreak\ {\footnotesize \NWtarget{nuweb70}{70}}$\,\rangle\equiv$
\vspace{-1ex}
\begin{list}{}{} \item
\mbox{}\verb@@\hbox{\sffamily\bfseries Id}\verb@ TEMPHeads(?all) = TEMPLegs(?all) * TEMPPinches(?all);@\\
\mbox{}\verb@@\hbox{\sffamily\bfseries Id}\verb@ TEMPTails(?all) = TEMPLegs(?all) * TEMPPinches(?all);@\\
\mbox{}\verb@@\hbox{\sffamily\bfseries Repeat}\verb@ @\hbox{\sffamily\bfseries Id}\verb@ TEMPLegs(?head, node(?all), ?tail) =@\\
\mbox{}\verb@   TEMPLegs(?head, ?all, ?tail);@\\
\mbox{}\verb@@\hbox{\sffamily\bfseries Repeat}\verb@ @\hbox{\sffamily\bfseries Id}\verb@ TEMPPinches(?head, node(?pinches, p1?), ?tail) =@\\
\mbox{}\verb@   TEMPPinches(?head, ?pinches, ?tail);@\\
\mbox{}\verb@@\hbox{\sffamily\bfseries Id}\verb@ TEMPLegs(p1?, ?all) = TEMPLegs(@\hbox{\sffamily\bfseries nargs}\verb@_(p1, ?all), p1, ?all);@\\
\mbox{}\verb@@\hbox{\sffamily\bfseries Id}\verb@ TEMPPinches(p1?, ?all) = TEMPPinches(nargs_(p1, ?all), p1, ?all);@\\
\mbox{}\verb@@\hbox{\sffamily\bfseries Id}\verb@ TEMPPinches() = TEMPPinches(0);@\\
\mbox{}\verb@@\hbox{\sffamily\bfseries Id}\verb@ TEMPLegs(`LEGS', ?all$legperm) = 1;@\\
\mbox{}\verb@@\hbox{\sffamily\bfseries Id}\verb@ TEMPLegs(cc0?, ?all) = 1;@\\
\mbox{}\verb@@\hbox{\sffamily\bfseries Id}\verb@ TEMPPinches({`LEGS'-`$loopsize'}, ?all$legpinches) = 1;@\\
\mbox{}\verb@@\hbox{\sffamily\bfseries Id}\verb@ TEMPPinches(cc0?, ?all) = 1;@\\
\mbox{}\verb@.@\hbox{\sffamily\bfseries sort}\verb@@\\
\mbox{}\verb@#@\hbox{\sffamily\bfseries ReDefine}\verb@ LEGPERMUTATION "`$legperm'"@\\
\mbox{}\verb@#@\hbox{\sffamily\bfseries ReDefine}\verb@ PINCHES "`$legpinches'"@{\NWsep}
\end{list}
\vspace{-1ex}
\footnotesize\addtolength{\baselineskip}{-1ex}
\begin{list}{}{\setlength{\itemsep}{-\parsep}\setlength{\itemindent}{-\leftmargin}}
\item \NWtxtMacroRefIn\ \NWlink{nuweb62}{62}.
\end{list}
\end{flushleft}
\subsubsection{The procedure \texttt{IntroduceRMomenta}}
The procedure \texttt{IntroduceRMomenta} replaces sums of
momenta that correspond to $r_i$ or $\Delta_{ij}=(r_i-r_j)$
or $q_i=p+r_i$ by a single vector.
These replacements are only done within
propagators. The loop over \texttt{Z} ensures that
different representations of the same sum of external
vectors are caught by the same set of substitution rules.
\begin{flushleft} \small \label{scrap71}
$\langle\,$define procedure \texttt{IntroduceRMomenta}\nobreak\ {\footnotesize \NWtarget{nuweb71}{71}}$\,\rangle\equiv$
\vspace{-1ex}
\begin{list}{}{} \item
\mbox{}\verb@#@\hbox{\sffamily\bfseries Procedure}\verb@ IntroduceRMomenta()@\\
\mbox{}\verb@@\hbox{$\langle\,$find $q_i$ vectors\nobreak\ {\footnotesize \NWlink{nuweb72}{72}}$\,\rangle$}\verb@@\\
\mbox{}\verb@@\\
\mbox{}\verb@#$zero = k1 + ... + k`LEGS';@\\
\mbox{}\verb@#@\hbox{\sffamily\bfseries Do}\verb@ Z={0,`$zero',-(`$zero')}@\\
\mbox{}\verb@   @\hbox{$\langle\,$find $r_i$ and $\Delta_{ij}$\nobreak\ {\footnotesize \NWlink{nuweb73}{73}}$\,\rangle$}\verb@@\\
\mbox{}\verb@#@\hbox{\sffamily\bfseries EndDo}\verb@@\\
\mbox{}\verb@#@\hbox{\sffamily\bfseries EndProcedure}\verb@@{\NWsep}
\end{list}
\vspace{-1ex}
\footnotesize\addtolength{\baselineskip}{-1ex}
\begin{list}{}{\setlength{\itemsep}{-\parsep}\setlength{\itemindent}{-\leftmargin}}
\item \NWtxtMacroRefIn\ \NWlink{nuweb61}{61}.
\end{list}
\end{flushleft}
The first set of replacements tries to match sums that correspond
to the vectors $q_i$. Only the two cases stemming from different
overall signs have to be considered.
\begin{flushleft} \small \label{scrap72}
$\langle\,$find $q_i$ vectors\nobreak\ {\footnotesize \NWtarget{nuweb72}{72}}$\,\rangle\equiv$
\vspace{-1ex}
\begin{list}{}{} \item
\mbox{}\verb@#@\hbox{\sffamily\bfseries Do}\verb@ i=1,`$loopsize'@\\
\mbox{}\verb@   @\hbox{\sffamily\bfseries Id}\verb@ ANYP?PropagatorFunction(ccFlavour?, iPROP?, p1 + (`$r`i''), @\\
\mbox{}\verb@         ccMass?, iFrom?, iTo?) =@\\
\mbox{}\verb@      ANYP(ccFlavour, iPROP, q`i', ccMass, iFrom, iTo);@\\
\mbox{}\verb@   @\hbox{\sffamily\bfseries Id}\verb@ ANYP?PropagatorFunction(ccFlavour?, iPROP?, -p1 - (`$r`i''), @\\
\mbox{}\verb@         ccMass?, iFrom?, iTo?) =@\\
\mbox{}\verb@      ANYP(ccFlavour, iPROP, -q`i', ccMass, iFrom, iTo);@\\
\mbox{}\verb@#@\hbox{\sffamily\bfseries EndDo}\verb@@{\NWsep}
\end{list}
\vspace{-1ex}
\footnotesize\addtolength{\baselineskip}{-1ex}
\begin{list}{}{\setlength{\itemsep}{-\parsep}\setlength{\itemindent}{-\leftmargin}}
\item \NWtxtMacroRefIn\ \NWlink{nuweb71}{71}.
\end{list}
\end{flushleft}
The substitution set inside the loop over \texttt{Z},
which runs over three different notations of the zero vector,
considers the patterns for $\Delta_{ij}$ and $\Delta_{ji}$
for $i<j$ only.
\begin{flushleft} \small \label{scrap73}
$\langle\,$find $r_i$ and $\Delta_{ij}$\nobreak\ {\footnotesize \NWtarget{nuweb73}{73}}$\,\rangle\equiv$
\vspace{-1ex}
\begin{list}{}{} \item
\mbox{}\verb@#@\hbox{\sffamily\bfseries Do}\verb@ i=1,{`$loopsize'-1}@\\
\mbox{}\verb@   #@\hbox{\sffamily\bfseries Do}\verb@ j={`i'+1}, `$loopsize'@\\
\mbox{}\verb@      @\hbox{\sffamily\bfseries Id}\verb@ ANYP?PropagatorFunction(ccFlavour?, iPROP?,@\\
\mbox{}\verb@            (`Z') + (`$r`i'') - (`$r`j''), @\\
\mbox{}\verb@            ccMass?, iFrom?, iTo?) =@\\
\mbox{}\verb@         ANYP(ccFlavour, iPROP, D`i'x`j', ccMass, iFrom, iTo);@\\
\mbox{}\verb@      @\hbox{\sffamily\bfseries Id}\verb@ ANYP?PropagatorFunction(ccFlavour?, iPROP?,@\\
\mbox{}\verb@            (`Z') + (`$r`j'') - (`$r`i''), @\\
\mbox{}\verb@            ccMass?, iFrom?, iTo?) =@\\
\mbox{}\verb@         ANYP(ccFlavour, iPROP, -D`i'x`j', ccMass, iFrom, iTo);@\\
\mbox{}\verb@   #@\hbox{\sffamily\bfseries EndDo}\verb@@\\
\mbox{}\verb@#@\hbox{\sffamily\bfseries EndDo}\verb@@{\NWsep}
\end{list}
\vspace{-1ex}
\footnotesize\addtolength{\baselineskip}{-1ex}
\begin{list}{}{\setlength{\itemsep}{-\parsep}\setlength{\itemindent}{-\leftmargin}}
\item \NWtxtMacroRefIn\ \NWlink{nuweb71}{71}.
\end{list}
\end{flushleft}
\subsubsection{The procedure \texttt{RemoveMetricTensors}}
As the name suggests the procedure \texttt{RemoveMetricTensors}
tries to replace all \person{Kronecker} deltas and metric
tensors that occur in the expression and do not require
special treatment.
\begin{flushleft} \small \label{scrap74}
$\langle\,$define procedure \texttt{RemoveMetricTensors}\nobreak\ {\footnotesize \NWtarget{nuweb74}{74}}$\,\rangle\equiv$
\vspace{-1ex}
\begin{list}{}{} \item
\mbox{}\verb@#@\hbox{\sffamily\bfseries Procedure}\verb@ RemoveMetricTensors()@\\
\mbox{}\verb@   @\hbox{\sffamily\bfseries Repeat}\verb@;@\\
\mbox{}\verb@      @\hbox{\sffamily\bfseries Id}\verb@ gTensor(n, i1?, i2?) * gTensor(n, i2?, i3?) =@\\
\mbox{}\verb@         gTensor(n, i1, i3);@\\
\mbox{}\verb@      @\hbox{\sffamily\bfseries Id}\verb@ gTensor(n, i1?, i2?) * MOMENTUM(n, k1?, i1?) =@\\
\mbox{}\verb@         MOMENTUM(n, k1, i2);@\\
\mbox{}\verb@      @\hbox{\sffamily\bfseries Id}\verb@ gTensor(n, i1?, i2?) * gg(n, i1?, is1?, is2?) =@\\
\mbox{}\verb@         gg(n, i2, is1, is2);@\\
\mbox{}\verb@      @\hbox{\sffamily\bfseries Id}\verb@ AdjointID(iA?, iB?) * T(i1?, i2?, iA?) =@\\
\mbox{}\verb@         T(i1, i2, iB);@\\
\mbox{}\verb@      @\hbox{\sffamily\bfseries Id}\verb@ AdjointID(iA?, iE?) * f(iA?, iB?, iC?) =@\\
\mbox{}\verb@         f(iE, iB, iC);@\\
\mbox{}\verb@      @\hbox{\sffamily\bfseries Id}\verb@ FundamentalID(i1?, i3?) * T(i1?, i2?, iA?) =@\\
\mbox{}\verb@         T(i3, i2, iA);@\\
\mbox{}\verb@      @\hbox{\sffamily\bfseries Id}\verb@ FundamentalID(i2?, i3?) * T(i1?, i2?, iA?) =@\\
\mbox{}\verb@         T(i1, i3, iA);@\\
\mbox{}\verb@   @\hbox{\sffamily\bfseries EndRepeat}\verb@;@\\
\mbox{}\verb@#@\hbox{\sffamily\bfseries EndProcedure}\verb@@{\NWsep}
\end{list}
\vspace{-1ex}
\footnotesize\addtolength{\baselineskip}{-1ex}
\begin{list}{}{\setlength{\itemsep}{-\parsep}\setlength{\itemindent}{-\leftmargin}}
\item \NWtxtMacroRefIn\ \NWlink{nuweb61}{61}.
\end{list}
\end{flushleft}
\subsubsection{Replacement of the Form Factors}
The procedure \texttt{findff} that scans for all form factors
that can occur in a certain diagram is probably the least
transparent of the procedures in this program.
For the understanding of the procedure it should be noted
that variables such as \texttt{fffound} which are
declared inside the procedure are in a local context.

The main program calls the procedure \texttt{recfind}
with the name of a form factor, the loop size and the
number of indices; a typical call would look like
``\texttt{\#call recfind(b53,5,1)}'' because the
form factor $B^{5,1}_j(S)$ has one index~$j$
and belongs to a topology of loop size~$5$.
The procedure \texttt{recfind} then calls
\texttt{recfind1} if the number of indices
is larger than one; it calls directly
to \texttt{findff} if there are no indices.
\begin{flushleft} \small \label{scrap75}
$\langle\,$define procedure \texttt{recfind}\nobreak\ {\footnotesize \NWtarget{nuweb75}{75}}$\,\rangle\equiv$
\vspace{-1ex}
\begin{list}{}{} \item
\mbox{}\verb@#@\hbox{\sffamily\bfseries Procedure}\verb@ recfind(name,l,counter)@\\
\mbox{}\verb@   #@\hbox{\sffamily\bfseries If}\verb@ `counter' > 0@\\
\mbox{}\verb@      #@\hbox{\sffamily\bfseries Call}\verb@ recfind1(`name',`l',`counter')@\\
\mbox{}\verb@   #@\hbox{\sffamily\bfseries Else}\verb@@\\
\mbox{}\verb@      #@\hbox{\sffamily\bfseries Call}\verb@ findff(`name')@\\
\mbox{}\verb@   #@\hbox{\sffamily\bfseries EndIf}\verb@@\\
\mbox{}\verb@#@\hbox{\sffamily\bfseries EndProcedure}\verb@@\\
\mbox{}\verb@@\hbox{$\langle\,$define procedure \texttt{recfind1}\nobreak\ {\footnotesize \NWlink{nuweb76}{76}}$\,\rangle$}\verb@@\\
\mbox{}\verb@@\hbox{$\langle\,$define procedure \texttt{findff}\nobreak\ {\footnotesize \NWlink{nuweb77}{77}}$\,\rangle$}\verb@@{\NWsep}
\end{list}
\vspace{-1ex}
\footnotesize\addtolength{\baselineskip}{-1ex}
\begin{list}{}{\setlength{\itemsep}{-\parsep}\setlength{\itemindent}{-\leftmargin}}
\item \NWtxtMacroRefIn\ \NWlink{nuweb61}{61}.
\end{list}
\end{flushleft}
The procedure \texttt{recfind1} adds an index
to the argument list and decreases the parameter
\texttt{counter} by one before it calls itself
recursively; if the counter reaches zero
the procedure \texttt{findff} is called to
actually do the substitution.
\begin{flushleft} \small \label{scrap76}
$\langle\,$define procedure \texttt{recfind1}\nobreak\ {\footnotesize \NWtarget{nuweb76}{76}}$\,\rangle\equiv$
\vspace{-1ex}
\begin{list}{}{} \item
\mbox{}\verb@#@\hbox{\sffamily\bfseries Procedure}\verb@ recfind1(name,l,counter,?args)@\\
\mbox{}\verb@   #@\hbox{\sffamily\bfseries Define}\verb@ newcounter "{`counter'-1}"@\\
\mbox{}\verb@   #@\hbox{\sffamily\bfseries If}\verb@ `counter' > 0@\\
\mbox{}\verb@      #@\hbox{\sffamily\bfseries Do}\verb@ i=1,`l'@\\
\mbox{}\verb@         #@\hbox{\sffamily\bfseries Call}\verb@ recfind1(`name',`l',`newcounter',`i',`?args')@\\
\mbox{}\verb@      #@\hbox{\sffamily\bfseries EndDo}\verb@@\\
\mbox{}\verb@   #@\hbox{\sffamily\bfseries Else}\verb@@\\
\mbox{}\verb@      #@\hbox{\sffamily\bfseries Call}\verb@ findff(`name',`?args')@\\
\mbox{}\verb@   #@\hbox{\sffamily\bfseries EndIf}\verb@@\\
\mbox{}\verb@#@\hbox{\sffamily\bfseries EndProcedure}\verb@@{\NWsep}
\end{list}
\vspace{-1ex}
\footnotesize\addtolength{\baselineskip}{-1ex}
\begin{list}{}{\setlength{\itemsep}{-\parsep}\setlength{\itemindent}{-\leftmargin}}
\item \NWtxtMacroRefIn\ \NWlink{nuweb75}{75}.
\end{list}
\end{flushleft}
Finally, the procedure \texttt{findff} builds a rewriting
rule for a given form factor \texttt{`name'(`?args')}
by communicating with the programme \texttt{golem.py}
to obtain a global symbol that represents its value
in the given topology.
\begin{flushleft} \small \label{scrap77}
$\langle\,$define procedure \texttt{findff}\nobreak\ {\footnotesize \NWtarget{nuweb77}{77}}$\,\rangle\equiv$
\vspace{-1ex}
\begin{list}{}{} \item
\mbox{}\verb@#$dummy = 0;@\\
\mbox{}\verb@#@\hbox{\sffamily\bfseries Procedure}\verb@ findff(name,?args)@\\
\mbox{}\verb@   #$dummy = {`$dummy'+1};@\\
\mbox{}\verb@   #@\hbox{\sffamily\bfseries Define}\verb@ fffound "0"@\\
\mbox{}\verb@@\\
\mbox{}\verb@   @\hbox{\sffamily\bfseries Id}\verb@ @\hbox{\sffamily\bfseries IfMatch}\verb@->lab`$dummy' `name'(`?args'`SNULL') =@\\
\mbox{}\verb@      `name'(`?args'`SNULL');@\\
\mbox{}\verb@   @\hbox{\sffamily\bfseries Goto}\verb@ lab`$dummy'fail;@\\
\mbox{}\verb@   @\hbox{\sffamily\bfseries Label}\verb@ lab`$dummy';@\\
\mbox{}\verb@      @\hbox{\sffamily\bfseries Redefine}\verb@ fffound, "1";@\\
\mbox{}\verb@   @\hbox{\sffamily\bfseries Label}\verb@ lab`$dummy'fail;@\\
\mbox{}\verb@   .@\hbox{\sffamily\bfseries sort}\verb@@\\
\mbox{}\verb@   #@\hbox{\sffamily\bfseries If}\verb@ `fffound'@\\
\mbox{}\verb@      #@\hbox{\sffamily\bfseries ToExternal}\verb@ "FF `?args'`name'\n"@\\
\mbox{}\verb@      #@\hbox{\sffamily\bfseries FromExternal}\verb@@\\
\mbox{}\verb@   #@\hbox{\sffamily\bfseries EndIf}\verb@@\\
\mbox{}\verb@#@\hbox{\sffamily\bfseries EndProcedure}\verb@@{\NWsep}
\end{list}
\vspace{-1ex}
\footnotesize\addtolength{\baselineskip}{-1ex}
\begin{list}{}{\setlength{\itemsep}{-\parsep}\setlength{\itemindent}{-\leftmargin}}
\item \NWtxtMacroRefIn\ \NWlink{nuweb75}{75}.
\end{list}
\end{flushleft}
\subsubsection{define procedure \texttt{kinematics}}
The procedure \texttt{kinematics} replaces dot-products
between momenta by \person{Mandelstam} variables.
As a first step it removes all momenta abbreviations
by their representation in external momenta.
\begin{flushleft} \small \label{scrap78}
$\langle\,$eliminate $r_i$ and $\Delta_{ij}$\nobreak\ {\footnotesize \NWtarget{nuweb78}{78}}$\,\rangle\equiv$
\vspace{-1ex}
\begin{list}{}{} \item
\mbox{}\verb@#@\hbox{\sffamily\bfseries If}\verb@ `LOOPS' == 1@\\
\mbox{}\verb@   #@\hbox{\sffamily\bfseries Do}\verb@ i=1,{`$loopsize'-1}@\\
\mbox{}\verb@      #@\hbox{\sffamily\bfseries Do}\verb@ j={`i'+1},`$loopsize'@\\
\mbox{}\verb@         @\hbox{\sffamily\bfseries Id}\verb@ D`i'x`j' = r`i' - r`j';@\\
\mbox{}\verb@      #@\hbox{\sffamily\bfseries EndDo}\verb@@\\
\mbox{}\verb@   #@\hbox{\sffamily\bfseries EndDo}\verb@@\\
\mbox{}\verb@   #@\hbox{\sffamily\bfseries Do}\verb@ i=1,{`$loopsize'}@\\
\mbox{}\verb@         @\hbox{\sffamily\bfseries Id}\verb@ r`i' = `$r`i'';@\\
\mbox{}\verb@   #@\hbox{\sffamily\bfseries EndDo}\verb@@\\
\mbox{}\verb@#@\hbox{\sffamily\bfseries EndIf}\verb@@{\NWsep}
\end{list}
\vspace{-1ex}
\footnotesize\addtolength{\baselineskip}{-1ex}
\begin{list}{}{\setlength{\itemsep}{-\parsep}\setlength{\itemindent}{-\leftmargin}}
\item \NWtxtMacroRefIn\ \NWlink{nuweb79}{79}.
\end{list}
\end{flushleft}
Then the \person{Mandelstam} variables are
plugged in and the on-shell conditions are applied.
\begin{flushleft} \small \label{scrap79}
$\langle\,$define procedure \texttt{kinematics}\nobreak\ {\footnotesize \NWtarget{nuweb79}{79}}$\,\rangle\equiv$
\vspace{-1ex}
\begin{list}{}{} \item
\mbox{}\verb@#@\hbox{\sffamily\bfseries Procedure}\verb@ kinematics()@\\
\mbox{}\verb@   @\hbox{$\langle\,$eliminate $r_i$ and $\Delta_{ij}$\nobreak\ {\footnotesize \NWlink{nuweb78}{78}}$\,\rangle$}\verb@@\\
\mbox{}\verb@   #@\hbox{\sffamily\bfseries Call}\verb@ mandelstam(`LEGPERMUTATION')@\\
\mbox{}\verb@   #@\hbox{\sffamily\bfseries Call}\verb@ onshell()@\\
\mbox{}\verb@#@\hbox{\sffamily\bfseries EndProcedure}\verb@ kinematics@{\NWsep}
\end{list}
\vspace{-1ex}
\footnotesize\addtolength{\baselineskip}{-1ex}
\begin{list}{}{\setlength{\itemsep}{-\parsep}\setlength{\itemindent}{-\leftmargin}}
\item \NWtxtMacroRefIn\ \NWlink{nuweb61}{61}.
\end{list}
\end{flushleft}
\subsubsection{The procedure \texttt{hProjectorSimplify}}
The procedure \texttt{hProjectorSimplify} orders a trace
with respect to the helicity projectors $\Pi_\pm$ it contains.
\begin{flushleft} \small \label{scrap80}
$\langle\,$define procedure \texttt{hProjectorSimplify}\nobreak\ {\footnotesize \NWtarget{nuweb80}{80}}$\,\rangle\equiv$
\vspace{-1ex}
\begin{list}{}{} \item
\mbox{}\verb@#@\hbox{\sffamily\bfseries Procedure}\verb@ hProjectorSimplify()@\\
\mbox{}\verb@   @\hbox{$\langle\,$shuffle projectors to the left\nobreak\ {\footnotesize \NWlink{nuweb81}{81}}$\,\rangle$}\verb@@\\
\mbox{}\verb@   @\hbox{$\langle\,$use projector properties\nobreak\ {\footnotesize \NWlink{nuweb82}{82}}$\,\rangle$}\verb@@\\
\mbox{}\verb@#@\hbox{\sffamily\bfseries EndProcedure}\verb@@{\NWsep}
\end{list}
\vspace{-1ex}
\footnotesize\addtolength{\baselineskip}{-1ex}
\begin{list}{}{\setlength{\itemsep}{-\parsep}\setlength{\itemindent}{-\leftmargin}}
\item \NWtxtMacroRefIn\ \NWlink{nuweb61}{61}.
\end{list}
\end{flushleft}
First the projectors are shuffled to the left of the trace.
If a trace contains more than one projector this ensures that
they are all adjacent to each other.
\begin{flushleft} \small \label{scrap81}
$\langle\,$shuffle projectors to the left\nobreak\ {\footnotesize \NWtarget{nuweb81}{81}}$\,\rangle\equiv$
\vspace{-1ex}
\begin{list}{}{} \item
\mbox{}\verb@@\hbox{\sffamily\bfseries Repeat}\verb@;@\\
\mbox{}\verb@   @\hbox{\sffamily\bfseries Repeat}\verb@ @\hbox{\sffamily\bfseries Id}\verb@ SpinorTrace(?head, vec?, hProjector(cc0?), ?tail) =@\\
\mbox{}\verb@      SpinorTrace(?head, hProjector(-cc0), vec, ?tail);@\\
\mbox{}\verb@   @\hbox{\sffamily\bfseries Id}\verb@ SpinorTrace(?head, iMu?, hProjector(cc0?), ?tail) =@\\
\mbox{}\verb@      SpinorTrace(?head, hProjector(-cc0), iMu, ?tail);@\\
\mbox{}\verb@@\hbox{\sffamily\bfseries EndRepeat}\verb@;@{\NWsep}
\end{list}
\vspace{-1ex}
\footnotesize\addtolength{\baselineskip}{-1ex}
\begin{list}{}{\setlength{\itemsep}{-\parsep}\setlength{\itemindent}{-\leftmargin}}
\item \NWtxtMacroRefIn\ \NWlink{nuweb80}{80}.
\end{list}
\end{flushleft}
In a second step the idempotence and the orthogonality of
the projectors are used to obtain a simplification.
\begin{flushleft} \small \label{scrap82}
$\langle\,$use projector properties\nobreak\ {\footnotesize \NWtarget{nuweb82}{82}}$\,\rangle\equiv$
\vspace{-1ex}
\begin{list}{}{} \item
\mbox{}\verb@@\hbox{\sffamily\bfseries Repeat}\verb@ @\hbox{\sffamily\bfseries Id}\verb@ SpinorTrace@\\
\mbox{}\verb@      (hProjector(cc0?), hProjector(cc0?), ?tail) =@\\
\mbox{}\verb@   SpinorTrace(hProjector(cc0), ?tail);@\\
\mbox{}\verb@@\hbox{\sffamily\bfseries Id}\verb@ SpinorTrace(hProjector(+1), hProjector(-1), ?tail) = 0;@\\
\mbox{}\verb@@\hbox{\sffamily\bfseries Id}\verb@ SpinorTrace(hProjector(-1), hProjector(+1), ?tail) = 0;@{\NWsep}
\end{list}
\vspace{-1ex}
\footnotesize\addtolength{\baselineskip}{-1ex}
\begin{list}{}{\setlength{\itemsep}{-\parsep}\setlength{\itemindent}{-\leftmargin}}
\item \NWtxtMacroRefIn\ \NWlink{nuweb80}{80}.
\end{list}
\end{flushleft}
\subsubsection{Procedure for the Colour Algebra}
The colour algebra is carried out in a very similar manner as
described in algorithm~\ref{alg:qcd-color:reduction}.
To speed the rewriting up the procedure brackets off
all factors that do not contain colour information.
\begin{flushleft} \small \label{scrap83}
$\langle\,$separate colour factor for efficiency\nobreak\ {\footnotesize \NWtarget{nuweb83}{83}}$\,\rangle\equiv$
\vspace{-1ex}
\begin{list}{}{} \item
\mbox{}\verb@@\hbox{\sffamily\bfseries Bracket}\verb@ AdjointID, FundamentalID, T, f;@\\
\mbox{}\verb@.@\hbox{\sffamily\bfseries sort}\verb@@\\
\mbox{}\verb@@\hbox{\sffamily\bfseries Collect}\verb@ ANYNonColor;@\\
\mbox{}\verb@@\hbox{\sffamily\bfseries Bracket}\verb@ AdjointID, FundamentalID, T, f;@\\
\mbox{}\verb@.@\hbox{\sffamily\bfseries sort}\verb@;@\\
\mbox{}\verb@@\hbox{\sffamily\bfseries Keep}\verb@ @\hbox{\sffamily\bfseries Brackets}\verb@;@{\NWsep}
\end{list}
\vspace{-1ex}
\footnotesize\addtolength{\baselineskip}{-1ex}
\begin{list}{}{\setlength{\itemsep}{-\parsep}\setlength{\itemindent}{-\leftmargin}}
\item \NWtxtMacroRefIn\ \NWlink{nuweb87}{87}.
\end{list}
\end{flushleft}
In the next step equation \eqref{eq:qcd-color:ftoTTT} is
exploited to eliminate all structure constants $f^{ABC}$.
The expression $2l+e-2$, where $l$ is the number of loops
and $e$ is the number of legs counts the maximum number
of structure constants that can occur in a diagram of that
complexity.
\begin{flushleft} \small \label{scrap84}
$\langle\,$eliminate $f^{ABD}$\nobreak\ {\footnotesize \NWtarget{nuweb84}{84}}$\,\rangle\equiv$
\vspace{-1ex}
\begin{list}{}{} \item
\mbox{}\verb@#@\hbox{\sffamily\bfseries Do}\verb@ i=1, {2*`LOOPS'+`LEGS'-2}@\\
\mbox{}\verb@   @\hbox{\sffamily\bfseries Id}\verb@ @\hbox{\sffamily\bfseries Once}\verb@ f(iADJ1?, iADJ2?, iADJ3?) =@\\
\mbox{}\verb@      -i_ * 1/TR * (@\\
\mbox{}\verb@       + T(i'i'T0, i'i'T1, iADJ1) *@\\
\mbox{}\verb@         T(i'i'T1, i'i'T2, iADJ2) *@\\
\mbox{}\verb@         T(i'i'T2, i'i'T0, iADJ3)@\\
\mbox{}\verb@       - T(i'i'T0, i'i'T1, iADJ3) *@\\
\mbox{}\verb@         T(i'i'T1, i'i'T2, iADJ2) *@\\
\mbox{}\verb@         T(i'i'T2, i'i'T0, iADJ1)@\\
\mbox{}\verb@      ); @\\
\mbox{}\verb@#@\hbox{\sffamily\bfseries EndDo}\verb@@{\NWsep}
\end{list}
\vspace{-1ex}
\footnotesize\addtolength{\baselineskip}{-1ex}
\begin{list}{}{\setlength{\itemsep}{-\parsep}\setlength{\itemindent}{-\leftmargin}}
\item \NWtxtMacroRefIn\ \NWlink{nuweb87}{87}.
\end{list}
\end{flushleft}
All internal gluon lines are removed from the colour
diagram using the completeness
relation~\eqref{eq:qcd-color:reduce}.
\begin{flushleft} \small
\begin{minipage}{\linewidth} \label{scrap85}
$\langle\,$completeness relation~\eqref{eq:qcd-color:reduce}\nobreak\ {\footnotesize \NWtarget{nuweb85}{85}}$\,\rangle\equiv$
\vspace{-1ex}
\begin{list}{}{} \item
\mbox{}\verb@@\hbox{\sffamily\bfseries Id}\verb@ T(i1?, i2?, iADJ?) * T(i3?, i4?, iADJ?) =@\\
\mbox{}\verb@   TR * (@\\
\mbox{}\verb@        FundamentalID(i1, i4) * FundamentalID(i2, i3)@\\
\mbox{}\verb@      - 1/dF * FundamentalID(i1, i2) * FundamentalID(i3, i4)@\\
\mbox{}\verb@   );@{\NWsep}
\end{list}
\vspace{-1ex}
\footnotesize\addtolength{\baselineskip}{-1ex}
\begin{list}{}{\setlength{\itemsep}{-\parsep}\setlength{\itemindent}{-\leftmargin}}
\item \NWtxtMacroRefIn\ \NWlink{nuweb87}{87}.
\end{list}
\end{minipage}\\[4ex]
\end{flushleft}
What remains are contractions of \person{Kronecker} deltas
and possibly tadpole graphs, which can be reduced by the
code below.
\begin{flushleft} \small \label{scrap86}
$\langle\,$contract colour deltas\nobreak\ {\footnotesize \NWtarget{nuweb86}{86}}$\,\rangle\equiv$
\vspace{-1ex}
\begin{list}{}{} \item
\mbox{}\verb@@\hbox{\sffamily\bfseries Repeat}\verb@ @\hbox{\sffamily\bfseries Id}\verb@ AdjointID(i1?, i2?) * AdjointID(i2?, i3?) =@\\
\mbox{}\verb@   AdjointID(i1, i3);@\\
\mbox{}\verb@@\hbox{\sffamily\bfseries Repeat}\verb@ @\hbox{\sffamily\bfseries Id}\verb@ FundamentalID(i1?, i2?) * FundamentalID(i2?, i3?) =@\\
\mbox{}\verb@         FundamentalID(i1, i3);@\\
\mbox{}\verb@@\hbox{\sffamily\bfseries Id}\verb@ AdjointID(iADJ1?, iADJ2?) * T(i1?, i2?, iADJ1?) =@\\
\mbox{}\verb@   T(i1, i2, iADJ2);@\\
\mbox{}\verb@@\hbox{\sffamily\bfseries Id}\verb@ FundamentalID(i1?, i3?) * T(i1?, i2?, iADJ1?) =@\\
\mbox{}\verb@   T(i3, i2, iADJ1);@\\
\mbox{}\verb@@\hbox{\sffamily\bfseries Id}\verb@ FundamentalID(i2?, i3?) * T(i1?, i2?, iADJ1?) =@\\
\mbox{}\verb@   T(i1, i3, iADJ1);@\\
\mbox{}\verb@@\hbox{\sffamily\bfseries Id}\verb@ AdjointID(iADJ1?, iADJ1?) = dA;@\\
\mbox{}\verb@@\hbox{\sffamily\bfseries Id}\verb@ FundamentalID(iADJ1?, iADJ1?) = dF;@\\
\mbox{}\verb@@\hbox{\sffamily\bfseries Id}\verb@ T(i1?, i1?, iADJ1?) = 0;@{\NWsep}
\end{list}
\vspace{-1ex}
\footnotesize\addtolength{\baselineskip}{-1ex}
\begin{list}{}{\setlength{\itemsep}{-\parsep}\setlength{\itemindent}{-\leftmargin}}
\item \NWtxtMacroRefIn\ \NWlink{nuweb87}{87}.
\end{list}
\end{flushleft}
It should be noted that the star-triangle relations
and the reduction of the quadratic \person{Casimir}
operators is not used explicitly, unlike stated in
Algorithm~\ref{alg:qcd-color:reduction} to maintain
the simplicity of the resulting algorithm.
The very last step just removes the function that
has been introduced earlier for efficiency reasons.
\begin{flushleft} \small \label{scrap87}
$\langle\,$define procedure \texttt{sunsimplify}\nobreak\ {\footnotesize \NWtarget{nuweb87}{87}}$\,\rangle\equiv$
\vspace{-1ex}
\begin{list}{}{} \item
\mbox{}\verb@#@\hbox{\sffamily\bfseries Procedure}\verb@ sunsimplify@\\
\mbox{}\verb@   @\hbox{$\langle\,$separate colour factor for efficiency\nobreak\ {\footnotesize \NWlink{nuweb83}{83}}$\,\rangle$}\verb@@\\
\mbox{}\verb@   @\hbox{$\langle\,$eliminate $f^{ABD}$\nobreak\ {\footnotesize \NWlink{nuweb84}{84}}$\,\rangle$}\verb@@\\
\mbox{}\verb@   @\hbox{\sffamily\bfseries Repeat}\verb@;@\\
\mbox{}\verb@      @\hbox{$\langle\,$completeness relation~\eqref{eq:qcd-color:reduce}\nobreak\ {\footnotesize \NWlink{nuweb85}{85}}$\,\rangle$}\verb@@\\
\mbox{}\verb@      @\hbox{$\langle\,$contract colour deltas\nobreak\ {\footnotesize \NWlink{nuweb86}{86}}$\,\rangle$}\verb@@\\
\mbox{}\verb@   @\hbox{\sffamily\bfseries EndRepeat}\verb@;@\\
\mbox{}\verb@.@\hbox{\sffamily\bfseries sort}\verb@@\\
\mbox{}\verb@   @\hbox{\sffamily\bfseries Id}\verb@ ANYNonColor(cc0?) = cc0;@\\
\mbox{}\verb@.@\hbox{\sffamily\bfseries sort}\verb@@\\
\mbox{}\verb@#@\hbox{\sffamily\bfseries EndProcedure}\verb@@{\NWsep}
\end{list}
\vspace{-1ex}
\footnotesize\addtolength{\baselineskip}{-1ex}
\begin{list}{}{\setlength{\itemsep}{-\parsep}\setlength{\itemindent}{-\leftmargin}}
\item \NWtxtMacroRefIn\ \NWlink{nuweb61}{61}.
\end{list}
\end{flushleft}
\section{Numerical Evaluation}
\label{sec:imp-fortran}
\input{imp-fortran}

\section{Index of Literate Programs}
\subsection{Index of Output Files}
\label{ssec:implementation:idx-files}

{\small\begin{list}{}{\setlength{\itemsep}{-\parsep}\setlength{\itemindent}{-\leftmargin}}
\item \verb@"colour.frm"@ {\footnotesize {\NWtxtDefBy} \NWlink{nuweb2}{2}.}
\item \verb@"preprocess.frm"@ {\footnotesize {\NWtxtDefBy} \NWlink{nuweb23}{23}.}
\item \verb@"symbols.h"@ {\footnotesize {\NWtxtDefBy} \NWlink{nuweb16}{16}.}
\end{list}}
\subsection{Index of Macros}
\label{ssec:implementation:idx-macros}

{\small\begin{list}{}{\setlength{\itemsep}{-\parsep}\setlength{\itemindent}{-\leftmargin}}
\item $\langle\,$$n$-dimensional spinor algebra\nobreak\ {\footnotesize \NWlink{nuweb41}{41}}$\,\rangle$ {\footnotesize {\NWtxtRefIn} \NWlink{nuweb53}{53}.}
\item $\langle\,$Build Symmetriser\nobreak\ {\footnotesize \NWlink{nuweb5}{5}}$\,\rangle$ {\footnotesize {\NWtxtRefIn} \NWlink{nuweb8}{8}.}
\item $\langle\,$Cut quark-lines\nobreak\ {\footnotesize \NWlink{nuweb10}{10}}$\,\rangle$ {\footnotesize {\NWtxtRefIn} \NWlink{nuweb8}{8}.}
\item $\langle\,$Insert $t_{ij}^g$\nobreak\ {\footnotesize \NWlink{nuweb11}{11}}$\,\rangle$ {\footnotesize {\NWtxtRefIn} \NWlink{nuweb8}{8}.}
\item $\langle\,$Insert a pair of cuts\nobreak\ {\footnotesize \NWlink{nuweb9}{9}}$\,\rangle$ {\footnotesize {\NWtxtRefIn} \NWlink{nuweb8}{8}.}
\item $\langle\,$Perform Insertions\nobreak\ {\footnotesize \NWlink{nuweb8}{8}}$\,\rangle$ {\footnotesize {\NWtxtRefIn} \NWlink{nuweb2}{2}.}
\item $\langle\,$Procedure definition \texttt{insertgluons}\nobreak\ {\footnotesize \NWlink{nuweb6}{6}}$\,\rangle$ {\footnotesize {\NWtxtRefIn} \NWlink{nuweb2}{2}.}
\item $\langle\,$Procedure definition \texttt{insertquarks}\nobreak\ {\footnotesize \NWlink{nuweb4}{4}}$\,\rangle$ {\footnotesize {\NWtxtRefIn} \NWlink{nuweb2}{2}.}
\item $\langle\,$Procedure definition \texttt{stripcoeff}\nobreak\ {\footnotesize \NWlink{nuweb7}{7}}$\,\rangle$ {\footnotesize {\NWtxtRefIn} \NWlink{nuweb2}{2}.}
\item $\langle\,$Process Specification\nobreak\ {\footnotesize \NWlink{nuweb1}{1}}$\,\rangle$ {\footnotesize {\NWtxtRefIn} \NWlink{nuweb2}{2}.}
\item $\langle\,$Simplify Result\nobreak\ {\footnotesize \NWlink{nuweb12}{12}}$\,\rangle$ {\footnotesize {\NWtxtRefIn} \NWlink{nuweb2}{2}.}
\item $\langle\,$Symbol Definitions\nobreak\ {\footnotesize \NWlink{nuweb3}{3}}$\,\rangle$ {\footnotesize {\NWtxtRefIn} \NWlink{nuweb2}{2}.}
\item $\langle\,$actual replacement of colour structure\nobreak\ {\footnotesize \NWlink{nuweb30}{30}}$\,\rangle$ {\footnotesize {\NWtxtRefIn} \NWlink{nuweb31}{31}.}
\item $\langle\,$calculate $\Delta_{i,i+1}$\nobreak\ {\footnotesize \NWlink{nuweb69}{69}}$\,\rangle$ {\footnotesize {\NWtxtRefIn} \NWlink{nuweb62}{62}.}
\item $\langle\,$carry out \person{Lorentz} algebra\nobreak\ {\footnotesize \NWlink{nuweb40}{40}}$\,\rangle$ {\footnotesize {\NWtxtRefIn} \NWlink{nuweb53}{53}.}
\item $\langle\,$check bounds on diagram number\nobreak\ {\footnotesize \NWlink{nuweb14}{14}}$\,\rangle$ {\footnotesize {\NWtxtRefIn} \NWlink{nuweb24}{24}.}
\item $\langle\,$check command line arguments\nobreak\ {\footnotesize \NWlink{nuweb13}{13}}$\,\rangle$ {\footnotesize {\NWtxtRefIn} \NWlink{nuweb23}{23}.}
\item $\langle\,$check communication channels\nobreak\ {\footnotesize \NWlink{nuweb15}{15}}$\,\rangle$ {\footnotesize {\NWtxtRefIn} \NWlink{nuweb23}{23}.}
\item $\langle\,$collect momenta in the loop\nobreak\ {\footnotesize \NWlink{nuweb65}{65}}$\,\rangle$ {\footnotesize {\NWtxtRefIn} \NWlink{nuweb62}{62}.}
\item $\langle\,$combine result\nobreak\ {\footnotesize \NWlink{nuweb60}{60}}$\,\rangle$ {\footnotesize {\NWtxtRefIn} \NWlink{nuweb56}{56}.}
\item $\langle\,$completeness relation~\eqref{eq:qcd-color:reduce}\nobreak\ {\footnotesize \NWlink{nuweb85}{85}}$\,\rangle$ {\footnotesize {\NWtxtRefIn} \NWlink{nuweb87}{87}.}
\item $\langle\,$construct tensor integral\nobreak\ {\footnotesize \NWlink{nuweb37}{37}}$\,\rangle$ {\footnotesize {\NWtxtRefIn} \NWlink{nuweb39}{39}.}
\item $\langle\,$contract colour deltas\nobreak\ {\footnotesize \NWlink{nuweb86}{86}}$\,\rangle$ {\footnotesize {\NWtxtRefIn} \NWlink{nuweb87}{87}.}
\item $\langle\,$create colour vector\nobreak\ {\footnotesize \NWlink{nuweb33}{33}}$\,\rangle$ {\footnotesize {\NWtxtRefIn} \NWlink{nuweb34}{34}.}
\item $\langle\,$deal with situation $\tr{\cdots\gamma^\mu\cdots\gamma_\mu\cdots}$\nobreak\ {\footnotesize \NWlink{nuweb46}{46}}$\,\rangle$ {\footnotesize {\NWtxtRefIn} \NWlink{nuweb45}{45}.}
\item $\langle\,$deal with situation $\tr{\cdots\gamma^\mu\cdots}\tr{\cdots\gamma_\mu\cdots}$\nobreak\ {\footnotesize \NWlink{nuweb47}{47}}$\,\rangle$ {\footnotesize {\NWtxtRefIn} \NWlink{nuweb45}{45}.}
\item $\langle\,$define auxiliary functions and symbols\nobreak\ {\footnotesize \NWlink{nuweb21}{21}}$\,\rangle$ {\footnotesize {\NWtxtRefIn} \NWlink{nuweb16}{16}.}
\item $\langle\,$define form factors\nobreak\ {\footnotesize \NWlink{nuweb22}{22}}$\,\rangle$ {\footnotesize {\NWtxtRefIn} \NWlink{nuweb16}{16}.}
\item $\langle\,$define procedure \texttt{IntroduceRMomenta}\nobreak\ {\footnotesize \NWlink{nuweb71}{71}}$\,\rangle$ {\footnotesize {\NWtxtRefIn} \NWlink{nuweb61}{61}.}
\item $\langle\,$define procedure \texttt{RemoveMetricTensors}\nobreak\ {\footnotesize \NWlink{nuweb74}{74}}$\,\rangle$ {\footnotesize {\NWtxtRefIn} \NWlink{nuweb61}{61}.}
\item $\langle\,$define procedure \texttt{TopologyInfo}\nobreak\ {\footnotesize \NWlink{nuweb62}{62}}$\,\rangle$ {\footnotesize {\NWtxtRefIn} \NWlink{nuweb61}{61}.}
\item $\langle\,$define procedure \texttt{findff}\nobreak\ {\footnotesize \NWlink{nuweb77}{77}}$\,\rangle$ {\footnotesize {\NWtxtRefIn} \NWlink{nuweb75}{75}.}
\item $\langle\,$define procedure \texttt{hProjectorSimplify}\nobreak\ {\footnotesize \NWlink{nuweb80}{80}}$\,\rangle$ {\footnotesize {\NWtxtRefIn} \NWlink{nuweb61}{61}.}
\item $\langle\,$define procedure \texttt{kinematics}\nobreak\ {\footnotesize \NWlink{nuweb79}{79}}$\,\rangle$ {\footnotesize {\NWtxtRefIn} \NWlink{nuweb61}{61}.}
\item $\langle\,$define procedure \texttt{recfind1}\nobreak\ {\footnotesize \NWlink{nuweb76}{76}}$\,\rangle$ {\footnotesize {\NWtxtRefIn} \NWlink{nuweb75}{75}.}
\item $\langle\,$define procedure \texttt{recfind}\nobreak\ {\footnotesize \NWlink{nuweb75}{75}}$\,\rangle$ {\footnotesize {\NWtxtRefIn} \NWlink{nuweb61}{61}.}
\item $\langle\,$define procedure \texttt{sunsimplify}\nobreak\ {\footnotesize \NWlink{nuweb87}{87}}$\,\rangle$ {\footnotesize {\NWtxtRefIn} \NWlink{nuweb61}{61}.}
\item $\langle\,$define procedures\nobreak\ {\footnotesize \NWlink{nuweb61}{61}}$\,\rangle$ {\footnotesize {\NWtxtRefIn} \NWlink{nuweb23}{23}.}
\item $\langle\,$define symbols for \person{Lorentz} and \person{Dirac} algebra\nobreak\ {\footnotesize \NWlink{nuweb18}{18}}$\,\rangle$ {\footnotesize {\NWtxtRefIn} \NWlink{nuweb16}{16}.}
\item $\langle\,$define symbols for colour algebra\nobreak\ {\footnotesize \NWlink{nuweb17}{17}}$\,\rangle$ {\footnotesize {\NWtxtRefIn} \NWlink{nuweb16}{16}.}
\item $\langle\,$define topological functions\nobreak\ {\footnotesize \NWlink{nuweb20}{20}}$\,\rangle$ {\footnotesize {\NWtxtRefIn} \NWlink{nuweb16}{16}.}
\item $\langle\,$define vectors\nobreak\ {\footnotesize \NWlink{nuweb19}{19}}$\,\rangle$ {\footnotesize {\NWtxtRefIn} \NWlink{nuweb16}{16}.}
\item $\langle\,$determine \texttt{\$loopsize} and $r_i$\nobreak\ {\footnotesize \NWlink{nuweb66}{66}}$\,\rangle$ {\footnotesize {\NWtxtRefIn} \NWlink{nuweb62}{62}.}
\item $\langle\,$determine graph topology\nobreak\ {\footnotesize \NWlink{nuweb28}{28}}$\,\rangle$ {\footnotesize {\NWtxtRefIn} \NWlink{nuweb53}{53}.}
\item $\langle\,$determine the helicities of the external particles\nobreak\ {\footnotesize \NWlink{nuweb25}{25}}$\,\rangle$ {\footnotesize {\NWtxtRefIn} \NWlink{nuweb23}{23}.}
\item $\langle\,$determine the permutation of the legs and the pinches\nobreak\ {\footnotesize \NWlink{nuweb70}{70}}$\,\rangle$ {\footnotesize {\NWtxtRefIn} \NWlink{nuweb62}{62}.}
\item $\langle\,$discriminate $v$ and $(v-z)$\nobreak\ {\footnotesize \NWlink{nuweb68}{68}}$\,\rangle$ {\footnotesize {\NWtxtRefIn} \NWlink{nuweb69}{69}.}
\item $\langle\,$eliminate $f^{ABD}$\nobreak\ {\footnotesize \NWlink{nuweb84}{84}}$\,\rangle$ {\footnotesize {\NWtxtRefIn} \NWlink{nuweb87}{87}.}
\item $\langle\,$eliminate $r_i$ and $\Delta_{ij}$\nobreak\ {\footnotesize \NWlink{nuweb78}{78}}$\,\rangle$ {\footnotesize {\NWtxtRefIn} \NWlink{nuweb79}{79}.}
\item $\langle\,$eliminate \person{Lorentz} indices\nobreak\ {\footnotesize \NWlink{nuweb45}{45}}$\,\rangle$ {\footnotesize {\NWtxtRefIn} \NWlink{nuweb53}{53}.}
\item $\langle\,$evaluate $(n-4)$-dimensional traces\nobreak\ {\footnotesize \NWlink{nuweb44}{44}}$\,\rangle$ {\footnotesize {\NWtxtRefIn} \NWlink{nuweb41}{41}.}
\item $\langle\,$evaluate colour vector numerically\nobreak\ {\footnotesize \NWlink{nuweb35}{35}}$\,\rangle$ {\footnotesize {\NWtxtRefIn} \NWlink{nuweb53}{53}.}
\item $\langle\,$find $q_i$ vectors\nobreak\ {\footnotesize \NWlink{nuweb72}{72}}$\,\rangle$ {\footnotesize {\NWtxtRefIn} \NWlink{nuweb71}{71}.}
\item $\langle\,$find $r_i$ and $\Delta_{ij}$\nobreak\ {\footnotesize \NWlink{nuweb73}{73}}$\,\rangle$ {\footnotesize {\NWtxtRefIn} \NWlink{nuweb71}{71}.}
\item $\langle\,$find next colour structure\nobreak\ {\footnotesize \NWlink{nuweb29}{29}}$\,\rangle$ {\footnotesize {\NWtxtRefIn} \NWlink{nuweb31}{31}.}
\item $\langle\,$initialise local variables\nobreak\ {\footnotesize \NWlink{nuweb59}{59}}$\,\rangle$ {\footnotesize {\NWtxtRefIn} \NWlink{nuweb56}{56}.}
\item $\langle\,$introduce \texttt{edge} and \texttt{node} functions\nobreak\ {\footnotesize \NWlink{nuweb63}{63}}$\,\rangle$ {\footnotesize {\NWtxtRefIn} \NWlink{nuweb62}{62}.}
\item $\langle\,$introduce form factors\nobreak\ {\footnotesize \NWlink{nuweb38}{38}}$\,\rangle$ {\footnotesize {\NWtxtRefIn} \NWlink{nuweb39}{39}.}
\item $\langle\,$introduce momenta $q_i$ and $r_i$ for one-loop processes\nobreak\ {\footnotesize \NWlink{nuweb27}{27}}$\,\rangle$ {\footnotesize {\NWtxtRefIn} \NWlink{nuweb53}{53}.}
\item $\langle\,$label colour structures\nobreak\ {\footnotesize \NWlink{nuweb31}{31}}$\,\rangle$ {\footnotesize {\NWtxtRefIn} \NWlink{nuweb32}{32}.}
\item $\langle\,$list all $\Delta_{i,i+1}$\nobreak\ {\footnotesize \NWlink{nuweb67}{67}}$\,\rangle$ {\footnotesize {\NWtxtRefIn} \NWlink{nuweb69}{69}.}
\item $\langle\,$make all momenta ingoing\nobreak\ {\footnotesize \NWlink{nuweb26}{26}}$\,\rangle$ {\footnotesize {\NWtxtRefIn} \NWlink{nuweb53}{53}.}
\item $\langle\,$move $\bar\gamma^\mu$ right\nobreak\ {\footnotesize \NWlink{nuweb42}{42}}$\,\rangle$ {\footnotesize {\NWtxtRefIn} \NWlink{nuweb41}{41}.}
\item $\langle\,$output section\nobreak\ {\footnotesize \NWlink{nuweb54}{54}}$\,\rangle$ {\footnotesize {\NWtxtRefIn} \NWlink{nuweb23}{23}.}
\item $\langle\,$perform integration\nobreak\ {\footnotesize \NWlink{nuweb39}{39}}$\,\rangle$ {\footnotesize {\NWtxtRefIn} \NWlink{nuweb53}{53}.}
\item $\langle\,$project on colour basis\nobreak\ {\footnotesize \NWlink{nuweb34}{34}}$\,\rangle$ {\footnotesize {\NWtxtRefIn} \NWlink{nuweb53}{53}.}
\item $\langle\,$read libraries and configuration\nobreak\ {\footnotesize \NWlink{nuweb24}{24}}$\,\rangle$ {\footnotesize {\NWtxtRefIn} \NWlink{nuweb23}{23}.}
\item $\langle\,$read propagator masses\nobreak\ {\footnotesize \NWlink{nuweb36}{36}}$\,\rangle$ {\footnotesize {\NWtxtRefIn} \NWlink{nuweb39}{39}.}
\item $\langle\,$remove \texttt{node} functions\nobreak\ {\footnotesize \NWlink{nuweb64}{64}}$\,\rangle$ {\footnotesize {\NWtxtRefIn} \NWlink{nuweb62}{62}.}
\item $\langle\,$replace form factors by symbols\nobreak\ {\footnotesize \NWlink{nuweb50}{50}}$\,\rangle$ {\footnotesize {\NWtxtRefIn} \NWlink{nuweb53}{53}.}
\item $\langle\,$replace spinor traces by constants\nobreak\ {\footnotesize \NWlink{nuweb51}{51}}$\,\rangle$ {\footnotesize {\NWtxtRefIn} \NWlink{nuweb53}{53}.}
\item $\langle\,$separate colour factor for efficiency\nobreak\ {\footnotesize \NWlink{nuweb83}{83}}$\,\rangle$ {\footnotesize {\NWtxtRefIn} \NWlink{nuweb87}{87}.}
\item $\langle\,$shuffle projectors to the left\nobreak\ {\footnotesize \NWlink{nuweb81}{81}}$\,\rangle$ {\footnotesize {\NWtxtRefIn} \NWlink{nuweb80}{80}.}
\item $\langle\,$simplification algorithm\nobreak\ {\footnotesize \NWlink{nuweb53}{53}}$\,\rangle$ {\footnotesize {\NWtxtRefIn} \NWlink{nuweb23}{23}.}
\item $\langle\,$split into colour structures\nobreak\ {\footnotesize \NWlink{nuweb32}{32}}$\,\rangle$ {\footnotesize {\NWtxtRefIn} \NWlink{nuweb53}{53}.}
\item $\langle\,$split traces\nobreak\ {\footnotesize \NWlink{nuweb43}{43}}$\,\rangle$ {\footnotesize {\NWtxtRefIn} \NWlink{nuweb41}{41}.}
\item $\langle\,$strip global factor\nobreak\ {\footnotesize \NWlink{nuweb52}{52}}$\,\rangle$ {\footnotesize {\NWtxtRefIn} \NWlink{nuweb53}{53}.}
\item $\langle\,$strip propagators\nobreak\ {\footnotesize \NWlink{nuweb48}{48}}$\,\rangle$ {\footnotesize {\NWtxtRefIn} \NWlink{nuweb53}{53}.}
\item $\langle\,$symmetrise form factors\nobreak\ {\footnotesize \NWlink{nuweb49}{49}}$\,\rangle$ {\footnotesize {\NWtxtRefIn} \NWlink{nuweb50}{50}.}
\item $\langle\,$use projector properties\nobreak\ {\footnotesize \NWlink{nuweb82}{82}}$\,\rangle$ {\footnotesize {\NWtxtRefIn} \NWlink{nuweb80}{80}.}
\item $\langle\,$write colour vector\nobreak\ {\footnotesize \NWlink{nuweb58}{58}}$\,\rangle$ {\footnotesize {\NWtxtRefIn} \NWlink{nuweb56}{56}.}
\item $\langle\,$write function for diagram\nobreak\ {\footnotesize \NWlink{nuweb56}{56}}$\,\rangle$ {\footnotesize {\NWtxtRefIn} \NWlink{nuweb54}{54}.}
\item $\langle\,$write header of \fortran{} module\nobreak\ {\footnotesize \NWlink{nuweb55}{55}}$\,\rangle$ {\footnotesize {\NWtxtRefIn} \NWlink{nuweb54}{54}.}
\item $\langle\,$write variable declarations\nobreak\ {\footnotesize \NWlink{nuweb57}{57}}$\,\rangle$ {\footnotesize {\NWtxtRefIn} \NWlink{nuweb56}{56}.}
\end{list}}
\subsection{Index of Identifiers}
\label{ssec:implementation:idx-identifiers}

{\small\begin{list}{}{\setlength{\itemsep}{-\parsep}\setlength{\itemindent}{-\leftmargin}}
\item \verb@$basis*@: (\underline{\NWlink{nuweb34}{34})}.
\item \verb@$legperm@: \underline{\NWlink{nuweb70}{70}}.
\item \verb@$legpinches@: \underline{\NWlink{nuweb70}{70}}.
\item \verb@$loopsize@: \NWlink{nuweb27}{27}\NWlink{nuweb28}{, 28}\NWlink{nuweb36}{, 36}\NWlink{nuweb37}{, 37}\NWlink{nuweb38}{, 38}\NWlink{nuweb39}{, 39}\NWlink{nuweb49}{, 49}\NWlink{nuweb50}{, 50}\NWlink{nuweb54}{, 54}\NWlink{nuweb62}{, 62}, \underline{\NWlink{nuweb66}{66}}\NWlink{nuweb67}{, 67}\NWlink{nuweb70}{, 70}\NWlink{nuweb72}{, 72}\NWlink{nuweb73}{, 73}\NWlink{nuweb78}{, 78}.
\item \verb@$mass*@: (\underline{\NWlink{nuweb36}{36})}.
\item \verb@$num@: \underline{\NWlink{nuweb2}{2}}.
\item \verb@$prefactor@: \underline{\NWlink{nuweb52}{52}}\NWlink{nuweb54}{, 54}.
\item \verb@$props@: \underline{\NWlink{nuweb48}{48}}\NWlink{nuweb54}{, 54}.
\item \verb@$r*@: (\underline{\NWlink{nuweb66}{66})}.
\item \verb@AdjointID@: \underline{\NWlink{nuweb17}{17}}\NWlink{nuweb74}{, 74}\NWlink{nuweb83}{, 83}\NWlink{nuweb86}{, 86}.
\item \verb@ANY*@: (\underline{\NWlink{nuweb16}{16})}.
\item \verb@ANYNonColor@: \underline{\NWlink{nuweb83}{83}}\NWlink{nuweb87}{, 87}.
\item \verb@aquarks@: \underline{\NWlink{nuweb3}{3}}\NWlink{nuweb9}{, 9}.
\item \verb@braket@: \underline{\NWlink{nuweb21}{21}}\NWlink{nuweb52}{, 52}\NWlink{nuweb54}{, 54}.
\item \verb@CA@: \underline{\NWlink{nuweb17}{17}}.
\item \verb@cc*@: (\underline{\NWlink{nuweb16}{16})}.
\item \verb@circle@: \underline{\NWlink{nuweb20}{20}}\NWlink{nuweb65}{, 65}.
\item \verb@color*@: (\underline{\NWlink{nuweb16}{16})}.
\item \verb@COLORBASIS@: \underline{\NWlink{nuweb21}{21}}\NWlink{nuweb33}{, 33}\NWlink{nuweb34}{, 34}.
\item \verb@colour@: \underline{\NWlink{nuweb1}{1}}\NWlink{nuweb16}{, 16}\NWlink{nuweb31}{, 31}\NWlink{nuweb32}{, 32}\NWlink{nuweb34}{, 34}\NWlink{nuweb53}{, 53}\NWlink{nuweb56}{, 56}\NWlink{nuweb87}{, 87}.
\item \verb@COUNTER@: \underline{\NWlink{nuweb54}{54}}\NWlink{nuweb57}{, 57}\NWlink{nuweb58}{, 58}\NWlink{nuweb60}{, 60}.
\item \verb@dA@: \underline{\NWlink{nuweb17}{17}}\NWlink{nuweb33}{, 33}\NWlink{nuweb35}{, 35}\NWlink{nuweb86}{, 86}.
\item \verb@DELTA@: \underline{\NWlink{nuweb21}{21}}\NWlink{nuweb39}{, 39}.
\item \verb@delta@: \underline{\NWlink{nuweb3}{3}}\NWlink{nuweb4}{, 4}\NWlink{nuweb5}{, 5}\NWlink{nuweb9}{, 9}\NWlink{nuweb10}{, 10}\NWlink{nuweb11}{, 11}\NWlink{nuweb12}{, 12}.
\item \verb@dF@: \underline{\NWlink{nuweb17}{17}}\NWlink{nuweb33}{, 33}\NWlink{nuweb35}{, 35}\NWlink{nuweb85}{, 85}\NWlink{nuweb86}{, 86}.
\item \verb@DIAG@: \underline{\NWlink{nuweb13}{13}}\NWlink{nuweb14}{, 14}\NWlink{nuweb23}{, 23}\NWlink{nuweb24}{, 24}\NWlink{nuweb32}{, 32}\NWlink{nuweb54}{, 54}\NWlink{nuweb56}{, 56}.
\item \verb@DIAGRAMCOUNT@: \underline{\NWlink{nuweb14}{14}}.
\item \verb@edge@: \underline{\NWlink{nuweb20}{20}}\NWlink{nuweb62}{, 62}\NWlink{nuweb63}{, 63}\NWlink{nuweb64}{, 64}\NWlink{nuweb65}{, 65}.
\item \verb@f@: \NWlink{nuweb7}{7}, \underline{\NWlink{nuweb17}{17}}\NWlink{nuweb33}{, 33}\NWlink{nuweb74}{, 74}\NWlink{nuweb83}{, 83}\NWlink{nuweb84}{, 84}\NWlink{nuweb87}{, 87}.
\item \verb@ff*@: (\underline{\NWlink{nuweb16}{16})}.
\item \verb@findff@: \NWlink{nuweb75}{75}\NWlink{nuweb76}{, 76}, \underline{\NWlink{nuweb77}{77}}.
\item \verb@FormFactors@: \underline{\NWlink{nuweb22}{22}}.
\item \verb@FundamentalID@: \underline{\NWlink{nuweb17}{17}}\NWlink{nuweb74}{, 74}\NWlink{nuweb83}{, 83}\NWlink{nuweb85}{, 85}\NWlink{nuweb86}{, 86}.
\item \verb@g@: \NWlink{nuweb3}{3}\NWlink{nuweb6}{, 6}\NWlink{nuweb8}{, 8}\NWlink{nuweb9}{, 9}\NWlink{nuweb11}{, 11}\NWlink{nuweb12}{, 12}, \underline{\NWlink{nuweb16}{16}}\NWlink{nuweb52}{, 52}\NWlink{nuweb63}{, 63}.
\item \verb@g*@: (\underline{\NWlink{nuweb3}{3})}.
\item \verb@HEL*@: (\underline{\NWlink{nuweb25}{25})}.
\item \verb@HELICITY@: \underline{\NWlink{nuweb13}{13}}\NWlink{nuweb23}{, 23}\NWlink{nuweb25}{, 25}\NWlink{nuweb54}{, 54}\NWlink{nuweb56}{, 56}.
\item \verb@hProjectorSimplify@: \NWlink{nuweb45}{45}\NWlink{nuweb61}{, 61}, \underline{\NWlink{nuweb80}{80}}.
\item \verb@i*@: (\underline{\NWlink{nuweb3}{3})}.
\item \verb@insertgluon@: \underline{\NWlink{nuweb3}{3}}\NWlink{nuweb5}{, 5}\NWlink{nuweb6}{, 6}\NWlink{nuweb9}{, 9}.
\item \verb@insertgluons@: \NWlink{nuweb1}{1}\NWlink{nuweb2}{, 2}, \underline{\NWlink{nuweb6}{6}}.
\item \verb@insertq@: \underline{\NWlink{nuweb3}{3}}\NWlink{nuweb9}{, 9}\NWlink{nuweb10}{, 10}.
\item \verb@insertquarks@: \NWlink{nuweb1}{1}\NWlink{nuweb2}{, 2}, \underline{\NWlink{nuweb4}{4}}.
\item \verb@insertt@: \underline{\NWlink{nuweb3}{3}}\NWlink{nuweb9}{, 9}\NWlink{nuweb10}{, 10}\NWlink{nuweb11}{, 11}.
\item \verb@IntroduceRMomenta@: \NWlink{nuweb27}{27}\NWlink{nuweb61}{, 61}, \underline{\NWlink{nuweb71}{71}}.
\item \verb@j*@: (\underline{\NWlink{nuweb3}{3})}.
\item \verb@kinematics@: \NWlink{nuweb40}{40}\NWlink{nuweb48}{, 48}\NWlink{nuweb61}{, 61}, \underline{\NWlink{nuweb79}{79}}.
\item \verb@k`i'@: \underline{\NWlink{nuweb19}{19}}\NWlink{nuweb26}{, 26}\NWlink{nuweb57}{, 57}\NWlink{nuweb59}{, 59}\NWlink{nuweb62}{, 62}\NWlink{nuweb63}{, 63}\NWlink{nuweb66}{, 66}.
\item \verb@LEGPERMUTATION@: \NWlink{nuweb28}{28}\NWlink{nuweb59}{, 59}\NWlink{nuweb62}{, 62}, \underline{\NWlink{nuweb70}{70}}\NWlink{nuweb79}{, 79}.
\item \verb@line@: \underline{\NWlink{nuweb3}{3}}\NWlink{nuweb12}{, 12}\NWlink{nuweb13}{, 13}\NWlink{nuweb23}{, 23}.
\item \verb@LOOPSIZE@: \underline{\NWlink{nuweb54}{54}}.
\item \verb@MAXLEGS@: \underline{\NWlink{nuweb19}{19}}.
\item \verb@MOMENTUM@: \underline{\NWlink{nuweb21}{21}}\NWlink{nuweb37}{, 37}\NWlink{nuweb39}{, 39}\NWlink{nuweb40}{, 40}\NWlink{nuweb74}{, 74}.
\item \verb@NCTEMP*@: (\underline{\NWlink{nuweb16}{16})}.
\item \verb@node@: \underline{\NWlink{nuweb20}{20}}\NWlink{nuweb62}{, 62}\NWlink{nuweb63}{, 63}\NWlink{nuweb64}{, 64}\NWlink{nuweb65}{, 65}\NWlink{nuweb66}{, 66}\NWlink{nuweb68}{, 68}\NWlink{nuweb69}{, 69}\NWlink{nuweb70}{, 70}.
\item \verb@nullarray@: \underline{\NWlink{nuweb21}{21}}.
\item \verb@p1@: \underline{\NWlink{nuweb19}{19}}\NWlink{nuweb64}{, 64}\NWlink{nuweb65}{, 65}\NWlink{nuweb68}{, 68}\NWlink{nuweb70}{, 70}\NWlink{nuweb72}{, 72}.
\item \verb@PINCHES@: \NWlink{nuweb28}{28}\NWlink{nuweb62}{, 62}, \underline{\NWlink{nuweb70}{70}}.
\item \verb@POW@: \underline{\NWlink{nuweb21}{21}}\NWlink{nuweb34}{, 34}\NWlink{nuweb48}{, 48}\NWlink{nuweb53}{, 53}.
\item \verb@PREFACTOR@: \underline{\NWlink{nuweb21}{21}}\NWlink{nuweb52}{, 52}.
\item \verb@prefactor@: \NWlink{nuweb52}{52}, \underline{\NWlink{nuweb54}{54}}\NWlink{nuweb57}{, 57}\NWlink{nuweb59}{, 59}\NWlink{nuweb60}{, 60}.
\item \verb@PREFIX@: \underline{\NWlink{nuweb13}{13}}\NWlink{nuweb23}{, 23}\NWlink{nuweb24}{, 24}\NWlink{nuweb54}{, 54}\NWlink{nuweb55}{, 55}\NWlink{nuweb56}{, 56}.
\item \verb@PROP@: \underline{\NWlink{nuweb21}{21}}\NWlink{nuweb36}{, 36}\NWlink{nuweb53}{, 53}.
\item \verb@props@: \NWlink{nuweb48}{48}, \underline{\NWlink{nuweb54}{54}}\NWlink{nuweb57}{, 57}\NWlink{nuweb59}{, 59}.
\item \verb@qGauge`i'@: (\underline{\NWlink{nuweb19}{19})}.
\item \verb@quarks@: \underline{\NWlink{nuweb3}{3}}\NWlink{nuweb9}{, 9}.
\item \verb@recfind@: \NWlink{nuweb50}{50}\NWlink{nuweb61}{, 61}, \underline{\NWlink{nuweb75}{75}}.
\item \verb@recfind1@: \NWlink{nuweb75}{75}, \underline{\NWlink{nuweb76}{76}}.
\item \verb@RemoveMetricTensors@: \NWlink{nuweb53}{53}\NWlink{nuweb61}{, 61}, \underline{\NWlink{nuweb74}{74}}.
\item \verb@SNULL@: \underline{\NWlink{nuweb21}{21}}\NWlink{nuweb77}{, 77}.
\item \verb@stripcoeff@: \NWlink{nuweb2}{2}\NWlink{nuweb5}{, 5}, \underline{\NWlink{nuweb7}{7}}\NWlink{nuweb12}{, 12}.
\item \verb@struct*@: (\underline{\NWlink{nuweb32}{32})}.
\item \verb@sunsimplify@: \NWlink{nuweb53}{53}\NWlink{nuweb61}{, 61}, \underline{\NWlink{nuweb87}{87}}.
\item \verb@T@: \NWlink{nuweb2}{2}, \underline{\NWlink{nuweb17}{17}}\NWlink{nuweb33}{, 33}\NWlink{nuweb74}{, 74}\NWlink{nuweb83}{, 83}\NWlink{nuweb84}{, 84}\NWlink{nuweb85}{, 85}\NWlink{nuweb86}{, 86}.
\item \verb@t@: \underline{\NWlink{nuweb3}{3}}\NWlink{nuweb11}{, 11}\NWlink{nuweb12}{, 12}.
\item \verb@TEMP*@: (\underline{\NWlink{nuweb16}{16})}.
\item \verb@TEMPHeads@: \underline{\NWlink{nuweb69}{69}}\NWlink{nuweb70}{, 70}.
\item \verb@TEMPLegs@: \underline{\NWlink{nuweb67}{67}}\NWlink{nuweb68}{, 68}\NWlink{nuweb69}{, 69}\NWlink{nuweb70}{, 70}.
\item \verb@TEMPTails@: \underline{\NWlink{nuweb69}{69}}\NWlink{nuweb70}{, 70}.
\item \verb@THEORY@: \underline{\NWlink{nuweb24}{24}}.
\item \verb@TI@: \underline{\NWlink{nuweb21}{21}}\NWlink{nuweb37}{, 37}.
\item \verb@TopologyInfo@: \NWlink{nuweb28}{28}\NWlink{nuweb61}{, 61}, \underline{\NWlink{nuweb62}{62}}.
\item \verb@TR@: \underline{\NWlink{nuweb17}{17}}\NWlink{nuweb33}{, 33}\NWlink{nuweb34}{, 34}\NWlink{nuweb35}{, 35}\NWlink{nuweb51}{, 51}\NWlink{nuweb84}{, 84}\NWlink{nuweb85}{, 85}.
\item \verb@tr@: \underline{\NWlink{nuweb3}{3}}\NWlink{nuweb12}{, 12}\NWlink{nuweb16}{, 16}\NWlink{nuweb45}{, 45}.
\item \verb@tr*@: (\underline{\NWlink{nuweb16}{16})}.
\item \verb@vec*@: (\underline{\NWlink{nuweb16}{16})}.
\end{list}}

%% file: imp-checks.tex
In the previous two sections I have discussed two methods
which help to write correct programs: A good documentation
as provided through Literate Programming helps to structure
the program and allows the reader to understand the program
bit by bit. Programming by Contract if implemented thoroughly
leads to clear interfaces between the components of a program
and very often also helps to debug the single components.

Another technique which originates in \emph{Extreme Programming}
is the idea of\index{unit test|see{test-driven development}}
\index{test-driven development}
\emph{Test-driven Development}~\cite{Beck:1994,Beck:2002}.
The programmer writes a test-case for every class or module.
In the original model\footnote{The concept was first discussed
for the language Smalltalk.} a test-case is a class with
at least three methods: \texttt{setUp()} generates the
data for the test-case to act on, \texttt{run()} runs the
test on the generated data and \texttt{tearDown()}
deallocates any resources held by the test-class.
This model has been adapted e.g. by the \texttt{JUnit}~\cite{Beck:JUnit}
testing framework for the programming language
\texttt{Java}~\cite{Gosling:Java} and similarly by the
module \texttt{unittest}~\cite{Purcell:unittest}
in \texttt{Python}. A slightly
different approach has been implemented by the \texttt{doctest}
module~\cite{Lang:doctest} which allows the programmer to place
simple tests directly
inside the documentation of a \texttt{Python} program; these
tests at the same time serve as examples about the usage of the
corresponding function, method or class. As an example we have
another look at the function \texttt{sort} that has been used
in the previous section about Programming by Contract.
\begin{Verbatim}[commandchars=\\\{\},numbers=left,frame=lines,framerule=2pt]
{\color{blue}def} sort(a):
    """Sort a list.

    pre: {\color{blue}isinstance}(a, {\color{blue}type}({\color{blue}list}))
    post[a]:
        # array size is unchanged
        {\color{blue}len}(a) == {\color{blue}len}(\_\_old\_\_.a)
        \ldots

    examples:
        {\color{red}>>> a = [5, 1, 3, 4, 2]}
        {\color{red}>>> sort(a)}
	{\color{red}>>>} {\color{blue}print} {\color{red}a}
	{\color{red}[1, 2, 3, 4, 5]}
    """
    \ldots
{\color{blue}import} {\color{red}doctest}
{\color{red}doctest.testmod()}
\end{Verbatim}
Lines 11--14 contain the test for the function \texttt{sort};
commands for tests
are marked as documentation lines that start with the
characters~\texttt{>>>} and are terminated by a blank line or the
end of the documentation string. The tests are checked by printing
out values which then are compared textually to the expected
output (line~14). For more involved tests the \texttt{unittest}
module is recommended where the setup of the data for the test
is separated from the test itself.

Although these techniques and concepts already detect many of
the possible errors there are also cases where even more checks
are needed. Especially for complicated amplitude calculations
in particle physics an established method is the implementation
of an alternative, redundant computation of the amplitude by
a second programmer. It is important that no untested code
is shared between the programmers, and even for trusted parts
of the code and third party contributions it is helpful to have
alternative implementations. For the
$u\bar{u}\rightarrow b\bar{b}s\bar{s}$
amplitude we produced two independent implementations that
also differed in the reduction method.
Table~\ref{tab:imp-checks:comparison} summarises the differences
between the two implementations. The numerical values of each
\person{Feynman} diagram for different phase space points have
been compiled by a script for all helicities and colour structures
and provided a regular and automated test tool during the code
development.
\begin{table}[hbtp]
\begin{center}
\begin{tabular}{l|ll}
& Implementation A & Implementation B \\
\hline
Diagram Generation & \qgraf & Mathematica/FeynArts \\
Simplification & \form & \form and \maple \\
Representation & numerical & analytic \\
   & form factors & basis integrals \\
Numerical Evaluation & \fortranXC & \maple
\end{tabular}
\end{center}
\caption{Comparison of two alternative implementations
of the $u\bar{u}\rightarrow b\bar{b}s\bar{s}$
amplitude. As many aspects as possible have been
chosen differently to ensure an effective error
detection.}
\label{tab:imp-checks:comparison}
\end{table}

%% file: imp-overview.tex
This section provides an overview over the interplay of the
components of the amplitude calculation before in the following
section the single program parts,
each of which stand for a phase in the generation of a \fortranXC code
for the efficient numerical computation of the amplitude.
\begin{figure}[hbtp]
\begin{center}
\includegraphics[width=\textwidth]{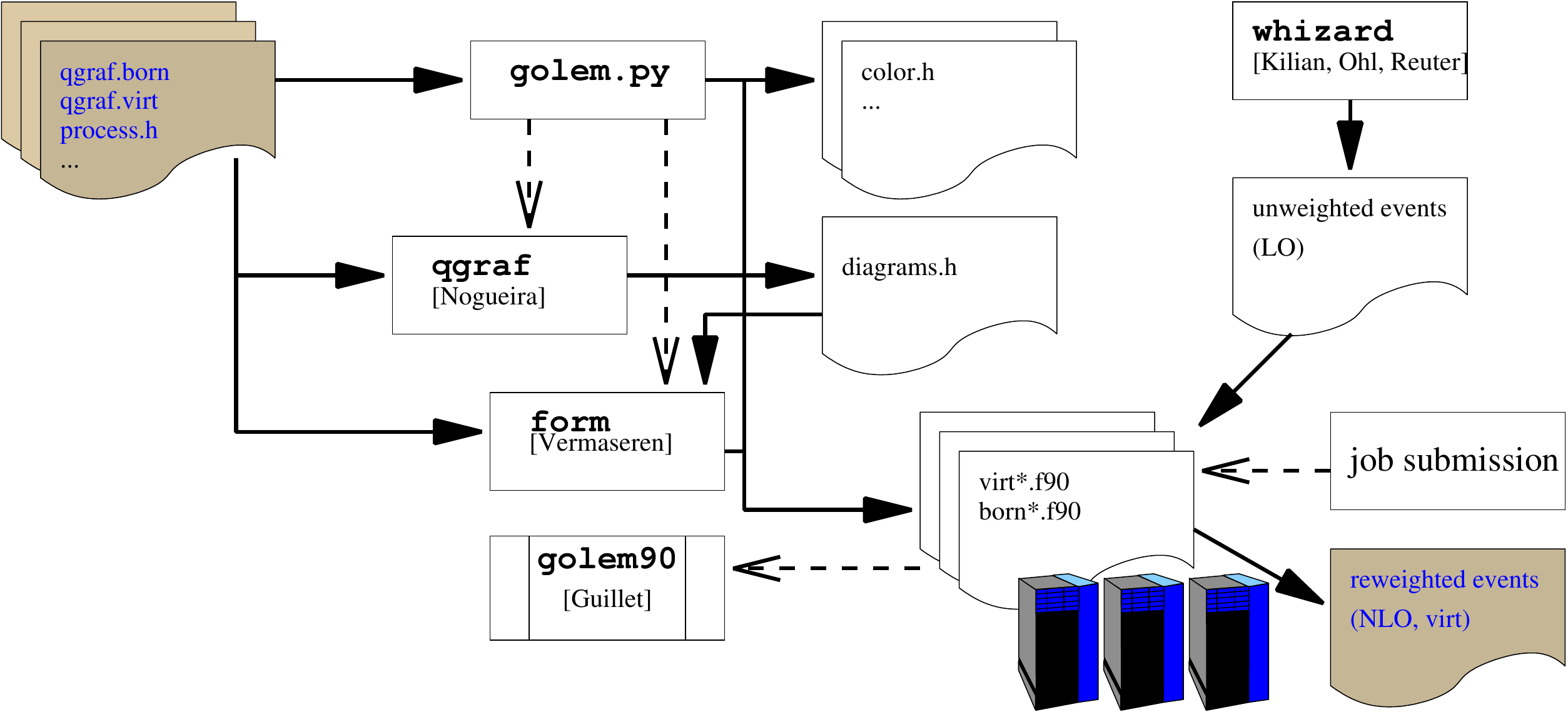}
\end{center}
\caption{Interplay between the components of the amplitude
computation in the implementation as described in this work.
Straight arrows denote data flow, dashed arrows stand for
control flow in the sense of a call graph.}
\label{fig:imp-overview:overview}
\end{figure}
Figure~\ref{fig:imp-overview:overview} shows the main components
and their interactions: the user specifies the process through
the control files. The diagram generator \qgraf reads in a
command file (\texttt{qgraf.born}, \texttt{qgraf.virt})
which has to be provided for each of the subprocesses.
A \texttt{Python} script (\texttt{golem.py}) is invoked and calls
\qgraf~\cite{Nogueira:1991ex} and
\form~\cite{Vermaseren:2000nd,Vermaseren:2002}
automatically with the correct parameters in order
to generate \fortran files for each diagram and the required helicity
projections.

The output of \qgraf{} (\texttt{diagrams.h}) contains all diagrams
as \form{} expressions and is processed by some \form{} program
(\texttt{preprocess.frm}). It generates a \fortranXC{} module for
each diagram in every helicity projection\footnote{The user specifies
which helicity projections are calculated directly and how to obtain
the remaining ones by parity transformation.}; the \person{Feynman}
diagrams are represented as a product of \person{Dirac} traces
and form factors as specified by
Equation~\eqref{eq:qcd-tensorred:formfactorrep}. A typical piece
of code would look like the following, which is the representation
of the all-plus helicity projection of the diagram shown in
Figure~\ref{fig:imp-qgraf:d167}:
\begin{Verbatim}[numbers=left,frame=lines,framerule=2pt,%
commandchars=\\\{\}]
module     virt167_63
   {\color{blue}! ... module imports ...}
   use virt_ff
   use virt_tr
   implicit none
contains
function     virt167h63(vecs) result(res)
   implicit none
   {\color{blue}! ... variable declarations ...}
   props = y12**2*y123*y56
   prefactor = 1/(braket(-k4,1,k1,-1))/(braket(-k6,1,k1,-1))/(braket(k1&
     & ,-1,-k3,1))/(braket(k1,-1,-k5,1))/(braket(k1,1,k4,-1))/(braket(&
     & k4,-1,k2,1))*g**6
   prefactor = prefactor / props
   basis1(1) = 1.0_ki/72.0_ki {\color{blue}! = dF**(-2)*TR**3}
   {\color{blue}! ... basis(2) .. basis(5) ...}
   basis1(6) =  - 1.0_ki/24.0_ki {\color{blue}! = -dF**(-1)*TR**3}
   result1 =  - 32*i_*ff48*tr4*tr16 - 32*i_*ff48*tr4*tr15 - 32*i_*ff48*&
     & tr4*tr2*tr17 - 32*i_*ff48*tr4*tr2*tr57 - 32*i_*ff48*tr4*tr56 - &
     & 32*i_*ff48*tr4*tr55 - 16*i_*ff47*tr28*tr30 - 16*i_*ff47*tr46*&
     & tr30 - 16*i_*ff47*tr6*tr28*tr1 - 16*i_*ff47*tr6*tr46*tr1 - 16*i_&
     & *ff47*tr4*tr6*tr30 - 16*i_*ff47*tr4*tr6**2*tr1 + 32*i_*eps*ff48*&
     & tr4*tr16 + 32*i_*eps*ff48*tr4*tr15 + 32*i_*eps*ff48*tr4*tr2*tr17&
     &  + 32*i_*eps*ff48*tr4*tr2*tr57 + 32*i_*eps*ff48*tr4*tr56 + 32*i_&
     & *eps*ff48*tr4*tr55
   do i = 1, 6
      res(i) = result1 * basis1(i)
      res(i) = res(i) * prefactor
   end do
end function virt167h63
end module virt167_63
\end{Verbatim}
The traces are computed once for each helicity in the module
\texttt{virt\_tr} and then recycled across all diagrams. The
form factors are the same for all helicities and are calculated
once per phase space point in the module \texttt{virt\_ff}.
The files \texttt{virt\_tr.f90}, \texttt{virt\_ff.f90} and a
interface for the summing all diagrams and producing the squared
matrix element including the \ac{ir} and \ac{uv} subtractions
are generated by the script \texttt{golem.py}.
So far everything is automated up to the point where the user
is left with a function \texttt{evaluate\_me2(vecs, alphas)}
that returns the matrix element squared for a given kinematics
and a given value of $\alpha_s$. The form factors are calculated
by the \texttt{golem90} library. The relevant piece of the
file \texttt{virt\_ff.f90} for the above diagram looks as follows
\begin{Verbatim}[numbers=left,frame=lines,framerule=2pt,%
commandchars=\\\{\}]
subroutine init_ff(vecs)
   {\color{blue}! ... other topologies ...}
   call allocation_s(6)
   call yvariables(k1,k2,k3,k4,k5,k6)
   {\color{blue}! ... initialize \(S\) ...}
   call allocate_cache(6)
   ff47 = a22(2,6,(/1,3,4,5/))
   ff1749 = a42(1,6,(/2,3/))
   ff48 = b22((/1,3,4,5/))
   ff1753 = b42((/2,3/))
   {\color{blue}! ...}
   call clear_cache()
   call deallocation_s()
   {\color{blue}! ... other topologies ...}
end subroutine init_ff
\end{Verbatim}
The \texttt{golem90} library uses a caching mechanism
because many of the form factors are calculated recursively;
lines 14 and~20 ensure that this cache is set up correctly.
The notation of the form factors is very mnemonic, e.g.
$\mathtt{a42(1,6,(/2,3/)}=A^{4,2}_{1,6}(S^{\{2,3\}})$.

The integrator that uses the matrix element is handwritten.
Automatising at that point is only of limited benefit since
the code very much depends on the observable the user wants to 
calculate.

%% file: imp-qgraf.tex
\begin{fmffile}{imp-qgraf-examples}
In the chosen approach which is based on \person{Feynman} diagrams
one of the first step in order to calculate an
amplitude at the given order in perturbation theory
is the generation of all contributing \person{Feynman} graphs.
The diagram generator \qgraf{}~\cite{Nogueira:1991ex} is a
fast and robust option which is easy to configure through
model files to determine the particle content and the interactions
of the physical model and through style files which control
the translation of the diagrams into formul\ae{} or programs.

For the calculation of the processes $u\bar{u}\rightarrow b\bar{b}s\bar{s}$
and $gg\rightarrow b\bar{b}s\bar{s}$ a model file for
\ac{sm}-\ac{qcd} has been implemented.
\begin{Verbatim}[frame=lines,framerule=2pt,%
commandchars=\\\{\},label=smqcd.model]
[ model = 'Standard Model QCD' ]
[ fmrules = 'smqcd' ]
{\color{blue}\% Propagators:}
[U, antiU,         -; PFUN='QuarkPropagator', FLAVOUR='1',
                      CHARGE=('+2/3', '-2/3'),
                      MASS=('emu', 'emu'),
                      IFUN=('u', 'vBAR'), OFUN=('uBAR', 'v')]
[D, antiD,         -; PFUN='QuarkPropagator', FLAVOUR='2',
                      CHARGE=('-1/3', '+1/3'),
                      MASS=('emd', 'emd'),
                      IFUN=('u', 'vBAR'), OFUN=('uBAR', 'v')]
\dots
[Ghost, antiGhost, -; PFUN='GhostPropagator', FLAVOUR='0',
                      CHARGE=('0', '0'),
                      MASS=('0', '0'),
                      IFUN=('ERR', 'ERR'), OFUN=('ERR', 'ERR')]
[glue, glue,       +; PFUN='GluonPropagator', FLAVOUR='0',
                      CHARGE=('0'),
                      MASS=('0'),
                      IFUN=('pol'), OFUN=('polCONJ')]
{\color{blue}\% Vertices:}
[glue, glue, glue;       VFUN='ThreeGluonVertex']
[glue, glue, glue, glue; VFUN='FourGluonVertex' ]
[antiU, U, glue;         VFUN='GluonQuarkVertex']
[antiD, D, glue;         VFUN='GluonQuarkVertex']
\dots
[antiGhost, Ghost, glue; VFUN='GluonGhostVertex']
\end{Verbatim}

Propagators have the format
\texttt{[$\langle\mathrm{field}_1\rangle$, $\langle\mathrm{field}_2\rangle$, %
$\langle\mathrm{sign}\rangle$; $\langle\mathrm{option}$, $\ldots$]}; the
$\langle\mathrm{sign}\rangle$ is the sign from the commutation relations of
that field, i.e. a minus for fermions and a plus for bosons. The options
after the semi-colon are user-defined functions. In this model file
\texttt{PFUN} has been used as the propagator function which is used in the
output, \texttt{IFUN} and \texttt{OFUN} are the names of the functions
for in- and outgoing legs. The electric charge \texttt{CHARGE} is not used
in the current implementation but one can use it to extend the model file
for the inclusion of \ac{qed} into future calculations. Similarly the
masses of the particles are provided by the field \texttt{MASS} but
are set to zero later in the calculation. Vertices have a similar
syntax: the interacting fields are given before the semi-colon and
can have a set of parameters after that. The only parameter for the vertices
is \texttt{VFUN}, the function that is used for a vertex in the output.

This model file also defines two constants \texttt{model}, which is
a description of the model, and \texttt{fmrules}. The latter is used
to include a corresponding \form{} file in the algebraic reduction that
will plug in the \person{Feynman} rules for the symbolic names given
in \texttt{PFUN}, \texttt{IFUN}, \texttt{OFUN} and \texttt{VFUN}.

The user selects a process by specifying the external particles in
a command file together with selection criteria for the diagrams.
The command file for the one-loop corrections to
$u\bar{u}\rightarrow b\bar{b}s\bar{s}$ is given below.
This file has to be called \texttt{qgraf.dat} and must reside
in the directory from which \qgraf{} is called.
\begin{Verbatim}[numbers=left,frame=lines,framerule=2pt,%
commandchars=\\\{\},label=qgraf.dat]
output = 'diagrams.h' ;
style  = 'form/form.sty' ;
model  = 'form/smqcd.model' ;

in     = U[k1], antiU[k2] ;
out    = B[k3], antiB[k4], S[k5], antiS[k6] ;
loops  = 1 ;
loop_momentum = p ;

options= onshell, notadpole ;
{\color{blue}\% no top loops:}
true = chord [ T, 0, 0 ] ;
\end{Verbatim}
Lines 1--8 are obligatory; lines 1--3 specify the output, style
and model file respectively. The parameters \texttt{in} and
\texttt{out} list the in- and out-going particles, where the
names are the $\langle\textrm{field}\rangle$ names in the model
file; in square brackets the user can add the names of the momenta
of these particles. For the one-loop correction \texttt{loops} is
set to one, for the tree-level amplitude one would have a zero in
its place. The value of the variable \texttt{loop\_momentum} prefixes
the loop momenta; as only up to one-loop corrections are considered
here the only loop momentum is~\texttt{p1}.
The \form code assumes that
the external momenta are called $\mathtt{k}_1$, \dots, $\mathtt{k}_n$,
and the loop momentum must be called \texttt{p1} (see line~9).

In the optional section of the command file, here lines~10--12,
restrictions can be applied to the diagram generation. The option
\texttt{onshell} discards all diagrams that have a self-energy
insertion on an external leg, \texttt{notadpole} suppresses the
generation of tadpole graphs; both diagram types
are zero in our renormalisation scheme for massless particles
and can be safely discarded.

Line~12 is to be understood as follows: \qgraf should only
include\footnote{exclude, if \texttt{true} was replaced by~\texttt{false}.}
diagrams which have exactly zero top-propagators running in a loop.
Since no top-quarks are in the initial or in the final state
this corresponds to excluding all top-loops.

The operator \texttt{chord[
$\langle\mathrm{field}\rangle$, 
$\langle\mathrm{min}\rangle$,
$\langle\mathrm{max}\rangle$]} tests if a diagram contains
at least $\langle\mathrm{min}\rangle$ but at most
$\langle\mathrm{max}\rangle$ propagators of a field
$\langle\mathrm{field}\rangle$ that belong to loops.
The opposite is the operator \texttt{bridge} that
tests for propagators \emph{not} belonging to loops.

\begin{figure}[hbtp]
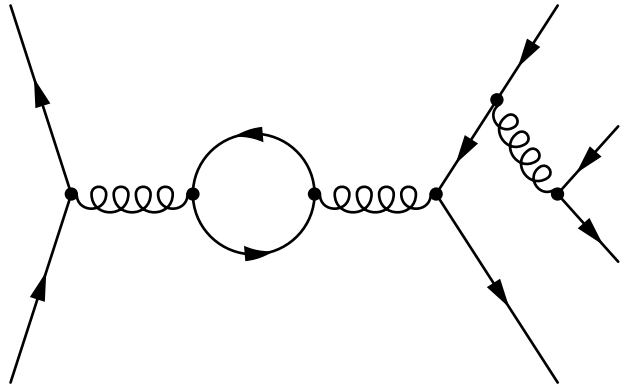

\begin{fmfchar*}(90,50)
\fmfleft{u,ubar}
\fmfright{b,s,sbar,bbar}
\fmf{fermion,label=$k_1$}{u,v1}\fmf{fermion,label=$k_2$}{v1,ubar}
\fmf{fermion,label=$k_4$}{bbar,v5}
\fmf{fermion}{v5,v4}
\fmf{fermion,tension=0.5,label=$k_3$}{v4,b}
\fmf{gluon}{v1,v2}
\fmf{gluon}{v3,v4}
\fmf{fermion,right,tension=0.5,label=$p_1$}{v2,v3}
\fmf{fermion,right,tension=0.5,label=$p_1-k_1-k_2$}{v3,v2}
\fmffreeze
\fmf{fermion,label=$k_6$,label.side=right}{sbar,v6}\fmf{fermion,label=$k_5$}{v6,s}
\fmf{phantom}{v4,v6}
\fmffreeze
\fmf{gluon}{v5,v6}
\fmfdotn{v}{6}
\fmfv{label=$v_1$}{v6}
\fmfv{label=$v_2$}{v1}
\fmfv{label=$v_3$,l.a=180}{v5}
\fmfv{label=$v_4$}{v4}
\fmfv{label=$v_5$,l.a=0}{v2}
\fmfv{label=$v_6$,l.a=180}{v3}
\end{fmfchar*}
\caption{\person{Feynman} diagram which illustrates the expression~\texttt{d167} 
described in the text. The vertices $v_i$ correspond to $\mathtt{iVERT}\langle i\rangle$
in the expression.}\label{fig:imp-qgraf:d167}
\end{figure}

The code below shows the expression which is created
for the diagram in Figure~\ref{fig:imp-qgraf:d167}.
\begin{Verbatim}[numbers=left,frame=lines,framerule=2pt,%
commandchars=\\\{\}]
{\color{blue}*---------- Diagram 167 ------------------}
{\color{blue}*--\#[ d167:}
{\color{blue}*}
Local `DIAGRAM'167 =
  - 1 *
   u(1, 1, k1, emu, i2r2) *
   vBAR(1, 2, k2, emu, i2r1) *
   uBAR(2, 1, k3, emb, i4r1) *
   v(2, 2, k4, emb, i3r2) *
   uBAR(3, 3, k5, ems, i1r1) *
   v(3, 4, k6, ems, i1r2) *
   GluonQuarkVertex(iVERT1,
      QuarkPropagator(3, iPROP\{2*6+(-6)\}, -k5, ems, i0r0, i1r1),
      QuarkPropagator(3, iPROP\{2*6+(-8)\}, -k6, ems, i0r0, i1r2),
      GluonPropagator(0, iPROP\{2*6+(1)\}, k5+k6, 0, i3r3, i1r3)) *
   GluonQuarkVertex(iVERT2,
      QuarkPropagator(1, iPROP\{2*6+(-3)\}, k2, emu, i0r0, i2r1),
      QuarkPropagator(1, iPROP\{2*6+(-1)\}, k1, emu, i0r0, i2r2),
      GluonPropagator(0, iPROP\{2*6+(2)\}, -k1-k2, 0, i5r3, i2r3)) *
   GluonQuarkVertex(iVERT3,
      QuarkPropagator(2, iPROP\{2*6+(3)\}, k4+k5+k6, emb, i4r2, i3r1),
      QuarkPropagator(2, iPROP\{2*6+(-4)\}, -k4, emb, i0r0, i3r2),
      GluonPropagator(0, iPROP\{2*6+(1)\}, -k5-k6, 0, i1r3, i3r3)) *
   GluonQuarkVertex(iVERT4,
      QuarkPropagator(2, iPROP\{2*6+(-2)\}, -k3, emb, i0r0, i4r1),
      QuarkPropagator(2, iPROP\{2*6+(3)\}, -k4-k5-k6, emb, i3r1, i4r2),
      GluonPropagator(0, iPROP\{2*6+(4)\}, k1+k2, 0, i6r3, i4r3)) *
   GluonQuarkVertex(iVERT5,
      QuarkPropagator(1, iPROP\{2*6+(6)\}, -p1, emu, i6r2, i5r1),
      QuarkPropagator(1, iPROP\{2*6+(5)\}, p1-k1-k2, emu, i6r1, i5r2),
      GluonPropagator(0, iPROP\{2*6+(2)\}, k1+k2, 0, i2r3, i5r3)) *
   GluonQuarkVertex(iVERT6,
      QuarkPropagator(1, iPROP\{2*6+(5)\}, -p1+k1+k2, emu, i5r2, i6r1),
      QuarkPropagator(1, iPROP\{2*6+(6)\}, p1, emu, i5r1, i6r2),
      GluonPropagator(0, iPROP\{2*6+(4)\}, -k1-k2, 0, i4r3, i6r3)) *
   GluonPropagator(0, iPROP\{2*6+(1)\}, k5+k6, 0, i3r3, i1r3) *
   GluonPropagator(0, iPROP\{2*6+(2)\}, -k1-k2, 0, i5r3, i2r3) *
   QuarkPropagator(2, iPROP\{2*6+(3)\}, -k4-k5-k6, emb, i3r1, i4r2) *
   GluonPropagator(0, iPROP\{2*6+(4)\}, k1+k2, 0, i6r3, i4r3) *
   QuarkPropagator(1, iPROP\{2*6+(5)\}, p1-k1-k2, emu, i6r1, i5r2) *
   QuarkPropagator(1, iPROP\{2*6+(6)\}, p1, emu, i5r1, i6r2)
;
#ifndef `LOOPS'
   #define LOOPS "1"
   #define LEGS "6"
#endif
{\color{blue}*--\#] d167:}
\end{Verbatim}

These expressions form the input of the \form{} program described
in Section~\ref{sec:imp-form}. The comments of the form
\texttt{*--\#[~d167:} and \texttt{*--\#]~d167:} have a special meaning
in \form{}: they form a so-called \emph{fold} and can be addressed
in an \texttt{\#include} statement to only include the lines
that are enclosed by these two lines. This feature is later used to
process one diagram at a time.
\end{fmffile}

%% file: imp-autocode.tex
One of the aims of this project, besides the calculation of
cross-sections for the \ac{lhc} was the automatising of
\ac{nlo} calculations in general. An important part of this
endeavour is the automatic code generation not only for the
\person{Feynman} diagrams but also for most of the auxiliary
parts of the code. Section~\ref{sec:imp-form} will show an
approach using \form{} to generate \fortranXC{} files for each
diagram. That \form{} program also generates a \texttt{Python}
file for each diagram containing information about its topology 
and all quantities which are to be managed by a global cache,
such as the form factors of the tensor integrals and the
spinor traces.

A full description of the program \texttt{golem.py} would
certainly go beyond the scope of this thesis and only be of
limited value for the reader. Therefore only selected algorithms
and concepts are presented in the following sections.
Some of the algorithms are valid for massless internal particles only.

\subsection{\person{Mandelstam} Variables}
In Section~\ref{ssec:qcd-diagramrep:mandelstam} of Chapter~\ref{chp:qcdnlo}
it has been
shown that all dot-products of two external momenta $k_i\cdot k_j$,
with $1\leq i,j \leq N$ can be expressed by a canonical set
of \person{Mandelstam} variables which correspond to
partitions of the set $\{1,\ldots,N\}$ into two subsets.
In principle this can be worked out once and for all for
each value of~$N$; it is, however, quite easy to automate
the generation of of these variables and the according
equations to translate between \person{Mandelstam} variables
and dot-products.

The function \texttt{sections} creates a list of all partitions
of a set \texttt{mom} of momenta into two subsets. The partition
is canonicalised and labelled according to the rules given in
Chapter~\ref{chp:qcdnlo}, Section~\ref{ssec:qcd-diagramrep:mandelstam}.
The function
\texttt{section\_name(i, j, n, prefix)} creates the according
names, e.g. \texttt{section\_name(2, 4, 6, "s")} would give
\texttt{s23} whereas \texttt{section\_name(4, 2, 6, "s")}
returns \texttt{s4561}.
\begin{Verbatim}[numbers=left,frame=lines,framerule=2pt,%
commandchars=\\\{\}]
def sections(mom, prefix):
   n = len(mom)
   result = []
   for i in range(1, n):
      for j in range(0, i):
         sets = [mom[j:i], mom[i:n] + mom[0:j]]
         if len(sets[0]) <= len(sets[1]):
            result.append([section_name(j, i, n, prefix), 
                  sets[0], sets[1]])
         else:
            result.append([section_name(i, j, n, prefix), 
                  sets[1], sets[0]])
   return result
\end{Verbatim}
The implementation ensures by the way the loops are nested
that in the case where both sets of the partition have equal
length the one starting with the lower index is chosen. In
the case of four external legs the following output can be
expected:
\begin{Verbatim}[commandchars=\\\{\}]
{\color{gray}>>>} lst = golem.sections(["k1", "k2", "k3", "k4"], 4, "s")
{\color{gray}>>>} print lst
[['s1', ['k1'], ['k2', 'k3', 'k4']], ['s12', ['k1', 'k2'], ['k3', 'k4']],
['s2', ['k2'], ['k3', 'k4', 'k1']], ['s4', ['k4'], ['k1', 'k2', 'k3']],
['s23', ['k2', 'k3'], ['k4', 'k1']], ['s3', ['k3'], ['k4', 'k1', 'k2']]]
\end{Verbatim}
It should be noted that the second element of each sublist contains
the range of momenta that define the name of the \person{Mandelstam} variable
and the third element contains the remaining momenta. Given this
table it is easy to produce code to compute the \person{Mandelstam}
variables numerically, as the following example shows:
\begin{Verbatim}[commandchars=\\\{\}]
{\color{gray}>>>} for line in lst:
{\color{gray}...}    print "\%s = square(\%s)" \% (line[0], "+".join(line[1]))
s1 = square(k1)
s12 = square(k1+k2)
\ldots
s3 = square(k3)
\end{Verbatim}

The reverse replacement, i.e. replacing dot products in favour
of \person{Mandelstam} variables can be achieved
using Equation~\eqref{eq:qcd-diagramrep:dot-to-mandelstam}.
A practical implementation of this equation requires some extra
care for the case $k_i\cdot k_j$ when $\vert i-j\vert\leq1$.
One can search for the according lists of momenta by the
second and third elements of each entry in \texttt{lst}
to reproduce the canonical names of the \person{Mandelstam}
variables.

\subsection{Heuristic Optimisation for Dirac Traces}
The two most time consuming parts in the calculation of
\ac{nlo} matrix elements in our case are the calculation
of the form factors and the calculation of the spinor
traces. Therefore one tries to reduce the amount of
computing time especially for these two topics.
All form factors are extracted from the amplitude
and identified if they appear in more than one diagram
such that they are calculated only once.

For the calculation of the \person{Dirac} traces
one has to choose between different approaches:
at the one end of the spectrum one could
expand out all traces to \person{Mandelstam} variables
and $\varepsilon$-tensors. This leads to huge expressions
for the \person{Feynman} diagrams and one faces technical
problems when compiling the resulting \fortran{} files.
On the other end of the spectrum one can evaluate all
traces numerically. The resulting expressions for the
\person{Feynman} diagrams become extremely compact
but one is left with a large number of different
traces that have to be evaluated. A set of identities
that reduce the number of traces without increasing
the number of terms is the following, where $\Gamma$
stands for a product of \person{Dirac} matrices:
\begin{subequations}
\begin{align}
\tr[\pm]{\Gamma\kslash[i]\kslash[i]}&\rightarrow k_i^2\tr[\pm]{\Gamma}\\
\tr[-]{\kslash\Gamma}&\rightarrow\tr[+]{\Gamma\kslash}\\
\tr[\pm]{\gamma_{a_1}^{\mu_1}\gamma_{a_2}^{\mu_2}\cdots\gamma_{a_n}^{\mu_n}}&=
\tr[\pm]{\gamma_{a_n}^{\mu_n}\cdots\gamma_{a_2}^{\mu_2}\gamma_{a_1}^{\mu_1}}\\
\tr[\pm]{\gamma_{a_1}^{\mu_1}\gamma_{a_2}^{\mu_2}\Gamma}&=
\tr[\pm]{\Gamma\gamma_{a_1}^{\mu_1}\gamma_{a_2}^{\mu_2}}
\end{align}
\end{subequations}
It is also advisable to include the rule
\begin{equation}
\tr{\Gamma}\rightarrow\tr[+]{\Gamma}+\tr[-]{\Gamma}
\end{equation}
although it doubles the number of terms in the expression.
For a lightlike vector~$k$ and an odd number of \person{Dirac}
matrices in $\Gamma^{(1)}$ one can use the additional relation
\begin{equation}
\tr[\pm]{\kslash\Gamma^{(1)}\kslash\Gamma^{(2)}}\rightarrow
\tr[\pm]{\kslash\Gamma^{(1)}}\tr[\pm]{\kslash\Gamma^{(2)}}
\text{.}
\end{equation}
Using the above set of replacements one can decrease the number
of traces that have to be computed already dramatically; in
the case of $u\bar{u}\rightarrow b\bar{b}d\bar{d}$ at \ac{nlo}
roughly 90\% of the trace calculations could be saved compared
to the case where no standardisation was applied.
In addition momentum conservation $k_N\rightarrow k_1+\ldots+k_{N-1}$
leads to an additional relation. It depends on
the process how much can be gained by this replacement for one
increases the number of terms on average by $(N-2)$ for the massless
case but at the same time one reduces the number of different traces
to be calculated.

The implementation uses external channels~\cite{Tentyukov:2006ys}
to establish a bi-directional communication between \form{}
and the \texttt{Python} program \texttt{golem.py} which
generates the according replacement rules on the fly for each
spinor trace. On the invocation of \form{} a pair of pipes
is generated using \texttt{Python}'s command \texttt{os.pipe()}.
A minimal version of a pipe communicating with form is
given below:
\begin{Verbatim}[numbers=left,frame=lines,framerule=2pt]
class FormPipe:
   def __init__(self, formfile):
      (r1, w1) = os.pipe()
      (r2, w2) = os.pipe()
      self._fds = [r1, w1, r2, w2]
      args = ["form", "-pipe", "%d,%d" % (r1, w2), formfile]
      self._proc = subprocess.Popen(args)
      self._in = r2
      self._out = w1
      self._pid = os.getpid()
      self._formPID = self.readLine().strip('\r\n')
      self.write("%s,%d\n" % (self._formPID, self._pid))
\end{Verbatim}
The last line is part of the protocol
as defined in~\cite{Tentyukov:2006ys}.

The methods for communicating with the pipe
simply act on the file descriptors \texttt{self.\_in}
and \texttt{self.\_out}. The method \texttt{readLine}
is for convenience; one has to be careful not to
read ahead because communication through pipes is
blocking and one easily creates a deadlock situation
where both processes wait for each other ad
infinitum.
\begin{Verbatim}[numbers=left,frame=lines,framerule=2pt]
def write(self, str):
   os.write(self._out, str)
   return self
def read(self, count=1):
   return os.read(self._in, count)
def close(self):
   for fd in self._fds:
      os.close(fd)
   self._fds = []
def readLine(self):
   s = os.read(self._in, 1)
   result = ""
   while len(s) == 1 and s != "\n":
      result += s
      s = os.read(self._in, 1)
   result += s
   return result
\end{Verbatim}

\subsection{Colour Correlation Matrices}
\begin{fmffile}{imp-ccorrel}
Since the treatment of the colour algebra is not
a computational issue for the processes that were
addressed in this work the implementation uses
the most simple colour basis rather than the most
efficient one: all gluons in colour space
are projected on a quark-antiquark pair
as described in Section~\ref{ssec:qcd-color:other-bases}
of Chapter~\ref{chp:repqcdamp}.
A basis is generated by all possible ways connecting
the quark with the antiquark lines. An efficient
non-recursive algorithm for generating all permutations
is the \person{Johnson}-\person{Trotter}
algorithm~\cite{Trotter,Johnson:Perm}.

The projection of the gluons on to quark
pairs as in Equation~\eqref{eq:qcd-color:fundamental-basis1},
\begin{equation}
\parbox[c][\height+1.5\baselineskip][c]{25mm}{
\begin{fmfchar}(25,10)
\fmfleft{g1}\fmfright{g2}
\fmf{gluon}{g1,g2}
\end{fmfchar}}
\rightarrow\frac{1}{\sqrt{T_R}}
\parbox[c][\height+1.5\baselineskip][c]{25mm}{
\begin{fmfchar}(25,10)
\fmfleft{p,q}\fmfright{g}
\fmf{gluon}{v,g}
\fmf{fermion,right=0.5}{p,v,q}
\fmfdot{v}
\end{fmfchar}}\text{,}
\end{equation}
also requires a change of the rules for the
insertion of a generator in the
\person{Catani}-\person{Seymour} dipole
subtraction~\cite{Catani:1996vz,Catani:2002hc}.
In the usual graphical notation the insertion
operator becomes
\begin{equation}
\parbox[c][\height+1.5\baselineskip][c]{25mm}{
\begin{fmfchar}(25,15)
\fmftop{g3}
\fmfleft{g1}\fmfright{g2}
\fmf{gluon}{g1,v,g2}
\fmfdot{v}
\fmffreeze
\fmf{gluon}{v,g3}
\end{fmfchar}}
\rightarrow\frac{1}{\sqrt{T_R}}
\parbox[c][\height+1.5\baselineskip][c]{25mm}{
\begin{fmfchar}(25,10)
\fmftop{g3}
\fmfleft{p,q}\fmfright{g}
\fmf{gluon}{v,w,g}
\fmf{fermion,right=0.4}{p,v,q}
\fmfdot{v,w}
\fmffreeze
\fmf{gluon}{w,g3}
\end{fmfchar}}
=\frac{1}{\sqrt{T_R}}\left(
\parbox[c][\height+1.5\baselineskip][c]{25mm}{
\begin{fmfchar}(25,10)
\fmftop{g3}
\fmfleft{p,q}\fmfright{g}
\fmf{gluon}{v,g}
\fmf{fermion}{p,v}
\fmf{fermion}{v,w,q}
\fmfdot{v,w}
\fmffreeze
\fmf{gluon}{g3,w}
\end{fmfchar}}
-%
\parbox[c][\height+1.5\baselineskip][c]{25mm}{
\begin{fmfchar}(25,10)
\fmfbottom{g3}
\fmfleft{p,q}\fmfright{g}
\fmf{gluon}{v,g}
\fmf{fermion}{p,w,v,q}
\fmfdot{v,w}
\fmffreeze
\fmf{gluon}{w,g3}
\end{fmfchar}}
\right)
\end{equation}

The current implementation automatises the
whole colour algebra including the generation
of the insertion operators for the infrared
regularisation\footnote{for massless partons only}
as defined in~\cite{Catani:1996vz}.
\end{fmffile}

%% file: appendix-ffactory.tex
\section{Translation of Tensor Integrals into Form Factors}
\label{sec:FormFactory.java}
This section describes a program that generates rewriting
rules for a \form{} program to translate from tensor integrals
into a form factor representation according to
Equation~\eqref{eq:qcd-tensorred:formfactorrep}.

The program assumes that tensor integrals are denoted as
\begin{equation}\label{eq:appendix-ffactory:translation01}
I_N^{d;\mu_1\mu_2\ldots\mu_R}(a_1,a_2,\ldots,a_R;S)=
\mathtt{TI}(d, N, R, r_{a_1}, \mu_1, r_{a_2}, \mu_2, \ldots, r_{a_R}, \mu_R)%
\text{.}
\end{equation}
The information about the matrix $S$ is to be kept elsewhere.
The program is written in \texttt{Java} and implemented as
a class called \texttt{FormFactory}.\footnote{All comments have
been stripped from the printed version of the program
and some of the
methods which are very similar to each other have been left out.}
The program expects three arguments which are the numbers $N$ and $R$
of Equation~\eqref{eq:appendix-ffactory:translation01} and the
name of the output file as the third argument.
\begin{Verbatim}
import java.io.*;
public class FormFactory {
   public static void main(String[] args) throws IOException {
      if(args.length == 3) {
         try {
            int legs =  Integer.parseInt(args[0]);
            int rank =  Integer.parseInt(args[1]);
            new FormFactory(legs, rank, args[2]);
         } catch(NumberFormatException ex) {
            System.err.println("Invalid numeric argument");
         }
      } else
         System.err.println("usage: java FormFactory <legs> <rank> <file name>");
   } // static method main

   // ... other methods ...

} // class FormFactory
\end{Verbatim}

The constructor serves as the main program. It opens the
output file and writes to it the left hand side
and, depending on the number of legs
and the rank of the integral,
the according terms of the right hand side: the $B$ and $C$
terms are only written if the number of legs is smaller than~6;
\begin{Verbatim}
public FormFactory(int legs, int rank, String filename) throws IOException {
   FileOutputStream theFile = new FileOutputStream(filename);
   PrintStream out = new PrintStream(theFile);

   this.generateLHS(legs, rank, out);
   this.generateFormFactorA(legs, rank, out);
   if(legs < 6) {
      if(rank >= 2)
         this.generateFormFactorB(legs, rank, out);
      if(rank >= 4)
         this.generateFormFactorC(legs, rank, out);
   }
   out.println(";");
   theFile.close();
}
\end{Verbatim}

The generation of the left hand side of the replacement
uses the triple-dot (``\texttt{...}'') operator of the \form{} preprocessor
rather than expanding out all arguments explicitly. The
angle brackets hereby ensure that both arguments, \texttt{q1}
and \texttt{i1} are incremented simultaneously.
\begin{Verbatim}
protected void generateLHS(int legs, int rank, PrintStream out) {
   out.print("id TI(" + legs + ", " + rank);
   switch(rank) {
   case 0: break;
   case 1: out.print(", q1?,i1?"); break;
   default:
      out.print(", <q1?,i1?>, ..., <q" + rank + "?,i" + rank + "?>");
   }
   out.println(") =");
}
\end{Verbatim}

The method \texttt{complementSet} computes the set theoretic
complement of a subset of $\{1,\ldots,\mathtt{n}\}$. The
routine assumes that the parameter \texttt{set} is already
in increasing order.
\begin{Verbatim}
protected static int[] complementSet(int[] set, int n) {
   int i = 0;
   int size = n - set.length;
   int[] result = new int[size];
   for(int k = 1; k <= n; ++k) {
      if((i < set.length) && (set[i] == k)) ++i;
      else result[k - i - 1] = k;
   }
   return result;
}
\end{Verbatim}

Another combinatorial algorithm that is required for the
generation of the translation formula enumerats all subsets
of the set $\{1,\ldots,\mathtt{n}\}$ that have $m\leq n$ elements.
The method \texttt{nextSelection} enumerates these subsets: the
first time the method has to be invoked with the array
\texttt{\{1, 2, \textrm{\dots}, m\}}, after that the method
must be called with the previous value of \texttt{set};
the next subset in the sequence is written in-place
to the argument \texttt{set}. If no more sets can be found the
method returns \texttt{false}.
\begin{Verbatim}
protected static boolean nextSelection(int[] set, int n) {
   int m = set.length;
   
   for(int i = m - 1; i >= 0; --i) {
      if(set[i] <= n - (m - i)) {
         for(int j = m - 1; j >= i; --j) {
            set[j] = set[i] + 1 + (j - i);
         } // for
         return true;
      } // if
   } // for
   return false;
}
\end{Verbatim}

All elements of the set $\{1,\ldots,\mathtt{n}\}^r$ are generated
by the method \texttt{nextCombination}; the calling conventions
are similar to the previous method. For the first call the
argument \texttt{lst} mut be initialized with
\texttt{\{1, 1, \textrm{\dots}, 1\}}.
\begin{Verbatim}
protected static boolean nextCombination(int[] lst, int n) {
	int m = lst.length;
	++ lst[m - 1];
	for(int i = m - 1; i >= 0; --i) {
		if(lst[i] > n) {
			lst[i] = 1; if(i > 0) ++lst[i - 1];
		}
		else return true;
	} // for
	return false;
}
\end{Verbatim}
Before the generation of the actual terms in the rewriting rule
is discussed two utility functions are introduced: the first
one, \texttt{printSymmetricTensor} prints the expression
$g^{i_1i_2}g^{i_3i_4}+g^{i_1i_3}g^{i_2i_4}+g^{i_1i_4}g^{i_2i_3}$
for a given set of indices $\{i_1,i_2,i_3,i_4\}$. The second
method, \texttt{printDelta} generate the \form{} equivalent
of the vector $\Delta_{ij}^{\mu_j}$.
\begin{Verbatim}
private void printSymmetricTensor(int[] G, PrintStream out) {
   out.print("(");
   out.print("gTensor(n,i" + G[0] + ",i" + G[1] + ")"); out.print("*");
   out.print("gTensor(n,i" + G[2] + ",i" + G[3] + ")"); out.print(" + ");
   out.print("gTensor(n,i" + G[0] + ",i" + G[2] + ")"); out.print("*");
   out.print("gTensor(n,i" + G[1] + ",i" + G[3] + ")"); out.print(" + ");
   out.print("gTensor(n,i" + G[0] + ",i" + G[3] + ")"); out.print("*");
   out.print("gTensor(n,i" + G[1] + ",i" + G[2] + ")");
   out.print(")");
}
private void printDelta(int i, int j, PrintStream out) {
   out.print("DELTA(r" + i + ",q" + j + ",i" + j + ")");
}
\end{Verbatim}

The form factors $A^{N,r}_{j_1,\ldots,j_r}$ occur under a
multiple sum over all $j_i$ with a coefficient
$[\Delta_{j_1\bul}^{\bul}\cdots\Delta_{j_r\bul}^{\bul}%
]_{a_1,\ldots,a_r}^{\mu_1\ldots\mu_r}$. To generate all
terms an array $\mathtt{J[]}=\{j_1,\ldots,j_r\}$
is generated for each possible combination of values
for the $j_i$ using the method \texttt{nextCombination}.
Only a single term is generated for the case~$r=0$, where
the tensor in front of the form factor is~$1$.
\begin{Verbatim}
protected void generateFormFactorA(int legs, int rank, PrintStream out) {
   if(rank > 0) {
      int[] J = new int[rank];
      for(int j = 0; j < rank; ++j) J[j] = 1;
      do {
         out.print("   + ");
         for(int j = 0; j < rank; ++j) {
            this.printDelta(J[j], j + 1, out); out.print(" * ");
         } // for
         out.print("a" + legs + "" + rank + "(");
         for(int j = 0; j < rank; ++j)
            out.print(Integer.toString(J[j]) + ",");
         out.println("`SNULL')");
      } while(nextCombination(J, legs));
   } else {
      out.print("   + ");
      out.println("a" + legs + "" + rank + "(`SNULL')");
   } // if
}
\end{Verbatim}

The form factor $B^{N,r}_{j_1,\ldots,j_{r-2}}$ has to take
into account the symmetrization over the additional $g^{\bul\bul}$.
The array \texttt{G} is filled with all possible selections of two
of the indices $\mu_1,\ldots,\mu_r$ using the method \texttt{nextSelection}.
The remaining indices are selected by the method \texttt{complementSet}
and distributed in the same manner as for the form factor~$A$.
\begin{Verbatim}
protected void generateFormFactorB(int legs, int rank, PrintStream out) {
   if(rank > 2) {
      int[] J = new int[rank - 2];
      int[] G = new int[2];
      for(int j = 0; j < rank - 2; ++j) J[j] = 1;
      for(int j = 0; j < 2; ++j) G[j] = j + 1;
      do {
         int[] C = complementSet(G, legs);
         do {
            out.print("   + ");
            out.print("gTensor(n,i" + G[0] + ",i" + G[1] + ")");
            out.print(" * ");
            for(int j = 0; j < rank - 2; ++j) {
               this.printDelta(J[j], C[j], out); out.print(" * ");
            } // for
            out.print("b" + legs + "" + rank + "(");
            for(int j = 0; j < rank - 2; ++j)
               out.print(Integer.toString(J[j]) + ",");
            out.println("`SNULL')");
         } while(nextCombination(J, legs));
      } while(nextSelection(G, rank));
   } else {
      out.print("   + "); out.print("gTensor(n, i1, i2) * ");
      out.println("b" + legs + "" + rank + "(`SNULL')");
   } // if
}
\end{Verbatim}

The overall structure of the method \texttt{generateFormFactorC}
is the same as the previous ones. The additional symmetrisation
over the two metric tensors $g^{\bul\bul}g^{\bul\bul}$ is done
explicitly by the method \texttt{printSymmetricTensor}.
\begin{Verbatim}
protected void generateFormFactorC(int legs, int rank, PrintStream out) {
   int[] G = new int[4];
   for(int j = 0; j < 4; ++j) G[j] = j + 1;
   if(rank > 4) {
      int[] J = new int[rank - 4];
      for(int j = 0; j < rank - 4; ++j) J[j] = 1;
      do {
         int[] C = complementSet(G, legs);
         do {
            out.print("   + ");
            this.printSymmetricTensor(G, out);
            out.print(" * ");
            for(int j = 0; j < rank - 4; ++j) {
               this.printDelta(J[j], C[j], out); out.print(" * ");
            } // for
            out.print("c" + legs + "" + rank + "(");
            for(int j = 0; j < rank - 4; ++j)
               out.print(Integer.toString(J[j]) + ",");
            out.println("`SNULL')");
         } while(nextCombination(J, legs));
      } while(nextSelection(G, rank));
   } else {
      out.print("   + "); this.printSymmetricTensor(G, out);
      out.print(" * "); out.println("c" + legs + "" + rank + "(`SNULL')");
   } // if
}
\end{Verbatim}

This concludes the program; it should be straight forward to modify
the program in order to suite different requirements if one uses another
computer algebra program or prefers another programming language. One of
the reasons for adding this program in the appendix is to give an unambiguous
specification of the meaning of the
brackets~$[\ldots]^{\mu_1\ldots\mu_r}_{a_1\ldots a_r}$.

%% file: imp-fortran.tex
\subsection{Introduction}
The numerical integration of a cross-section can be computationally
very expensive when many particles are in the final state and the
amplitude consists of millions of terms. The previous sections
described how \fortranXC{}~\cite{Fortran90standard}
code for the expression for an amplitude
can be generated automatically at the one-loop level in a way that
spurious calculations are kept to a minimum.

In this approach most steps are carried out numerically, and algebraic
simplifications are done where necessary and where the simplification
is immediate; lengthy simplifications where cancellations only happen
in the very end have been avoided since the success of such calculations
very often depends on human intervention at many steps and automation
is only possible to a limited degree; conversely, a fully analytic
calculation can lead to more compact expressions and to faster code.

In the rest of this section the calculation of the matrix element is
considered only as far as the evaluation of the form factors goes.
Apart from that it is considered as a black box that has been explained
in the previous sections, and for the following discussion it is enough
to assume the existence of the function \texttt{evaluate\_me2(vecs, alphas)}
that returns the squared matrix element including \ac{ir} and \ac{uv}
counterterms and therefore returns a finite, real result.

Section~\ref{ssec:imp-fortran:overview} gives an overview over the
structure of a phase space integrator that uses the reweighting
procedure as described in Equations
\eqref{eq:qcd-psintegral:reweighting01}
and~\eqref{eq:qcd-psintegral:reweighting01}.
Although \fortran{} is not famous for object-oriented design,
Section~\ref{ssec:imp-fortran:oop90} shows for the example of the
form-factor type how \fortranXC{} provides object-oriented principles,
which have proved useful in the context of a direct translation from
expression obtained from computer algebra programs into \fortran{} code.
Section~\ref{ssec:imp-fortran:golem90} introduces the \texttt{Golem90}
library in detail with the focus on the calculation of the
form factors. A short review of parallelisation methods and the
description of \ac{ecdf}, the cluster on which the calculations have
been performed conclude this chapter.

\subsection{Overview}
\label{ssec:imp-fortran:overview}
In Section~\ref{ssec:qcd-psintegral:nlo} of Chapter~\ref{chp:qcdnlo}
I have introduce a method for integrating the \ac{nlo} corrections
of a process over a phase space.~\cite{Binoth:2008gx}

This method has been implemented by
reading unweighted \ac{lo} events that have been created using
\texttt{Whizard}~\cite{Kilian:2007gr,Moretti:2001zz,Whizard1.0}
and are stored in the \texttt{HepEvent} format~\cite{Bambah:1989ab};
to each of the events is attached the local $K$-factor as defined
in Equation~\eqref{eq:qcd-psintegral:reweighting02}, which defines
our reweighting procedure. Apart from
the additional entry the output is still in \texttt{HepEvent}
format to allow for an easy interchange with other packages;
In the current version the analysis of the reweighted events
is done with custom programmes.

\subsection{Object-Oriented Design in \fortranXC}
\label{ssec:imp-fortran:oop90}
Although \fortran{} is not renowned for its object-oriented
capabilities it does support many of the language constructs
one would expect from an object-oriented programming language.
The authors of~\cite{Polymorphism90} argue that
\textit{``Fortran 90 clearly has some language features
which are useful for \acs{oop} (derived types, modules, generic interfaces),
but clearly lacks some others (inheritance, run-time polymorphism)''}
and give recipes how to emulate the latter features; thereby they
extend the earlier work of~\cite{Dupee:F90-OOP}.

For the current implementation of the \texttt{Golem90} library
certain object-oriented concepts are applied in representation
of the form factors. The expansion of an arbitrary one-loop integral
in dimensional regularisation starts at $\varepsilon^{-2}$, where
the dimension is $d=4-2\varepsilon$. Neglecting all positive powers
of $\varepsilon$ the integral can therefore be represented by three
coefficients which are independent of~$\varepsilon$,
\begin{equation}
I(\{p_i\})=\frac{A(\{p_i\})}{\varepsilon^2}
+\frac{B(\{p_i\})}{\varepsilon}+C(\{p_i\})
+{\mathcal O}(\varepsilon)\text{.}
\end{equation}
This expansion can be directly mapped onto a derived type in \fortranXC:
\begin{Verbatim}[frame=lines,framerule=2pt,%
commandchars=\\\{\},label=type form\_factor]
type form_factor
   complex(ki) :: a, b, c
end type form_factor
\end{Verbatim}
The parameter \texttt{ki} specifies the precision and is defined
in another module. The interface to the routines implementing
the form factors, e.g. the routine for the
form factor~$A^{3,2}_{j_1j_2}(S)$ is declared below.
\begin{Verbatim}[frame=lines,framerule=2pt,%
commandchars=\\\{\},label=function a32]
interface
   function     a32(j1,j2,S)
      integer, intent (in) :: j1,j2
      integer, intent (in), dimension(:) :: S
      type(form_factor) :: a32
   end function a32
end interface
\end{Verbatim}
The argument~\texttt{S} is a list of pinches, i.e. the form factor
$A^{3,2}_{j_1,j_2}(S^{1,2})$ translates to the function call
\texttt{a32(j1,j2,(/1,2/))}. Up to this point the only advantage
of a derived type over an array of complex numbers is the improved
type safety. The main justification of using derived types in this
example is the possibility of overloading operators. The reason is
that the \fortran{} code is generated from a computer algebra program
which should translate expressions like
\begin{displaymath}
2A^{3,2}_{4,3}(S^{\{1,2\}})-3\varepsilon B^{3,2}(S^{\{1,2\}})
\end{displaymath}
as straight-forward as possible both to keep the computer algebra code
simple and to maintain the readability of the output; we aim for
an output like \texttt{2*ff1-3*eps*ff2}, where \texttt{ff1} and
\texttt{ff2} are defined as the above form factors
\texttt{a32(\hspace{0em}4,\hspace{0em}3,\hspace{0em}(/1,\hspace{0em}2/))}
and \texttt{b32(\hspace{0em}(/1,{\hspace{0em}}2/))} respectively.
The symbol \texttt{eps} is of
the type \texttt{epsilon\_type} which is defined below and represents
positive powers of~$\varepsilon$ with an optional complex coefficient.
\begin{Verbatim}[frame=lines,framerule=2pt,%
commandchars=\\\{\},label=type epsilon\_type]
type epsilon_type
   complex(ki) :: coefficient
   integer :: power
end type epsilon_type
\end{Verbatim}

In order to allow the above expression in \fortranXC{} one has to overload
the according operators, which below is shown in two examples. For a fully
working implementation all possible combination of operators and types
that may appear in an expression have to be defined. One basic operation
that has to be defined for the form factors is the addition of two
form factors. In an \texttt{interface} declaration the operator~$+$
is overloaded by the function \texttt{add\_ff\_ff} which is defined
later in the module.
\begin{Verbatim}[frame=lines,framerule=2pt,%
commandchars=\\\{\},label=type function add\_ff\_ff]
interface operator(+)
   module procedure add_ff_ff
end interface
\ldots
pure function add_ff_ff(ff1, ff2) result(r)
	type(form_factor), intent(in) :: ff1, ff2
	type(form_factor) :: r
	r%a = ff1%a + ff2%a
	r%b = ff1%b + ff2%b
	r%c = ff1%c + ff2%c
end function add_ff_ff
\end{Verbatim}

As a second example the multiplication of $\varepsilon^n$
with a form factor is implemented.
\begin{Verbatim}[frame=lines,framerule=2pt,%
commandchars=\\\{\},label=type function mul\_ff\_eps]
interface operator(*)
   module procedure mul_ff_eps
end interface
\ldots
pure function mul_ff_eps(ff, x) result(r)
   type(epsilon_type), intent(in) :: x
   type(form_factor), intent(in) :: ff
   type(form_factor) :: r
   if (x%power >= 3) then
      r = 0.0_ki
   elseif (X%power == 2) then
      r%a = 0.0_ki 
      r%b = 0.0_ki 
      r%c = x%coefficient * r%a
   else
      r%a = 0.0_ki
      r%b = x%coefficient * ff%a
      r%c = x%coefficient * ff%b
   end if
end function mul_ff_eps
\end{Verbatim}

It should be noted that the above setup cannot evaluate an expression
like $(\varepsilon+2\varepsilon^2)B^{3,2}(S)$ although it is
mathematically equivalent to its expanded form, unless one overloads
the operator $+$ for the signature
(\texttt{epsilon\_type}, \texttt{epsilon\_type}). Therefore care has
to be taken when the expression is generated by the computer algebra
programme.

\subsection{Tensor-Integrals in \texttt{Golem90}}
\label{ssec:imp-fortran:golem90}
\texttt{Golem90}~\cite{Golem90:2008inprep} is a \fortranXC{} library
that defines an interface to the form factors that are defined
through Equation~\eqref{eq:qcd-tensorred:formfactorrep}. It maps
the form factors
$A^{N,r}_{j_1,\ldots,j_r}(S)$,
$B^{N,r}_{j_1,\ldots,j_{r-2}}(S)$ and
$C^{N,r}_{j_1,\ldots,j_{r-4}}(S)$ onto functions
\texttt{a}$\langle N\rangle\langle r\rangle$\texttt{($j_1$,\dots $j_r$)},
\texttt{b}$\langle N\rangle\langle r\rangle$\texttt{($j_1$,\dots $j_{r-2}$)}
and
\texttt{c}$\langle N\rangle\langle r\rangle$\texttt{($j_1$,\dots $j_{r-4}$)}
respectively. It should be noted, that the form factors $B^{6,r}$
and $C^{6,r}$ are not implemented. Instead, they are absorbed through
relation~\eqref{eq:qcd-tensorred:completeness}.

As explained in Sections \ref{sec:scalarred} and~\ref{sec:tensorred} of
Chapter~\ref{chp:qcdnlo} it is always possible to reduce the tensor integrals
down to an extended integral basis~$\mathcal{I}_\mathcal{N}$ as defined
in Equation~\eqref{eq:qcd-diagramrep:IN} without introducing inverse
\person{Gram} determinants. This basis contains integrals with non-trivial
numerators; their evaluation allows for different methods.
The phase space can be separated by a parameter $\Lambda$ which
discriminates
\begin{align*}
\text{the bulk of the phase space}&\;\text{where}\;
\frac{\det{G}}{\det{S}}>\Lambda\\
\text{from the critical phase space}&\;\text{where}\;
\frac{\det{G}}{\det{S}}<\Lambda\text{.}
\end{align*}
In the bulk region one can use 
the equations given in Chapter~\ref{chp:qcdnlo},
Section~\ref{ssec:qcd-tensorred-ibp:threepoint}
and similar relations for the four-point functions
in order to reduce the integral set~$\mathcal{I}_\mathcal{N}$
to the purely scalar integral set~$\mathcal{I}_\mathcal{S}$.
In the critical phase space an analytic reduction would
introduce numerical instabilities through inverse \person{Gram}
determinants. For many of the three-point functions integral-free
representations can be found analytically without a reduction
in the above sense. The remaining integrals
in the \texttt{Golem90} library are solved by numerical integration.

\subsection{Numerical Integration of the Basis Functions}
As an example we shall consider the box-integrals in $n+2$ and $n+4$
dimensions,
\begin{align}\label{eq:imp-fortran:box6}
I_4^{n+2}&=\Gamma(1+\varepsilon)\int_0\!\!\diff[4]z\,\delta_z%
\frac{z_1^{\alpha_1}z_2^{\alpha_2}z_3^{\alpha_3}z_4^{\alpha_4}}{%
\left(-\frac12z^\transposed Sz-i\delta\right)^{1+\varepsilon}}%
\quad\text{and}\\
\label{eq:imp-fortran:box8}
I_4^{n+4}&=\frac{\Gamma(1+\varepsilon)}{\varepsilon}%
\int_0\!\!\diff[4]z\,\delta_z%
\frac{z_1^{\alpha_1}z_2^{\alpha_2}z_3^{\alpha_3}z_4^{\alpha_4}}{%
\left(-\frac12z^\transposed Sz-i\delta\right)^{\varepsilon}}\text{.}
\end{align}
As all propagators and $k_1$ are assumed to be massless,
the matrix $S$ reads
\begin{equation}
S=\left(\begin{matrix}
0&s_2&s_{23}&0\\
s_2&0&s_3&s_{12}\\
s_{23}&s_3&0&s_4\\
0&s_{12}&s_4&0
\end{matrix}\right)\text{.}
\end{equation}
We are only interested in the $\varepsilon$ expansion of the
integrals and hence can write them as
\begin{align}
I_4^{n+2}&=\int_0\!\!\diff[4]z\,\delta_z%
\frac{z_1^{\alpha_1}z_2^{\alpha_2}z_3^{\alpha_3}z_4^{\alpha_4}}{%
-\frac12z^{\transposed}Sz-i\delta}%
+{\mathcal O}(\varepsilon)\quad\text{and}\\
I_4^{n+4}&=\frac{\Gamma(1+\varepsilon)C}{\varepsilon}
-\int_0\!\!\diff[4]z\,\delta_z%
z_1^{\alpha_1}z_2^{\alpha_2}z_3^{\alpha_3}z_4^{\alpha_4}
\ln\left(-\frac12z^\transposed Sz-i\delta\right)+{\mathcal O}(\varepsilon)%
\text{,}
\end{align}
where the constant $C$ can be derived from
Equations~\eqref{eq:qcd-dimreg:fppoly01}
and~\eqref{eq:qcd-dimreg:solvePolyloop},
\begin{equation}
C=P_{\alpha_1,\alpha_2,\alpha_3,\alpha_4}=
\frac{\alpha_1!\alpha_2!\alpha_3!\alpha_4!}{%
(3+\alpha_1+\alpha_2+\alpha_3+\alpha_4)!}\text{.}
\end{equation}
In the remaining integrals the $\delta$-function can be eliminated
by the transformation,
\begin{equation}
z=\left(\begin{matrix}z_1\\z_2\\z_3\\z_4\end{matrix}\right)
=w\cdot\left(\begin{matrix}(1-z)\\xyz\\xy(1-z)\\x(1-y)\end{matrix}\right)
\end{equation}
that leads to the \person{Jacobi}an factor ${\mathcal J}=-w^3x^2y$;
the $\delta$ function becomes $\delta_z=\delta(1-w)$.
In terms of the new variables the denominators of
Equation~\eqref{eq:imp-fortran:box6}
and~\eqref{eq:imp-fortran:box8} factorise into
\begin{equation}
-\frac12z^{\transposed}Sz=-xy\left[a_1xy+a_2 x+a_3\right]\text{,}
\end{equation}
with $a_1$, $a_2$ and $a_3$ being quadratic polynomials in $z$.
The integrals over $x$ and $y$ can be worked out analytically, in
general leading to an integral like~\cite{Golem90:2008inprep}
\begin{equation}
I=\int_0^1\!\!\diff{z}\frac{f(z)\ln[f(z)]-g(z)\ln[g(z)]}{h(z)[f(z)-g(z)]}
\end{equation}
with $f$, $g$ and~$h$ being polynomials in~$z$.
This remaining integral is solved numerically by the
\person{Gauss}-\person{Kronrod}
quadrature~\cite{Kronrod:1964,Patterson:1968} after applying the
contour deformation $z=u\pm u(1-u)$, which avoids crossing
the cut of the logarithms in~$I$.~\cite{Golem90:2008inprep}

\subsection{Distributed and Parallel Computing}
The problem of a Monte Carlo integration is very often described
as \emph{embarrassingly parallel}. This essentially means that
the problem can be split into subprocesses which require no or
virtually no communication. Such problems are very well suited
for computation on clusters and have no need for specialised
hardware. On the other hand the evaluation of each helicity
amplitude can be split into the computation of single \person{Feynman}
diagrams. In this case the diagrams share a pool of data, i.e.
form factors and spinor traces which have to be computed once
per helicity amplitude, but do not modify any of the data.
In this case it would be natural and desirable to distribute
the amplitude calculation across multiple processors with shared memory.
To achieve this second kind of parallelisation one can take advantage
of nodes equipped with multiple cores.

\subsubsection{Message Passing with \acs{mpi}} A Monte Carlo integrator
for our purpose can be split into two parts,
a stochastic source of phase space points and an integrand function.
In \ac{nlo} calculations typically the generation of the phase space
points is relatively cheap compared to the evaluation of the integrand.

\begin{figure}[hbtp]
\includegraphics[scale=0.5]{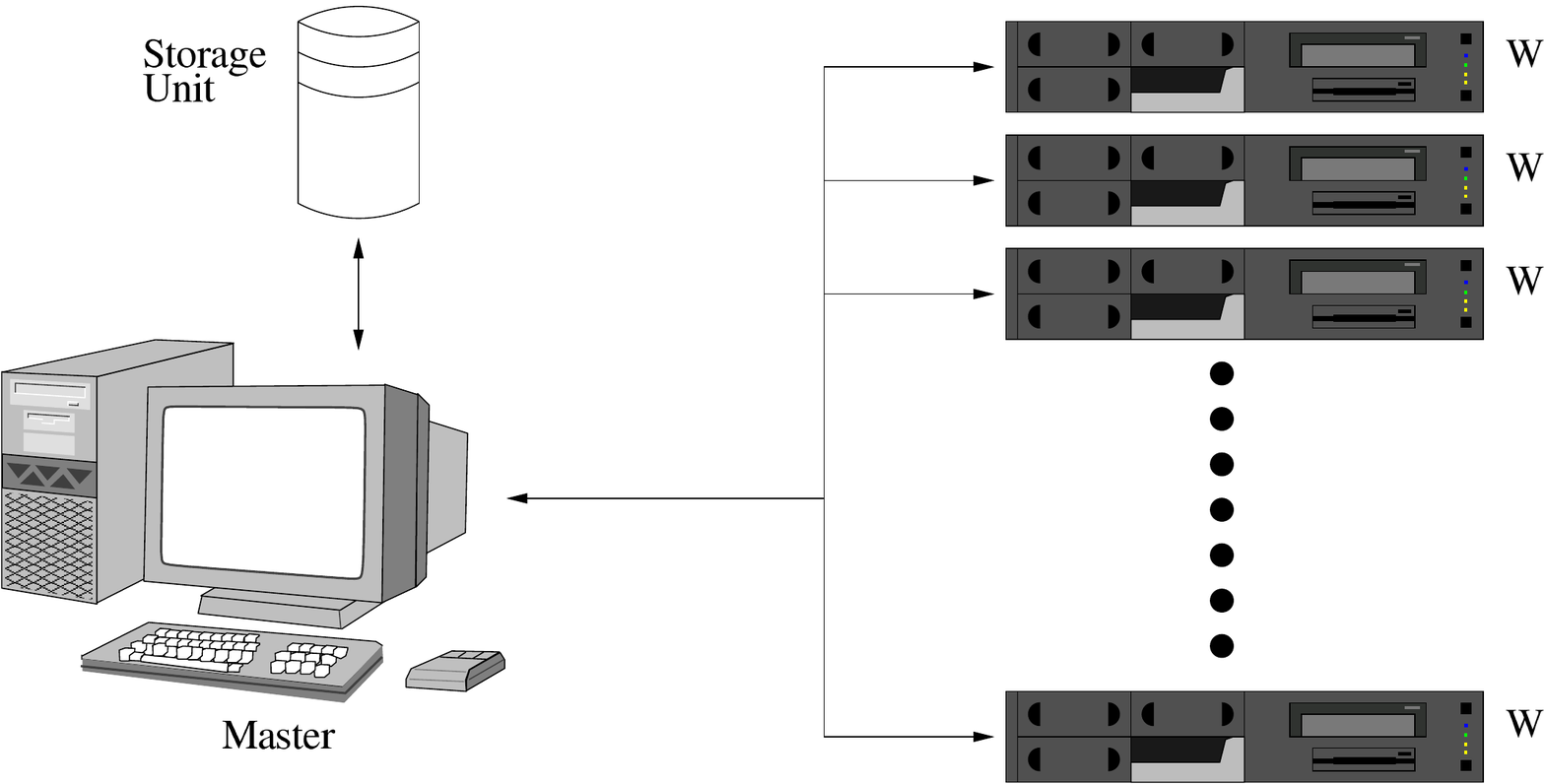}
\caption{Monte Carlo integration on a cluster: the \emph{Master}
supplies each \emph{Worker} with phase-space points and collects the
results. In this scenario the \emph{Master} is the only process that
needs access to I/O devices.}\label{fig:imp-fortran:MCMPI}
\end{figure}

Although Monte Carlo integration is an example where $N$ processes
can work in parallel with \emph{no} communication at all there are
two reasons why a message-passing implementation can be advantageous
over the completely independent approach:
\begin{itemize}
\item Only one of the processes needs access to the storage device.
Since the phase space points in our case were pregenerated and are
stored in a file the integrator needs disk access both for reading
and writing. Large numbers of parallel processes accessing the same
disk at the same time, however, can destabilise the system and decrease
the performance. Therefore the one-to-$n$ topology as shown in
Figure~\ref{fig:imp-fortran:MCMPI} has been used; the alternative of
using the local disks of each nodes involves additional file transfers
which can complicate the data handling unnecessarily.
\item It cannot be guaranteed that a single evaluation of the integrand
always takes the same time. At critical phase-space points the integrals
can converge much slower than for the bulk of the phase space. The
\emph{Master} process in Figure~\ref{fig:imp-fortran:MCMPI} handles
these situations dynamically and guarantees an optimal distribution
of the workload such the idle time of the processors is minimised.
\end{itemize}
However, these benefits do not come for free. More care and effort
by the programmer is required for the \ac{mpi} implementation than
for the sequential approach. A second drawback is the lower failure
tolerance if one of the processes drops out.

\subsubsection{Shared Memory Parallel Computing with \texttt{openMP}}
Shared memory parallel programming comes into play when the parallel
entities of a program need to update shared parts of the memory. It
is typically easier to implement but errors due to improper synchronisation
of the parallel threads are more likely. Most programming languages
require the use of explicit multi-threading to make use of shared
memory parallelism. In order to hide these technicalities from the
programmer \texttt{openMP} has been introduced as a standard interface
for parallelisation through multi-threading~\cite{openMP}. Rather
than forcing the programmer to set up threads, \texttt{openMP} allows
to place instructions to tell the compiler where parallelisation is
possible and how to deal with global data regions.

The use of \texttt{openMP} has not been built into the current
version of the code as the structure of the
\texttt{Golem90} library makes heavy use of global but private
variables. Therefore a parallelisation of the calculation of the
form factors using \texttt{openMP} appears to be very difficult.

Applying \texttt{openMP} to other less time-consuming sections
like the calculation of the spinor traces does not lead to significant
improvements; when applied to rather inexpensive code fragments
the overhead of creating threads even increases the run time and cannot be
compensated for by distributing the workload to multiple processors.

\subsection{The \acs{ecdf}~Cluster}
\label{ssec:imp-fortran:ecdf}
The University of Edinburgh runs the \acf{ecdf},
which has been used for the numerical integration of the cross section.
Part of it is a cluster consisting of
``128 worker nodes each with two dual core CPUs,
and 118 worker nodes each with two quad core CPUs,
giving a total of 1456 cores.
The worker nodes are interconnected by gigabit Ethernet networking.''%
\footnote{\url{http://www.is.ed.ac.uk/ecdf/eddie.shtml}, 6~August~2008}
Only part of the cluster has been used subject to availability, which
usually was between 100 and 200 nodes.

The job submission is controlled by the Sun Grid Engine and
message passing has been realised through the \texttt{openMPI}
implementation. The source file have been compiled with the
Intel \fortran{} compiler which optimises the code for the
underlying architecture\footnote{Intel Xeon 5160,
\unit{3}{\giga\hertz} (dual core nodes)
and Intel Xeon 5450, \unit{3}{\giga\hertz} (quad core nodes)}.

%% file: appendix-minor.tex
%\section{Software}
%\label{qpp:software}
%\input{appendix-software}
\phantomsection
\chapter*{Acronyms}
\label{app:acro}
\input{acro}

%% file: acro.tex
\begin{acronym}[\hspace{5em}]
\acro{bsm}[BSM]{Beyond Standard Model}
\acro{cas}[CAS]{Computer Algebra System}
\acro{cms}[CMS]{Compact Muon Solenoid}
\acro{dred}[DR]{Dimensional Reduction}
\acro{dreg}[DReg]{Dimensional Regularization}
\acro{ecdf}[ECDF]{Edinburgh Computing and Data Facility}
\acro{gln}[\gln]{General Linear Group}
\acro{hc}[h{.}c{.}]{hermitian conjugate}
\acro{ir}[IR]{infrared}
\acro{lep}[LEP]{Large Electron Positron Collider}
\acro{lhc}[LHC]{Large Hadron Collider}
\acro{lo}[LO]{Leading Order}
\acro{lsp}[LSP]{Lightest Supersymmetric Particle}
\acro{lr}[LR]{\person{Littlewood}-\person{Richardson}}
\acro{mpi}[MPI]{Message Passing Interface}
\acro{ms}[{\ensuremath{\mathrm{MS}}}]{Minimal Subtraction}
\acro{msbar}[\hbox{\ensuremath{\overline{\mathrm{MS}}}}]{%
   Modified Minimal Subtraction}
\acro{mssm}[MSSM]{Minimal Supersymmetric Standard Model}
\acro{ndr}[NDR]{Na{\"i}ve Dimensional Reduction}
\acro{nlo}[NLO]{Next to Leading Order}
\acro{nnlo}[NNLO]{Next to Next to Leading Order}
\acro{oop}[OOP]{Object Oriented Programming}
\acro{pdf}[PDF]{Parton Distribution Function}
\acro{qcd}[QCD]{Quantum Chromodynamics}
\acro{qed}[QED]{Quantum Electrodynamics}
\acro{shp}[SHP]{Spinor Helicity Projection}
\acro{sm}[SM]{Standard~Model}
\acro{sln}[\sln]{Special Linear Group}
\acro{son}[\son]{Special Orthogonal Group}
\acro{spn}[\spn]{Symplectic Group}
\acro{sun}[\sun]{Special Unitary Group}
\acro{sym}[\Sym]{Symmetric Group}
\acro{tho}['tHo]{\person{'t~Hooft}-\person{Veltman}}
\acro{uv}[UV]{ultraviolet}
\acro{wvdw}[WvdW]{\person{Weyl}-\person{van~der~Waerden}}
\end{acronym}

%% file: very-end.tex
\phantomsection
\begin{bibunit}[hepdoi]
\makeatletter
\def\thebibliography#1{\chapter*{List of Publications}
%\@mkboth%
%	{PUBLICATIONS}{PUBLICATIONS}}
   This work has contributed to the publications listed below:
   \list%
	{[\arabic{enumi}]}{\settowidth\labelwidth{[#1]}\leftmargin\labelwidth
\advance\leftmargin\labelsep
\usecounter{enumi}}
\def\newblock{\hskip .11em plus .33em minus .07em}
\sloppy\clubpenalty4000\widowpenalty4000
\sfcode`\.=1000\relax}
\makeatother

\nocite{Bern:2008ef}
\nocite{Binoth:2008bj}
\nocite{Binoth:2006mf}
\nocite{Binoth:2008inprep}
\nocite{Golem90:2008inprep}
\putbib[dissert]
\end{bibunit}

\clearpage
\phantomsection
\pagestyle{empty}
\chapter*{Acknowledgements}
I want to thank my supervisors Thomas Binoth ans Tony Kennedy
for their support and guidance during my postgraduate studies.
Parts of the results of this work were achieved through the
support of the \texttt{Golem} collaboration: Jean-Philippe Guillet
took the main responsibility for the \texttt{Golem90} library,
Alberto Guffanti and J{\"u}rgen Reuter calculated
the real emission contribution of the
$u\bar{u}\rightarrow b\bar{b}b\bar{b}$ amplitude and the
generated the \ac{lo} event files which served as an input
for my integrated results.

Part of my research was accomplished at at the LAPPTH in Annecy,
the Fermilab in Batavia and the GGI in Florence; I wish to thank
all three hosts for inviting me to their institutes,
which offered an inspiring environment and great hospitality.
I also want to thank the Scottish Universities Physics Alliance
who founded my research through a studentship.

This work has made use of the resources provided by the
\acf{ecdf}\footnote{\url{http://www.ecdf.ed.ac.uk/}}.
The \acs{ecdf} is partially supported by the
eDIKT initiative\footnote{\url{http://www.edikt.org}}.

The \person{Feynman} diagrams in this thesis have been drawn
using the \texttt{FeynMF} package~\cite{Ohl:1995kr}.

\clearpage
\phantomsection
\printindex
\clearpage
\remarks